\documentclass[twocolumn,prl,aps]{revtex4-2}
\usepackage{amsmath}
\usepackage{tensor}
\usepackage{physics}
\usepackage[pdftex]{graphicx}
\usepackage{subfigure}
\usepackage{float}
\usepackage{amssymb} 
\usepackage{xcolor}
\usepackage{hyperref}

\usepackage{etoolbox}
\apptocmd{\thebibliography}{\raggedright}{}{}

\begin{document}
\author{
Enda Xiao$^1$,
Hao Ma$^2$,
Matthew S. Bryan$^2$,
Lyuwen Fu$^3$,
J. Matthew Mann$^4$,
Barry Winn$^5$,
Douglas L. Abernathy$^5$,
Raphaël P. Hermann$^2$,
Amey R. Khanolkar$^6$,
Cody A. Dennett$^6$,
David H. Hurley$^6$,
Michael E. Manley$^2$,
Chris A. Marianetti$^3$
}

\title{Validating First-Principles Phonon Lifetimes  via Inelastic Neutron Scattering}

\address{$^1$Department of Chemistry, Columbia University, New York, New York 10027, USA}
\address{$^2$Materials Science and Technology Division, Oak Ridge National Laboratory, Oak Ridge, Tennessee 37831, USA}
\address{$^3$Department of Applied Physics and Applied Mathematics, Columbia University, New York, New York 10027, USA}
\address{$^4$Air Force Research Laboratory, Sensors Directorate, Wright-Patterson AFB, OH 45433, USA}
\address{$^5$Neutron Scattering Division, Oak Ridge National Laboratory, Oak Ridge, Tennessee 37831, USA}
\address{$^6$Materials Science and Engineering Department, Idaho National Laboratory, Idaho Falls, ID 83415, USA}
\date{\today}

\begin{abstract}
Phonon lifetimes are a key component of quasiparticle theories of transport,
yet first-principles lifetimes are rarely directly compared to inelastic neutron
scattering (INS) results. Existing comparisons show discrepancies even at
temperatures where perturbation theory is expected to be reliable. In this work, we
demonstrate that the reciprocal space voxel ($q$-voxel), which is the finite
region in reciprocal space required in INS data analysis, must be explicitly
accounted for within theory in order to draw a meaningful comparison.
We demonstrate accurate predictions of peak widths of the scattering function
when accounting for the $q$-voxel in CaF$_2$ and ThO$_2$. Passing this test
implies high fidelity of the phonon interactions and the approximations used to
compute the Green's function, serving as critical benchmark of theory, and
indicating that other material properties should be accurately predicted; which
we demonstrate for thermal conductivity.

\end{abstract}

Notice of Copyright This manuscript has been authored by UT-Battelle, LLC under Contract No. DE-AC05-00OR22725 with the U.S. Department of Energy. The United States Government retains and the publisher, by accepting the article for publication, acknowledges that the United States Government retains a non-exclusive, paid-up, irrevocable, world-wide license to publish or reproduce the published form of this manuscript, or allow others to do so, for United States Government purposes. The Department of Energy will provide public access to these results of federally sponsored research in accordance with the DOE Public Access Plan (http://energy.gov/downloads/doe-public-access-plan).

\clearpage 
\maketitle

When computing anharmonic vibrational properties from first-principles, various approximations 
are employed, making it challenging to assess the integrity of any single observable
as compared to experiment (e.g. thermal conductivity). 
Comparing a 
$q$-space resolved observable (e.g. the scattering function) inherently 
provides
a large number of comparisons,
offering a very stringent test. 
While verification of the harmonic
vibrational first-principles Hamiltonian is a standard practice, the same
cannot be said for the anharmonic vibrational Hamiltonian and subsequent approximations which 
are used to compute observalbes.  
Anharmonic terms result in finite phonon lifetimes yielding finite
widths of the peaks in the scattering function.  A direct comparison of peak
width between theory and INS is rare, and existing studies reveal anomalous discrepancies. A recent study on
Si notes the large discrepancy between theory and experiment, leading the
authors to only compare the relative change as a function of
temperature\cite{kim2020temperature}. A study on Al at high temperatures finds that perturbation theory
does not reliably predict the experimental peak width, 
and their first-principles
molecular dynamics simulations often differ from experiment by a factor of two\cite{glensk2019phonon}.

Here we show that a proper comparison between the theoretical and
experimental scattering function requires  an explicit accounting
for the finite region probed in reciprocal space, which is referred to as the $q$-voxel.
Due to the flux-limited nature of INS, there is a minimum
region of $q$-space which can be sampled while maintaining sufficient statistics, 
and therefore there is a minimum $q$-voxel size below which INS cannot probe.
Depending on the measurement type,  the $q$-voxel shape is either set by the instrument configuration (for triple-axis) or defined post measurement in the analysis of large volumes of data (for time-of-flight). This work will focus on the latter since the large data volumes allow for a more comprehensive assessment across many zones and with varying $q$-voxels. While there is no formal standard for choosing a $q$-voxel, 
it is typical to choose the $q$-voxel size and shape
based on the shape of the phonon dispersion surface; selecting a smaller dimension along
directions with steeper dispersion and larger dimensions along directions with flatter dispersion.

In the present work, we demonstrate the critical role of the $q$-voxel using two fluorite structured materials, ThO$_2$ and
CaF$_2$, showing excellent agreement between peakwidths within a $q$-voxel
obtained from INS and 
computed from perturbation theory based on the first-principles phonons and cubic phonon
interactions. Successful agreement validates both the
anhmarmonic Hamiltonian and the level of theory being used to evaluate the scattering function.  Given our
successful peakwidth predictions, it is expected that thermal conductivity predicted using the same
anhmarmonic Hamiltonian and the Boltzmann transport equation should faithfully
describe experimental measurements in some temperature regime, 
which is demonstrated for 
CaF$_2$ and ThO$_2$.

Time-of-flight INS measurements were performed using the
ARCS\cite{abernathy2012design} and HYSPEC\cite{shapiro2006hyspec} instruments
at the Spallation Neutron Source. Thermal conductivity measurements were made on single crystal ThO$_2$ using spatial domain thermoreflectance (SDTR) via the methods described in~\cite{Dennett2021}.
First-principles
calculations were performed using density functional theory (DFT) with the strongly constrained
and appropriately normed (SCAN) functional\cite{sun_strongly_2015}, and phonons and
cubic phonon interactions were computed using the lone and bundled irreducible
derivative approaches\cite{fu_group_2019,supmat}. Detailed information about experimental and computational methods are included in supplemental material \cite{supmat}. 

In order to compute the phonon linewidths and the one-phonon scattering function $S_1(\mathbf{Q},\omega)$,
the single-particle phonon Green's function can be approximately 
evaluated using the phonons and cubic phonon interactions within leading order perturbation
theory\cite{Kokkedee1962374,Maradudin19622589} (see Eq. 2 in \cite{supmat}). 
The scattering function $S_1(\mathbf{Q},\omega)$ is evaluated precisely at
$\mathbf{Q}$, but in INS experiments
a finite $q$-voxel must be
chosen to provide sufficient counting statistics.
This $q$-voxel can
be accounted for theoretically by integrating $S_1(\mathbf{Q},\omega)$ 
over the $q$-voxel, resulting in 
\begin{equation} S_1^{\text{vox}}(\mathbf{Q},\omega) =
\frac{1}{\Omega_{\text{vox}}}
\int\limits_{\text{vox}} d^3Q' \hspace{2mm} S_
1(\mathbf{Q}',\omega), 
\end{equation} 
where $\Omega_{\text{vox}}$ is the reciprocal space volume of the $q$-voxel.
For clarity, we refer to $S_1(\mathbf{Q},\omega)$ as the $q$-point scattering function and
$S_1^{\text{vox}}(\mathbf{Q},\omega)$ as the $q$-voxel scattering function.

\begin{figure}[htbp]
\centering
\includegraphics[width=0.9\linewidth]{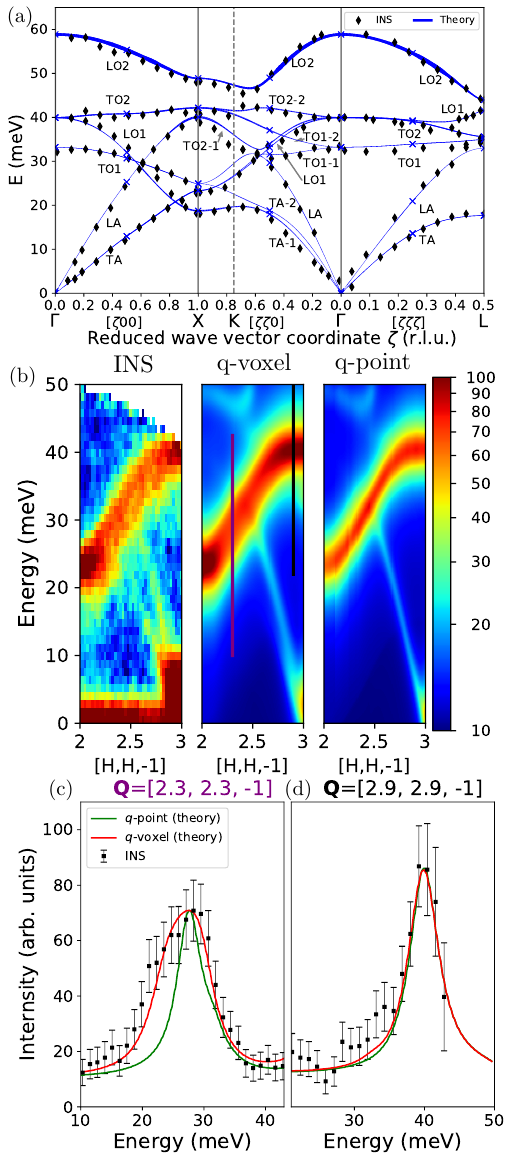}
\caption{
(a) Phonon dispersion of CaF$_2$. Black points are INS measurements at $T=300$K; blue crosses are computed using DFT (SCAN), blue lines are a Fourier interpolation, and the line width is proportional to the computed phonon linewidth.
(b) Color contour plots of the scattering function as a function of energy and
$\mathbf Q$ obtained using INS, theoretical $q$-voxel, and theoretical $q$-point
results.  The $q$-voxel dimensions are 0.025 r.l.u.
along the [H,H,0] direction (i.e. the dispersion direction) and relaxed to 0.2 r.l.u  along the
orthogonal directions [0,0,L] 
and [H,-H,0].
(c, d) Scattering function as a function of energy at selected $Q$-points,
corresponding to purple and black lines in the color contour plots. 
The red and
green curves are theoretical $q$-voxel and $q$-point scattering functions,
respectively; black points are INS measurements.}
\label{fig:cuts}
\end{figure}

We now proceed to evaluate the $q$-point and $q$-voxel scattering function in
CaF$_2$ and ThO$_2$, and compare them to INS measurements.  Given that CaF$_2$
and ThO$_2$ are  band insulators, standard implementations of DFT are expected
to perform well in terms of describing the ground state properties. 
As anticipated, the computed phonon spectrum is in good agreement with the scattering function 
peak positions obtained from INS
at ambient temperature for both CaF$_2$ (see Fig. \ref{fig:cuts},
panel $a$) and ThO$_2$ (see Fig. S2 in \cite{supmat}). 
While the peak locations agree well, it is interesting to directly compare the respective scattering
functions via contour plots along  a path through $q$-space (see Fig.\ref{fig:cuts}, panel $b$). 
For a direct comparison, the INS instrumental energy resolution is accounted for in the theoretical result\cite{abernathy2012design}.
We find that the theoretical and INS $q$-voxel scattering functions are in reasonable agreement, while 
they have nontrivial differences with the $q$-point scattering function in certain regions. The theoretical $q$-voxel
scattering function even recovers subtle features of the INS, such as the presence of the TA1 band (see panel $a$
for naming convention),
which is forbidden in the $q$-point scattering function for the path shown. 
In order to illustrate potential differences between the $q$-point and $q$-voxel
scattering function, both large and small, we present the scattering function at two $Q$ points as a
function of energy around the LO1 band, which scatters strongly (see Fig.\ref{fig:cuts}, panels $c$ and $d$).
As shown, the $Q$ which is closer to the zone center (panel $d$) only shows negligible differences between
the $q$-point and $q$-voxel scattering functions, while the other $Q$ (panel $c$) shows good agreement between the
theoretical and experimental $q$-voxel scattering function, but a substantial difference with the 
$q$-point scattering function; and this difference may be attributed to the large slope in the latter case.
The preceding examples already illustrate that one can only meaningfully compare theory and experiment
if $q$-voxel quantities are being employed. 

\begin{figure}[htbp]
\centering
\subfigure{
\includegraphics[width=0.46\textwidth]{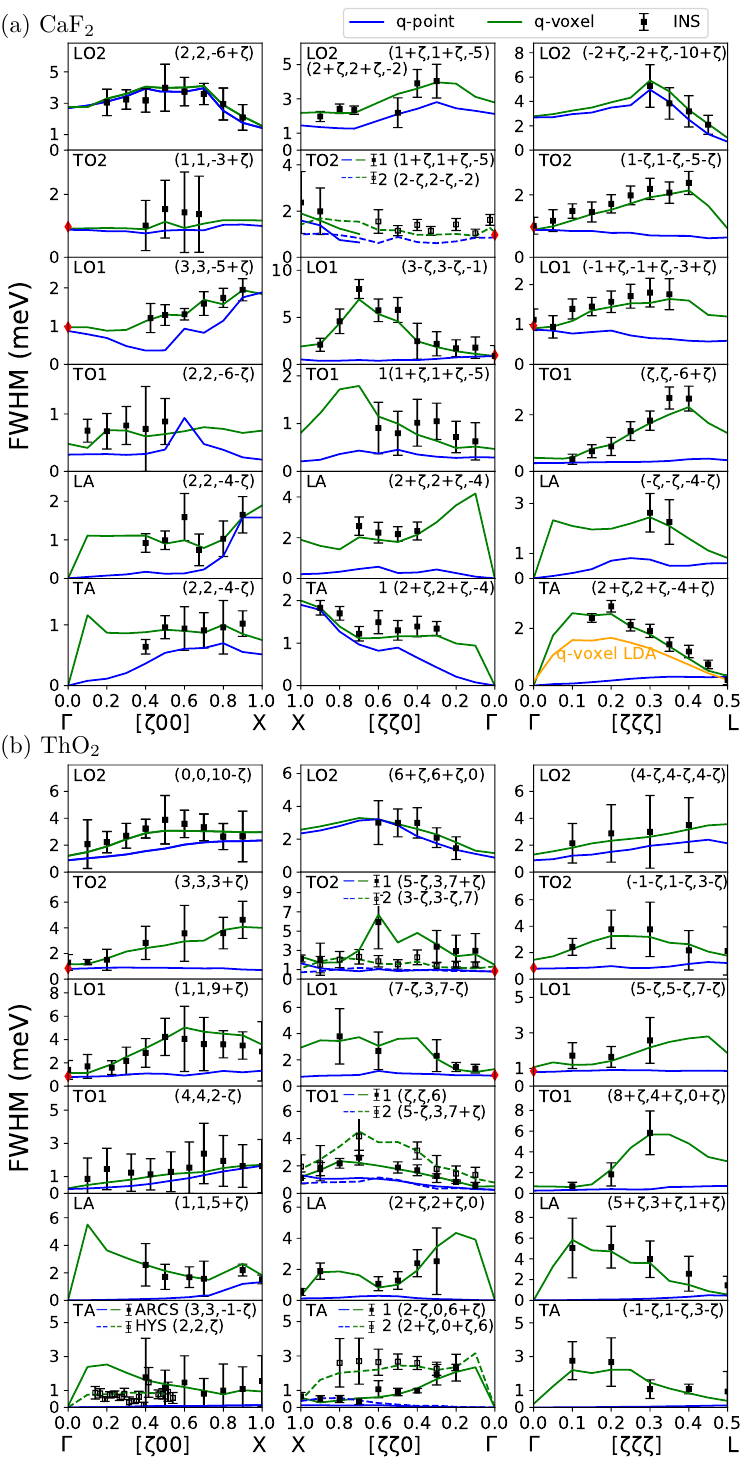}
}
\caption{ FWHM's of the scattering function peaks as a function of $q$ in various zones for CaF$_2$ (panel $a$) and ThO$_2$ (panel $b$) at
$T=300$ K. The DFT (SCAN) $q$-point and $q$-voxel results are shown as blue and green
lines, respectively; INS results are shown as black points.
Certain panels contain multiple modes (see legend).  
The $q$-voxel dimensions are reported
in supplemental material\cite{supmat}. Previous Raman measurements are denoted by a red diamond\cite{elliott_experimental_1978,rao_raman_2014}. 
}
\label{fig:figure_FWHM}
\end{figure}

Having illustrated that accounting for the $q$-voxel scattering function can be critical to 
describing experiments, we now proceed to comprehensively quantify the differences across the Brillouin zone
for both CaF$_2$ and ThO$_2$.
In Fig.\ref{fig:figure_FWHM}, we compare the full width at half maximum (FWHM)
of each peak obtained from the INS $q$-voxel scattering function, the theoretical $q$-voxel scattering
function, and the theoretical $q$-point scattering function (see \cite{supmat} for $\mathbf{Q}$ and voxel sizes). 
Following standard INS conventions, the energy resolution is removed from the INS scattering function peak width\cite{abernathy2012design, bruce_analytical_2000}.
The INS and theoretical $q$-voxel FWHM results are in favorable agreement
across all modes and $q$-paths, while there is substantial difference with the
FWHM obtained from the theoretical $q$-point scattering function in most cases. 
The acoustic
modes, which are highly relevant for thermal conductivity, 
 show a strong difference between $q$-voxel and $q$-point
FWHM values. 
We also compare our theoretical results at the $\Gamma$ point to the Raman measurements 
of the 
T$_{2g}$ widths in CaF$_2$ \cite{elliott_experimental_1978} and ThO$_2$ \cite{rao_raman_2014},
indicated with a red diamond, demonstrating good agreement with the $q$-point result; which is to be expected. 
In summary, we have shown that it is critical to employ the $q$-voxel in order to validate
a first-principles theory using INS.

\begin{figure}[htbp]
\centering
\includegraphics[width=0.45\textwidth]{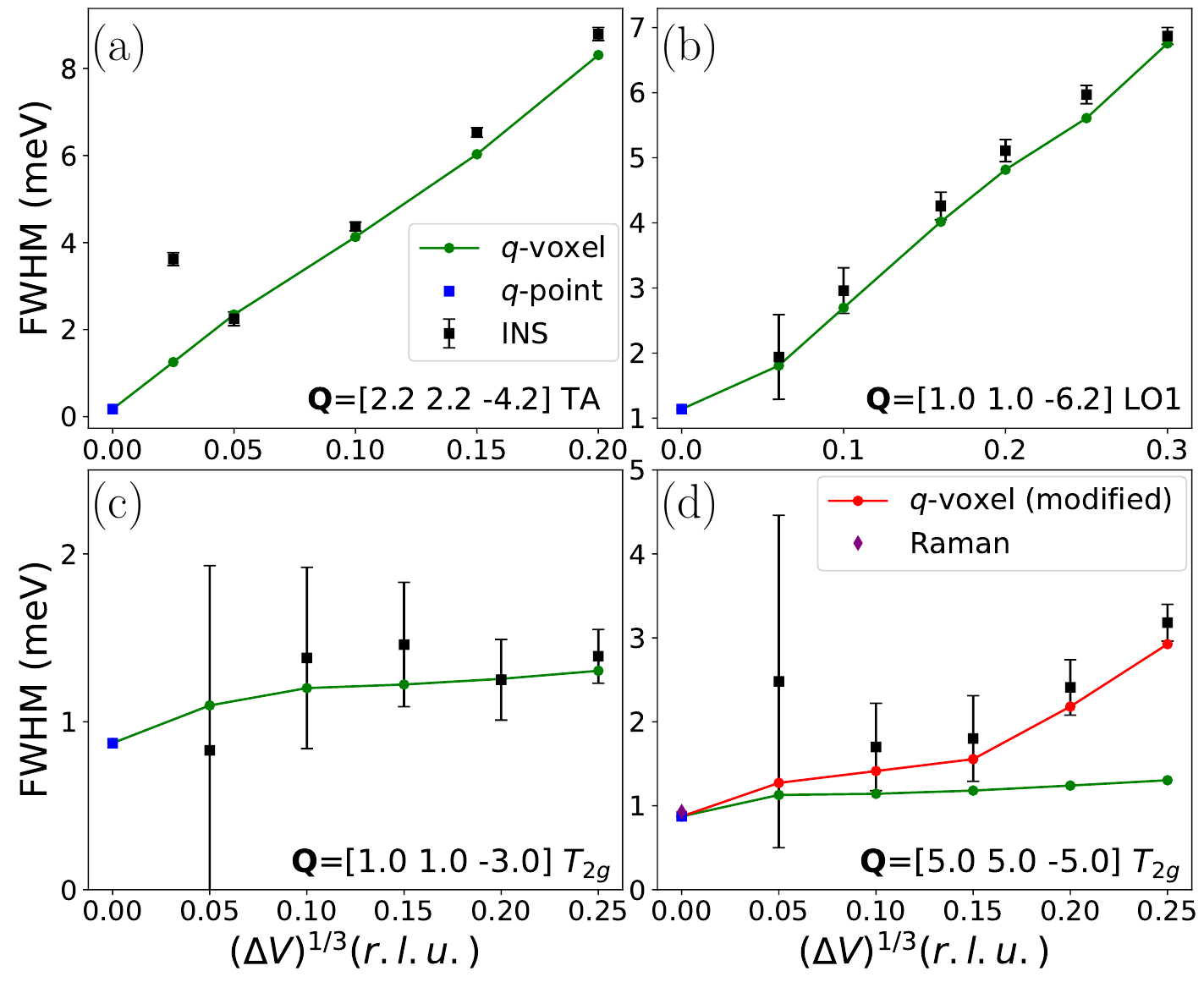}
\caption{ 
 The FWHM as a function of the cubic $q$-voxel dimension in 
 CaF$_2$ at $T=300$ K for three modes: the TA mode at $\mathbf{Q}=$[2.2,  2.2,
 -4.2] (panel $a$),  the LO1 mode at mode at $\mathbf{Q}=$[1.0, 1.0, 6.2] (panel $b$), and the T$_{2g}$ mode at
 $\mathbf{Q}=$[1.0, 1.0, -3.0] (panel $c$) and $\mathbf{Q}=$[5.0, 5.0, -5.0] (panel $d$).  
See text for explanation of $q$-voxel (modified) result (red line).
The Raman measurement of the 
T$_{2g}$ width in CaF$_2$ by Elliott \textit{et al.}\cite{elliott_experimental_1978} is shown as a purple diamond.
 }
\label{fig:voxel_size3}
\end{figure}

In the preceding analysis, we demonstrated that theory and experiment are in
good agreement when using the same $q$-voxel, and that the results can be
substantially different from the $q$-point results. Now we explore the effect
of voxel size in the case of a cubic voxel, and examine the possibility of
extrapolating the $q$-voxel results to zero voxel size in order to recover the
$q$-point results, which would allow INS experiments to independently obtain
$q$-point peakwidths.  The $q$-voxel size used in INS data analysis is a
compromise between having good counting statistics (larger volume) and
minimizing contamination from neighboring regions in reciprocal space (smaller
volume).  The effects of the $q$-voxel size on the experimental FWHM of CaF$_2$ is shown in
Fig.\ref{fig:voxel_size3} for three different modes: the TA mode at
$\mathbf{Q}=$[2.2,  2.2,  -4.2],  the LO1 mode at mode at $\mathbf{Q}=$[1.0, 1.0,
6.2], and the T$_{2g}$ mode (i.e. LO1+TO2) at $\mathbf{Q}=$[1.0, 1.0, -3.0] and $\mathbf{Q}=$[5.0,
5.0, -5.0].  
With the exception of the T$_{2g}$ mode at $\mathbf{Q}=$[5.0, 5.0, -5.0] (i.e. Fig.\ref{fig:voxel_size3}, panel $d$),
the theoretical results are in good agreement with experiment for a sufficiently large
voxel size; 
and the exception can be attributed to the overlap of the T$_{2g}$ peak with
the TO1 peak, which corrupts the fitting process for the experimental data
which must account for the energy resolution of the instrument.  To make a
direct comparison, we convolve the theoretical results with the energy
resolution, and then perform the identical fitting process used for the
experimental results where the energy resolution is removed (see red line in
Fig.\ref{fig:voxel_size3}, panel $d$).  This modified theoretical result now
agrees well with experiment for large $q$-voxel sizes, and recovers the usual
theoretical $q$-voxel results at small $q$-voxel sizes. Our analysis resolves a
previous anomoly in the literature, explaining why the INS peak width of the
T$_{2g}$ mode was found to be more than twice that of Raman
measurements\cite{schmalzl2003lattice}.  Nonetheless, the simplest solution to
this problem is to choose a more favorable zone, as shown in
Fig.\ref{fig:voxel_size3}, panel $c$. In all the preceding examples, we see
that sufficiently small voxel sizes lead to poor results for the experimental
case given the poor counting statistics, as expected.  Furthermore, it appears
that there is sufficient uncertainty within the experimental measurements which
would preclude the possibility of extrapolating to zero voxel size to obtain
the peakwidth purely from experiment.  The effects on non-cubic voxels are
explored in Supplementary Material\cite{supmat} (see Section VII).

\begin{figure}[h]
\centering
\includegraphics[width=0.48\textwidth]{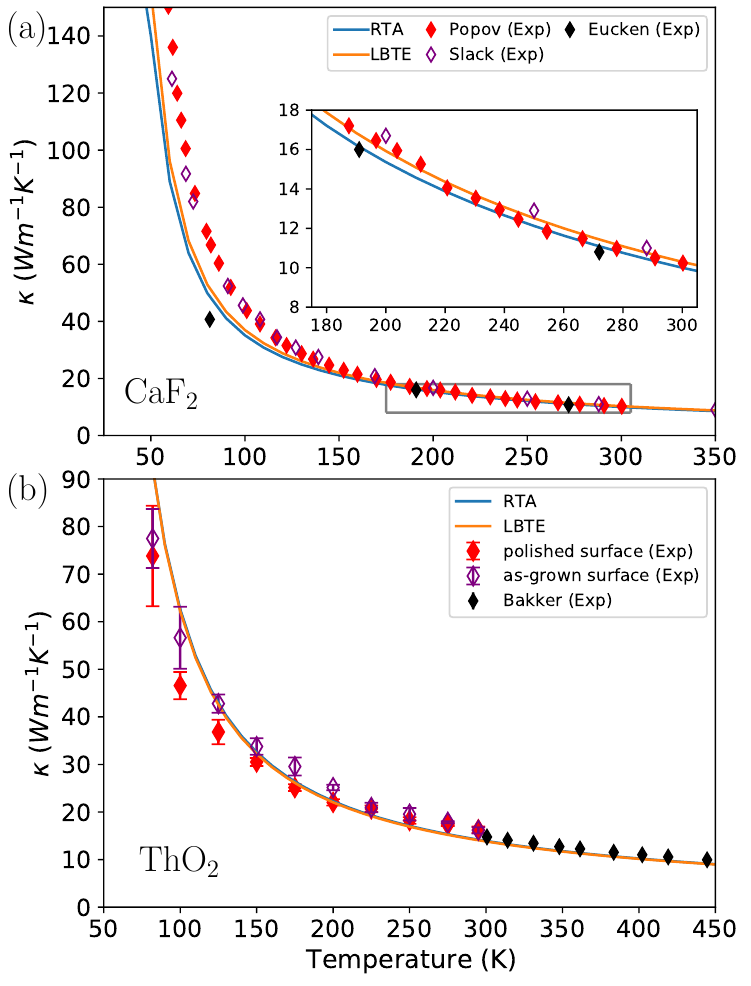} 
\caption{ 
Thermal conductivity $\kappa$ as a function of temperature in  CaF$_2$ (panel $a$) and ThO$_2$ (panel $b$). 
Our DFT (SCAN) results are denoted with orange and blue lines for LBTE and RTA solutions, respectively.
Experimental measurements on CaF$_2$ by Slack\cite{slack1961thermal}, Eucken\cite{eucken_uber_1911}, and Popov \textit{et al.}\cite{popov_thermal_2014,digitfn} are shown, in addition to 
measurements in  ThO$_2$ by Bakker \textit{et al.}\cite{bakker_critical_nodate} and our own (with one sample having a polished surface \cite{supmat}).
}
\label{fig:thermal_condu}
\end{figure}

We have demonstrated that the $q$-voxel peakwidths from theory and experiment are in good agreement at room temperature.
This agreement serves as a verification of the quality of our cubic phonon interactions, in addition to the level of perturbation
theory used to construct the scattering function. The former is an indirect assessment of the quality of the approximation to the 
exchange-correlation
energy used within DFT, and it should be emphasized that the SCAN functional is critical to such good agreement; whereas the local density approximation \cite{Perdew19815048} (LDA)
produces substantial deviations 
(see Fig. \ref{fig:voxel_size3}, panel $a$, orange line, and  
\cite{supmat}, Section VIII).
We can now
predict other quantities, such as thermal conductivity, and anticipate robust results. 
The thermal conductivity can be computed by solving the linearized Boltzmann
transport equation (LBTE) \cite{broido_lattice_2005,broido_thermal_2012,chaput_direct_2013}.  
Within the 
relaxation time approximation (RTA), the LBTE solution is obtained as an explicit function of the
phonon spectrum and the phonon linewidths.
The RTA is sometimes an excellent approximation to the LBTE solution, and previous work has demonstrated that this is
the case in CaF$_2$ \cite{qi_lattice_2016}; and we reach the same conclusion in both CaF$_2$ and ThO$_2$.
Therefore, we expect our predicted thermal conductivity to be very robust, and we evaluate both 
CaF$_2$ and ThO$_2$. 

In the case of CaF$_2$ (Fig.\ref{fig:thermal_condu}, panel $a$), our
predictions are in good agreement with all available experimental data
\cite{eucken_uber_1911,slack1961thermal,popov_thermal_2014} in the temperature
range of 200 K to 300 K.  
At low
temperatures, the results of both Slack\cite{slack1961thermal} and Popov 
\textit{et al}. \cite{popov_thermal_2014} are somewhat higher than our predictions, 
though the single measurement at $T=77K$ by
Eucken\cite{eucken_uber_1911} is below our result.  In the case of ThO$_2$
(Fig.\ref{fig:thermal_condu}, panel $b$), our predicted thermal conductivity is
in good agreement with the data of Bakker \textit{et al}. \cite{bakker_critical_nodate},
which extends from 300 K to 450 K, and reasonable comparison is found with our
measurements from 82 K to 295 K.  The higher conductivity predicted at low
temperatures as compared with our experiments is likely due to small, native
impurity concentrations resulting from the hydrothermal growth process.  Our
laser based measurements show nontrivial variability at low temperatures based
on the surface condition of the sample, and there is a small difference with
the data of Bakker \textit{et al}. near room temperature. It is difficult to assess
which experimental results are more reliable. 
Comparison with previous first-principles calculations is provided in Supplementary
Material for both CaF$_2$ and ThO$_2$ \cite{supmat}.

In summary, we have shown that the $q$-voxel of the INS measured scattering
function must be accounted for when comparing to theoretical predictions, elucidating
why INS peak widths  had not been well matched to predictions until now.  Given that the $q$-voxel can be straightforwardly implemented within
theory, INS is thus elevated to a refined judge of all ingredients of a
quasiparticle theory of phonons, whether ab-initio or empirical. 
The $q$-voxel should be carefully
considered in the design of future INS instruments, with the possible goal of allowing
INS to independently extrapolate to the $q$-point limit. 

\section*{acknowledgement}
INS measurements by M.S.B., H. M., and M.E.M., first-principles calculations by E.X., L.F., and C.A.M., 
crystal growth by J.M.M., and thermal conductivity measurements by C.A.D., A.K., and D.H. were supported by the Center for Thermal Energy Transport under Irradiation, an Energy Frontier Research Center funded by the U.S. Department of Energy (DOE), Office of Science, United States, Office of Basic Energy Sciences. 
Neutron scattering data acquisition and provision of CaF2 by RPH was supported by the DOE Office of Science, Basic Energy Science, Materials Science and Engineering Division. 
Portions of this research used resources at the Spallation Neutron Source, a U.S. DOE Office of Science User Facility operated by the Oak Ridge National Laboratory.  This research used resources of the National Energy Research Scientific Computing Center, a DOE Office of Science User Facility supported by the Office of Science of the U.S. Department of Energy under Contract No. DE-AC02-05CH11231. The formulation and encoding of linewidths via irreducible derivatives by L.F. and C.A.M. was supported by the grant DE-SC0016507 funded by the U.S. Department of Energy, Office of Science.

*E.X., H.M. and M.S.B.  contributed equally to this work.

\clearpage

\bibliography{main}
\end{document}


\renewcommand*{\thefigure}{S\arabic{figure}}
\renewcommand*{\thetable}{S\arabic{table}}

\author{
Enda Xiao$^1$,
Hao Ma$^2$,
Matthew S. Bryan$^2$,
Lyuwen Fu$^3$,
J. Matthew Mann$^4$,
Barry Winn$^5$,
Douglas L. Abernathy$^5$,
Raphaël P. Hermann$^2$,
Amey R. Khanolkar$^6$,
Cody A. Dennett$^6$,
David H. Hurley$^6$,
Michael E. Manley$^2$,
Chris A. Marianetti$^3$
}

\title{Validating First Principles Phonon Lifetimes  via Inelastic Neutron Scattering [Supplemental Material]} 

\address{$^1$Department of Chemistry, Columbia University, New York, New York 10027, USA}
\address{$^2$Materials Science and Technology Division, Oak Ridge National Laboratory, Oak Ridge, Tennessee 37831, USA}
\address{$^3$Department of Applied Physics and Applied Mathematics, Columbia University, New York, New York 10027, USA}
\address{$^4$Air Force Research Laboratory, Sensors Directorate, Wright-Patterson AFB, OH 45433, USA}
\address{$^5$Neutron Scattering Division, Oak Ridge National Laboratory, Oak Ridge, Tennessee 37831, USA}
\address{$^6$Materials Science and Engineering Department, Idaho National Laboratory, Idaho Falls, ID 83415, USA}

\date{\today}

\maketitle

\section{INS experimetal details}

The ThO$_2$ sample was grown by hydrothermal growth\cite{mann2010hydrothermal}
and produced a 1.48 g crystal, which was measured at
ARCS\cite{abernathy2012design} (E$_i$ = 50, 120 meV) and
HYSPEC\cite{shapiro2006hyspec} (E$_i$ = 17 meV). Further details of the ThO$_2$ crystal and ARCS
measurements details have been reported previously\cite{bryan2020nonlinear}.
The CaF$_2$ crystal was purchased from
United Crystals Inc, and it was grown using the Czochralski technique; resulting in
a 52 g single crystal which we then measured at ARCS
(E$_i$ = 60, 120 meV). Both instruments measure the dynamic structure factor
$S(\mathbf Q, \omega)$. The ARCS and HYSPEC energy resolution
functions\cite{lin2019energy} were used in fitting the phonon peaks, and
reported widths are the intrinsic full-width half-maximum (FWHM) values that
have been corrected for the instrument contribution. All measurements reported
here are taken at $T=300$ K. The ARCS instrument, in particular, measures a large
volume in $\mathbf Q$ and $E$ which contains many Brillouin zones, and the data
analysis allows for an adjustable $q$-voxel size in all crystallographic
directions on both ARCS and HYSPEC.

\section{T\lowercase{h}O$_2$ thermal conductivity measurement}
The thermal conductivity of two 1~mm-scale ThO$_2$ single crystals from the same hydrothermal synthesis~\cite{mann2010hydrothermal} were measured using spatial domain thermoreflectance (SDTR)~\cite{Dennett2021}. One crystal was measured on an as-grown \{001\} crystal face identified by the facet morphology, receiving only chemical cleaning once removed from the growth chamber. A second with a \{001\} surface facet was mechanically polished using an Ultrapol Advanced polisher with a 0.1~{\textmu}m diamond pad from UltraTec. Both crystals were mounted to copper blocks using first high-conductivity silver paste and then epoxy resin to ensure good thermal and mechanical contact for cryogenic thermal conductivity measurements. 

SDTR measurements were carried out in a liquid-nitrogen cooled optical cryostat (Cryo Industries model XEM) over a temperature range of 80-295~K using a sample management protocol described in detail by Dennett and coworkers~\cite{Dennett2021}. The estimated temperature uncertainty for samples of these geometries in this apparatus is less than 3~K across the temperature range under consideration, including local laser heating effects. In SDTR, a transient local temperature variation is induced by using an intensity-modulated CW laser with a wavelength of 660~nm, and detected by using a constant-intensity CW laser with a wavelength of 532~nm through the thermoreflectance effect. A 50$\times$ objective lens is used to focus both heating and probing lasers. The convolved radius of the laser spots at the sample surface is approximately 2~{\textmu}m, with an optical power of $\sim$1~mW and $\sim$0.3~mW for the heating and probe lasers, respectively. Samples are coated with a 60~nm gold film as an optical and thermal transducer~\cite{Hurley2015}. Scans of the thermal phase lag as a function of a 20~{\textmu}m scan distance are collected at three modulation frequencies in the range 50--100~kHz at a minimum of five spatially-varying surface locations at each temperature. Thermal diffusivity is optimized from the multi-frequency, multi-layer response using a three-dimensional thermal transport model -- and independent measurements of all thermal parameters except the bulk crystal diffusivity -- computed using the method of thermal quadrupoles~\cite{Hurley2015,Maillet2000}. The measured thermal diffusivity is converted to thermal conductivity using independently-measured values of the low-temperature heat capacity of hydrothermally-grown ThO$_2$ as reported in~\cite{Dennett2020}.

\section{Computational details}

DFT calculations with the strongly constrained and appropriately normed
functional (SCAN) \cite{sun_strongly_2015} were performed using the projector
augmented wave (PAW) method \cite{blochl_improved_1994,Kresse19991758}, as implemented in the Vienna
ab initio simulation package (VASP)
\cite{Kresse1993558,Kresse199414251,Kresse199615,Kresse199611169}. The local density approximation (LDA)\cite{Perdew19815048} was also employed for select comparison with the SCAN functional for CaF$_2$.  A plane
wave basis with an energy cutoff of 600 eV was employed, along with a $k$-point density
consistent with a centered
k-point mesh of 20$\times$20$\times$20 in the primitive unit cell. All k-point integrations were done using
tetrahedron method with Blöchl corrections. The DFT energies were converged to
within $10^{-6}$ eV, while ionic relaxations were converged to within $10^{-5}$
eV.  

The structure was relaxed yielding a lattice parameter of 5.592 {\AA} for
ThO$_2$ and 5.448 {\AA} for CaF$_2$.  The measured coefficient of linear thermal expansion (CLTE) for
ThO$_2$ is $7.8 \times 10^{-6} K^{-1} $ at $T=300$K \cite{momin_high_1991} and
the CLTE for CaF$_2$ is \cite{schumann_thermal_1984} is:
\begin{align*} \alpha(T)=&1.885 \cdot 10^{-5}+1.67 \cdot
10^{-8}\left(T-T_{0}\right) + 5.5 \cdot 10^{-12}\left(T-T_{0}\right)^{2}
\end{align*} where $T_0$ = 273.15K, which is roughly 2.5 times larger than that of ThO$_2$ at room temperature. 
Given the relatively small thermal expansion of ThO$_2$,
we assume that the quadratic and cubic terms do not
change significantly in the temperature range of 0K to 300K.  Thus, for ThO$_2$ the quadratic and cubic force constants (FCs)
computed at the relaxed structure are used in all computations.  For CaF$_2$, quadratic
and  cubic FCs are computed at the fully relaxed volume, denoted $V_{0K}$, and the volume at 300K as dictated by the
experimental CLTE, denoted $V_{300K}$. All results generated at $T=300$K use the 
quadratic and cubic force constants from $V_{300K}$. For computations of thermal conductivity
as a function of temperature, all results in Figure 4a in the main manuscript use a linear interpolation of the quadratic and cubic force constants from $V_{0K}$ and  $V_{300K}$. To understand how the volume dependence of the
the quadratic and cubic force constants affect the thermal conductivity, 
we explore all permutations of quadratic and cubic force constants from $V_{0K}$ and $V_{300K}$ in Fig. \ref{fig:expansion}; 
demonstrating that the changes in the quadratic and cubic force constants largely cancel within the 
computation of thermal conductivity (see Sec. \ref{sec:influence_therm_exp} for details).

\begin{figure*}[h]
\centering
\includegraphics[width=0.8\textwidth]{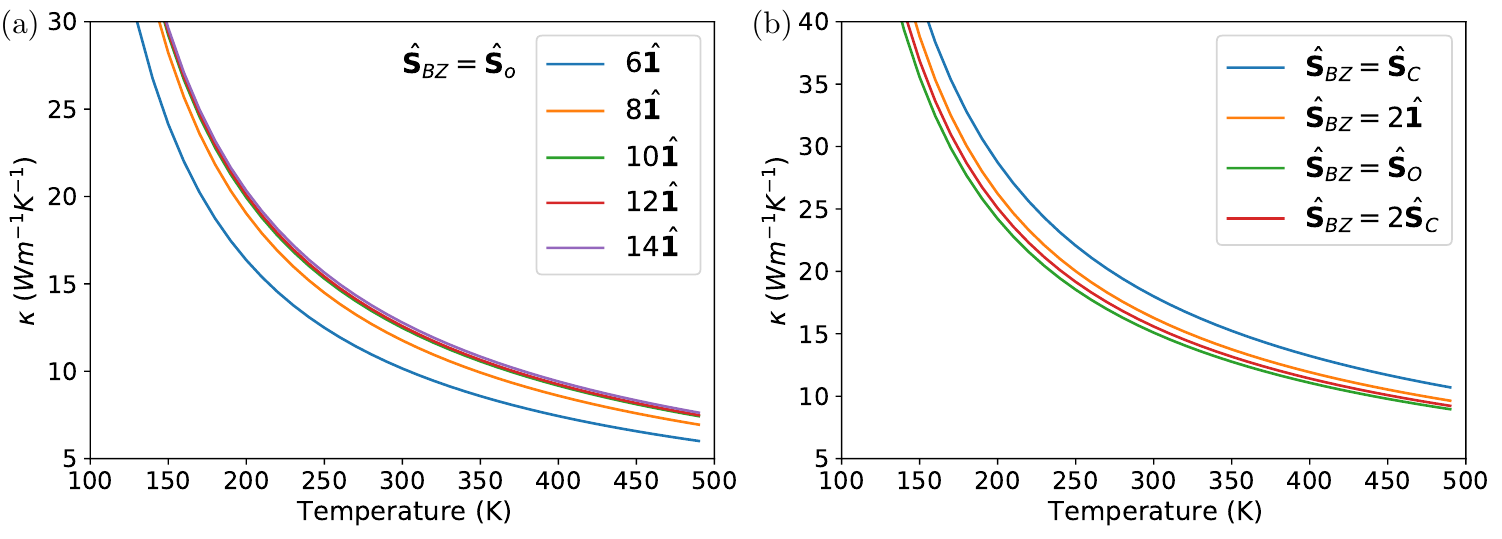}
\caption{Thermal conductivity of ThO$_2$ computed from DFT (SCAN) within the RTA. (Panel $a$) 
Results from different interpolation grid densities.
(Panel $b$) Results from cubic force constants extracted
from supercells of increasing size: $\hat{\mathbf{S}}_{C}$ (4 primitive cells), $2\hat{\mathbf1}$ (8 primitive cells),  $\hat{\mathbf{S}}_{O}$ (16 primitive cells), and $2\hat{\mathbf{S}}_{C}$ (32 primitive cells). 
An interpolation grid of $12\hat1$ is used in all cases.
}
\label{fig:cubic_supa}
\end{figure*}

The phonons and cubic phonon interactions are computed using the lone and
bundled irreducible derivative approaches (LID, BID)\cite{fu_group_2019}, respectively
(see \cite{fu_group_2019} for notation and further detail). Force constants are
obtained by Fourier transforming the phonon interactions and using Wigner-Seitz
packing.  We account for the dipole-dipole interactions at 2nd order when
Fourier interpolating using the standard approach\cite{gonze_interatomic_1994,
gonze_dynamical_1997}.  The  dielectric constants computed by SCAN for CaF$_2$ and ThO$_2$ are 2.161 and 4.482. Born effective charges are 2.263, -1.168, 5.317 and -2.673 for Ca, F, Th, and O in CaF$_2$ and ThO$_2$.

For the face-centered cubic lattice, we choose lattice vectors given in a row stacked matrix as
\begin{align}\label{eq:rocksa}
  \lmat =  \frac{a}{2}
    \begin{bmatrix}
      0 & 1 & 1 \\[0.2em]
      1 & 0 & 1 \\[0.2em]
      1 & 1 & 0
    \end{bmatrix}.
\end{align}
Three classes of supercells, denoted $\hat{\mathbf{S}}_{BZ}$, are used to define the discretization of the phonon interactions computed from first-principles:
$n\hat{\mathbf1}$ (i.e., uniform supercells),  $n\hat{\mathbf{S}}_{C}=n(\hat{\mathbf{J}}-2\hat{\mathbf{1}})$,
and  $n\hat{\mathbf{S}}_{O}=n(4\hat{\mathbf{1}}-\hat{\mathbf{J}})$; where $n$
is a positive integer, $\hat{\mathbf{1}}$ is the $3\times3$ identity matrix,
and $\hat{\mathbf{J}}$ is a $3\times3$ matrix with each element being 1. 
The second and third order irreducible derivatives were computed for 
$\hat{\mathbf{S}}_{BZ}=4\hat{\mathbf1}$ (containing 64 primitive cells) and $\hat{\mathbf{S}}_{BZ}=\hat{\mathbf{S}}_{O}$ (containing 16 primitive cells), respectively.
Up to
10 finite difference $\Delta$ were evaluated for a given measurement, such that
robust error tails could be constructed and used to extrapolate to zero
discretization.  While the BID method only requires the absolute minimum number
of measurements as required by group theory, we doubled this minumum number in
order to remove the possibility of contamination due to a defective
measurement.

When evaluating phonon linewidths via perturbation theory, or solving the LBTE, 
all quantities are Fourier interpolated to supercells of $12\hat{\mathbf1}$; and
all integrations of the Dirac delta function are performed using the tetrahedron method\cite{blochl_improved_1994}.
We found that denser grids beyond $12\hat{\mathbf1}$ are not necessary,
as illustrated in Fig. \ref{fig:cubic_supa}, panel $a$. Another important question is convergence of the
results with respect to the size of $\hat{\mathbf{S}}_{BZ}$ used to compute
cubic phonon interactions, which sets the overall range of the cubic force
constants.  We explicitly tested this in the case of ThO$_2$ by evaluating the
thermal conductivity using cubic interactions from an $\hat{\mathbf{S}}_{BZ}$
of $\hat{\mathbf{S}}_{C}$, 2$\hat{\mathbf1}$,  $\hat{\mathbf{S}}_{O}$, and $2\hat{\mathbf{S}}_{C}$ 
(see Fig. \ref{fig:cubic_supa}, panel $b$). At room temperature, results beyond $\hat{\mathbf{S}}_{C}$ show small differences,
with 2$\hat{\mathbf1}$ being less than 5\% greater than $2\hat{\mathbf{S}}_{C}$ and
$\hat{\mathbf{S}}_{O}$ being less than 3\% smaller than $2\hat{\mathbf{S}}_{C}$.
Overall, the difference between $\hat{\mathbf{S}}_{O}$ and $2\hat{\mathbf{S}}_{C}$
is relatively small at all temperatures. Therefore, we use $\hat{\mathbf{S}}_{O}$ for cubic interactions in all 
computations in the main manuscript.

\section{Phonon Dispersion of Thoria}

\begin{figure*}[h]
\centering
\includegraphics[width=0.6\textwidth]{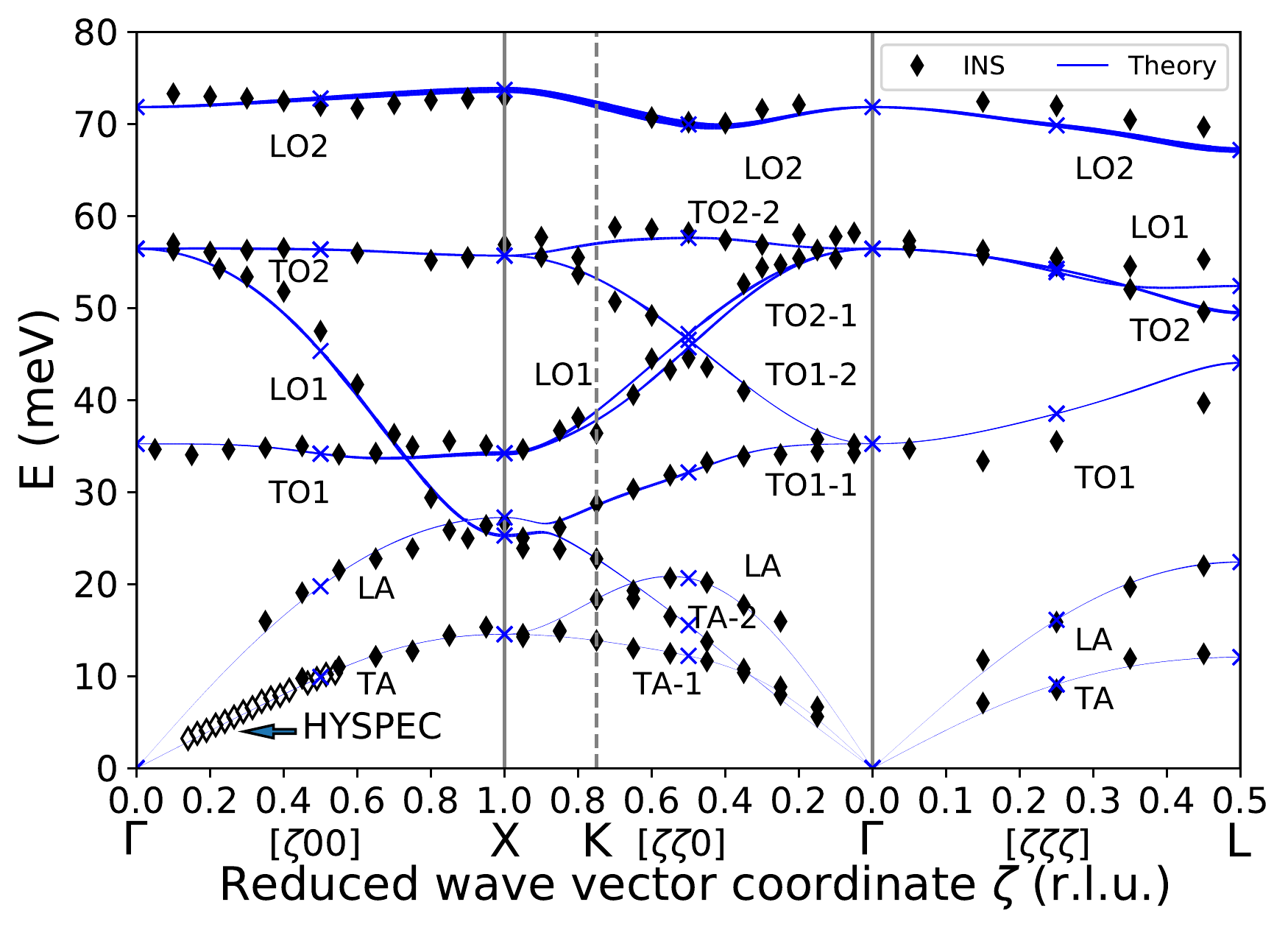}
\caption{ 
Phonon dispersion of ThO$_2$. Black points are INS measurements; blue crosses are computed from DFT (SCAN),
blue lines are a Fourier interpolation, and the line width is proportional to the computed phonon linewidth.}
\label{fig:dispersion}
\end{figure*}

\newpage

\section{Thermal conductivity of CaF$_2$ and ThO$_2$ }

\begin{figure}[h]
\centering
\includegraphics[width=0.8\textwidth]{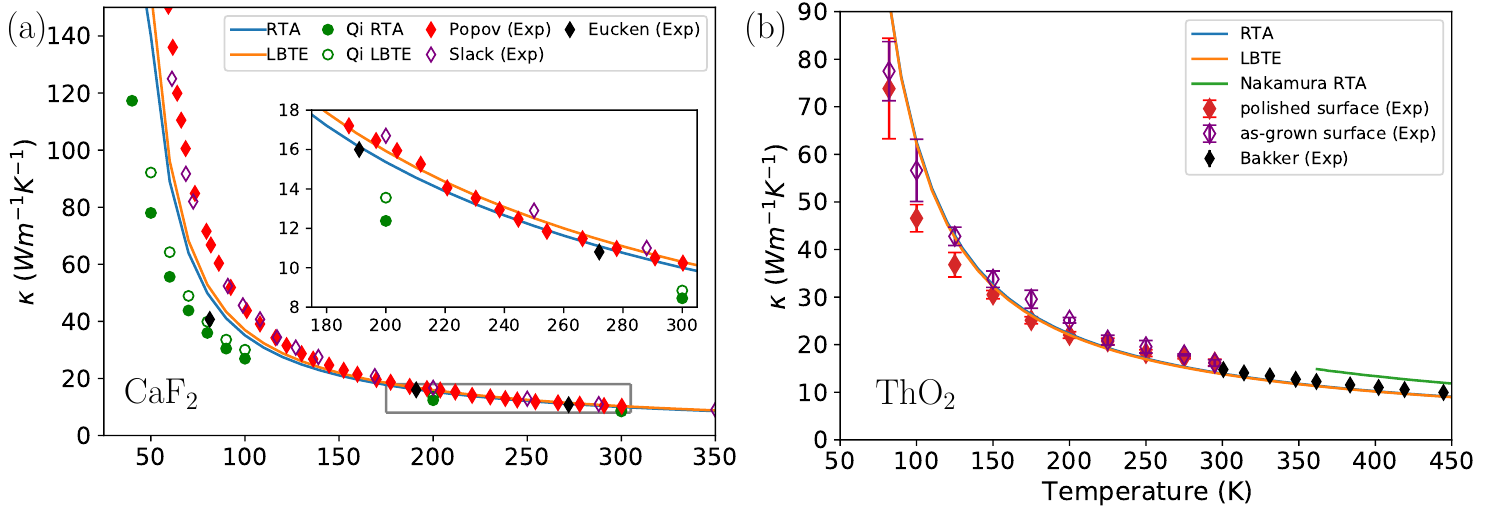}
\caption{ 
Thermal conductivity $\kappa$ as a function of temperature in  CaF$_2$ (panel $a$) and ThO$_2$ (panel $b$). 
Our DFT (SCAN) results are denoted with orange and blue lines for LBTE and RTA solutions, respectively.
Experimental measurements on CaF$_2$ by Slack\cite{slack1961thermal}, Eucken\cite{eucken_uber_1911}, and Popov \textit{et al.}\cite{popov_thermal_2014,digitfn} are shown, in addition to 
measurements in  ThO$_2$ by Bakker \textit{et al.}\cite{bakker_critical_nodate} and our own (with one sample having a polished surface \cite{supmat}).
Previous DFT results for CaF$_2$ by Qi \textit{et al.} (GGA) \cite{qi_lattice_2016} and ThO$_2$ by Nakamura
\textit{et al.} (LDA) \cite{nakamura_first-principles_2019} are also included.
}
\label{fig:thermal_condu}
\end{figure}

\section{$q$-voxel dimension information }
\begin{figure*}[h]
\centering
\includegraphics[width=0.4\textwidth]{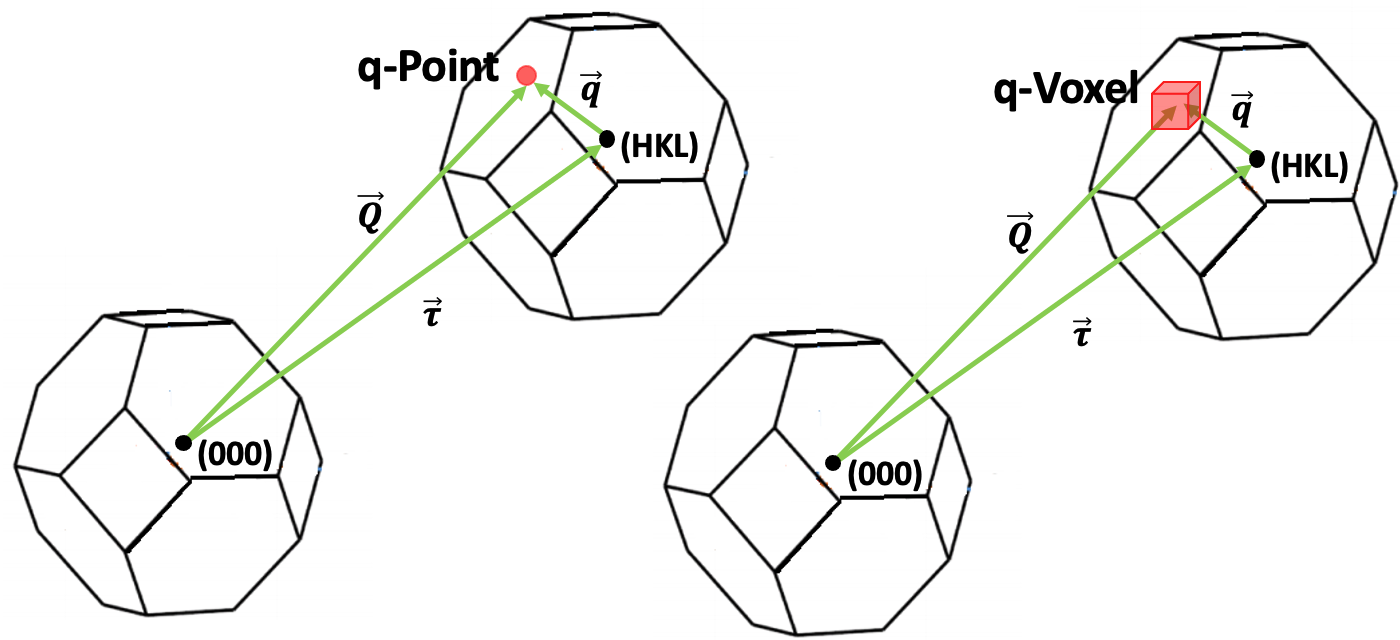}
\caption{ A schematic illustrating the difference between the $q$-point and
$q$-voxel.  The $q$-point is a single point in reciprocal space
while the $q$-voxel is a finite region that is used for the
INS data analysis.  }
\label{fig:schematic}
\end{figure*}

Here we provide the dimensions of all $q$-voxels used for CaF$_2$ and ThO$_2$ in the main manuscript.
A schematic of a $q$-voxel is shown in Fig. \ref{fig:schematic} for clarity.
The CaF$_2$ $q$-voxel dimensions used in Fig. 2 of the main manuscript are listed in
Tables \ref{table:caf2_001}, \ref{table:caf2_011}, and \ref{table:caf2_111}.
In all of the aforementioned tables, the first column specifies the phonon branch; 
the second column specifies a line segment in reciprocal space; 
the third, fourth, and fifth columns specify the three voxel dimensions;
and the last column specifies the incident energy. 
The ThO$_2$ $q$-voxel dimensions used in Fig. 2 in the main manuscript are presented in the same
format in Tables \ref{table:tho2_001}, \ref{table:tho2_011}, and \ref{table:tho2_111}. 
\newpage
\begin{table}[hp]
\begin{tabular}{|c|c|c|c|c|c|}
\hline
(001) direction & path                                  & $\Delta_{[0,0,L] }$ & $\Delta_{[H,H,0]}$ & $\Delta_{[H,\Bar{H},0]}$ & Ei      \\ \hline
TA              & {[}2,2,-4{]}$\rightarrow${[}2,2,-5{]} & 0.025               & 0.1                & 0.1                 & 60 meV  \\ \hline
LA              & {[}2,2,-4{]}$\rightarrow${[}2,2,-5{]} & 0.025               & 0.1                & 0.1                 & 60 meV  \\ \hline
TO1             & {[}2,2,-6{]}$\rightarrow${[}2,2,-7{]} & 0.025               & 0.1                & 0.1                 & 60 meV  \\ \hline
LO1             & {[}3,3,-5{]}$\rightarrow${[}3,3,-4{]} & 0.025               & 0.1                & 0.1                 & 60 meV  \\ \hline
TO2             & {[}1,1,-3{]}$\rightarrow${[}1,1,-2{]} & 0.025               & 0.1                & 0.2                 & 60 meV  \\ \hline
LO2             & {[}2,2,-6{]}$\rightarrow${[}2,2,-5{]} & 0.025               & 0.2                & 0.2                 & 120 meV \\ \hline
\end{tabular}

\caption{Voxel information for $[\zeta00]$ direction in CaF$_2$. \label{table:caf2_001} }

\vspace{0.5cm}

\begin{tabular}{|c|c|c|c|c|c|}
\hline
(110) direction & path                                      & $\Delta_{[H,H,0]}$ & $\Delta_{[0,0,L]}$ & $\Delta_{[H,\Bar{H},0]}$ & Ei      \\ \hline
TA              & {[}2,2,-4{]}$\rightarrow${[}3,3,-4{]}     & 0.025               & 0.1                 & 0.2                  & 60 meV  \\ \hline
LA              & {[}2,2,-4{]}$\rightarrow${[}3,3,-4{]}     & 0.025               & 0.1                 & 0.2                  & 60 meV  \\ \hline
TO1 1           & {[}1,1,-5{]}$\rightarrow${[}2,2,-5{]}     & 0.025               & 0.1                 & 0.2                  & 60 meV  \\ \hline
LO1             & {[}3,3,-1{]}$\rightarrow${[}2,2,-1{]}     & 0.025               & 0.2                 & 0.2                  & 120 meV \\ \hline
TO2 1           & {[}1.7,1.7,-5{]}$\rightarrow${[}2,2,-5{]} & 0.025               & 0.1                 & 0.2                  & 60 meV  \\ \hline
TO2 2           & {[}2,2,-2{]}$\rightarrow${[}1,1,-2{]}     & 0.025               & 0.1                 & 0.1                  & 60 meV  \\ \hline
LO2 (half)      & {[}1,1,-5{]}$\rightarrow${[}1.5,1.5,-5{]} & 0.025               & 0.1                 & 0.2                  & 60 meV  \\ \hline
LO2 (half)      & {[}2.5,2.5,-2{]}$\rightarrow${[}3,3,-2{]} & 0.025               & 0.2                 & 0.2                  & 120 meV \\ \hline
\end{tabular}
\caption{Voxel information for $[\zeta\zeta0]$ direction in CaF$_2$. \label{table:caf2_011} }

\vspace{0.5cm}

\begin{tabular}{|c|c|c|c|c|c|}
\hline
(111) direction & path                                     & $\Delta_{[H,H,H]}$ & $\Delta_{[0,0,L]}$ & $\Delta_{[H,\Bar{H},0]}$ & Ei      \\ \hline
TA              & {[}2,2,-4{]}$\rightarrow${[}2.5,2.5,-3.5{]}      & 0.025              & 0.1                & 0.1                 & 60 meV  \\ \hline
LA              & {[}0,0,-4{]}$\rightarrow${[}-0.5,-0.5,-4.5{]}    & 0.025              & 0.1                & 0.1                 & 60 meV  \\ \hline
TO1             & {[}0,0,-6{]}$\rightarrow${[}0.5,0.5,-5.5{]}      & 0.025              & 0.1                & 0.1                 & 60 meV  \\ \hline
LO1             & {[}-1,-1,-3{]}$\rightarrow${[}-0.5,-0.5,-2.5{]}  & 0.025              & 0.1                & 0.2                 & 60 meV  \\ \hline
TO2             & {[}1,1,-5{]}$\rightarrow${[}0.5,0.5,-5.5{]}      & 0.025              & 0.1                & 0.1                 & 60 meV  \\ \hline
LO2             & {[}-2,-2,-10{]}$\rightarrow${[}-1.5,-1.5,-9.5{]} & 0.025              & 0.2                & 0.2                 & 120 meV \\ \hline
\end{tabular}
\caption{Voxel information for $[\zeta\zeta\zeta]$ direction in CaF$_2$. \label{table:caf2_111} }
\end{table}

\begin{table}[hp]
\begin{tabular}{|c|c|c|c|c|c|}
\hline
(001) direction & path                                 & $\Delta_{[0,0,L] }$ & $\Delta_{[H,H,0]}$ & $\Delta_{[H,\Bar{H},0]}$ & Ei      \\ \hline
LA              & {[}1,1,5{]}$\rightarrow${[}1,1,-6{]} & 0.1                 & 0.3                & 0.3                 & 50meV   \\ \hline
TA              & {[}3,3,-1{]}$\rightarrow${[}3,3,0{]} & 0.1                 & 0.2                & 0.2                 & 50 meV  \\ \hline
TO1             & {[}4,4,2{]}$\rightarrow${[}4,4,1{]}  & 0.1                 & 0.1                & 0.1                 & 50 meV  \\ \hline
TO2             & {[}3,3,3{]}$\rightarrow${[}3,3,4{]}  & 0.1                 & 0.3                & 0.4                 & 120 meV \\ \hline
LO1             & {[}1,1,9{]}$\rightarrow${[}1,1,10{]} & 0.1                 & 0.4                & 0.4                 & 120 meV \\ \hline
LO2             & {[}0,0,10{]}$\rightarrow${[}0,0,9{]} & 0.1                 & 0.4                & 0.4                 & 120 meV \\ \hline
\end{tabular}
\caption{Voxel information for $[\zeta00]$ direction in ThO$_2$. \label{table:tho2_001} }

\vspace{0.5cm}

\begin{tabular}{|c|c|c|c|c|c|}
\hline
(110) direction & path                                                     & $\Delta_{[H,H,0]}$  & $\Delta_{[0,0,L]}$ & $\Delta_{[H,\Bar{H},0]}$         & Ei      \\ \hline
LA              & {[}2,2,0{]}$\rightarrow${[}3,3,-0{]}                     & 0.1                 & 0.2                & 0.3                         & 50meV   \\ \hline
TO1\_1          & {[}0,0,6{]}$\rightarrow${[}1,1,6{]}                   & 0.1                 & 0.1                & 0.2                         & 50 meV  \\ \hline
TO2\_2          & {[}3,3,7{]}$\rightarrow${[}2,2,7{]}                   & 0.1                 & 0.4                & 0.4                         & 120 meV \\ \hline
LO2             & \multicolumn{1}{l|}{{[}6,6,0{]}$\rightarrow${[}7,7,0{]}}       & 0.1                 & 0.4                & 0.4                         & 120 meV \\ \hline
(110) direction & path                                                     & $\Delta_{[H,H,0]}$  & $\Delta_{[0,0,L]}$ & $\Delta_{[\Bar{H},H,0]}$         & Ei      \\ \hline
TA2             & {[}2,0,6{]}$\rightarrow${[}1,-1,6{]}                     & 0.1                 & 0.2                & 0.4                         & 50 meV  \\ \hline
(110) direction & path                                                     & $\Delta_{[\Bar{H},0,H]}$ & $\Delta_{[0,L,0]}$ & $\Delta_{ [H,0,H]}$         & Ei      \\ \hline
TA1             & {[}2,0,6{]}$\rightarrow${[}1,0,7{]}                      & 0.1                 & 0.2                & 0.2                         & 50 meV  \\ \hline
(110) direction & path                                                     & $\Delta_{[\Bar{H},0,H]}$ & $\Delta_{[H,H,H]}$ & $\Delta_{[\frac{1}{2}H,\Bar{H},\frac{1}{2}H]}$   & Ei      \\ \hline
TO1\_2          & {[}1,1,7{]}$\rightarrow${[}2,1,6{]}                   & 0.1                 & 0.2                & 0.4                         & 120 meV \\ \hline
(110) direction & path                                                     & $\Delta_{[\Bar{H},0,H]}$ & $\Delta_{[0,L,0]}$ & $\Delta_{ [H,0,H]}$         & Ei      \\ \hline
TO2\_1          & {[}5,3,7{]}$\rightarrow${[}4,3,8{]}                    & 0.1                 & 0.4                & 0.4                         & 120 meV \\ \hline
(110) direction & path                                                     & $\Delta_{[H,0,H]}$  & $\Delta_{[0,L,0]}$ & $\Delta_{ [\Bar{H},0,H]}$        & Ei      \\ \hline
LO1             & {[}7,3,7{]}$\rightarrow${[}6,3,6{]}                      & 0.1                 & 0.2                & 0.4                         & 120 meV \\ \hline
\end{tabular}
\caption{Voxel information for $[\zeta\zeta0]$ direction in ThO$_2$. \label{table:tho2_011} }
\vspace{0.5cm}

\begin{tabular}{|c|c|c|c|c|c|}
\hline
(111) direction & path                                        & $\Delta_{[H,H,H]}$ & $\Delta_{[\frac{1}{2}H,\Bar{H},\frac{1}{2}H]}$    & $\Delta_{[\Bar{H},0,H]}$ & Ei      \\ \hline
LA              & {[}4,3,1{]}$\rightarrow${[}4.5,3.5,1.5{]}   & 0.1                & 0.2                          & 0.2                 & 50 meV  \\ \hline
TA              & {[}3,1,-1{]}$\rightarrow${[}2.5,0.5,-1.5{]} & 0.1                & 0.2                          & 0.2                 & 50 meV  \\ \hline
TO1             & {[}8,4,0{]}$\rightarrow${[}7.5,3.5,-0.5{]}  & 0.1                & 0.4                          & 0.2                 & 120 meV \\ \hline
TO2             & {[}-1,1,3{]}$\rightarrow${[}-1.5,0.5,2.5{]} & 0.1                & 0.4                          & 0.2                 & 120 meV \\ \hline
LO2             & {[}4,4,4{]}$\rightarrow${[}3.5,3.5,3.5{]}   & 0.1                & 0.4                          & 0.4                 & 120 meV \\ \hline
(111) direction & path                                        & $\Delta_{[H,H,H]}$ & $\Delta_{[\frac{1}{3}H,\frac{1}{3}H,\frac{2}{3} \Bar{H}]}$ & $\Delta_{[\Bar{H},0,H]}$ & Ei      \\ \hline
LO1             & {[}5,5,7{]}$\rightarrow${[}4.5,4.5,6.5{]}   & 0.1                & 0.4                          & 0.4                 & 120 meV \\ \hline
\end{tabular}
\caption{Voxel information for $[\zeta\zeta\zeta]$ direction in ThO$_2$. \label{table:tho2_111} }
\end{table}

\newpage

\section{Effect of varying different $q$-voxel dimensions}
\begin{figure}[h]
\centering
\includegraphics[width=0.8\textwidth]{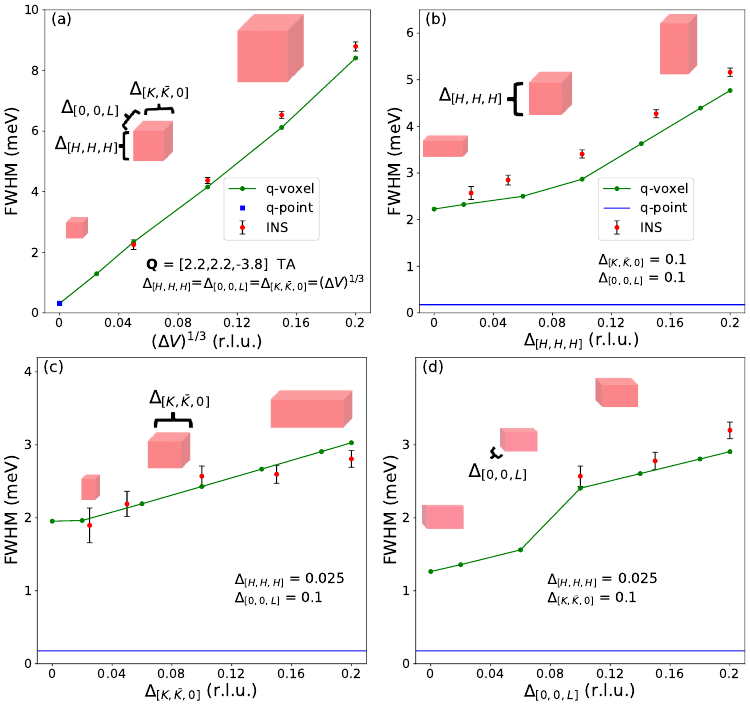}
\caption{ The effect of the $q$-voxel size and shape on the experimental and
theoretical $q$-voxel scattering function peak, full width half maximum (FWHM), for CaF$_2$ at 300K.  
The FWHM 
of the transverse acoustic (TA) phonon at $\mathbf{Q}$ = [2.2, 2.2, -3.8] as a
function of (a) cube root of $q$-voxel volume in reciprocal lattice units
(r.l.u.), (b) dimension along $[H,H,H]$, (c)  $[K,\Bar{K},0]$, and
(d) $[0,0,L]$. In panels $b-d$,  the other dimensions are held fixed, as indicated
in the legend. The blue point in panel $a$ and the blue line in panels $b-d$
indicate the theoretical $q$-point result. }
\label{fig:voxel_size3}
\end{figure}

The $q$-voxel is normally chosen as a rectangular cuboid, and therefore it has three independent dimensions.
In the main manuscript, we explored how the peak width changed as a function of a cubic $q$-voxel,
demonstrating that one cannot reliably extrapolate to the $q$-point limit. Here we evaulate
how the peak width changes while varying a single $q$-voxel dimension, with the limiting $q$-voxel shape being
a rectangle instead of a point. Such an analysis is useful as a further comparison between theory and
experiment, in addition to demonstrating the quantitive effects of different $q$-voxel dimensions.
We focus on the scattering function peak width as a function of the $q$-voxel shape for the TA phonon at $\mathbf{Q}$ = [2.2,2.2, -3.8] (See Fig.\ref{fig:voxel_size3}).
Panel $a$ shows the result of changing the dimension of a cubic voxel, which is the same
result from Fig. 3 in the main manuscript.
In panels
$b-d$, the $q$-voxel is stretched along the [H,H,H], [K,-K,0], and [0,0,L] direction,
respectively,  while  the other dimensions are held fixed.  We show that the scattering
function FWHMs measured using different $q$-voxel shapes are well captured 
by theory.
In panels $b-d$, when squeezing
one dimension to zero and keeping the others fixed, the FWHM does not approach
the $q$-point value since the $q$-voxel scattering function is still obtained by summing
over a plane. This anaylsis validates the reasonable supposition that the
$q$-voxel dimensions should be reported  for any experimental result.

\section{Influence of the DFT Functional on computed Scattering function }
\begin{figure*}[h]
\centering
\includegraphics[width=0.9\textwidth]{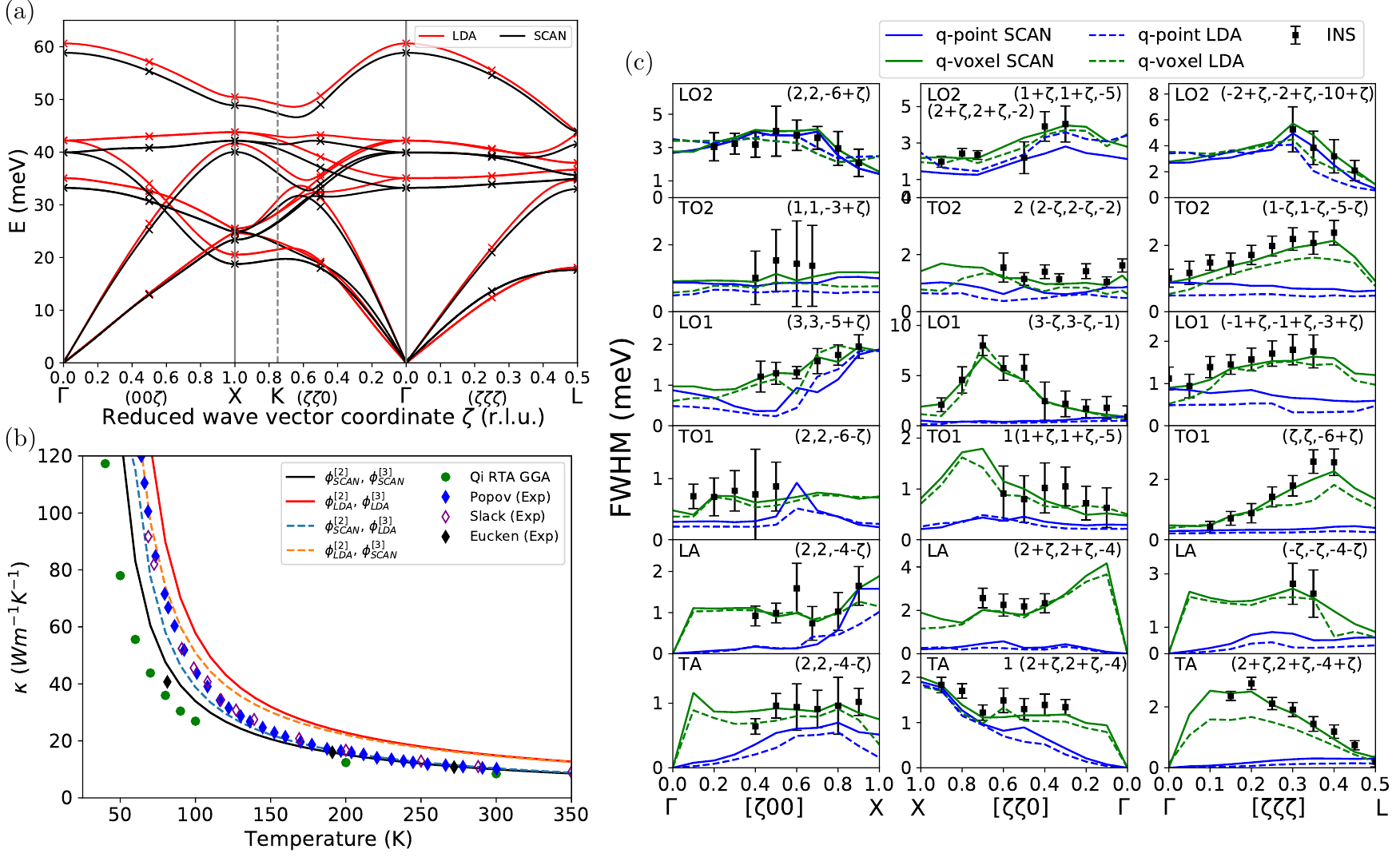}
\caption{
(a) Phonon dispersion of CaF$_2$ computed using LDA and SCAN at $V_{300K}$. Crosses are computed using DFT and lines are a Fourier interpolation.
(b) Thermal conductivity of CaF$_2$ computed within the RTA using all permutations of phonons and
cubic phonon interactions computed from the LDA and SCAN functionals at $V_{300K}$. Previous theoretical work using GGA \cite{qi_lattice_2016} and experimental measurements by Slack\cite{slack1961thermal}, Eucken\cite{eucken_uber_1911}, and Popov \textit{et al.}\cite{popov_thermal_2014,digitfn} are shown for comparison.
(c) FWHMs of the scattering function peaks for CaF$_2$ at 300 K computed using the LDA and SCAN functionals. The $q$-point and $q$-voxel FWHM values are shown as blue and green lines; results for SCAN and LDA are shown by solid and dashed lines, respectively; INS results are shown as black points. }
\label{fig:LDA}
\end{figure*}

In the main manuscript, all calculations are done using the SCAN functional.
Here we explore how the phonons and the $q$-voxel scattering function
peak widths change when using the LDA functional.  The relaxed lattice parameter
of CaF$_2$ within LDA is 5.330 {\AA}, in agreement with previous work
\cite{bader_teraflops_2003}. The second and third order derivatives (i.e.  the phonons
and cubic phonon interactions) computed using LDA and SCAN are denoted
$\phi^{[2]}_{LDA}$ , $\phi^{[3]}_{LDA}$, $\phi^{[2]}_{SCAN}$ and $\phi^{[3]}_{SCAN}$. 
The LDA phonons are scaled to slightly higher
frequencies relative to SCAN, which is expected given that LDA underpredicts
the volume (see Fig. \ref{fig:LDA}, panel $a$).  
Experiments have found the lattice parameter to be 5.466 {\AA} at $T=298 K$\cite{schumann_thermal_1984} and 5.46 {\AA} at $T=296 K$\cite{Hazen:a20511}. 

For the FWHM of the scattering
function peaks, there can be a drastic difference in both the $q$-point and
$q$-voxel results when comparing LDA and SCAN (see Fig. \ref{fig:LDA}, panel
$c$).  For example, for the TA mode along $[\zeta\zeta\zeta]$ direction
(bottom right subplot), the $q$-voxel result for SCAN is in outstanding
agreement with the experimental $q$-voxel result, while the LDA 
$q$-voxel result is notably smaller.
Therefore, we see that INS $q$-voxel scattering function
peak widths can serve as a critical judge of the exchange-correlation energy
functional used within DFT.
Given that the $q$-point FWHM obtained by LDA is smaller than SCAN in most cases,
especially the acoustic modes, the thermal conductivity will be larger within LDA
as compared to SCAN (see Fig. \ref{fig:LDA}, panel $b$).

\section{Influence of thermal expansion on thermal conductivity and Scattering function }
\label{sec:influence_therm_exp}
\begin{figure*}[h]
\centering
\includegraphics[width=0.9\textwidth]{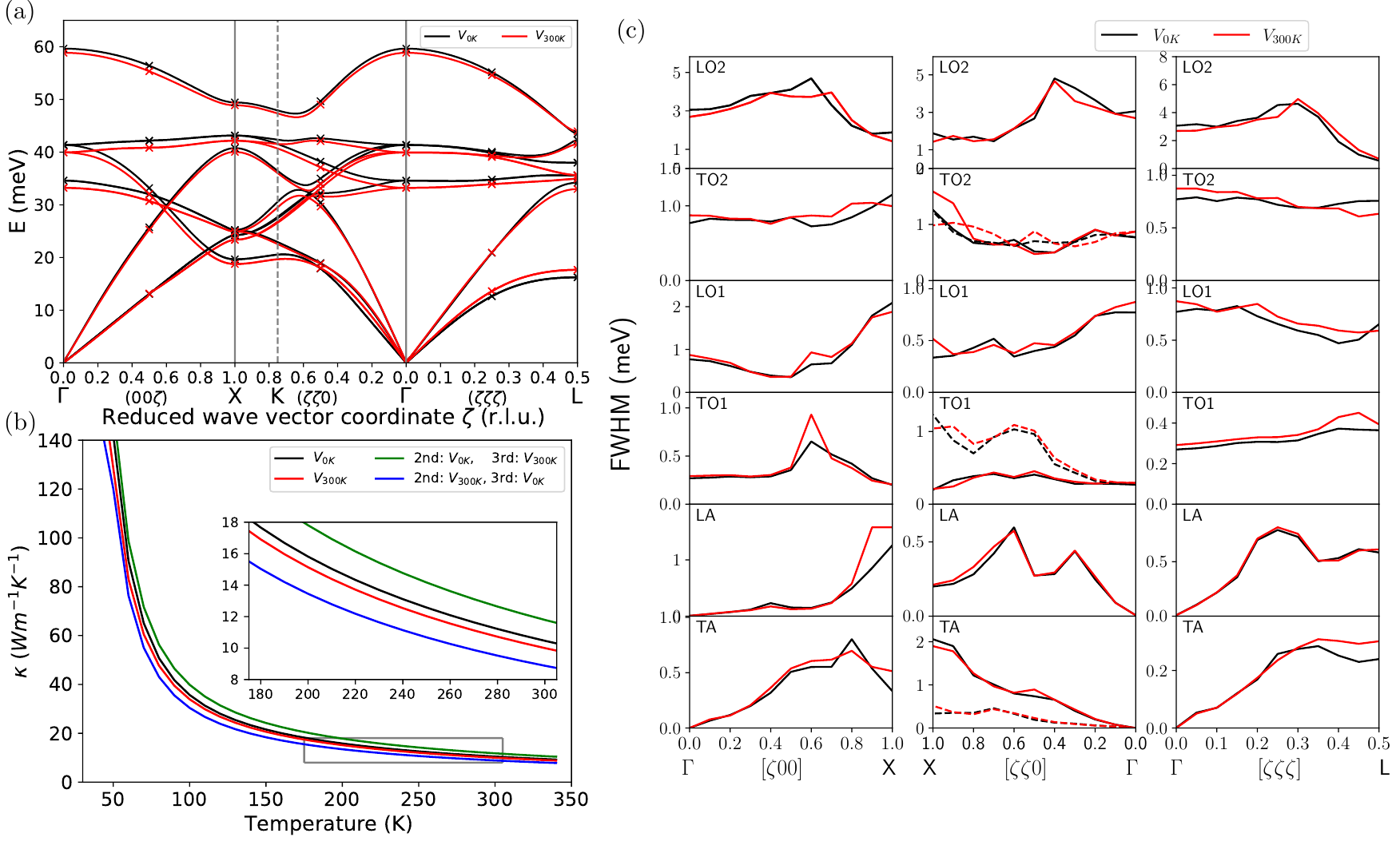}
\caption{
(a) Phonon dispersion of CaF$_2$ computed using DFT (SCAN) at the fully relaxed volume
(denoted $V_{0K}$) and the volume at 300K as dictated by the CLTE (denoted
$V_{300K}$).  Crosses are computed using DFT and lines are a Fourier
interpolation.  (b) Thermal conductivity within the RTA for CaF$_2$ computed using all permutations of phonons and
cubic phonon interactions from $V_{0K}$ and $V_{300K}$.
(c) A plot of $2\Gamma$ for CaF$_2$ using phonons and phonon interactions from $V_{0K}$ and $V_{300K}$. 
In panels which contain two modes, mode 1 and mode 2 (see Fig. 1a in the main manuscript) are shown as solid and dashed lines.  }
\label{fig:expansion}
\end{figure*}

The displacement derivatives of the Born-Oppenheimer potential are a function of volume. Given
that experiments are performed under conditions of constant pressure and temperature, the volume 
is not fixed. If a crystal has a substantial CLTE, this must be accounted for. Therefore, for the case of
CaF$_2$, we compute the second and third order derivatives (i.e. the phonons and cubic phonon interactions) 
at the fully relaxed volume, denoted $V_{0K}$, and the volume at T=300 K as dictated by the CLTE,
denoted $V_{300K}$. Here we explore
how important these changes are, starting with the phonons (see Fig. \ref{fig:expansion}, panel $a$). 
As shown, the changes in the phonons at these two different volumes is small, but nonzero. The changes
in the phonon linewidths $2\Gamma$, which reflect both the changes in the phonons and the cubic phonon 
interactions, is also relatively small (see Fig. \ref{fig:expansion}, panel $c$); though this is 
somewhat fortuitous given that the change in the phonon frequencies are offset by the changes in the cubic
phonon interactions. Finally, we compare the thermal conductivity using all permutations of $V_{0K}$
and $V_{300K}$ phonons and cubic phonon interactions (see Fig. \ref{fig:expansion}, panel $b$). 
The pure $V_{0K}$ result is similar to the $V_{300K}$ result due to the aforementioned 
cancellation, while a nontrivial change can be seen when using combinations of the two results. 
The similarity
bewteen the pure $V_{0K}$ and $V_{300K}$ implies that the use of any interpolation between these parameter
sets to account for thermal expansion will not have any appreciable effect on the thermal conductivity.

\section{One-phonon Scattering Function via perturbation theory}
%
The 1-phonon scattering function can be obtained using perturbation theory in terms of the phonons
and the cubic phonon interactions as \cite{Maradudin19622589} 
\begin{align}
S_1(\boldsymbol{Q},\omega) &=\frac{\hbar N}{2\pi} \sum\limits_{d d^{\prime}}  
\frac{b_d b_{d^{\prime}} }{\sqrt{M_dM_{d^{\prime}}}}
e^{-W_d} e^{-W_{d^{\prime}}} 
e^{-i\boldsymbol{Q}\cdot (\boldsymbol{x}_{d} - \boldsymbol{x}_{d^{\prime}}) } \sum\limits_{j} \frac{[\boldsymbol{Q} \cdot \mathbf{e}_{\mathbf{q} j d}][\boldsymbol{Q} \cdot \mathbf{e}_{-\mathbf{q} j d^{\prime}}]}{\omega_{\mathbf{q} j}} \\
&\times\frac{1}{1-e^{-\beta \hbar \omega}}
\left[\frac{ \Gamma_{\boldsymbol{q}j}(\omega)}{(\omega + \omega_{\boldsymbol{q} j} +\Delta_{\boldsymbol{q}j}(\omega)  )^2 +  \Gamma_{\boldsymbol{q}j}(\omega)^2 }  \nonumber  
+ \frac{ \Gamma_{\boldsymbol{q},j}(\omega) }{(-\omega + \omega_{\boldsymbol{q} j} +\Delta_{\boldsymbol{q},j}(\omega)  )^2 +  \Gamma_{\boldsymbol{q}j}(\omega)^2 } \right]
\end{align}
where $\mathbf{Q} = \mathbf{q} + \mathbf{G}$ and $\mathbf{G}$ is a reciprocal lattice vector;
$b_d$,
$M_d$, $W_d$, and $x_d$ denote the scattering length, mass, Debye-Waller factor,
and equilibrium position within unit cell of atom $d$;
$\mathbf{e}_{\mathbf{q} j d}$ denotes the polarization vector of phonon
${\boldsymbol{q},j}$ and atom $d$; $\omega_{\mathbf{q} j}$ denotes the phonon
frequency of phonon ${\boldsymbol{q},j}$; $\beta=(kT)^{-1}$ where $k$
and $T$ are  the Boltzmann's constant and temperature; $\Delta$ and $\Gamma$
are the real and imaginary part of self energy,
${\Sigma}_{\boldsymbol{q}j}(\omega)  = \Delta_{\boldsymbol{q}j}(\omega) +
i\Gamma_{\boldsymbol{q}j}(\omega)$ \cite{Maradudin19622589}. 
The phonon linewidth for a given ${\boldsymbol{q},j}$ is $2\Gamma_{\boldsymbol{q},j}(\omega_{\boldsymbol{q},j})$.


%

\newpage
\section{Space Group Irreducible Derivatives}
\label{sec:irrderivinfo}
This section contains all information necessary to specify the irreducible derivatives. 

\begin{table*}[hp]
\centering
\caption{The labels of $\mathbf{q}$ points, which are used in tables \ref{table:caf2_2_0}-\ref{table:caf2_3_300_lda}.}
\begin{ruledtabular}
\begin{tabular*}{\textwidth}{cccccc}
$ {\Gamma} = {\left( 0 0 0\right) }$ & 
\\[0.4em] 
$ {L} = {\left( \frac{1}{2} 0 0\right) }$ & 
$ {L_{1}} = {\left( \frac{1}{2} \frac{1}{2} \frac{1}{2}\right) }$ & 
$ {L_{2}} = {\left( 0 \frac{1}{2} 0\right) }$ & 
$ {L_{3}} = {\left( 0 0 \frac{1}{2}\right) }$ & 
\\[0.4em] 
$ {X} = {\left( \frac{1}{2} \frac{1}{2} 0\right) }$ & 
$ {X_{1}} = {\left( 0 \frac{1}{2} \frac{1}{2}\right) }$ & 
$ {X_{2}} = {\left( \frac{1}{2} 0 \frac{1}{2}\right) }$ & 
\\[0.4em] 
$ {\Delta} = {\left( \frac{1}{4} \frac{1}{4} 0\right) }$ & 
$ {\bar{\Delta}} = {\left( \frac{3}{4} \frac{3}{4} 0\right) }$ & 
$ {\Delta_{1}} = {\left( 0 \frac{1}{4} \frac{1}{4}\right) }$ & 
$ {\bar{\Delta}_{1}} = {\left( 0 \frac{3}{4} \frac{3}{4}\right) }$ & 
$ {\Delta_{2}} = {\left( \frac{3}{4} 0 \frac{3}{4}\right) }$ & 
$ {\bar{\Delta}_{2}} = {\left( \frac{1}{4} 0 \frac{1}{4}\right) }$
\\[0.4em]
$ {W} = {\left( \frac{1}{4} \frac{3}{4} \frac{1}{2}\right) }$ & 
$ {\bar{W}} = {\left( \frac{3}{4} \frac{1}{4} \frac{1}{2}\right) }$ & 
$ {W_{1}} = {\left( \frac{1}{2} \frac{1}{4} \frac{3}{4}\right) }$ & 
$ {\bar{W}_{1}} = {\left( \frac{1}{2} \frac{3}{4} \frac{1}{4}\right) }$ & 
$ {W_{2}} = {\left( \frac{1}{4} \frac{1}{2} \frac{3}{4}\right) }$ & 
$ {\bar{W}_{2}} = {\left( \frac{3}{4} \frac{1}{2} \frac{1}{4}\right) }$
\\[0.4em]
$ {A} = {\left( \frac{1}{4} \frac{3}{4} 0\right) }$ & 
$ {\bar{A}} = {\left( \frac{3}{4} \frac{1}{4} 0\right) }$ & 
$ {A_{1}} = {\left( \frac{1}{2} \frac{3}{4} \frac{3}{4}\right) }$ & 
$ {\bar{A}_{1}} = {\left( \frac{1}{2} \frac{1}{4} \frac{1}{4}\right) }$ & 
$ {A_{2}} = {\left( \frac{1}{4} 0 \frac{3}{4}\right) }$ & 
$ {\bar{A}_{2}} = {\left( \frac{3}{4} 0 \frac{1}{4}\right) }$\\[0.4em] 
$ {A_{3}} = {\left( 0 \frac{3}{4} \frac{1}{4}\right) }$ & 
$ {\bar{A}_{3}} = {\left( 0 \frac{1}{4} \frac{3}{4}\right) }$ & 
$ {A_{4}} = {\left( \frac{3}{4} \frac{1}{2} \frac{3}{4}\right) }$ & 
$ {\bar{A}_{4}} = {\left( \frac{1}{4} \frac{1}{2} \frac{1}{4}\right) }$ & 
$ {A_{5}} = {\left( \frac{1}{4} \frac{1}{4} \frac{1}{2}\right) }$ & 
$ {\bar{A}_{5}} = {\left( \frac{3}{4} \frac{3}{4} \frac{1}{2}\right) }$
\\[0.4em] 
$ {\Lambda} = {\left( \frac{1}{4} 0 0\right) }$ & 
$ {\bar{\Lambda}} = {\left( \frac{3}{4} 0 0\right) }$ & 
$ {\Lambda_{1}} = {\left( 0 \frac{3}{4} 0\right) }$ & 
$ {\bar{\Lambda}_{1}} = {\left( 0 \frac{1}{4} 0\right) }$ & 
$ {\Lambda_{2}} = {\left( 0 0 \frac{3}{4}\right) }$ & 
$ {\bar{\Lambda}_{2}} = {\left( 0 0 \frac{1}{4}\right) }$\\[0.4em] 
$ {\Lambda_{3}} = {\left( \frac{1}{4} \frac{1}{4} \frac{1}{4}\right) }$ & 
$ {\bar{\Lambda}_{3}} = {\left( \frac{3}{4} \frac{3}{4} \frac{3}{4}\right) }$ & 
\\[0.4em] 
$ {B} = {\left( \frac{1}{4} \frac{1}{2} 0\right) }$ & 
$ {\bar{B}} = {\left( \frac{3}{4} \frac{1}{2} 0\right) }$ & 
$ {B_{1}} = {\left( \frac{1}{2} \frac{3}{4} \frac{1}{2}\right) }$ & 
$ {\bar{B}_{1}} = {\left( \frac{1}{2} \frac{1}{4} \frac{1}{2}\right) }$ & 
$ {B_{2}} = {\left( \frac{3}{4} \frac{3}{4} \frac{1}{4}\right) }$ & 
$ {\bar{B}_{2}} = {\left( \frac{1}{4} \frac{1}{4} \frac{3}{4}\right) }$\\[0.4em] 
$ {B_{3}} = {\left( \frac{1}{4} \frac{3}{4} \frac{3}{4}\right) }$ & 
$ {\bar{B}_{3}} = {\left( \frac{3}{4} \frac{1}{4} \frac{1}{4}\right) }$ & 
$ {B_{4}} = {\left( \frac{1}{2} \frac{1}{2} \frac{1}{4}\right) }$ & 
$ {\bar{B}_{4}} = {\left( \frac{1}{2} \frac{1}{2} \frac{3}{4}\right) }$ & 
$ {B_{5}} = {\left( \frac{3}{4} \frac{1}{4} \frac{3}{4}\right) }$ & 
$ {\bar{B}_{5}} = {\left( \frac{1}{4} \frac{3}{4} \frac{1}{4}\right) }$\\[0.4em] 
$ {B_{6}} = {\left( \frac{1}{4} \frac{1}{2} \frac{1}{2}\right) }$ & 
$ {\bar{B}_{6}} = {\left( \frac{3}{4} \frac{1}{2} \frac{1}{2}\right) }$ & 
$ {B_{7}} = {\left( \frac{1}{2} \frac{3}{4} 0\right) }$ & 
$ {\bar{B}_{7}} = {\left( \frac{1}{2} \frac{1}{4} 0\right) }$ & 
$ {B_{8}} = {\left( \frac{1}{4} 0 \frac{1}{2}\right) }$ & 
$ {\bar{B}_{8}} = {\left( \frac{3}{4} 0 \frac{1}{2}\right) }$\\[0.4em] 
$ {B_{9}} = {\left( \frac{1}{2} 0 \frac{1}{4}\right) }$ & 
$ {\bar{B}_{9}} = {\left( \frac{1}{2} 0 \frac{3}{4}\right) }$ & 
$ {B_{10}} = {\left( 0 \frac{1}{4} \frac{1}{2}\right) }$ & 
$ {\bar{B}_{10}} = {\left( 0 \frac{3}{4} \frac{1}{2}\right) }$ & 
$ {B_{11}} = {\left( 0 \frac{1}{2} \frac{1}{4}\right) }$ & 
$ {\bar{B}_{11}} = {\left( 0 \frac{1}{2} \frac{3}{4}\right) }$
\\[0.4em] 

\end{tabular*}
\end{ruledtabular}
%

\end{table*}

\begin{longtable*}{cccccccccc}
\caption{All second order irreducible derivatives for CaF$_2$ at the fully relaxed volume. All results were computed using the LID approach,  $4\hat{\mathbf1}$ supercell, and SCAN functional. The units are eV/\AA$^2$.\label{table:caf2_2_0}} \\
\hline\hline 
Derivative & Value & Derivative & Value & Derivative & Value & Derivative & Value & Derivative & Value \\
\hline\\[-0.9em]
\endfirsthead
\hline\hline
Derivative & Value & Derivative & Value & Derivative & Value & Derivative & Value & Derivative & Value  \\
\hline\\[-0.9em]
\endhead
\hline\hline
\endfoot
                                                                          $d\indices*{*_{\Gamma}^{T_{2g}^{}}_{\Gamma}^{T_{2g}^{}}}$ &   $7.7748$ &                                                                         $d\indices*{*_{\Gamma}^{T_{1u}^{}}_{\Gamma}^{T_{1u}^{}}}$ &   $8.3779$ &                                                                                   $d\indices*{*_{L}^{A_{1g}^{}}_{L}^{A_{1g}^{}}}$ &   $8.1278$ &                                                                                   $d\indices*{*_{L}^{E_{g}^{}}_{L}^{E_{g}^{}}}$ &   $5.7516$ &                                                                               $d\indices*{*_{L}^{A_{2u}^{}}_{L}^{A_{2u}^{}}}$ &  $15.3524$ \\[0.4em]
                                                   $d\indices*{*_{L}^{A_{2u}^{}}_{L}^{\tensor*[^{1}]{\hspace{-0.2em}A}{}_{2u}^{}}}$ &   $2.3443$ &                 $d\indices*{*_{L}^{\tensor*[^{1}]{\hspace{-0.2em}A}{}_{2u}^{}}_{L}^{\tensor*[^{1}]{\hspace{-0.2em}A}{}_{2u}^{}}}$ &   $6.6428$ &                                                                                     $d\indices*{*_{L}^{E_{u}^{}}_{L}^{E_{u}^{}}}$ &   $5.9754$ &                                                  $d\indices*{*_{L}^{E_{u}^{}}_{L}^{\tensor*[^{1}]{\hspace{-0.2em}E}{}_{u}^{}}}$ & $- 3.5849$ &               $d\indices*{*_{L}^{\tensor*[^{1}]{\hspace{-0.2em}E}{}_{u}^{}}_{L}^{\tensor*[^{1}]{\hspace{-0.2em}E}{}_{u}^{}}}$ &   $4.9288$ \\[0.4em]
                                                                                    $d\indices*{*_{X}^{A_{1g}^{}}_{X}^{A_{1g}^{}}}$ &  $11.1101$ &                                                                                     $d\indices*{*_{X}^{E_{g}^{}}_{X}^{E_{g}^{}}}$ &   $2.6728$ &                                                                                   $d\indices*{*_{X}^{A_{2u}^{}}_{X}^{A_{2u}^{}}}$ &  $15.9373$ &                                                                                 $d\indices*{*_{X}^{B_{1u}^{}}_{X}^{B_{1u}^{}}}$ &   $1.7560$ &                                                                                 $d\indices*{*_{X}^{E_{u}^{}}_{X}^{E_{u}^{}}}$ &   $7.9770$ \\[0.4em]
                                                     $d\indices*{*_{X}^{E_{u}^{}}_{X}^{\tensor*[^{1}]{\hspace{-0.2em}E}{}_{u}^{}}}$ & $- 2.2887$ &                   $d\indices*{*_{X}^{\tensor*[^{1}]{\hspace{-0.2em}E}{}_{u}^{}}_{X}^{\tensor*[^{1}]{\hspace{-0.2em}E}{}_{u}^{}}}$ &   $7.1280$ &                                                                               $d\indices*{*_{A}^{A_{1}^{}}_{\bar{A}}^{A_{1}^{}}}$ &  $13.5875$ &                                            $d\indices*{*_{A}^{A_{1}^{}}_{\bar{A}}^{\tensor*[^{1}]{\hspace{-0.2em}A}{}_{1}^{}}}$ & $- 2.0420$ &                                          $d\indices*{*_{A}^{A_{1}^{}}_{\bar{A}}^{\tensor*[^{2}]{\hspace{-0.2em}A}{}_{1}^{}}}$ &   $4.2516$ \\[0.4em]
              $d\indices*{*_{A}^{\tensor*[^{1}]{\hspace{-0.2em}A}{}_{1}^{}}_{\bar{A}}^{\tensor*[^{1}]{\hspace{-0.2em}A}{}_{1}^{}}}$ &   $8.7209$ &             $d\indices*{*_{A}^{\tensor*[^{1}]{\hspace{-0.2em}A}{}_{1}^{}}_{\bar{A}}^{\tensor*[^{2}]{\hspace{-0.2em}A}{}_{1}^{}}}$ & $- 0.5105$ &             $d\indices*{*_{A}^{\tensor*[^{2}]{\hspace{-0.2em}A}{}_{1}^{}}_{\bar{A}}^{\tensor*[^{2}]{\hspace{-0.2em}A}{}_{1}^{}}}$ &   $8.5115$ &                                                                             $d\indices*{*_{A}^{A_{2}^{}}_{\bar{A}}^{A_{2}^{}}}$ &   $5.2514$ &                                                                           $d\indices*{*_{A}^{B_{1}^{}}_{\bar{A}}^{B_{1}^{}}}$ &   $2.4534$ \\[0.4em]
                                               $d\indices*{*_{A}^{B_{1}^{}}_{\bar{A}}^{\tensor*[^{1}]{\hspace{-0.2em}B}{}_{1}^{}}}$ & $- 2.1815$ &                                              $d\indices*{*_{A}^{B_{1}^{}}_{\bar{A}}^{\tensor*[^{2}]{\hspace{-0.2em}B}{}_{1}^{}}}$ &   $0.2604$ &             $d\indices*{*_{A}^{\tensor*[^{1}]{\hspace{-0.2em}B}{}_{1}^{}}_{\bar{A}}^{\tensor*[^{1}]{\hspace{-0.2em}B}{}_{1}^{}}}$ &   $8.3852$ &           $d\indices*{*_{A}^{\tensor*[^{1}]{\hspace{-0.2em}B}{}_{1}^{}}_{\bar{A}}^{\tensor*[^{2}]{\hspace{-0.2em}B}{}_{1}^{}}}$ &   $0.8503$ &         $d\indices*{*_{A}^{\tensor*[^{2}]{\hspace{-0.2em}B}{}_{1}^{}}_{\bar{A}}^{\tensor*[^{2}]{\hspace{-0.2em}B}{}_{1}^{}}}$ &   $5.2992$ \\[0.4em]
                                                                                $d\indices*{*_{A}^{B_{2}^{}}_{\bar{A}}^{B_{2}^{}}}$ &   $5.9484$ &                                              $d\indices*{*_{A}^{B_{2}^{}}_{\bar{A}}^{\tensor*[^{1}]{\hspace{-0.2em}B}{}_{2}^{}}}$ & $- 3.0883$ &             $d\indices*{*_{A}^{\tensor*[^{1}]{\hspace{-0.2em}B}{}_{2}^{}}_{\bar{A}}^{\tensor*[^{1}]{\hspace{-0.2em}B}{}_{2}^{}}}$ &   $5.4462$ &                                                                             $d\indices*{*_{W}^{A_{1}^{}}_{\bar{W}}^{A_{1}^{}}}$ &  $10.1514$ &                                                                           $d\indices*{*_{W}^{B_{1}^{}}_{\bar{W}}^{B_{1}^{}}}$ &   $6.9876$ \\[0.4em]
                                               $d\indices*{*_{W}^{B_{1}^{}}_{\bar{W}}^{\tensor*[^{1}]{\hspace{-0.2em}B}{}_{1}^{}}}$ & $- 1.9748$ &             $d\indices*{*_{W}^{\tensor*[^{1}]{\hspace{-0.2em}B}{}_{1}^{}}_{\bar{W}}^{\tensor*[^{1}]{\hspace{-0.2em}B}{}_{1}^{}}}$ &   $8.8277$ &                                                                               $d\indices*{*_{W}^{A_{2}^{}}_{\bar{W}}^{A_{2}^{}}}$ &   $2.1341$ &                                                                             $d\indices*{*_{W}^{B_{2}^{}}_{\bar{W}}^{B_{2}^{}}}$ &   $2.2791$ &                                                                             $d\indices*{*_{W}^{E_{}^{}}_{\bar{W}}^{E_{}^{}}}$ &  $10.6246$ \\[0.4em]
                                                 $d\indices*{*_{W}^{E_{}^{}}_{\bar{W}}^{\tensor*[^{1}]{\hspace{-0.2em}E}{}_{}^{}}}$ &   $0.8281$ &               $d\indices*{*_{W}^{\tensor*[^{1}]{\hspace{-0.2em}E}{}_{}^{}}_{\bar{W}}^{\tensor*[^{1}]{\hspace{-0.2em}E}{}_{}^{}}}$ &   $5.5497$ &                                                                   $d\indices*{*_{\Lambda}^{A_{1}^{}}_{\bar{\Lambda}}^{A_{1}^{}}}$ &  $15.8491$ &                                $d\indices*{*_{\Lambda}^{A_{1}^{}}_{\bar{\Lambda}}^{\tensor*[^{1}]{\hspace{-0.2em}A}{}_{1}^{}}}$ & $- 8.4452$ &                              $d\indices*{*_{\Lambda}^{A_{1}^{}}_{\bar{\Lambda}}^{\tensor*[^{2}]{\hspace{-0.2em}A}{}_{1}^{}}}$ &   $1.5504$ \\[0.4em]
  $d\indices*{*_{\Lambda}^{\tensor*[^{1}]{\hspace{-0.2em}A}{}_{1}^{}}_{\bar{\Lambda}}^{\tensor*[^{1}]{\hspace{-0.2em}A}{}_{1}^{}}}$ &   $8.3371$ & $d\indices*{*_{\Lambda}^{\tensor*[^{1}]{\hspace{-0.2em}A}{}_{1}^{}}_{\bar{\Lambda}}^{\tensor*[^{2}]{\hspace{-0.2em}A}{}_{1}^{}}}$ & $- 0.0308$ & $d\indices*{*_{\Lambda}^{\tensor*[^{2}]{\hspace{-0.2em}A}{}_{1}^{}}_{\bar{\Lambda}}^{\tensor*[^{2}]{\hspace{-0.2em}A}{}_{1}^{}}}$ &   $7.2777$ &                                                                   $d\indices*{*_{\Lambda}^{E_{}^{}}_{\bar{\Lambda}}^{E_{}^{}}}$ &   $5.7023$ &                                $d\indices*{*_{\Lambda}^{E_{}^{}}_{\bar{\Lambda}}^{\tensor*[^{1}]{\hspace{-0.2em}E}{}_{}^{}}}$ & $- 2.7351$ \\[0.4em]
                                     $d\indices*{*_{\Lambda}^{E_{}^{}}_{\bar{\Lambda}}^{\tensor*[^{2}]{\hspace{-0.2em}E}{}_{}^{}}}$ & $- 2.1881$ &   $d\indices*{*_{\Lambda}^{\tensor*[^{1}]{\hspace{-0.2em}E}{}_{}^{}}_{\bar{\Lambda}}^{\tensor*[^{1}]{\hspace{-0.2em}E}{}_{}^{}}}$ &   $4.3342$ &   $d\indices*{*_{\Lambda}^{\tensor*[^{1}]{\hspace{-0.2em}E}{}_{}^{}}_{\bar{\Lambda}}^{\tensor*[^{2}]{\hspace{-0.2em}E}{}_{}^{}}}$ & $- 1.4310$ & $d\indices*{*_{\Lambda}^{\tensor*[^{2}]{\hspace{-0.2em}E}{}_{}^{}}_{\bar{\Lambda}}^{\tensor*[^{2}]{\hspace{-0.2em}E}{}_{}^{}}}$ &   $6.4073$ &                                                                             $d\indices*{*_{B}^{A_{}^{}}_{\bar{B}}^{A_{}^{}}}$ &  $12.4558$ \\[0.4em]
                                                 $d\indices*{*_{B}^{A_{}^{}}_{\bar{B}}^{\tensor*[^{1}]{\hspace{-0.2em}A}{}_{}^{}}}$ & $- 3.1695$ &                                                $d\indices*{*_{B}^{A_{}^{}}_{\bar{B}}^{\tensor*[^{2}]{\hspace{-0.2em}A}{}_{}^{}}}$ & $- 2.2244$ &                                                $d\indices*{*_{B}^{A_{}^{}}_{\bar{B}}^{\tensor*[^{3}]{\hspace{-0.2em}A}{}_{}^{}}}$ & $- 0.0641$ &                                              $d\indices*{*_{B}^{A_{}^{}}_{\bar{B}}^{\tensor*[^{4}]{\hspace{-0.2em}A}{}_{}^{}}}$ & $- 0.6986$ &                                            $d\indices*{*_{B}^{A_{}^{}}_{\bar{B}}^{\tensor*[^{5}]{\hspace{-0.2em}A}{}_{}^{}}}$ &   $0.5838$ \\[0.4em]
                $d\indices*{*_{B}^{\tensor*[^{1}]{\hspace{-0.2em}A}{}_{}^{}}_{\bar{B}}^{\tensor*[^{1}]{\hspace{-0.2em}A}{}_{}^{}}}$ &   $8.5334$ &               $d\indices*{*_{B}^{\tensor*[^{1}]{\hspace{-0.2em}A}{}_{}^{}}_{\bar{B}}^{\tensor*[^{2}]{\hspace{-0.2em}A}{}_{}^{}}}$ & $- 0.3670$ &               $d\indices*{*_{B}^{\tensor*[^{1}]{\hspace{-0.2em}A}{}_{}^{}}_{\bar{B}}^{\tensor*[^{3}]{\hspace{-0.2em}A}{}_{}^{}}}$ &   $0.5734$ &             $d\indices*{*_{B}^{\tensor*[^{1}]{\hspace{-0.2em}A}{}_{}^{}}_{\bar{B}}^{\tensor*[^{4}]{\hspace{-0.2em}A}{}_{}^{}}}$ &   $0.6925$ &           $d\indices*{*_{B}^{\tensor*[^{1}]{\hspace{-0.2em}A}{}_{}^{}}_{\bar{B}}^{\tensor*[^{5}]{\hspace{-0.2em}A}{}_{}^{}}}$ & $- 2.4602$ \\[0.4em]
                $d\indices*{*_{B}^{\tensor*[^{2}]{\hspace{-0.2em}A}{}_{}^{}}_{\bar{B}}^{\tensor*[^{2}]{\hspace{-0.2em}A}{}_{}^{}}}$ &   $9.1756$ &               $d\indices*{*_{B}^{\tensor*[^{2}]{\hspace{-0.2em}A}{}_{}^{}}_{\bar{B}}^{\tensor*[^{3}]{\hspace{-0.2em}A}{}_{}^{}}}$ &   $1.1844$ &               $d\indices*{*_{B}^{\tensor*[^{2}]{\hspace{-0.2em}A}{}_{}^{}}_{\bar{B}}^{\tensor*[^{4}]{\hspace{-0.2em}A}{}_{}^{}}}$ &   $1.1655$ &             $d\indices*{*_{B}^{\tensor*[^{2}]{\hspace{-0.2em}A}{}_{}^{}}_{\bar{B}}^{\tensor*[^{5}]{\hspace{-0.2em}A}{}_{}^{}}}$ &   $0.1551$ &           $d\indices*{*_{B}^{\tensor*[^{3}]{\hspace{-0.2em}A}{}_{}^{}}_{\bar{B}}^{\tensor*[^{3}]{\hspace{-0.2em}A}{}_{}^{}}}$ &   $4.3134$ \\[0.4em]
                $d\indices*{*_{B}^{\tensor*[^{3}]{\hspace{-0.2em}A}{}_{}^{}}_{\bar{B}}^{\tensor*[^{4}]{\hspace{-0.2em}A}{}_{}^{}}}$ & $- 1.4677$ &               $d\indices*{*_{B}^{\tensor*[^{3}]{\hspace{-0.2em}A}{}_{}^{}}_{\bar{B}}^{\tensor*[^{5}]{\hspace{-0.2em}A}{}_{}^{}}}$ & $- 0.1277$ &               $d\indices*{*_{B}^{\tensor*[^{4}]{\hspace{-0.2em}A}{}_{}^{}}_{\bar{B}}^{\tensor*[^{4}]{\hspace{-0.2em}A}{}_{}^{}}}$ &   $7.4591$ &             $d\indices*{*_{B}^{\tensor*[^{4}]{\hspace{-0.2em}A}{}_{}^{}}_{\bar{B}}^{\tensor*[^{5}]{\hspace{-0.2em}A}{}_{}^{}}}$ & $- 0.4298$ &           $d\indices*{*_{B}^{\tensor*[^{5}]{\hspace{-0.2em}A}{}_{}^{}}_{\bar{B}}^{\tensor*[^{5}]{\hspace{-0.2em}A}{}_{}^{}}}$ &   $4.0646$ \\[0.4em]
                                                                                  $d\indices*{*_{B}^{B_{}^{}}_{\bar{B}}^{B_{}^{}}}$ &   $6.2854$ &                                                $d\indices*{*_{B}^{B_{}^{}}_{\bar{B}}^{\tensor*[^{1}]{\hspace{-0.2em}B}{}_{}^{}}}$ & $- 2.2101$ &                                                $d\indices*{*_{B}^{B_{}^{}}_{\bar{B}}^{\tensor*[^{2}]{\hspace{-0.2em}B}{}_{}^{}}}$ & $- 1.6289$ &             $d\indices*{*_{B}^{\tensor*[^{1}]{\hspace{-0.2em}B}{}_{}^{}}_{\bar{B}}^{\tensor*[^{1}]{\hspace{-0.2em}B}{}_{}^{}}}$ &   $3.7678$ &           $d\indices*{*_{B}^{\tensor*[^{1}]{\hspace{-0.2em}B}{}_{}^{}}_{\bar{B}}^{\tensor*[^{2}]{\hspace{-0.2em}B}{}_{}^{}}}$ &   $1.1154$ \\[0.4em]
                $d\indices*{*_{B}^{\tensor*[^{2}]{\hspace{-0.2em}B}{}_{}^{}}_{\bar{B}}^{\tensor*[^{2}]{\hspace{-0.2em}B}{}_{}^{}}}$ &   $6.8078$ &                                                                     $d\indices*{*_{\Delta}^{A_{1}^{}}_{\bar{\Delta}}^{A_{1}^{}}}$ &  $16.3126$ &                                    $d\indices*{*_{\Delta}^{A_{1}^{}}_{\bar{\Delta}}^{\tensor*[^{1}]{\hspace{-0.2em}A}{}_{1}^{}}}$ & $- 8.1932$ & $d\indices*{*_{\Delta}^{\tensor*[^{1}]{\hspace{-0.2em}A}{}_{1}^{}}_{\bar{\Delta}}^{\tensor*[^{1}]{\hspace{-0.2em}A}{}_{1}^{}}}$ &   $9.7413$ &                                                                 $d\indices*{*_{\Delta}^{B_{2}^{}}_{\bar{\Delta}}^{B_{2}^{}}}$ &   $5.0054$ \\[0.4em]
                                                                        $d\indices*{*_{\Delta}^{E_{}^{}}_{\bar{\Delta}}^{E_{}^{}}}$ &   $6.2883$ &                                      $d\indices*{*_{\Delta}^{E_{}^{}}_{\bar{\Delta}}^{\tensor*[^{1}]{\hspace{-0.2em}E}{}_{}^{}}}$ &   $2.7609$ &                                      $d\indices*{*_{\Delta}^{E_{}^{}}_{\bar{\Delta}}^{\tensor*[^{2}]{\hspace{-0.2em}E}{}_{}^{}}}$ & $- 1.6287$ &   $d\indices*{*_{\Delta}^{\tensor*[^{1}]{\hspace{-0.2em}E}{}_{}^{}}_{\bar{\Delta}}^{\tensor*[^{1}]{\hspace{-0.2em}E}{}_{}^{}}}$ &   $2.7002$ & $d\indices*{*_{\Delta}^{\tensor*[^{1}]{\hspace{-0.2em}E}{}_{}^{}}_{\bar{\Delta}}^{\tensor*[^{2}]{\hspace{-0.2em}E}{}_{}^{}}}$ &   $0.3821$ \\[0.4em]
      $d\indices*{*_{\Delta}^{\tensor*[^{2}]{\hspace{-0.2em}E}{}_{}^{}}_{\bar{\Delta}}^{\tensor*[^{2}]{\hspace{-0.2em}E}{}_{}^{}}}$ &   $7.8346$ &                                                                                                                                   &            &                                                                                                                                   &            &                                                                                                                                 &            &                                                                                                                               &             \\
%
%
%

\end{longtable*}

\begin{longtable*}{cccccccccc}
\caption{All the third order irreducible derivatives for CaF$_2$ at the fully relaxed volume.  All results were computed using the BID approach, $\hat{\mathbf{S}}_{O}$ supercell, and SCAN functional.  The units are eV/\AA$^3$.\label{table:caf2_3_0}} \\
\hline\hline 
Derivative & Value & Derivative & Value & Derivative & Value & Derivative & Value & Derivative & Value\\
\hline\\[-0.9em]
\endfirsthead
\hline\hline
Derivative & Value & Derivative & Value & Derivative & Value & Derivative & Value  & Derivative & Value\\
\hline\\[-0.9em]
\endhead
\hline\hline
\endfoot
                                                                                                                 $d\indices*{*_{\Gamma}^{T_{2g}^{}}_{\Gamma}^{T_{2g}^{}}_{\Gamma}^{T_{2g}^{}}}$ & $- 64.4191$ &                                                                                                                $d\indices*{*_{\Gamma}^{T_{1u}^{}}_{\Gamma}^{T_{1u}^{}}_{\Gamma}^{T_{2g}^{}}}$ & $- 50.5341$ &                                                                                                                     $d\indices*{*_{X}^{A_{1g}^{}}_{\bar{A}}^{A_{1}^{}}_{\bar{A}}^{A_{1}^{}}}$ &   $17.1481$ &                                                                                                                      $d\indices*{*_{X}^{E_{u}^{}}_{\bar{A}}^{A_{1}^{}}_{\bar{A}}^{A_{1}^{}}}$ &    $0.2890$ &                                                                                     $d\indices*{*_{X}^{\tensor*[^{1}]{\hspace{-0.2em}E}{}_{u}^{}}_{\bar{A}}^{A_{1}^{}}_{\bar{A}}^{A_{1}^{}}}$ &  $- 4.5606$ \\[0.4em]
                                                                                     $d\indices*{*_{X}^{A_{1g}^{}}_{\bar{A}}^{A_{1}^{}}_{\bar{A}}^{\tensor*[^{1}]{\hspace{-0.2em}A}{}_{1}^{}}}$ &  $- 7.8596$ &                                                                                     $d\indices*{*_{X}^{E_{u}^{}}_{\bar{A}}^{A_{1}^{}}_{\bar{A}}^{\tensor*[^{1}]{\hspace{-0.2em}A}{}_{1}^{}}}$ &  $- 1.8020$ &                                                    $d\indices*{*_{X}^{\tensor*[^{1}]{\hspace{-0.2em}E}{}_{u}^{}}_{\bar{A}}^{A_{1}^{}}_{\bar{A}}^{\tensor*[^{1}]{\hspace{-0.2em}A}{}_{1}^{}}}$ &    $0.0510$ &                                                                                    $d\indices*{*_{X}^{A_{1g}^{}}_{\bar{A}}^{A_{1}^{}}_{\bar{A}}^{\tensor*[^{2}]{\hspace{-0.2em}A}{}_{1}^{}}}$ &  $- 0.0628$ &                                                                                     $d\indices*{*_{X}^{E_{u}^{}}_{\bar{A}}^{A_{1}^{}}_{\bar{A}}^{\tensor*[^{2}]{\hspace{-0.2em}A}{}_{1}^{}}}$ &   $10.6578$ \\[0.4em]
                                                     $d\indices*{*_{X}^{\tensor*[^{1}]{\hspace{-0.2em}E}{}_{u}^{}}_{\bar{A}}^{A_{1}^{}}_{\bar{A}}^{\tensor*[^{2}]{\hspace{-0.2em}A}{}_{1}^{}}}$ & $- 17.1636$ &                                                                                                                      $d\indices*{*_{X}^{E_{g}^{}}_{\bar{A}}^{A_{1}^{}}_{\bar{A}}^{A_{2}^{}}}$ &    $0.9989$ &                                                                                                                      $d\indices*{*_{X}^{E_{g}^{}}_{\bar{A}}^{A_{1}^{}}_{\bar{A}}^{B_{1}^{}}}$ &    $0.7063$ &                                                                                                                     $d\indices*{*_{X}^{A_{2u}^{}}_{\bar{A}}^{A_{1}^{}}_{\bar{A}}^{B_{1}^{}}}$ &   $13.3452$ &                                                                                                                     $d\indices*{*_{X}^{B_{1u}^{}}_{\bar{A}}^{A_{1}^{}}_{\bar{A}}^{B_{1}^{}}}$ &    $8.4931$ \\[0.4em]
                                                                                      $d\indices*{*_{X}^{E_{g}^{}}_{\bar{A}}^{A_{1}^{}}_{\bar{A}}^{\tensor*[^{1}]{\hspace{-0.2em}B}{}_{1}^{}}}$ &   $12.8477$ &                                                                                    $d\indices*{*_{X}^{A_{2u}^{}}_{\bar{A}}^{A_{1}^{}}_{\bar{A}}^{\tensor*[^{1}]{\hspace{-0.2em}B}{}_{1}^{}}}$ &    $6.7236$ &                                                                                    $d\indices*{*_{X}^{B_{1u}^{}}_{\bar{A}}^{A_{1}^{}}_{\bar{A}}^{\tensor*[^{1}]{\hspace{-0.2em}B}{}_{1}^{}}}$ &    $0.4564$ &                                                                                     $d\indices*{*_{X}^{E_{g}^{}}_{\bar{A}}^{A_{1}^{}}_{\bar{A}}^{\tensor*[^{2}]{\hspace{-0.2em}B}{}_{1}^{}}}$ &   $12.3939$ &                                                                                    $d\indices*{*_{X}^{A_{2u}^{}}_{\bar{A}}^{A_{1}^{}}_{\bar{A}}^{\tensor*[^{2}]{\hspace{-0.2em}B}{}_{1}^{}}}$ &    $0.3806$ \\[0.4em]
                                                                                     $d\indices*{*_{X}^{B_{1u}^{}}_{\bar{A}}^{A_{1}^{}}_{\bar{A}}^{\tensor*[^{2}]{\hspace{-0.2em}B}{}_{1}^{}}}$ &  $- 2.0362$ &                                                                                                                      $d\indices*{*_{X}^{E_{u}^{}}_{\bar{A}}^{A_{1}^{}}_{\bar{A}}^{B_{2}^{}}}$ &    $0.4795$ &                                                                                     $d\indices*{*_{X}^{\tensor*[^{1}]{\hspace{-0.2em}E}{}_{u}^{}}_{\bar{A}}^{A_{1}^{}}_{\bar{A}}^{B_{2}^{}}}$ &  $- 2.5452$ &                                                                                     $d\indices*{*_{X}^{E_{u}^{}}_{\bar{A}}^{A_{1}^{}}_{\bar{A}}^{\tensor*[^{1}]{\hspace{-0.2em}B}{}_{2}^{}}}$ &  $- 0.1799$ &                                                    $d\indices*{*_{X}^{\tensor*[^{1}]{\hspace{-0.2em}E}{}_{u}^{}}_{\bar{A}}^{A_{1}^{}}_{\bar{A}}^{\tensor*[^{1}]{\hspace{-0.2em}B}{}_{2}^{}}}$ &    $1.7827$ \\[0.4em]
                                                    $d\indices*{*_{X}^{A_{1g}^{}}_{\bar{A}}^{\tensor*[^{1}]{\hspace{-0.2em}A}{}_{1}^{}}_{\bar{A}}^{\tensor*[^{1}]{\hspace{-0.2em}A}{}_{1}^{}}}$ &  $- 1.1111$ &                                                    $d\indices*{*_{X}^{E_{u}^{}}_{\bar{A}}^{\tensor*[^{1}]{\hspace{-0.2em}A}{}_{1}^{}}_{\bar{A}}^{\tensor*[^{1}]{\hspace{-0.2em}A}{}_{1}^{}}}$ &    $1.4395$ &                   $d\indices*{*_{X}^{\tensor*[^{1}]{\hspace{-0.2em}E}{}_{u}^{}}_{\bar{A}}^{\tensor*[^{1}]{\hspace{-0.2em}A}{}_{1}^{}}_{\bar{A}}^{\tensor*[^{1}]{\hspace{-0.2em}A}{}_{1}^{}}}$ &  $- 1.6452$ &                                                   $d\indices*{*_{X}^{A_{1g}^{}}_{\bar{A}}^{\tensor*[^{1}]{\hspace{-0.2em}A}{}_{1}^{}}_{\bar{A}}^{\tensor*[^{2}]{\hspace{-0.2em}A}{}_{1}^{}}}$ &  $- 0.7719$ &                                                    $d\indices*{*_{X}^{E_{u}^{}}_{\bar{A}}^{\tensor*[^{1}]{\hspace{-0.2em}A}{}_{1}^{}}_{\bar{A}}^{\tensor*[^{2}]{\hspace{-0.2em}A}{}_{1}^{}}}$ &  $- 6.1235$ \\[0.4em]
                    $d\indices*{*_{X}^{\tensor*[^{1}]{\hspace{-0.2em}E}{}_{u}^{}}_{\bar{A}}^{\tensor*[^{1}]{\hspace{-0.2em}A}{}_{1}^{}}_{\bar{A}}^{\tensor*[^{2}]{\hspace{-0.2em}A}{}_{1}^{}}}$ &    $7.4674$ &                                                                                     $d\indices*{*_{X}^{E_{g}^{}}_{\bar{A}}^{\tensor*[^{1}]{\hspace{-0.2em}A}{}_{1}^{}}_{\bar{A}}^{A_{2}^{}}}$ &  $- 7.2149$ &                                                                                     $d\indices*{*_{X}^{E_{g}^{}}_{\bar{A}}^{\tensor*[^{1}]{\hspace{-0.2em}A}{}_{1}^{}}_{\bar{A}}^{B_{1}^{}}}$ &  $- 1.6754$ &                                                                                    $d\indices*{*_{X}^{A_{2u}^{}}_{\bar{A}}^{\tensor*[^{1}]{\hspace{-0.2em}A}{}_{1}^{}}_{\bar{A}}^{B_{1}^{}}}$ & $- 10.4595$ &                                                                                    $d\indices*{*_{X}^{B_{1u}^{}}_{\bar{A}}^{\tensor*[^{1}]{\hspace{-0.2em}A}{}_{1}^{}}_{\bar{A}}^{B_{1}^{}}}$ &    $0.6680$ \\[0.4em]
                                                     $d\indices*{*_{X}^{E_{g}^{}}_{\bar{A}}^{\tensor*[^{1}]{\hspace{-0.2em}A}{}_{1}^{}}_{\bar{A}}^{\tensor*[^{1}]{\hspace{-0.2em}B}{}_{1}^{}}}$ &  $- 6.1592$ &                                                   $d\indices*{*_{X}^{A_{2u}^{}}_{\bar{A}}^{\tensor*[^{1}]{\hspace{-0.2em}A}{}_{1}^{}}_{\bar{A}}^{\tensor*[^{1}]{\hspace{-0.2em}B}{}_{1}^{}}}$ &    $4.0347$ &                                                   $d\indices*{*_{X}^{B_{1u}^{}}_{\bar{A}}^{\tensor*[^{1}]{\hspace{-0.2em}A}{}_{1}^{}}_{\bar{A}}^{\tensor*[^{1}]{\hspace{-0.2em}B}{}_{1}^{}}}$ &  $- 4.6676$ &                                                    $d\indices*{*_{X}^{E_{g}^{}}_{\bar{A}}^{\tensor*[^{1}]{\hspace{-0.2em}A}{}_{1}^{}}_{\bar{A}}^{\tensor*[^{2}]{\hspace{-0.2em}B}{}_{1}^{}}}$ &  $- 6.8507$ &                                                   $d\indices*{*_{X}^{A_{2u}^{}}_{\bar{A}}^{\tensor*[^{1}]{\hspace{-0.2em}A}{}_{1}^{}}_{\bar{A}}^{\tensor*[^{2}]{\hspace{-0.2em}B}{}_{1}^{}}}$ &    $0.9071$ \\[0.4em]
                                                    $d\indices*{*_{X}^{B_{1u}^{}}_{\bar{A}}^{\tensor*[^{1}]{\hspace{-0.2em}A}{}_{1}^{}}_{\bar{A}}^{\tensor*[^{2}]{\hspace{-0.2em}B}{}_{1}^{}}}$ &    $3.7090$ &                                                                                     $d\indices*{*_{X}^{E_{u}^{}}_{\bar{A}}^{\tensor*[^{1}]{\hspace{-0.2em}A}{}_{1}^{}}_{\bar{A}}^{B_{2}^{}}}$ &    $6.6677$ &                                                    $d\indices*{*_{X}^{\tensor*[^{1}]{\hspace{-0.2em}E}{}_{u}^{}}_{\bar{A}}^{\tensor*[^{1}]{\hspace{-0.2em}A}{}_{1}^{}}_{\bar{A}}^{B_{2}^{}}}$ & $- 10.0459$ &                                                    $d\indices*{*_{X}^{E_{u}^{}}_{\bar{A}}^{\tensor*[^{1}]{\hspace{-0.2em}A}{}_{1}^{}}_{\bar{A}}^{\tensor*[^{1}]{\hspace{-0.2em}B}{}_{2}^{}}}$ &  $- 5.8458$ &                   $d\indices*{*_{X}^{\tensor*[^{1}]{\hspace{-0.2em}E}{}_{u}^{}}_{\bar{A}}^{\tensor*[^{1}]{\hspace{-0.2em}A}{}_{1}^{}}_{\bar{A}}^{\tensor*[^{1}]{\hspace{-0.2em}B}{}_{2}^{}}}$ &    $6.2296$ \\[0.4em]
                                                    $d\indices*{*_{X}^{A_{1g}^{}}_{\bar{A}}^{\tensor*[^{2}]{\hspace{-0.2em}A}{}_{1}^{}}_{\bar{A}}^{\tensor*[^{2}]{\hspace{-0.2em}A}{}_{1}^{}}}$ &  $- 8.2551$ &                                                    $d\indices*{*_{X}^{E_{u}^{}}_{\bar{A}}^{\tensor*[^{2}]{\hspace{-0.2em}A}{}_{1}^{}}_{\bar{A}}^{\tensor*[^{2}]{\hspace{-0.2em}A}{}_{1}^{}}}$ &  $- 0.3846$ &                   $d\indices*{*_{X}^{\tensor*[^{1}]{\hspace{-0.2em}E}{}_{u}^{}}_{\bar{A}}^{\tensor*[^{2}]{\hspace{-0.2em}A}{}_{1}^{}}_{\bar{A}}^{\tensor*[^{2}]{\hspace{-0.2em}A}{}_{1}^{}}}$ &  $- 0.8417$ &                                                                                     $d\indices*{*_{X}^{E_{g}^{}}_{\bar{A}}^{\tensor*[^{2}]{\hspace{-0.2em}A}{}_{1}^{}}_{\bar{A}}^{A_{2}^{}}}$ &    $1.0560$ &                                                                                     $d\indices*{*_{X}^{E_{g}^{}}_{\bar{A}}^{\tensor*[^{2}]{\hspace{-0.2em}A}{}_{1}^{}}_{\bar{A}}^{B_{1}^{}}}$ &    $8.3289$ \\[0.4em]
                                                                                     $d\indices*{*_{X}^{A_{2u}^{}}_{\bar{A}}^{\tensor*[^{2}]{\hspace{-0.2em}A}{}_{1}^{}}_{\bar{A}}^{B_{1}^{}}}$ &    $0.4388$ &                                                                                    $d\indices*{*_{X}^{B_{1u}^{}}_{\bar{A}}^{\tensor*[^{2}]{\hspace{-0.2em}A}{}_{1}^{}}_{\bar{A}}^{B_{1}^{}}}$ &    $1.9792$ &                                                    $d\indices*{*_{X}^{E_{g}^{}}_{\bar{A}}^{\tensor*[^{2}]{\hspace{-0.2em}A}{}_{1}^{}}_{\bar{A}}^{\tensor*[^{1}]{\hspace{-0.2em}B}{}_{1}^{}}}$ &    $0.7308$ &                                                   $d\indices*{*_{X}^{A_{2u}^{}}_{\bar{A}}^{\tensor*[^{2}]{\hspace{-0.2em}A}{}_{1}^{}}_{\bar{A}}^{\tensor*[^{1}]{\hspace{-0.2em}B}{}_{1}^{}}}$ & $- 11.1259$ &                                                   $d\indices*{*_{X}^{B_{1u}^{}}_{\bar{A}}^{\tensor*[^{2}]{\hspace{-0.2em}A}{}_{1}^{}}_{\bar{A}}^{\tensor*[^{1}]{\hspace{-0.2em}B}{}_{1}^{}}}$ &  $- 9.1157$ \\[0.4em]
                                                     $d\indices*{*_{X}^{E_{g}^{}}_{\bar{A}}^{\tensor*[^{2}]{\hspace{-0.2em}A}{}_{1}^{}}_{\bar{A}}^{\tensor*[^{2}]{\hspace{-0.2em}B}{}_{1}^{}}}$ &    $1.6755$ &                                                   $d\indices*{*_{X}^{A_{2u}^{}}_{\bar{A}}^{\tensor*[^{2}]{\hspace{-0.2em}A}{}_{1}^{}}_{\bar{A}}^{\tensor*[^{2}]{\hspace{-0.2em}B}{}_{1}^{}}}$ & $- 11.3920$ &                                                   $d\indices*{*_{X}^{B_{1u}^{}}_{\bar{A}}^{\tensor*[^{2}]{\hspace{-0.2em}A}{}_{1}^{}}_{\bar{A}}^{\tensor*[^{2}]{\hspace{-0.2em}B}{}_{1}^{}}}$ &  $- 9.5286$ &                                                                                     $d\indices*{*_{X}^{E_{u}^{}}_{\bar{A}}^{\tensor*[^{2}]{\hspace{-0.2em}A}{}_{1}^{}}_{\bar{A}}^{B_{2}^{}}}$ &  $- 2.1299$ &                                                    $d\indices*{*_{X}^{\tensor*[^{1}]{\hspace{-0.2em}E}{}_{u}^{}}_{\bar{A}}^{\tensor*[^{2}]{\hspace{-0.2em}A}{}_{1}^{}}_{\bar{A}}^{B_{2}^{}}}$ &  $- 3.0902$ \\[0.4em]
                                                     $d\indices*{*_{X}^{E_{u}^{}}_{\bar{A}}^{\tensor*[^{2}]{\hspace{-0.2em}A}{}_{1}^{}}_{\bar{A}}^{\tensor*[^{1}]{\hspace{-0.2em}B}{}_{2}^{}}}$ &    $5.6261$ &                   $d\indices*{*_{X}^{\tensor*[^{1}]{\hspace{-0.2em}E}{}_{u}^{}}_{\bar{A}}^{\tensor*[^{2}]{\hspace{-0.2em}A}{}_{1}^{}}_{\bar{A}}^{\tensor*[^{1}]{\hspace{-0.2em}B}{}_{2}^{}}}$ &    $0.4506$ &                                                                                                                     $d\indices*{*_{X}^{A_{1g}^{}}_{\bar{A}}^{A_{2}^{}}_{\bar{A}}^{A_{2}^{}}}$ &    $8.7541$ &                                                                                                                      $d\indices*{*_{X}^{E_{u}^{}}_{\bar{A}}^{A_{2}^{}}_{\bar{A}}^{A_{2}^{}}}$ &    $6.6974$ &                                                                                     $d\indices*{*_{X}^{\tensor*[^{1}]{\hspace{-0.2em}E}{}_{u}^{}}_{\bar{A}}^{A_{2}^{}}_{\bar{A}}^{A_{2}^{}}}$ &  $- 0.0021$ \\[0.4em]
                                                                                                                       $d\indices*{*_{X}^{E_{u}^{}}_{\bar{A}}^{A_{2}^{}}_{\bar{A}}^{B_{1}^{}}}$ &    $6.3874$ &                                                                                     $d\indices*{*_{X}^{\tensor*[^{1}]{\hspace{-0.2em}E}{}_{u}^{}}_{\bar{A}}^{A_{2}^{}}_{\bar{A}}^{B_{1}^{}}}$ &  $- 7.1773$ &                                                                                     $d\indices*{*_{X}^{E_{u}^{}}_{\bar{A}}^{A_{2}^{}}_{\bar{A}}^{\tensor*[^{1}]{\hspace{-0.2em}B}{}_{1}^{}}}$ &  $- 7.2970$ &                                                    $d\indices*{*_{X}^{\tensor*[^{1}]{\hspace{-0.2em}E}{}_{u}^{}}_{\bar{A}}^{A_{2}^{}}_{\bar{A}}^{\tensor*[^{1}]{\hspace{-0.2em}B}{}_{1}^{}}}$ &    $9.3667$ &                                                                                     $d\indices*{*_{X}^{E_{u}^{}}_{\bar{A}}^{A_{2}^{}}_{\bar{A}}^{\tensor*[^{2}]{\hspace{-0.2em}B}{}_{1}^{}}}$ &  $- 5.4315$ \\[0.4em]
                                                     $d\indices*{*_{X}^{\tensor*[^{1}]{\hspace{-0.2em}E}{}_{u}^{}}_{\bar{A}}^{A_{2}^{}}_{\bar{A}}^{\tensor*[^{2}]{\hspace{-0.2em}B}{}_{1}^{}}}$ &  $- 1.2195$ &                                                                                                                      $d\indices*{*_{X}^{E_{g}^{}}_{\bar{A}}^{A_{2}^{}}_{\bar{A}}^{B_{2}^{}}}$ &  $- 0.3982$ &                                                                                                                     $d\indices*{*_{X}^{A_{2u}^{}}_{\bar{A}}^{A_{2}^{}}_{\bar{A}}^{B_{2}^{}}}$ &   $17.1782$ &                                                                                                                     $d\indices*{*_{X}^{B_{1u}^{}}_{\bar{A}}^{A_{2}^{}}_{\bar{A}}^{B_{2}^{}}}$ & $- 13.0666$ &                                                                                     $d\indices*{*_{X}^{E_{g}^{}}_{\bar{A}}^{A_{2}^{}}_{\bar{A}}^{\tensor*[^{1}]{\hspace{-0.2em}B}{}_{2}^{}}}$ &  $- 0.5302$ \\[0.4em]
                                                                                     $d\indices*{*_{X}^{A_{2u}^{}}_{\bar{A}}^{A_{2}^{}}_{\bar{A}}^{\tensor*[^{1}]{\hspace{-0.2em}B}{}_{2}^{}}}$ & $- 10.5151$ &                                                                                    $d\indices*{*_{X}^{B_{1u}^{}}_{\bar{A}}^{A_{2}^{}}_{\bar{A}}^{\tensor*[^{1}]{\hspace{-0.2em}B}{}_{2}^{}}}$ &    $9.1719$ &                                                                                                                     $d\indices*{*_{X}^{A_{1g}^{}}_{\bar{A}}^{B_{1}^{}}_{\bar{A}}^{B_{1}^{}}}$ &  $- 0.3415$ &                                                                                                                      $d\indices*{*_{X}^{E_{u}^{}}_{\bar{A}}^{B_{1}^{}}_{\bar{A}}^{B_{1}^{}}}$ &  $- 1.4778$ &                                                                                     $d\indices*{*_{X}^{\tensor*[^{1}]{\hspace{-0.2em}E}{}_{u}^{}}_{\bar{A}}^{B_{1}^{}}_{\bar{A}}^{B_{1}^{}}}$ &    $0.3416$ \\[0.4em]
                                                                                     $d\indices*{*_{X}^{A_{1g}^{}}_{\bar{A}}^{B_{1}^{}}_{\bar{A}}^{\tensor*[^{1}]{\hspace{-0.2em}B}{}_{1}^{}}}$ &    $2.0157$ &                                                                                     $d\indices*{*_{X}^{E_{u}^{}}_{\bar{A}}^{B_{1}^{}}_{\bar{A}}^{\tensor*[^{1}]{\hspace{-0.2em}B}{}_{1}^{}}}$ &    $4.3366$ &                                                    $d\indices*{*_{X}^{\tensor*[^{1}]{\hspace{-0.2em}E}{}_{u}^{}}_{\bar{A}}^{B_{1}^{}}_{\bar{A}}^{\tensor*[^{1}]{\hspace{-0.2em}B}{}_{1}^{}}}$ &  $- 7.4685$ &                                                                                    $d\indices*{*_{X}^{A_{1g}^{}}_{\bar{A}}^{B_{1}^{}}_{\bar{A}}^{\tensor*[^{2}]{\hspace{-0.2em}B}{}_{1}^{}}}$ &  $- 1.8540$ &                                                                                     $d\indices*{*_{X}^{E_{u}^{}}_{\bar{A}}^{B_{1}^{}}_{\bar{A}}^{\tensor*[^{2}]{\hspace{-0.2em}B}{}_{1}^{}}}$ &    $6.0218$ \\[0.4em]
                                                     $d\indices*{*_{X}^{\tensor*[^{1}]{\hspace{-0.2em}E}{}_{u}^{}}_{\bar{A}}^{B_{1}^{}}_{\bar{A}}^{\tensor*[^{2}]{\hspace{-0.2em}B}{}_{1}^{}}}$ &  $- 7.3675$ &                                                                                                                      $d\indices*{*_{X}^{E_{g}^{}}_{\bar{A}}^{B_{2}^{}}_{\bar{A}}^{B_{1}^{}}}$ &  $- 7.3664$ &                                                                                     $d\indices*{*_{X}^{E_{g}^{}}_{\bar{A}}^{\tensor*[^{1}]{\hspace{-0.2em}B}{}_{2}^{}}_{\bar{A}}^{B_{1}^{}}}$ &    $6.3358$ &                                                   $d\indices*{*_{X}^{A_{1g}^{}}_{\bar{A}}^{\tensor*[^{1}]{\hspace{-0.2em}B}{}_{1}^{}}_{\bar{A}}^{\tensor*[^{1}]{\hspace{-0.2em}B}{}_{1}^{}}}$ & $- 10.6891$ &                                                    $d\indices*{*_{X}^{E_{u}^{}}_{\bar{A}}^{\tensor*[^{1}]{\hspace{-0.2em}B}{}_{1}^{}}_{\bar{A}}^{\tensor*[^{1}]{\hspace{-0.2em}B}{}_{1}^{}}}$ &  $- 0.1116$ \\[0.4em]
                    $d\indices*{*_{X}^{\tensor*[^{1}]{\hspace{-0.2em}E}{}_{u}^{}}_{\bar{A}}^{\tensor*[^{1}]{\hspace{-0.2em}B}{}_{1}^{}}_{\bar{A}}^{\tensor*[^{1}]{\hspace{-0.2em}B}{}_{1}^{}}}$ &  $- 5.5166$ &                                                   $d\indices*{*_{X}^{A_{1g}^{}}_{\bar{A}}^{\tensor*[^{1}]{\hspace{-0.2em}B}{}_{1}^{}}_{\bar{A}}^{\tensor*[^{2}]{\hspace{-0.2em}B}{}_{1}^{}}}$ &  $- 6.9156$ &                                                    $d\indices*{*_{X}^{E_{u}^{}}_{\bar{A}}^{\tensor*[^{1}]{\hspace{-0.2em}B}{}_{1}^{}}_{\bar{A}}^{\tensor*[^{2}]{\hspace{-0.2em}B}{}_{1}^{}}}$ &  $- 0.7560$ &                   $d\indices*{*_{X}^{\tensor*[^{1}]{\hspace{-0.2em}E}{}_{u}^{}}_{\bar{A}}^{\tensor*[^{1}]{\hspace{-0.2em}B}{}_{1}^{}}_{\bar{A}}^{\tensor*[^{2}]{\hspace{-0.2em}B}{}_{1}^{}}}$ &    $0.3361$ &                                                                                     $d\indices*{*_{X}^{E_{g}^{}}_{\bar{A}}^{B_{2}^{}}_{\bar{A}}^{\tensor*[^{1}]{\hspace{-0.2em}B}{}_{1}^{}}}$ &   $11.0061$ \\[0.4em]
                                                     $d\indices*{*_{X}^{E_{g}^{}}_{\bar{A}}^{\tensor*[^{1}]{\hspace{-0.2em}B}{}_{2}^{}}_{\bar{A}}^{\tensor*[^{1}]{\hspace{-0.2em}B}{}_{1}^{}}}$ &  $- 5.8041$ &                                                   $d\indices*{*_{X}^{A_{1g}^{}}_{\bar{A}}^{\tensor*[^{2}]{\hspace{-0.2em}B}{}_{1}^{}}_{\bar{A}}^{\tensor*[^{2}]{\hspace{-0.2em}B}{}_{1}^{}}}$ &  $- 9.5020$ &                                                    $d\indices*{*_{X}^{E_{u}^{}}_{\bar{A}}^{\tensor*[^{2}]{\hspace{-0.2em}B}{}_{1}^{}}_{\bar{A}}^{\tensor*[^{2}]{\hspace{-0.2em}B}{}_{1}^{}}}$ &    $2.8561$ &                   $d\indices*{*_{X}^{\tensor*[^{1}]{\hspace{-0.2em}E}{}_{u}^{}}_{\bar{A}}^{\tensor*[^{2}]{\hspace{-0.2em}B}{}_{1}^{}}_{\bar{A}}^{\tensor*[^{2}]{\hspace{-0.2em}B}{}_{1}^{}}}$ &    $0.2217$ &                                                                                     $d\indices*{*_{X}^{E_{g}^{}}_{\bar{A}}^{B_{2}^{}}_{\bar{A}}^{\tensor*[^{2}]{\hspace{-0.2em}B}{}_{1}^{}}}$ &  $- 2.2155$ \\[0.4em]
                                                     $d\indices*{*_{X}^{E_{g}^{}}_{\bar{A}}^{\tensor*[^{1}]{\hspace{-0.2em}B}{}_{2}^{}}_{\bar{A}}^{\tensor*[^{2}]{\hspace{-0.2em}B}{}_{1}^{}}}$ &    $1.1445$ &                                                                                                                     $d\indices*{*_{X}^{A_{1g}^{}}_{\bar{A}}^{B_{2}^{}}_{\bar{A}}^{B_{2}^{}}}$ &    $9.5946$ &                                                                                                                      $d\indices*{*_{X}^{E_{u}^{}}_{\bar{A}}^{B_{2}^{}}_{\bar{A}}^{B_{2}^{}}}$ &    $0.5217$ &                                                                                     $d\indices*{*_{X}^{\tensor*[^{1}]{\hspace{-0.2em}E}{}_{u}^{}}_{\bar{A}}^{B_{2}^{}}_{\bar{A}}^{B_{2}^{}}}$ &  $- 0.3315$ &                                                                                    $d\indices*{*_{X}^{A_{1g}^{}}_{\bar{A}}^{B_{2}^{}}_{\bar{A}}^{\tensor*[^{1}]{\hspace{-0.2em}B}{}_{2}^{}}}$ &  $- 8.8277$ \\[0.4em]
                                                                                      $d\indices*{*_{X}^{E_{u}^{}}_{\bar{A}}^{B_{2}^{}}_{\bar{A}}^{\tensor*[^{1}]{\hspace{-0.2em}B}{}_{2}^{}}}$ &    $0.4438$ &                                                    $d\indices*{*_{X}^{\tensor*[^{1}]{\hspace{-0.2em}E}{}_{u}^{}}_{\bar{A}}^{B_{2}^{}}_{\bar{A}}^{\tensor*[^{1}]{\hspace{-0.2em}B}{}_{2}^{}}}$ &  $- 0.2755$ &                                                   $d\indices*{*_{X}^{A_{1g}^{}}_{\bar{A}}^{\tensor*[^{1}]{\hspace{-0.2em}B}{}_{2}^{}}_{\bar{A}}^{\tensor*[^{1}]{\hspace{-0.2em}B}{}_{2}^{}}}$ &   $10.3247$ &                                                    $d\indices*{*_{X}^{E_{u}^{}}_{\bar{A}}^{\tensor*[^{1}]{\hspace{-0.2em}B}{}_{2}^{}}_{\bar{A}}^{\tensor*[^{1}]{\hspace{-0.2em}B}{}_{2}^{}}}$ &  $- 6.5938$ &                   $d\indices*{*_{X}^{\tensor*[^{1}]{\hspace{-0.2em}E}{}_{u}^{}}_{\bar{A}}^{\tensor*[^{1}]{\hspace{-0.2em}B}{}_{2}^{}}_{\bar{A}}^{\tensor*[^{1}]{\hspace{-0.2em}B}{}_{2}^{}}}$ &  $- 0.4910$ \\[0.4em]
                                                                                                                             $d\indices*{*_{\Gamma}^{T_{2g}^{}}_{X}^{E_{g}^{}}_{X}^{E_{g}^{}}}$ &   $20.9195$ &                                                                                                                          $d\indices*{*_{\Gamma}^{T_{2g}^{}}_{X}^{B_{1u}^{}}_{X}^{A_{2u}^{}}}$ &   $11.8667$ &                                                                                                                            $d\indices*{*_{\Gamma}^{T_{2g}^{}}_{X}^{E_{u}^{}}_{X}^{E_{u}^{}}}$ & $- 13.2732$ &                                                                                           $d\indices*{*_{\Gamma}^{T_{2g}^{}}_{X}^{E_{u}^{}}_{X}^{\tensor*[^{1}]{\hspace{-0.2em}E}{}_{u}^{}}}$ &   $10.2409$ &                                                          $d\indices*{*_{\Gamma}^{T_{2g}^{}}_{X}^{\tensor*[^{1}]{\hspace{-0.2em}E}{}_{u}^{}}_{X}^{\tensor*[^{1}]{\hspace{-0.2em}E}{}_{u}^{}}}$ & $- 27.8611$ \\[0.4em]
                                                                                                                            $d\indices*{*_{\Gamma}^{T_{2g}^{}}_{X}^{A_{1g}^{}}_{X}^{E_{g}^{}}}$ & $- 11.5450$ &                                                                                                                           $d\indices*{*_{\Gamma}^{T_{2g}^{}}_{X}^{A_{2u}^{}}_{X}^{E_{u}^{}}}$ &   $10.3394$ &                                                                                          $d\indices*{*_{\Gamma}^{T_{2g}^{}}_{X}^{A_{2u}^{}}_{X}^{\tensor*[^{1}]{\hspace{-0.2em}E}{}_{u}^{}}}$ & $- 19.5053$ &                                                                                                                           $d\indices*{*_{\Gamma}^{T_{2g}^{}}_{X}^{B_{1u}^{}}_{X}^{E_{u}^{}}}$ &  $- 8.6407$ &                                                                                          $d\indices*{*_{\Gamma}^{T_{2g}^{}}_{X}^{B_{1u}^{}}_{X}^{\tensor*[^{1}]{\hspace{-0.2em}E}{}_{u}^{}}}$ &   $10.9421$ \\[0.4em]
                                                                                                                            $d\indices*{*_{\Gamma}^{T_{1u}^{}}_{X}^{A_{1g}^{}}_{X}^{E_{u}^{}}}$ & $- 15.6541$ &                                                                                          $d\indices*{*_{\Gamma}^{T_{1u}^{}}_{X}^{A_{1g}^{}}_{X}^{\tensor*[^{1}]{\hspace{-0.2em}E}{}_{u}^{}}}$ &   $17.8462$ &                                                                                                                           $d\indices*{*_{\Gamma}^{T_{1u}^{}}_{X}^{B_{1u}^{}}_{X}^{E_{g}^{}}}$ & $- 15.2877$ &                                                                                                                           $d\indices*{*_{\Gamma}^{T_{1u}^{}}_{X}^{E_{g}^{}}_{X}^{A_{2u}^{}}}$ &   $21.9969$ &                                                                                                                          $d\indices*{*_{\Gamma}^{T_{1u}^{}}_{X}^{A_{1g}^{}}_{X}^{A_{2u}^{}}}$ &   $19.4232$ \\[0.4em]
                                                                                                                             $d\indices*{*_{\Gamma}^{T_{1u}^{}}_{X}^{E_{g}^{}}_{X}^{E_{u}^{}}}$ & $- 13.2825$ &                                                                                           $d\indices*{*_{\Gamma}^{T_{1u}^{}}_{X}^{E_{g}^{}}_{X}^{\tensor*[^{1}]{\hspace{-0.2em}E}{}_{u}^{}}}$ &   $19.2363$ &                                                                                                                      $d\indices*{*_{\Gamma}^{T_{2g}^{}}_{A}^{A_{1}^{}}_{\bar{A}}^{A_{1}^{}}}$ &   $20.3346$ &                                                                                     $d\indices*{*_{\Gamma}^{T_{2g}^{}}_{A}^{A_{1}^{}}_{\bar{A}}^{\tensor*[^{1}]{\hspace{-0.2em}A}{}_{1}^{}}}$ &  $- 6.2875$ &                                                                                     $d\indices*{*_{\Gamma}^{T_{2g}^{}}_{A}^{A_{1}^{}}_{\bar{A}}^{\tensor*[^{2}]{\hspace{-0.2em}A}{}_{1}^{}}}$ &    $2.5564$ \\[0.4em]
                                                     $d\indices*{*_{\Gamma}^{T_{2g}^{}}_{A}^{\tensor*[^{1}]{\hspace{-0.2em}A}{}_{1}^{}}_{\bar{A}}^{\tensor*[^{1}]{\hspace{-0.2em}A}{}_{1}^{}}}$ &    $0.9174$ &                                                    $d\indices*{*_{\Gamma}^{T_{2g}^{}}_{A}^{\tensor*[^{1}]{\hspace{-0.2em}A}{}_{1}^{}}_{\bar{A}}^{\tensor*[^{2}]{\hspace{-0.2em}A}{}_{1}^{}}}$ &  $- 2.3305$ &                                                    $d\indices*{*_{\Gamma}^{T_{2g}^{}}_{A}^{\tensor*[^{2}]{\hspace{-0.2em}A}{}_{1}^{}}_{\bar{A}}^{\tensor*[^{2}]{\hspace{-0.2em}A}{}_{1}^{}}}$ &   $11.0079$ &                                                                                                                      $d\indices*{*_{\Gamma}^{T_{2g}^{}}_{A}^{A_{2}^{}}_{\bar{A}}^{A_{2}^{}}}$ & $- 10.2759$ &                                                                                                                      $d\indices*{*_{\Gamma}^{T_{2g}^{}}_{A}^{B_{1}^{}}_{\bar{A}}^{B_{1}^{}}}$ &  $- 1.6105$ \\[0.4em]
                                                                                      $d\indices*{*_{\Gamma}^{T_{2g}^{}}_{A}^{B_{1}^{}}_{\bar{A}}^{\tensor*[^{1}]{\hspace{-0.2em}B}{}_{1}^{}}}$ &    $5.0038$ &                                                                                     $d\indices*{*_{\Gamma}^{T_{2g}^{}}_{A}^{B_{1}^{}}_{\bar{A}}^{\tensor*[^{2}]{\hspace{-0.2em}B}{}_{1}^{}}}$ &  $- 4.4113$ &                                                    $d\indices*{*_{\Gamma}^{T_{2g}^{}}_{A}^{\tensor*[^{1}]{\hspace{-0.2em}B}{}_{1}^{}}_{\bar{A}}^{\tensor*[^{1}]{\hspace{-0.2em}B}{}_{1}^{}}}$ &    $1.1840$ &                                                    $d\indices*{*_{\Gamma}^{T_{2g}^{}}_{A}^{\tensor*[^{1}]{\hspace{-0.2em}B}{}_{1}^{}}_{\bar{A}}^{\tensor*[^{2}]{\hspace{-0.2em}B}{}_{1}^{}}}$ &    $6.2417$ &                                                    $d\indices*{*_{\Gamma}^{T_{2g}^{}}_{A}^{\tensor*[^{2}]{\hspace{-0.2em}B}{}_{1}^{}}_{\bar{A}}^{\tensor*[^{2}]{\hspace{-0.2em}B}{}_{1}^{}}}$ &    $8.0407$ \\[0.4em]
                                                                                                                       $d\indices*{*_{\Gamma}^{T_{2g}^{}}_{A}^{B_{2}^{}}_{\bar{A}}^{B_{2}^{}}}$ & $- 18.2706$ &                                                                                     $d\indices*{*_{\Gamma}^{T_{2g}^{}}_{A}^{B_{2}^{}}_{\bar{A}}^{\tensor*[^{1}]{\hspace{-0.2em}B}{}_{2}^{}}}$ &   $10.4297$ &                                                    $d\indices*{*_{\Gamma}^{T_{2g}^{}}_{A}^{\tensor*[^{1}]{\hspace{-0.2em}B}{}_{2}^{}}_{\bar{A}}^{\tensor*[^{1}]{\hspace{-0.2em}B}{}_{2}^{}}}$ &  $- 8.3232$ &                                                                                                                      $d\indices*{*_{\Gamma}^{T_{2g}^{}}_{A}^{A_{1}^{}}_{\bar{A}}^{A_{2}^{}}}$ &    $0.0764$ &                                                                                                                      $d\indices*{*_{\Gamma}^{T_{2g}^{}}_{A}^{A_{1}^{}}_{\bar{A}}^{B_{1}^{}}}$ &    $0.5353$ \\[0.4em]
                                                                                      $d\indices*{*_{\Gamma}^{T_{2g}^{}}_{A}^{A_{1}^{}}_{\bar{A}}^{\tensor*[^{1}]{\hspace{-0.2em}B}{}_{1}^{}}}$ & $- 14.8122$ &                                                                                     $d\indices*{*_{\Gamma}^{T_{2g}^{}}_{A}^{A_{1}^{}}_{\bar{A}}^{\tensor*[^{2}]{\hspace{-0.2em}B}{}_{1}^{}}}$ & $- 11.2975$ &                                                                                     $d\indices*{*_{\Gamma}^{T_{2g}^{}}_{A}^{\tensor*[^{1}]{\hspace{-0.2em}A}{}_{1}^{}}_{\bar{A}}^{A_{2}^{}}}$ &    $7.3660$ &                                                                                     $d\indices*{*_{\Gamma}^{T_{2g}^{}}_{A}^{\tensor*[^{1}]{\hspace{-0.2em}A}{}_{1}^{}}_{\bar{A}}^{B_{1}^{}}}$ &  $- 1.5259$ &                                                    $d\indices*{*_{\Gamma}^{T_{2g}^{}}_{A}^{\tensor*[^{1}]{\hspace{-0.2em}A}{}_{1}^{}}_{\bar{A}}^{\tensor*[^{1}]{\hspace{-0.2em}B}{}_{1}^{}}}$ &    $3.7734$ \\[0.4em]
                                                     $d\indices*{*_{\Gamma}^{T_{2g}^{}}_{A}^{\tensor*[^{1}]{\hspace{-0.2em}A}{}_{1}^{}}_{\bar{A}}^{\tensor*[^{2}]{\hspace{-0.2em}B}{}_{1}^{}}}$ &    $7.1488$ &                                                                                     $d\indices*{*_{\Gamma}^{T_{2g}^{}}_{A}^{\tensor*[^{2}]{\hspace{-0.2em}A}{}_{1}^{}}_{\bar{A}}^{A_{2}^{}}}$ &  $- 5.4573$ &                                                                                     $d\indices*{*_{\Gamma}^{T_{2g}^{}}_{A}^{\tensor*[^{2}]{\hspace{-0.2em}A}{}_{1}^{}}_{\bar{A}}^{B_{1}^{}}}$ &    $7.9851$ &                                                    $d\indices*{*_{\Gamma}^{T_{2g}^{}}_{A}^{\tensor*[^{2}]{\hspace{-0.2em}A}{}_{1}^{}}_{\bar{A}}^{\tensor*[^{1}]{\hspace{-0.2em}B}{}_{1}^{}}}$ &  $- 0.1178$ &                                                    $d\indices*{*_{\Gamma}^{T_{2g}^{}}_{A}^{\tensor*[^{2}]{\hspace{-0.2em}A}{}_{1}^{}}_{\bar{A}}^{\tensor*[^{2}]{\hspace{-0.2em}B}{}_{1}^{}}}$ &    $1.3923$ \\[0.4em]
                                                                                                                       $d\indices*{*_{\Gamma}^{T_{2g}^{}}_{A}^{A_{2}^{}}_{\bar{A}}^{B_{2}^{}}}$ &    $2.6007$ &                                                                                     $d\indices*{*_{\Gamma}^{T_{2g}^{}}_{A}^{A_{2}^{}}_{\bar{A}}^{\tensor*[^{1}]{\hspace{-0.2em}B}{}_{2}^{}}}$ &  $- 5.9591$ &                                                                                                                      $d\indices*{*_{\Gamma}^{T_{2g}^{}}_{A}^{B_{1}^{}}_{\bar{A}}^{B_{2}^{}}}$ &  $- 7.3570$ &                                                                                     $d\indices*{*_{\Gamma}^{T_{2g}^{}}_{A}^{B_{1}^{}}_{\bar{A}}^{\tensor*[^{1}]{\hspace{-0.2em}B}{}_{2}^{}}}$ &    $6.6007$ &                                                                                     $d\indices*{*_{\Gamma}^{T_{2g}^{}}_{A}^{\tensor*[^{1}]{\hspace{-0.2em}B}{}_{1}^{}}_{\bar{A}}^{B_{2}^{}}}$ &    $9.8166$ \\[0.4em]
                                                     $d\indices*{*_{\Gamma}^{T_{2g}^{}}_{A}^{\tensor*[^{1}]{\hspace{-0.2em}B}{}_{1}^{}}_{\bar{A}}^{\tensor*[^{1}]{\hspace{-0.2em}B}{}_{2}^{}}}$ &  $- 8.2643$ &                                                                                     $d\indices*{*_{\Gamma}^{T_{2g}^{}}_{A}^{\tensor*[^{2}]{\hspace{-0.2em}B}{}_{1}^{}}_{\bar{A}}^{B_{2}^{}}}$ &    $0.3217$ &                                                    $d\indices*{*_{\Gamma}^{T_{2g}^{}}_{A}^{\tensor*[^{2}]{\hspace{-0.2em}B}{}_{1}^{}}_{\bar{A}}^{\tensor*[^{1}]{\hspace{-0.2em}B}{}_{2}^{}}}$ &  $- 5.5527$ &                                                                                     $d\indices*{*_{\Gamma}^{T_{1u}^{}}_{A}^{A_{1}^{}}_{\bar{A}}^{\tensor*[^{1}]{\hspace{-0.2em}A}{}_{1}^{}}}$ &    $1.6539$ &                                                                                     $d\indices*{*_{\Gamma}^{T_{1u}^{}}_{A}^{A_{1}^{}}_{\bar{A}}^{\tensor*[^{2}]{\hspace{-0.2em}A}{}_{1}^{}}}$ &   $21.6267$ \\[0.4em]
                                                                                                                       $d\indices*{*_{\Gamma}^{T_{1u}^{}}_{A}^{A_{1}^{}}_{\bar{A}}^{B_{2}^{}}}$ &  $- 0.9589$ &                                                                                     $d\indices*{*_{\Gamma}^{T_{1u}^{}}_{A}^{A_{1}^{}}_{\bar{A}}^{\tensor*[^{1}]{\hspace{-0.2em}B}{}_{2}^{}}}$ &  $- 1.1783$ &                                                    $d\indices*{*_{\Gamma}^{T_{1u}^{}}_{A}^{\tensor*[^{1}]{\hspace{-0.2em}A}{}_{1}^{}}_{\bar{A}}^{\tensor*[^{2}]{\hspace{-0.2em}A}{}_{1}^{}}}$ & $- 11.6152$ &                                                                                     $d\indices*{*_{\Gamma}^{T_{1u}^{}}_{A}^{\tensor*[^{1}]{\hspace{-0.2em}A}{}_{1}^{}}_{\bar{A}}^{B_{2}^{}}}$ & $- 12.3283$ &                                                    $d\indices*{*_{\Gamma}^{T_{1u}^{}}_{A}^{\tensor*[^{1}]{\hspace{-0.2em}A}{}_{1}^{}}_{\bar{A}}^{\tensor*[^{1}]{\hspace{-0.2em}B}{}_{2}^{}}}$ &    $8.2490$ \\[0.4em]
                                                                                      $d\indices*{*_{\Gamma}^{T_{1u}^{}}_{A}^{\tensor*[^{2}]{\hspace{-0.2em}A}{}_{1}^{}}_{\bar{A}}^{B_{2}^{}}}$ &  $- 1.8116$ &                                                    $d\indices*{*_{\Gamma}^{T_{1u}^{}}_{A}^{\tensor*[^{2}]{\hspace{-0.2em}A}{}_{1}^{}}_{\bar{A}}^{\tensor*[^{1}]{\hspace{-0.2em}B}{}_{2}^{}}}$ &    $0.7354$ &                                                                                                                      $d\indices*{*_{\Gamma}^{T_{1u}^{}}_{A}^{A_{2}^{}}_{\bar{A}}^{B_{1}^{}}}$ &    $9.4261$ &                                                                                     $d\indices*{*_{\Gamma}^{T_{1u}^{}}_{A}^{A_{2}^{}}_{\bar{A}}^{\tensor*[^{1}]{\hspace{-0.2em}B}{}_{1}^{}}}$ & $- 12.6424$ &                                                                                     $d\indices*{*_{\Gamma}^{T_{1u}^{}}_{A}^{A_{2}^{}}_{\bar{A}}^{\tensor*[^{2}]{\hspace{-0.2em}B}{}_{1}^{}}}$ &    $0.9681$ \\[0.4em]
                                                                                      $d\indices*{*_{\Gamma}^{T_{1u}^{}}_{A}^{B_{1}^{}}_{\bar{A}}^{\tensor*[^{1}]{\hspace{-0.2em}B}{}_{1}^{}}}$ &    $9.3349$ &                                                                                     $d\indices*{*_{\Gamma}^{T_{1u}^{}}_{A}^{B_{1}^{}}_{\bar{A}}^{\tensor*[^{2}]{\hspace{-0.2em}B}{}_{1}^{}}}$ &   $10.1546$ &                                                    $d\indices*{*_{\Gamma}^{T_{1u}^{}}_{A}^{\tensor*[^{1}]{\hspace{-0.2em}B}{}_{1}^{}}_{\bar{A}}^{\tensor*[^{2}]{\hspace{-0.2em}B}{}_{1}^{}}}$ &    $1.5624$ &                                                                                     $d\indices*{*_{\Gamma}^{T_{1u}^{}}_{A}^{B_{2}^{}}_{\bar{A}}^{\tensor*[^{1}]{\hspace{-0.2em}B}{}_{2}^{}}}$ &    $1.0833$ &                                                                                                                      $d\indices*{*_{\Gamma}^{T_{1u}^{}}_{A}^{A_{1}^{}}_{\bar{A}}^{B_{1}^{}}}$ &   $14.5205$ \\[0.4em]
                                                                                      $d\indices*{*_{\Gamma}^{T_{1u}^{}}_{A}^{A_{1}^{}}_{\bar{A}}^{\tensor*[^{1}]{\hspace{-0.2em}B}{}_{1}^{}}}$ &    $1.9915$ &                                                                                     $d\indices*{*_{\Gamma}^{T_{1u}^{}}_{A}^{A_{1}^{}}_{\bar{A}}^{\tensor*[^{2}]{\hspace{-0.2em}B}{}_{1}^{}}}$ &    $0.7735$ &                                                                                     $d\indices*{*_{\Gamma}^{T_{1u}^{}}_{A}^{\tensor*[^{1}]{\hspace{-0.2em}A}{}_{1}^{}}_{\bar{A}}^{B_{1}^{}}}$ &  $- 0.2457$ &                                                    $d\indices*{*_{\Gamma}^{T_{1u}^{}}_{A}^{\tensor*[^{1}]{\hspace{-0.2em}A}{}_{1}^{}}_{\bar{A}}^{\tensor*[^{1}]{\hspace{-0.2em}B}{}_{1}^{}}}$ &  $- 9.8948$ &                                                    $d\indices*{*_{\Gamma}^{T_{1u}^{}}_{A}^{\tensor*[^{1}]{\hspace{-0.2em}A}{}_{1}^{}}_{\bar{A}}^{\tensor*[^{2}]{\hspace{-0.2em}B}{}_{1}^{}}}$ &    $0.3134$ \\[0.4em]
                                                                                      $d\indices*{*_{\Gamma}^{T_{1u}^{}}_{A}^{\tensor*[^{2}]{\hspace{-0.2em}A}{}_{1}^{}}_{\bar{A}}^{B_{1}^{}}}$ &    $3.0872$ &                                                    $d\indices*{*_{\Gamma}^{T_{1u}^{}}_{A}^{\tensor*[^{2}]{\hspace{-0.2em}A}{}_{1}^{}}_{\bar{A}}^{\tensor*[^{1}]{\hspace{-0.2em}B}{}_{1}^{}}}$ &   $11.4120$ &                                                    $d\indices*{*_{\Gamma}^{T_{1u}^{}}_{A}^{\tensor*[^{2}]{\hspace{-0.2em}A}{}_{1}^{}}_{\bar{A}}^{\tensor*[^{2}]{\hspace{-0.2em}B}{}_{1}^{}}}$ &   $15.2482$ &                                                                                                                      $d\indices*{*_{\Gamma}^{T_{1u}^{}}_{A}^{A_{2}^{}}_{\bar{A}}^{B_{2}^{}}}$ &   $16.3685$ &                                                                                     $d\indices*{*_{\Gamma}^{T_{1u}^{}}_{A}^{A_{2}^{}}_{\bar{A}}^{\tensor*[^{1}]{\hspace{-0.2em}B}{}_{2}^{}}}$ & $- 13.8115$ \\[0.4em]
                                                                                                                $d\indices*{*_{A_{3}}^{A_{1}^{}}_{X_{2}}^{A_{1g}^{}}_{\bar{A}_{1}}^{A_{1}^{}}}$ &  $- 1.9567$ &                                                                                                               $d\indices*{*_{A_{3}}^{A_{1}^{}}_{X_{2}}^{A_{2u}^{}}_{\bar{A}_{1}}^{A_{1}^{}}}$ &  $- 3.5637$ &                                                                              $d\indices*{*_{A_{3}}^{A_{1}^{}}_{X_{2}}^{A_{1g}^{}}_{\bar{A}_{1}}^{\tensor*[^{1}]{\hspace{-0.2em}A}{}_{1}^{}}}$ &    $1.3445$ &                                                                               $d\indices*{*_{A_{3}}^{A_{1}^{}}_{X_{2}}^{E_{g}^{}}_{\bar{A}_{1}}^{\tensor*[^{1}]{\hspace{-0.2em}A}{}_{1}^{}}}$ &  $- 0.4098$ &                                                                              $d\indices*{*_{A_{3}}^{A_{1}^{}}_{X_{2}}^{A_{2u}^{}}_{\bar{A}_{1}}^{\tensor*[^{1}]{\hspace{-0.2em}A}{}_{1}^{}}}$ & $- 11.1423$ \\[0.4em]
                                                                                $d\indices*{*_{A_{3}}^{A_{1}^{}}_{X_{2}}^{E_{u}^{}}_{\bar{A}_{1}}^{\tensor*[^{1}]{\hspace{-0.2em}A}{}_{1}^{}}}$ &  $- 6.8310$ &                                              $d\indices*{*_{A_{3}}^{A_{1}^{}}_{X_{2}}^{\tensor*[^{1}]{\hspace{-0.2em}E}{}_{u}^{}}_{\bar{A}_{1}}^{\tensor*[^{1}]{\hspace{-0.2em}A}{}_{1}^{}}}$ &   $12.5873$ &                                                                              $d\indices*{*_{A_{3}}^{A_{1}^{}}_{X_{2}}^{A_{1g}^{}}_{\bar{A}_{1}}^{\tensor*[^{2}]{\hspace{-0.2em}A}{}_{1}^{}}}$ &  $- 1.1799$ &                                                                               $d\indices*{*_{A_{3}}^{A_{1}^{}}_{X_{2}}^{E_{g}^{}}_{\bar{A}_{1}}^{\tensor*[^{2}]{\hspace{-0.2em}A}{}_{1}^{}}}$ &  $- 0.4718$ &                                                                              $d\indices*{*_{A_{3}}^{A_{1}^{}}_{X_{2}}^{A_{2u}^{}}_{\bar{A}_{1}}^{\tensor*[^{2}]{\hspace{-0.2em}A}{}_{1}^{}}}$ &  $- 3.2535$ \\[0.4em]
                                                                                $d\indices*{*_{A_{3}}^{A_{1}^{}}_{X_{2}}^{E_{u}^{}}_{\bar{A}_{1}}^{\tensor*[^{2}]{\hspace{-0.2em}A}{}_{1}^{}}}$ &  $- 0.7207$ &                                              $d\indices*{*_{A_{3}}^{A_{1}^{}}_{X_{2}}^{\tensor*[^{1}]{\hspace{-0.2em}E}{}_{u}^{}}_{\bar{A}_{1}}^{\tensor*[^{2}]{\hspace{-0.2em}A}{}_{1}^{}}}$ &    $3.4585$ &                                                                                                                $d\indices*{*_{A_{3}}^{A_{1}^{}}_{X_{2}}^{E_{g}^{}}_{\bar{A}_{1}}^{A_{2}^{}}}$ &    $0.6043$ &                                                                                                               $d\indices*{*_{A_{3}}^{A_{1}^{}}_{X_{2}}^{B_{1u}^{}}_{\bar{A}_{1}}^{A_{2}^{}}}$ & $- 12.9566$ &                                                                                                                $d\indices*{*_{A_{3}}^{A_{1}^{}}_{X_{2}}^{E_{u}^{}}_{\bar{A}_{1}}^{A_{2}^{}}}$ &   $10.9341$ \\[0.4em]
                                                                                $d\indices*{*_{A_{3}}^{A_{1}^{}}_{X_{2}}^{\tensor*[^{1}]{\hspace{-0.2em}E}{}_{u}^{}}_{\bar{A}_{1}}^{A_{2}^{}}}$ & $- 18.5719$ &                                                                                                                $d\indices*{*_{A_{3}}^{A_{1}^{}}_{X_{2}}^{E_{g}^{}}_{\bar{A}_{1}}^{B_{1}^{}}}$ &  $- 9.0607$ &                                                                                                               $d\indices*{*_{A_{3}}^{A_{1}^{}}_{X_{2}}^{B_{1u}^{}}_{\bar{A}_{1}}^{B_{1}^{}}}$ &  $- 1.2721$ &                                                                                                                $d\indices*{*_{A_{3}}^{A_{1}^{}}_{X_{2}}^{E_{u}^{}}_{\bar{A}_{1}}^{B_{1}^{}}}$ &    $1.5111$ &                                                                               $d\indices*{*_{A_{3}}^{A_{1}^{}}_{X_{2}}^{\tensor*[^{1}]{\hspace{-0.2em}E}{}_{u}^{}}_{\bar{A}_{1}}^{B_{1}^{}}}$ &  $- 0.1622$ \\[0.4em]
                                                                                $d\indices*{*_{A_{3}}^{A_{1}^{}}_{X_{2}}^{E_{g}^{}}_{\bar{A}_{1}}^{\tensor*[^{1}]{\hspace{-0.2em}B}{}_{1}^{}}}$ &   $12.3323$ &                                                                              $d\indices*{*_{A_{3}}^{A_{1}^{}}_{X_{2}}^{B_{1u}^{}}_{\bar{A}_{1}}^{\tensor*[^{1}]{\hspace{-0.2em}B}{}_{1}^{}}}$ &    $0.4179$ &                                                                               $d\indices*{*_{A_{3}}^{A_{1}^{}}_{X_{2}}^{E_{u}^{}}_{\bar{A}_{1}}^{\tensor*[^{1}]{\hspace{-0.2em}B}{}_{1}^{}}}$ &    $0.3862$ &                                              $d\indices*{*_{A_{3}}^{A_{1}^{}}_{X_{2}}^{\tensor*[^{1}]{\hspace{-0.2em}E}{}_{u}^{}}_{\bar{A}_{1}}^{\tensor*[^{1}]{\hspace{-0.2em}B}{}_{1}^{}}}$ &  $- 3.1696$ &                                                                               $d\indices*{*_{A_{3}}^{A_{1}^{}}_{X_{2}}^{E_{g}^{}}_{\bar{A}_{1}}^{\tensor*[^{2}]{\hspace{-0.2em}B}{}_{1}^{}}}$ &  $- 1.3005$ \\[0.4em]
                                                                               $d\indices*{*_{A_{3}}^{A_{1}^{}}_{X_{2}}^{B_{1u}^{}}_{\bar{A}_{1}}^{\tensor*[^{2}]{\hspace{-0.2em}B}{}_{1}^{}}}$ &  $- 0.0007$ &                                                                               $d\indices*{*_{A_{3}}^{A_{1}^{}}_{X_{2}}^{E_{u}^{}}_{\bar{A}_{1}}^{\tensor*[^{2}]{\hspace{-0.2em}B}{}_{1}^{}}}$ &    $0.5873$ &                                              $d\indices*{*_{A_{3}}^{A_{1}^{}}_{X_{2}}^{\tensor*[^{1}]{\hspace{-0.2em}E}{}_{u}^{}}_{\bar{A}_{1}}^{\tensor*[^{2}]{\hspace{-0.2em}B}{}_{1}^{}}}$ &    $0.0793$ &                                                                                                               $d\indices*{*_{A_{3}}^{A_{1}^{}}_{X_{2}}^{A_{1g}^{}}_{\bar{A}_{1}}^{B_{2}^{}}}$ & $- 14.5308$ &                                                                                                                $d\indices*{*_{A_{3}}^{A_{1}^{}}_{X_{2}}^{E_{g}^{}}_{\bar{A}_{1}}^{B_{2}^{}}}$ & $- 16.3691$ \\[0.4em]
                                                                                                                $d\indices*{*_{A_{3}}^{A_{1}^{}}_{X_{2}}^{A_{2u}^{}}_{\bar{A}_{1}}^{B_{2}^{}}}$ &    $0.6673$ &                                                                                                                $d\indices*{*_{A_{3}}^{A_{1}^{}}_{X_{2}}^{E_{u}^{}}_{\bar{A}_{1}}^{B_{2}^{}}}$ &    $1.2884$ &                                                                               $d\indices*{*_{A_{3}}^{A_{1}^{}}_{X_{2}}^{\tensor*[^{1}]{\hspace{-0.2em}E}{}_{u}^{}}_{\bar{A}_{1}}^{B_{2}^{}}}$ &  $- 3.2236$ &                                                                              $d\indices*{*_{A_{3}}^{A_{1}^{}}_{X_{2}}^{A_{1g}^{}}_{\bar{A}_{1}}^{\tensor*[^{1}]{\hspace{-0.2em}B}{}_{2}^{}}}$ &   $10.6070$ &                                                                               $d\indices*{*_{A_{3}}^{A_{1}^{}}_{X_{2}}^{E_{g}^{}}_{\bar{A}_{1}}^{\tensor*[^{1}]{\hspace{-0.2em}B}{}_{2}^{}}}$ &   $12.6620$ \\[0.4em]
                                                                               $d\indices*{*_{A_{3}}^{A_{1}^{}}_{X_{2}}^{A_{2u}^{}}_{\bar{A}_{1}}^{\tensor*[^{1}]{\hspace{-0.2em}B}{}_{2}^{}}}$ &    $2.0512$ &                                                                               $d\indices*{*_{A_{3}}^{A_{1}^{}}_{X_{2}}^{E_{u}^{}}_{\bar{A}_{1}}^{\tensor*[^{1}]{\hspace{-0.2em}B}{}_{2}^{}}}$ &  $- 1.3823$ &                                              $d\indices*{*_{A_{3}}^{A_{1}^{}}_{X_{2}}^{\tensor*[^{1}]{\hspace{-0.2em}E}{}_{u}^{}}_{\bar{A}_{1}}^{\tensor*[^{1}]{\hspace{-0.2em}B}{}_{2}^{}}}$ &    $0.1008$ &                                             $d\indices*{*_{A_{3}}^{\tensor*[^{1}]{\hspace{-0.2em}A}{}_{1}^{}}_{X_{2}}^{A_{1g}^{}}_{\bar{A}_{1}}^{\tensor*[^{1}]{\hspace{-0.2em}A}{}_{1}^{}}}$ &  $- 1.2286$ &                                             $d\indices*{*_{A_{3}}^{\tensor*[^{1}]{\hspace{-0.2em}A}{}_{1}^{}}_{X_{2}}^{A_{2u}^{}}_{\bar{A}_{1}}^{\tensor*[^{1}]{\hspace{-0.2em}A}{}_{1}^{}}}$ &    $9.1384$ \\[0.4em]
                                              $d\indices*{*_{A_{3}}^{\tensor*[^{1}]{\hspace{-0.2em}A}{}_{1}^{}}_{X_{2}}^{A_{1g}^{}}_{\bar{A}_{1}}^{\tensor*[^{2}]{\hspace{-0.2em}A}{}_{1}^{}}}$ &    $6.7613$ &                                              $d\indices*{*_{A_{3}}^{\tensor*[^{1}]{\hspace{-0.2em}A}{}_{1}^{}}_{X_{2}}^{E_{g}^{}}_{\bar{A}_{1}}^{\tensor*[^{2}]{\hspace{-0.2em}A}{}_{1}^{}}}$ &    $7.3620$ &                                             $d\indices*{*_{A_{3}}^{\tensor*[^{1}]{\hspace{-0.2em}A}{}_{1}^{}}_{X_{2}}^{A_{2u}^{}}_{\bar{A}_{1}}^{\tensor*[^{2}]{\hspace{-0.2em}A}{}_{1}^{}}}$ &  $- 0.4544$ &                                              $d\indices*{*_{A_{3}}^{\tensor*[^{1}]{\hspace{-0.2em}A}{}_{1}^{}}_{X_{2}}^{E_{u}^{}}_{\bar{A}_{1}}^{\tensor*[^{2}]{\hspace{-0.2em}A}{}_{1}^{}}}$ &    $1.7750$ &             $d\indices*{*_{A_{3}}^{\tensor*[^{1}]{\hspace{-0.2em}A}{}_{1}^{}}_{X_{2}}^{\tensor*[^{1}]{\hspace{-0.2em}E}{}_{u}^{}}_{\bar{A}_{1}}^{\tensor*[^{2}]{\hspace{-0.2em}A}{}_{1}^{}}}$ &  $- 0.0789$ \\[0.4em]
                                                                                $d\indices*{*_{A_{3}}^{\tensor*[^{1}]{\hspace{-0.2em}A}{}_{1}^{}}_{X_{2}}^{E_{g}^{}}_{\bar{A}_{1}}^{A_{2}^{}}}$ &  $- 2.1364$ &                                                                              $d\indices*{*_{A_{3}}^{\tensor*[^{1}]{\hspace{-0.2em}A}{}_{1}^{}}_{X_{2}}^{B_{1u}^{}}_{\bar{A}_{1}}^{A_{2}^{}}}$ &    $7.1363$ &                                                                               $d\indices*{*_{A_{3}}^{\tensor*[^{1}]{\hspace{-0.2em}A}{}_{1}^{}}_{X_{2}}^{E_{u}^{}}_{\bar{A}_{1}}^{A_{2}^{}}}$ &  $- 6.6482$ &                                              $d\indices*{*_{A_{3}}^{\tensor*[^{1}]{\hspace{-0.2em}A}{}_{1}^{}}_{X_{2}}^{\tensor*[^{1}]{\hspace{-0.2em}E}{}_{u}^{}}_{\bar{A}_{1}}^{A_{2}^{}}}$ &    $6.1929$ &                                                                               $d\indices*{*_{A_{3}}^{\tensor*[^{1}]{\hspace{-0.2em}A}{}_{1}^{}}_{X_{2}}^{E_{g}^{}}_{\bar{A}_{1}}^{B_{1}^{}}}$ &  $- 0.7046$ \\[0.4em]
                                                                               $d\indices*{*_{A_{3}}^{\tensor*[^{1}]{\hspace{-0.2em}A}{}_{1}^{}}_{X_{2}}^{B_{1u}^{}}_{\bar{A}_{1}}^{B_{1}^{}}}$ &    $0.3086$ &                                                                               $d\indices*{*_{A_{3}}^{\tensor*[^{1}]{\hspace{-0.2em}A}{}_{1}^{}}_{X_{2}}^{E_{u}^{}}_{\bar{A}_{1}}^{B_{1}^{}}}$ &  $- 0.8478$ &                                              $d\indices*{*_{A_{3}}^{\tensor*[^{1}]{\hspace{-0.2em}A}{}_{1}^{}}_{X_{2}}^{\tensor*[^{1}]{\hspace{-0.2em}E}{}_{u}^{}}_{\bar{A}_{1}}^{B_{1}^{}}}$ &    $1.7442$ &                                              $d\indices*{*_{A_{3}}^{\tensor*[^{1}]{\hspace{-0.2em}A}{}_{1}^{}}_{X_{2}}^{E_{g}^{}}_{\bar{A}_{1}}^{\tensor*[^{1}]{\hspace{-0.2em}B}{}_{1}^{}}}$ &  $- 5.0930$ &                                             $d\indices*{*_{A_{3}}^{\tensor*[^{1}]{\hspace{-0.2em}A}{}_{1}^{}}_{X_{2}}^{B_{1u}^{}}_{\bar{A}_{1}}^{\tensor*[^{1}]{\hspace{-0.2em}B}{}_{1}^{}}}$ &    $5.4147$ \\[0.4em]
                                               $d\indices*{*_{A_{3}}^{\tensor*[^{1}]{\hspace{-0.2em}A}{}_{1}^{}}_{X_{2}}^{E_{u}^{}}_{\bar{A}_{1}}^{\tensor*[^{1}]{\hspace{-0.2em}B}{}_{1}^{}}}$ &    $4.6447$ &             $d\indices*{*_{A_{3}}^{\tensor*[^{1}]{\hspace{-0.2em}A}{}_{1}^{}}_{X_{2}}^{\tensor*[^{1}]{\hspace{-0.2em}E}{}_{u}^{}}_{\bar{A}_{1}}^{\tensor*[^{1}]{\hspace{-0.2em}B}{}_{1}^{}}}$ &  $- 8.0581$ &                                              $d\indices*{*_{A_{3}}^{\tensor*[^{1}]{\hspace{-0.2em}A}{}_{1}^{}}_{X_{2}}^{E_{g}^{}}_{\bar{A}_{1}}^{\tensor*[^{2}]{\hspace{-0.2em}B}{}_{1}^{}}}$ &    $1.2156$ &                                             $d\indices*{*_{A_{3}}^{\tensor*[^{1}]{\hspace{-0.2em}A}{}_{1}^{}}_{X_{2}}^{B_{1u}^{}}_{\bar{A}_{1}}^{\tensor*[^{2}]{\hspace{-0.2em}B}{}_{1}^{}}}$ &    $6.7192$ &                                              $d\indices*{*_{A_{3}}^{\tensor*[^{1}]{\hspace{-0.2em}A}{}_{1}^{}}_{X_{2}}^{E_{u}^{}}_{\bar{A}_{1}}^{\tensor*[^{2}]{\hspace{-0.2em}B}{}_{1}^{}}}$ &    $5.8043$ \\[0.4em]
              $d\indices*{*_{A_{3}}^{\tensor*[^{1}]{\hspace{-0.2em}A}{}_{1}^{}}_{X_{2}}^{\tensor*[^{1}]{\hspace{-0.2em}E}{}_{u}^{}}_{\bar{A}_{1}}^{\tensor*[^{2}]{\hspace{-0.2em}B}{}_{1}^{}}}$ &  $- 7.4634$ &                                                                              $d\indices*{*_{A_{3}}^{\tensor*[^{1}]{\hspace{-0.2em}A}{}_{1}^{}}_{X_{2}}^{A_{1g}^{}}_{\bar{A}_{1}}^{B_{2}^{}}}$ &    $7.8530$ &                                                                               $d\indices*{*_{A_{3}}^{\tensor*[^{1}]{\hspace{-0.2em}A}{}_{1}^{}}_{X_{2}}^{E_{g}^{}}_{\bar{A}_{1}}^{B_{2}^{}}}$ &    $7.8117$ &                                                                              $d\indices*{*_{A_{3}}^{\tensor*[^{1}]{\hspace{-0.2em}A}{}_{1}^{}}_{X_{2}}^{A_{2u}^{}}_{\bar{A}_{1}}^{B_{2}^{}}}$ &  $- 0.3806$ &                                                                               $d\indices*{*_{A_{3}}^{\tensor*[^{1}]{\hspace{-0.2em}A}{}_{1}^{}}_{X_{2}}^{E_{u}^{}}_{\bar{A}_{1}}^{B_{2}^{}}}$ &  $- 0.2979$ \\[0.4em]
                                               $d\indices*{*_{A_{3}}^{\tensor*[^{1}]{\hspace{-0.2em}A}{}_{1}^{}}_{X_{2}}^{\tensor*[^{1}]{\hspace{-0.2em}E}{}_{u}^{}}_{\bar{A}_{1}}^{B_{2}^{}}}$ &  $- 0.3700$ &                                             $d\indices*{*_{A_{3}}^{\tensor*[^{1}]{\hspace{-0.2em}A}{}_{1}^{}}_{X_{2}}^{A_{1g}^{}}_{\bar{A}_{1}}^{\tensor*[^{1}]{\hspace{-0.2em}B}{}_{2}^{}}}$ &  $- 5.7432$ &                                              $d\indices*{*_{A_{3}}^{\tensor*[^{1}]{\hspace{-0.2em}A}{}_{1}^{}}_{X_{2}}^{E_{g}^{}}_{\bar{A}_{1}}^{\tensor*[^{1}]{\hspace{-0.2em}B}{}_{2}^{}}}$ &  $- 6.0377$ &                                             $d\indices*{*_{A_{3}}^{\tensor*[^{1}]{\hspace{-0.2em}A}{}_{1}^{}}_{X_{2}}^{A_{2u}^{}}_{\bar{A}_{1}}^{\tensor*[^{1}]{\hspace{-0.2em}B}{}_{2}^{}}}$ &    $0.1776$ &                                              $d\indices*{*_{A_{3}}^{\tensor*[^{1}]{\hspace{-0.2em}A}{}_{1}^{}}_{X_{2}}^{E_{u}^{}}_{\bar{A}_{1}}^{\tensor*[^{1}]{\hspace{-0.2em}B}{}_{2}^{}}}$ &    $1.1954$ \\[0.4em]
              $d\indices*{*_{A_{3}}^{\tensor*[^{1}]{\hspace{-0.2em}A}{}_{1}^{}}_{X_{2}}^{\tensor*[^{1}]{\hspace{-0.2em}E}{}_{u}^{}}_{\bar{A}_{1}}^{\tensor*[^{1}]{\hspace{-0.2em}B}{}_{2}^{}}}$ &  $- 1.6853$ &                                             $d\indices*{*_{A_{3}}^{\tensor*[^{2}]{\hspace{-0.2em}A}{}_{1}^{}}_{X_{2}}^{A_{1g}^{}}_{\bar{A}_{1}}^{\tensor*[^{2}]{\hspace{-0.2em}A}{}_{1}^{}}}$ &  $- 4.4681$ &                                             $d\indices*{*_{A_{3}}^{\tensor*[^{2}]{\hspace{-0.2em}A}{}_{1}^{}}_{X_{2}}^{A_{2u}^{}}_{\bar{A}_{1}}^{\tensor*[^{2}]{\hspace{-0.2em}A}{}_{1}^{}}}$ &  $- 3.0011$ &                                                                               $d\indices*{*_{A_{3}}^{\tensor*[^{2}]{\hspace{-0.2em}A}{}_{1}^{}}_{X_{2}}^{E_{g}^{}}_{\bar{A}_{1}}^{A_{2}^{}}}$ &   $10.2886$ &                                                                              $d\indices*{*_{A_{3}}^{\tensor*[^{2}]{\hspace{-0.2em}A}{}_{1}^{}}_{X_{2}}^{B_{1u}^{}}_{\bar{A}_{1}}^{A_{2}^{}}}$ &  $- 2.8219$ \\[0.4em]
                                                                                $d\indices*{*_{A_{3}}^{\tensor*[^{2}]{\hspace{-0.2em}A}{}_{1}^{}}_{X_{2}}^{E_{u}^{}}_{\bar{A}_{1}}^{A_{2}^{}}}$ &  $- 3.2072$ &                                              $d\indices*{*_{A_{3}}^{\tensor*[^{2}]{\hspace{-0.2em}A}{}_{1}^{}}_{X_{2}}^{\tensor*[^{1}]{\hspace{-0.2em}E}{}_{u}^{}}_{\bar{A}_{1}}^{A_{2}^{}}}$ &  $- 2.2822$ &                                                                               $d\indices*{*_{A_{3}}^{\tensor*[^{2}]{\hspace{-0.2em}A}{}_{1}^{}}_{X_{2}}^{E_{g}^{}}_{\bar{A}_{1}}^{B_{1}^{}}}$ &  $- 1.6615$ &                                                                              $d\indices*{*_{A_{3}}^{\tensor*[^{2}]{\hspace{-0.2em}A}{}_{1}^{}}_{X_{2}}^{B_{1u}^{}}_{\bar{A}_{1}}^{B_{1}^{}}}$ &  $- 7.3345$ &                                                                               $d\indices*{*_{A_{3}}^{\tensor*[^{2}]{\hspace{-0.2em}A}{}_{1}^{}}_{X_{2}}^{E_{u}^{}}_{\bar{A}_{1}}^{B_{1}^{}}}$ &    $7.3033$ \\[0.4em]
                                               $d\indices*{*_{A_{3}}^{\tensor*[^{2}]{\hspace{-0.2em}A}{}_{1}^{}}_{X_{2}}^{\tensor*[^{1}]{\hspace{-0.2em}E}{}_{u}^{}}_{\bar{A}_{1}}^{B_{1}^{}}}$ &  $- 7.2934$ &                                              $d\indices*{*_{A_{3}}^{\tensor*[^{2}]{\hspace{-0.2em}A}{}_{1}^{}}_{X_{2}}^{E_{g}^{}}_{\bar{A}_{1}}^{\tensor*[^{1}]{\hspace{-0.2em}B}{}_{1}^{}}}$ &    $1.6119$ &                                             $d\indices*{*_{A_{3}}^{\tensor*[^{2}]{\hspace{-0.2em}A}{}_{1}^{}}_{X_{2}}^{B_{1u}^{}}_{\bar{A}_{1}}^{\tensor*[^{1}]{\hspace{-0.2em}B}{}_{1}^{}}}$ &    $9.0130$ &                                              $d\indices*{*_{A_{3}}^{\tensor*[^{2}]{\hspace{-0.2em}A}{}_{1}^{}}_{X_{2}}^{E_{u}^{}}_{\bar{A}_{1}}^{\tensor*[^{1}]{\hspace{-0.2em}B}{}_{1}^{}}}$ &  $- 8.0522$ &             $d\indices*{*_{A_{3}}^{\tensor*[^{2}]{\hspace{-0.2em}A}{}_{1}^{}}_{X_{2}}^{\tensor*[^{1}]{\hspace{-0.2em}E}{}_{u}^{}}_{\bar{A}_{1}}^{\tensor*[^{1}]{\hspace{-0.2em}B}{}_{1}^{}}}$ &    $9.4100$ \\[0.4em]
                                               $d\indices*{*_{A_{3}}^{\tensor*[^{2}]{\hspace{-0.2em}A}{}_{1}^{}}_{X_{2}}^{E_{g}^{}}_{\bar{A}_{1}}^{\tensor*[^{2}]{\hspace{-0.2em}B}{}_{1}^{}}}$ &  $- 0.3514$ &                                             $d\indices*{*_{A_{3}}^{\tensor*[^{2}]{\hspace{-0.2em}A}{}_{1}^{}}_{X_{2}}^{B_{1u}^{}}_{\bar{A}_{1}}^{\tensor*[^{2}]{\hspace{-0.2em}B}{}_{1}^{}}}$ &  $- 0.9513$ &                                              $d\indices*{*_{A_{3}}^{\tensor*[^{2}]{\hspace{-0.2em}A}{}_{1}^{}}_{X_{2}}^{E_{u}^{}}_{\bar{A}_{1}}^{\tensor*[^{2}]{\hspace{-0.2em}B}{}_{1}^{}}}$ &  $- 4.8328$ &             $d\indices*{*_{A_{3}}^{\tensor*[^{2}]{\hspace{-0.2em}A}{}_{1}^{}}_{X_{2}}^{\tensor*[^{1}]{\hspace{-0.2em}E}{}_{u}^{}}_{\bar{A}_{1}}^{\tensor*[^{2}]{\hspace{-0.2em}B}{}_{1}^{}}}$ &    $0.3175$ &                                                                              $d\indices*{*_{A_{3}}^{\tensor*[^{2}]{\hspace{-0.2em}A}{}_{1}^{}}_{X_{2}}^{A_{1g}^{}}_{\bar{A}_{1}}^{B_{2}^{}}}$ &    $1.1107$ \\[0.4em]
                                                                                $d\indices*{*_{A_{3}}^{\tensor*[^{2}]{\hspace{-0.2em}A}{}_{1}^{}}_{X_{2}}^{E_{g}^{}}_{\bar{A}_{1}}^{B_{2}^{}}}$ &  $- 1.6240$ &                                                                              $d\indices*{*_{A_{3}}^{\tensor*[^{2}]{\hspace{-0.2em}A}{}_{1}^{}}_{X_{2}}^{A_{2u}^{}}_{\bar{A}_{1}}^{B_{2}^{}}}$ &   $17.7457$ &                                                                               $d\indices*{*_{A_{3}}^{\tensor*[^{2}]{\hspace{-0.2em}A}{}_{1}^{}}_{X_{2}}^{E_{u}^{}}_{\bar{A}_{1}}^{B_{2}^{}}}$ &   $11.1160$ &                                              $d\indices*{*_{A_{3}}^{\tensor*[^{2}]{\hspace{-0.2em}A}{}_{1}^{}}_{X_{2}}^{\tensor*[^{1}]{\hspace{-0.2em}E}{}_{u}^{}}_{\bar{A}_{1}}^{B_{2}^{}}}$ & $- 15.7543$ &                                             $d\indices*{*_{A_{3}}^{\tensor*[^{2}]{\hspace{-0.2em}A}{}_{1}^{}}_{X_{2}}^{A_{1g}^{}}_{\bar{A}_{1}}^{\tensor*[^{1}]{\hspace{-0.2em}B}{}_{2}^{}}}$ &  $- 2.7226$ \\[0.4em]
                                               $d\indices*{*_{A_{3}}^{\tensor*[^{2}]{\hspace{-0.2em}A}{}_{1}^{}}_{X_{2}}^{E_{g}^{}}_{\bar{A}_{1}}^{\tensor*[^{1}]{\hspace{-0.2em}B}{}_{2}^{}}}$ &    $3.6748$ &                                             $d\indices*{*_{A_{3}}^{\tensor*[^{2}]{\hspace{-0.2em}A}{}_{1}^{}}_{X_{2}}^{A_{2u}^{}}_{\bar{A}_{1}}^{\tensor*[^{1}]{\hspace{-0.2em}B}{}_{2}^{}}}$ & $- 10.9055$ &                                              $d\indices*{*_{A_{3}}^{\tensor*[^{2}]{\hspace{-0.2em}A}{}_{1}^{}}_{X_{2}}^{E_{u}^{}}_{\bar{A}_{1}}^{\tensor*[^{1}]{\hspace{-0.2em}B}{}_{2}^{}}}$ &  $- 9.1977$ &             $d\indices*{*_{A_{3}}^{\tensor*[^{2}]{\hspace{-0.2em}A}{}_{1}^{}}_{X_{2}}^{\tensor*[^{1}]{\hspace{-0.2em}E}{}_{u}^{}}_{\bar{A}_{1}}^{\tensor*[^{1}]{\hspace{-0.2em}B}{}_{2}^{}}}$ &   $10.4305$ &                                                                                                               $d\indices*{*_{A_{3}}^{A_{2}^{}}_{X_{2}}^{A_{1g}^{}}_{\bar{A}_{1}}^{A_{2}^{}}}$ &    $6.2445$ \\[0.4em]
                                                                                                                $d\indices*{*_{A_{3}}^{A_{2}^{}}_{X_{2}}^{A_{2u}^{}}_{\bar{A}_{1}}^{A_{2}^{}}}$ &  $- 0.6480$ &                                                                                                               $d\indices*{*_{A_{3}}^{A_{2}^{}}_{X_{2}}^{A_{1g}^{}}_{\bar{A}_{1}}^{B_{1}^{}}}$ &  $- 7.2126$ &                                                                                                                $d\indices*{*_{A_{3}}^{A_{2}^{}}_{X_{2}}^{E_{g}^{}}_{\bar{A}_{1}}^{B_{1}^{}}}$ &  $- 6.9332$ &                                                                                                               $d\indices*{*_{A_{3}}^{A_{2}^{}}_{X_{2}}^{A_{2u}^{}}_{\bar{A}_{1}}^{B_{1}^{}}}$ &  $- 0.8136$ &                                                                                                                $d\indices*{*_{A_{3}}^{A_{2}^{}}_{X_{2}}^{E_{u}^{}}_{\bar{A}_{1}}^{B_{1}^{}}}$ &  $- 0.8478$ \\[0.4em]
                                                                                $d\indices*{*_{A_{3}}^{A_{2}^{}}_{X_{2}}^{\tensor*[^{1}]{\hspace{-0.2em}E}{}_{u}^{}}_{\bar{A}_{1}}^{B_{1}^{}}}$ &  $- 0.7353$ &                                                                              $d\indices*{*_{A_{3}}^{A_{2}^{}}_{X_{2}}^{A_{1g}^{}}_{\bar{A}_{1}}^{\tensor*[^{1}]{\hspace{-0.2em}B}{}_{1}^{}}}$ &    $0.8222$ &                                                                               $d\indices*{*_{A_{3}}^{A_{2}^{}}_{X_{2}}^{E_{g}^{}}_{\bar{A}_{1}}^{\tensor*[^{1}]{\hspace{-0.2em}B}{}_{1}^{}}}$ &  $- 1.3058$ &                                                                              $d\indices*{*_{A_{3}}^{A_{2}^{}}_{X_{2}}^{A_{2u}^{}}_{\bar{A}_{1}}^{\tensor*[^{1}]{\hspace{-0.2em}B}{}_{1}^{}}}$ & $- 13.7299$ &                                                                               $d\indices*{*_{A_{3}}^{A_{2}^{}}_{X_{2}}^{E_{u}^{}}_{\bar{A}_{1}}^{\tensor*[^{1}]{\hspace{-0.2em}B}{}_{1}^{}}}$ &    $6.5163$ \\[0.4em]
                                               $d\indices*{*_{A_{3}}^{A_{2}^{}}_{X_{2}}^{\tensor*[^{1}]{\hspace{-0.2em}E}{}_{u}^{}}_{\bar{A}_{1}}^{\tensor*[^{1}]{\hspace{-0.2em}B}{}_{1}^{}}}$ & $- 11.4729$ &                                                                              $d\indices*{*_{A_{3}}^{A_{2}^{}}_{X_{2}}^{A_{1g}^{}}_{\bar{A}_{1}}^{\tensor*[^{2}]{\hspace{-0.2em}B}{}_{1}^{}}}$ &  $- 4.7821$ &                                                                               $d\indices*{*_{A_{3}}^{A_{2}^{}}_{X_{2}}^{E_{g}^{}}_{\bar{A}_{1}}^{\tensor*[^{2}]{\hspace{-0.2em}B}{}_{1}^{}}}$ &    $2.4477$ &                                                                              $d\indices*{*_{A_{3}}^{A_{2}^{}}_{X_{2}}^{A_{2u}^{}}_{\bar{A}_{1}}^{\tensor*[^{2}]{\hspace{-0.2em}B}{}_{1}^{}}}$ & $- 11.1467$ &                                                                               $d\indices*{*_{A_{3}}^{A_{2}^{}}_{X_{2}}^{E_{u}^{}}_{\bar{A}_{1}}^{\tensor*[^{2}]{\hspace{-0.2em}B}{}_{1}^{}}}$ &    $8.9774$ \\[0.4em]
                                               $d\indices*{*_{A_{3}}^{A_{2}^{}}_{X_{2}}^{\tensor*[^{1}]{\hspace{-0.2em}E}{}_{u}^{}}_{\bar{A}_{1}}^{\tensor*[^{2}]{\hspace{-0.2em}B}{}_{1}^{}}}$ &  $- 9.8507$ &                                                                                                                $d\indices*{*_{A_{3}}^{B_{2}^{}}_{X_{2}}^{E_{g}^{}}_{\bar{A}_{1}}^{A_{2}^{}}}$ &  $- 0.8084$ &                                                                                                               $d\indices*{*_{A_{3}}^{B_{2}^{}}_{X_{2}}^{B_{1u}^{}}_{\bar{A}_{1}}^{A_{2}^{}}}$ &    $1.0429$ &                                                                                                                $d\indices*{*_{A_{3}}^{B_{2}^{}}_{X_{2}}^{E_{u}^{}}_{\bar{A}_{1}}^{A_{2}^{}}}$ &  $- 1.0920$ &                                                                               $d\indices*{*_{A_{3}}^{B_{2}^{}}_{X_{2}}^{\tensor*[^{1}]{\hspace{-0.2em}E}{}_{u}^{}}_{\bar{A}_{1}}^{A_{2}^{}}}$ &    $0.4012$ \\[0.4em]
                                                                                $d\indices*{*_{A_{3}}^{\tensor*[^{1}]{\hspace{-0.2em}B}{}_{2}^{}}_{X_{2}}^{E_{g}^{}}_{\bar{A}_{1}}^{A_{2}^{}}}$ &  $- 0.1189$ &                                                                              $d\indices*{*_{A_{3}}^{\tensor*[^{1}]{\hspace{-0.2em}B}{}_{2}^{}}_{X_{2}}^{B_{1u}^{}}_{\bar{A}_{1}}^{A_{2}^{}}}$ &  $- 0.6936$ &                                                                               $d\indices*{*_{A_{3}}^{\tensor*[^{1}]{\hspace{-0.2em}B}{}_{2}^{}}_{X_{2}}^{E_{u}^{}}_{\bar{A}_{1}}^{A_{2}^{}}}$ &    $6.1035$ &                                              $d\indices*{*_{A_{3}}^{\tensor*[^{1}]{\hspace{-0.2em}B}{}_{2}^{}}_{X_{2}}^{\tensor*[^{1}]{\hspace{-0.2em}E}{}_{u}^{}}_{\bar{A}_{1}}^{A_{2}^{}}}$ &  $- 0.4577$ &                                                                                                               $d\indices*{*_{A_{3}}^{B_{1}^{}}_{X_{2}}^{A_{1g}^{}}_{\bar{A}_{1}}^{B_{1}^{}}}$ &  $- 0.4988$ \\[0.4em]
                                                                                                                $d\indices*{*_{A_{3}}^{B_{1}^{}}_{X_{2}}^{A_{2u}^{}}_{\bar{A}_{1}}^{B_{1}^{}}}$ &   $12.5794$ &                                                                              $d\indices*{*_{A_{3}}^{B_{1}^{}}_{X_{2}}^{A_{1g}^{}}_{\bar{A}_{1}}^{\tensor*[^{1}]{\hspace{-0.2em}B}{}_{1}^{}}}$ &  $- 5.4304$ &                                                                               $d\indices*{*_{A_{3}}^{B_{1}^{}}_{X_{2}}^{E_{g}^{}}_{\bar{A}_{1}}^{\tensor*[^{1}]{\hspace{-0.2em}B}{}_{1}^{}}}$ &  $- 5.4534$ &                                                                              $d\indices*{*_{A_{3}}^{B_{1}^{}}_{X_{2}}^{A_{2u}^{}}_{\bar{A}_{1}}^{\tensor*[^{1}]{\hspace{-0.2em}B}{}_{1}^{}}}$ & $- 10.3899$ &                                                                               $d\indices*{*_{A_{3}}^{B_{1}^{}}_{X_{2}}^{E_{u}^{}}_{\bar{A}_{1}}^{\tensor*[^{1}]{\hspace{-0.2em}B}{}_{1}^{}}}$ &  $- 3.8418$ \\[0.4em]
                                               $d\indices*{*_{A_{3}}^{B_{1}^{}}_{X_{2}}^{\tensor*[^{1}]{\hspace{-0.2em}E}{}_{u}^{}}_{\bar{A}_{1}}^{\tensor*[^{1}]{\hspace{-0.2em}B}{}_{1}^{}}}$ &    $7.9384$ &                                                                              $d\indices*{*_{A_{3}}^{B_{1}^{}}_{X_{2}}^{A_{1g}^{}}_{\bar{A}_{1}}^{\tensor*[^{2}]{\hspace{-0.2em}B}{}_{1}^{}}}$ &  $- 7.2028$ &                                                                               $d\indices*{*_{A_{3}}^{B_{1}^{}}_{X_{2}}^{E_{g}^{}}_{\bar{A}_{1}}^{\tensor*[^{2}]{\hspace{-0.2em}B}{}_{1}^{}}}$ &  $- 6.8020$ &                                                                              $d\indices*{*_{A_{3}}^{B_{1}^{}}_{X_{2}}^{A_{2u}^{}}_{\bar{A}_{1}}^{\tensor*[^{2}]{\hspace{-0.2em}B}{}_{1}^{}}}$ &    $0.2979$ &                                                                               $d\indices*{*_{A_{3}}^{B_{1}^{}}_{X_{2}}^{E_{u}^{}}_{\bar{A}_{1}}^{\tensor*[^{2}]{\hspace{-0.2em}B}{}_{1}^{}}}$ &    $1.7656$ \\[0.4em]
                                               $d\indices*{*_{A_{3}}^{B_{1}^{}}_{X_{2}}^{\tensor*[^{1}]{\hspace{-0.2em}E}{}_{u}^{}}_{\bar{A}_{1}}^{\tensor*[^{2}]{\hspace{-0.2em}B}{}_{1}^{}}}$ &  $- 1.2598$ &                                                                                                                $d\indices*{*_{A_{3}}^{B_{2}^{}}_{X_{2}}^{E_{g}^{}}_{\bar{A}_{1}}^{B_{1}^{}}}$ &  $- 1.0541$ &                                                                                                               $d\indices*{*_{A_{3}}^{B_{2}^{}}_{X_{2}}^{B_{1u}^{}}_{\bar{A}_{1}}^{B_{1}^{}}}$ &    $9.5349$ &                                                                                                                $d\indices*{*_{A_{3}}^{B_{2}^{}}_{X_{2}}^{E_{u}^{}}_{\bar{A}_{1}}^{B_{1}^{}}}$ &    $6.7580$ &                                                                               $d\indices*{*_{A_{3}}^{B_{2}^{}}_{X_{2}}^{\tensor*[^{1}]{\hspace{-0.2em}E}{}_{u}^{}}_{\bar{A}_{1}}^{B_{1}^{}}}$ & $- 10.3672$ \\[0.4em]
                                                                                $d\indices*{*_{A_{3}}^{\tensor*[^{1}]{\hspace{-0.2em}B}{}_{2}^{}}_{X_{2}}^{E_{g}^{}}_{\bar{A}_{1}}^{B_{1}^{}}}$ &    $1.8816$ &                                                                              $d\indices*{*_{A_{3}}^{\tensor*[^{1}]{\hspace{-0.2em}B}{}_{2}^{}}_{X_{2}}^{B_{1u}^{}}_{\bar{A}_{1}}^{B_{1}^{}}}$ &  $- 6.3017$ &                                                                               $d\indices*{*_{A_{3}}^{\tensor*[^{1}]{\hspace{-0.2em}B}{}_{2}^{}}_{X_{2}}^{E_{u}^{}}_{\bar{A}_{1}}^{B_{1}^{}}}$ &  $- 5.6426$ &                                              $d\indices*{*_{A_{3}}^{\tensor*[^{1}]{\hspace{-0.2em}B}{}_{2}^{}}_{X_{2}}^{\tensor*[^{1}]{\hspace{-0.2em}E}{}_{u}^{}}_{\bar{A}_{1}}^{B_{1}^{}}}$ &    $7.4949$ &                                             $d\indices*{*_{A_{3}}^{\tensor*[^{1}]{\hspace{-0.2em}B}{}_{1}^{}}_{X_{2}}^{A_{1g}^{}}_{\bar{A}_{1}}^{\tensor*[^{1}]{\hspace{-0.2em}B}{}_{1}^{}}}$ &   $14.5569$ \\[0.4em]
                                              $d\indices*{*_{A_{3}}^{\tensor*[^{1}]{\hspace{-0.2em}B}{}_{1}^{}}_{X_{2}}^{A_{2u}^{}}_{\bar{A}_{1}}^{\tensor*[^{1}]{\hspace{-0.2em}B}{}_{1}^{}}}$ &  $- 3.6726$ &                                             $d\indices*{*_{A_{3}}^{\tensor*[^{1}]{\hspace{-0.2em}B}{}_{1}^{}}_{X_{2}}^{A_{1g}^{}}_{\bar{A}_{1}}^{\tensor*[^{2}]{\hspace{-0.2em}B}{}_{1}^{}}}$ &    $8.3510$ &                                              $d\indices*{*_{A_{3}}^{\tensor*[^{1}]{\hspace{-0.2em}B}{}_{1}^{}}_{X_{2}}^{E_{g}^{}}_{\bar{A}_{1}}^{\tensor*[^{2}]{\hspace{-0.2em}B}{}_{1}^{}}}$ &    $9.2612$ &                                             $d\indices*{*_{A_{3}}^{\tensor*[^{1}]{\hspace{-0.2em}B}{}_{1}^{}}_{X_{2}}^{A_{2u}^{}}_{\bar{A}_{1}}^{\tensor*[^{2}]{\hspace{-0.2em}B}{}_{1}^{}}}$ &    $1.3871$ &                                              $d\indices*{*_{A_{3}}^{\tensor*[^{1}]{\hspace{-0.2em}B}{}_{1}^{}}_{X_{2}}^{E_{u}^{}}_{\bar{A}_{1}}^{\tensor*[^{2}]{\hspace{-0.2em}B}{}_{1}^{}}}$ &  $- 0.0294$ \\[0.4em]
              $d\indices*{*_{A_{3}}^{\tensor*[^{1}]{\hspace{-0.2em}B}{}_{1}^{}}_{X_{2}}^{\tensor*[^{1}]{\hspace{-0.2em}E}{}_{u}^{}}_{\bar{A}_{1}}^{\tensor*[^{2}]{\hspace{-0.2em}B}{}_{1}^{}}}$ &  $- 0.4684$ &                                                                               $d\indices*{*_{A_{3}}^{B_{2}^{}}_{X_{2}}^{E_{g}^{}}_{\bar{A}_{1}}^{\tensor*[^{1}]{\hspace{-0.2em}B}{}_{1}^{}}}$ &   $10.3725$ &                                                                              $d\indices*{*_{A_{3}}^{B_{2}^{}}_{X_{2}}^{B_{1u}^{}}_{\bar{A}_{1}}^{\tensor*[^{1}]{\hspace{-0.2em}B}{}_{1}^{}}}$ &    $0.9973$ &                                                                               $d\indices*{*_{A_{3}}^{B_{2}^{}}_{X_{2}}^{E_{u}^{}}_{\bar{A}_{1}}^{\tensor*[^{1}]{\hspace{-0.2em}B}{}_{1}^{}}}$ &    $0.1448$ &                                              $d\indices*{*_{A_{3}}^{B_{2}^{}}_{X_{2}}^{\tensor*[^{1}]{\hspace{-0.2em}E}{}_{u}^{}}_{\bar{A}_{1}}^{\tensor*[^{1}]{\hspace{-0.2em}B}{}_{1}^{}}}$ &  $- 4.6448$ \\[0.4em]
                                               $d\indices*{*_{A_{3}}^{\tensor*[^{1}]{\hspace{-0.2em}B}{}_{2}^{}}_{X_{2}}^{E_{g}^{}}_{\bar{A}_{1}}^{\tensor*[^{1}]{\hspace{-0.2em}B}{}_{1}^{}}}$ &  $- 7.6974$ &                                             $d\indices*{*_{A_{3}}^{\tensor*[^{1}]{\hspace{-0.2em}B}{}_{2}^{}}_{X_{2}}^{B_{1u}^{}}_{\bar{A}_{1}}^{\tensor*[^{1}]{\hspace{-0.2em}B}{}_{1}^{}}}$ &    $0.4129$ &                                              $d\indices*{*_{A_{3}}^{\tensor*[^{1}]{\hspace{-0.2em}B}{}_{2}^{}}_{X_{2}}^{E_{u}^{}}_{\bar{A}_{1}}^{\tensor*[^{1}]{\hspace{-0.2em}B}{}_{1}^{}}}$ &  $- 1.4578$ &             $d\indices*{*_{A_{3}}^{\tensor*[^{1}]{\hspace{-0.2em}B}{}_{2}^{}}_{X_{2}}^{\tensor*[^{1}]{\hspace{-0.2em}E}{}_{u}^{}}_{\bar{A}_{1}}^{\tensor*[^{1}]{\hspace{-0.2em}B}{}_{1}^{}}}$ &    $0.4376$ &                                             $d\indices*{*_{A_{3}}^{\tensor*[^{2}]{\hspace{-0.2em}B}{}_{1}^{}}_{X_{2}}^{A_{1g}^{}}_{\bar{A}_{1}}^{\tensor*[^{2}]{\hspace{-0.2em}B}{}_{1}^{}}}$ &    $3.4043$ \\[0.4em]
                                              $d\indices*{*_{A_{3}}^{\tensor*[^{2}]{\hspace{-0.2em}B}{}_{1}^{}}_{X_{2}}^{A_{2u}^{}}_{\bar{A}_{1}}^{\tensor*[^{2}]{\hspace{-0.2em}B}{}_{1}^{}}}$ &  $- 0.5991$ &                                                                               $d\indices*{*_{A_{3}}^{B_{2}^{}}_{X_{2}}^{E_{g}^{}}_{\bar{A}_{1}}^{\tensor*[^{2}]{\hspace{-0.2em}B}{}_{1}^{}}}$ &   $11.0510$ &                                                                              $d\indices*{*_{A_{3}}^{B_{2}^{}}_{X_{2}}^{B_{1u}^{}}_{\bar{A}_{1}}^{\tensor*[^{2}]{\hspace{-0.2em}B}{}_{1}^{}}}$ &  $- 1.0824$ &                                                                               $d\indices*{*_{A_{3}}^{B_{2}^{}}_{X_{2}}^{E_{u}^{}}_{\bar{A}_{1}}^{\tensor*[^{2}]{\hspace{-0.2em}B}{}_{1}^{}}}$ &    $0.8619$ &                                              $d\indices*{*_{A_{3}}^{B_{2}^{}}_{X_{2}}^{\tensor*[^{1}]{\hspace{-0.2em}E}{}_{u}^{}}_{\bar{A}_{1}}^{\tensor*[^{2}]{\hspace{-0.2em}B}{}_{1}^{}}}$ &    $1.2201$ \\[0.4em]
                                               $d\indices*{*_{A_{3}}^{\tensor*[^{1}]{\hspace{-0.2em}B}{}_{2}^{}}_{X_{2}}^{E_{g}^{}}_{\bar{A}_{1}}^{\tensor*[^{2}]{\hspace{-0.2em}B}{}_{1}^{}}}$ &  $- 9.7728$ &                                             $d\indices*{*_{A_{3}}^{\tensor*[^{1}]{\hspace{-0.2em}B}{}_{2}^{}}_{X_{2}}^{B_{1u}^{}}_{\bar{A}_{1}}^{\tensor*[^{2}]{\hspace{-0.2em}B}{}_{1}^{}}}$ &    $1.4277$ &                                              $d\indices*{*_{A_{3}}^{\tensor*[^{1}]{\hspace{-0.2em}B}{}_{2}^{}}_{X_{2}}^{E_{u}^{}}_{\bar{A}_{1}}^{\tensor*[^{2}]{\hspace{-0.2em}B}{}_{1}^{}}}$ &  $- 4.9178$ &             $d\indices*{*_{A_{3}}^{\tensor*[^{1}]{\hspace{-0.2em}B}{}_{2}^{}}_{X_{2}}^{\tensor*[^{1}]{\hspace{-0.2em}E}{}_{u}^{}}_{\bar{A}_{1}}^{\tensor*[^{2}]{\hspace{-0.2em}B}{}_{1}^{}}}$ &    $1.0596$ &                                                                                                               $d\indices*{*_{A_{3}}^{B_{2}^{}}_{X_{2}}^{A_{1g}^{}}_{\bar{A}_{1}}^{B_{2}^{}}}$ &    $0.6491$ \\[0.4em]
                                                                                                                $d\indices*{*_{A_{3}}^{B_{2}^{}}_{X_{2}}^{A_{2u}^{}}_{\bar{A}_{1}}^{B_{2}^{}}}$ &  $- 3.1510$ &                                                                              $d\indices*{*_{A_{3}}^{B_{2}^{}}_{X_{2}}^{A_{1g}^{}}_{\bar{A}_{1}}^{\tensor*[^{1}]{\hspace{-0.2em}B}{}_{2}^{}}}$ &    $1.0030$ &                                                                               $d\indices*{*_{A_{3}}^{B_{2}^{}}_{X_{2}}^{E_{g}^{}}_{\bar{A}_{1}}^{\tensor*[^{1}]{\hspace{-0.2em}B}{}_{2}^{}}}$ &    $0.2366$ &                                                                              $d\indices*{*_{A_{3}}^{B_{2}^{}}_{X_{2}}^{A_{2u}^{}}_{\bar{A}_{1}}^{\tensor*[^{1}]{\hspace{-0.2em}B}{}_{2}^{}}}$ &  $- 0.5381$ &                                                                               $d\indices*{*_{A_{3}}^{B_{2}^{}}_{X_{2}}^{E_{u}^{}}_{\bar{A}_{1}}^{\tensor*[^{1}]{\hspace{-0.2em}B}{}_{2}^{}}}$ &  $- 0.8124$ \\[0.4em]
                                               $d\indices*{*_{A_{3}}^{B_{2}^{}}_{X_{2}}^{\tensor*[^{1}]{\hspace{-0.2em}E}{}_{u}^{}}_{\bar{A}_{1}}^{\tensor*[^{1}]{\hspace{-0.2em}B}{}_{2}^{}}}$ &    $0.0884$ &                                             $d\indices*{*_{A_{3}}^{\tensor*[^{1}]{\hspace{-0.2em}B}{}_{2}^{}}_{X_{2}}^{A_{1g}^{}}_{\bar{A}_{1}}^{\tensor*[^{1}]{\hspace{-0.2em}B}{}_{2}^{}}}$ &  $- 6.1114$ &                                             $d\indices*{*_{A_{3}}^{\tensor*[^{1}]{\hspace{-0.2em}B}{}_{2}^{}}_{X_{2}}^{A_{2u}^{}}_{\bar{A}_{1}}^{\tensor*[^{1}]{\hspace{-0.2em}B}{}_{2}^{}}}$ &    $2.5651$ &                                                                                                          $d\indices*{*_{A_{5}}^{A_{1}^{}}_{\bar{A}_{4}}^{A_{1}^{}}_{\bar{A}_{1}}^{A_{1}^{}}}$ &    $0.1742$ &                                                                         $d\indices*{*_{A_{5}}^{A_{1}^{}}_{\bar{A}_{4}}^{A_{1}^{}}_{\bar{A}_{1}}^{\tensor*[^{1}]{\hspace{-0.2em}A}{}_{1}^{}}}$ &    $1.3642$ \\[0.4em]
                                                                          $d\indices*{*_{A_{5}}^{A_{1}^{}}_{\bar{A}_{4}}^{A_{1}^{}}_{\bar{A}_{1}}^{\tensor*[^{2}]{\hspace{-0.2em}A}{}_{1}^{}}}$ &   $12.4116$ &                                                                                                          $d\indices*{*_{A_{5}}^{A_{1}^{}}_{\bar{A}_{4}}^{A_{1}^{}}_{\bar{A}_{1}}^{B_{1}^{}}}$ &    $8.7332$ &                                                                         $d\indices*{*_{A_{5}}^{A_{1}^{}}_{\bar{A}_{4}}^{A_{1}^{}}_{\bar{A}_{1}}^{\tensor*[^{1}]{\hspace{-0.2em}B}{}_{1}^{}}}$ &    $2.7576$ &                                                                         $d\indices*{*_{A_{5}}^{A_{1}^{}}_{\bar{A}_{4}}^{A_{1}^{}}_{\bar{A}_{1}}^{\tensor*[^{2}]{\hspace{-0.2em}B}{}_{1}^{}}}$ &  $- 0.0865$ &                                        $d\indices*{*_{A_{5}}^{A_{1}^{}}_{\bar{A}_{4}}^{\tensor*[^{1}]{\hspace{-0.2em}A}{}_{1}^{}}_{\bar{A}_{1}}^{\tensor*[^{1}]{\hspace{-0.2em}A}{}_{1}^{}}}$ &    $0.1244$ \\[0.4em]
                                         $d\indices*{*_{A_{5}}^{A_{1}^{}}_{\bar{A}_{4}}^{\tensor*[^{1}]{\hspace{-0.2em}A}{}_{1}^{}}_{\bar{A}_{1}}^{\tensor*[^{2}]{\hspace{-0.2em}A}{}_{1}^{}}}$ &  $- 4.6882$ &                                                                         $d\indices*{*_{A_{5}}^{A_{1}^{}}_{\bar{A}_{4}}^{\tensor*[^{1}]{\hspace{-0.2em}A}{}_{1}^{}}_{\bar{A}_{1}}^{A_{2}^{}}}$ &    $0.4568$ &                                                                         $d\indices*{*_{A_{5}}^{A_{1}^{}}_{\bar{A}_{4}}^{\tensor*[^{1}]{\hspace{-0.2em}A}{}_{1}^{}}_{\bar{A}_{1}}^{B_{1}^{}}}$ &  $- 8.4546$ &                                        $d\indices*{*_{A_{5}}^{A_{1}^{}}_{\bar{A}_{4}}^{\tensor*[^{1}]{\hspace{-0.2em}A}{}_{1}^{}}_{\bar{A}_{1}}^{\tensor*[^{1}]{\hspace{-0.2em}B}{}_{1}^{}}}$ &    $4.9321$ &                                        $d\indices*{*_{A_{5}}^{A_{1}^{}}_{\bar{A}_{4}}^{\tensor*[^{1}]{\hspace{-0.2em}A}{}_{1}^{}}_{\bar{A}_{1}}^{\tensor*[^{2}]{\hspace{-0.2em}B}{}_{1}^{}}}$ &    $0.4719$ \\[0.4em]
                                                                          $d\indices*{*_{A_{5}}^{A_{1}^{}}_{\bar{A}_{4}}^{\tensor*[^{1}]{\hspace{-0.2em}A}{}_{1}^{}}_{\bar{A}_{1}}^{B_{2}^{}}}$ &  $- 8.2941$ &                                        $d\indices*{*_{A_{5}}^{A_{1}^{}}_{\bar{A}_{4}}^{\tensor*[^{1}]{\hspace{-0.2em}A}{}_{1}^{}}_{\bar{A}_{1}}^{\tensor*[^{1}]{\hspace{-0.2em}B}{}_{2}^{}}}$ &    $5.5210$ &                                        $d\indices*{*_{A_{5}}^{A_{1}^{}}_{\bar{A}_{4}}^{\tensor*[^{2}]{\hspace{-0.2em}A}{}_{1}^{}}_{\bar{A}_{1}}^{\tensor*[^{2}]{\hspace{-0.2em}A}{}_{1}^{}}}$ &    $0.7735$ &                                                                         $d\indices*{*_{A_{5}}^{A_{1}^{}}_{\bar{A}_{4}}^{\tensor*[^{2}]{\hspace{-0.2em}A}{}_{1}^{}}_{\bar{A}_{1}}^{A_{2}^{}}}$ &    $0.1026$ &                                                                         $d\indices*{*_{A_{5}}^{A_{1}^{}}_{\bar{A}_{4}}^{\tensor*[^{2}]{\hspace{-0.2em}A}{}_{1}^{}}_{\bar{A}_{1}}^{B_{1}^{}}}$ &    $0.3846$ \\[0.4em]
                                         $d\indices*{*_{A_{5}}^{A_{1}^{}}_{\bar{A}_{4}}^{\tensor*[^{2}]{\hspace{-0.2em}A}{}_{1}^{}}_{\bar{A}_{1}}^{\tensor*[^{1}]{\hspace{-0.2em}B}{}_{1}^{}}}$ &  $- 6.9581$ &                                        $d\indices*{*_{A_{5}}^{A_{1}^{}}_{\bar{A}_{4}}^{\tensor*[^{2}]{\hspace{-0.2em}A}{}_{1}^{}}_{\bar{A}_{1}}^{\tensor*[^{2}]{\hspace{-0.2em}B}{}_{1}^{}}}$ &  $- 9.1737$ &                                                                         $d\indices*{*_{A_{5}}^{A_{1}^{}}_{\bar{A}_{4}}^{\tensor*[^{2}]{\hspace{-0.2em}A}{}_{1}^{}}_{\bar{A}_{1}}^{B_{2}^{}}}$ &  $- 2.1323$ &                                        $d\indices*{*_{A_{5}}^{A_{1}^{}}_{\bar{A}_{4}}^{\tensor*[^{2}]{\hspace{-0.2em}A}{}_{1}^{}}_{\bar{A}_{1}}^{\tensor*[^{1}]{\hspace{-0.2em}B}{}_{2}^{}}}$ &    $0.2321$ &                                                                                                          $d\indices*{*_{A_{5}}^{A_{1}^{}}_{\bar{A}_{4}}^{A_{2}^{}}_{\bar{A}_{1}}^{A_{2}^{}}}$ &  $- 0.8179$ \\[0.4em]
                                                                                                           $d\indices*{*_{A_{5}}^{A_{1}^{}}_{\bar{A}_{4}}^{A_{2}^{}}_{\bar{A}_{1}}^{B_{1}^{}}}$ &    $5.6248$ &                                                                         $d\indices*{*_{A_{5}}^{A_{1}^{}}_{\bar{A}_{4}}^{A_{2}^{}}_{\bar{A}_{1}}^{\tensor*[^{1}]{\hspace{-0.2em}B}{}_{1}^{}}}$ &  $- 8.1766$ &                                                                         $d\indices*{*_{A_{5}}^{A_{1}^{}}_{\bar{A}_{4}}^{A_{2}^{}}_{\bar{A}_{1}}^{\tensor*[^{2}]{\hspace{-0.2em}B}{}_{1}^{}}}$ &    $0.5028$ &                                                                                                          $d\indices*{*_{A_{5}}^{A_{1}^{}}_{\bar{A}_{4}}^{A_{2}^{}}_{\bar{A}_{1}}^{B_{2}^{}}}$ & $- 12.2678$ &                                                                         $d\indices*{*_{A_{5}}^{A_{1}^{}}_{\bar{A}_{4}}^{A_{2}^{}}_{\bar{A}_{1}}^{\tensor*[^{1}]{\hspace{-0.2em}B}{}_{2}^{}}}$ &    $7.6296$ \\[0.4em]
                                                                                                           $d\indices*{*_{A_{5}}^{A_{1}^{}}_{\bar{A}_{4}}^{B_{1}^{}}_{\bar{A}_{1}}^{B_{1}^{}}}$ &    $0.7914$ &                                                                         $d\indices*{*_{A_{5}}^{A_{1}^{}}_{\bar{A}_{4}}^{B_{1}^{}}_{\bar{A}_{1}}^{\tensor*[^{1}]{\hspace{-0.2em}B}{}_{1}^{}}}$ &    $6.4567$ &                                                                         $d\indices*{*_{A_{5}}^{A_{1}^{}}_{\bar{A}_{4}}^{B_{1}^{}}_{\bar{A}_{1}}^{\tensor*[^{2}]{\hspace{-0.2em}B}{}_{1}^{}}}$ &    $6.3422$ &                                                                                                          $d\indices*{*_{A_{5}}^{A_{1}^{}}_{\bar{A}_{4}}^{B_{2}^{}}_{\bar{A}_{1}}^{B_{1}^{}}}$ &    $0.1795$ &                                                                         $d\indices*{*_{A_{5}}^{A_{1}^{}}_{\bar{A}_{4}}^{\tensor*[^{1}]{\hspace{-0.2em}B}{}_{2}^{}}_{\bar{A}_{1}}^{B_{1}^{}}}$ &    $0.8947$ \\[0.4em]
                                         $d\indices*{*_{A_{5}}^{A_{1}^{}}_{\bar{A}_{4}}^{\tensor*[^{1}]{\hspace{-0.2em}B}{}_{1}^{}}_{\bar{A}_{1}}^{\tensor*[^{1}]{\hspace{-0.2em}B}{}_{1}^{}}}$ &    $2.2026$ &                                        $d\indices*{*_{A_{5}}^{A_{1}^{}}_{\bar{A}_{4}}^{\tensor*[^{1}]{\hspace{-0.2em}B}{}_{1}^{}}_{\bar{A}_{1}}^{\tensor*[^{2}]{\hspace{-0.2em}B}{}_{1}^{}}}$ &    $0.3772$ &                                                                         $d\indices*{*_{A_{5}}^{A_{1}^{}}_{\bar{A}_{4}}^{B_{2}^{}}_{\bar{A}_{1}}^{\tensor*[^{1}]{\hspace{-0.2em}B}{}_{1}^{}}}$ &  $- 0.0186$ &                                        $d\indices*{*_{A_{5}}^{A_{1}^{}}_{\bar{A}_{4}}^{\tensor*[^{1}]{\hspace{-0.2em}B}{}_{2}^{}}_{\bar{A}_{1}}^{\tensor*[^{1}]{\hspace{-0.2em}B}{}_{1}^{}}}$ &    $1.2308$ &                                        $d\indices*{*_{A_{5}}^{A_{1}^{}}_{\bar{A}_{4}}^{\tensor*[^{2}]{\hspace{-0.2em}B}{}_{1}^{}}_{\bar{A}_{1}}^{\tensor*[^{2}]{\hspace{-0.2em}B}{}_{1}^{}}}$ &  $- 1.5175$ \\[0.4em]
                                                                          $d\indices*{*_{A_{5}}^{A_{1}^{}}_{\bar{A}_{4}}^{B_{2}^{}}_{\bar{A}_{1}}^{\tensor*[^{2}]{\hspace{-0.2em}B}{}_{1}^{}}}$ &    $0.0601$ &                                        $d\indices*{*_{A_{5}}^{A_{1}^{}}_{\bar{A}_{4}}^{\tensor*[^{1}]{\hspace{-0.2em}B}{}_{2}^{}}_{\bar{A}_{1}}^{\tensor*[^{2}]{\hspace{-0.2em}B}{}_{1}^{}}}$ &  $- 0.5115$ &                                                                                                          $d\indices*{*_{A_{5}}^{A_{1}^{}}_{\bar{A}_{4}}^{B_{2}^{}}_{\bar{A}_{1}}^{B_{2}^{}}}$ &  $- 0.7401$ &                                                                         $d\indices*{*_{A_{5}}^{A_{1}^{}}_{\bar{A}_{4}}^{B_{2}^{}}_{\bar{A}_{1}}^{\tensor*[^{1}]{\hspace{-0.2em}B}{}_{2}^{}}}$ &    $0.9256$ &                                        $d\indices*{*_{A_{5}}^{A_{1}^{}}_{\bar{A}_{4}}^{\tensor*[^{1}]{\hspace{-0.2em}B}{}_{2}^{}}_{\bar{A}_{1}}^{\tensor*[^{1}]{\hspace{-0.2em}B}{}_{2}^{}}}$ &  $- 1.0077$ \\[0.4em]
        $d\indices*{*_{A_{5}}^{\tensor*[^{1}]{\hspace{-0.2em}A}{}_{1}^{}}_{\bar{A}_{4}}^{\tensor*[^{1}]{\hspace{-0.2em}A}{}_{1}^{}}_{\bar{A}_{1}}^{\tensor*[^{1}]{\hspace{-0.2em}A}{}_{1}^{}}}$ &  $- 0.3719$ &       $d\indices*{*_{A_{5}}^{\tensor*[^{1}]{\hspace{-0.2em}A}{}_{1}^{}}_{\bar{A}_{4}}^{\tensor*[^{1}]{\hspace{-0.2em}A}{}_{1}^{}}_{\bar{A}_{1}}^{\tensor*[^{2}]{\hspace{-0.2em}A}{}_{1}^{}}}$ &    $0.3674$ &                                        $d\indices*{*_{A_{5}}^{\tensor*[^{1}]{\hspace{-0.2em}A}{}_{1}^{}}_{\bar{A}_{4}}^{\tensor*[^{1}]{\hspace{-0.2em}A}{}_{1}^{}}_{\bar{A}_{1}}^{B_{1}^{}}}$ &    $9.9220$ &       $d\indices*{*_{A_{5}}^{\tensor*[^{1}]{\hspace{-0.2em}A}{}_{1}^{}}_{\bar{A}_{4}}^{\tensor*[^{1}]{\hspace{-0.2em}A}{}_{1}^{}}_{\bar{A}_{1}}^{\tensor*[^{1}]{\hspace{-0.2em}B}{}_{1}^{}}}$ &  $- 5.8812$ &       $d\indices*{*_{A_{5}}^{\tensor*[^{1}]{\hspace{-0.2em}A}{}_{1}^{}}_{\bar{A}_{4}}^{\tensor*[^{1}]{\hspace{-0.2em}A}{}_{1}^{}}_{\bar{A}_{1}}^{\tensor*[^{2}]{\hspace{-0.2em}B}{}_{1}^{}}}$ &  $- 1.5818$ \\[0.4em]
        $d\indices*{*_{A_{5}}^{\tensor*[^{1}]{\hspace{-0.2em}A}{}_{1}^{}}_{\bar{A}_{4}}^{\tensor*[^{2}]{\hspace{-0.2em}A}{}_{1}^{}}_{\bar{A}_{1}}^{\tensor*[^{2}]{\hspace{-0.2em}A}{}_{1}^{}}}$ &    $2.1758$ &                                        $d\indices*{*_{A_{5}}^{\tensor*[^{1}]{\hspace{-0.2em}A}{}_{1}^{}}_{\bar{A}_{4}}^{\tensor*[^{2}]{\hspace{-0.2em}A}{}_{1}^{}}_{\bar{A}_{1}}^{A_{2}^{}}}$ &  $- 6.0437$ &                                        $d\indices*{*_{A_{5}}^{\tensor*[^{1}]{\hspace{-0.2em}A}{}_{1}^{}}_{\bar{A}_{4}}^{\tensor*[^{2}]{\hspace{-0.2em}A}{}_{1}^{}}_{\bar{A}_{1}}^{B_{1}^{}}}$ &  $- 1.0055$ &       $d\indices*{*_{A_{5}}^{\tensor*[^{1}]{\hspace{-0.2em}A}{}_{1}^{}}_{\bar{A}_{4}}^{\tensor*[^{2}]{\hspace{-0.2em}A}{}_{1}^{}}_{\bar{A}_{1}}^{\tensor*[^{1}]{\hspace{-0.2em}B}{}_{1}^{}}}$ &    $3.8205$ &       $d\indices*{*_{A_{5}}^{\tensor*[^{1}]{\hspace{-0.2em}A}{}_{1}^{}}_{\bar{A}_{4}}^{\tensor*[^{2}]{\hspace{-0.2em}A}{}_{1}^{}}_{\bar{A}_{1}}^{\tensor*[^{2}]{\hspace{-0.2em}B}{}_{1}^{}}}$ &    $4.7179$ \\[0.4em]
                                         $d\indices*{*_{A_{5}}^{\tensor*[^{1}]{\hspace{-0.2em}A}{}_{1}^{}}_{\bar{A}_{4}}^{\tensor*[^{2}]{\hspace{-0.2em}A}{}_{1}^{}}_{\bar{A}_{1}}^{B_{2}^{}}}$ &  $- 0.3262$ &       $d\indices*{*_{A_{5}}^{\tensor*[^{1}]{\hspace{-0.2em}A}{}_{1}^{}}_{\bar{A}_{4}}^{\tensor*[^{2}]{\hspace{-0.2em}A}{}_{1}^{}}_{\bar{A}_{1}}^{\tensor*[^{1}]{\hspace{-0.2em}B}{}_{2}^{}}}$ &    $4.3167$ &                                                                         $d\indices*{*_{A_{5}}^{\tensor*[^{1}]{\hspace{-0.2em}A}{}_{1}^{}}_{\bar{A}_{4}}^{A_{2}^{}}_{\bar{A}_{1}}^{A_{2}^{}}}$ &    $7.5207$ &                                                                         $d\indices*{*_{A_{5}}^{\tensor*[^{1}]{\hspace{-0.2em}A}{}_{1}^{}}_{\bar{A}_{4}}^{A_{2}^{}}_{\bar{A}_{1}}^{B_{1}^{}}}$ &    $0.0445$ &                                        $d\indices*{*_{A_{5}}^{\tensor*[^{1}]{\hspace{-0.2em}A}{}_{1}^{}}_{\bar{A}_{4}}^{A_{2}^{}}_{\bar{A}_{1}}^{\tensor*[^{1}]{\hspace{-0.2em}B}{}_{1}^{}}}$ &    $3.6572$ \\[0.4em]
                                         $d\indices*{*_{A_{5}}^{\tensor*[^{1}]{\hspace{-0.2em}A}{}_{1}^{}}_{\bar{A}_{4}}^{A_{2}^{}}_{\bar{A}_{1}}^{\tensor*[^{2}]{\hspace{-0.2em}B}{}_{1}^{}}}$ &    $4.2530$ &                                                                         $d\indices*{*_{A_{5}}^{\tensor*[^{1}]{\hspace{-0.2em}A}{}_{1}^{}}_{\bar{A}_{4}}^{A_{2}^{}}_{\bar{A}_{1}}^{B_{2}^{}}}$ &    $4.7230$ &                                        $d\indices*{*_{A_{5}}^{\tensor*[^{1}]{\hspace{-0.2em}A}{}_{1}^{}}_{\bar{A}_{4}}^{A_{2}^{}}_{\bar{A}_{1}}^{\tensor*[^{1}]{\hspace{-0.2em}B}{}_{2}^{}}}$ &  $- 3.7514$ &                                                                         $d\indices*{*_{A_{5}}^{\tensor*[^{1}]{\hspace{-0.2em}A}{}_{1}^{}}_{\bar{A}_{4}}^{B_{1}^{}}_{\bar{A}_{1}}^{B_{1}^{}}}$ &    $0.3475$ &                                        $d\indices*{*_{A_{5}}^{\tensor*[^{1}]{\hspace{-0.2em}A}{}_{1}^{}}_{\bar{A}_{4}}^{B_{1}^{}}_{\bar{A}_{1}}^{\tensor*[^{1}]{\hspace{-0.2em}B}{}_{1}^{}}}$ &    $0.3794$ \\[0.4em]
                                         $d\indices*{*_{A_{5}}^{\tensor*[^{1}]{\hspace{-0.2em}A}{}_{1}^{}}_{\bar{A}_{4}}^{B_{1}^{}}_{\bar{A}_{1}}^{\tensor*[^{2}]{\hspace{-0.2em}B}{}_{1}^{}}}$ &    $0.3131$ &                                                                         $d\indices*{*_{A_{5}}^{\tensor*[^{1}]{\hspace{-0.2em}A}{}_{1}^{}}_{\bar{A}_{4}}^{B_{2}^{}}_{\bar{A}_{1}}^{B_{1}^{}}}$ &  $- 1.0558$ &                                        $d\indices*{*_{A_{5}}^{\tensor*[^{1}]{\hspace{-0.2em}A}{}_{1}^{}}_{\bar{A}_{4}}^{\tensor*[^{1}]{\hspace{-0.2em}B}{}_{2}^{}}_{\bar{A}_{1}}^{B_{1}^{}}}$ &  $- 0.3316$ &       $d\indices*{*_{A_{5}}^{\tensor*[^{1}]{\hspace{-0.2em}A}{}_{1}^{}}_{\bar{A}_{4}}^{\tensor*[^{1}]{\hspace{-0.2em}B}{}_{1}^{}}_{\bar{A}_{1}}^{\tensor*[^{1}]{\hspace{-0.2em}B}{}_{1}^{}}}$ &  $- 6.2669$ &       $d\indices*{*_{A_{5}}^{\tensor*[^{1}]{\hspace{-0.2em}A}{}_{1}^{}}_{\bar{A}_{4}}^{\tensor*[^{1}]{\hspace{-0.2em}B}{}_{1}^{}}_{\bar{A}_{1}}^{\tensor*[^{2}]{\hspace{-0.2em}B}{}_{1}^{}}}$ &  $- 3.2965$ \\[0.4em]
                                         $d\indices*{*_{A_{5}}^{\tensor*[^{1}]{\hspace{-0.2em}A}{}_{1}^{}}_{\bar{A}_{4}}^{B_{2}^{}}_{\bar{A}_{1}}^{\tensor*[^{1}]{\hspace{-0.2em}B}{}_{1}^{}}}$ &  $- 6.1459$ &       $d\indices*{*_{A_{5}}^{\tensor*[^{1}]{\hspace{-0.2em}A}{}_{1}^{}}_{\bar{A}_{4}}^{\tensor*[^{1}]{\hspace{-0.2em}B}{}_{2}^{}}_{\bar{A}_{1}}^{\tensor*[^{1}]{\hspace{-0.2em}B}{}_{1}^{}}}$ &    $3.6016$ &       $d\indices*{*_{A_{5}}^{\tensor*[^{1}]{\hspace{-0.2em}A}{}_{1}^{}}_{\bar{A}_{4}}^{\tensor*[^{2}]{\hspace{-0.2em}B}{}_{1}^{}}_{\bar{A}_{1}}^{\tensor*[^{2}]{\hspace{-0.2em}B}{}_{1}^{}}}$ &  $- 2.1745$ &                                        $d\indices*{*_{A_{5}}^{\tensor*[^{1}]{\hspace{-0.2em}A}{}_{1}^{}}_{\bar{A}_{4}}^{B_{2}^{}}_{\bar{A}_{1}}^{\tensor*[^{2}]{\hspace{-0.2em}B}{}_{1}^{}}}$ &  $- 5.1911$ &       $d\indices*{*_{A_{5}}^{\tensor*[^{1}]{\hspace{-0.2em}A}{}_{1}^{}}_{\bar{A}_{4}}^{\tensor*[^{1}]{\hspace{-0.2em}B}{}_{2}^{}}_{\bar{A}_{1}}^{\tensor*[^{2}]{\hspace{-0.2em}B}{}_{1}^{}}}$ &    $4.0541$ \\[0.4em]
                                                                          $d\indices*{*_{A_{5}}^{\tensor*[^{1}]{\hspace{-0.2em}A}{}_{1}^{}}_{\bar{A}_{4}}^{B_{2}^{}}_{\bar{A}_{1}}^{B_{2}^{}}}$ &    $0.0192$ &                                        $d\indices*{*_{A_{5}}^{\tensor*[^{1}]{\hspace{-0.2em}A}{}_{1}^{}}_{\bar{A}_{4}}^{B_{2}^{}}_{\bar{A}_{1}}^{\tensor*[^{1}]{\hspace{-0.2em}B}{}_{2}^{}}}$ &    $0.4362$ &       $d\indices*{*_{A_{5}}^{\tensor*[^{1}]{\hspace{-0.2em}A}{}_{1}^{}}_{\bar{A}_{4}}^{\tensor*[^{1}]{\hspace{-0.2em}B}{}_{2}^{}}_{\bar{A}_{1}}^{\tensor*[^{1}]{\hspace{-0.2em}B}{}_{2}^{}}}$ &  $- 6.4996$ &       $d\indices*{*_{A_{5}}^{\tensor*[^{2}]{\hspace{-0.2em}A}{}_{1}^{}}_{\bar{A}_{4}}^{\tensor*[^{2}]{\hspace{-0.2em}A}{}_{1}^{}}_{\bar{A}_{1}}^{\tensor*[^{2}]{\hspace{-0.2em}A}{}_{1}^{}}}$ &  $- 9.3907$ &                                        $d\indices*{*_{A_{5}}^{\tensor*[^{2}]{\hspace{-0.2em}A}{}_{1}^{}}_{\bar{A}_{4}}^{\tensor*[^{2}]{\hspace{-0.2em}A}{}_{1}^{}}_{\bar{A}_{1}}^{B_{1}^{}}}$ &  $- 4.5030$ \\[0.4em]
        $d\indices*{*_{A_{5}}^{\tensor*[^{2}]{\hspace{-0.2em}A}{}_{1}^{}}_{\bar{A}_{4}}^{\tensor*[^{2}]{\hspace{-0.2em}A}{}_{1}^{}}_{\bar{A}_{1}}^{\tensor*[^{1}]{\hspace{-0.2em}B}{}_{1}^{}}}$ &    $1.2290$ &       $d\indices*{*_{A_{5}}^{\tensor*[^{2}]{\hspace{-0.2em}A}{}_{1}^{}}_{\bar{A}_{4}}^{\tensor*[^{2}]{\hspace{-0.2em}A}{}_{1}^{}}_{\bar{A}_{1}}^{\tensor*[^{2}]{\hspace{-0.2em}B}{}_{1}^{}}}$ &  $- 1.3159$ &                                                                         $d\indices*{*_{A_{5}}^{\tensor*[^{2}]{\hspace{-0.2em}A}{}_{1}^{}}_{\bar{A}_{4}}^{A_{2}^{}}_{\bar{A}_{1}}^{A_{2}^{}}}$ &  $- 7.3613$ &                                                                         $d\indices*{*_{A_{5}}^{\tensor*[^{2}]{\hspace{-0.2em}A}{}_{1}^{}}_{\bar{A}_{4}}^{A_{2}^{}}_{\bar{A}_{1}}^{B_{1}^{}}}$ &    $1.5430$ &                                        $d\indices*{*_{A_{5}}^{\tensor*[^{2}]{\hspace{-0.2em}A}{}_{1}^{}}_{\bar{A}_{4}}^{A_{2}^{}}_{\bar{A}_{1}}^{\tensor*[^{1}]{\hspace{-0.2em}B}{}_{1}^{}}}$ &  $- 0.8319$ \\[0.4em]
                                         $d\indices*{*_{A_{5}}^{\tensor*[^{2}]{\hspace{-0.2em}A}{}_{1}^{}}_{\bar{A}_{4}}^{A_{2}^{}}_{\bar{A}_{1}}^{\tensor*[^{2}]{\hspace{-0.2em}B}{}_{1}^{}}}$ &  $- 0.1318$ &                                                                         $d\indices*{*_{A_{5}}^{\tensor*[^{2}]{\hspace{-0.2em}A}{}_{1}^{}}_{\bar{A}_{4}}^{A_{2}^{}}_{\bar{A}_{1}}^{B_{2}^{}}}$ &  $- 0.8570$ &                                        $d\indices*{*_{A_{5}}^{\tensor*[^{2}]{\hspace{-0.2em}A}{}_{1}^{}}_{\bar{A}_{4}}^{A_{2}^{}}_{\bar{A}_{1}}^{\tensor*[^{1}]{\hspace{-0.2em}B}{}_{2}^{}}}$ &    $2.6112$ &                                                                         $d\indices*{*_{A_{5}}^{\tensor*[^{2}]{\hspace{-0.2em}A}{}_{1}^{}}_{\bar{A}_{4}}^{B_{1}^{}}_{\bar{A}_{1}}^{B_{1}^{}}}$ &    $0.6960$ &                                        $d\indices*{*_{A_{5}}^{\tensor*[^{2}]{\hspace{-0.2em}A}{}_{1}^{}}_{\bar{A}_{4}}^{B_{1}^{}}_{\bar{A}_{1}}^{\tensor*[^{1}]{\hspace{-0.2em}B}{}_{1}^{}}}$ &    $4.7523$ \\[0.4em]
                                         $d\indices*{*_{A_{5}}^{\tensor*[^{2}]{\hspace{-0.2em}A}{}_{1}^{}}_{\bar{A}_{4}}^{B_{1}^{}}_{\bar{A}_{1}}^{\tensor*[^{2}]{\hspace{-0.2em}B}{}_{1}^{}}}$ &    $1.2283$ &                                                                         $d\indices*{*_{A_{5}}^{\tensor*[^{2}]{\hspace{-0.2em}A}{}_{1}^{}}_{\bar{A}_{4}}^{B_{2}^{}}_{\bar{A}_{1}}^{B_{1}^{}}}$ &  $- 6.2679$ &                                        $d\indices*{*_{A_{5}}^{\tensor*[^{2}]{\hspace{-0.2em}A}{}_{1}^{}}_{\bar{A}_{4}}^{\tensor*[^{1}]{\hspace{-0.2em}B}{}_{2}^{}}_{\bar{A}_{1}}^{B_{1}^{}}}$ &    $5.0251$ &       $d\indices*{*_{A_{5}}^{\tensor*[^{2}]{\hspace{-0.2em}A}{}_{1}^{}}_{\bar{A}_{4}}^{\tensor*[^{1}]{\hspace{-0.2em}B}{}_{1}^{}}_{\bar{A}_{1}}^{\tensor*[^{1}]{\hspace{-0.2em}B}{}_{1}^{}}}$ & $- 12.9480$ &       $d\indices*{*_{A_{5}}^{\tensor*[^{2}]{\hspace{-0.2em}A}{}_{1}^{}}_{\bar{A}_{4}}^{\tensor*[^{1}]{\hspace{-0.2em}B}{}_{1}^{}}_{\bar{A}_{1}}^{\tensor*[^{2}]{\hspace{-0.2em}B}{}_{1}^{}}}$ &  $- 4.9044$ \\[0.4em]
                                         $d\indices*{*_{A_{5}}^{\tensor*[^{2}]{\hspace{-0.2em}A}{}_{1}^{}}_{\bar{A}_{4}}^{B_{2}^{}}_{\bar{A}_{1}}^{\tensor*[^{1}]{\hspace{-0.2em}B}{}_{1}^{}}}$ &    $6.7721$ &       $d\indices*{*_{A_{5}}^{\tensor*[^{2}]{\hspace{-0.2em}A}{}_{1}^{}}_{\bar{A}_{4}}^{\tensor*[^{1}]{\hspace{-0.2em}B}{}_{2}^{}}_{\bar{A}_{1}}^{\tensor*[^{1}]{\hspace{-0.2em}B}{}_{1}^{}}}$ &  $- 5.0219$ &       $d\indices*{*_{A_{5}}^{\tensor*[^{2}]{\hspace{-0.2em}A}{}_{1}^{}}_{\bar{A}_{4}}^{\tensor*[^{2}]{\hspace{-0.2em}B}{}_{1}^{}}_{\bar{A}_{1}}^{\tensor*[^{2}]{\hspace{-0.2em}B}{}_{1}^{}}}$ &  $- 7.0715$ &                                        $d\indices*{*_{A_{5}}^{\tensor*[^{2}]{\hspace{-0.2em}A}{}_{1}^{}}_{\bar{A}_{4}}^{B_{2}^{}}_{\bar{A}_{1}}^{\tensor*[^{2}]{\hspace{-0.2em}B}{}_{1}^{}}}$ &  $- 0.7473$ &       $d\indices*{*_{A_{5}}^{\tensor*[^{2}]{\hspace{-0.2em}A}{}_{1}^{}}_{\bar{A}_{4}}^{\tensor*[^{1}]{\hspace{-0.2em}B}{}_{2}^{}}_{\bar{A}_{1}}^{\tensor*[^{2}]{\hspace{-0.2em}B}{}_{1}^{}}}$ &    $0.3591$ \\[0.4em]
                                                                          $d\indices*{*_{A_{5}}^{\tensor*[^{2}]{\hspace{-0.2em}A}{}_{1}^{}}_{\bar{A}_{4}}^{B_{2}^{}}_{\bar{A}_{1}}^{B_{2}^{}}}$ & $- 11.7522$ &                                        $d\indices*{*_{A_{5}}^{\tensor*[^{2}]{\hspace{-0.2em}A}{}_{1}^{}}_{\bar{A}_{4}}^{B_{2}^{}}_{\bar{A}_{1}}^{\tensor*[^{1}]{\hspace{-0.2em}B}{}_{2}^{}}}$ &    $7.5284$ &       $d\indices*{*_{A_{5}}^{\tensor*[^{2}]{\hspace{-0.2em}A}{}_{1}^{}}_{\bar{A}_{4}}^{\tensor*[^{1}]{\hspace{-0.2em}B}{}_{2}^{}}_{\bar{A}_{1}}^{\tensor*[^{1}]{\hspace{-0.2em}B}{}_{2}^{}}}$ &  $- 6.0740$ &                                                                                                          $d\indices*{*_{A_{5}}^{A_{2}^{}}_{\bar{A}_{4}}^{A_{2}^{}}_{\bar{A}_{1}}^{B_{1}^{}}}$ &    $4.3430$ &                                                                         $d\indices*{*_{A_{5}}^{A_{2}^{}}_{\bar{A}_{4}}^{A_{2}^{}}_{\bar{A}_{1}}^{\tensor*[^{1}]{\hspace{-0.2em}B}{}_{1}^{}}}$ &  $- 0.6644$ \\[0.4em]
                                                                          $d\indices*{*_{A_{5}}^{A_{2}^{}}_{\bar{A}_{4}}^{A_{2}^{}}_{\bar{A}_{1}}^{\tensor*[^{2}]{\hspace{-0.2em}B}{}_{1}^{}}}$ &  $- 2.0919$ &                                                                         $d\indices*{*_{A_{5}}^{A_{2}^{}}_{\bar{A}_{4}}^{B_{1}^{}}_{\bar{A}_{1}}^{\tensor*[^{1}]{\hspace{-0.2em}B}{}_{1}^{}}}$ &    $4.1441$ &                                                                         $d\indices*{*_{A_{5}}^{A_{2}^{}}_{\bar{A}_{4}}^{B_{1}^{}}_{\bar{A}_{1}}^{\tensor*[^{2}]{\hspace{-0.2em}B}{}_{1}^{}}}$ &    $4.4356$ &                                                                                                          $d\indices*{*_{A_{5}}^{A_{2}^{}}_{\bar{A}_{4}}^{B_{2}^{}}_{\bar{A}_{1}}^{B_{1}^{}}}$ &  $- 1.0262$ &                                                                         $d\indices*{*_{A_{5}}^{A_{2}^{}}_{\bar{A}_{4}}^{\tensor*[^{1}]{\hspace{-0.2em}B}{}_{2}^{}}_{\bar{A}_{1}}^{B_{1}^{}}}$ &    $0.2311$ \\[0.4em]
                                         $d\indices*{*_{A_{5}}^{A_{2}^{}}_{\bar{A}_{4}}^{\tensor*[^{1}]{\hspace{-0.2em}B}{}_{1}^{}}_{\bar{A}_{1}}^{\tensor*[^{2}]{\hspace{-0.2em}B}{}_{1}^{}}}$ &  $- 6.8250$ &                                                                         $d\indices*{*_{A_{5}}^{A_{2}^{}}_{\bar{A}_{4}}^{B_{2}^{}}_{\bar{A}_{1}}^{\tensor*[^{1}]{\hspace{-0.2em}B}{}_{1}^{}}}$ &    $8.6642$ &                                        $d\indices*{*_{A_{5}}^{A_{2}^{}}_{\bar{A}_{4}}^{\tensor*[^{1}]{\hspace{-0.2em}B}{}_{2}^{}}_{\bar{A}_{1}}^{\tensor*[^{1}]{\hspace{-0.2em}B}{}_{1}^{}}}$ &  $- 5.3262$ &                                                                         $d\indices*{*_{A_{5}}^{A_{2}^{}}_{\bar{A}_{4}}^{B_{2}^{}}_{\bar{A}_{1}}^{\tensor*[^{2}]{\hspace{-0.2em}B}{}_{1}^{}}}$ &    $8.4716$ &                                        $d\indices*{*_{A_{5}}^{A_{2}^{}}_{\bar{A}_{4}}^{\tensor*[^{1}]{\hspace{-0.2em}B}{}_{2}^{}}_{\bar{A}_{1}}^{\tensor*[^{2}]{\hspace{-0.2em}B}{}_{1}^{}}}$ &  $- 6.2593$ \\[0.4em]
                                                                          $d\indices*{*_{A_{5}}^{A_{2}^{}}_{\bar{A}_{4}}^{B_{2}^{}}_{\bar{A}_{1}}^{\tensor*[^{1}]{\hspace{-0.2em}B}{}_{2}^{}}}$ &  $- 0.6310$ &                                                                                                          $d\indices*{*_{A_{5}}^{B_{1}^{}}_{\bar{A}_{4}}^{B_{1}^{}}_{\bar{A}_{1}}^{B_{1}^{}}}$ &   $10.9179$ &                                                                         $d\indices*{*_{A_{5}}^{B_{1}^{}}_{\bar{A}_{4}}^{B_{1}^{}}_{\bar{A}_{1}}^{\tensor*[^{1}]{\hspace{-0.2em}B}{}_{1}^{}}}$ &  $- 6.7567$ &                                                                         $d\indices*{*_{A_{5}}^{B_{1}^{}}_{\bar{A}_{4}}^{B_{1}^{}}_{\bar{A}_{1}}^{\tensor*[^{2}]{\hspace{-0.2em}B}{}_{1}^{}}}$ &  $- 0.4612$ &                                        $d\indices*{*_{A_{5}}^{B_{1}^{}}_{\bar{A}_{4}}^{\tensor*[^{1}]{\hspace{-0.2em}B}{}_{1}^{}}_{\bar{A}_{1}}^{\tensor*[^{1}]{\hspace{-0.2em}B}{}_{1}^{}}}$ &  $- 0.8155$ \\[0.4em]
                                         $d\indices*{*_{A_{5}}^{B_{1}^{}}_{\bar{A}_{4}}^{\tensor*[^{1}]{\hspace{-0.2em}B}{}_{1}^{}}_{\bar{A}_{1}}^{\tensor*[^{2}]{\hspace{-0.2em}B}{}_{1}^{}}}$ &  $- 4.2854$ &                                                                         $d\indices*{*_{A_{5}}^{B_{2}^{}}_{\bar{A}_{4}}^{B_{1}^{}}_{\bar{A}_{1}}^{\tensor*[^{1}]{\hspace{-0.2em}B}{}_{1}^{}}}$ &  $- 5.8654$ &                                        $d\indices*{*_{A_{5}}^{\tensor*[^{1}]{\hspace{-0.2em}B}{}_{2}^{}}_{\bar{A}_{4}}^{B_{1}^{}}_{\bar{A}_{1}}^{\tensor*[^{1}]{\hspace{-0.2em}B}{}_{1}^{}}}$ &    $4.3714$ &                                        $d\indices*{*_{A_{5}}^{B_{1}^{}}_{\bar{A}_{4}}^{\tensor*[^{2}]{\hspace{-0.2em}B}{}_{1}^{}}_{\bar{A}_{1}}^{\tensor*[^{2}]{\hspace{-0.2em}B}{}_{1}^{}}}$ &  $- 4.9717$ &                                                                         $d\indices*{*_{A_{5}}^{B_{2}^{}}_{\bar{A}_{4}}^{B_{1}^{}}_{\bar{A}_{1}}^{\tensor*[^{2}]{\hspace{-0.2em}B}{}_{1}^{}}}$ &    $2.6861$ \\[0.4em]
                                         $d\indices*{*_{A_{5}}^{\tensor*[^{1}]{\hspace{-0.2em}B}{}_{2}^{}}_{\bar{A}_{4}}^{B_{1}^{}}_{\bar{A}_{1}}^{\tensor*[^{2}]{\hspace{-0.2em}B}{}_{1}^{}}}$ &  $- 0.8733$ &                                                                                                          $d\indices*{*_{A_{5}}^{B_{2}^{}}_{\bar{A}_{4}}^{B_{2}^{}}_{\bar{A}_{1}}^{B_{1}^{}}}$ &    $7.3082$ &                                                                         $d\indices*{*_{A_{5}}^{B_{2}^{}}_{\bar{A}_{4}}^{\tensor*[^{1}]{\hspace{-0.2em}B}{}_{2}^{}}_{\bar{A}_{1}}^{B_{1}^{}}}$ &  $- 4.3852$ &                                        $d\indices*{*_{A_{5}}^{\tensor*[^{1}]{\hspace{-0.2em}B}{}_{2}^{}}_{\bar{A}_{4}}^{\tensor*[^{1}]{\hspace{-0.2em}B}{}_{2}^{}}_{\bar{A}_{1}}^{B_{1}^{}}}$ &    $4.1340$ &       $d\indices*{*_{A_{5}}^{\tensor*[^{1}]{\hspace{-0.2em}B}{}_{1}^{}}_{\bar{A}_{4}}^{\tensor*[^{1}]{\hspace{-0.2em}B}{}_{1}^{}}_{\bar{A}_{1}}^{\tensor*[^{1}]{\hspace{-0.2em}B}{}_{1}^{}}}$ &  $- 1.0674$ \\[0.4em]
        $d\indices*{*_{A_{5}}^{\tensor*[^{1}]{\hspace{-0.2em}B}{}_{1}^{}}_{\bar{A}_{4}}^{\tensor*[^{1}]{\hspace{-0.2em}B}{}_{1}^{}}_{\bar{A}_{1}}^{\tensor*[^{2}]{\hspace{-0.2em}B}{}_{1}^{}}}$ &  $- 1.3268$ &       $d\indices*{*_{A_{5}}^{\tensor*[^{1}]{\hspace{-0.2em}B}{}_{1}^{}}_{\bar{A}_{4}}^{\tensor*[^{2}]{\hspace{-0.2em}B}{}_{1}^{}}_{\bar{A}_{1}}^{\tensor*[^{2}]{\hspace{-0.2em}B}{}_{1}^{}}}$ &    $1.1572$ &                                        $d\indices*{*_{A_{5}}^{B_{2}^{}}_{\bar{A}_{4}}^{\tensor*[^{1}]{\hspace{-0.2em}B}{}_{1}^{}}_{\bar{A}_{1}}^{\tensor*[^{2}]{\hspace{-0.2em}B}{}_{1}^{}}}$ &  $- 0.2777$ &       $d\indices*{*_{A_{5}}^{\tensor*[^{1}]{\hspace{-0.2em}B}{}_{2}^{}}_{\bar{A}_{4}}^{\tensor*[^{1}]{\hspace{-0.2em}B}{}_{1}^{}}_{\bar{A}_{1}}^{\tensor*[^{2}]{\hspace{-0.2em}B}{}_{1}^{}}}$ &    $0.5533$ &                                                                         $d\indices*{*_{A_{5}}^{B_{2}^{}}_{\bar{A}_{4}}^{B_{2}^{}}_{\bar{A}_{1}}^{\tensor*[^{1}]{\hspace{-0.2em}B}{}_{1}^{}}}$ &    $1.0769$ \\[0.4em]
                                         $d\indices*{*_{A_{5}}^{B_{2}^{}}_{\bar{A}_{4}}^{\tensor*[^{1}]{\hspace{-0.2em}B}{}_{2}^{}}_{\bar{A}_{1}}^{\tensor*[^{1}]{\hspace{-0.2em}B}{}_{1}^{}}}$ &  $- 1.0676$ &       $d\indices*{*_{A_{5}}^{\tensor*[^{1}]{\hspace{-0.2em}B}{}_{2}^{}}_{\bar{A}_{4}}^{\tensor*[^{1}]{\hspace{-0.2em}B}{}_{2}^{}}_{\bar{A}_{1}}^{\tensor*[^{1}]{\hspace{-0.2em}B}{}_{1}^{}}}$ &  $- 0.7442$ &       $d\indices*{*_{A_{5}}^{\tensor*[^{2}]{\hspace{-0.2em}B}{}_{1}^{}}_{\bar{A}_{4}}^{\tensor*[^{2}]{\hspace{-0.2em}B}{}_{1}^{}}_{\bar{A}_{1}}^{\tensor*[^{2}]{\hspace{-0.2em}B}{}_{1}^{}}}$ &    $4.6697$ &                                                                         $d\indices*{*_{A_{5}}^{B_{2}^{}}_{\bar{A}_{4}}^{B_{2}^{}}_{\bar{A}_{1}}^{\tensor*[^{2}]{\hspace{-0.2em}B}{}_{1}^{}}}$ &  $- 0.8982$ &                                        $d\indices*{*_{A_{5}}^{B_{2}^{}}_{\bar{A}_{4}}^{\tensor*[^{1}]{\hspace{-0.2em}B}{}_{2}^{}}_{\bar{A}_{1}}^{\tensor*[^{2}]{\hspace{-0.2em}B}{}_{1}^{}}}$ &    $0.2613$ \\[0.4em]
        $d\indices*{*_{A_{5}}^{\tensor*[^{1}]{\hspace{-0.2em}B}{}_{2}^{}}_{\bar{A}_{4}}^{\tensor*[^{1}]{\hspace{-0.2em}B}{}_{2}^{}}_{\bar{A}_{1}}^{\tensor*[^{2}]{\hspace{-0.2em}B}{}_{1}^{}}}$ &  $- 1.7851$ &                                                                                                                       $d\indices*{*_{X_{1}}^{A_{1g}^{}}_{X_{2}}^{A_{1g}^{}}_{X}^{A_{1g}^{}}}$ &  $- 9.0724$ &                                                                                                                         $d\indices*{*_{X_{1}}^{A_{1g}^{}}_{X_{2}}^{E_{g}^{}}_{X}^{E_{g}^{}}}$ &    $9.8360$ &                                                                                                                        $d\indices*{*_{X_{1}}^{A_{1g}^{}}_{X_{2}}^{A_{2u}^{}}_{X}^{E_{u}^{}}}$ &   $11.5765$ &                                                                                       $d\indices*{*_{X_{1}}^{A_{1g}^{}}_{X_{2}}^{A_{2u}^{}}_{X}^{\tensor*[^{1}]{\hspace{-0.2em}E}{}_{u}^{}}}$ & $- 14.2932$ \\[0.4em]
                                                                                                                        $d\indices*{*_{X_{1}}^{A_{1g}^{}}_{X_{2}}^{B_{1u}^{}}_{X}^{B_{1u}^{}}}$ &  $- 9.7139$ &                                                                                                                         $d\indices*{*_{X_{1}}^{A_{1g}^{}}_{X_{2}}^{E_{u}^{}}_{X}^{E_{u}^{}}}$ &  $- 7.2834$ &                                                                                        $d\indices*{*_{X_{1}}^{A_{1g}^{}}_{X_{2}}^{E_{u}^{}}_{X}^{\tensor*[^{1}]{\hspace{-0.2em}E}{}_{u}^{}}}$ &   $11.7033$ &                                                       $d\indices*{*_{X_{1}}^{A_{1g}^{}}_{X_{2}}^{\tensor*[^{1}]{\hspace{-0.2em}E}{}_{u}^{}}_{X}^{\tensor*[^{1}]{\hspace{-0.2em}E}{}_{u}^{}}}$ & $- 15.1883$ &                                                                                                                          $d\indices*{*_{X_{1}}^{E_{g}^{}}_{X_{2}}^{E_{g}^{}}_{X}^{E_{g}^{}}}$ &  $- 9.6244$ \\[0.4em]
                                                                                                                          $d\indices*{*_{X_{1}}^{A_{2u}^{}}_{X_{2}}^{E_{g}^{}}_{X}^{E_{u}^{}}}$ &   $11.2415$ &                                                                                        $d\indices*{*_{X_{1}}^{A_{2u}^{}}_{X_{2}}^{E_{g}^{}}_{X}^{\tensor*[^{1}]{\hspace{-0.2em}E}{}_{u}^{}}}$ & $- 21.2686$ &                                                                                                                        $d\indices*{*_{X_{1}}^{B_{1u}^{}}_{X_{2}}^{A_{2u}^{}}_{X}^{E_{g}^{}}}$ &   $13.7190$ &                                                                                                                         $d\indices*{*_{X_{1}}^{B_{1u}^{}}_{X_{2}}^{E_{g}^{}}_{X}^{E_{u}^{}}}$ &    $8.1487$ &                                                                                                                          $d\indices*{*_{X_{1}}^{E_{g}^{}}_{X_{2}}^{E_{u}^{}}_{X}^{E_{u}^{}}}$ &  $- 8.2609$ \\[0.4em]
                                                                        $\tensor*[^{1}]{d}{}\indices*{*_{X_{1}}^{E_{g}^{}}_{X_{2}}^{E_{u}^{}}_{X}^{\tensor*[^{1}]{\hspace{-0.2em}E}{}_{u}^{}}}$ &    $9.9274$ &                                                                                        $d\indices*{*_{X_{1}}^{B_{1u}^{}}_{X_{2}}^{E_{g}^{}}_{X}^{\tensor*[^{1}]{\hspace{-0.2em}E}{}_{u}^{}}}$ & $- 11.6635$ &                                                                       $\tensor*[^{0}]{d}{}\indices*{*_{X_{1}}^{E_{g}^{}}_{X_{2}}^{E_{u}^{}}_{X}^{\tensor*[^{1}]{\hspace{-0.2em}E}{}_{u}^{}}}$ &    $9.7093$ &                                                        $d\indices*{*_{X_{1}}^{E_{g}^{}}_{X_{2}}^{\tensor*[^{1}]{\hspace{-0.2em}E}{}_{u}^{}}_{X}^{\tensor*[^{1}]{\hspace{-0.2em}E}{}_{u}^{}}}$ & $- 15.2820$ &                                                                                                    $d\indices*{*_{\bar{A}_{1}}^{A_{1}^{}}_{\bar{A}_{2}}^{A_{1}^{}}_{\bar{A}_{5}}^{A_{1}^{}}}$ &    $7.6793$ \\[0.4em]
                                                                    $d\indices*{*_{\bar{A}_{1}}^{A_{1}^{}}_{\bar{A}_{2}}^{A_{1}^{}}_{\bar{A}_{5}}^{\tensor*[^{1}]{\hspace{-0.2em}A}{}_{1}^{}}}$ &    $9.5571$ &                                                                   $d\indices*{*_{\bar{A}_{1}}^{A_{1}^{}}_{\bar{A}_{2}}^{A_{1}^{}}_{\bar{A}_{5}}^{\tensor*[^{2}]{\hspace{-0.2em}A}{}_{1}^{}}}$ &    $1.9766$ &                                                                                                    $d\indices*{*_{\bar{A}_{1}}^{A_{1}^{}}_{\bar{A}_{2}}^{A_{1}^{}}_{\bar{A}_{5}}^{A_{2}^{}}}$ &   $14.0428$ &                                  $d\indices*{*_{\bar{A}_{1}}^{A_{1}^{}}_{\bar{A}_{2}}^{\tensor*[^{1}]{\hspace{-0.2em}A}{}_{1}^{}}_{\bar{A}_{5}}^{\tensor*[^{1}]{\hspace{-0.2em}A}{}_{1}^{}}}$ &  $- 7.1252$ &                                  $d\indices*{*_{\bar{A}_{1}}^{A_{1}^{}}_{\bar{A}_{2}}^{\tensor*[^{1}]{\hspace{-0.2em}A}{}_{1}^{}}_{\bar{A}_{5}}^{\tensor*[^{2}]{\hspace{-0.2em}A}{}_{1}^{}}}$ &    $0.8471$ \\[0.4em]
                                                                    $d\indices*{*_{\bar{A}_{1}}^{A_{1}^{}}_{\bar{A}_{2}}^{\tensor*[^{1}]{\hspace{-0.2em}A}{}_{1}^{}}_{\bar{A}_{5}}^{A_{2}^{}}}$ &  $- 5.5616$ &                                                                   $d\indices*{*_{\bar{A}_{1}}^{A_{1}^{}}_{\bar{A}_{2}}^{\tensor*[^{1}]{\hspace{-0.2em}A}{}_{1}^{}}_{\bar{A}_{5}}^{B_{1}^{}}}$ &  $- 0.1393$ &                                  $d\indices*{*_{\bar{A}_{1}}^{A_{1}^{}}_{\bar{A}_{2}}^{\tensor*[^{1}]{\hspace{-0.2em}A}{}_{1}^{}}_{\bar{A}_{5}}^{\tensor*[^{1}]{\hspace{-0.2em}B}{}_{1}^{}}}$ &  $- 5.7606$ &                                  $d\indices*{*_{\bar{A}_{1}}^{A_{1}^{}}_{\bar{A}_{2}}^{\tensor*[^{1}]{\hspace{-0.2em}A}{}_{1}^{}}_{\bar{A}_{5}}^{\tensor*[^{2}]{\hspace{-0.2em}B}{}_{1}^{}}}$ &  $- 5.6977$ &                                                                   $d\indices*{*_{\bar{A}_{1}}^{A_{1}^{}}_{\bar{A}_{2}}^{\tensor*[^{1}]{\hspace{-0.2em}A}{}_{1}^{}}_{\bar{A}_{5}}^{B_{2}^{}}}$ &    $1.3976$ \\[0.4em]
                                   $d\indices*{*_{\bar{A}_{1}}^{A_{1}^{}}_{\bar{A}_{2}}^{\tensor*[^{1}]{\hspace{-0.2em}A}{}_{1}^{}}_{\bar{A}_{5}}^{\tensor*[^{1}]{\hspace{-0.2em}B}{}_{2}^{}}}$ &  $- 1.0023$ &                                  $d\indices*{*_{\bar{A}_{1}}^{A_{1}^{}}_{\bar{A}_{2}}^{\tensor*[^{2}]{\hspace{-0.2em}A}{}_{1}^{}}_{\bar{A}_{5}}^{\tensor*[^{2}]{\hspace{-0.2em}A}{}_{1}^{}}}$ &    $0.1432$ &                                                                   $d\indices*{*_{\bar{A}_{1}}^{A_{1}^{}}_{\bar{A}_{2}}^{\tensor*[^{2}]{\hspace{-0.2em}A}{}_{1}^{}}_{\bar{A}_{5}}^{A_{2}^{}}}$ &    $1.5668$ &                                                                   $d\indices*{*_{\bar{A}_{1}}^{A_{1}^{}}_{\bar{A}_{2}}^{\tensor*[^{2}]{\hspace{-0.2em}A}{}_{1}^{}}_{\bar{A}_{5}}^{B_{1}^{}}}$ &  $- 5.6923$ &                                  $d\indices*{*_{\bar{A}_{1}}^{A_{1}^{}}_{\bar{A}_{2}}^{\tensor*[^{2}]{\hspace{-0.2em}A}{}_{1}^{}}_{\bar{A}_{5}}^{\tensor*[^{1}]{\hspace{-0.2em}B}{}_{1}^{}}}$ &    $8.2829$ \\[0.4em]
                                   $d\indices*{*_{\bar{A}_{1}}^{A_{1}^{}}_{\bar{A}_{2}}^{\tensor*[^{2}]{\hspace{-0.2em}A}{}_{1}^{}}_{\bar{A}_{5}}^{\tensor*[^{2}]{\hspace{-0.2em}B}{}_{1}^{}}}$ &  $- 0.3054$ &                                                                   $d\indices*{*_{\bar{A}_{1}}^{A_{1}^{}}_{\bar{A}_{2}}^{\tensor*[^{2}]{\hspace{-0.2em}A}{}_{1}^{}}_{\bar{A}_{5}}^{B_{2}^{}}}$ &   $11.5849$ &                                  $d\indices*{*_{\bar{A}_{1}}^{A_{1}^{}}_{\bar{A}_{2}}^{\tensor*[^{2}]{\hspace{-0.2em}A}{}_{1}^{}}_{\bar{A}_{5}}^{\tensor*[^{1}]{\hspace{-0.2em}B}{}_{2}^{}}}$ &  $- 8.9217$ &                                                                                                    $d\indices*{*_{\bar{A}_{1}}^{A_{1}^{}}_{\bar{A}_{2}}^{A_{2}^{}}_{\bar{A}_{5}}^{A_{2}^{}}}$ &    $1.3468$ &                                                                                                    $d\indices*{*_{\bar{A}_{1}}^{A_{1}^{}}_{\bar{A}_{2}}^{A_{2}^{}}_{\bar{A}_{5}}^{B_{1}^{}}}$ &  $- 0.8454$ \\[0.4em]
                                                                    $d\indices*{*_{\bar{A}_{1}}^{A_{1}^{}}_{\bar{A}_{2}}^{A_{2}^{}}_{\bar{A}_{5}}^{\tensor*[^{1}]{\hspace{-0.2em}B}{}_{1}^{}}}$ &    $8.8838$ &                                                                   $d\indices*{*_{\bar{A}_{1}}^{A_{1}^{}}_{\bar{A}_{2}}^{A_{2}^{}}_{\bar{A}_{5}}^{\tensor*[^{2}]{\hspace{-0.2em}B}{}_{1}^{}}}$ &    $6.9853$ &                                                                                                    $d\indices*{*_{\bar{A}_{1}}^{A_{1}^{}}_{\bar{A}_{2}}^{A_{2}^{}}_{\bar{A}_{5}}^{B_{2}^{}}}$ &  $- 0.9379$ &                                                                   $d\indices*{*_{\bar{A}_{1}}^{A_{1}^{}}_{\bar{A}_{2}}^{A_{2}^{}}_{\bar{A}_{5}}^{\tensor*[^{1}]{\hspace{-0.2em}B}{}_{2}^{}}}$ &    $0.3867$ &                                                                                                    $d\indices*{*_{\bar{A}_{1}}^{A_{1}^{}}_{\bar{A}_{2}}^{B_{1}^{}}_{\bar{A}_{5}}^{B_{1}^{}}}$ &  $- 9.1808$ \\[0.4em]
                                                                    $d\indices*{*_{\bar{A}_{1}}^{A_{1}^{}}_{\bar{A}_{2}}^{B_{1}^{}}_{\bar{A}_{5}}^{\tensor*[^{1}]{\hspace{-0.2em}B}{}_{1}^{}}}$ &    $6.4321$ &                                                                   $d\indices*{*_{\bar{A}_{1}}^{A_{1}^{}}_{\bar{A}_{2}}^{B_{1}^{}}_{\bar{A}_{5}}^{\tensor*[^{2}]{\hspace{-0.2em}B}{}_{1}^{}}}$ &  $- 0.6464$ &                                                                                                    $d\indices*{*_{\bar{A}_{1}}^{A_{1}^{}}_{\bar{A}_{2}}^{B_{2}^{}}_{\bar{A}_{5}}^{B_{1}^{}}}$ &  $- 8.3970$ &                                                                   $d\indices*{*_{\bar{A}_{1}}^{A_{1}^{}}_{\bar{A}_{2}}^{\tensor*[^{1}]{\hspace{-0.2em}B}{}_{2}^{}}_{\bar{A}_{5}}^{B_{1}^{}}}$ &    $6.0526$ &                                  $d\indices*{*_{\bar{A}_{1}}^{A_{1}^{}}_{\bar{A}_{2}}^{\tensor*[^{1}]{\hspace{-0.2em}B}{}_{1}^{}}_{\bar{A}_{5}}^{\tensor*[^{1}]{\hspace{-0.2em}B}{}_{1}^{}}}$ &    $8.2724$ \\[0.4em]
                                   $d\indices*{*_{\bar{A}_{1}}^{A_{1}^{}}_{\bar{A}_{2}}^{\tensor*[^{1}]{\hspace{-0.2em}B}{}_{1}^{}}_{\bar{A}_{5}}^{\tensor*[^{2}]{\hspace{-0.2em}B}{}_{1}^{}}}$ &    $0.5648$ &                                                                   $d\indices*{*_{\bar{A}_{1}}^{A_{1}^{}}_{\bar{A}_{2}}^{B_{2}^{}}_{\bar{A}_{5}}^{\tensor*[^{1}]{\hspace{-0.2em}B}{}_{1}^{}}}$ &  $- 4.0932$ &                                  $d\indices*{*_{\bar{A}_{1}}^{A_{1}^{}}_{\bar{A}_{2}}^{\tensor*[^{1}]{\hspace{-0.2em}B}{}_{2}^{}}_{\bar{A}_{5}}^{\tensor*[^{1}]{\hspace{-0.2em}B}{}_{1}^{}}}$ &    $0.5501$ &                                  $d\indices*{*_{\bar{A}_{1}}^{A_{1}^{}}_{\bar{A}_{2}}^{\tensor*[^{2}]{\hspace{-0.2em}B}{}_{1}^{}}_{\bar{A}_{5}}^{\tensor*[^{2}]{\hspace{-0.2em}B}{}_{1}^{}}}$ &  $- 0.3518$ &                                                                   $d\indices*{*_{\bar{A}_{1}}^{A_{1}^{}}_{\bar{A}_{2}}^{B_{2}^{}}_{\bar{A}_{5}}^{\tensor*[^{2}]{\hspace{-0.2em}B}{}_{1}^{}}}$ &    $0.0779$ \\[0.4em]
                                   $d\indices*{*_{\bar{A}_{1}}^{A_{1}^{}}_{\bar{A}_{2}}^{\tensor*[^{1}]{\hspace{-0.2em}B}{}_{2}^{}}_{\bar{A}_{5}}^{\tensor*[^{2}]{\hspace{-0.2em}B}{}_{1}^{}}}$ &  $- 0.0213$ &                                                                                                    $d\indices*{*_{\bar{A}_{1}}^{A_{1}^{}}_{\bar{A}_{2}}^{B_{2}^{}}_{\bar{A}_{5}}^{B_{2}^{}}}$ &  $- 4.4805$ &                                                                   $d\indices*{*_{\bar{A}_{1}}^{A_{1}^{}}_{\bar{A}_{2}}^{B_{2}^{}}_{\bar{A}_{5}}^{\tensor*[^{1}]{\hspace{-0.2em}B}{}_{2}^{}}}$ &    $0.7508$ &                                  $d\indices*{*_{\bar{A}_{1}}^{A_{1}^{}}_{\bar{A}_{2}}^{\tensor*[^{1}]{\hspace{-0.2em}B}{}_{2}^{}}_{\bar{A}_{5}}^{\tensor*[^{1}]{\hspace{-0.2em}B}{}_{2}^{}}}$ &    $0.1038$ & $d\indices*{*_{\bar{A}_{1}}^{\tensor*[^{1}]{\hspace{-0.2em}A}{}_{1}^{}}_{\bar{A}_{2}}^{\tensor*[^{1}]{\hspace{-0.2em}A}{}_{1}^{}}_{\bar{A}_{5}}^{\tensor*[^{1}]{\hspace{-0.2em}A}{}_{1}^{}}}$ &    $9.4232$ \\[0.4em]
  $d\indices*{*_{\bar{A}_{1}}^{\tensor*[^{1}]{\hspace{-0.2em}A}{}_{1}^{}}_{\bar{A}_{2}}^{\tensor*[^{1}]{\hspace{-0.2em}A}{}_{1}^{}}_{\bar{A}_{5}}^{\tensor*[^{2}]{\hspace{-0.2em}A}{}_{1}^{}}}$ &  $- 2.8697$ &                                  $d\indices*{*_{\bar{A}_{1}}^{\tensor*[^{1}]{\hspace{-0.2em}A}{}_{1}^{}}_{\bar{A}_{2}}^{\tensor*[^{1}]{\hspace{-0.2em}A}{}_{1}^{}}_{\bar{A}_{5}}^{A_{2}^{}}}$ &    $0.4487$ & $d\indices*{*_{\bar{A}_{1}}^{\tensor*[^{1}]{\hspace{-0.2em}A}{}_{1}^{}}_{\bar{A}_{2}}^{\tensor*[^{2}]{\hspace{-0.2em}A}{}_{1}^{}}_{\bar{A}_{5}}^{\tensor*[^{2}]{\hspace{-0.2em}A}{}_{1}^{}}}$ &    $5.4444$ &                                  $d\indices*{*_{\bar{A}_{1}}^{\tensor*[^{1}]{\hspace{-0.2em}A}{}_{1}^{}}_{\bar{A}_{2}}^{\tensor*[^{2}]{\hspace{-0.2em}A}{}_{1}^{}}_{\bar{A}_{5}}^{A_{2}^{}}}$ &    $3.4094$ &                                  $d\indices*{*_{\bar{A}_{1}}^{\tensor*[^{1}]{\hspace{-0.2em}A}{}_{1}^{}}_{\bar{A}_{2}}^{\tensor*[^{2}]{\hspace{-0.2em}A}{}_{1}^{}}_{\bar{A}_{5}}^{B_{1}^{}}}$ &  $- 0.9161$ \\[0.4em]
  $d\indices*{*_{\bar{A}_{1}}^{\tensor*[^{1}]{\hspace{-0.2em}A}{}_{1}^{}}_{\bar{A}_{2}}^{\tensor*[^{2}]{\hspace{-0.2em}A}{}_{1}^{}}_{\bar{A}_{5}}^{\tensor*[^{1}]{\hspace{-0.2em}B}{}_{1}^{}}}$ &  $- 2.9189$ & $d\indices*{*_{\bar{A}_{1}}^{\tensor*[^{1}]{\hspace{-0.2em}A}{}_{1}^{}}_{\bar{A}_{2}}^{\tensor*[^{2}]{\hspace{-0.2em}A}{}_{1}^{}}_{\bar{A}_{5}}^{\tensor*[^{2}]{\hspace{-0.2em}B}{}_{1}^{}}}$ &  $- 2.9224$ &                                  $d\indices*{*_{\bar{A}_{1}}^{\tensor*[^{1}]{\hspace{-0.2em}A}{}_{1}^{}}_{\bar{A}_{2}}^{\tensor*[^{2}]{\hspace{-0.2em}A}{}_{1}^{}}_{\bar{A}_{5}}^{B_{2}^{}}}$ &  $- 6.6485$ & $d\indices*{*_{\bar{A}_{1}}^{\tensor*[^{1}]{\hspace{-0.2em}A}{}_{1}^{}}_{\bar{A}_{2}}^{\tensor*[^{2}]{\hspace{-0.2em}A}{}_{1}^{}}_{\bar{A}_{5}}^{\tensor*[^{1}]{\hspace{-0.2em}B}{}_{2}^{}}}$ &    $4.7543$ &                                                                   $d\indices*{*_{\bar{A}_{1}}^{\tensor*[^{1}]{\hspace{-0.2em}A}{}_{1}^{}}_{\bar{A}_{2}}^{A_{2}^{}}_{\bar{A}_{5}}^{A_{2}^{}}}$ &  $- 6.2035$ \\[0.4em]
                                                                    $d\indices*{*_{\bar{A}_{1}}^{\tensor*[^{1}]{\hspace{-0.2em}A}{}_{1}^{}}_{\bar{A}_{2}}^{A_{2}^{}}_{\bar{A}_{5}}^{B_{1}^{}}}$ &    $1.3446$ &                                  $d\indices*{*_{\bar{A}_{1}}^{\tensor*[^{1}]{\hspace{-0.2em}A}{}_{1}^{}}_{\bar{A}_{2}}^{A_{2}^{}}_{\bar{A}_{5}}^{\tensor*[^{1}]{\hspace{-0.2em}B}{}_{1}^{}}}$ &  $- 3.6841$ &                                  $d\indices*{*_{\bar{A}_{1}}^{\tensor*[^{1}]{\hspace{-0.2em}A}{}_{1}^{}}_{\bar{A}_{2}}^{A_{2}^{}}_{\bar{A}_{5}}^{\tensor*[^{2}]{\hspace{-0.2em}B}{}_{1}^{}}}$ &  $- 3.9054$ &                                                                   $d\indices*{*_{\bar{A}_{1}}^{\tensor*[^{1}]{\hspace{-0.2em}A}{}_{1}^{}}_{\bar{A}_{2}}^{A_{2}^{}}_{\bar{A}_{5}}^{B_{2}^{}}}$ &    $0.8543$ &                                  $d\indices*{*_{\bar{A}_{1}}^{\tensor*[^{1}]{\hspace{-0.2em}A}{}_{1}^{}}_{\bar{A}_{2}}^{A_{2}^{}}_{\bar{A}_{5}}^{\tensor*[^{1}]{\hspace{-0.2em}B}{}_{2}^{}}}$ &  $- 6.5843$ \\[0.4em]
                                                                    $d\indices*{*_{\bar{A}_{1}}^{\tensor*[^{1}]{\hspace{-0.2em}A}{}_{1}^{}}_{\bar{A}_{2}}^{B_{1}^{}}_{\bar{A}_{5}}^{B_{1}^{}}}$ &    $8.7415$ &                                  $d\indices*{*_{\bar{A}_{1}}^{\tensor*[^{1}]{\hspace{-0.2em}A}{}_{1}^{}}_{\bar{A}_{2}}^{B_{1}^{}}_{\bar{A}_{5}}^{\tensor*[^{1}]{\hspace{-0.2em}B}{}_{1}^{}}}$ &  $- 4.5269$ &                                  $d\indices*{*_{\bar{A}_{1}}^{\tensor*[^{1}]{\hspace{-0.2em}A}{}_{1}^{}}_{\bar{A}_{2}}^{B_{1}^{}}_{\bar{A}_{5}}^{\tensor*[^{2}]{\hspace{-0.2em}B}{}_{1}^{}}}$ &  $- 0.9188$ &                                                                   $d\indices*{*_{\bar{A}_{1}}^{\tensor*[^{1}]{\hspace{-0.2em}A}{}_{1}^{}}_{\bar{A}_{2}}^{B_{2}^{}}_{\bar{A}_{5}}^{B_{1}^{}}}$ &  $- 0.0129$ &                                  $d\indices*{*_{\bar{A}_{1}}^{\tensor*[^{1}]{\hspace{-0.2em}A}{}_{1}^{}}_{\bar{A}_{2}}^{\tensor*[^{1}]{\hspace{-0.2em}B}{}_{2}^{}}_{\bar{A}_{5}}^{B_{1}^{}}}$ &    $0.7883$ \\[0.4em]
  $d\indices*{*_{\bar{A}_{1}}^{\tensor*[^{1}]{\hspace{-0.2em}A}{}_{1}^{}}_{\bar{A}_{2}}^{\tensor*[^{1}]{\hspace{-0.2em}B}{}_{1}^{}}_{\bar{A}_{5}}^{\tensor*[^{1}]{\hspace{-0.2em}B}{}_{1}^{}}}$ &    $7.7019$ & $d\indices*{*_{\bar{A}_{1}}^{\tensor*[^{1}]{\hspace{-0.2em}A}{}_{1}^{}}_{\bar{A}_{2}}^{\tensor*[^{1}]{\hspace{-0.2em}B}{}_{1}^{}}_{\bar{A}_{5}}^{\tensor*[^{2}]{\hspace{-0.2em}B}{}_{1}^{}}}$ &    $2.6681$ &                                  $d\indices*{*_{\bar{A}_{1}}^{\tensor*[^{1}]{\hspace{-0.2em}A}{}_{1}^{}}_{\bar{A}_{2}}^{B_{2}^{}}_{\bar{A}_{5}}^{\tensor*[^{1}]{\hspace{-0.2em}B}{}_{1}^{}}}$ &    $5.7393$ & $d\indices*{*_{\bar{A}_{1}}^{\tensor*[^{1}]{\hspace{-0.2em}A}{}_{1}^{}}_{\bar{A}_{2}}^{\tensor*[^{1}]{\hspace{-0.2em}B}{}_{2}^{}}_{\bar{A}_{5}}^{\tensor*[^{1}]{\hspace{-0.2em}B}{}_{1}^{}}}$ &  $- 3.5189$ & $d\indices*{*_{\bar{A}_{1}}^{\tensor*[^{1}]{\hspace{-0.2em}A}{}_{1}^{}}_{\bar{A}_{2}}^{\tensor*[^{2}]{\hspace{-0.2em}B}{}_{1}^{}}_{\bar{A}_{5}}^{\tensor*[^{2}]{\hspace{-0.2em}B}{}_{1}^{}}}$ &    $3.9365$ \\[0.4em]
                                   $d\indices*{*_{\bar{A}_{1}}^{\tensor*[^{1}]{\hspace{-0.2em}A}{}_{1}^{}}_{\bar{A}_{2}}^{B_{2}^{}}_{\bar{A}_{5}}^{\tensor*[^{2}]{\hspace{-0.2em}B}{}_{1}^{}}}$ &    $1.4590$ & $d\indices*{*_{\bar{A}_{1}}^{\tensor*[^{1}]{\hspace{-0.2em}A}{}_{1}^{}}_{\bar{A}_{2}}^{\tensor*[^{1}]{\hspace{-0.2em}B}{}_{2}^{}}_{\bar{A}_{5}}^{\tensor*[^{2}]{\hspace{-0.2em}B}{}_{1}^{}}}$ &  $- 5.8948$ &                                                                   $d\indices*{*_{\bar{A}_{1}}^{\tensor*[^{1}]{\hspace{-0.2em}A}{}_{1}^{}}_{\bar{A}_{2}}^{B_{2}^{}}_{\bar{A}_{5}}^{B_{2}^{}}}$ &  $- 7.3051$ &                                  $d\indices*{*_{\bar{A}_{1}}^{\tensor*[^{1}]{\hspace{-0.2em}A}{}_{1}^{}}_{\bar{A}_{2}}^{B_{2}^{}}_{\bar{A}_{5}}^{\tensor*[^{1}]{\hspace{-0.2em}B}{}_{2}^{}}}$ &    $5.7474$ & $d\indices*{*_{\bar{A}_{1}}^{\tensor*[^{1}]{\hspace{-0.2em}A}{}_{1}^{}}_{\bar{A}_{2}}^{\tensor*[^{1}]{\hspace{-0.2em}B}{}_{2}^{}}_{\bar{A}_{5}}^{\tensor*[^{1}]{\hspace{-0.2em}B}{}_{2}^{}}}$ &  $- 2.1379$ \\[0.4em]
  $d\indices*{*_{\bar{A}_{1}}^{\tensor*[^{2}]{\hspace{-0.2em}A}{}_{1}^{}}_{\bar{A}_{2}}^{\tensor*[^{2}]{\hspace{-0.2em}A}{}_{1}^{}}_{\bar{A}_{5}}^{\tensor*[^{2}]{\hspace{-0.2em}A}{}_{1}^{}}}$ &    $1.8715$ &                                  $d\indices*{*_{\bar{A}_{1}}^{\tensor*[^{2}]{\hspace{-0.2em}A}{}_{1}^{}}_{\bar{A}_{2}}^{\tensor*[^{2}]{\hspace{-0.2em}A}{}_{1}^{}}_{\bar{A}_{5}}^{A_{2}^{}}}$ &    $7.8308$ &                                                                   $d\indices*{*_{\bar{A}_{1}}^{\tensor*[^{2}]{\hspace{-0.2em}A}{}_{1}^{}}_{\bar{A}_{2}}^{A_{2}^{}}_{\bar{A}_{5}}^{A_{2}^{}}}$ &    $1.3216$ &                                                                   $d\indices*{*_{\bar{A}_{1}}^{\tensor*[^{2}]{\hspace{-0.2em}A}{}_{1}^{}}_{\bar{A}_{2}}^{A_{2}^{}}_{\bar{A}_{5}}^{B_{1}^{}}}$ &  $- 5.1385$ &                                  $d\indices*{*_{\bar{A}_{1}}^{\tensor*[^{2}]{\hspace{-0.2em}A}{}_{1}^{}}_{\bar{A}_{2}}^{A_{2}^{}}_{\bar{A}_{5}}^{\tensor*[^{1}]{\hspace{-0.2em}B}{}_{1}^{}}}$ &    $2.0773$ \\[0.4em]
                                   $d\indices*{*_{\bar{A}_{1}}^{\tensor*[^{2}]{\hspace{-0.2em}A}{}_{1}^{}}_{\bar{A}_{2}}^{A_{2}^{}}_{\bar{A}_{5}}^{\tensor*[^{2}]{\hspace{-0.2em}B}{}_{1}^{}}}$ &    $3.4644$ &                                                                   $d\indices*{*_{\bar{A}_{1}}^{\tensor*[^{2}]{\hspace{-0.2em}A}{}_{1}^{}}_{\bar{A}_{2}}^{A_{2}^{}}_{\bar{A}_{5}}^{B_{2}^{}}}$ &  $- 0.3383$ &                                  $d\indices*{*_{\bar{A}_{1}}^{\tensor*[^{2}]{\hspace{-0.2em}A}{}_{1}^{}}_{\bar{A}_{2}}^{A_{2}^{}}_{\bar{A}_{5}}^{\tensor*[^{1}]{\hspace{-0.2em}B}{}_{2}^{}}}$ &  $- 0.7120$ &                                                                   $d\indices*{*_{\bar{A}_{1}}^{\tensor*[^{2}]{\hspace{-0.2em}A}{}_{1}^{}}_{\bar{A}_{2}}^{B_{1}^{}}_{\bar{A}_{5}}^{B_{1}^{}}}$ &  $- 1.1434$ &                                  $d\indices*{*_{\bar{A}_{1}}^{\tensor*[^{2}]{\hspace{-0.2em}A}{}_{1}^{}}_{\bar{A}_{2}}^{B_{1}^{}}_{\bar{A}_{5}}^{\tensor*[^{1}]{\hspace{-0.2em}B}{}_{1}^{}}}$ &  $- 3.3577$ \\[0.4em]
                                   $d\indices*{*_{\bar{A}_{1}}^{\tensor*[^{2}]{\hspace{-0.2em}A}{}_{1}^{}}_{\bar{A}_{2}}^{B_{1}^{}}_{\bar{A}_{5}}^{\tensor*[^{2}]{\hspace{-0.2em}B}{}_{1}^{}}}$ &  $- 4.8747$ &                                                                   $d\indices*{*_{\bar{A}_{1}}^{\tensor*[^{2}]{\hspace{-0.2em}A}{}_{1}^{}}_{\bar{A}_{2}}^{B_{2}^{}}_{\bar{A}_{5}}^{B_{1}^{}}}$ &  $- 1.9700$ &                                  $d\indices*{*_{\bar{A}_{1}}^{\tensor*[^{2}]{\hspace{-0.2em}A}{}_{1}^{}}_{\bar{A}_{2}}^{\tensor*[^{1}]{\hspace{-0.2em}B}{}_{2}^{}}_{\bar{A}_{5}}^{B_{1}^{}}}$ &    $1.4885$ & $d\indices*{*_{\bar{A}_{1}}^{\tensor*[^{2}]{\hspace{-0.2em}A}{}_{1}^{}}_{\bar{A}_{2}}^{\tensor*[^{1}]{\hspace{-0.2em}B}{}_{1}^{}}_{\bar{A}_{5}}^{\tensor*[^{1}]{\hspace{-0.2em}B}{}_{1}^{}}}$ &   $13.2047$ & $d\indices*{*_{\bar{A}_{1}}^{\tensor*[^{2}]{\hspace{-0.2em}A}{}_{1}^{}}_{\bar{A}_{2}}^{\tensor*[^{1}]{\hspace{-0.2em}B}{}_{1}^{}}_{\bar{A}_{5}}^{\tensor*[^{2}]{\hspace{-0.2em}B}{}_{1}^{}}}$ &    $4.4920$ \\[0.4em]
                                   $d\indices*{*_{\bar{A}_{1}}^{\tensor*[^{2}]{\hspace{-0.2em}A}{}_{1}^{}}_{\bar{A}_{2}}^{B_{2}^{}}_{\bar{A}_{5}}^{\tensor*[^{1}]{\hspace{-0.2em}B}{}_{1}^{}}}$ &  $- 7.1326$ & $d\indices*{*_{\bar{A}_{1}}^{\tensor*[^{2}]{\hspace{-0.2em}A}{}_{1}^{}}_{\bar{A}_{2}}^{\tensor*[^{1}]{\hspace{-0.2em}B}{}_{2}^{}}_{\bar{A}_{5}}^{\tensor*[^{1}]{\hspace{-0.2em}B}{}_{1}^{}}}$ &    $3.8474$ & $d\indices*{*_{\bar{A}_{1}}^{\tensor*[^{2}]{\hspace{-0.2em}A}{}_{1}^{}}_{\bar{A}_{2}}^{\tensor*[^{2}]{\hspace{-0.2em}B}{}_{1}^{}}_{\bar{A}_{5}}^{\tensor*[^{2}]{\hspace{-0.2em}B}{}_{1}^{}}}$ &  $- 2.2173$ &                                  $d\indices*{*_{\bar{A}_{1}}^{\tensor*[^{2}]{\hspace{-0.2em}A}{}_{1}^{}}_{\bar{A}_{2}}^{B_{2}^{}}_{\bar{A}_{5}}^{\tensor*[^{2}]{\hspace{-0.2em}B}{}_{1}^{}}}$ &  $- 7.8460$ & $d\indices*{*_{\bar{A}_{1}}^{\tensor*[^{2}]{\hspace{-0.2em}A}{}_{1}^{}}_{\bar{A}_{2}}^{\tensor*[^{1}]{\hspace{-0.2em}B}{}_{2}^{}}_{\bar{A}_{5}}^{\tensor*[^{2}]{\hspace{-0.2em}B}{}_{1}^{}}}$ &    $6.5430$ \\[0.4em]
                                                                    $d\indices*{*_{\bar{A}_{1}}^{\tensor*[^{2}]{\hspace{-0.2em}A}{}_{1}^{}}_{\bar{A}_{2}}^{B_{2}^{}}_{\bar{A}_{5}}^{B_{2}^{}}}$ &  $- 0.4528$ &                                  $d\indices*{*_{\bar{A}_{1}}^{\tensor*[^{2}]{\hspace{-0.2em}A}{}_{1}^{}}_{\bar{A}_{2}}^{B_{2}^{}}_{\bar{A}_{5}}^{\tensor*[^{1}]{\hspace{-0.2em}B}{}_{2}^{}}}$ &  $- 0.1446$ & $d\indices*{*_{\bar{A}_{1}}^{\tensor*[^{2}]{\hspace{-0.2em}A}{}_{1}^{}}_{\bar{A}_{2}}^{\tensor*[^{1}]{\hspace{-0.2em}B}{}_{2}^{}}_{\bar{A}_{5}}^{\tensor*[^{1}]{\hspace{-0.2em}B}{}_{2}^{}}}$ &    $0.0879$ &                                                                                                    $d\indices*{*_{\bar{A}_{1}}^{A_{2}^{}}_{\bar{A}_{2}}^{A_{2}^{}}_{\bar{A}_{5}}^{A_{2}^{}}}$ &    $7.6629$ &                                                                                                    $d\indices*{*_{\bar{A}_{1}}^{A_{2}^{}}_{\bar{A}_{2}}^{B_{1}^{}}_{\bar{A}_{5}}^{B_{1}^{}}}$ &    $0.7485$ \\[0.4em]
                                                                    $d\indices*{*_{\bar{A}_{1}}^{A_{2}^{}}_{\bar{A}_{2}}^{B_{1}^{}}_{\bar{A}_{5}}^{\tensor*[^{1}]{\hspace{-0.2em}B}{}_{1}^{}}}$ &    $3.6249$ &                                                                   $d\indices*{*_{\bar{A}_{1}}^{A_{2}^{}}_{\bar{A}_{2}}^{B_{1}^{}}_{\bar{A}_{5}}^{\tensor*[^{2}]{\hspace{-0.2em}B}{}_{1}^{}}}$ &  $- 1.3859$ &                                                                                                    $d\indices*{*_{\bar{A}_{1}}^{A_{2}^{}}_{\bar{A}_{2}}^{B_{2}^{}}_{\bar{A}_{5}}^{B_{1}^{}}}$ &  $- 5.6611$ &                                                                   $d\indices*{*_{\bar{A}_{1}}^{A_{2}^{}}_{\bar{A}_{2}}^{\tensor*[^{1}]{\hspace{-0.2em}B}{}_{2}^{}}_{\bar{A}_{5}}^{B_{1}^{}}}$ &    $4.8699$ &                                  $d\indices*{*_{\bar{A}_{1}}^{A_{2}^{}}_{\bar{A}_{2}}^{\tensor*[^{1}]{\hspace{-0.2em}B}{}_{1}^{}}_{\bar{A}_{5}}^{\tensor*[^{1}]{\hspace{-0.2em}B}{}_{1}^{}}}$ &    $1.0346$ \\[0.4em]
                                   $d\indices*{*_{\bar{A}_{1}}^{A_{2}^{}}_{\bar{A}_{2}}^{\tensor*[^{1}]{\hspace{-0.2em}B}{}_{1}^{}}_{\bar{A}_{5}}^{\tensor*[^{2}]{\hspace{-0.2em}B}{}_{1}^{}}}$ &    $5.6857$ &                                                                   $d\indices*{*_{\bar{A}_{1}}^{A_{2}^{}}_{\bar{A}_{2}}^{B_{2}^{}}_{\bar{A}_{5}}^{\tensor*[^{1}]{\hspace{-0.2em}B}{}_{1}^{}}}$ &    $9.2277$ &                                  $d\indices*{*_{\bar{A}_{1}}^{A_{2}^{}}_{\bar{A}_{2}}^{\tensor*[^{1}]{\hspace{-0.2em}B}{}_{2}^{}}_{\bar{A}_{5}}^{\tensor*[^{1}]{\hspace{-0.2em}B}{}_{1}^{}}}$ &  $- 5.0935$ &                                  $d\indices*{*_{\bar{A}_{1}}^{A_{2}^{}}_{\bar{A}_{2}}^{\tensor*[^{2}]{\hspace{-0.2em}B}{}_{1}^{}}_{\bar{A}_{5}}^{\tensor*[^{2}]{\hspace{-0.2em}B}{}_{1}^{}}}$ &    $7.0185$ &                                                                   $d\indices*{*_{\bar{A}_{1}}^{A_{2}^{}}_{\bar{A}_{2}}^{B_{2}^{}}_{\bar{A}_{5}}^{\tensor*[^{2}]{\hspace{-0.2em}B}{}_{1}^{}}}$ &    $0.0925$ \\[0.4em]
                                   $d\indices*{*_{\bar{A}_{1}}^{A_{2}^{}}_{\bar{A}_{2}}^{\tensor*[^{1}]{\hspace{-0.2em}B}{}_{2}^{}}_{\bar{A}_{5}}^{\tensor*[^{2}]{\hspace{-0.2em}B}{}_{1}^{}}}$ &    $1.1929$ &                                                                                                    $d\indices*{*_{\bar{A}_{1}}^{A_{2}^{}}_{\bar{A}_{2}}^{B_{2}^{}}_{\bar{A}_{5}}^{B_{2}^{}}}$ &   $11.0551$ &                                                                   $d\indices*{*_{\bar{A}_{1}}^{A_{2}^{}}_{\bar{A}_{2}}^{B_{2}^{}}_{\bar{A}_{5}}^{\tensor*[^{1}]{\hspace{-0.2em}B}{}_{2}^{}}}$ &  $- 7.7656$ &                                  $d\indices*{*_{\bar{A}_{1}}^{A_{2}^{}}_{\bar{A}_{2}}^{\tensor*[^{1}]{\hspace{-0.2em}B}{}_{2}^{}}_{\bar{A}_{5}}^{\tensor*[^{1}]{\hspace{-0.2em}B}{}_{2}^{}}}$ &    $5.6475$ &                                  $d\indices*{*_{\bar{A}_{1}}^{B_{1}^{}}_{\bar{A}_{2}}^{\tensor*[^{1}]{\hspace{-0.2em}B}{}_{1}^{}}_{\bar{A}_{5}}^{\tensor*[^{2}]{\hspace{-0.2em}B}{}_{1}^{}}}$ &  $- 4.8940$ \\[0.4em]
                                                                    $d\indices*{*_{\bar{A}_{1}}^{B_{2}^{}}_{\bar{A}_{2}}^{B_{1}^{}}_{\bar{A}_{5}}^{\tensor*[^{1}]{\hspace{-0.2em}B}{}_{1}^{}}}$ &    $6.2930$ &                                  $d\indices*{*_{\bar{A}_{1}}^{\tensor*[^{1}]{\hspace{-0.2em}B}{}_{2}^{}}_{\bar{A}_{2}}^{B_{1}^{}}_{\bar{A}_{5}}^{\tensor*[^{1}]{\hspace{-0.2em}B}{}_{1}^{}}}$ &  $- 4.2143$ &                                                                   $d\indices*{*_{\bar{A}_{1}}^{B_{2}^{}}_{\bar{A}_{2}}^{B_{1}^{}}_{\bar{A}_{5}}^{\tensor*[^{2}]{\hspace{-0.2em}B}{}_{1}^{}}}$ &    $5.2756$ &                                  $d\indices*{*_{\bar{A}_{1}}^{\tensor*[^{1}]{\hspace{-0.2em}B}{}_{2}^{}}_{\bar{A}_{2}}^{B_{1}^{}}_{\bar{A}_{5}}^{\tensor*[^{2}]{\hspace{-0.2em}B}{}_{1}^{}}}$ &  $- 3.9898$ &                                                                   $d\indices*{*_{\bar{A}_{1}}^{B_{2}^{}}_{\bar{A}_{2}}^{\tensor*[^{1}]{\hspace{-0.2em}B}{}_{2}^{}}_{\bar{A}_{5}}^{B_{1}^{}}}$ &    $1.5417$ \\[0.4em]
                                   $d\indices*{*_{\bar{A}_{1}}^{B_{2}^{}}_{\bar{A}_{2}}^{\tensor*[^{1}]{\hspace{-0.2em}B}{}_{1}^{}}_{\bar{A}_{5}}^{\tensor*[^{2}]{\hspace{-0.2em}B}{}_{1}^{}}}$ &    $0.8993$ & $d\indices*{*_{\bar{A}_{1}}^{\tensor*[^{1}]{\hspace{-0.2em}B}{}_{2}^{}}_{\bar{A}_{2}}^{\tensor*[^{1}]{\hspace{-0.2em}B}{}_{1}^{}}_{\bar{A}_{5}}^{\tensor*[^{2}]{\hspace{-0.2em}B}{}_{1}^{}}}$ &    $0.0127$ &                                  $d\indices*{*_{\bar{A}_{1}}^{B_{2}^{}}_{\bar{A}_{2}}^{\tensor*[^{1}]{\hspace{-0.2em}B}{}_{2}^{}}_{\bar{A}_{5}}^{\tensor*[^{1}]{\hspace{-0.2em}B}{}_{1}^{}}}$ &  $- 0.3837$ &                                  $d\indices*{*_{\bar{A}_{1}}^{B_{2}^{}}_{\bar{A}_{2}}^{\tensor*[^{1}]{\hspace{-0.2em}B}{}_{2}^{}}_{\bar{A}_{5}}^{\tensor*[^{2}]{\hspace{-0.2em}B}{}_{1}^{}}}$ &    $0.0527$ &                                                                                                                                                                                               &              \\
%
%
%

\end{longtable*}

\begin{longtable*}{cccccccccc}
\caption{All the second order irreducible derivatives for CaF$_2$ at $V_{300K}$. All results were computed using the LID approach,  $4\hat{\mathbf1}$ supercell, and SCAN functional. The units are eV/\AA$^2$.\label{table:caf2_2_300}} \\
\hline\hline 
Derivative & Value & Derivative & Value & Derivative & Value & Derivative & Value & Derivative & Value \\
\hline\\[-0.9em]
\endfirsthead
\hline\hline
Derivative & Value & Derivative & Value & Derivative & Value & Derivative & Value & Derivative & Value  \\
\hline\\[-0.9em]
\endhead
\hline\hline
\endfoot
                                                                          $d\indices*{*_{\Gamma}^{T_{2g}^{}}_{\Gamma}^{T_{2g}^{}}}$ &   $7.4731$ &                                                                         $d\indices*{*_{\Gamma}^{T_{1u}^{}}_{\Gamma}^{T_{1u}^{}}}$ &   $7.9105$ &                                                                                   $d\indices*{*_{L}^{A_{1g}^{}}_{L}^{A_{1g}^{}}}$ &   $7.9248$ &                                                                                   $d\indices*{*_{L}^{E_{g}^{}}_{L}^{E_{g}^{}}}$ &   $5.5266$ &                                                                               $d\indices*{*_{L}^{A_{2u}^{}}_{L}^{A_{2u}^{}}}$ &  $14.9915$ \\[0.4em]
                                                   $d\indices*{*_{L}^{A_{2u}^{}}_{L}^{\tensor*[^{1}]{\hspace{-0.2em}A}{}_{2u}^{}}}$ &   $2.8370$ &                 $d\indices*{*_{L}^{\tensor*[^{1}]{\hspace{-0.2em}A}{}_{2u}^{}}_{L}^{\tensor*[^{1}]{\hspace{-0.2em}A}{}_{2u}^{}}}$ &   $6.4590$ &                                                                                     $d\indices*{*_{L}^{E_{u}^{}}_{L}^{E_{u}^{}}}$ &   $5.6389$ &                                                  $d\indices*{*_{L}^{E_{u}^{}}_{L}^{\tensor*[^{1}]{\hspace{-0.2em}E}{}_{u}^{}}}$ & $- 2.6972$ &               $d\indices*{*_{L}^{\tensor*[^{1}]{\hspace{-0.2em}E}{}_{u}^{}}_{L}^{\tensor*[^{1}]{\hspace{-0.2em}E}{}_{u}^{}}}$ &   $4.7885$ \\[0.4em]
                                                                                    $d\indices*{*_{X}^{A_{1g}^{}}_{X}^{A_{1g}^{}}}$ &  $10.7809$ &                                                                                     $d\indices*{*_{X}^{E_{g}^{}}_{X}^{E_{g}^{}}}$ &   $2.4879$ &                                                                                   $d\indices*{*_{X}^{A_{2u}^{}}_{X}^{A_{2u}^{}}}$ &  $15.2647$ &                                                                                 $d\indices*{*_{X}^{B_{1u}^{}}_{X}^{B_{1u}^{}}}$ &   $1.6104$ &                                                                                 $d\indices*{*_{X}^{E_{u}^{}}_{X}^{E_{u}^{}}}$ &   $7.6950$ \\[0.4em]
                                                     $d\indices*{*_{X}^{E_{u}^{}}_{X}^{\tensor*[^{1}]{\hspace{-0.2em}E}{}_{u}^{}}}$ & $- 1.9737$ &                   $d\indices*{*_{X}^{\tensor*[^{1}]{\hspace{-0.2em}E}{}_{u}^{}}_{X}^{\tensor*[^{1}]{\hspace{-0.2em}E}{}_{u}^{}}}$ &   $6.6912$ &                                                                               $d\indices*{*_{A}^{A_{1}^{}}_{\bar{A}}^{A_{1}^{}}}$ &  $12.9743$ &                                            $d\indices*{*_{A}^{A_{1}^{}}_{\bar{A}}^{\tensor*[^{1}]{\hspace{-0.2em}A}{}_{1}^{}}}$ & $- 1.7957$ &                                          $d\indices*{*_{A}^{A_{1}^{}}_{\bar{A}}^{\tensor*[^{2}]{\hspace{-0.2em}A}{}_{1}^{}}}$ &   $4.2220$ \\[0.4em]
              $d\indices*{*_{A}^{\tensor*[^{1}]{\hspace{-0.2em}A}{}_{1}^{}}_{\bar{A}}^{\tensor*[^{1}]{\hspace{-0.2em}A}{}_{1}^{}}}$ &   $8.4159$ &             $d\indices*{*_{A}^{\tensor*[^{1}]{\hspace{-0.2em}A}{}_{1}^{}}_{\bar{A}}^{\tensor*[^{2}]{\hspace{-0.2em}A}{}_{1}^{}}}$ & $- 0.5108$ &             $d\indices*{*_{A}^{\tensor*[^{2}]{\hspace{-0.2em}A}{}_{1}^{}}_{\bar{A}}^{\tensor*[^{2}]{\hspace{-0.2em}A}{}_{1}^{}}}$ &   $8.2441$ &                                                                             $d\indices*{*_{A}^{A_{2}^{}}_{\bar{A}}^{A_{2}^{}}}$ &   $4.9873$ &                                                                           $d\indices*{*_{A}^{B_{1}^{}}_{\bar{A}}^{B_{1}^{}}}$ &   $2.2962$ \\[0.4em]
                                               $d\indices*{*_{A}^{B_{1}^{}}_{\bar{A}}^{\tensor*[^{1}]{\hspace{-0.2em}B}{}_{1}^{}}}$ & $- 1.9542$ &                                              $d\indices*{*_{A}^{B_{1}^{}}_{\bar{A}}^{\tensor*[^{2}]{\hspace{-0.2em}B}{}_{1}^{}}}$ &   $0.2329$ &             $d\indices*{*_{A}^{\tensor*[^{1}]{\hspace{-0.2em}B}{}_{1}^{}}_{\bar{A}}^{\tensor*[^{1}]{\hspace{-0.2em}B}{}_{1}^{}}}$ &   $8.1321$ &           $d\indices*{*_{A}^{\tensor*[^{1}]{\hspace{-0.2em}B}{}_{1}^{}}_{\bar{A}}^{\tensor*[^{2}]{\hspace{-0.2em}B}{}_{1}^{}}}$ &   $0.6518$ &         $d\indices*{*_{A}^{\tensor*[^{2}]{\hspace{-0.2em}B}{}_{1}^{}}_{\bar{A}}^{\tensor*[^{2}]{\hspace{-0.2em}B}{}_{1}^{}}}$ &   $5.0824$ \\[0.4em]
                                                                                $d\indices*{*_{A}^{B_{2}^{}}_{\bar{A}}^{B_{2}^{}}}$ &   $5.5234$ &                                              $d\indices*{*_{A}^{B_{2}^{}}_{\bar{A}}^{\tensor*[^{1}]{\hspace{-0.2em}B}{}_{2}^{}}}$ & $- 2.8067$ &             $d\indices*{*_{A}^{\tensor*[^{1}]{\hspace{-0.2em}B}{}_{2}^{}}_{\bar{A}}^{\tensor*[^{1}]{\hspace{-0.2em}B}{}_{2}^{}}}$ &   $5.2296$ &                                                                             $d\indices*{*_{W}^{A_{1}^{}}_{\bar{W}}^{A_{1}^{}}}$ &   $9.8104$ &                                                                           $d\indices*{*_{W}^{B_{1}^{}}_{\bar{W}}^{B_{1}^{}}}$ &   $6.8257$ \\[0.4em]
                                               $d\indices*{*_{W}^{B_{1}^{}}_{\bar{W}}^{\tensor*[^{1}]{\hspace{-0.2em}B}{}_{1}^{}}}$ & $- 1.8176$ &             $d\indices*{*_{W}^{\tensor*[^{1}]{\hspace{-0.2em}B}{}_{1}^{}}_{\bar{W}}^{\tensor*[^{1}]{\hspace{-0.2em}B}{}_{1}^{}}}$ &   $8.5005$ &                                                                               $d\indices*{*_{W}^{A_{2}^{}}_{\bar{W}}^{A_{2}^{}}}$ &   $1.9343$ &                                                                             $d\indices*{*_{W}^{B_{2}^{}}_{\bar{W}}^{B_{2}^{}}}$ &   $2.0837$ &                                                                             $d\indices*{*_{W}^{E_{}^{}}_{\bar{W}}^{E_{}^{}}}$ &   $9.7196$ \\[0.4em]
                                                 $d\indices*{*_{W}^{E_{}^{}}_{\bar{W}}^{\tensor*[^{1}]{\hspace{-0.2em}E}{}_{}^{}}}$ &   $0.6485$ &               $d\indices*{*_{W}^{\tensor*[^{1}]{\hspace{-0.2em}E}{}_{}^{}}_{\bar{W}}^{\tensor*[^{1}]{\hspace{-0.2em}E}{}_{}^{}}}$ &   $5.3027$ &                                                                   $d\indices*{*_{\Lambda}^{A_{1}^{}}_{\bar{\Lambda}}^{A_{1}^{}}}$ &  $15.5124$ &                                $d\indices*{*_{\Lambda}^{A_{1}^{}}_{\bar{\Lambda}}^{\tensor*[^{1}]{\hspace{-0.2em}A}{}_{1}^{}}}$ & $- 8.1053$ &                              $d\indices*{*_{\Lambda}^{A_{1}^{}}_{\bar{\Lambda}}^{\tensor*[^{2}]{\hspace{-0.2em}A}{}_{1}^{}}}$ &   $1.6489$ \\[0.4em]
  $d\indices*{*_{\Lambda}^{\tensor*[^{1}]{\hspace{-0.2em}A}{}_{1}^{}}_{\bar{\Lambda}}^{\tensor*[^{1}]{\hspace{-0.2em}A}{}_{1}^{}}}$ &   $8.0653$ & $d\indices*{*_{\Lambda}^{\tensor*[^{1}]{\hspace{-0.2em}A}{}_{1}^{}}_{\bar{\Lambda}}^{\tensor*[^{2}]{\hspace{-0.2em}A}{}_{1}^{}}}$ & $- 0.0414$ & $d\indices*{*_{\Lambda}^{\tensor*[^{2}]{\hspace{-0.2em}A}{}_{1}^{}}_{\bar{\Lambda}}^{\tensor*[^{2}]{\hspace{-0.2em}A}{}_{1}^{}}}$ &   $6.9719$ &                                                                   $d\indices*{*_{\Lambda}^{E_{}^{}}_{\bar{\Lambda}}^{E_{}^{}}}$ &   $5.5596$ &                                $d\indices*{*_{\Lambda}^{E_{}^{}}_{\bar{\Lambda}}^{\tensor*[^{1}]{\hspace{-0.2em}E}{}_{}^{}}}$ & $- 2.5395$ \\[0.4em]
                                     $d\indices*{*_{\Lambda}^{E_{}^{}}_{\bar{\Lambda}}^{\tensor*[^{2}]{\hspace{-0.2em}E}{}_{}^{}}}$ & $- 1.9734$ &   $d\indices*{*_{\Lambda}^{\tensor*[^{1}]{\hspace{-0.2em}E}{}_{}^{}}_{\bar{\Lambda}}^{\tensor*[^{1}]{\hspace{-0.2em}E}{}_{}^{}}}$ &   $4.1831$ &   $d\indices*{*_{\Lambda}^{\tensor*[^{1}]{\hspace{-0.2em}E}{}_{}^{}}_{\bar{\Lambda}}^{\tensor*[^{2}]{\hspace{-0.2em}E}{}_{}^{}}}$ & $- 1.3602$ & $d\indices*{*_{\Lambda}^{\tensor*[^{2}]{\hspace{-0.2em}E}{}_{}^{}}_{\bar{\Lambda}}^{\tensor*[^{2}]{\hspace{-0.2em}E}{}_{}^{}}}$ &   $6.1936$ &                                                                             $d\indices*{*_{B}^{A_{}^{}}_{\bar{B}}^{A_{}^{}}}$ &  $12.1704$ \\[0.4em]
                                                 $d\indices*{*_{B}^{A_{}^{}}_{\bar{B}}^{\tensor*[^{1}]{\hspace{-0.2em}A}{}_{}^{}}}$ & $- 3.0435$ &                                                $d\indices*{*_{B}^{A_{}^{}}_{\bar{B}}^{\tensor*[^{2}]{\hspace{-0.2em}A}{}_{}^{}}}$ & $- 2.3818$ &                                                $d\indices*{*_{B}^{A_{}^{}}_{\bar{B}}^{\tensor*[^{3}]{\hspace{-0.2em}A}{}_{}^{}}}$ & $- 0.2155$ &                                              $d\indices*{*_{B}^{A_{}^{}}_{\bar{B}}^{\tensor*[^{4}]{\hspace{-0.2em}A}{}_{}^{}}}$ & $- 0.6536$ &                                            $d\indices*{*_{B}^{A_{}^{}}_{\bar{B}}^{\tensor*[^{5}]{\hspace{-0.2em}A}{}_{}^{}}}$ &   $0.5875$ \\[0.4em]
                $d\indices*{*_{B}^{\tensor*[^{1}]{\hspace{-0.2em}A}{}_{}^{}}_{\bar{B}}^{\tensor*[^{1}]{\hspace{-0.2em}A}{}_{}^{}}}$ &   $8.2471$ &               $d\indices*{*_{B}^{\tensor*[^{1}]{\hspace{-0.2em}A}{}_{}^{}}_{\bar{B}}^{\tensor*[^{2}]{\hspace{-0.2em}A}{}_{}^{}}}$ & $- 0.2187$ &               $d\indices*{*_{B}^{\tensor*[^{1}]{\hspace{-0.2em}A}{}_{}^{}}_{\bar{B}}^{\tensor*[^{3}]{\hspace{-0.2em}A}{}_{}^{}}}$ &   $0.5827$ &             $d\indices*{*_{B}^{\tensor*[^{1}]{\hspace{-0.2em}A}{}_{}^{}}_{\bar{B}}^{\tensor*[^{4}]{\hspace{-0.2em}A}{}_{}^{}}}$ &   $0.6808$ &           $d\indices*{*_{B}^{\tensor*[^{1}]{\hspace{-0.2em}A}{}_{}^{}}_{\bar{B}}^{\tensor*[^{5}]{\hspace{-0.2em}A}{}_{}^{}}}$ & $- 2.4138$ \\[0.4em]
                $d\indices*{*_{B}^{\tensor*[^{2}]{\hspace{-0.2em}A}{}_{}^{}}_{\bar{B}}^{\tensor*[^{2}]{\hspace{-0.2em}A}{}_{}^{}}}$ &   $8.4952$ &               $d\indices*{*_{B}^{\tensor*[^{2}]{\hspace{-0.2em}A}{}_{}^{}}_{\bar{B}}^{\tensor*[^{3}]{\hspace{-0.2em}A}{}_{}^{}}}$ &   $1.1847$ &               $d\indices*{*_{B}^{\tensor*[^{2}]{\hspace{-0.2em}A}{}_{}^{}}_{\bar{B}}^{\tensor*[^{4}]{\hspace{-0.2em}A}{}_{}^{}}}$ &   $0.9624$ &             $d\indices*{*_{B}^{\tensor*[^{2}]{\hspace{-0.2em}A}{}_{}^{}}_{\bar{B}}^{\tensor*[^{5}]{\hspace{-0.2em}A}{}_{}^{}}}$ &   $0.1423$ &           $d\indices*{*_{B}^{\tensor*[^{3}]{\hspace{-0.2em}A}{}_{}^{}}_{\bar{B}}^{\tensor*[^{3}]{\hspace{-0.2em}A}{}_{}^{}}}$ &   $4.0666$ \\[0.4em]
                $d\indices*{*_{B}^{\tensor*[^{3}]{\hspace{-0.2em}A}{}_{}^{}}_{\bar{B}}^{\tensor*[^{4}]{\hspace{-0.2em}A}{}_{}^{}}}$ & $- 1.4342$ &               $d\indices*{*_{B}^{\tensor*[^{3}]{\hspace{-0.2em}A}{}_{}^{}}_{\bar{B}}^{\tensor*[^{5}]{\hspace{-0.2em}A}{}_{}^{}}}$ & $- 0.1299$ &               $d\indices*{*_{B}^{\tensor*[^{4}]{\hspace{-0.2em}A}{}_{}^{}}_{\bar{B}}^{\tensor*[^{4}]{\hspace{-0.2em}A}{}_{}^{}}}$ &   $7.1570$ &             $d\indices*{*_{B}^{\tensor*[^{4}]{\hspace{-0.2em}A}{}_{}^{}}_{\bar{B}}^{\tensor*[^{5}]{\hspace{-0.2em}A}{}_{}^{}}}$ & $- 0.4604$ &           $d\indices*{*_{B}^{\tensor*[^{5}]{\hspace{-0.2em}A}{}_{}^{}}_{\bar{B}}^{\tensor*[^{5}]{\hspace{-0.2em}A}{}_{}^{}}}$ &   $3.8407$ \\[0.4em]
                                                                                  $d\indices*{*_{B}^{B_{}^{}}_{\bar{B}}^{B_{}^{}}}$ &   $5.8619$ &                                                $d\indices*{*_{B}^{B_{}^{}}_{\bar{B}}^{\tensor*[^{1}]{\hspace{-0.2em}B}{}_{}^{}}}$ & $- 2.0140$ &                                                $d\indices*{*_{B}^{B_{}^{}}_{\bar{B}}^{\tensor*[^{2}]{\hspace{-0.2em}B}{}_{}^{}}}$ & $- 1.4143$ &             $d\indices*{*_{B}^{\tensor*[^{1}]{\hspace{-0.2em}B}{}_{}^{}}_{\bar{B}}^{\tensor*[^{1}]{\hspace{-0.2em}B}{}_{}^{}}}$ &   $3.5727$ &           $d\indices*{*_{B}^{\tensor*[^{1}]{\hspace{-0.2em}B}{}_{}^{}}_{\bar{B}}^{\tensor*[^{2}]{\hspace{-0.2em}B}{}_{}^{}}}$ &   $1.1140$ \\[0.4em]
                $d\indices*{*_{B}^{\tensor*[^{2}]{\hspace{-0.2em}B}{}_{}^{}}_{\bar{B}}^{\tensor*[^{2}]{\hspace{-0.2em}B}{}_{}^{}}}$ &   $6.5289$ &                                                                     $d\indices*{*_{\Delta}^{A_{1}^{}}_{\bar{\Delta}}^{A_{1}^{}}}$ &  $15.7240$ &                                    $d\indices*{*_{\Delta}^{A_{1}^{}}_{\bar{\Delta}}^{\tensor*[^{1}]{\hspace{-0.2em}A}{}_{1}^{}}}$ & $- 7.8933$ & $d\indices*{*_{\Delta}^{\tensor*[^{1}]{\hspace{-0.2em}A}{}_{1}^{}}_{\bar{\Delta}}^{\tensor*[^{1}]{\hspace{-0.2em}A}{}_{1}^{}}}$ &   $9.4158$ &                                                                 $d\indices*{*_{\Delta}^{B_{2}^{}}_{\bar{\Delta}}^{B_{2}^{}}}$ &   $4.7192$ \\[0.4em]
                                                                        $d\indices*{*_{\Delta}^{E_{}^{}}_{\bar{\Delta}}^{E_{}^{}}}$ &   $5.8969$ &                                      $d\indices*{*_{\Delta}^{E_{}^{}}_{\bar{\Delta}}^{\tensor*[^{1}]{\hspace{-0.2em}E}{}_{}^{}}}$ &   $2.5718$ &                                      $d\indices*{*_{\Delta}^{E_{}^{}}_{\bar{\Delta}}^{\tensor*[^{2}]{\hspace{-0.2em}E}{}_{}^{}}}$ & $- 1.4026$ &   $d\indices*{*_{\Delta}^{\tensor*[^{1}]{\hspace{-0.2em}E}{}_{}^{}}_{\bar{\Delta}}^{\tensor*[^{1}]{\hspace{-0.2em}E}{}_{}^{}}}$ &   $2.5146$ & $d\indices*{*_{\Delta}^{\tensor*[^{1}]{\hspace{-0.2em}E}{}_{}^{}}_{\bar{\Delta}}^{\tensor*[^{2}]{\hspace{-0.2em}E}{}_{}^{}}}$ &   $0.3473$ \\[0.4em]
      $d\indices*{*_{\Delta}^{\tensor*[^{2}]{\hspace{-0.2em}E}{}_{}^{}}_{\bar{\Delta}}^{\tensor*[^{2}]{\hspace{-0.2em}E}{}_{}^{}}}$ &   $7.4990$ &                                                                                                                                   &            &                                                                                                                                   &            &                                                                                                                                 &            &                                                                                                                               &             \\
%
%
%

\end{longtable*}

\begin{longtable*}{cccccccccc}
\caption{All the third order space group irreducible derivatives for CaF$_2$ computed at  $V_{300K}$. All results were computed using the BID approach, $\hat{\mathbf{S}}_{O}$ supercell, and SCAN functional.  The units are eV/\AA$^3$.\label{table:caf2_3_300}} \\
\hline\hline 
Derivative & Value & Derivative & Value & Derivative & Value & Derivative & Value & Derivative & Value\\
\hline\\[-0.9em]
\endfirsthead
\hline\hline
Derivative & Value & Derivative & Value & Derivative & Value & Derivative & Value  & Derivative & Value\\
\hline\\[-0.9em]
\endhead
\hline\hline
\endfoot
                                                                                                                 $d\indices*{*_{\Gamma}^{T_{2g}^{}}_{\Gamma}^{T_{2g}^{}}_{\Gamma}^{T_{2g}^{}}}$ & $- 50.1487$ &                                                                                                                $d\indices*{*_{\Gamma}^{T_{1u}^{}}_{\Gamma}^{T_{1u}^{}}_{\Gamma}^{T_{2g}^{}}}$ & $- 38.7711$ &                                                                                                                     $d\indices*{*_{X}^{A_{1g}^{}}_{\bar{A}}^{A_{1}^{}}_{\bar{A}}^{A_{1}^{}}}$ &   $15.4230$ &                                                                                                                      $d\indices*{*_{X}^{E_{u}^{}}_{\bar{A}}^{A_{1}^{}}_{\bar{A}}^{A_{1}^{}}}$ &    $0.5837$ &                                                                                     $d\indices*{*_{X}^{\tensor*[^{1}]{\hspace{-0.2em}E}{}_{u}^{}}_{\bar{A}}^{A_{1}^{}}_{\bar{A}}^{A_{1}^{}}}$ &  $- 4.8662$ \\[0.4em]
                                                                                     $d\indices*{*_{X}^{A_{1g}^{}}_{\bar{A}}^{A_{1}^{}}_{\bar{A}}^{\tensor*[^{1}]{\hspace{-0.2em}A}{}_{1}^{}}}$ &  $- 7.4309$ &                                                                                     $d\indices*{*_{X}^{E_{u}^{}}_{\bar{A}}^{A_{1}^{}}_{\bar{A}}^{\tensor*[^{1}]{\hspace{-0.2em}A}{}_{1}^{}}}$ &  $- 1.4351$ &                                                    $d\indices*{*_{X}^{\tensor*[^{1}]{\hspace{-0.2em}E}{}_{u}^{}}_{\bar{A}}^{A_{1}^{}}_{\bar{A}}^{\tensor*[^{1}]{\hspace{-0.2em}A}{}_{1}^{}}}$ &  $- 0.0615$ &                                                                                    $d\indices*{*_{X}^{A_{1g}^{}}_{\bar{A}}^{A_{1}^{}}_{\bar{A}}^{\tensor*[^{2}]{\hspace{-0.2em}A}{}_{1}^{}}}$ &    $0.1255$ &                                                                                     $d\indices*{*_{X}^{E_{u}^{}}_{\bar{A}}^{A_{1}^{}}_{\bar{A}}^{\tensor*[^{2}]{\hspace{-0.2em}A}{}_{1}^{}}}$ &   $10.4760$ \\[0.4em]
                                                     $d\indices*{*_{X}^{\tensor*[^{1}]{\hspace{-0.2em}E}{}_{u}^{}}_{\bar{A}}^{A_{1}^{}}_{\bar{A}}^{\tensor*[^{2}]{\hspace{-0.2em}A}{}_{1}^{}}}$ & $- 16.5210$ &                                                                                                                      $d\indices*{*_{X}^{E_{g}^{}}_{\bar{A}}^{A_{1}^{}}_{\bar{A}}^{A_{2}^{}}}$ &    $0.6123$ &                                                                                                                      $d\indices*{*_{X}^{E_{g}^{}}_{\bar{A}}^{A_{1}^{}}_{\bar{A}}^{B_{1}^{}}}$ &    $0.7216$ &                                                                                                                     $d\indices*{*_{X}^{A_{2u}^{}}_{\bar{A}}^{A_{1}^{}}_{\bar{A}}^{B_{1}^{}}}$ &   $13.6174$ &                                                                                                                     $d\indices*{*_{X}^{B_{1u}^{}}_{\bar{A}}^{A_{1}^{}}_{\bar{A}}^{B_{1}^{}}}$ &    $8.2678$ \\[0.4em]
                                                                                      $d\indices*{*_{X}^{E_{g}^{}}_{\bar{A}}^{A_{1}^{}}_{\bar{A}}^{\tensor*[^{1}]{\hspace{-0.2em}B}{}_{1}^{}}}$ &   $11.7703$ &                                                                                    $d\indices*{*_{X}^{A_{2u}^{}}_{\bar{A}}^{A_{1}^{}}_{\bar{A}}^{\tensor*[^{1}]{\hspace{-0.2em}B}{}_{1}^{}}}$ &    $6.6847$ &                                                                                    $d\indices*{*_{X}^{B_{1u}^{}}_{\bar{A}}^{A_{1}^{}}_{\bar{A}}^{\tensor*[^{1}]{\hspace{-0.2em}B}{}_{1}^{}}}$ &    $0.6152$ &                                                                                     $d\indices*{*_{X}^{E_{g}^{}}_{\bar{A}}^{A_{1}^{}}_{\bar{A}}^{\tensor*[^{2}]{\hspace{-0.2em}B}{}_{1}^{}}}$ &   $12.0473$ &                                                                                    $d\indices*{*_{X}^{A_{2u}^{}}_{\bar{A}}^{A_{1}^{}}_{\bar{A}}^{\tensor*[^{2}]{\hspace{-0.2em}B}{}_{1}^{}}}$ &  $- 0.9395$ \\[0.4em]
                                                                                     $d\indices*{*_{X}^{B_{1u}^{}}_{\bar{A}}^{A_{1}^{}}_{\bar{A}}^{\tensor*[^{2}]{\hspace{-0.2em}B}{}_{1}^{}}}$ &  $- 2.0655$ &                                                                                                                      $d\indices*{*_{X}^{E_{u}^{}}_{\bar{A}}^{A_{1}^{}}_{\bar{A}}^{B_{2}^{}}}$ &    $0.4435$ &                                                                                     $d\indices*{*_{X}^{\tensor*[^{1}]{\hspace{-0.2em}E}{}_{u}^{}}_{\bar{A}}^{A_{1}^{}}_{\bar{A}}^{B_{2}^{}}}$ &  $- 2.4584$ &                                                                                     $d\indices*{*_{X}^{E_{u}^{}}_{\bar{A}}^{A_{1}^{}}_{\bar{A}}^{\tensor*[^{1}]{\hspace{-0.2em}B}{}_{2}^{}}}$ &  $- 0.0196$ &                                                    $d\indices*{*_{X}^{\tensor*[^{1}]{\hspace{-0.2em}E}{}_{u}^{}}_{\bar{A}}^{A_{1}^{}}_{\bar{A}}^{\tensor*[^{1}]{\hspace{-0.2em}B}{}_{2}^{}}}$ &    $1.7433$ \\[0.4em]
                                                    $d\indices*{*_{X}^{A_{1g}^{}}_{\bar{A}}^{\tensor*[^{1}]{\hspace{-0.2em}A}{}_{1}^{}}_{\bar{A}}^{\tensor*[^{1}]{\hspace{-0.2em}A}{}_{1}^{}}}$ &  $- 0.9067$ &                                                    $d\indices*{*_{X}^{E_{u}^{}}_{\bar{A}}^{\tensor*[^{1}]{\hspace{-0.2em}A}{}_{1}^{}}_{\bar{A}}^{\tensor*[^{1}]{\hspace{-0.2em}A}{}_{1}^{}}}$ &    $1.1862$ &                   $d\indices*{*_{X}^{\tensor*[^{1}]{\hspace{-0.2em}E}{}_{u}^{}}_{\bar{A}}^{\tensor*[^{1}]{\hspace{-0.2em}A}{}_{1}^{}}_{\bar{A}}^{\tensor*[^{1}]{\hspace{-0.2em}A}{}_{1}^{}}}$ &  $- 1.1771$ &                                                   $d\indices*{*_{X}^{A_{1g}^{}}_{\bar{A}}^{\tensor*[^{1}]{\hspace{-0.2em}A}{}_{1}^{}}_{\bar{A}}^{\tensor*[^{2}]{\hspace{-0.2em}A}{}_{1}^{}}}$ &  $- 0.6848$ &                                                    $d\indices*{*_{X}^{E_{u}^{}}_{\bar{A}}^{\tensor*[^{1}]{\hspace{-0.2em}A}{}_{1}^{}}_{\bar{A}}^{\tensor*[^{2}]{\hspace{-0.2em}A}{}_{1}^{}}}$ &  $- 5.7444$ \\[0.4em]
                    $d\indices*{*_{X}^{\tensor*[^{1}]{\hspace{-0.2em}E}{}_{u}^{}}_{\bar{A}}^{\tensor*[^{1}]{\hspace{-0.2em}A}{}_{1}^{}}_{\bar{A}}^{\tensor*[^{2}]{\hspace{-0.2em}A}{}_{1}^{}}}$ &    $7.1117$ &                                                                                     $d\indices*{*_{X}^{E_{g}^{}}_{\bar{A}}^{\tensor*[^{1}]{\hspace{-0.2em}A}{}_{1}^{}}_{\bar{A}}^{A_{2}^{}}}$ &  $- 6.8516$ &                                                                                     $d\indices*{*_{X}^{E_{g}^{}}_{\bar{A}}^{\tensor*[^{1}]{\hspace{-0.2em}A}{}_{1}^{}}_{\bar{A}}^{B_{1}^{}}}$ &  $- 1.5404$ &                                                                                    $d\indices*{*_{X}^{A_{2u}^{}}_{\bar{A}}^{\tensor*[^{1}]{\hspace{-0.2em}A}{}_{1}^{}}_{\bar{A}}^{B_{1}^{}}}$ & $- 10.4068$ &                                                                                    $d\indices*{*_{X}^{B_{1u}^{}}_{\bar{A}}^{\tensor*[^{1}]{\hspace{-0.2em}A}{}_{1}^{}}_{\bar{A}}^{B_{1}^{}}}$ &    $0.7237$ \\[0.4em]
                                                     $d\indices*{*_{X}^{E_{g}^{}}_{\bar{A}}^{\tensor*[^{1}]{\hspace{-0.2em}A}{}_{1}^{}}_{\bar{A}}^{\tensor*[^{1}]{\hspace{-0.2em}B}{}_{1}^{}}}$ &  $- 5.3417$ &                                                   $d\indices*{*_{X}^{A_{2u}^{}}_{\bar{A}}^{\tensor*[^{1}]{\hspace{-0.2em}A}{}_{1}^{}}_{\bar{A}}^{\tensor*[^{1}]{\hspace{-0.2em}B}{}_{1}^{}}}$ &    $4.4766$ &                                                   $d\indices*{*_{X}^{B_{1u}^{}}_{\bar{A}}^{\tensor*[^{1}]{\hspace{-0.2em}A}{}_{1}^{}}_{\bar{A}}^{\tensor*[^{1}]{\hspace{-0.2em}B}{}_{1}^{}}}$ &  $- 4.5757$ &                                                    $d\indices*{*_{X}^{E_{g}^{}}_{\bar{A}}^{\tensor*[^{1}]{\hspace{-0.2em}A}{}_{1}^{}}_{\bar{A}}^{\tensor*[^{2}]{\hspace{-0.2em}B}{}_{1}^{}}}$ &  $- 6.3256$ &                                                   $d\indices*{*_{X}^{A_{2u}^{}}_{\bar{A}}^{\tensor*[^{1}]{\hspace{-0.2em}A}{}_{1}^{}}_{\bar{A}}^{\tensor*[^{2}]{\hspace{-0.2em}B}{}_{1}^{}}}$ &    $0.8095$ \\[0.4em]
                                                    $d\indices*{*_{X}^{B_{1u}^{}}_{\bar{A}}^{\tensor*[^{1}]{\hspace{-0.2em}A}{}_{1}^{}}_{\bar{A}}^{\tensor*[^{2}]{\hspace{-0.2em}B}{}_{1}^{}}}$ &    $3.7982$ &                                                                                     $d\indices*{*_{X}^{E_{u}^{}}_{\bar{A}}^{\tensor*[^{1}]{\hspace{-0.2em}A}{}_{1}^{}}_{\bar{A}}^{B_{2}^{}}}$ &    $6.2224$ &                                                    $d\indices*{*_{X}^{\tensor*[^{1}]{\hspace{-0.2em}E}{}_{u}^{}}_{\bar{A}}^{\tensor*[^{1}]{\hspace{-0.2em}A}{}_{1}^{}}_{\bar{A}}^{B_{2}^{}}}$ &  $- 9.5957$ &                                                    $d\indices*{*_{X}^{E_{u}^{}}_{\bar{A}}^{\tensor*[^{1}]{\hspace{-0.2em}A}{}_{1}^{}}_{\bar{A}}^{\tensor*[^{1}]{\hspace{-0.2em}B}{}_{2}^{}}}$ &  $- 5.5658$ &                   $d\indices*{*_{X}^{\tensor*[^{1}]{\hspace{-0.2em}E}{}_{u}^{}}_{\bar{A}}^{\tensor*[^{1}]{\hspace{-0.2em}A}{}_{1}^{}}_{\bar{A}}^{\tensor*[^{1}]{\hspace{-0.2em}B}{}_{2}^{}}}$ &    $6.1909$ \\[0.4em]
                                                    $d\indices*{*_{X}^{A_{1g}^{}}_{\bar{A}}^{\tensor*[^{2}]{\hspace{-0.2em}A}{}_{1}^{}}_{\bar{A}}^{\tensor*[^{2}]{\hspace{-0.2em}A}{}_{1}^{}}}$ &  $- 8.4349$ &                                                    $d\indices*{*_{X}^{E_{u}^{}}_{\bar{A}}^{\tensor*[^{2}]{\hspace{-0.2em}A}{}_{1}^{}}_{\bar{A}}^{\tensor*[^{2}]{\hspace{-0.2em}A}{}_{1}^{}}}$ &  $- 0.0925$ &                   $d\indices*{*_{X}^{\tensor*[^{1}]{\hspace{-0.2em}E}{}_{u}^{}}_{\bar{A}}^{\tensor*[^{2}]{\hspace{-0.2em}A}{}_{1}^{}}_{\bar{A}}^{\tensor*[^{2}]{\hspace{-0.2em}A}{}_{1}^{}}}$ &  $- 0.8376$ &                                                                                     $d\indices*{*_{X}^{E_{g}^{}}_{\bar{A}}^{\tensor*[^{2}]{\hspace{-0.2em}A}{}_{1}^{}}_{\bar{A}}^{A_{2}^{}}}$ &    $0.8302$ &                                                                                     $d\indices*{*_{X}^{E_{g}^{}}_{\bar{A}}^{\tensor*[^{2}]{\hspace{-0.2em}A}{}_{1}^{}}_{\bar{A}}^{B_{1}^{}}}$ &    $7.7856$ \\[0.4em]
                                                                                     $d\indices*{*_{X}^{A_{2u}^{}}_{\bar{A}}^{\tensor*[^{2}]{\hspace{-0.2em}A}{}_{1}^{}}_{\bar{A}}^{B_{1}^{}}}$ &    $0.2842$ &                                                                                    $d\indices*{*_{X}^{B_{1u}^{}}_{\bar{A}}^{\tensor*[^{2}]{\hspace{-0.2em}A}{}_{1}^{}}_{\bar{A}}^{B_{1}^{}}}$ &    $1.9981$ &                                                    $d\indices*{*_{X}^{E_{g}^{}}_{\bar{A}}^{\tensor*[^{2}]{\hspace{-0.2em}A}{}_{1}^{}}_{\bar{A}}^{\tensor*[^{1}]{\hspace{-0.2em}B}{}_{1}^{}}}$ &    $0.5587$ &                                                   $d\indices*{*_{X}^{A_{2u}^{}}_{\bar{A}}^{\tensor*[^{2}]{\hspace{-0.2em}A}{}_{1}^{}}_{\bar{A}}^{\tensor*[^{1}]{\hspace{-0.2em}B}{}_{1}^{}}}$ & $- 11.0852$ &                                                   $d\indices*{*_{X}^{B_{1u}^{}}_{\bar{A}}^{\tensor*[^{2}]{\hspace{-0.2em}A}{}_{1}^{}}_{\bar{A}}^{\tensor*[^{1}]{\hspace{-0.2em}B}{}_{1}^{}}}$ &  $- 8.2978$ \\[0.4em]
                                                     $d\indices*{*_{X}^{E_{g}^{}}_{\bar{A}}^{\tensor*[^{2}]{\hspace{-0.2em}A}{}_{1}^{}}_{\bar{A}}^{\tensor*[^{2}]{\hspace{-0.2em}B}{}_{1}^{}}}$ &    $1.5041$ &                                                   $d\indices*{*_{X}^{A_{2u}^{}}_{\bar{A}}^{\tensor*[^{2}]{\hspace{-0.2em}A}{}_{1}^{}}_{\bar{A}}^{\tensor*[^{2}]{\hspace{-0.2em}B}{}_{1}^{}}}$ & $- 10.2805$ &                                                   $d\indices*{*_{X}^{B_{1u}^{}}_{\bar{A}}^{\tensor*[^{2}]{\hspace{-0.2em}A}{}_{1}^{}}_{\bar{A}}^{\tensor*[^{2}]{\hspace{-0.2em}B}{}_{1}^{}}}$ &  $- 8.5504$ &                                                                                     $d\indices*{*_{X}^{E_{u}^{}}_{\bar{A}}^{\tensor*[^{2}]{\hspace{-0.2em}A}{}_{1}^{}}_{\bar{A}}^{B_{2}^{}}}$ &  $- 1.5548$ &                                                    $d\indices*{*_{X}^{\tensor*[^{1}]{\hspace{-0.2em}E}{}_{u}^{}}_{\bar{A}}^{\tensor*[^{2}]{\hspace{-0.2em}A}{}_{1}^{}}_{\bar{A}}^{B_{2}^{}}}$ &  $- 3.3680$ \\[0.4em]
                                                     $d\indices*{*_{X}^{E_{u}^{}}_{\bar{A}}^{\tensor*[^{2}]{\hspace{-0.2em}A}{}_{1}^{}}_{\bar{A}}^{\tensor*[^{1}]{\hspace{-0.2em}B}{}_{2}^{}}}$ &    $5.4956$ &                   $d\indices*{*_{X}^{\tensor*[^{1}]{\hspace{-0.2em}E}{}_{u}^{}}_{\bar{A}}^{\tensor*[^{2}]{\hspace{-0.2em}A}{}_{1}^{}}_{\bar{A}}^{\tensor*[^{1}]{\hspace{-0.2em}B}{}_{2}^{}}}$ &    $0.4260$ &                                                                                                                     $d\indices*{*_{X}^{A_{1g}^{}}_{\bar{A}}^{A_{2}^{}}_{\bar{A}}^{A_{2}^{}}}$ &    $8.9114$ &                                                                                                                      $d\indices*{*_{X}^{E_{u}^{}}_{\bar{A}}^{A_{2}^{}}_{\bar{A}}^{A_{2}^{}}}$ &    $6.1979$ &                                                                                     $d\indices*{*_{X}^{\tensor*[^{1}]{\hspace{-0.2em}E}{}_{u}^{}}_{\bar{A}}^{A_{2}^{}}_{\bar{A}}^{A_{2}^{}}}$ &  $- 0.5530$ \\[0.4em]
                                                                                                                       $d\indices*{*_{X}^{E_{u}^{}}_{\bar{A}}^{A_{2}^{}}_{\bar{A}}^{B_{1}^{}}}$ &    $6.0354$ &                                                                                     $d\indices*{*_{X}^{\tensor*[^{1}]{\hspace{-0.2em}E}{}_{u}^{}}_{\bar{A}}^{A_{2}^{}}_{\bar{A}}^{B_{1}^{}}}$ &  $- 6.9740$ &                                                                                     $d\indices*{*_{X}^{E_{u}^{}}_{\bar{A}}^{A_{2}^{}}_{\bar{A}}^{\tensor*[^{1}]{\hspace{-0.2em}B}{}_{1}^{}}}$ &  $- 7.1518$ &                                                    $d\indices*{*_{X}^{\tensor*[^{1}]{\hspace{-0.2em}E}{}_{u}^{}}_{\bar{A}}^{A_{2}^{}}_{\bar{A}}^{\tensor*[^{1}]{\hspace{-0.2em}B}{}_{1}^{}}}$ &    $8.2155$ &                                                                                     $d\indices*{*_{X}^{E_{u}^{}}_{\bar{A}}^{A_{2}^{}}_{\bar{A}}^{\tensor*[^{2}]{\hspace{-0.2em}B}{}_{1}^{}}}$ &  $- 5.2379$ \\[0.4em]
                                                     $d\indices*{*_{X}^{\tensor*[^{1}]{\hspace{-0.2em}E}{}_{u}^{}}_{\bar{A}}^{A_{2}^{}}_{\bar{A}}^{\tensor*[^{2}]{\hspace{-0.2em}B}{}_{1}^{}}}$ &  $- 1.3464$ &                                                                                                                      $d\indices*{*_{X}^{E_{g}^{}}_{\bar{A}}^{A_{2}^{}}_{\bar{A}}^{B_{2}^{}}}$ &  $- 0.6657$ &                                                                                                                     $d\indices*{*_{X}^{A_{2u}^{}}_{\bar{A}}^{A_{2}^{}}_{\bar{A}}^{B_{2}^{}}}$ &   $15.9194$ &                                                                                                                     $d\indices*{*_{X}^{B_{1u}^{}}_{\bar{A}}^{A_{2}^{}}_{\bar{A}}^{B_{2}^{}}}$ & $- 12.4465$ &                                                                                     $d\indices*{*_{X}^{E_{g}^{}}_{\bar{A}}^{A_{2}^{}}_{\bar{A}}^{\tensor*[^{1}]{\hspace{-0.2em}B}{}_{2}^{}}}$ &  $- 0.5752$ \\[0.4em]
                                                                                     $d\indices*{*_{X}^{A_{2u}^{}}_{\bar{A}}^{A_{2}^{}}_{\bar{A}}^{\tensor*[^{1}]{\hspace{-0.2em}B}{}_{2}^{}}}$ &  $- 9.4810$ &                                                                                    $d\indices*{*_{X}^{B_{1u}^{}}_{\bar{A}}^{A_{2}^{}}_{\bar{A}}^{\tensor*[^{1}]{\hspace{-0.2em}B}{}_{2}^{}}}$ &    $8.4151$ &                                                                                                                     $d\indices*{*_{X}^{A_{1g}^{}}_{\bar{A}}^{B_{1}^{}}_{\bar{A}}^{B_{1}^{}}}$ &    $0.1829$ &                                                                                                                      $d\indices*{*_{X}^{E_{u}^{}}_{\bar{A}}^{B_{1}^{}}_{\bar{A}}^{B_{1}^{}}}$ &  $- 1.6625$ &                                                                                     $d\indices*{*_{X}^{\tensor*[^{1}]{\hspace{-0.2em}E}{}_{u}^{}}_{\bar{A}}^{B_{1}^{}}_{\bar{A}}^{B_{1}^{}}}$ &    $0.5989$ \\[0.4em]
                                                                                     $d\indices*{*_{X}^{A_{1g}^{}}_{\bar{A}}^{B_{1}^{}}_{\bar{A}}^{\tensor*[^{1}]{\hspace{-0.2em}B}{}_{1}^{}}}$ &    $1.9947$ &                                                                                     $d\indices*{*_{X}^{E_{u}^{}}_{\bar{A}}^{B_{1}^{}}_{\bar{A}}^{\tensor*[^{1}]{\hspace{-0.2em}B}{}_{1}^{}}}$ &    $4.1588$ &                                                    $d\indices*{*_{X}^{\tensor*[^{1}]{\hspace{-0.2em}E}{}_{u}^{}}_{\bar{A}}^{B_{1}^{}}_{\bar{A}}^{\tensor*[^{1}]{\hspace{-0.2em}B}{}_{1}^{}}}$ &  $- 6.6274$ &                                                                                    $d\indices*{*_{X}^{A_{1g}^{}}_{\bar{A}}^{B_{1}^{}}_{\bar{A}}^{\tensor*[^{2}]{\hspace{-0.2em}B}{}_{1}^{}}}$ &  $- 1.6492$ &                                                                                     $d\indices*{*_{X}^{E_{u}^{}}_{\bar{A}}^{B_{1}^{}}_{\bar{A}}^{\tensor*[^{2}]{\hspace{-0.2em}B}{}_{1}^{}}}$ &    $5.3757$ \\[0.4em]
                                                     $d\indices*{*_{X}^{\tensor*[^{1}]{\hspace{-0.2em}E}{}_{u}^{}}_{\bar{A}}^{B_{1}^{}}_{\bar{A}}^{\tensor*[^{2}]{\hspace{-0.2em}B}{}_{1}^{}}}$ &  $- 6.8985$ &                                                                                                                      $d\indices*{*_{X}^{E_{g}^{}}_{\bar{A}}^{B_{2}^{}}_{\bar{A}}^{B_{1}^{}}}$ &  $- 7.1161$ &                                                                                     $d\indices*{*_{X}^{E_{g}^{}}_{\bar{A}}^{\tensor*[^{1}]{\hspace{-0.2em}B}{}_{2}^{}}_{\bar{A}}^{B_{1}^{}}}$ &    $6.0104$ &                                                   $d\indices*{*_{X}^{A_{1g}^{}}_{\bar{A}}^{\tensor*[^{1}]{\hspace{-0.2em}B}{}_{1}^{}}_{\bar{A}}^{\tensor*[^{1}]{\hspace{-0.2em}B}{}_{1}^{}}}$ & $- 11.2566$ &                                                    $d\indices*{*_{X}^{E_{u}^{}}_{\bar{A}}^{\tensor*[^{1}]{\hspace{-0.2em}B}{}_{1}^{}}_{\bar{A}}^{\tensor*[^{1}]{\hspace{-0.2em}B}{}_{1}^{}}}$ &    $0.5051$ \\[0.4em]
                    $d\indices*{*_{X}^{\tensor*[^{1}]{\hspace{-0.2em}E}{}_{u}^{}}_{\bar{A}}^{\tensor*[^{1}]{\hspace{-0.2em}B}{}_{1}^{}}_{\bar{A}}^{\tensor*[^{1}]{\hspace{-0.2em}B}{}_{1}^{}}}$ &  $- 5.2429$ &                                                   $d\indices*{*_{X}^{A_{1g}^{}}_{\bar{A}}^{\tensor*[^{1}]{\hspace{-0.2em}B}{}_{1}^{}}_{\bar{A}}^{\tensor*[^{2}]{\hspace{-0.2em}B}{}_{1}^{}}}$ &  $- 6.3947$ &                                                    $d\indices*{*_{X}^{E_{u}^{}}_{\bar{A}}^{\tensor*[^{1}]{\hspace{-0.2em}B}{}_{1}^{}}_{\bar{A}}^{\tensor*[^{2}]{\hspace{-0.2em}B}{}_{1}^{}}}$ &  $- 0.8411$ &                   $d\indices*{*_{X}^{\tensor*[^{1}]{\hspace{-0.2em}E}{}_{u}^{}}_{\bar{A}}^{\tensor*[^{1}]{\hspace{-0.2em}B}{}_{1}^{}}_{\bar{A}}^{\tensor*[^{2}]{\hspace{-0.2em}B}{}_{1}^{}}}$ &    $0.8576$ &                                                                                     $d\indices*{*_{X}^{E_{g}^{}}_{\bar{A}}^{B_{2}^{}}_{\bar{A}}^{\tensor*[^{1}]{\hspace{-0.2em}B}{}_{1}^{}}}$ &   $10.7154$ \\[0.4em]
                                                     $d\indices*{*_{X}^{E_{g}^{}}_{\bar{A}}^{\tensor*[^{1}]{\hspace{-0.2em}B}{}_{2}^{}}_{\bar{A}}^{\tensor*[^{1}]{\hspace{-0.2em}B}{}_{1}^{}}}$ &  $- 5.5312$ &                                                   $d\indices*{*_{X}^{A_{1g}^{}}_{\bar{A}}^{\tensor*[^{2}]{\hspace{-0.2em}B}{}_{1}^{}}_{\bar{A}}^{\tensor*[^{2}]{\hspace{-0.2em}B}{}_{1}^{}}}$ &  $- 9.8849$ &                                                    $d\indices*{*_{X}^{E_{u}^{}}_{\bar{A}}^{\tensor*[^{2}]{\hspace{-0.2em}B}{}_{1}^{}}_{\bar{A}}^{\tensor*[^{2}]{\hspace{-0.2em}B}{}_{1}^{}}}$ &    $2.5289$ &                   $d\indices*{*_{X}^{\tensor*[^{1}]{\hspace{-0.2em}E}{}_{u}^{}}_{\bar{A}}^{\tensor*[^{2}]{\hspace{-0.2em}B}{}_{1}^{}}_{\bar{A}}^{\tensor*[^{2}]{\hspace{-0.2em}B}{}_{1}^{}}}$ &    $0.5472$ &                                                                                     $d\indices*{*_{X}^{E_{g}^{}}_{\bar{A}}^{B_{2}^{}}_{\bar{A}}^{\tensor*[^{2}]{\hspace{-0.2em}B}{}_{1}^{}}}$ &  $- 1.8645$ \\[0.4em]
                                                     $d\indices*{*_{X}^{E_{g}^{}}_{\bar{A}}^{\tensor*[^{1}]{\hspace{-0.2em}B}{}_{2}^{}}_{\bar{A}}^{\tensor*[^{2}]{\hspace{-0.2em}B}{}_{1}^{}}}$ &    $1.0226$ &                                                                                                                     $d\indices*{*_{X}^{A_{1g}^{}}_{\bar{A}}^{B_{2}^{}}_{\bar{A}}^{B_{2}^{}}}$ &    $8.8610$ &                                                                                                                      $d\indices*{*_{X}^{E_{u}^{}}_{\bar{A}}^{B_{2}^{}}_{\bar{A}}^{B_{2}^{}}}$ &    $0.4921$ &                                                                                     $d\indices*{*_{X}^{\tensor*[^{1}]{\hspace{-0.2em}E}{}_{u}^{}}_{\bar{A}}^{B_{2}^{}}_{\bar{A}}^{B_{2}^{}}}$ &    $0.0334$ &                                                                                    $d\indices*{*_{X}^{A_{1g}^{}}_{\bar{A}}^{B_{2}^{}}_{\bar{A}}^{\tensor*[^{1}]{\hspace{-0.2em}B}{}_{2}^{}}}$ &  $- 8.9431$ \\[0.4em]
                                                                                      $d\indices*{*_{X}^{E_{u}^{}}_{\bar{A}}^{B_{2}^{}}_{\bar{A}}^{\tensor*[^{1}]{\hspace{-0.2em}B}{}_{2}^{}}}$ &    $0.2633$ &                                                    $d\indices*{*_{X}^{\tensor*[^{1}]{\hspace{-0.2em}E}{}_{u}^{}}_{\bar{A}}^{B_{2}^{}}_{\bar{A}}^{\tensor*[^{1}]{\hspace{-0.2em}B}{}_{2}^{}}}$ &  $- 0.5939$ &                                                   $d\indices*{*_{X}^{A_{1g}^{}}_{\bar{A}}^{\tensor*[^{1}]{\hspace{-0.2em}B}{}_{2}^{}}_{\bar{A}}^{\tensor*[^{1}]{\hspace{-0.2em}B}{}_{2}^{}}}$ &    $9.3366$ &                                                    $d\indices*{*_{X}^{E_{u}^{}}_{\bar{A}}^{\tensor*[^{1}]{\hspace{-0.2em}B}{}_{2}^{}}_{\bar{A}}^{\tensor*[^{1}]{\hspace{-0.2em}B}{}_{2}^{}}}$ &  $- 6.2745$ &                   $d\indices*{*_{X}^{\tensor*[^{1}]{\hspace{-0.2em}E}{}_{u}^{}}_{\bar{A}}^{\tensor*[^{1}]{\hspace{-0.2em}B}{}_{2}^{}}_{\bar{A}}^{\tensor*[^{1}]{\hspace{-0.2em}B}{}_{2}^{}}}$ &  $- 0.5277$ \\[0.4em]
                                                                                                                             $d\indices*{*_{\Gamma}^{T_{2g}^{}}_{X}^{E_{g}^{}}_{X}^{E_{g}^{}}}$ &   $21.4028$ &                                                                                                                          $d\indices*{*_{\Gamma}^{T_{2g}^{}}_{X}^{B_{1u}^{}}_{X}^{A_{2u}^{}}}$ &   $11.4780$ &                                                                                                                            $d\indices*{*_{\Gamma}^{T_{2g}^{}}_{X}^{E_{u}^{}}_{X}^{E_{u}^{}}}$ & $- 14.2406$ &                                                                                           $d\indices*{*_{\Gamma}^{T_{2g}^{}}_{X}^{E_{u}^{}}_{X}^{\tensor*[^{1}]{\hspace{-0.2em}E}{}_{u}^{}}}$ &    $9.8211$ &                                                          $d\indices*{*_{\Gamma}^{T_{2g}^{}}_{X}^{\tensor*[^{1}]{\hspace{-0.2em}E}{}_{u}^{}}_{X}^{\tensor*[^{1}]{\hspace{-0.2em}E}{}_{u}^{}}}$ & $- 26.4481$ \\[0.4em]
                                                                                                                            $d\indices*{*_{\Gamma}^{T_{2g}^{}}_{X}^{A_{1g}^{}}_{X}^{E_{g}^{}}}$ & $- 10.5475$ &                                                                                                                           $d\indices*{*_{\Gamma}^{T_{2g}^{}}_{X}^{A_{2u}^{}}_{X}^{E_{u}^{}}}$ &   $10.2813$ &                                                                                          $d\indices*{*_{\Gamma}^{T_{2g}^{}}_{X}^{A_{2u}^{}}_{X}^{\tensor*[^{1}]{\hspace{-0.2em}E}{}_{u}^{}}}$ & $- 18.9499$ &                                                                                                                           $d\indices*{*_{\Gamma}^{T_{2g}^{}}_{X}^{B_{1u}^{}}_{X}^{E_{u}^{}}}$ &  $- 8.3174$ &                                                                                          $d\indices*{*_{\Gamma}^{T_{2g}^{}}_{X}^{B_{1u}^{}}_{X}^{\tensor*[^{1}]{\hspace{-0.2em}E}{}_{u}^{}}}$ &    $9.5593$ \\[0.4em]
                                                                                                                            $d\indices*{*_{\Gamma}^{T_{1u}^{}}_{X}^{A_{1g}^{}}_{X}^{E_{u}^{}}}$ & $- 14.8725$ &                                                                                          $d\indices*{*_{\Gamma}^{T_{1u}^{}}_{X}^{A_{1g}^{}}_{X}^{\tensor*[^{1}]{\hspace{-0.2em}E}{}_{u}^{}}}$ &   $17.6106$ &                                                                                                                           $d\indices*{*_{\Gamma}^{T_{1u}^{}}_{X}^{B_{1u}^{}}_{X}^{E_{g}^{}}}$ & $- 13.8487$ &                                                                                                                           $d\indices*{*_{\Gamma}^{T_{1u}^{}}_{X}^{E_{g}^{}}_{X}^{A_{2u}^{}}}$ &   $20.1671$ &                                                                                                                          $d\indices*{*_{\Gamma}^{T_{1u}^{}}_{X}^{A_{1g}^{}}_{X}^{A_{2u}^{}}}$ &   $17.3681$ \\[0.4em]
                                                                                                                             $d\indices*{*_{\Gamma}^{T_{1u}^{}}_{X}^{E_{g}^{}}_{X}^{E_{u}^{}}}$ & $- 12.5897$ &                                                                                           $d\indices*{*_{\Gamma}^{T_{1u}^{}}_{X}^{E_{g}^{}}_{X}^{\tensor*[^{1}]{\hspace{-0.2em}E}{}_{u}^{}}}$ &   $17.7092$ &                                                                                                                      $d\indices*{*_{\Gamma}^{T_{2g}^{}}_{A}^{A_{1}^{}}_{\bar{A}}^{A_{1}^{}}}$ &   $18.3325$ &                                                                                     $d\indices*{*_{\Gamma}^{T_{2g}^{}}_{A}^{A_{1}^{}}_{\bar{A}}^{\tensor*[^{1}]{\hspace{-0.2em}A}{}_{1}^{}}}$ &  $- 6.2474$ &                                                                                     $d\indices*{*_{\Gamma}^{T_{2g}^{}}_{A}^{A_{1}^{}}_{\bar{A}}^{\tensor*[^{2}]{\hspace{-0.2em}A}{}_{1}^{}}}$ &    $2.5723$ \\[0.4em]
                                                     $d\indices*{*_{\Gamma}^{T_{2g}^{}}_{A}^{\tensor*[^{1}]{\hspace{-0.2em}A}{}_{1}^{}}_{\bar{A}}^{\tensor*[^{1}]{\hspace{-0.2em}A}{}_{1}^{}}}$ &    $0.8814$ &                                                    $d\indices*{*_{\Gamma}^{T_{2g}^{}}_{A}^{\tensor*[^{1}]{\hspace{-0.2em}A}{}_{1}^{}}_{\bar{A}}^{\tensor*[^{2}]{\hspace{-0.2em}A}{}_{1}^{}}}$ &  $- 1.6607$ &                                                    $d\indices*{*_{\Gamma}^{T_{2g}^{}}_{A}^{\tensor*[^{2}]{\hspace{-0.2em}A}{}_{1}^{}}_{\bar{A}}^{\tensor*[^{2}]{\hspace{-0.2em}A}{}_{1}^{}}}$ &   $10.6971$ &                                                                                                                      $d\indices*{*_{\Gamma}^{T_{2g}^{}}_{A}^{A_{2}^{}}_{\bar{A}}^{A_{2}^{}}}$ &  $- 8.5764$ &                                                                                                                      $d\indices*{*_{\Gamma}^{T_{2g}^{}}_{A}^{B_{1}^{}}_{\bar{A}}^{B_{1}^{}}}$ &  $- 1.9350$ \\[0.4em]
                                                                                      $d\indices*{*_{\Gamma}^{T_{2g}^{}}_{A}^{B_{1}^{}}_{\bar{A}}^{\tensor*[^{1}]{\hspace{-0.2em}B}{}_{1}^{}}}$ &    $3.9463$ &                                                                                     $d\indices*{*_{\Gamma}^{T_{2g}^{}}_{A}^{B_{1}^{}}_{\bar{A}}^{\tensor*[^{2}]{\hspace{-0.2em}B}{}_{1}^{}}}$ &  $- 4.3564$ &                                                    $d\indices*{*_{\Gamma}^{T_{2g}^{}}_{A}^{\tensor*[^{1}]{\hspace{-0.2em}B}{}_{1}^{}}_{\bar{A}}^{\tensor*[^{1}]{\hspace{-0.2em}B}{}_{1}^{}}}$ &  $- 0.2902$ &                                                    $d\indices*{*_{\Gamma}^{T_{2g}^{}}_{A}^{\tensor*[^{1}]{\hspace{-0.2em}B}{}_{1}^{}}_{\bar{A}}^{\tensor*[^{2}]{\hspace{-0.2em}B}{}_{1}^{}}}$ &    $6.6803$ &                                                    $d\indices*{*_{\Gamma}^{T_{2g}^{}}_{A}^{\tensor*[^{2}]{\hspace{-0.2em}B}{}_{1}^{}}_{\bar{A}}^{\tensor*[^{2}]{\hspace{-0.2em}B}{}_{1}^{}}}$ &    $8.0156$ \\[0.4em]
                                                                                                                       $d\indices*{*_{\Gamma}^{T_{2g}^{}}_{A}^{B_{2}^{}}_{\bar{A}}^{B_{2}^{}}}$ & $- 16.9287$ &                                                                                     $d\indices*{*_{\Gamma}^{T_{2g}^{}}_{A}^{B_{2}^{}}_{\bar{A}}^{\tensor*[^{1}]{\hspace{-0.2em}B}{}_{2}^{}}}$ &    $9.7516$ &                                                    $d\indices*{*_{\Gamma}^{T_{2g}^{}}_{A}^{\tensor*[^{1}]{\hspace{-0.2em}B}{}_{2}^{}}_{\bar{A}}^{\tensor*[^{1}]{\hspace{-0.2em}B}{}_{2}^{}}}$ &  $- 6.8532$ &                                                                                                                      $d\indices*{*_{\Gamma}^{T_{2g}^{}}_{A}^{A_{1}^{}}_{\bar{A}}^{A_{2}^{}}}$ &  $- 0.0626$ &                                                                                                                      $d\indices*{*_{\Gamma}^{T_{2g}^{}}_{A}^{A_{1}^{}}_{\bar{A}}^{B_{1}^{}}}$ &    $0.3910$ \\[0.4em]
                                                                                      $d\indices*{*_{\Gamma}^{T_{2g}^{}}_{A}^{A_{1}^{}}_{\bar{A}}^{\tensor*[^{1}]{\hspace{-0.2em}B}{}_{1}^{}}}$ & $- 13.1652$ &                                                                                     $d\indices*{*_{\Gamma}^{T_{2g}^{}}_{A}^{A_{1}^{}}_{\bar{A}}^{\tensor*[^{2}]{\hspace{-0.2em}B}{}_{1}^{}}}$ & $- 10.9096$ &                                                                                     $d\indices*{*_{\Gamma}^{T_{2g}^{}}_{A}^{\tensor*[^{1}]{\hspace{-0.2em}A}{}_{1}^{}}_{\bar{A}}^{A_{2}^{}}}$ &    $6.7225$ &                                                                                     $d\indices*{*_{\Gamma}^{T_{2g}^{}}_{A}^{\tensor*[^{1}]{\hspace{-0.2em}A}{}_{1}^{}}_{\bar{A}}^{B_{1}^{}}}$ &  $- 1.1552$ &                                                    $d\indices*{*_{\Gamma}^{T_{2g}^{}}_{A}^{\tensor*[^{1}]{\hspace{-0.2em}A}{}_{1}^{}}_{\bar{A}}^{\tensor*[^{1}]{\hspace{-0.2em}B}{}_{1}^{}}}$ &    $3.8424$ \\[0.4em]
                                                     $d\indices*{*_{\Gamma}^{T_{2g}^{}}_{A}^{\tensor*[^{1}]{\hspace{-0.2em}A}{}_{1}^{}}_{\bar{A}}^{\tensor*[^{2}]{\hspace{-0.2em}B}{}_{1}^{}}}$ &    $6.5291$ &                                                                                     $d\indices*{*_{\Gamma}^{T_{2g}^{}}_{A}^{\tensor*[^{2}]{\hspace{-0.2em}A}{}_{1}^{}}_{\bar{A}}^{A_{2}^{}}}$ &  $- 4.5386$ &                                                                                     $d\indices*{*_{\Gamma}^{T_{2g}^{}}_{A}^{\tensor*[^{2}]{\hspace{-0.2em}A}{}_{1}^{}}_{\bar{A}}^{B_{1}^{}}}$ &    $7.7981$ &                                                    $d\indices*{*_{\Gamma}^{T_{2g}^{}}_{A}^{\tensor*[^{2}]{\hspace{-0.2em}A}{}_{1}^{}}_{\bar{A}}^{\tensor*[^{1}]{\hspace{-0.2em}B}{}_{1}^{}}}$ &  $- 0.5790$ &                                                    $d\indices*{*_{\Gamma}^{T_{2g}^{}}_{A}^{\tensor*[^{2}]{\hspace{-0.2em}A}{}_{1}^{}}_{\bar{A}}^{\tensor*[^{2}]{\hspace{-0.2em}B}{}_{1}^{}}}$ &    $0.7226$ \\[0.4em]
                                                                                                                       $d\indices*{*_{\Gamma}^{T_{2g}^{}}_{A}^{A_{2}^{}}_{\bar{A}}^{B_{2}^{}}}$ &    $2.4884$ &                                                                                     $d\indices*{*_{\Gamma}^{T_{2g}^{}}_{A}^{A_{2}^{}}_{\bar{A}}^{\tensor*[^{1}]{\hspace{-0.2em}B}{}_{2}^{}}}$ &  $- 5.9427$ &                                                                                                                      $d\indices*{*_{\Gamma}^{T_{2g}^{}}_{A}^{B_{1}^{}}_{\bar{A}}^{B_{2}^{}}}$ &  $- 7.2866$ &                                                                                     $d\indices*{*_{\Gamma}^{T_{2g}^{}}_{A}^{B_{1}^{}}_{\bar{A}}^{\tensor*[^{1}]{\hspace{-0.2em}B}{}_{2}^{}}}$ &    $5.9482$ &                                                                                     $d\indices*{*_{\Gamma}^{T_{2g}^{}}_{A}^{\tensor*[^{1}]{\hspace{-0.2em}B}{}_{1}^{}}_{\bar{A}}^{B_{2}^{}}}$ &    $8.8460$ \\[0.4em]
                                                     $d\indices*{*_{\Gamma}^{T_{2g}^{}}_{A}^{\tensor*[^{1}]{\hspace{-0.2em}B}{}_{1}^{}}_{\bar{A}}^{\tensor*[^{1}]{\hspace{-0.2em}B}{}_{2}^{}}}$ &  $- 7.5685$ &                                                                                     $d\indices*{*_{\Gamma}^{T_{2g}^{}}_{A}^{\tensor*[^{2}]{\hspace{-0.2em}B}{}_{1}^{}}_{\bar{A}}^{B_{2}^{}}}$ &    $0.4889$ &                                                    $d\indices*{*_{\Gamma}^{T_{2g}^{}}_{A}^{\tensor*[^{2}]{\hspace{-0.2em}B}{}_{1}^{}}_{\bar{A}}^{\tensor*[^{1}]{\hspace{-0.2em}B}{}_{2}^{}}}$ &  $- 5.3958$ &                                                                                     $d\indices*{*_{\Gamma}^{T_{1u}^{}}_{A}^{A_{1}^{}}_{\bar{A}}^{\tensor*[^{1}]{\hspace{-0.2em}A}{}_{1}^{}}}$ &    $0.8961$ &                                                                                     $d\indices*{*_{\Gamma}^{T_{1u}^{}}_{A}^{A_{1}^{}}_{\bar{A}}^{\tensor*[^{2}]{\hspace{-0.2em}A}{}_{1}^{}}}$ &   $20.1154$ \\[0.4em]
                                                                                                                       $d\indices*{*_{\Gamma}^{T_{1u}^{}}_{A}^{A_{1}^{}}_{\bar{A}}^{B_{2}^{}}}$ &  $- 0.6180$ &                                                                                     $d\indices*{*_{\Gamma}^{T_{1u}^{}}_{A}^{A_{1}^{}}_{\bar{A}}^{\tensor*[^{1}]{\hspace{-0.2em}B}{}_{2}^{}}}$ &  $- 1.0426$ &                                                    $d\indices*{*_{\Gamma}^{T_{1u}^{}}_{A}^{\tensor*[^{1}]{\hspace{-0.2em}A}{}_{1}^{}}_{\bar{A}}^{\tensor*[^{2}]{\hspace{-0.2em}A}{}_{1}^{}}}$ & $- 11.1751$ &                                                                                     $d\indices*{*_{\Gamma}^{T_{1u}^{}}_{A}^{\tensor*[^{1}]{\hspace{-0.2em}A}{}_{1}^{}}_{\bar{A}}^{B_{2}^{}}}$ & $- 11.1196$ &                                                    $d\indices*{*_{\Gamma}^{T_{1u}^{}}_{A}^{\tensor*[^{1}]{\hspace{-0.2em}A}{}_{1}^{}}_{\bar{A}}^{\tensor*[^{1}]{\hspace{-0.2em}B}{}_{2}^{}}}$ &    $7.1850$ \\[0.4em]
                                                                                      $d\indices*{*_{\Gamma}^{T_{1u}^{}}_{A}^{\tensor*[^{2}]{\hspace{-0.2em}A}{}_{1}^{}}_{\bar{A}}^{B_{2}^{}}}$ &  $- 0.8239$ &                                                    $d\indices*{*_{\Gamma}^{T_{1u}^{}}_{A}^{\tensor*[^{2}]{\hspace{-0.2em}A}{}_{1}^{}}_{\bar{A}}^{\tensor*[^{1}]{\hspace{-0.2em}B}{}_{2}^{}}}$ &    $0.8297$ &                                                                                                                      $d\indices*{*_{\Gamma}^{T_{1u}^{}}_{A}^{A_{2}^{}}_{\bar{A}}^{B_{1}^{}}}$ &    $9.1770$ &                                                                                     $d\indices*{*_{\Gamma}^{T_{1u}^{}}_{A}^{A_{2}^{}}_{\bar{A}}^{\tensor*[^{1}]{\hspace{-0.2em}B}{}_{1}^{}}}$ & $- 12.3899$ &                                                                                     $d\indices*{*_{\Gamma}^{T_{1u}^{}}_{A}^{A_{2}^{}}_{\bar{A}}^{\tensor*[^{2}]{\hspace{-0.2em}B}{}_{1}^{}}}$ &    $0.2641$ \\[0.4em]
                                                                                      $d\indices*{*_{\Gamma}^{T_{1u}^{}}_{A}^{B_{1}^{}}_{\bar{A}}^{\tensor*[^{1}]{\hspace{-0.2em}B}{}_{1}^{}}}$ &    $8.7010$ &                                                                                     $d\indices*{*_{\Gamma}^{T_{1u}^{}}_{A}^{B_{1}^{}}_{\bar{A}}^{\tensor*[^{2}]{\hspace{-0.2em}B}{}_{1}^{}}}$ &   $10.2567$ &                                                    $d\indices*{*_{\Gamma}^{T_{1u}^{}}_{A}^{\tensor*[^{1}]{\hspace{-0.2em}B}{}_{1}^{}}_{\bar{A}}^{\tensor*[^{2}]{\hspace{-0.2em}B}{}_{1}^{}}}$ &    $1.0384$ &                                                                                     $d\indices*{*_{\Gamma}^{T_{1u}^{}}_{A}^{B_{2}^{}}_{\bar{A}}^{\tensor*[^{1}]{\hspace{-0.2em}B}{}_{2}^{}}}$ &    $0.8561$ &                                                                                                                      $d\indices*{*_{\Gamma}^{T_{1u}^{}}_{A}^{A_{1}^{}}_{\bar{A}}^{B_{1}^{}}}$ &   $14.2093$ \\[0.4em]
                                                                                      $d\indices*{*_{\Gamma}^{T_{1u}^{}}_{A}^{A_{1}^{}}_{\bar{A}}^{\tensor*[^{1}]{\hspace{-0.2em}B}{}_{1}^{}}}$ &    $2.4120$ &                                                                                     $d\indices*{*_{\Gamma}^{T_{1u}^{}}_{A}^{A_{1}^{}}_{\bar{A}}^{\tensor*[^{2}]{\hspace{-0.2em}B}{}_{1}^{}}}$ &  $- 0.1510$ &                                                                                     $d\indices*{*_{\Gamma}^{T_{1u}^{}}_{A}^{\tensor*[^{1}]{\hspace{-0.2em}A}{}_{1}^{}}_{\bar{A}}^{B_{1}^{}}}$ &    $0.2873$ &                                                    $d\indices*{*_{\Gamma}^{T_{1u}^{}}_{A}^{\tensor*[^{1}]{\hspace{-0.2em}A}{}_{1}^{}}_{\bar{A}}^{\tensor*[^{1}]{\hspace{-0.2em}B}{}_{1}^{}}}$ &  $- 9.6343$ &                                                    $d\indices*{*_{\Gamma}^{T_{1u}^{}}_{A}^{\tensor*[^{1}]{\hspace{-0.2em}A}{}_{1}^{}}_{\bar{A}}^{\tensor*[^{2}]{\hspace{-0.2em}B}{}_{1}^{}}}$ &    $0.6573$ \\[0.4em]
                                                                                      $d\indices*{*_{\Gamma}^{T_{1u}^{}}_{A}^{\tensor*[^{2}]{\hspace{-0.2em}A}{}_{1}^{}}_{\bar{A}}^{B_{1}^{}}}$ &    $2.2228$ &                                                    $d\indices*{*_{\Gamma}^{T_{1u}^{}}_{A}^{\tensor*[^{2}]{\hspace{-0.2em}A}{}_{1}^{}}_{\bar{A}}^{\tensor*[^{1}]{\hspace{-0.2em}B}{}_{1}^{}}}$ &   $11.0350$ &                                                    $d\indices*{*_{\Gamma}^{T_{1u}^{}}_{A}^{\tensor*[^{2}]{\hspace{-0.2em}A}{}_{1}^{}}_{\bar{A}}^{\tensor*[^{2}]{\hspace{-0.2em}B}{}_{1}^{}}}$ &   $15.1386$ &                                                                                                                      $d\indices*{*_{\Gamma}^{T_{1u}^{}}_{A}^{A_{2}^{}}_{\bar{A}}^{B_{2}^{}}}$ &   $16.6010$ &                                                                                     $d\indices*{*_{\Gamma}^{T_{1u}^{}}_{A}^{A_{2}^{}}_{\bar{A}}^{\tensor*[^{1}]{\hspace{-0.2em}B}{}_{2}^{}}}$ & $- 13.2186$ \\[0.4em]
                                                                                                                $d\indices*{*_{A_{3}}^{A_{1}^{}}_{X_{2}}^{A_{1g}^{}}_{\bar{A}_{1}}^{A_{1}^{}}}$ &  $- 1.2955$ &                                                                                                               $d\indices*{*_{A_{3}}^{A_{1}^{}}_{X_{2}}^{A_{2u}^{}}_{\bar{A}_{1}}^{A_{1}^{}}}$ &  $- 1.5835$ &                                                                              $d\indices*{*_{A_{3}}^{A_{1}^{}}_{X_{2}}^{A_{1g}^{}}_{\bar{A}_{1}}^{\tensor*[^{1}]{\hspace{-0.2em}A}{}_{1}^{}}}$ &    $1.2732$ &                                                                               $d\indices*{*_{A_{3}}^{A_{1}^{}}_{X_{2}}^{E_{g}^{}}_{\bar{A}_{1}}^{\tensor*[^{1}]{\hspace{-0.2em}A}{}_{1}^{}}}$ &  $- 0.2866$ &                                                                              $d\indices*{*_{A_{3}}^{A_{1}^{}}_{X_{2}}^{A_{2u}^{}}_{\bar{A}_{1}}^{\tensor*[^{1}]{\hspace{-0.2em}A}{}_{1}^{}}}$ & $- 11.0080$ \\[0.4em]
                                                                                $d\indices*{*_{A_{3}}^{A_{1}^{}}_{X_{2}}^{E_{u}^{}}_{\bar{A}_{1}}^{\tensor*[^{1}]{\hspace{-0.2em}A}{}_{1}^{}}}$ &  $- 6.4160$ &                                              $d\indices*{*_{A_{3}}^{A_{1}^{}}_{X_{2}}^{\tensor*[^{1}]{\hspace{-0.2em}E}{}_{u}^{}}_{\bar{A}_{1}}^{\tensor*[^{1}]{\hspace{-0.2em}A}{}_{1}^{}}}$ &   $11.1299$ &                                                                              $d\indices*{*_{A_{3}}^{A_{1}^{}}_{X_{2}}^{A_{1g}^{}}_{\bar{A}_{1}}^{\tensor*[^{2}]{\hspace{-0.2em}A}{}_{1}^{}}}$ &  $- 0.9583$ &                                                                               $d\indices*{*_{A_{3}}^{A_{1}^{}}_{X_{2}}^{E_{g}^{}}_{\bar{A}_{1}}^{\tensor*[^{2}]{\hspace{-0.2em}A}{}_{1}^{}}}$ &  $- 0.5213$ &                                                                              $d\indices*{*_{A_{3}}^{A_{1}^{}}_{X_{2}}^{A_{2u}^{}}_{\bar{A}_{1}}^{\tensor*[^{2}]{\hspace{-0.2em}A}{}_{1}^{}}}$ &  $- 3.3748$ \\[0.4em]
                                                                                $d\indices*{*_{A_{3}}^{A_{1}^{}}_{X_{2}}^{E_{u}^{}}_{\bar{A}_{1}}^{\tensor*[^{2}]{\hspace{-0.2em}A}{}_{1}^{}}}$ &  $- 0.4903$ &                                              $d\indices*{*_{A_{3}}^{A_{1}^{}}_{X_{2}}^{\tensor*[^{1}]{\hspace{-0.2em}E}{}_{u}^{}}_{\bar{A}_{1}}^{\tensor*[^{2}]{\hspace{-0.2em}A}{}_{1}^{}}}$ &    $2.8745$ &                                                                                                                $d\indices*{*_{A_{3}}^{A_{1}^{}}_{X_{2}}^{E_{g}^{}}_{\bar{A}_{1}}^{A_{2}^{}}}$ &    $0.1857$ &                                                                                                               $d\indices*{*_{A_{3}}^{A_{1}^{}}_{X_{2}}^{B_{1u}^{}}_{\bar{A}_{1}}^{A_{2}^{}}}$ & $- 13.4983$ &                                                                                                                $d\indices*{*_{A_{3}}^{A_{1}^{}}_{X_{2}}^{E_{u}^{}}_{\bar{A}_{1}}^{A_{2}^{}}}$ &    $9.9672$ \\[0.4em]
                                                                                $d\indices*{*_{A_{3}}^{A_{1}^{}}_{X_{2}}^{\tensor*[^{1}]{\hspace{-0.2em}E}{}_{u}^{}}_{\bar{A}_{1}}^{A_{2}^{}}}$ & $- 17.3649$ &                                                                                                                $d\indices*{*_{A_{3}}^{A_{1}^{}}_{X_{2}}^{E_{g}^{}}_{\bar{A}_{1}}^{B_{1}^{}}}$ &  $- 8.1953$ &                                                                                                               $d\indices*{*_{A_{3}}^{A_{1}^{}}_{X_{2}}^{B_{1u}^{}}_{\bar{A}_{1}}^{B_{1}^{}}}$ &  $- 0.7430$ &                                                                                                                $d\indices*{*_{A_{3}}^{A_{1}^{}}_{X_{2}}^{E_{u}^{}}_{\bar{A}_{1}}^{B_{1}^{}}}$ &    $2.2600$ &                                                                               $d\indices*{*_{A_{3}}^{A_{1}^{}}_{X_{2}}^{\tensor*[^{1}]{\hspace{-0.2em}E}{}_{u}^{}}_{\bar{A}_{1}}^{B_{1}^{}}}$ &    $0.0002$ \\[0.4em]
                                                                                $d\indices*{*_{A_{3}}^{A_{1}^{}}_{X_{2}}^{E_{g}^{}}_{\bar{A}_{1}}^{\tensor*[^{1}]{\hspace{-0.2em}B}{}_{1}^{}}}$ &   $10.5565$ &                                                                              $d\indices*{*_{A_{3}}^{A_{1}^{}}_{X_{2}}^{B_{1u}^{}}_{\bar{A}_{1}}^{\tensor*[^{1}]{\hspace{-0.2em}B}{}_{1}^{}}}$ &  $- 0.0808$ &                                                                               $d\indices*{*_{A_{3}}^{A_{1}^{}}_{X_{2}}^{E_{u}^{}}_{\bar{A}_{1}}^{\tensor*[^{1}]{\hspace{-0.2em}B}{}_{1}^{}}}$ &    $0.1341$ &                                              $d\indices*{*_{A_{3}}^{A_{1}^{}}_{X_{2}}^{\tensor*[^{1}]{\hspace{-0.2em}E}{}_{u}^{}}_{\bar{A}_{1}}^{\tensor*[^{1}]{\hspace{-0.2em}B}{}_{1}^{}}}$ &  $- 2.2094$ &                                                                               $d\indices*{*_{A_{3}}^{A_{1}^{}}_{X_{2}}^{E_{g}^{}}_{\bar{A}_{1}}^{\tensor*[^{2}]{\hspace{-0.2em}B}{}_{1}^{}}}$ &  $- 0.9078$ \\[0.4em]
                                                                               $d\indices*{*_{A_{3}}^{A_{1}^{}}_{X_{2}}^{B_{1u}^{}}_{\bar{A}_{1}}^{\tensor*[^{2}]{\hspace{-0.2em}B}{}_{1}^{}}}$ &  $- 0.1682$ &                                                                               $d\indices*{*_{A_{3}}^{A_{1}^{}}_{X_{2}}^{E_{u}^{}}_{\bar{A}_{1}}^{\tensor*[^{2}]{\hspace{-0.2em}B}{}_{1}^{}}}$ &    $0.4878$ &                                              $d\indices*{*_{A_{3}}^{A_{1}^{}}_{X_{2}}^{\tensor*[^{1}]{\hspace{-0.2em}E}{}_{u}^{}}_{\bar{A}_{1}}^{\tensor*[^{2}]{\hspace{-0.2em}B}{}_{1}^{}}}$ &    $0.2145$ &                                                                                                               $d\indices*{*_{A_{3}}^{A_{1}^{}}_{X_{2}}^{A_{1g}^{}}_{\bar{A}_{1}}^{B_{2}^{}}}$ & $- 13.7236$ &                                                                                                                $d\indices*{*_{A_{3}}^{A_{1}^{}}_{X_{2}}^{E_{g}^{}}_{\bar{A}_{1}}^{B_{2}^{}}}$ & $- 15.4023$ \\[0.4em]
                                                                                                                $d\indices*{*_{A_{3}}^{A_{1}^{}}_{X_{2}}^{A_{2u}^{}}_{\bar{A}_{1}}^{B_{2}^{}}}$ &    $1.6961$ &                                                                                                                $d\indices*{*_{A_{3}}^{A_{1}^{}}_{X_{2}}^{E_{u}^{}}_{\bar{A}_{1}}^{B_{2}^{}}}$ &    $1.8238$ &                                                                               $d\indices*{*_{A_{3}}^{A_{1}^{}}_{X_{2}}^{\tensor*[^{1}]{\hspace{-0.2em}E}{}_{u}^{}}_{\bar{A}_{1}}^{B_{2}^{}}}$ &  $- 3.5616$ &                                                                              $d\indices*{*_{A_{3}}^{A_{1}^{}}_{X_{2}}^{A_{1g}^{}}_{\bar{A}_{1}}^{\tensor*[^{1}]{\hspace{-0.2em}B}{}_{2}^{}}}$ &    $9.8717$ &                                                                               $d\indices*{*_{A_{3}}^{A_{1}^{}}_{X_{2}}^{E_{g}^{}}_{\bar{A}_{1}}^{\tensor*[^{1}]{\hspace{-0.2em}B}{}_{2}^{}}}$ &   $11.8561$ \\[0.4em]
                                                                               $d\indices*{*_{A_{3}}^{A_{1}^{}}_{X_{2}}^{A_{2u}^{}}_{\bar{A}_{1}}^{\tensor*[^{1}]{\hspace{-0.2em}B}{}_{2}^{}}}$ &    $1.6440$ &                                                                               $d\indices*{*_{A_{3}}^{A_{1}^{}}_{X_{2}}^{E_{u}^{}}_{\bar{A}_{1}}^{\tensor*[^{1}]{\hspace{-0.2em}B}{}_{2}^{}}}$ &  $- 0.8676$ &                                              $d\indices*{*_{A_{3}}^{A_{1}^{}}_{X_{2}}^{\tensor*[^{1}]{\hspace{-0.2em}E}{}_{u}^{}}_{\bar{A}_{1}}^{\tensor*[^{1}]{\hspace{-0.2em}B}{}_{2}^{}}}$ &  $- 0.4211$ &                                             $d\indices*{*_{A_{3}}^{\tensor*[^{1}]{\hspace{-0.2em}A}{}_{1}^{}}_{X_{2}}^{A_{1g}^{}}_{\bar{A}_{1}}^{\tensor*[^{1}]{\hspace{-0.2em}A}{}_{1}^{}}}$ &  $- 0.9124$ &                                             $d\indices*{*_{A_{3}}^{\tensor*[^{1}]{\hspace{-0.2em}A}{}_{1}^{}}_{X_{2}}^{A_{2u}^{}}_{\bar{A}_{1}}^{\tensor*[^{1}]{\hspace{-0.2em}A}{}_{1}^{}}}$ &    $7.9774$ \\[0.4em]
                                              $d\indices*{*_{A_{3}}^{\tensor*[^{1}]{\hspace{-0.2em}A}{}_{1}^{}}_{X_{2}}^{A_{1g}^{}}_{\bar{A}_{1}}^{\tensor*[^{2}]{\hspace{-0.2em}A}{}_{1}^{}}}$ &    $6.8434$ &                                              $d\indices*{*_{A_{3}}^{\tensor*[^{1}]{\hspace{-0.2em}A}{}_{1}^{}}_{X_{2}}^{E_{g}^{}}_{\bar{A}_{1}}^{\tensor*[^{2}]{\hspace{-0.2em}A}{}_{1}^{}}}$ &    $6.5739$ &                                             $d\indices*{*_{A_{3}}^{\tensor*[^{1}]{\hspace{-0.2em}A}{}_{1}^{}}_{X_{2}}^{A_{2u}^{}}_{\bar{A}_{1}}^{\tensor*[^{2}]{\hspace{-0.2em}A}{}_{1}^{}}}$ &  $- 0.9048$ &                                              $d\indices*{*_{A_{3}}^{\tensor*[^{1}]{\hspace{-0.2em}A}{}_{1}^{}}_{X_{2}}^{E_{u}^{}}_{\bar{A}_{1}}^{\tensor*[^{2}]{\hspace{-0.2em}A}{}_{1}^{}}}$ &    $1.6302$ &             $d\indices*{*_{A_{3}}^{\tensor*[^{1}]{\hspace{-0.2em}A}{}_{1}^{}}_{X_{2}}^{\tensor*[^{1}]{\hspace{-0.2em}E}{}_{u}^{}}_{\bar{A}_{1}}^{\tensor*[^{2}]{\hspace{-0.2em}A}{}_{1}^{}}}$ &  $- 0.9155$ \\[0.4em]
                                                                                $d\indices*{*_{A_{3}}^{\tensor*[^{1}]{\hspace{-0.2em}A}{}_{1}^{}}_{X_{2}}^{E_{g}^{}}_{\bar{A}_{1}}^{A_{2}^{}}}$ &  $- 2.2886$ &                                                                              $d\indices*{*_{A_{3}}^{\tensor*[^{1}]{\hspace{-0.2em}A}{}_{1}^{}}_{X_{2}}^{B_{1u}^{}}_{\bar{A}_{1}}^{A_{2}^{}}}$ &    $5.8356$ &                                                                               $d\indices*{*_{A_{3}}^{\tensor*[^{1}]{\hspace{-0.2em}A}{}_{1}^{}}_{X_{2}}^{E_{u}^{}}_{\bar{A}_{1}}^{A_{2}^{}}}$ &  $- 6.2681$ &                                              $d\indices*{*_{A_{3}}^{\tensor*[^{1}]{\hspace{-0.2em}A}{}_{1}^{}}_{X_{2}}^{\tensor*[^{1}]{\hspace{-0.2em}E}{}_{u}^{}}_{\bar{A}_{1}}^{A_{2}^{}}}$ &    $4.9668$ &                                                                               $d\indices*{*_{A_{3}}^{\tensor*[^{1}]{\hspace{-0.2em}A}{}_{1}^{}}_{X_{2}}^{E_{g}^{}}_{\bar{A}_{1}}^{B_{1}^{}}}$ &  $- 1.1600$ \\[0.4em]
                                                                               $d\indices*{*_{A_{3}}^{\tensor*[^{1}]{\hspace{-0.2em}A}{}_{1}^{}}_{X_{2}}^{B_{1u}^{}}_{\bar{A}_{1}}^{B_{1}^{}}}$ &    $0.6697$ &                                                                               $d\indices*{*_{A_{3}}^{\tensor*[^{1}]{\hspace{-0.2em}A}{}_{1}^{}}_{X_{2}}^{E_{u}^{}}_{\bar{A}_{1}}^{B_{1}^{}}}$ &  $- 0.9158$ &                                              $d\indices*{*_{A_{3}}^{\tensor*[^{1}]{\hspace{-0.2em}A}{}_{1}^{}}_{X_{2}}^{\tensor*[^{1}]{\hspace{-0.2em}E}{}_{u}^{}}_{\bar{A}_{1}}^{B_{1}^{}}}$ &    $1.0472$ &                                              $d\indices*{*_{A_{3}}^{\tensor*[^{1}]{\hspace{-0.2em}A}{}_{1}^{}}_{X_{2}}^{E_{g}^{}}_{\bar{A}_{1}}^{\tensor*[^{1}]{\hspace{-0.2em}B}{}_{1}^{}}}$ &  $- 5.2418$ &                                             $d\indices*{*_{A_{3}}^{\tensor*[^{1}]{\hspace{-0.2em}A}{}_{1}^{}}_{X_{2}}^{B_{1u}^{}}_{\bar{A}_{1}}^{\tensor*[^{1}]{\hspace{-0.2em}B}{}_{1}^{}}}$ &    $5.4490$ \\[0.4em]
                                               $d\indices*{*_{A_{3}}^{\tensor*[^{1}]{\hspace{-0.2em}A}{}_{1}^{}}_{X_{2}}^{E_{u}^{}}_{\bar{A}_{1}}^{\tensor*[^{1}]{\hspace{-0.2em}B}{}_{1}^{}}}$ &    $5.6771$ &             $d\indices*{*_{A_{3}}^{\tensor*[^{1}]{\hspace{-0.2em}A}{}_{1}^{}}_{X_{2}}^{\tensor*[^{1}]{\hspace{-0.2em}E}{}_{u}^{}}_{\bar{A}_{1}}^{\tensor*[^{1}]{\hspace{-0.2em}B}{}_{1}^{}}}$ &  $- 7.4219$ &                                              $d\indices*{*_{A_{3}}^{\tensor*[^{1}]{\hspace{-0.2em}A}{}_{1}^{}}_{X_{2}}^{E_{g}^{}}_{\bar{A}_{1}}^{\tensor*[^{2}]{\hspace{-0.2em}B}{}_{1}^{}}}$ &    $1.1131$ &                                             $d\indices*{*_{A_{3}}^{\tensor*[^{1}]{\hspace{-0.2em}A}{}_{1}^{}}_{X_{2}}^{B_{1u}^{}}_{\bar{A}_{1}}^{\tensor*[^{2}]{\hspace{-0.2em}B}{}_{1}^{}}}$ &    $7.3205$ &                                              $d\indices*{*_{A_{3}}^{\tensor*[^{1}]{\hspace{-0.2em}A}{}_{1}^{}}_{X_{2}}^{E_{u}^{}}_{\bar{A}_{1}}^{\tensor*[^{2}]{\hspace{-0.2em}B}{}_{1}^{}}}$ &    $5.6671$ \\[0.4em]
              $d\indices*{*_{A_{3}}^{\tensor*[^{1}]{\hspace{-0.2em}A}{}_{1}^{}}_{X_{2}}^{\tensor*[^{1}]{\hspace{-0.2em}E}{}_{u}^{}}_{\bar{A}_{1}}^{\tensor*[^{2}]{\hspace{-0.2em}B}{}_{1}^{}}}$ &  $- 7.0615$ &                                                                              $d\indices*{*_{A_{3}}^{\tensor*[^{1}]{\hspace{-0.2em}A}{}_{1}^{}}_{X_{2}}^{A_{1g}^{}}_{\bar{A}_{1}}^{B_{2}^{}}}$ &    $7.7617$ &                                                                               $d\indices*{*_{A_{3}}^{\tensor*[^{1}]{\hspace{-0.2em}A}{}_{1}^{}}_{X_{2}}^{E_{g}^{}}_{\bar{A}_{1}}^{B_{2}^{}}}$ &    $7.3547$ &                                                                              $d\indices*{*_{A_{3}}^{\tensor*[^{1}]{\hspace{-0.2em}A}{}_{1}^{}}_{X_{2}}^{A_{2u}^{}}_{\bar{A}_{1}}^{B_{2}^{}}}$ &    $0.0612$ &                                                                               $d\indices*{*_{A_{3}}^{\tensor*[^{1}]{\hspace{-0.2em}A}{}_{1}^{}}_{X_{2}}^{E_{u}^{}}_{\bar{A}_{1}}^{B_{2}^{}}}$ &  $- 0.5441$ \\[0.4em]
                                               $d\indices*{*_{A_{3}}^{\tensor*[^{1}]{\hspace{-0.2em}A}{}_{1}^{}}_{X_{2}}^{\tensor*[^{1}]{\hspace{-0.2em}E}{}_{u}^{}}_{\bar{A}_{1}}^{B_{2}^{}}}$ &  $- 1.5695$ &                                             $d\indices*{*_{A_{3}}^{\tensor*[^{1}]{\hspace{-0.2em}A}{}_{1}^{}}_{X_{2}}^{A_{1g}^{}}_{\bar{A}_{1}}^{\tensor*[^{1}]{\hspace{-0.2em}B}{}_{2}^{}}}$ &  $- 4.8751$ &                                              $d\indices*{*_{A_{3}}^{\tensor*[^{1}]{\hspace{-0.2em}A}{}_{1}^{}}_{X_{2}}^{E_{g}^{}}_{\bar{A}_{1}}^{\tensor*[^{1}]{\hspace{-0.2em}B}{}_{2}^{}}}$ &  $- 5.8189$ &                                             $d\indices*{*_{A_{3}}^{\tensor*[^{1}]{\hspace{-0.2em}A}{}_{1}^{}}_{X_{2}}^{A_{2u}^{}}_{\bar{A}_{1}}^{\tensor*[^{1}]{\hspace{-0.2em}B}{}_{2}^{}}}$ &  $- 0.1256$ &                                              $d\indices*{*_{A_{3}}^{\tensor*[^{1}]{\hspace{-0.2em}A}{}_{1}^{}}_{X_{2}}^{E_{u}^{}}_{\bar{A}_{1}}^{\tensor*[^{1}]{\hspace{-0.2em}B}{}_{2}^{}}}$ &    $0.9747$ \\[0.4em]
              $d\indices*{*_{A_{3}}^{\tensor*[^{1}]{\hspace{-0.2em}A}{}_{1}^{}}_{X_{2}}^{\tensor*[^{1}]{\hspace{-0.2em}E}{}_{u}^{}}_{\bar{A}_{1}}^{\tensor*[^{1}]{\hspace{-0.2em}B}{}_{2}^{}}}$ &  $- 0.7220$ &                                             $d\indices*{*_{A_{3}}^{\tensor*[^{2}]{\hspace{-0.2em}A}{}_{1}^{}}_{X_{2}}^{A_{1g}^{}}_{\bar{A}_{1}}^{\tensor*[^{2}]{\hspace{-0.2em}A}{}_{1}^{}}}$ &  $- 3.8518$ &                                             $d\indices*{*_{A_{3}}^{\tensor*[^{2}]{\hspace{-0.2em}A}{}_{1}^{}}_{X_{2}}^{A_{2u}^{}}_{\bar{A}_{1}}^{\tensor*[^{2}]{\hspace{-0.2em}A}{}_{1}^{}}}$ &  $- 2.9444$ &                                                                               $d\indices*{*_{A_{3}}^{\tensor*[^{2}]{\hspace{-0.2em}A}{}_{1}^{}}_{X_{2}}^{E_{g}^{}}_{\bar{A}_{1}}^{A_{2}^{}}}$ &    $9.1599$ &                                                                              $d\indices*{*_{A_{3}}^{\tensor*[^{2}]{\hspace{-0.2em}A}{}_{1}^{}}_{X_{2}}^{B_{1u}^{}}_{\bar{A}_{1}}^{A_{2}^{}}}$ &  $- 2.7691$ \\[0.4em]
                                                                                $d\indices*{*_{A_{3}}^{\tensor*[^{2}]{\hspace{-0.2em}A}{}_{1}^{}}_{X_{2}}^{E_{u}^{}}_{\bar{A}_{1}}^{A_{2}^{}}}$ &  $- 3.0303$ &                                              $d\indices*{*_{A_{3}}^{\tensor*[^{2}]{\hspace{-0.2em}A}{}_{1}^{}}_{X_{2}}^{\tensor*[^{1}]{\hspace{-0.2em}E}{}_{u}^{}}_{\bar{A}_{1}}^{A_{2}^{}}}$ &  $- 1.8831$ &                                                                               $d\indices*{*_{A_{3}}^{\tensor*[^{2}]{\hspace{-0.2em}A}{}_{1}^{}}_{X_{2}}^{E_{g}^{}}_{\bar{A}_{1}}^{B_{1}^{}}}$ &  $- 1.4030$ &                                                                              $d\indices*{*_{A_{3}}^{\tensor*[^{2}]{\hspace{-0.2em}A}{}_{1}^{}}_{X_{2}}^{B_{1u}^{}}_{\bar{A}_{1}}^{B_{1}^{}}}$ &  $- 7.4358$ &                                                                               $d\indices*{*_{A_{3}}^{\tensor*[^{2}]{\hspace{-0.2em}A}{}_{1}^{}}_{X_{2}}^{E_{u}^{}}_{\bar{A}_{1}}^{B_{1}^{}}}$ &    $6.9694$ \\[0.4em]
                                               $d\indices*{*_{A_{3}}^{\tensor*[^{2}]{\hspace{-0.2em}A}{}_{1}^{}}_{X_{2}}^{\tensor*[^{1}]{\hspace{-0.2em}E}{}_{u}^{}}_{\bar{A}_{1}}^{B_{1}^{}}}$ &  $- 6.9993$ &                                              $d\indices*{*_{A_{3}}^{\tensor*[^{2}]{\hspace{-0.2em}A}{}_{1}^{}}_{X_{2}}^{E_{g}^{}}_{\bar{A}_{1}}^{\tensor*[^{1}]{\hspace{-0.2em}B}{}_{1}^{}}}$ &    $1.9265$ &                                             $d\indices*{*_{A_{3}}^{\tensor*[^{2}]{\hspace{-0.2em}A}{}_{1}^{}}_{X_{2}}^{B_{1u}^{}}_{\bar{A}_{1}}^{\tensor*[^{1}]{\hspace{-0.2em}B}{}_{1}^{}}}$ &    $8.3652$ &                                              $d\indices*{*_{A_{3}}^{\tensor*[^{2}]{\hspace{-0.2em}A}{}_{1}^{}}_{X_{2}}^{E_{u}^{}}_{\bar{A}_{1}}^{\tensor*[^{1}]{\hspace{-0.2em}B}{}_{1}^{}}}$ &  $- 7.5997$ &             $d\indices*{*_{A_{3}}^{\tensor*[^{2}]{\hspace{-0.2em}A}{}_{1}^{}}_{X_{2}}^{\tensor*[^{1}]{\hspace{-0.2em}E}{}_{u}^{}}_{\bar{A}_{1}}^{\tensor*[^{1}]{\hspace{-0.2em}B}{}_{1}^{}}}$ &    $7.3510$ \\[0.4em]
                                               $d\indices*{*_{A_{3}}^{\tensor*[^{2}]{\hspace{-0.2em}A}{}_{1}^{}}_{X_{2}}^{E_{g}^{}}_{\bar{A}_{1}}^{\tensor*[^{2}]{\hspace{-0.2em}B}{}_{1}^{}}}$ &  $- 0.5347$ &                                             $d\indices*{*_{A_{3}}^{\tensor*[^{2}]{\hspace{-0.2em}A}{}_{1}^{}}_{X_{2}}^{B_{1u}^{}}_{\bar{A}_{1}}^{\tensor*[^{2}]{\hspace{-0.2em}B}{}_{1}^{}}}$ &  $- 0.9181$ &                                              $d\indices*{*_{A_{3}}^{\tensor*[^{2}]{\hspace{-0.2em}A}{}_{1}^{}}_{X_{2}}^{E_{u}^{}}_{\bar{A}_{1}}^{\tensor*[^{2}]{\hspace{-0.2em}B}{}_{1}^{}}}$ &  $- 4.2795$ &             $d\indices*{*_{A_{3}}^{\tensor*[^{2}]{\hspace{-0.2em}A}{}_{1}^{}}_{X_{2}}^{\tensor*[^{1}]{\hspace{-0.2em}E}{}_{u}^{}}_{\bar{A}_{1}}^{\tensor*[^{2}]{\hspace{-0.2em}B}{}_{1}^{}}}$ &  $- 0.0548$ &                                                                              $d\indices*{*_{A_{3}}^{\tensor*[^{2}]{\hspace{-0.2em}A}{}_{1}^{}}_{X_{2}}^{A_{1g}^{}}_{\bar{A}_{1}}^{B_{2}^{}}}$ &    $0.8159$ \\[0.4em]
                                                                                $d\indices*{*_{A_{3}}^{\tensor*[^{2}]{\hspace{-0.2em}A}{}_{1}^{}}_{X_{2}}^{E_{g}^{}}_{\bar{A}_{1}}^{B_{2}^{}}}$ &  $- 1.8459$ &                                                                              $d\indices*{*_{A_{3}}^{\tensor*[^{2}]{\hspace{-0.2em}A}{}_{1}^{}}_{X_{2}}^{A_{2u}^{}}_{\bar{A}_{1}}^{B_{2}^{}}}$ &   $16.8424$ &                                                                               $d\indices*{*_{A_{3}}^{\tensor*[^{2}]{\hspace{-0.2em}A}{}_{1}^{}}_{X_{2}}^{E_{u}^{}}_{\bar{A}_{1}}^{B_{2}^{}}}$ &   $10.0107$ &                                              $d\indices*{*_{A_{3}}^{\tensor*[^{2}]{\hspace{-0.2em}A}{}_{1}^{}}_{X_{2}}^{\tensor*[^{1}]{\hspace{-0.2em}E}{}_{u}^{}}_{\bar{A}_{1}}^{B_{2}^{}}}$ & $- 14.4125$ &                                             $d\indices*{*_{A_{3}}^{\tensor*[^{2}]{\hspace{-0.2em}A}{}_{1}^{}}_{X_{2}}^{A_{1g}^{}}_{\bar{A}_{1}}^{\tensor*[^{1}]{\hspace{-0.2em}B}{}_{2}^{}}}$ &  $- 2.1783$ \\[0.4em]
                                               $d\indices*{*_{A_{3}}^{\tensor*[^{2}]{\hspace{-0.2em}A}{}_{1}^{}}_{X_{2}}^{E_{g}^{}}_{\bar{A}_{1}}^{\tensor*[^{1}]{\hspace{-0.2em}B}{}_{2}^{}}}$ &    $3.2086$ &                                             $d\indices*{*_{A_{3}}^{\tensor*[^{2}]{\hspace{-0.2em}A}{}_{1}^{}}_{X_{2}}^{A_{2u}^{}}_{\bar{A}_{1}}^{\tensor*[^{1}]{\hspace{-0.2em}B}{}_{2}^{}}}$ & $- 10.3698$ &                                              $d\indices*{*_{A_{3}}^{\tensor*[^{2}]{\hspace{-0.2em}A}{}_{1}^{}}_{X_{2}}^{E_{u}^{}}_{\bar{A}_{1}}^{\tensor*[^{1}]{\hspace{-0.2em}B}{}_{2}^{}}}$ &  $- 8.6552$ &             $d\indices*{*_{A_{3}}^{\tensor*[^{2}]{\hspace{-0.2em}A}{}_{1}^{}}_{X_{2}}^{\tensor*[^{1}]{\hspace{-0.2em}E}{}_{u}^{}}_{\bar{A}_{1}}^{\tensor*[^{1}]{\hspace{-0.2em}B}{}_{2}^{}}}$ &    $9.7198$ &                                                                                                               $d\indices*{*_{A_{3}}^{A_{2}^{}}_{X_{2}}^{A_{1g}^{}}_{\bar{A}_{1}}^{A_{2}^{}}}$ &    $6.2857$ \\[0.4em]
                                                                                                                $d\indices*{*_{A_{3}}^{A_{2}^{}}_{X_{2}}^{A_{2u}^{}}_{\bar{A}_{1}}^{A_{2}^{}}}$ &    $0.3302$ &                                                                                                               $d\indices*{*_{A_{3}}^{A_{2}^{}}_{X_{2}}^{A_{1g}^{}}_{\bar{A}_{1}}^{B_{1}^{}}}$ &  $- 6.2894$ &                                                                                                                $d\indices*{*_{A_{3}}^{A_{2}^{}}_{X_{2}}^{E_{g}^{}}_{\bar{A}_{1}}^{B_{1}^{}}}$ &  $- 5.8045$ &                                                                                                               $d\indices*{*_{A_{3}}^{A_{2}^{}}_{X_{2}}^{A_{2u}^{}}_{\bar{A}_{1}}^{B_{1}^{}}}$ &  $- 1.0300$ &                                                                                                                $d\indices*{*_{A_{3}}^{A_{2}^{}}_{X_{2}}^{E_{u}^{}}_{\bar{A}_{1}}^{B_{1}^{}}}$ &  $- 0.3274$ \\[0.4em]
                                                                                $d\indices*{*_{A_{3}}^{A_{2}^{}}_{X_{2}}^{\tensor*[^{1}]{\hspace{-0.2em}E}{}_{u}^{}}_{\bar{A}_{1}}^{B_{1}^{}}}$ &  $- 0.2852$ &                                                                              $d\indices*{*_{A_{3}}^{A_{2}^{}}_{X_{2}}^{A_{1g}^{}}_{\bar{A}_{1}}^{\tensor*[^{1}]{\hspace{-0.2em}B}{}_{1}^{}}}$ &    $0.2560$ &                                                                               $d\indices*{*_{A_{3}}^{A_{2}^{}}_{X_{2}}^{E_{g}^{}}_{\bar{A}_{1}}^{\tensor*[^{1}]{\hspace{-0.2em}B}{}_{1}^{}}}$ &  $- 1.7962$ &                                                                              $d\indices*{*_{A_{3}}^{A_{2}^{}}_{X_{2}}^{A_{2u}^{}}_{\bar{A}_{1}}^{\tensor*[^{1}]{\hspace{-0.2em}B}{}_{1}^{}}}$ & $- 13.6715$ &                                                                               $d\indices*{*_{A_{3}}^{A_{2}^{}}_{X_{2}}^{E_{u}^{}}_{\bar{A}_{1}}^{\tensor*[^{1}]{\hspace{-0.2em}B}{}_{1}^{}}}$ &    $5.6913$ \\[0.4em]
                                               $d\indices*{*_{A_{3}}^{A_{2}^{}}_{X_{2}}^{\tensor*[^{1}]{\hspace{-0.2em}E}{}_{u}^{}}_{\bar{A}_{1}}^{\tensor*[^{1}]{\hspace{-0.2em}B}{}_{1}^{}}}$ & $- 10.5519$ &                                                                              $d\indices*{*_{A_{3}}^{A_{2}^{}}_{X_{2}}^{A_{1g}^{}}_{\bar{A}_{1}}^{\tensor*[^{2}]{\hspace{-0.2em}B}{}_{1}^{}}}$ &  $- 4.5510$ &                                                                               $d\indices*{*_{A_{3}}^{A_{2}^{}}_{X_{2}}^{E_{g}^{}}_{\bar{A}_{1}}^{\tensor*[^{2}]{\hspace{-0.2em}B}{}_{1}^{}}}$ &    $1.7370$ &                                                                              $d\indices*{*_{A_{3}}^{A_{2}^{}}_{X_{2}}^{A_{2u}^{}}_{\bar{A}_{1}}^{\tensor*[^{2}]{\hspace{-0.2em}B}{}_{1}^{}}}$ & $- 10.3328$ &                                                                               $d\indices*{*_{A_{3}}^{A_{2}^{}}_{X_{2}}^{E_{u}^{}}_{\bar{A}_{1}}^{\tensor*[^{2}]{\hspace{-0.2em}B}{}_{1}^{}}}$ &    $8.0722$ \\[0.4em]
                                               $d\indices*{*_{A_{3}}^{A_{2}^{}}_{X_{2}}^{\tensor*[^{1}]{\hspace{-0.2em}E}{}_{u}^{}}_{\bar{A}_{1}}^{\tensor*[^{2}]{\hspace{-0.2em}B}{}_{1}^{}}}$ &  $- 8.5372$ &                                                                                                                $d\indices*{*_{A_{3}}^{B_{2}^{}}_{X_{2}}^{E_{g}^{}}_{\bar{A}_{1}}^{A_{2}^{}}}$ &  $- 0.8470$ &                                                                                                               $d\indices*{*_{A_{3}}^{B_{2}^{}}_{X_{2}}^{B_{1u}^{}}_{\bar{A}_{1}}^{A_{2}^{}}}$ &    $0.6476$ &                                                                                                                $d\indices*{*_{A_{3}}^{B_{2}^{}}_{X_{2}}^{E_{u}^{}}_{\bar{A}_{1}}^{A_{2}^{}}}$ &  $- 0.7884$ &                                                                               $d\indices*{*_{A_{3}}^{B_{2}^{}}_{X_{2}}^{\tensor*[^{1}]{\hspace{-0.2em}E}{}_{u}^{}}_{\bar{A}_{1}}^{A_{2}^{}}}$ &    $0.8607$ \\[0.4em]
                                                                                $d\indices*{*_{A_{3}}^{\tensor*[^{1}]{\hspace{-0.2em}B}{}_{2}^{}}_{X_{2}}^{E_{g}^{}}_{\bar{A}_{1}}^{A_{2}^{}}}$ &  $- 0.1728$ &                                                                              $d\indices*{*_{A_{3}}^{\tensor*[^{1}]{\hspace{-0.2em}B}{}_{2}^{}}_{X_{2}}^{B_{1u}^{}}_{\bar{A}_{1}}^{A_{2}^{}}}$ &  $- 0.5064$ &                                                                               $d\indices*{*_{A_{3}}^{\tensor*[^{1}]{\hspace{-0.2em}B}{}_{2}^{}}_{X_{2}}^{E_{u}^{}}_{\bar{A}_{1}}^{A_{2}^{}}}$ &    $6.2568$ &                                              $d\indices*{*_{A_{3}}^{\tensor*[^{1}]{\hspace{-0.2em}B}{}_{2}^{}}_{X_{2}}^{\tensor*[^{1}]{\hspace{-0.2em}E}{}_{u}^{}}_{\bar{A}_{1}}^{A_{2}^{}}}$ &    $0.0439$ &                                                                                                               $d\indices*{*_{A_{3}}^{B_{1}^{}}_{X_{2}}^{A_{1g}^{}}_{\bar{A}_{1}}^{B_{1}^{}}}$ &  $- 1.0857$ \\[0.4em]
                                                                                                                $d\indices*{*_{A_{3}}^{B_{1}^{}}_{X_{2}}^{A_{2u}^{}}_{\bar{A}_{1}}^{B_{1}^{}}}$ &   $12.5755$ &                                                                              $d\indices*{*_{A_{3}}^{B_{1}^{}}_{X_{2}}^{A_{1g}^{}}_{\bar{A}_{1}}^{\tensor*[^{1}]{\hspace{-0.2em}B}{}_{1}^{}}}$ &  $- 5.2087$ &                                                                               $d\indices*{*_{A_{3}}^{B_{1}^{}}_{X_{2}}^{E_{g}^{}}_{\bar{A}_{1}}^{\tensor*[^{1}]{\hspace{-0.2em}B}{}_{1}^{}}}$ &  $- 5.2729$ &                                                                              $d\indices*{*_{A_{3}}^{B_{1}^{}}_{X_{2}}^{A_{2u}^{}}_{\bar{A}_{1}}^{\tensor*[^{1}]{\hspace{-0.2em}B}{}_{1}^{}}}$ &  $- 9.2973$ &                                                                               $d\indices*{*_{A_{3}}^{B_{1}^{}}_{X_{2}}^{E_{u}^{}}_{\bar{A}_{1}}^{\tensor*[^{1}]{\hspace{-0.2em}B}{}_{1}^{}}}$ &  $- 4.2572$ \\[0.4em]
                                               $d\indices*{*_{A_{3}}^{B_{1}^{}}_{X_{2}}^{\tensor*[^{1}]{\hspace{-0.2em}E}{}_{u}^{}}_{\bar{A}_{1}}^{\tensor*[^{1}]{\hspace{-0.2em}B}{}_{1}^{}}}$ &    $7.9850$ &                                                                              $d\indices*{*_{A_{3}}^{B_{1}^{}}_{X_{2}}^{A_{1g}^{}}_{\bar{A}_{1}}^{\tensor*[^{2}]{\hspace{-0.2em}B}{}_{1}^{}}}$ &  $- 6.8302$ &                                                                               $d\indices*{*_{A_{3}}^{B_{1}^{}}_{X_{2}}^{E_{g}^{}}_{\bar{A}_{1}}^{\tensor*[^{2}]{\hspace{-0.2em}B}{}_{1}^{}}}$ &  $- 6.3375$ &                                                                              $d\indices*{*_{A_{3}}^{B_{1}^{}}_{X_{2}}^{A_{2u}^{}}_{\bar{A}_{1}}^{\tensor*[^{2}]{\hspace{-0.2em}B}{}_{1}^{}}}$ &  $- 0.1345$ &                                                                               $d\indices*{*_{A_{3}}^{B_{1}^{}}_{X_{2}}^{E_{u}^{}}_{\bar{A}_{1}}^{\tensor*[^{2}]{\hspace{-0.2em}B}{}_{1}^{}}}$ &    $1.4586$ \\[0.4em]
                                               $d\indices*{*_{A_{3}}^{B_{1}^{}}_{X_{2}}^{\tensor*[^{1}]{\hspace{-0.2em}E}{}_{u}^{}}_{\bar{A}_{1}}^{\tensor*[^{2}]{\hspace{-0.2em}B}{}_{1}^{}}}$ &  $- 0.8179$ &                                                                                                                $d\indices*{*_{A_{3}}^{B_{2}^{}}_{X_{2}}^{E_{g}^{}}_{\bar{A}_{1}}^{B_{1}^{}}}$ &  $- 0.7967$ &                                                                                                               $d\indices*{*_{A_{3}}^{B_{2}^{}}_{X_{2}}^{B_{1u}^{}}_{\bar{A}_{1}}^{B_{1}^{}}}$ &    $9.5348$ &                                                                                                                $d\indices*{*_{A_{3}}^{B_{2}^{}}_{X_{2}}^{E_{u}^{}}_{\bar{A}_{1}}^{B_{1}^{}}}$ &    $5.7157$ &                                                                               $d\indices*{*_{A_{3}}^{B_{2}^{}}_{X_{2}}^{\tensor*[^{1}]{\hspace{-0.2em}E}{}_{u}^{}}_{\bar{A}_{1}}^{B_{1}^{}}}$ &  $- 9.7988$ \\[0.4em]
                                                                                $d\indices*{*_{A_{3}}^{\tensor*[^{1}]{\hspace{-0.2em}B}{}_{2}^{}}_{X_{2}}^{E_{g}^{}}_{\bar{A}_{1}}^{B_{1}^{}}}$ &    $1.8338$ &                                                                              $d\indices*{*_{A_{3}}^{\tensor*[^{1}]{\hspace{-0.2em}B}{}_{2}^{}}_{X_{2}}^{B_{1u}^{}}_{\bar{A}_{1}}^{B_{1}^{}}}$ &  $- 5.7015$ &                                                                               $d\indices*{*_{A_{3}}^{\tensor*[^{1}]{\hspace{-0.2em}B}{}_{2}^{}}_{X_{2}}^{E_{u}^{}}_{\bar{A}_{1}}^{B_{1}^{}}}$ &  $- 5.1679$ &                                              $d\indices*{*_{A_{3}}^{\tensor*[^{1}]{\hspace{-0.2em}B}{}_{2}^{}}_{X_{2}}^{\tensor*[^{1}]{\hspace{-0.2em}E}{}_{u}^{}}_{\bar{A}_{1}}^{B_{1}^{}}}$ &    $6.2222$ &                                             $d\indices*{*_{A_{3}}^{\tensor*[^{1}]{\hspace{-0.2em}B}{}_{1}^{}}_{X_{2}}^{A_{1g}^{}}_{\bar{A}_{1}}^{\tensor*[^{1}]{\hspace{-0.2em}B}{}_{1}^{}}}$ &   $13.2145$ \\[0.4em]
                                              $d\indices*{*_{A_{3}}^{\tensor*[^{1}]{\hspace{-0.2em}B}{}_{1}^{}}_{X_{2}}^{A_{2u}^{}}_{\bar{A}_{1}}^{\tensor*[^{1}]{\hspace{-0.2em}B}{}_{1}^{}}}$ &  $- 3.5000$ &                                             $d\indices*{*_{A_{3}}^{\tensor*[^{1}]{\hspace{-0.2em}B}{}_{1}^{}}_{X_{2}}^{A_{1g}^{}}_{\bar{A}_{1}}^{\tensor*[^{2}]{\hspace{-0.2em}B}{}_{1}^{}}}$ &    $7.6676$ &                                              $d\indices*{*_{A_{3}}^{\tensor*[^{1}]{\hspace{-0.2em}B}{}_{1}^{}}_{X_{2}}^{E_{g}^{}}_{\bar{A}_{1}}^{\tensor*[^{2}]{\hspace{-0.2em}B}{}_{1}^{}}}$ &    $8.1967$ &                                             $d\indices*{*_{A_{3}}^{\tensor*[^{1}]{\hspace{-0.2em}B}{}_{1}^{}}_{X_{2}}^{A_{2u}^{}}_{\bar{A}_{1}}^{\tensor*[^{2}]{\hspace{-0.2em}B}{}_{1}^{}}}$ &    $1.3561$ &                                              $d\indices*{*_{A_{3}}^{\tensor*[^{1}]{\hspace{-0.2em}B}{}_{1}^{}}_{X_{2}}^{E_{u}^{}}_{\bar{A}_{1}}^{\tensor*[^{2}]{\hspace{-0.2em}B}{}_{1}^{}}}$ &  $- 0.0473$ \\[0.4em]
              $d\indices*{*_{A_{3}}^{\tensor*[^{1}]{\hspace{-0.2em}B}{}_{1}^{}}_{X_{2}}^{\tensor*[^{1}]{\hspace{-0.2em}E}{}_{u}^{}}_{\bar{A}_{1}}^{\tensor*[^{2}]{\hspace{-0.2em}B}{}_{1}^{}}}$ &    $0.1571$ &                                                                               $d\indices*{*_{A_{3}}^{B_{2}^{}}_{X_{2}}^{E_{g}^{}}_{\bar{A}_{1}}^{\tensor*[^{1}]{\hspace{-0.2em}B}{}_{1}^{}}}$ &    $9.1222$ &                                                                              $d\indices*{*_{A_{3}}^{B_{2}^{}}_{X_{2}}^{B_{1u}^{}}_{\bar{A}_{1}}^{\tensor*[^{1}]{\hspace{-0.2em}B}{}_{1}^{}}}$ &    $1.0555$ &                                                                               $d\indices*{*_{A_{3}}^{B_{2}^{}}_{X_{2}}^{E_{u}^{}}_{\bar{A}_{1}}^{\tensor*[^{1}]{\hspace{-0.2em}B}{}_{1}^{}}}$ &  $- 0.3543$ &                                              $d\indices*{*_{A_{3}}^{B_{2}^{}}_{X_{2}}^{\tensor*[^{1}]{\hspace{-0.2em}E}{}_{u}^{}}_{\bar{A}_{1}}^{\tensor*[^{1}]{\hspace{-0.2em}B}{}_{1}^{}}}$ &  $- 4.6639$ \\[0.4em]
                                               $d\indices*{*_{A_{3}}^{\tensor*[^{1}]{\hspace{-0.2em}B}{}_{2}^{}}_{X_{2}}^{E_{g}^{}}_{\bar{A}_{1}}^{\tensor*[^{1}]{\hspace{-0.2em}B}{}_{1}^{}}}$ &  $- 7.3217$ &                                             $d\indices*{*_{A_{3}}^{\tensor*[^{1}]{\hspace{-0.2em}B}{}_{2}^{}}_{X_{2}}^{B_{1u}^{}}_{\bar{A}_{1}}^{\tensor*[^{1}]{\hspace{-0.2em}B}{}_{1}^{}}}$ &    $0.7520$ &                                              $d\indices*{*_{A_{3}}^{\tensor*[^{1}]{\hspace{-0.2em}B}{}_{2}^{}}_{X_{2}}^{E_{u}^{}}_{\bar{A}_{1}}^{\tensor*[^{1}]{\hspace{-0.2em}B}{}_{1}^{}}}$ &  $- 1.5109$ &             $d\indices*{*_{A_{3}}^{\tensor*[^{1}]{\hspace{-0.2em}B}{}_{2}^{}}_{X_{2}}^{\tensor*[^{1}]{\hspace{-0.2em}E}{}_{u}^{}}_{\bar{A}_{1}}^{\tensor*[^{1}]{\hspace{-0.2em}B}{}_{1}^{}}}$ &    $1.0281$ &                                             $d\indices*{*_{A_{3}}^{\tensor*[^{2}]{\hspace{-0.2em}B}{}_{1}^{}}_{X_{2}}^{A_{1g}^{}}_{\bar{A}_{1}}^{\tensor*[^{2}]{\hspace{-0.2em}B}{}_{1}^{}}}$ &    $3.2812$ \\[0.4em]
                                              $d\indices*{*_{A_{3}}^{\tensor*[^{2}]{\hspace{-0.2em}B}{}_{1}^{}}_{X_{2}}^{A_{2u}^{}}_{\bar{A}_{1}}^{\tensor*[^{2}]{\hspace{-0.2em}B}{}_{1}^{}}}$ &    $0.5353$ &                                                                               $d\indices*{*_{A_{3}}^{B_{2}^{}}_{X_{2}}^{E_{g}^{}}_{\bar{A}_{1}}^{\tensor*[^{2}]{\hspace{-0.2em}B}{}_{1}^{}}}$ &   $10.1869$ &                                                                              $d\indices*{*_{A_{3}}^{B_{2}^{}}_{X_{2}}^{B_{1u}^{}}_{\bar{A}_{1}}^{\tensor*[^{2}]{\hspace{-0.2em}B}{}_{1}^{}}}$ &  $- 0.8993$ &                                                                               $d\indices*{*_{A_{3}}^{B_{2}^{}}_{X_{2}}^{E_{u}^{}}_{\bar{A}_{1}}^{\tensor*[^{2}]{\hspace{-0.2em}B}{}_{1}^{}}}$ &    $1.6277$ &                                              $d\indices*{*_{A_{3}}^{B_{2}^{}}_{X_{2}}^{\tensor*[^{1}]{\hspace{-0.2em}E}{}_{u}^{}}_{\bar{A}_{1}}^{\tensor*[^{2}]{\hspace{-0.2em}B}{}_{1}^{}}}$ &    $1.7570$ \\[0.4em]
                                               $d\indices*{*_{A_{3}}^{\tensor*[^{1}]{\hspace{-0.2em}B}{}_{2}^{}}_{X_{2}}^{E_{g}^{}}_{\bar{A}_{1}}^{\tensor*[^{2}]{\hspace{-0.2em}B}{}_{1}^{}}}$ &  $- 9.9289$ &                                             $d\indices*{*_{A_{3}}^{\tensor*[^{1}]{\hspace{-0.2em}B}{}_{2}^{}}_{X_{2}}^{B_{1u}^{}}_{\bar{A}_{1}}^{\tensor*[^{2}]{\hspace{-0.2em}B}{}_{1}^{}}}$ &    $1.4588$ &                                              $d\indices*{*_{A_{3}}^{\tensor*[^{1}]{\hspace{-0.2em}B}{}_{2}^{}}_{X_{2}}^{E_{u}^{}}_{\bar{A}_{1}}^{\tensor*[^{2}]{\hspace{-0.2em}B}{}_{1}^{}}}$ &  $- 4.5742$ &             $d\indices*{*_{A_{3}}^{\tensor*[^{1}]{\hspace{-0.2em}B}{}_{2}^{}}_{X_{2}}^{\tensor*[^{1}]{\hspace{-0.2em}E}{}_{u}^{}}_{\bar{A}_{1}}^{\tensor*[^{2}]{\hspace{-0.2em}B}{}_{1}^{}}}$ &    $1.5913$ &                                                                                                               $d\indices*{*_{A_{3}}^{B_{2}^{}}_{X_{2}}^{A_{1g}^{}}_{\bar{A}_{1}}^{B_{2}^{}}}$ &    $0.4428$ \\[0.4em]
                                                                                                                $d\indices*{*_{A_{3}}^{B_{2}^{}}_{X_{2}}^{A_{2u}^{}}_{\bar{A}_{1}}^{B_{2}^{}}}$ &  $- 2.6626$ &                                                                              $d\indices*{*_{A_{3}}^{B_{2}^{}}_{X_{2}}^{A_{1g}^{}}_{\bar{A}_{1}}^{\tensor*[^{1}]{\hspace{-0.2em}B}{}_{2}^{}}}$ &    $0.8355$ &                                                                               $d\indices*{*_{A_{3}}^{B_{2}^{}}_{X_{2}}^{E_{g}^{}}_{\bar{A}_{1}}^{\tensor*[^{1}]{\hspace{-0.2em}B}{}_{2}^{}}}$ &    $0.4024$ &                                                                              $d\indices*{*_{A_{3}}^{B_{2}^{}}_{X_{2}}^{A_{2u}^{}}_{\bar{A}_{1}}^{\tensor*[^{1}]{\hspace{-0.2em}B}{}_{2}^{}}}$ &  $- 0.4607$ &                                                                               $d\indices*{*_{A_{3}}^{B_{2}^{}}_{X_{2}}^{E_{u}^{}}_{\bar{A}_{1}}^{\tensor*[^{1}]{\hspace{-0.2em}B}{}_{2}^{}}}$ &  $- 1.0907$ \\[0.4em]
                                               $d\indices*{*_{A_{3}}^{B_{2}^{}}_{X_{2}}^{\tensor*[^{1}]{\hspace{-0.2em}E}{}_{u}^{}}_{\bar{A}_{1}}^{\tensor*[^{1}]{\hspace{-0.2em}B}{}_{2}^{}}}$ &    $0.0971$ &                                             $d\indices*{*_{A_{3}}^{\tensor*[^{1}]{\hspace{-0.2em}B}{}_{2}^{}}_{X_{2}}^{A_{1g}^{}}_{\bar{A}_{1}}^{\tensor*[^{1}]{\hspace{-0.2em}B}{}_{2}^{}}}$ &  $- 6.0181$ &                                             $d\indices*{*_{A_{3}}^{\tensor*[^{1}]{\hspace{-0.2em}B}{}_{2}^{}}_{X_{2}}^{A_{2u}^{}}_{\bar{A}_{1}}^{\tensor*[^{1}]{\hspace{-0.2em}B}{}_{2}^{}}}$ &    $0.4058$ &                                                                                                          $d\indices*{*_{A_{5}}^{A_{1}^{}}_{\bar{A}_{4}}^{A_{1}^{}}_{\bar{A}_{1}}^{A_{1}^{}}}$ &  $- 1.2180$ &                                                                         $d\indices*{*_{A_{5}}^{A_{1}^{}}_{\bar{A}_{4}}^{A_{1}^{}}_{\bar{A}_{1}}^{\tensor*[^{1}]{\hspace{-0.2em}A}{}_{1}^{}}}$ &    $1.2712$ \\[0.4em]
                                                                          $d\indices*{*_{A_{5}}^{A_{1}^{}}_{\bar{A}_{4}}^{A_{1}^{}}_{\bar{A}_{1}}^{\tensor*[^{2}]{\hspace{-0.2em}A}{}_{1}^{}}}$ &   $11.8528$ &                                                                                                          $d\indices*{*_{A_{5}}^{A_{1}^{}}_{\bar{A}_{4}}^{A_{1}^{}}_{\bar{A}_{1}}^{B_{1}^{}}}$ &    $8.3068$ &                                                                         $d\indices*{*_{A_{5}}^{A_{1}^{}}_{\bar{A}_{4}}^{A_{1}^{}}_{\bar{A}_{1}}^{\tensor*[^{1}]{\hspace{-0.2em}B}{}_{1}^{}}}$ &    $3.5155$ &                                                                         $d\indices*{*_{A_{5}}^{A_{1}^{}}_{\bar{A}_{4}}^{A_{1}^{}}_{\bar{A}_{1}}^{\tensor*[^{2}]{\hspace{-0.2em}B}{}_{1}^{}}}$ &  $- 0.2277$ &                                        $d\indices*{*_{A_{5}}^{A_{1}^{}}_{\bar{A}_{4}}^{\tensor*[^{1}]{\hspace{-0.2em}A}{}_{1}^{}}_{\bar{A}_{1}}^{\tensor*[^{1}]{\hspace{-0.2em}A}{}_{1}^{}}}$ &  $- 0.8191$ \\[0.4em]
                                         $d\indices*{*_{A_{5}}^{A_{1}^{}}_{\bar{A}_{4}}^{\tensor*[^{1}]{\hspace{-0.2em}A}{}_{1}^{}}_{\bar{A}_{1}}^{\tensor*[^{2}]{\hspace{-0.2em}A}{}_{1}^{}}}$ &  $- 3.9673$ &                                                                         $d\indices*{*_{A_{5}}^{A_{1}^{}}_{\bar{A}_{4}}^{\tensor*[^{1}]{\hspace{-0.2em}A}{}_{1}^{}}_{\bar{A}_{1}}^{A_{2}^{}}}$ &    $0.7464$ &                                                                         $d\indices*{*_{A_{5}}^{A_{1}^{}}_{\bar{A}_{4}}^{\tensor*[^{1}]{\hspace{-0.2em}A}{}_{1}^{}}_{\bar{A}_{1}}^{B_{1}^{}}}$ &  $- 7.8286$ &                                        $d\indices*{*_{A_{5}}^{A_{1}^{}}_{\bar{A}_{4}}^{\tensor*[^{1}]{\hspace{-0.2em}A}{}_{1}^{}}_{\bar{A}_{1}}^{\tensor*[^{1}]{\hspace{-0.2em}B}{}_{1}^{}}}$ &    $4.2899$ &                                        $d\indices*{*_{A_{5}}^{A_{1}^{}}_{\bar{A}_{4}}^{\tensor*[^{1}]{\hspace{-0.2em}A}{}_{1}^{}}_{\bar{A}_{1}}^{\tensor*[^{2}]{\hspace{-0.2em}B}{}_{1}^{}}}$ &    $0.3002$ \\[0.4em]
                                                                          $d\indices*{*_{A_{5}}^{A_{1}^{}}_{\bar{A}_{4}}^{\tensor*[^{1}]{\hspace{-0.2em}A}{}_{1}^{}}_{\bar{A}_{1}}^{B_{2}^{}}}$ &  $- 7.8807$ &                                        $d\indices*{*_{A_{5}}^{A_{1}^{}}_{\bar{A}_{4}}^{\tensor*[^{1}]{\hspace{-0.2em}A}{}_{1}^{}}_{\bar{A}_{1}}^{\tensor*[^{1}]{\hspace{-0.2em}B}{}_{2}^{}}}$ &    $5.2314$ &                                        $d\indices*{*_{A_{5}}^{A_{1}^{}}_{\bar{A}_{4}}^{\tensor*[^{2}]{\hspace{-0.2em}A}{}_{1}^{}}_{\bar{A}_{1}}^{\tensor*[^{2}]{\hspace{-0.2em}A}{}_{1}^{}}}$ &    $1.5567$ &                                                                         $d\indices*{*_{A_{5}}^{A_{1}^{}}_{\bar{A}_{4}}^{\tensor*[^{2}]{\hspace{-0.2em}A}{}_{1}^{}}_{\bar{A}_{1}}^{A_{2}^{}}}$ &  $- 0.1058$ &                                                                         $d\indices*{*_{A_{5}}^{A_{1}^{}}_{\bar{A}_{4}}^{\tensor*[^{2}]{\hspace{-0.2em}A}{}_{1}^{}}_{\bar{A}_{1}}^{B_{1}^{}}}$ &    $0.3597$ \\[0.4em]
                                         $d\indices*{*_{A_{5}}^{A_{1}^{}}_{\bar{A}_{4}}^{\tensor*[^{2}]{\hspace{-0.2em}A}{}_{1}^{}}_{\bar{A}_{1}}^{\tensor*[^{1}]{\hspace{-0.2em}B}{}_{1}^{}}}$ &  $- 5.9474$ &                                        $d\indices*{*_{A_{5}}^{A_{1}^{}}_{\bar{A}_{4}}^{\tensor*[^{2}]{\hspace{-0.2em}A}{}_{1}^{}}_{\bar{A}_{1}}^{\tensor*[^{2}]{\hspace{-0.2em}B}{}_{1}^{}}}$ &  $- 7.8854$ &                                                                         $d\indices*{*_{A_{5}}^{A_{1}^{}}_{\bar{A}_{4}}^{\tensor*[^{2}]{\hspace{-0.2em}A}{}_{1}^{}}_{\bar{A}_{1}}^{B_{2}^{}}}$ &  $- 2.1159$ &                                        $d\indices*{*_{A_{5}}^{A_{1}^{}}_{\bar{A}_{4}}^{\tensor*[^{2}]{\hspace{-0.2em}A}{}_{1}^{}}_{\bar{A}_{1}}^{\tensor*[^{1}]{\hspace{-0.2em}B}{}_{2}^{}}}$ &  $- 0.0717$ &                                                                                                          $d\indices*{*_{A_{5}}^{A_{1}^{}}_{\bar{A}_{4}}^{A_{2}^{}}_{\bar{A}_{1}}^{A_{2}^{}}}$ &  $- 0.6711$ \\[0.4em]
                                                                                                           $d\indices*{*_{A_{5}}^{A_{1}^{}}_{\bar{A}_{4}}^{A_{2}^{}}_{\bar{A}_{1}}^{B_{1}^{}}}$ &    $5.2485$ &                                                                         $d\indices*{*_{A_{5}}^{A_{1}^{}}_{\bar{A}_{4}}^{A_{2}^{}}_{\bar{A}_{1}}^{\tensor*[^{1}]{\hspace{-0.2em}B}{}_{1}^{}}}$ &  $- 7.9296$ &                                                                         $d\indices*{*_{A_{5}}^{A_{1}^{}}_{\bar{A}_{4}}^{A_{2}^{}}_{\bar{A}_{1}}^{\tensor*[^{2}]{\hspace{-0.2em}B}{}_{1}^{}}}$ &    $0.3326$ &                                                                                                          $d\indices*{*_{A_{5}}^{A_{1}^{}}_{\bar{A}_{4}}^{A_{2}^{}}_{\bar{A}_{1}}^{B_{2}^{}}}$ & $- 11.0896$ &                                                                         $d\indices*{*_{A_{5}}^{A_{1}^{}}_{\bar{A}_{4}}^{A_{2}^{}}_{\bar{A}_{1}}^{\tensor*[^{1}]{\hspace{-0.2em}B}{}_{2}^{}}}$ &    $6.9269$ \\[0.4em]
                                                                                                           $d\indices*{*_{A_{5}}^{A_{1}^{}}_{\bar{A}_{4}}^{B_{1}^{}}_{\bar{A}_{1}}^{B_{1}^{}}}$ &    $1.0534$ &                                                                         $d\indices*{*_{A_{5}}^{A_{1}^{}}_{\bar{A}_{4}}^{B_{1}^{}}_{\bar{A}_{1}}^{\tensor*[^{1}]{\hspace{-0.2em}B}{}_{1}^{}}}$ &    $6.2502$ &                                                                         $d\indices*{*_{A_{5}}^{A_{1}^{}}_{\bar{A}_{4}}^{B_{1}^{}}_{\bar{A}_{1}}^{\tensor*[^{2}]{\hspace{-0.2em}B}{}_{1}^{}}}$ &    $5.9836$ &                                                                                                          $d\indices*{*_{A_{5}}^{A_{1}^{}}_{\bar{A}_{4}}^{B_{2}^{}}_{\bar{A}_{1}}^{B_{1}^{}}}$ &  $- 0.1458$ &                                                                         $d\indices*{*_{A_{5}}^{A_{1}^{}}_{\bar{A}_{4}}^{\tensor*[^{1}]{\hspace{-0.2em}B}{}_{2}^{}}_{\bar{A}_{1}}^{B_{1}^{}}}$ &    $0.8022$ \\[0.4em]
                                         $d\indices*{*_{A_{5}}^{A_{1}^{}}_{\bar{A}_{4}}^{\tensor*[^{1}]{\hspace{-0.2em}B}{}_{1}^{}}_{\bar{A}_{1}}^{\tensor*[^{1}]{\hspace{-0.2em}B}{}_{1}^{}}}$ &    $2.2250$ &                                        $d\indices*{*_{A_{5}}^{A_{1}^{}}_{\bar{A}_{4}}^{\tensor*[^{1}]{\hspace{-0.2em}B}{}_{1}^{}}_{\bar{A}_{1}}^{\tensor*[^{2}]{\hspace{-0.2em}B}{}_{1}^{}}}$ &  $- 0.0427$ &                                                                         $d\indices*{*_{A_{5}}^{A_{1}^{}}_{\bar{A}_{4}}^{B_{2}^{}}_{\bar{A}_{1}}^{\tensor*[^{1}]{\hspace{-0.2em}B}{}_{1}^{}}}$ &  $- 0.0263$ &                                        $d\indices*{*_{A_{5}}^{A_{1}^{}}_{\bar{A}_{4}}^{\tensor*[^{1}]{\hspace{-0.2em}B}{}_{2}^{}}_{\bar{A}_{1}}^{\tensor*[^{1}]{\hspace{-0.2em}B}{}_{1}^{}}}$ &    $0.9808$ &                                        $d\indices*{*_{A_{5}}^{A_{1}^{}}_{\bar{A}_{4}}^{\tensor*[^{2}]{\hspace{-0.2em}B}{}_{1}^{}}_{\bar{A}_{1}}^{\tensor*[^{2}]{\hspace{-0.2em}B}{}_{1}^{}}}$ &  $- 2.5601$ \\[0.4em]
                                                                          $d\indices*{*_{A_{5}}^{A_{1}^{}}_{\bar{A}_{4}}^{B_{2}^{}}_{\bar{A}_{1}}^{\tensor*[^{2}]{\hspace{-0.2em}B}{}_{1}^{}}}$ &    $0.0054$ &                                        $d\indices*{*_{A_{5}}^{A_{1}^{}}_{\bar{A}_{4}}^{\tensor*[^{1}]{\hspace{-0.2em}B}{}_{2}^{}}_{\bar{A}_{1}}^{\tensor*[^{2}]{\hspace{-0.2em}B}{}_{1}^{}}}$ &  $- 0.1213$ &                                                                                                          $d\indices*{*_{A_{5}}^{A_{1}^{}}_{\bar{A}_{4}}^{B_{2}^{}}_{\bar{A}_{1}}^{B_{2}^{}}}$ &  $- 0.2245$ &                                                                         $d\indices*{*_{A_{5}}^{A_{1}^{}}_{\bar{A}_{4}}^{B_{2}^{}}_{\bar{A}_{1}}^{\tensor*[^{1}]{\hspace{-0.2em}B}{}_{2}^{}}}$ &    $0.7437$ &                                        $d\indices*{*_{A_{5}}^{A_{1}^{}}_{\bar{A}_{4}}^{\tensor*[^{1}]{\hspace{-0.2em}B}{}_{2}^{}}_{\bar{A}_{1}}^{\tensor*[^{1}]{\hspace{-0.2em}B}{}_{2}^{}}}$ &  $- 0.2069$ \\[0.4em]
        $d\indices*{*_{A_{5}}^{\tensor*[^{1}]{\hspace{-0.2em}A}{}_{1}^{}}_{\bar{A}_{4}}^{\tensor*[^{1}]{\hspace{-0.2em}A}{}_{1}^{}}_{\bar{A}_{1}}^{\tensor*[^{1}]{\hspace{-0.2em}A}{}_{1}^{}}}$ &  $- 0.4742$ &       $d\indices*{*_{A_{5}}^{\tensor*[^{1}]{\hspace{-0.2em}A}{}_{1}^{}}_{\bar{A}_{4}}^{\tensor*[^{1}]{\hspace{-0.2em}A}{}_{1}^{}}_{\bar{A}_{1}}^{\tensor*[^{2}]{\hspace{-0.2em}A}{}_{1}^{}}}$ &    $0.2038$ &                                        $d\indices*{*_{A_{5}}^{\tensor*[^{1}]{\hspace{-0.2em}A}{}_{1}^{}}_{\bar{A}_{4}}^{\tensor*[^{1}]{\hspace{-0.2em}A}{}_{1}^{}}_{\bar{A}_{1}}^{B_{1}^{}}}$ &    $8.7607$ &       $d\indices*{*_{A_{5}}^{\tensor*[^{1}]{\hspace{-0.2em}A}{}_{1}^{}}_{\bar{A}_{4}}^{\tensor*[^{1}]{\hspace{-0.2em}A}{}_{1}^{}}_{\bar{A}_{1}}^{\tensor*[^{1}]{\hspace{-0.2em}B}{}_{1}^{}}}$ &  $- 5.5207$ &       $d\indices*{*_{A_{5}}^{\tensor*[^{1}]{\hspace{-0.2em}A}{}_{1}^{}}_{\bar{A}_{4}}^{\tensor*[^{1}]{\hspace{-0.2em}A}{}_{1}^{}}_{\bar{A}_{1}}^{\tensor*[^{2}]{\hspace{-0.2em}B}{}_{1}^{}}}$ &  $- 1.0313$ \\[0.4em]
        $d\indices*{*_{A_{5}}^{\tensor*[^{1}]{\hspace{-0.2em}A}{}_{1}^{}}_{\bar{A}_{4}}^{\tensor*[^{2}]{\hspace{-0.2em}A}{}_{1}^{}}_{\bar{A}_{1}}^{\tensor*[^{2}]{\hspace{-0.2em}A}{}_{1}^{}}}$ &    $1.5366$ &                                        $d\indices*{*_{A_{5}}^{\tensor*[^{1}]{\hspace{-0.2em}A}{}_{1}^{}}_{\bar{A}_{4}}^{\tensor*[^{2}]{\hspace{-0.2em}A}{}_{1}^{}}_{\bar{A}_{1}}^{A_{2}^{}}}$ &  $- 5.4372$ &                                        $d\indices*{*_{A_{5}}^{\tensor*[^{1}]{\hspace{-0.2em}A}{}_{1}^{}}_{\bar{A}_{4}}^{\tensor*[^{2}]{\hspace{-0.2em}A}{}_{1}^{}}_{\bar{A}_{1}}^{B_{1}^{}}}$ &  $- 0.9677$ &       $d\indices*{*_{A_{5}}^{\tensor*[^{1}]{\hspace{-0.2em}A}{}_{1}^{}}_{\bar{A}_{4}}^{\tensor*[^{2}]{\hspace{-0.2em}A}{}_{1}^{}}_{\bar{A}_{1}}^{\tensor*[^{1}]{\hspace{-0.2em}B}{}_{1}^{}}}$ &    $3.0410$ &       $d\indices*{*_{A_{5}}^{\tensor*[^{1}]{\hspace{-0.2em}A}{}_{1}^{}}_{\bar{A}_{4}}^{\tensor*[^{2}]{\hspace{-0.2em}A}{}_{1}^{}}_{\bar{A}_{1}}^{\tensor*[^{2}]{\hspace{-0.2em}B}{}_{1}^{}}}$ &    $4.1743$ \\[0.4em]
                                         $d\indices*{*_{A_{5}}^{\tensor*[^{1}]{\hspace{-0.2em}A}{}_{1}^{}}_{\bar{A}_{4}}^{\tensor*[^{2}]{\hspace{-0.2em}A}{}_{1}^{}}_{\bar{A}_{1}}^{B_{2}^{}}}$ &    $0.0366$ &       $d\indices*{*_{A_{5}}^{\tensor*[^{1}]{\hspace{-0.2em}A}{}_{1}^{}}_{\bar{A}_{4}}^{\tensor*[^{2}]{\hspace{-0.2em}A}{}_{1}^{}}_{\bar{A}_{1}}^{\tensor*[^{1}]{\hspace{-0.2em}B}{}_{2}^{}}}$ &    $3.5753$ &                                                                         $d\indices*{*_{A_{5}}^{\tensor*[^{1}]{\hspace{-0.2em}A}{}_{1}^{}}_{\bar{A}_{4}}^{A_{2}^{}}_{\bar{A}_{1}}^{A_{2}^{}}}$ &    $6.9017$ &                                                                         $d\indices*{*_{A_{5}}^{\tensor*[^{1}]{\hspace{-0.2em}A}{}_{1}^{}}_{\bar{A}_{4}}^{A_{2}^{}}_{\bar{A}_{1}}^{B_{1}^{}}}$ &  $- 0.0037$ &                                        $d\indices*{*_{A_{5}}^{\tensor*[^{1}]{\hspace{-0.2em}A}{}_{1}^{}}_{\bar{A}_{4}}^{A_{2}^{}}_{\bar{A}_{1}}^{\tensor*[^{1}]{\hspace{-0.2em}B}{}_{1}^{}}}$ &    $3.3010$ \\[0.4em]
                                         $d\indices*{*_{A_{5}}^{\tensor*[^{1}]{\hspace{-0.2em}A}{}_{1}^{}}_{\bar{A}_{4}}^{A_{2}^{}}_{\bar{A}_{1}}^{\tensor*[^{2}]{\hspace{-0.2em}B}{}_{1}^{}}}$ &    $3.5157$ &                                                                         $d\indices*{*_{A_{5}}^{\tensor*[^{1}]{\hspace{-0.2em}A}{}_{1}^{}}_{\bar{A}_{4}}^{A_{2}^{}}_{\bar{A}_{1}}^{B_{2}^{}}}$ &    $4.0172$ &                                        $d\indices*{*_{A_{5}}^{\tensor*[^{1}]{\hspace{-0.2em}A}{}_{1}^{}}_{\bar{A}_{4}}^{A_{2}^{}}_{\bar{A}_{1}}^{\tensor*[^{1}]{\hspace{-0.2em}B}{}_{2}^{}}}$ &  $- 2.8986$ &                                                                         $d\indices*{*_{A_{5}}^{\tensor*[^{1}]{\hspace{-0.2em}A}{}_{1}^{}}_{\bar{A}_{4}}^{B_{1}^{}}_{\bar{A}_{1}}^{B_{1}^{}}}$ &    $0.0546$ &                                        $d\indices*{*_{A_{5}}^{\tensor*[^{1}]{\hspace{-0.2em}A}{}_{1}^{}}_{\bar{A}_{4}}^{B_{1}^{}}_{\bar{A}_{1}}^{\tensor*[^{1}]{\hspace{-0.2em}B}{}_{1}^{}}}$ &    $0.1834$ \\[0.4em]
                                         $d\indices*{*_{A_{5}}^{\tensor*[^{1}]{\hspace{-0.2em}A}{}_{1}^{}}_{\bar{A}_{4}}^{B_{1}^{}}_{\bar{A}_{1}}^{\tensor*[^{2}]{\hspace{-0.2em}B}{}_{1}^{}}}$ &    $0.6745$ &                                                                         $d\indices*{*_{A_{5}}^{\tensor*[^{1}]{\hspace{-0.2em}A}{}_{1}^{}}_{\bar{A}_{4}}^{B_{2}^{}}_{\bar{A}_{1}}^{B_{1}^{}}}$ &  $- 1.5735$ &                                        $d\indices*{*_{A_{5}}^{\tensor*[^{1}]{\hspace{-0.2em}A}{}_{1}^{}}_{\bar{A}_{4}}^{\tensor*[^{1}]{\hspace{-0.2em}B}{}_{2}^{}}_{\bar{A}_{1}}^{B_{1}^{}}}$ &  $- 0.1240$ &       $d\indices*{*_{A_{5}}^{\tensor*[^{1}]{\hspace{-0.2em}A}{}_{1}^{}}_{\bar{A}_{4}}^{\tensor*[^{1}]{\hspace{-0.2em}B}{}_{1}^{}}_{\bar{A}_{1}}^{\tensor*[^{1}]{\hspace{-0.2em}B}{}_{1}^{}}}$ &  $- 5.2935$ &       $d\indices*{*_{A_{5}}^{\tensor*[^{1}]{\hspace{-0.2em}A}{}_{1}^{}}_{\bar{A}_{4}}^{\tensor*[^{1}]{\hspace{-0.2em}B}{}_{1}^{}}_{\bar{A}_{1}}^{\tensor*[^{2}]{\hspace{-0.2em}B}{}_{1}^{}}}$ &  $- 2.7915$ \\[0.4em]
                                         $d\indices*{*_{A_{5}}^{\tensor*[^{1}]{\hspace{-0.2em}A}{}_{1}^{}}_{\bar{A}_{4}}^{B_{2}^{}}_{\bar{A}_{1}}^{\tensor*[^{1}]{\hspace{-0.2em}B}{}_{1}^{}}}$ &  $- 6.1291$ &       $d\indices*{*_{A_{5}}^{\tensor*[^{1}]{\hspace{-0.2em}A}{}_{1}^{}}_{\bar{A}_{4}}^{\tensor*[^{1}]{\hspace{-0.2em}B}{}_{2}^{}}_{\bar{A}_{1}}^{\tensor*[^{1}]{\hspace{-0.2em}B}{}_{1}^{}}}$ &    $3.4112$ &       $d\indices*{*_{A_{5}}^{\tensor*[^{1}]{\hspace{-0.2em}A}{}_{1}^{}}_{\bar{A}_{4}}^{\tensor*[^{2}]{\hspace{-0.2em}B}{}_{1}^{}}_{\bar{A}_{1}}^{\tensor*[^{2}]{\hspace{-0.2em}B}{}_{1}^{}}}$ &  $- 1.7902$ &                                        $d\indices*{*_{A_{5}}^{\tensor*[^{1}]{\hspace{-0.2em}A}{}_{1}^{}}_{\bar{A}_{4}}^{B_{2}^{}}_{\bar{A}_{1}}^{\tensor*[^{2}]{\hspace{-0.2em}B}{}_{1}^{}}}$ &  $- 3.9895$ &       $d\indices*{*_{A_{5}}^{\tensor*[^{1}]{\hspace{-0.2em}A}{}_{1}^{}}_{\bar{A}_{4}}^{\tensor*[^{1}]{\hspace{-0.2em}B}{}_{2}^{}}_{\bar{A}_{1}}^{\tensor*[^{2}]{\hspace{-0.2em}B}{}_{1}^{}}}$ &    $3.6452$ \\[0.4em]
                                                                          $d\indices*{*_{A_{5}}^{\tensor*[^{1}]{\hspace{-0.2em}A}{}_{1}^{}}_{\bar{A}_{4}}^{B_{2}^{}}_{\bar{A}_{1}}^{B_{2}^{}}}$ &  $- 0.4896$ &                                        $d\indices*{*_{A_{5}}^{\tensor*[^{1}]{\hspace{-0.2em}A}{}_{1}^{}}_{\bar{A}_{4}}^{B_{2}^{}}_{\bar{A}_{1}}^{\tensor*[^{1}]{\hspace{-0.2em}B}{}_{2}^{}}}$ &    $0.9971$ &       $d\indices*{*_{A_{5}}^{\tensor*[^{1}]{\hspace{-0.2em}A}{}_{1}^{}}_{\bar{A}_{4}}^{\tensor*[^{1}]{\hspace{-0.2em}B}{}_{2}^{}}_{\bar{A}_{1}}^{\tensor*[^{1}]{\hspace{-0.2em}B}{}_{2}^{}}}$ &  $- 5.3448$ &       $d\indices*{*_{A_{5}}^{\tensor*[^{2}]{\hspace{-0.2em}A}{}_{1}^{}}_{\bar{A}_{4}}^{\tensor*[^{2}]{\hspace{-0.2em}A}{}_{1}^{}}_{\bar{A}_{1}}^{\tensor*[^{2}]{\hspace{-0.2em}A}{}_{1}^{}}}$ &  $- 8.4972$ &                                        $d\indices*{*_{A_{5}}^{\tensor*[^{2}]{\hspace{-0.2em}A}{}_{1}^{}}_{\bar{A}_{4}}^{\tensor*[^{2}]{\hspace{-0.2em}A}{}_{1}^{}}_{\bar{A}_{1}}^{B_{1}^{}}}$ &  $- 4.0347$ \\[0.4em]
        $d\indices*{*_{A_{5}}^{\tensor*[^{2}]{\hspace{-0.2em}A}{}_{1}^{}}_{\bar{A}_{4}}^{\tensor*[^{2}]{\hspace{-0.2em}A}{}_{1}^{}}_{\bar{A}_{1}}^{\tensor*[^{1}]{\hspace{-0.2em}B}{}_{1}^{}}}$ &    $0.7317$ &       $d\indices*{*_{A_{5}}^{\tensor*[^{2}]{\hspace{-0.2em}A}{}_{1}^{}}_{\bar{A}_{4}}^{\tensor*[^{2}]{\hspace{-0.2em}A}{}_{1}^{}}_{\bar{A}_{1}}^{\tensor*[^{2}]{\hspace{-0.2em}B}{}_{1}^{}}}$ &  $- 0.9137$ &                                                                         $d\indices*{*_{A_{5}}^{\tensor*[^{2}]{\hspace{-0.2em}A}{}_{1}^{}}_{\bar{A}_{4}}^{A_{2}^{}}_{\bar{A}_{1}}^{A_{2}^{}}}$ &  $- 6.2821$ &                                                                         $d\indices*{*_{A_{5}}^{\tensor*[^{2}]{\hspace{-0.2em}A}{}_{1}^{}}_{\bar{A}_{4}}^{A_{2}^{}}_{\bar{A}_{1}}^{B_{1}^{}}}$ &    $1.8218$ &                                        $d\indices*{*_{A_{5}}^{\tensor*[^{2}]{\hspace{-0.2em}A}{}_{1}^{}}_{\bar{A}_{4}}^{A_{2}^{}}_{\bar{A}_{1}}^{\tensor*[^{1}]{\hspace{-0.2em}B}{}_{1}^{}}}$ &  $- 1.0210$ \\[0.4em]
                                         $d\indices*{*_{A_{5}}^{\tensor*[^{2}]{\hspace{-0.2em}A}{}_{1}^{}}_{\bar{A}_{4}}^{A_{2}^{}}_{\bar{A}_{1}}^{\tensor*[^{2}]{\hspace{-0.2em}B}{}_{1}^{}}}$ &  $- 0.2313$ &                                                                         $d\indices*{*_{A_{5}}^{\tensor*[^{2}]{\hspace{-0.2em}A}{}_{1}^{}}_{\bar{A}_{4}}^{A_{2}^{}}_{\bar{A}_{1}}^{B_{2}^{}}}$ &  $- 0.5009$ &                                        $d\indices*{*_{A_{5}}^{\tensor*[^{2}]{\hspace{-0.2em}A}{}_{1}^{}}_{\bar{A}_{4}}^{A_{2}^{}}_{\bar{A}_{1}}^{\tensor*[^{1}]{\hspace{-0.2em}B}{}_{2}^{}}}$ &    $2.9052$ &                                                                         $d\indices*{*_{A_{5}}^{\tensor*[^{2}]{\hspace{-0.2em}A}{}_{1}^{}}_{\bar{A}_{4}}^{B_{1}^{}}_{\bar{A}_{1}}^{B_{1}^{}}}$ &    $0.6868$ &                                        $d\indices*{*_{A_{5}}^{\tensor*[^{2}]{\hspace{-0.2em}A}{}_{1}^{}}_{\bar{A}_{4}}^{B_{1}^{}}_{\bar{A}_{1}}^{\tensor*[^{1}]{\hspace{-0.2em}B}{}_{1}^{}}}$ &    $4.2485$ \\[0.4em]
                                         $d\indices*{*_{A_{5}}^{\tensor*[^{2}]{\hspace{-0.2em}A}{}_{1}^{}}_{\bar{A}_{4}}^{B_{1}^{}}_{\bar{A}_{1}}^{\tensor*[^{2}]{\hspace{-0.2em}B}{}_{1}^{}}}$ &    $1.3323$ &                                                                         $d\indices*{*_{A_{5}}^{\tensor*[^{2}]{\hspace{-0.2em}A}{}_{1}^{}}_{\bar{A}_{4}}^{B_{2}^{}}_{\bar{A}_{1}}^{B_{1}^{}}}$ &  $- 5.1920$ &                                        $d\indices*{*_{A_{5}}^{\tensor*[^{2}]{\hspace{-0.2em}A}{}_{1}^{}}_{\bar{A}_{4}}^{\tensor*[^{1}]{\hspace{-0.2em}B}{}_{2}^{}}_{\bar{A}_{1}}^{B_{1}^{}}}$ &    $4.9328$ &       $d\indices*{*_{A_{5}}^{\tensor*[^{2}]{\hspace{-0.2em}A}{}_{1}^{}}_{\bar{A}_{4}}^{\tensor*[^{1}]{\hspace{-0.2em}B}{}_{1}^{}}_{\bar{A}_{1}}^{\tensor*[^{1}]{\hspace{-0.2em}B}{}_{1}^{}}}$ & $- 12.0043$ &       $d\indices*{*_{A_{5}}^{\tensor*[^{2}]{\hspace{-0.2em}A}{}_{1}^{}}_{\bar{A}_{4}}^{\tensor*[^{1}]{\hspace{-0.2em}B}{}_{1}^{}}_{\bar{A}_{1}}^{\tensor*[^{2}]{\hspace{-0.2em}B}{}_{1}^{}}}$ &  $- 4.8426$ \\[0.4em]
                                         $d\indices*{*_{A_{5}}^{\tensor*[^{2}]{\hspace{-0.2em}A}{}_{1}^{}}_{\bar{A}_{4}}^{B_{2}^{}}_{\bar{A}_{1}}^{\tensor*[^{1}]{\hspace{-0.2em}B}{}_{1}^{}}}$ &    $6.8707$ &       $d\indices*{*_{A_{5}}^{\tensor*[^{2}]{\hspace{-0.2em}A}{}_{1}^{}}_{\bar{A}_{4}}^{\tensor*[^{1}]{\hspace{-0.2em}B}{}_{2}^{}}_{\bar{A}_{1}}^{\tensor*[^{1}]{\hspace{-0.2em}B}{}_{1}^{}}}$ &  $- 4.4868$ &       $d\indices*{*_{A_{5}}^{\tensor*[^{2}]{\hspace{-0.2em}A}{}_{1}^{}}_{\bar{A}_{4}}^{\tensor*[^{2}]{\hspace{-0.2em}B}{}_{1}^{}}_{\bar{A}_{1}}^{\tensor*[^{2}]{\hspace{-0.2em}B}{}_{1}^{}}}$ &  $- 6.1261$ &                                        $d\indices*{*_{A_{5}}^{\tensor*[^{2}]{\hspace{-0.2em}A}{}_{1}^{}}_{\bar{A}_{4}}^{B_{2}^{}}_{\bar{A}_{1}}^{\tensor*[^{2}]{\hspace{-0.2em}B}{}_{1}^{}}}$ &  $- 0.6881$ &       $d\indices*{*_{A_{5}}^{\tensor*[^{2}]{\hspace{-0.2em}A}{}_{1}^{}}_{\bar{A}_{4}}^{\tensor*[^{1}]{\hspace{-0.2em}B}{}_{2}^{}}_{\bar{A}_{1}}^{\tensor*[^{2}]{\hspace{-0.2em}B}{}_{1}^{}}}$ &    $0.6070$ \\[0.4em]
                                                                          $d\indices*{*_{A_{5}}^{\tensor*[^{2}]{\hspace{-0.2em}A}{}_{1}^{}}_{\bar{A}_{4}}^{B_{2}^{}}_{\bar{A}_{1}}^{B_{2}^{}}}$ & $- 10.9010$ &                                        $d\indices*{*_{A_{5}}^{\tensor*[^{2}]{\hspace{-0.2em}A}{}_{1}^{}}_{\bar{A}_{4}}^{B_{2}^{}}_{\bar{A}_{1}}^{\tensor*[^{1}]{\hspace{-0.2em}B}{}_{2}^{}}}$ &    $6.8644$ &       $d\indices*{*_{A_{5}}^{\tensor*[^{2}]{\hspace{-0.2em}A}{}_{1}^{}}_{\bar{A}_{4}}^{\tensor*[^{1}]{\hspace{-0.2em}B}{}_{2}^{}}_{\bar{A}_{1}}^{\tensor*[^{1}]{\hspace{-0.2em}B}{}_{2}^{}}}$ &  $- 4.6029$ &                                                                                                          $d\indices*{*_{A_{5}}^{A_{2}^{}}_{\bar{A}_{4}}^{A_{2}^{}}_{\bar{A}_{1}}^{B_{1}^{}}}$ &    $3.9607$ &                                                                         $d\indices*{*_{A_{5}}^{A_{2}^{}}_{\bar{A}_{4}}^{A_{2}^{}}_{\bar{A}_{1}}^{\tensor*[^{1}]{\hspace{-0.2em}B}{}_{1}^{}}}$ &  $- 0.7628$ \\[0.4em]
                                                                          $d\indices*{*_{A_{5}}^{A_{2}^{}}_{\bar{A}_{4}}^{A_{2}^{}}_{\bar{A}_{1}}^{\tensor*[^{2}]{\hspace{-0.2em}B}{}_{1}^{}}}$ &  $- 2.5045$ &                                                                         $d\indices*{*_{A_{5}}^{A_{2}^{}}_{\bar{A}_{4}}^{B_{1}^{}}_{\bar{A}_{1}}^{\tensor*[^{1}]{\hspace{-0.2em}B}{}_{1}^{}}}$ &    $3.3235$ &                                                                         $d\indices*{*_{A_{5}}^{A_{2}^{}}_{\bar{A}_{4}}^{B_{1}^{}}_{\bar{A}_{1}}^{\tensor*[^{2}]{\hspace{-0.2em}B}{}_{1}^{}}}$ &    $3.6163$ &                                                                                                          $d\indices*{*_{A_{5}}^{A_{2}^{}}_{\bar{A}_{4}}^{B_{2}^{}}_{\bar{A}_{1}}^{B_{1}^{}}}$ &  $- 1.1434$ &                                                                         $d\indices*{*_{A_{5}}^{A_{2}^{}}_{\bar{A}_{4}}^{\tensor*[^{1}]{\hspace{-0.2em}B}{}_{2}^{}}_{\bar{A}_{1}}^{B_{1}^{}}}$ &  $- 0.0333$ \\[0.4em]
                                         $d\indices*{*_{A_{5}}^{A_{2}^{}}_{\bar{A}_{4}}^{\tensor*[^{1}]{\hspace{-0.2em}B}{}_{1}^{}}_{\bar{A}_{1}}^{\tensor*[^{2}]{\hspace{-0.2em}B}{}_{1}^{}}}$ &  $- 6.3640$ &                                                                         $d\indices*{*_{A_{5}}^{A_{2}^{}}_{\bar{A}_{4}}^{B_{2}^{}}_{\bar{A}_{1}}^{\tensor*[^{1}]{\hspace{-0.2em}B}{}_{1}^{}}}$ &    $7.2756$ &                                        $d\indices*{*_{A_{5}}^{A_{2}^{}}_{\bar{A}_{4}}^{\tensor*[^{1}]{\hspace{-0.2em}B}{}_{2}^{}}_{\bar{A}_{1}}^{\tensor*[^{1}]{\hspace{-0.2em}B}{}_{1}^{}}}$ &  $- 4.1125$ &                                                                         $d\indices*{*_{A_{5}}^{A_{2}^{}}_{\bar{A}_{4}}^{B_{2}^{}}_{\bar{A}_{1}}^{\tensor*[^{2}]{\hspace{-0.2em}B}{}_{1}^{}}}$ &    $7.4686$ &                                        $d\indices*{*_{A_{5}}^{A_{2}^{}}_{\bar{A}_{4}}^{\tensor*[^{1}]{\hspace{-0.2em}B}{}_{2}^{}}_{\bar{A}_{1}}^{\tensor*[^{2}]{\hspace{-0.2em}B}{}_{1}^{}}}$ &  $- 5.6518$ \\[0.4em]
                                                                          $d\indices*{*_{A_{5}}^{A_{2}^{}}_{\bar{A}_{4}}^{B_{2}^{}}_{\bar{A}_{1}}^{\tensor*[^{1}]{\hspace{-0.2em}B}{}_{2}^{}}}$ &  $- 0.3667$ &                                                                                                          $d\indices*{*_{A_{5}}^{B_{1}^{}}_{\bar{A}_{4}}^{B_{1}^{}}_{\bar{A}_{1}}^{B_{1}^{}}}$ &    $9.5444$ &                                                                         $d\indices*{*_{A_{5}}^{B_{1}^{}}_{\bar{A}_{4}}^{B_{1}^{}}_{\bar{A}_{1}}^{\tensor*[^{1}]{\hspace{-0.2em}B}{}_{1}^{}}}$ &  $- 5.4346$ &                                                                         $d\indices*{*_{A_{5}}^{B_{1}^{}}_{\bar{A}_{4}}^{B_{1}^{}}_{\bar{A}_{1}}^{\tensor*[^{2}]{\hspace{-0.2em}B}{}_{1}^{}}}$ &  $- 0.0850$ &                                        $d\indices*{*_{A_{5}}^{B_{1}^{}}_{\bar{A}_{4}}^{\tensor*[^{1}]{\hspace{-0.2em}B}{}_{1}^{}}_{\bar{A}_{1}}^{\tensor*[^{1}]{\hspace{-0.2em}B}{}_{1}^{}}}$ &  $- 0.5345$ \\[0.4em]
                                         $d\indices*{*_{A_{5}}^{B_{1}^{}}_{\bar{A}_{4}}^{\tensor*[^{1}]{\hspace{-0.2em}B}{}_{1}^{}}_{\bar{A}_{1}}^{\tensor*[^{2}]{\hspace{-0.2em}B}{}_{1}^{}}}$ &  $- 3.6385$ &                                                                         $d\indices*{*_{A_{5}}^{B_{2}^{}}_{\bar{A}_{4}}^{B_{1}^{}}_{\bar{A}_{1}}^{\tensor*[^{1}]{\hspace{-0.2em}B}{}_{1}^{}}}$ &  $- 4.9560$ &                                        $d\indices*{*_{A_{5}}^{\tensor*[^{1}]{\hspace{-0.2em}B}{}_{2}^{}}_{\bar{A}_{4}}^{B_{1}^{}}_{\bar{A}_{1}}^{\tensor*[^{1}]{\hspace{-0.2em}B}{}_{1}^{}}}$ &    $3.7214$ &                                        $d\indices*{*_{A_{5}}^{B_{1}^{}}_{\bar{A}_{4}}^{\tensor*[^{2}]{\hspace{-0.2em}B}{}_{1}^{}}_{\bar{A}_{1}}^{\tensor*[^{2}]{\hspace{-0.2em}B}{}_{1}^{}}}$ &  $- 3.7257$ &                                                                         $d\indices*{*_{A_{5}}^{B_{2}^{}}_{\bar{A}_{4}}^{B_{1}^{}}_{\bar{A}_{1}}^{\tensor*[^{2}]{\hspace{-0.2em}B}{}_{1}^{}}}$ &    $2.5955$ \\[0.4em]
                                         $d\indices*{*_{A_{5}}^{\tensor*[^{1}]{\hspace{-0.2em}B}{}_{2}^{}}_{\bar{A}_{4}}^{B_{1}^{}}_{\bar{A}_{1}}^{\tensor*[^{2}]{\hspace{-0.2em}B}{}_{1}^{}}}$ &  $- 1.2080$ &                                                                                                          $d\indices*{*_{A_{5}}^{B_{2}^{}}_{\bar{A}_{4}}^{B_{2}^{}}_{\bar{A}_{1}}^{B_{1}^{}}}$ &    $6.1164$ &                                                                         $d\indices*{*_{A_{5}}^{B_{2}^{}}_{\bar{A}_{4}}^{\tensor*[^{1}]{\hspace{-0.2em}B}{}_{2}^{}}_{\bar{A}_{1}}^{B_{1}^{}}}$ &  $- 4.3035$ &                                        $d\indices*{*_{A_{5}}^{\tensor*[^{1}]{\hspace{-0.2em}B}{}_{2}^{}}_{\bar{A}_{4}}^{\tensor*[^{1}]{\hspace{-0.2em}B}{}_{2}^{}}_{\bar{A}_{1}}^{B_{1}^{}}}$ &    $3.1406$ &       $d\indices*{*_{A_{5}}^{\tensor*[^{1}]{\hspace{-0.2em}B}{}_{1}^{}}_{\bar{A}_{4}}^{\tensor*[^{1}]{\hspace{-0.2em}B}{}_{1}^{}}_{\bar{A}_{1}}^{\tensor*[^{1}]{\hspace{-0.2em}B}{}_{1}^{}}}$ &  $- 2.1334$ \\[0.4em]
        $d\indices*{*_{A_{5}}^{\tensor*[^{1}]{\hspace{-0.2em}B}{}_{1}^{}}_{\bar{A}_{4}}^{\tensor*[^{1}]{\hspace{-0.2em}B}{}_{1}^{}}_{\bar{A}_{1}}^{\tensor*[^{2}]{\hspace{-0.2em}B}{}_{1}^{}}}$ &  $- 1.6363$ &       $d\indices*{*_{A_{5}}^{\tensor*[^{1}]{\hspace{-0.2em}B}{}_{1}^{}}_{\bar{A}_{4}}^{\tensor*[^{2}]{\hspace{-0.2em}B}{}_{1}^{}}_{\bar{A}_{1}}^{\tensor*[^{2}]{\hspace{-0.2em}B}{}_{1}^{}}}$ &    $0.6380$ &                                        $d\indices*{*_{A_{5}}^{B_{2}^{}}_{\bar{A}_{4}}^{\tensor*[^{1}]{\hspace{-0.2em}B}{}_{1}^{}}_{\bar{A}_{1}}^{\tensor*[^{2}]{\hspace{-0.2em}B}{}_{1}^{}}}$ &  $- 0.3343$ &       $d\indices*{*_{A_{5}}^{\tensor*[^{1}]{\hspace{-0.2em}B}{}_{2}^{}}_{\bar{A}_{4}}^{\tensor*[^{1}]{\hspace{-0.2em}B}{}_{1}^{}}_{\bar{A}_{1}}^{\tensor*[^{2}]{\hspace{-0.2em}B}{}_{1}^{}}}$ &  $- 0.1322$ &                                                                         $d\indices*{*_{A_{5}}^{B_{2}^{}}_{\bar{A}_{4}}^{B_{2}^{}}_{\bar{A}_{1}}^{\tensor*[^{1}]{\hspace{-0.2em}B}{}_{1}^{}}}$ &    $2.5162$ \\[0.4em]
                                         $d\indices*{*_{A_{5}}^{B_{2}^{}}_{\bar{A}_{4}}^{\tensor*[^{1}]{\hspace{-0.2em}B}{}_{2}^{}}_{\bar{A}_{1}}^{\tensor*[^{1}]{\hspace{-0.2em}B}{}_{1}^{}}}$ &  $- 0.9317$ &       $d\indices*{*_{A_{5}}^{\tensor*[^{1}]{\hspace{-0.2em}B}{}_{2}^{}}_{\bar{A}_{4}}^{\tensor*[^{1}]{\hspace{-0.2em}B}{}_{2}^{}}_{\bar{A}_{1}}^{\tensor*[^{1}]{\hspace{-0.2em}B}{}_{1}^{}}}$ &  $- 0.9248$ &       $d\indices*{*_{A_{5}}^{\tensor*[^{2}]{\hspace{-0.2em}B}{}_{1}^{}}_{\bar{A}_{4}}^{\tensor*[^{2}]{\hspace{-0.2em}B}{}_{1}^{}}_{\bar{A}_{1}}^{\tensor*[^{2}]{\hspace{-0.2em}B}{}_{1}^{}}}$ &    $4.3170$ &                                                                         $d\indices*{*_{A_{5}}^{B_{2}^{}}_{\bar{A}_{4}}^{B_{2}^{}}_{\bar{A}_{1}}^{\tensor*[^{2}]{\hspace{-0.2em}B}{}_{1}^{}}}$ &    $0.0676$ &                                        $d\indices*{*_{A_{5}}^{B_{2}^{}}_{\bar{A}_{4}}^{\tensor*[^{1}]{\hspace{-0.2em}B}{}_{2}^{}}_{\bar{A}_{1}}^{\tensor*[^{2}]{\hspace{-0.2em}B}{}_{1}^{}}}$ &    $0.7355$ \\[0.4em]
        $d\indices*{*_{A_{5}}^{\tensor*[^{1}]{\hspace{-0.2em}B}{}_{2}^{}}_{\bar{A}_{4}}^{\tensor*[^{1}]{\hspace{-0.2em}B}{}_{2}^{}}_{\bar{A}_{1}}^{\tensor*[^{2}]{\hspace{-0.2em}B}{}_{1}^{}}}$ &  $- 1.5079$ &                                                                                                                       $d\indices*{*_{X_{1}}^{A_{1g}^{}}_{X_{2}}^{A_{1g}^{}}_{X}^{A_{1g}^{}}}$ &  $- 9.5916$ &                                                                                                                         $d\indices*{*_{X_{1}}^{A_{1g}^{}}_{X_{2}}^{E_{g}^{}}_{X}^{E_{g}^{}}}$ &    $8.7236$ &                                                                                                                        $d\indices*{*_{X_{1}}^{A_{1g}^{}}_{X_{2}}^{A_{2u}^{}}_{X}^{E_{u}^{}}}$ &   $11.1533$ &                                                                                       $d\indices*{*_{X_{1}}^{A_{1g}^{}}_{X_{2}}^{A_{2u}^{}}_{X}^{\tensor*[^{1}]{\hspace{-0.2em}E}{}_{u}^{}}}$ & $- 13.5804$ \\[0.4em]
                                                                                                                        $d\indices*{*_{X_{1}}^{A_{1g}^{}}_{X_{2}}^{B_{1u}^{}}_{X}^{B_{1u}^{}}}$ &  $- 8.8709$ &                                                                                                                         $d\indices*{*_{X_{1}}^{A_{1g}^{}}_{X_{2}}^{E_{u}^{}}_{X}^{E_{u}^{}}}$ &  $- 7.2007$ &                                                                                        $d\indices*{*_{X_{1}}^{A_{1g}^{}}_{X_{2}}^{E_{u}^{}}_{X}^{\tensor*[^{1}]{\hspace{-0.2em}E}{}_{u}^{}}}$ &   $10.8015$ &                                                       $d\indices*{*_{X_{1}}^{A_{1g}^{}}_{X_{2}}^{\tensor*[^{1}]{\hspace{-0.2em}E}{}_{u}^{}}_{X}^{\tensor*[^{1}]{\hspace{-0.2em}E}{}_{u}^{}}}$ & $- 15.7199$ &                                                                                                                          $d\indices*{*_{X_{1}}^{E_{g}^{}}_{X_{2}}^{E_{g}^{}}_{X}^{E_{g}^{}}}$ &  $- 8.5261$ \\[0.4em]
                                                                                                                          $d\indices*{*_{X_{1}}^{A_{2u}^{}}_{X_{2}}^{E_{g}^{}}_{X}^{E_{u}^{}}}$ &   $11.1083$ &                                                                                        $d\indices*{*_{X_{1}}^{A_{2u}^{}}_{X_{2}}^{E_{g}^{}}_{X}^{\tensor*[^{1}]{\hspace{-0.2em}E}{}_{u}^{}}}$ & $- 20.4775$ &                                                                                                                        $d\indices*{*_{X_{1}}^{B_{1u}^{}}_{X_{2}}^{A_{2u}^{}}_{X}^{E_{g}^{}}}$ &   $12.4779$ &                                                                                                                         $d\indices*{*_{X_{1}}^{B_{1u}^{}}_{X_{2}}^{E_{g}^{}}_{X}^{E_{u}^{}}}$ &    $8.1954$ &                                                                                                                          $d\indices*{*_{X_{1}}^{E_{g}^{}}_{X_{2}}^{E_{u}^{}}_{X}^{E_{u}^{}}}$ &  $- 7.9308$ \\[0.4em]
                                                                        $\tensor*[^{1}]{d}{}\indices*{*_{X_{1}}^{E_{g}^{}}_{X_{2}}^{E_{u}^{}}_{X}^{\tensor*[^{1}]{\hspace{-0.2em}E}{}_{u}^{}}}$ &    $9.6645$ &                                                                                        $d\indices*{*_{X_{1}}^{B_{1u}^{}}_{X_{2}}^{E_{g}^{}}_{X}^{\tensor*[^{1}]{\hspace{-0.2em}E}{}_{u}^{}}}$ & $- 10.9011$ &                                                                       $\tensor*[^{0}]{d}{}\indices*{*_{X_{1}}^{E_{g}^{}}_{X_{2}}^{E_{u}^{}}_{X}^{\tensor*[^{1}]{\hspace{-0.2em}E}{}_{u}^{}}}$ &    $8.8866$ &                                                        $d\indices*{*_{X_{1}}^{E_{g}^{}}_{X_{2}}^{\tensor*[^{1}]{\hspace{-0.2em}E}{}_{u}^{}}_{X}^{\tensor*[^{1}]{\hspace{-0.2em}E}{}_{u}^{}}}$ & $- 14.2874$ &                                                                                                    $d\indices*{*_{\bar{A}_{1}}^{A_{1}^{}}_{\bar{A}_{2}}^{A_{1}^{}}_{\bar{A}_{5}}^{A_{1}^{}}}$ &    $6.9553$ \\[0.4em]
                                                                    $d\indices*{*_{\bar{A}_{1}}^{A_{1}^{}}_{\bar{A}_{2}}^{A_{1}^{}}_{\bar{A}_{5}}^{\tensor*[^{1}]{\hspace{-0.2em}A}{}_{1}^{}}}$ &    $9.5724$ &                                                                   $d\indices*{*_{\bar{A}_{1}}^{A_{1}^{}}_{\bar{A}_{2}}^{A_{1}^{}}_{\bar{A}_{5}}^{\tensor*[^{2}]{\hspace{-0.2em}A}{}_{1}^{}}}$ &    $2.4962$ &                                                                                                    $d\indices*{*_{\bar{A}_{1}}^{A_{1}^{}}_{\bar{A}_{2}}^{A_{1}^{}}_{\bar{A}_{5}}^{A_{2}^{}}}$ &   $12.3097$ &                                  $d\indices*{*_{\bar{A}_{1}}^{A_{1}^{}}_{\bar{A}_{2}}^{\tensor*[^{1}]{\hspace{-0.2em}A}{}_{1}^{}}_{\bar{A}_{5}}^{\tensor*[^{1}]{\hspace{-0.2em}A}{}_{1}^{}}}$ &  $- 6.4750$ &                                  $d\indices*{*_{\bar{A}_{1}}^{A_{1}^{}}_{\bar{A}_{2}}^{\tensor*[^{1}]{\hspace{-0.2em}A}{}_{1}^{}}_{\bar{A}_{5}}^{\tensor*[^{2}]{\hspace{-0.2em}A}{}_{1}^{}}}$ &    $1.4239$ \\[0.4em]
                                                                    $d\indices*{*_{\bar{A}_{1}}^{A_{1}^{}}_{\bar{A}_{2}}^{\tensor*[^{1}]{\hspace{-0.2em}A}{}_{1}^{}}_{\bar{A}_{5}}^{A_{2}^{}}}$ &  $- 5.1168$ &                                                                   $d\indices*{*_{\bar{A}_{1}}^{A_{1}^{}}_{\bar{A}_{2}}^{\tensor*[^{1}]{\hspace{-0.2em}A}{}_{1}^{}}_{\bar{A}_{5}}^{B_{1}^{}}}$ &  $- 0.3453$ &                                  $d\indices*{*_{\bar{A}_{1}}^{A_{1}^{}}_{\bar{A}_{2}}^{\tensor*[^{1}]{\hspace{-0.2em}A}{}_{1}^{}}_{\bar{A}_{5}}^{\tensor*[^{1}]{\hspace{-0.2em}B}{}_{1}^{}}}$ &  $- 4.8654$ &                                  $d\indices*{*_{\bar{A}_{1}}^{A_{1}^{}}_{\bar{A}_{2}}^{\tensor*[^{1}]{\hspace{-0.2em}A}{}_{1}^{}}_{\bar{A}_{5}}^{\tensor*[^{2}]{\hspace{-0.2em}B}{}_{1}^{}}}$ &  $- 5.4169$ &                                                                   $d\indices*{*_{\bar{A}_{1}}^{A_{1}^{}}_{\bar{A}_{2}}^{\tensor*[^{1}]{\hspace{-0.2em}A}{}_{1}^{}}_{\bar{A}_{5}}^{B_{2}^{}}}$ &    $1.5593$ \\[0.4em]
                                   $d\indices*{*_{\bar{A}_{1}}^{A_{1}^{}}_{\bar{A}_{2}}^{\tensor*[^{1}]{\hspace{-0.2em}A}{}_{1}^{}}_{\bar{A}_{5}}^{\tensor*[^{1}]{\hspace{-0.2em}B}{}_{2}^{}}}$ &  $- 1.2103$ &                                  $d\indices*{*_{\bar{A}_{1}}^{A_{1}^{}}_{\bar{A}_{2}}^{\tensor*[^{2}]{\hspace{-0.2em}A}{}_{1}^{}}_{\bar{A}_{5}}^{\tensor*[^{2}]{\hspace{-0.2em}A}{}_{1}^{}}}$ &    $0.0548$ &                                                                   $d\indices*{*_{\bar{A}_{1}}^{A_{1}^{}}_{\bar{A}_{2}}^{\tensor*[^{2}]{\hspace{-0.2em}A}{}_{1}^{}}_{\bar{A}_{5}}^{A_{2}^{}}}$ &    $2.2221$ &                                                                   $d\indices*{*_{\bar{A}_{1}}^{A_{1}^{}}_{\bar{A}_{2}}^{\tensor*[^{2}]{\hspace{-0.2em}A}{}_{1}^{}}_{\bar{A}_{5}}^{B_{1}^{}}}$ &  $- 5.1183$ &                                  $d\indices*{*_{\bar{A}_{1}}^{A_{1}^{}}_{\bar{A}_{2}}^{\tensor*[^{2}]{\hspace{-0.2em}A}{}_{1}^{}}_{\bar{A}_{5}}^{\tensor*[^{1}]{\hspace{-0.2em}B}{}_{1}^{}}}$ &    $7.5614$ \\[0.4em]
                                   $d\indices*{*_{\bar{A}_{1}}^{A_{1}^{}}_{\bar{A}_{2}}^{\tensor*[^{2}]{\hspace{-0.2em}A}{}_{1}^{}}_{\bar{A}_{5}}^{\tensor*[^{2}]{\hspace{-0.2em}B}{}_{1}^{}}}$ &  $- 0.7094$ &                                                                   $d\indices*{*_{\bar{A}_{1}}^{A_{1}^{}}_{\bar{A}_{2}}^{\tensor*[^{2}]{\hspace{-0.2em}A}{}_{1}^{}}_{\bar{A}_{5}}^{B_{2}^{}}}$ &   $10.9585$ &                                  $d\indices*{*_{\bar{A}_{1}}^{A_{1}^{}}_{\bar{A}_{2}}^{\tensor*[^{2}]{\hspace{-0.2em}A}{}_{1}^{}}_{\bar{A}_{5}}^{\tensor*[^{1}]{\hspace{-0.2em}B}{}_{2}^{}}}$ &  $- 7.9840$ &                                                                                                    $d\indices*{*_{\bar{A}_{1}}^{A_{1}^{}}_{\bar{A}_{2}}^{A_{2}^{}}_{\bar{A}_{5}}^{A_{2}^{}}}$ &    $0.8886$ &                                                                                                    $d\indices*{*_{\bar{A}_{1}}^{A_{1}^{}}_{\bar{A}_{2}}^{A_{2}^{}}_{\bar{A}_{5}}^{B_{1}^{}}}$ &  $- 0.6930$ \\[0.4em]
                                                                    $d\indices*{*_{\bar{A}_{1}}^{A_{1}^{}}_{\bar{A}_{2}}^{A_{2}^{}}_{\bar{A}_{5}}^{\tensor*[^{1}]{\hspace{-0.2em}B}{}_{1}^{}}}$ &    $7.9989$ &                                                                   $d\indices*{*_{\bar{A}_{1}}^{A_{1}^{}}_{\bar{A}_{2}}^{A_{2}^{}}_{\bar{A}_{5}}^{\tensor*[^{2}]{\hspace{-0.2em}B}{}_{1}^{}}}$ &    $5.7846$ &                                                                                                    $d\indices*{*_{\bar{A}_{1}}^{A_{1}^{}}_{\bar{A}_{2}}^{A_{2}^{}}_{\bar{A}_{5}}^{B_{2}^{}}}$ &  $- 1.0079$ &                                                                   $d\indices*{*_{\bar{A}_{1}}^{A_{1}^{}}_{\bar{A}_{2}}^{A_{2}^{}}_{\bar{A}_{5}}^{\tensor*[^{1}]{\hspace{-0.2em}B}{}_{2}^{}}}$ &    $0.3713$ &                                                                                                    $d\indices*{*_{\bar{A}_{1}}^{A_{1}^{}}_{\bar{A}_{2}}^{B_{1}^{}}_{\bar{A}_{5}}^{B_{1}^{}}}$ &  $- 8.9108$ \\[0.4em]
                                                                    $d\indices*{*_{\bar{A}_{1}}^{A_{1}^{}}_{\bar{A}_{2}}^{B_{1}^{}}_{\bar{A}_{5}}^{\tensor*[^{1}]{\hspace{-0.2em}B}{}_{1}^{}}}$ &    $5.5933$ &                                                                   $d\indices*{*_{\bar{A}_{1}}^{A_{1}^{}}_{\bar{A}_{2}}^{B_{1}^{}}_{\bar{A}_{5}}^{\tensor*[^{2}]{\hspace{-0.2em}B}{}_{1}^{}}}$ &  $- 0.3984$ &                                                                                                    $d\indices*{*_{\bar{A}_{1}}^{A_{1}^{}}_{\bar{A}_{2}}^{B_{2}^{}}_{\bar{A}_{5}}^{B_{1}^{}}}$ &  $- 8.0415$ &                                                                   $d\indices*{*_{\bar{A}_{1}}^{A_{1}^{}}_{\bar{A}_{2}}^{\tensor*[^{1}]{\hspace{-0.2em}B}{}_{2}^{}}_{\bar{A}_{5}}^{B_{1}^{}}}$ &    $5.7925$ &                                  $d\indices*{*_{\bar{A}_{1}}^{A_{1}^{}}_{\bar{A}_{2}}^{\tensor*[^{1}]{\hspace{-0.2em}B}{}_{1}^{}}_{\bar{A}_{5}}^{\tensor*[^{1}]{\hspace{-0.2em}B}{}_{1}^{}}}$ &    $7.7429$ \\[0.4em]
                                   $d\indices*{*_{\bar{A}_{1}}^{A_{1}^{}}_{\bar{A}_{2}}^{\tensor*[^{1}]{\hspace{-0.2em}B}{}_{1}^{}}_{\bar{A}_{5}}^{\tensor*[^{2}]{\hspace{-0.2em}B}{}_{1}^{}}}$ &    $0.4598$ &                                                                   $d\indices*{*_{\bar{A}_{1}}^{A_{1}^{}}_{\bar{A}_{2}}^{B_{2}^{}}_{\bar{A}_{5}}^{\tensor*[^{1}]{\hspace{-0.2em}B}{}_{1}^{}}}$ &  $- 4.4905$ &                                  $d\indices*{*_{\bar{A}_{1}}^{A_{1}^{}}_{\bar{A}_{2}}^{\tensor*[^{1}]{\hspace{-0.2em}B}{}_{2}^{}}_{\bar{A}_{5}}^{\tensor*[^{1}]{\hspace{-0.2em}B}{}_{1}^{}}}$ &    $1.0846$ &                                  $d\indices*{*_{\bar{A}_{1}}^{A_{1}^{}}_{\bar{A}_{2}}^{\tensor*[^{2}]{\hspace{-0.2em}B}{}_{1}^{}}_{\bar{A}_{5}}^{\tensor*[^{2}]{\hspace{-0.2em}B}{}_{1}^{}}}$ &    $0.1625$ &                                                                   $d\indices*{*_{\bar{A}_{1}}^{A_{1}^{}}_{\bar{A}_{2}}^{B_{2}^{}}_{\bar{A}_{5}}^{\tensor*[^{2}]{\hspace{-0.2em}B}{}_{1}^{}}}$ &    $0.3734$ \\[0.4em]
                                   $d\indices*{*_{\bar{A}_{1}}^{A_{1}^{}}_{\bar{A}_{2}}^{\tensor*[^{1}]{\hspace{-0.2em}B}{}_{2}^{}}_{\bar{A}_{5}}^{\tensor*[^{2}]{\hspace{-0.2em}B}{}_{1}^{}}}$ &    $0.0815$ &                                                                                                    $d\indices*{*_{\bar{A}_{1}}^{A_{1}^{}}_{\bar{A}_{2}}^{B_{2}^{}}_{\bar{A}_{5}}^{B_{2}^{}}}$ &  $- 3.6135$ &                                                                   $d\indices*{*_{\bar{A}_{1}}^{A_{1}^{}}_{\bar{A}_{2}}^{B_{2}^{}}_{\bar{A}_{5}}^{\tensor*[^{1}]{\hspace{-0.2em}B}{}_{2}^{}}}$ &    $0.4362$ &                                  $d\indices*{*_{\bar{A}_{1}}^{A_{1}^{}}_{\bar{A}_{2}}^{\tensor*[^{1}]{\hspace{-0.2em}B}{}_{2}^{}}_{\bar{A}_{5}}^{\tensor*[^{1}]{\hspace{-0.2em}B}{}_{2}^{}}}$ &    $0.1472$ & $d\indices*{*_{\bar{A}_{1}}^{\tensor*[^{1}]{\hspace{-0.2em}A}{}_{1}^{}}_{\bar{A}_{2}}^{\tensor*[^{1}]{\hspace{-0.2em}A}{}_{1}^{}}_{\bar{A}_{5}}^{\tensor*[^{1}]{\hspace{-0.2em}A}{}_{1}^{}}}$ &    $8.1677$ \\[0.4em]
  $d\indices*{*_{\bar{A}_{1}}^{\tensor*[^{1}]{\hspace{-0.2em}A}{}_{1}^{}}_{\bar{A}_{2}}^{\tensor*[^{1}]{\hspace{-0.2em}A}{}_{1}^{}}_{\bar{A}_{5}}^{\tensor*[^{2}]{\hspace{-0.2em}A}{}_{1}^{}}}$ &  $- 3.5888$ &                                  $d\indices*{*_{\bar{A}_{1}}^{\tensor*[^{1}]{\hspace{-0.2em}A}{}_{1}^{}}_{\bar{A}_{2}}^{\tensor*[^{1}]{\hspace{-0.2em}A}{}_{1}^{}}_{\bar{A}_{5}}^{A_{2}^{}}}$ &    $0.2551$ & $d\indices*{*_{\bar{A}_{1}}^{\tensor*[^{1}]{\hspace{-0.2em}A}{}_{1}^{}}_{\bar{A}_{2}}^{\tensor*[^{2}]{\hspace{-0.2em}A}{}_{1}^{}}_{\bar{A}_{5}}^{\tensor*[^{2}]{\hspace{-0.2em}A}{}_{1}^{}}}$ &    $5.2736$ &                                  $d\indices*{*_{\bar{A}_{1}}^{\tensor*[^{1}]{\hspace{-0.2em}A}{}_{1}^{}}_{\bar{A}_{2}}^{\tensor*[^{2}]{\hspace{-0.2em}A}{}_{1}^{}}_{\bar{A}_{5}}^{A_{2}^{}}}$ &    $2.8990$ &                                  $d\indices*{*_{\bar{A}_{1}}^{\tensor*[^{1}]{\hspace{-0.2em}A}{}_{1}^{}}_{\bar{A}_{2}}^{\tensor*[^{2}]{\hspace{-0.2em}A}{}_{1}^{}}_{\bar{A}_{5}}^{B_{1}^{}}}$ &  $- 0.7111$ \\[0.4em]
  $d\indices*{*_{\bar{A}_{1}}^{\tensor*[^{1}]{\hspace{-0.2em}A}{}_{1}^{}}_{\bar{A}_{2}}^{\tensor*[^{2}]{\hspace{-0.2em}A}{}_{1}^{}}_{\bar{A}_{5}}^{\tensor*[^{1}]{\hspace{-0.2em}B}{}_{1}^{}}}$ &  $- 2.7744$ & $d\indices*{*_{\bar{A}_{1}}^{\tensor*[^{1}]{\hspace{-0.2em}A}{}_{1}^{}}_{\bar{A}_{2}}^{\tensor*[^{2}]{\hspace{-0.2em}A}{}_{1}^{}}_{\bar{A}_{5}}^{\tensor*[^{2}]{\hspace{-0.2em}B}{}_{1}^{}}}$ &  $- 3.1698$ &                                  $d\indices*{*_{\bar{A}_{1}}^{\tensor*[^{1}]{\hspace{-0.2em}A}{}_{1}^{}}_{\bar{A}_{2}}^{\tensor*[^{2}]{\hspace{-0.2em}A}{}_{1}^{}}_{\bar{A}_{5}}^{B_{2}^{}}}$ &  $- 5.3910$ & $d\indices*{*_{\bar{A}_{1}}^{\tensor*[^{1}]{\hspace{-0.2em}A}{}_{1}^{}}_{\bar{A}_{2}}^{\tensor*[^{2}]{\hspace{-0.2em}A}{}_{1}^{}}_{\bar{A}_{5}}^{\tensor*[^{1}]{\hspace{-0.2em}B}{}_{2}^{}}}$ &    $4.2268$ &                                                                   $d\indices*{*_{\bar{A}_{1}}^{\tensor*[^{1}]{\hspace{-0.2em}A}{}_{1}^{}}_{\bar{A}_{2}}^{A_{2}^{}}_{\bar{A}_{5}}^{A_{2}^{}}}$ &  $- 5.7935$ \\[0.4em]
                                                                    $d\indices*{*_{\bar{A}_{1}}^{\tensor*[^{1}]{\hspace{-0.2em}A}{}_{1}^{}}_{\bar{A}_{2}}^{A_{2}^{}}_{\bar{A}_{5}}^{B_{1}^{}}}$ &    $1.0343$ &                                  $d\indices*{*_{\bar{A}_{1}}^{\tensor*[^{1}]{\hspace{-0.2em}A}{}_{1}^{}}_{\bar{A}_{2}}^{A_{2}^{}}_{\bar{A}_{5}}^{\tensor*[^{1}]{\hspace{-0.2em}B}{}_{1}^{}}}$ &  $- 2.9833$ &                                  $d\indices*{*_{\bar{A}_{1}}^{\tensor*[^{1}]{\hspace{-0.2em}A}{}_{1}^{}}_{\bar{A}_{2}}^{A_{2}^{}}_{\bar{A}_{5}}^{\tensor*[^{2}]{\hspace{-0.2em}B}{}_{1}^{}}}$ &  $- 3.4970$ &                                                                   $d\indices*{*_{\bar{A}_{1}}^{\tensor*[^{1}]{\hspace{-0.2em}A}{}_{1}^{}}_{\bar{A}_{2}}^{A_{2}^{}}_{\bar{A}_{5}}^{B_{2}^{}}}$ &    $1.0971$ &                                  $d\indices*{*_{\bar{A}_{1}}^{\tensor*[^{1}]{\hspace{-0.2em}A}{}_{1}^{}}_{\bar{A}_{2}}^{A_{2}^{}}_{\bar{A}_{5}}^{\tensor*[^{1}]{\hspace{-0.2em}B}{}_{2}^{}}}$ &  $- 6.1434$ \\[0.4em]
                                                                    $d\indices*{*_{\bar{A}_{1}}^{\tensor*[^{1}]{\hspace{-0.2em}A}{}_{1}^{}}_{\bar{A}_{2}}^{B_{1}^{}}_{\bar{A}_{5}}^{B_{1}^{}}}$ &    $8.2434$ &                                  $d\indices*{*_{\bar{A}_{1}}^{\tensor*[^{1}]{\hspace{-0.2em}A}{}_{1}^{}}_{\bar{A}_{2}}^{B_{1}^{}}_{\bar{A}_{5}}^{\tensor*[^{1}]{\hspace{-0.2em}B}{}_{1}^{}}}$ &  $- 4.1076$ &                                  $d\indices*{*_{\bar{A}_{1}}^{\tensor*[^{1}]{\hspace{-0.2em}A}{}_{1}^{}}_{\bar{A}_{2}}^{B_{1}^{}}_{\bar{A}_{5}}^{\tensor*[^{2}]{\hspace{-0.2em}B}{}_{1}^{}}}$ &  $- 1.1598$ &                                                                   $d\indices*{*_{\bar{A}_{1}}^{\tensor*[^{1}]{\hspace{-0.2em}A}{}_{1}^{}}_{\bar{A}_{2}}^{B_{2}^{}}_{\bar{A}_{5}}^{B_{1}^{}}}$ &  $- 0.1595$ &                                  $d\indices*{*_{\bar{A}_{1}}^{\tensor*[^{1}]{\hspace{-0.2em}A}{}_{1}^{}}_{\bar{A}_{2}}^{\tensor*[^{1}]{\hspace{-0.2em}B}{}_{2}^{}}_{\bar{A}_{5}}^{B_{1}^{}}}$ &    $0.5979$ \\[0.4em]
  $d\indices*{*_{\bar{A}_{1}}^{\tensor*[^{1}]{\hspace{-0.2em}A}{}_{1}^{}}_{\bar{A}_{2}}^{\tensor*[^{1}]{\hspace{-0.2em}B}{}_{1}^{}}_{\bar{A}_{5}}^{\tensor*[^{1}]{\hspace{-0.2em}B}{}_{1}^{}}}$ &    $5.8817$ & $d\indices*{*_{\bar{A}_{1}}^{\tensor*[^{1}]{\hspace{-0.2em}A}{}_{1}^{}}_{\bar{A}_{2}}^{\tensor*[^{1}]{\hspace{-0.2em}B}{}_{1}^{}}_{\bar{A}_{5}}^{\tensor*[^{2}]{\hspace{-0.2em}B}{}_{1}^{}}}$ &    $2.5494$ &                                  $d\indices*{*_{\bar{A}_{1}}^{\tensor*[^{1}]{\hspace{-0.2em}A}{}_{1}^{}}_{\bar{A}_{2}}^{B_{2}^{}}_{\bar{A}_{5}}^{\tensor*[^{1}]{\hspace{-0.2em}B}{}_{1}^{}}}$ &    $5.4220$ & $d\indices*{*_{\bar{A}_{1}}^{\tensor*[^{1}]{\hspace{-0.2em}A}{}_{1}^{}}_{\bar{A}_{2}}^{\tensor*[^{1}]{\hspace{-0.2em}B}{}_{2}^{}}_{\bar{A}_{5}}^{\tensor*[^{1}]{\hspace{-0.2em}B}{}_{1}^{}}}$ &  $- 3.3936$ & $d\indices*{*_{\bar{A}_{1}}^{\tensor*[^{1}]{\hspace{-0.2em}A}{}_{1}^{}}_{\bar{A}_{2}}^{\tensor*[^{2}]{\hspace{-0.2em}B}{}_{1}^{}}_{\bar{A}_{5}}^{\tensor*[^{2}]{\hspace{-0.2em}B}{}_{1}^{}}}$ &    $3.5179$ \\[0.4em]
                                   $d\indices*{*_{\bar{A}_{1}}^{\tensor*[^{1}]{\hspace{-0.2em}A}{}_{1}^{}}_{\bar{A}_{2}}^{B_{2}^{}}_{\bar{A}_{5}}^{\tensor*[^{2}]{\hspace{-0.2em}B}{}_{1}^{}}}$ &    $1.6017$ & $d\indices*{*_{\bar{A}_{1}}^{\tensor*[^{1}]{\hspace{-0.2em}A}{}_{1}^{}}_{\bar{A}_{2}}^{\tensor*[^{1}]{\hspace{-0.2em}B}{}_{2}^{}}_{\bar{A}_{5}}^{\tensor*[^{2}]{\hspace{-0.2em}B}{}_{1}^{}}}$ &  $- 4.9596$ &                                                                   $d\indices*{*_{\bar{A}_{1}}^{\tensor*[^{1}]{\hspace{-0.2em}A}{}_{1}^{}}_{\bar{A}_{2}}^{B_{2}^{}}_{\bar{A}_{5}}^{B_{2}^{}}}$ &  $- 7.3609$ &                                  $d\indices*{*_{\bar{A}_{1}}^{\tensor*[^{1}]{\hspace{-0.2em}A}{}_{1}^{}}_{\bar{A}_{2}}^{B_{2}^{}}_{\bar{A}_{5}}^{\tensor*[^{1}]{\hspace{-0.2em}B}{}_{2}^{}}}$ &    $4.8193$ & $d\indices*{*_{\bar{A}_{1}}^{\tensor*[^{1}]{\hspace{-0.2em}A}{}_{1}^{}}_{\bar{A}_{2}}^{\tensor*[^{1}]{\hspace{-0.2em}B}{}_{2}^{}}_{\bar{A}_{5}}^{\tensor*[^{1}]{\hspace{-0.2em}B}{}_{2}^{}}}$ &  $- 2.6042$ \\[0.4em]
  $d\indices*{*_{\bar{A}_{1}}^{\tensor*[^{2}]{\hspace{-0.2em}A}{}_{1}^{}}_{\bar{A}_{2}}^{\tensor*[^{2}]{\hspace{-0.2em}A}{}_{1}^{}}_{\bar{A}_{5}}^{\tensor*[^{2}]{\hspace{-0.2em}A}{}_{1}^{}}}$ &    $1.6921$ &                                  $d\indices*{*_{\bar{A}_{1}}^{\tensor*[^{2}]{\hspace{-0.2em}A}{}_{1}^{}}_{\bar{A}_{2}}^{\tensor*[^{2}]{\hspace{-0.2em}A}{}_{1}^{}}_{\bar{A}_{5}}^{A_{2}^{}}}$ &    $7.9069$ &                                                                   $d\indices*{*_{\bar{A}_{1}}^{\tensor*[^{2}]{\hspace{-0.2em}A}{}_{1}^{}}_{\bar{A}_{2}}^{A_{2}^{}}_{\bar{A}_{5}}^{A_{2}^{}}}$ &    $1.2874$ &                                                                   $d\indices*{*_{\bar{A}_{1}}^{\tensor*[^{2}]{\hspace{-0.2em}A}{}_{1}^{}}_{\bar{A}_{2}}^{A_{2}^{}}_{\bar{A}_{5}}^{B_{1}^{}}}$ &  $- 4.7949$ &                                  $d\indices*{*_{\bar{A}_{1}}^{\tensor*[^{2}]{\hspace{-0.2em}A}{}_{1}^{}}_{\bar{A}_{2}}^{A_{2}^{}}_{\bar{A}_{5}}^{\tensor*[^{1}]{\hspace{-0.2em}B}{}_{1}^{}}}$ &    $1.6361$ \\[0.4em]
                                   $d\indices*{*_{\bar{A}_{1}}^{\tensor*[^{2}]{\hspace{-0.2em}A}{}_{1}^{}}_{\bar{A}_{2}}^{A_{2}^{}}_{\bar{A}_{5}}^{\tensor*[^{2}]{\hspace{-0.2em}B}{}_{1}^{}}}$ &    $3.7299$ &                                                                   $d\indices*{*_{\bar{A}_{1}}^{\tensor*[^{2}]{\hspace{-0.2em}A}{}_{1}^{}}_{\bar{A}_{2}}^{A_{2}^{}}_{\bar{A}_{5}}^{B_{2}^{}}}$ &  $- 0.0457$ &                                  $d\indices*{*_{\bar{A}_{1}}^{\tensor*[^{2}]{\hspace{-0.2em}A}{}_{1}^{}}_{\bar{A}_{2}}^{A_{2}^{}}_{\bar{A}_{5}}^{\tensor*[^{1}]{\hspace{-0.2em}B}{}_{2}^{}}}$ &  $- 0.0576$ &                                                                   $d\indices*{*_{\bar{A}_{1}}^{\tensor*[^{2}]{\hspace{-0.2em}A}{}_{1}^{}}_{\bar{A}_{2}}^{B_{1}^{}}_{\bar{A}_{5}}^{B_{1}^{}}}$ &  $- 1.3269$ &                                  $d\indices*{*_{\bar{A}_{1}}^{\tensor*[^{2}]{\hspace{-0.2em}A}{}_{1}^{}}_{\bar{A}_{2}}^{B_{1}^{}}_{\bar{A}_{5}}^{\tensor*[^{1}]{\hspace{-0.2em}B}{}_{1}^{}}}$ &  $- 3.2413$ \\[0.4em]
                                   $d\indices*{*_{\bar{A}_{1}}^{\tensor*[^{2}]{\hspace{-0.2em}A}{}_{1}^{}}_{\bar{A}_{2}}^{B_{1}^{}}_{\bar{A}_{5}}^{\tensor*[^{2}]{\hspace{-0.2em}B}{}_{1}^{}}}$ &  $- 4.4891$ &                                                                   $d\indices*{*_{\bar{A}_{1}}^{\tensor*[^{2}]{\hspace{-0.2em}A}{}_{1}^{}}_{\bar{A}_{2}}^{B_{2}^{}}_{\bar{A}_{5}}^{B_{1}^{}}}$ &  $- 2.2687$ &                                  $d\indices*{*_{\bar{A}_{1}}^{\tensor*[^{2}]{\hspace{-0.2em}A}{}_{1}^{}}_{\bar{A}_{2}}^{\tensor*[^{1}]{\hspace{-0.2em}B}{}_{2}^{}}_{\bar{A}_{5}}^{B_{1}^{}}}$ &    $1.4441$ & $d\indices*{*_{\bar{A}_{1}}^{\tensor*[^{2}]{\hspace{-0.2em}A}{}_{1}^{}}_{\bar{A}_{2}}^{\tensor*[^{1}]{\hspace{-0.2em}B}{}_{1}^{}}_{\bar{A}_{5}}^{\tensor*[^{1}]{\hspace{-0.2em}B}{}_{1}^{}}}$ &   $13.3139$ & $d\indices*{*_{\bar{A}_{1}}^{\tensor*[^{2}]{\hspace{-0.2em}A}{}_{1}^{}}_{\bar{A}_{2}}^{\tensor*[^{1}]{\hspace{-0.2em}B}{}_{1}^{}}_{\bar{A}_{5}}^{\tensor*[^{2}]{\hspace{-0.2em}B}{}_{1}^{}}}$ &    $4.2784$ \\[0.4em]
                                   $d\indices*{*_{\bar{A}_{1}}^{\tensor*[^{2}]{\hspace{-0.2em}A}{}_{1}^{}}_{\bar{A}_{2}}^{B_{2}^{}}_{\bar{A}_{5}}^{\tensor*[^{1}]{\hspace{-0.2em}B}{}_{1}^{}}}$ &  $- 6.2771$ & $d\indices*{*_{\bar{A}_{1}}^{\tensor*[^{2}]{\hspace{-0.2em}A}{}_{1}^{}}_{\bar{A}_{2}}^{\tensor*[^{1}]{\hspace{-0.2em}B}{}_{2}^{}}_{\bar{A}_{5}}^{\tensor*[^{1}]{\hspace{-0.2em}B}{}_{1}^{}}}$ &    $4.2212$ & $d\indices*{*_{\bar{A}_{1}}^{\tensor*[^{2}]{\hspace{-0.2em}A}{}_{1}^{}}_{\bar{A}_{2}}^{\tensor*[^{2}]{\hspace{-0.2em}B}{}_{1}^{}}_{\bar{A}_{5}}^{\tensor*[^{2}]{\hspace{-0.2em}B}{}_{1}^{}}}$ &  $- 2.4243$ &                                  $d\indices*{*_{\bar{A}_{1}}^{\tensor*[^{2}]{\hspace{-0.2em}A}{}_{1}^{}}_{\bar{A}_{2}}^{B_{2}^{}}_{\bar{A}_{5}}^{\tensor*[^{2}]{\hspace{-0.2em}B}{}_{1}^{}}}$ &  $- 7.2147$ & $d\indices*{*_{\bar{A}_{1}}^{\tensor*[^{2}]{\hspace{-0.2em}A}{}_{1}^{}}_{\bar{A}_{2}}^{\tensor*[^{1}]{\hspace{-0.2em}B}{}_{2}^{}}_{\bar{A}_{5}}^{\tensor*[^{2}]{\hspace{-0.2em}B}{}_{1}^{}}}$ &    $6.0620$ \\[0.4em]
                                                                    $d\indices*{*_{\bar{A}_{1}}^{\tensor*[^{2}]{\hspace{-0.2em}A}{}_{1}^{}}_{\bar{A}_{2}}^{B_{2}^{}}_{\bar{A}_{5}}^{B_{2}^{}}}$ &  $- 1.6293$ &                                  $d\indices*{*_{\bar{A}_{1}}^{\tensor*[^{2}]{\hspace{-0.2em}A}{}_{1}^{}}_{\bar{A}_{2}}^{B_{2}^{}}_{\bar{A}_{5}}^{\tensor*[^{1}]{\hspace{-0.2em}B}{}_{2}^{}}}$ &  $- 0.5297$ & $d\indices*{*_{\bar{A}_{1}}^{\tensor*[^{2}]{\hspace{-0.2em}A}{}_{1}^{}}_{\bar{A}_{2}}^{\tensor*[^{1}]{\hspace{-0.2em}B}{}_{2}^{}}_{\bar{A}_{5}}^{\tensor*[^{1}]{\hspace{-0.2em}B}{}_{2}^{}}}$ &    $0.3611$ &                                                                                                    $d\indices*{*_{\bar{A}_{1}}^{A_{2}^{}}_{\bar{A}_{2}}^{A_{2}^{}}_{\bar{A}_{5}}^{A_{2}^{}}}$ &    $7.3647$ &                                                                                                    $d\indices*{*_{\bar{A}_{1}}^{A_{2}^{}}_{\bar{A}_{2}}^{B_{1}^{}}_{\bar{A}_{5}}^{B_{1}^{}}}$ &    $1.0134$ \\[0.4em]
                                                                    $d\indices*{*_{\bar{A}_{1}}^{A_{2}^{}}_{\bar{A}_{2}}^{B_{1}^{}}_{\bar{A}_{5}}^{\tensor*[^{1}]{\hspace{-0.2em}B}{}_{1}^{}}}$ &    $3.7407$ &                                                                   $d\indices*{*_{\bar{A}_{1}}^{A_{2}^{}}_{\bar{A}_{2}}^{B_{1}^{}}_{\bar{A}_{5}}^{\tensor*[^{2}]{\hspace{-0.2em}B}{}_{1}^{}}}$ &  $- 1.5954$ &                                                                                                    $d\indices*{*_{\bar{A}_{1}}^{A_{2}^{}}_{\bar{A}_{2}}^{B_{2}^{}}_{\bar{A}_{5}}^{B_{1}^{}}}$ &  $- 4.8702$ &                                                                   $d\indices*{*_{\bar{A}_{1}}^{A_{2}^{}}_{\bar{A}_{2}}^{\tensor*[^{1}]{\hspace{-0.2em}B}{}_{2}^{}}_{\bar{A}_{5}}^{B_{1}^{}}}$ &    $4.8695$ &                                  $d\indices*{*_{\bar{A}_{1}}^{A_{2}^{}}_{\bar{A}_{2}}^{\tensor*[^{1}]{\hspace{-0.2em}B}{}_{1}^{}}_{\bar{A}_{5}}^{\tensor*[^{1}]{\hspace{-0.2em}B}{}_{1}^{}}}$ &    $1.5120$ \\[0.4em]
                                   $d\indices*{*_{\bar{A}_{1}}^{A_{2}^{}}_{\bar{A}_{2}}^{\tensor*[^{1}]{\hspace{-0.2em}B}{}_{1}^{}}_{\bar{A}_{5}}^{\tensor*[^{2}]{\hspace{-0.2em}B}{}_{1}^{}}}$ &    $5.3742$ &                                                                   $d\indices*{*_{\bar{A}_{1}}^{A_{2}^{}}_{\bar{A}_{2}}^{B_{2}^{}}_{\bar{A}_{5}}^{\tensor*[^{1}]{\hspace{-0.2em}B}{}_{1}^{}}}$ &    $8.2880$ &                                  $d\indices*{*_{\bar{A}_{1}}^{A_{2}^{}}_{\bar{A}_{2}}^{\tensor*[^{1}]{\hspace{-0.2em}B}{}_{2}^{}}_{\bar{A}_{5}}^{\tensor*[^{1}]{\hspace{-0.2em}B}{}_{1}^{}}}$ &  $- 4.1776$ &                                  $d\indices*{*_{\bar{A}_{1}}^{A_{2}^{}}_{\bar{A}_{2}}^{\tensor*[^{2}]{\hspace{-0.2em}B}{}_{1}^{}}_{\bar{A}_{5}}^{\tensor*[^{2}]{\hspace{-0.2em}B}{}_{1}^{}}}$ &    $6.5404$ &                                                                   $d\indices*{*_{\bar{A}_{1}}^{A_{2}^{}}_{\bar{A}_{2}}^{B_{2}^{}}_{\bar{A}_{5}}^{\tensor*[^{2}]{\hspace{-0.2em}B}{}_{1}^{}}}$ &  $- 0.4468$ \\[0.4em]
                                   $d\indices*{*_{\bar{A}_{1}}^{A_{2}^{}}_{\bar{A}_{2}}^{\tensor*[^{1}]{\hspace{-0.2em}B}{}_{2}^{}}_{\bar{A}_{5}}^{\tensor*[^{2}]{\hspace{-0.2em}B}{}_{1}^{}}}$ &    $0.8437$ &                                                                                                    $d\indices*{*_{\bar{A}_{1}}^{A_{2}^{}}_{\bar{A}_{2}}^{B_{2}^{}}_{\bar{A}_{5}}^{B_{2}^{}}}$ &   $10.5046$ &                                                                   $d\indices*{*_{\bar{A}_{1}}^{A_{2}^{}}_{\bar{A}_{2}}^{B_{2}^{}}_{\bar{A}_{5}}^{\tensor*[^{1}]{\hspace{-0.2em}B}{}_{2}^{}}}$ &  $- 7.0084$ &                                  $d\indices*{*_{\bar{A}_{1}}^{A_{2}^{}}_{\bar{A}_{2}}^{\tensor*[^{1}]{\hspace{-0.2em}B}{}_{2}^{}}_{\bar{A}_{5}}^{\tensor*[^{1}]{\hspace{-0.2em}B}{}_{2}^{}}}$ &    $4.6520$ &                                  $d\indices*{*_{\bar{A}_{1}}^{B_{1}^{}}_{\bar{A}_{2}}^{\tensor*[^{1}]{\hspace{-0.2em}B}{}_{1}^{}}_{\bar{A}_{5}}^{\tensor*[^{2}]{\hspace{-0.2em}B}{}_{1}^{}}}$ &  $- 4.3444$ \\[0.4em]
                                                                    $d\indices*{*_{\bar{A}_{1}}^{B_{2}^{}}_{\bar{A}_{2}}^{B_{1}^{}}_{\bar{A}_{5}}^{\tensor*[^{1}]{\hspace{-0.2em}B}{}_{1}^{}}}$ &    $5.1531$ &                                  $d\indices*{*_{\bar{A}_{1}}^{\tensor*[^{1}]{\hspace{-0.2em}B}{}_{2}^{}}_{\bar{A}_{2}}^{B_{1}^{}}_{\bar{A}_{5}}^{\tensor*[^{1}]{\hspace{-0.2em}B}{}_{1}^{}}}$ &  $- 3.8848$ &                                                                   $d\indices*{*_{\bar{A}_{1}}^{B_{2}^{}}_{\bar{A}_{2}}^{B_{1}^{}}_{\bar{A}_{5}}^{\tensor*[^{2}]{\hspace{-0.2em}B}{}_{1}^{}}}$ &    $5.1203$ &                                  $d\indices*{*_{\bar{A}_{1}}^{\tensor*[^{1}]{\hspace{-0.2em}B}{}_{2}^{}}_{\bar{A}_{2}}^{B_{1}^{}}_{\bar{A}_{5}}^{\tensor*[^{2}]{\hspace{-0.2em}B}{}_{1}^{}}}$ &  $- 3.6242$ &                                                                   $d\indices*{*_{\bar{A}_{1}}^{B_{2}^{}}_{\bar{A}_{2}}^{\tensor*[^{1}]{\hspace{-0.2em}B}{}_{2}^{}}_{\bar{A}_{5}}^{B_{1}^{}}}$ &    $1.4608$ \\[0.4em]
                                   $d\indices*{*_{\bar{A}_{1}}^{B_{2}^{}}_{\bar{A}_{2}}^{\tensor*[^{1}]{\hspace{-0.2em}B}{}_{1}^{}}_{\bar{A}_{5}}^{\tensor*[^{2}]{\hspace{-0.2em}B}{}_{1}^{}}}$ &    $0.7216$ & $d\indices*{*_{\bar{A}_{1}}^{\tensor*[^{1}]{\hspace{-0.2em}B}{}_{2}^{}}_{\bar{A}_{2}}^{\tensor*[^{1}]{\hspace{-0.2em}B}{}_{1}^{}}_{\bar{A}_{5}}^{\tensor*[^{2}]{\hspace{-0.2em}B}{}_{1}^{}}}$ &  $- 0.0810$ &                                  $d\indices*{*_{\bar{A}_{1}}^{B_{2}^{}}_{\bar{A}_{2}}^{\tensor*[^{1}]{\hspace{-0.2em}B}{}_{2}^{}}_{\bar{A}_{5}}^{\tensor*[^{1}]{\hspace{-0.2em}B}{}_{1}^{}}}$ &  $- 0.3419$ &                                  $d\indices*{*_{\bar{A}_{1}}^{B_{2}^{}}_{\bar{A}_{2}}^{\tensor*[^{1}]{\hspace{-0.2em}B}{}_{2}^{}}_{\bar{A}_{5}}^{\tensor*[^{2}]{\hspace{-0.2em}B}{}_{1}^{}}}$ &  $- 0.2718$ &                                                                                                                                                                                               &              \\
%
%
%

\end{longtable*}

\begin{longtable*}{cccccccccc}
\caption{All second order irreducible derivatives for ThO$_2$. All results were computed using the LID approach, $4\hat{\mathbf1}$ supercell, and SCAN functional. The units are eV/\AA$^2$. \label{table:tho2_2}} \\
\hline\hline 
Derivative & Value & Derivative & Value & Derivative & Value & Derivative & Value & Derivative & Value \\
\hline\\[-0.9em]
\endfirsthead
\hline\hline
Derivative & Value & Derivative & Value & Derivative & Value & Derivative & Value & Derivative & Value   \\
\hline\\[-0.9em]
\endhead
\hline\hline
\endfoot
                                                                          $d\indices*{*_{\Gamma}^{T_{2g}^{}}_{\Gamma}^{T_{2g}^{}}}$ &  $12.1990$ &                                                                       $d\indices*{*_{\Gamma}^{T_{1u}^{}}_{\Gamma}^{T_{1u}^{}}}$ &   $12.5163$ &                                                                       $d\indices*{*_{L}^{A_{1g}^{}}_{L}^{A_{1g}^{}}}$ &  $17.2149$ &                                                                                     $d\indices*{*_{L}^{E_{g}^{}}_{L}^{E_{g}^{}}}$ &    $9.3760$ &                                                                                   $d\indices*{*_{L}^{A_{2u}^{}}_{L}^{A_{2u}^{}}}$ &  $35.5237$ \\[0.4em]
                                                   $d\indices*{*_{L}^{A_{2u}^{}}_{L}^{\tensor*[^{1}]{\hspace{-0.2em}A}{}_{2u}^{}}}$ &   $7.8665$ &               $d\indices*{*_{L}^{\tensor*[^{1}]{\hspace{-0.2em}A}{}_{2u}^{}}_{L}^{\tensor*[^{1}]{\hspace{-0.2em}A}{}_{2u}^{}}}$ &    $9.9096$ &                                                                         $d\indices*{*_{L}^{E_{u}^{}}_{L}^{E_{u}^{}}}$ &   $9.4382$ &                                                    $d\indices*{*_{L}^{E_{u}^{}}_{L}^{\tensor*[^{1}]{\hspace{-0.2em}E}{}_{u}^{}}}$ &  $- 3.0224$ &                   $d\indices*{*_{L}^{\tensor*[^{1}]{\hspace{-0.2em}E}{}_{u}^{}}_{L}^{\tensor*[^{1}]{\hspace{-0.2em}E}{}_{u}^{}}}$ &   $7.3315$ \\[0.4em]
                                                                    $d\indices*{*_{\Lambda}^{A_{1}^{}}_{\bar{\Lambda}}^{A_{1}^{}}}$ &  $35.2643$ &                                $d\indices*{*_{\Lambda}^{A_{1}^{}}_{\bar{\Lambda}}^{\tensor*[^{1}]{\hspace{-0.2em}A}{}_{1}^{}}}$ & $- 17.4190$ &                      $d\indices*{*_{\Lambda}^{A_{1}^{}}_{\bar{\Lambda}}^{\tensor*[^{2}]{\hspace{-0.2em}A}{}_{1}^{}}}$ &   $4.8629$ & $d\indices*{*_{\Lambda}^{\tensor*[^{1}]{\hspace{-0.2em}A}{}_{1}^{}}_{\bar{\Lambda}}^{\tensor*[^{1}]{\hspace{-0.2em}A}{}_{1}^{}}}$ &   $17.1959$ & $d\indices*{*_{\Lambda}^{\tensor*[^{1}]{\hspace{-0.2em}A}{}_{1}^{}}_{\bar{\Lambda}}^{\tensor*[^{2}]{\hspace{-0.2em}A}{}_{1}^{}}}$ & $- 0.2206$ \\[0.4em]
  $d\indices*{*_{\Lambda}^{\tensor*[^{2}]{\hspace{-0.2em}A}{}_{1}^{}}_{\bar{\Lambda}}^{\tensor*[^{2}]{\hspace{-0.2em}A}{}_{1}^{}}}$ &  $11.0067$ &                                                                   $d\indices*{*_{\Lambda}^{E_{}^{}}_{\bar{\Lambda}}^{E_{}^{}}}$ &    $8.9241$ &                        $d\indices*{*_{\Lambda}^{E_{}^{}}_{\bar{\Lambda}}^{\tensor*[^{1}]{\hspace{-0.2em}E}{}_{}^{}}}$ & $- 4.1097$ &                                    $d\indices*{*_{\Lambda}^{E_{}^{}}_{\bar{\Lambda}}^{\tensor*[^{2}]{\hspace{-0.2em}E}{}_{}^{}}}$ &  $- 2.2106$ &   $d\indices*{*_{\Lambda}^{\tensor*[^{1}]{\hspace{-0.2em}E}{}_{}^{}}_{\bar{\Lambda}}^{\tensor*[^{1}]{\hspace{-0.2em}E}{}_{}^{}}}$ &   $6.8261$ \\[0.4em]
    $d\indices*{*_{\Lambda}^{\tensor*[^{1}]{\hspace{-0.2em}E}{}_{}^{}}_{\bar{\Lambda}}^{\tensor*[^{2}]{\hspace{-0.2em}E}{}_{}^{}}}$ & $- 2.5362$ & $d\indices*{*_{\Lambda}^{\tensor*[^{2}]{\hspace{-0.2em}E}{}_{}^{}}_{\bar{\Lambda}}^{\tensor*[^{2}]{\hspace{-0.2em}E}{}_{}^{}}}$ &    $9.8200$ &                                                                       $d\indices*{*_{X}^{A_{1g}^{}}_{X}^{A_{1g}^{}}}$ &  $20.7290$ &                                                                                     $d\indices*{*_{X}^{E_{g}^{}}_{X}^{E_{g}^{}}}$ &    $4.4730$ &                                                                                   $d\indices*{*_{X}^{A_{2u}^{}}_{X}^{A_{2u}^{}}}$ &  $41.0700$ \\[0.4em]
                                                                                    $d\indices*{*_{X}^{B_{1u}^{}}_{X}^{B_{1u}^{}}}$ &   $2.4343$ &                                                                                   $d\indices*{*_{X}^{E_{u}^{}}_{X}^{E_{u}^{}}}$ &   $11.8687$ &                                        $d\indices*{*_{X}^{E_{u}^{}}_{X}^{\tensor*[^{1}]{\hspace{-0.2em}E}{}_{u}^{}}}$ & $- 0.4391$ &                   $d\indices*{*_{X}^{\tensor*[^{1}]{\hspace{-0.2em}E}{}_{u}^{}}_{X}^{\tensor*[^{1}]{\hspace{-0.2em}E}{}_{u}^{}}}$ &   $11.7599$ &                                                                                 $d\indices*{*_{B}^{A_{}^{}}_{\bar{B}}^{A_{}^{}}}$ &  $30.4708$ \\[0.4em]
                                                 $d\indices*{*_{B}^{A_{}^{}}_{\bar{B}}^{\tensor*[^{1}]{\hspace{-0.2em}A}{}_{}^{}}}$ & $- 6.0319$ &                                              $d\indices*{*_{B}^{A_{}^{}}_{\bar{B}}^{\tensor*[^{2}]{\hspace{-0.2em}A}{}_{}^{}}}$ &  $- 6.4816$ &                                    $d\indices*{*_{B}^{A_{}^{}}_{\bar{B}}^{\tensor*[^{3}]{\hspace{-0.2em}A}{}_{}^{}}}$ & $- 2.3498$ &                                                $d\indices*{*_{B}^{A_{}^{}}_{\bar{B}}^{\tensor*[^{4}]{\hspace{-0.2em}A}{}_{}^{}}}$ &    $1.5032$ &                                                $d\indices*{*_{B}^{A_{}^{}}_{\bar{B}}^{\tensor*[^{5}]{\hspace{-0.2em}A}{}_{}^{}}}$ & $- 2.1761$ \\[0.4em]
                $d\indices*{*_{B}^{\tensor*[^{1}]{\hspace{-0.2em}A}{}_{}^{}}_{\bar{B}}^{\tensor*[^{1}]{\hspace{-0.2em}A}{}_{}^{}}}$ &  $15.2502$ &             $d\indices*{*_{B}^{\tensor*[^{1}]{\hspace{-0.2em}A}{}_{}^{}}_{\bar{B}}^{\tensor*[^{2}]{\hspace{-0.2em}A}{}_{}^{}}}$ &    $1.1007$ &   $d\indices*{*_{B}^{\tensor*[^{1}]{\hspace{-0.2em}A}{}_{}^{}}_{\bar{B}}^{\tensor*[^{3}]{\hspace{-0.2em}A}{}_{}^{}}}$ &   $0.8918$ &               $d\indices*{*_{B}^{\tensor*[^{1}]{\hspace{-0.2em}A}{}_{}^{}}_{\bar{B}}^{\tensor*[^{4}]{\hspace{-0.2em}A}{}_{}^{}}}$ &  $- 5.0103$ &               $d\indices*{*_{B}^{\tensor*[^{1}]{\hspace{-0.2em}A}{}_{}^{}}_{\bar{B}}^{\tensor*[^{5}]{\hspace{-0.2em}A}{}_{}^{}}}$ &   $2.9225$ \\[0.4em]
                $d\indices*{*_{B}^{\tensor*[^{2}]{\hspace{-0.2em}A}{}_{}^{}}_{\bar{B}}^{\tensor*[^{2}]{\hspace{-0.2em}A}{}_{}^{}}}$ &  $18.8181$ &             $d\indices*{*_{B}^{\tensor*[^{2}]{\hspace{-0.2em}A}{}_{}^{}}_{\bar{B}}^{\tensor*[^{3}]{\hspace{-0.2em}A}{}_{}^{}}}$ &    $3.1393$ &   $d\indices*{*_{B}^{\tensor*[^{2}]{\hspace{-0.2em}A}{}_{}^{}}_{\bar{B}}^{\tensor*[^{4}]{\hspace{-0.2em}A}{}_{}^{}}}$ &   $0.5820$ &               $d\indices*{*_{B}^{\tensor*[^{2}]{\hspace{-0.2em}A}{}_{}^{}}_{\bar{B}}^{\tensor*[^{5}]{\hspace{-0.2em}A}{}_{}^{}}}$ &  $- 0.5627$ &               $d\indices*{*_{B}^{\tensor*[^{3}]{\hspace{-0.2em}A}{}_{}^{}}_{\bar{B}}^{\tensor*[^{3}]{\hspace{-0.2em}A}{}_{}^{}}}$ &   $6.6890$ \\[0.4em]
                $d\indices*{*_{B}^{\tensor*[^{3}]{\hspace{-0.2em}A}{}_{}^{}}_{\bar{B}}^{\tensor*[^{4}]{\hspace{-0.2em}A}{}_{}^{}}}$ & $- 0.3181$ &             $d\indices*{*_{B}^{\tensor*[^{3}]{\hspace{-0.2em}A}{}_{}^{}}_{\bar{B}}^{\tensor*[^{5}]{\hspace{-0.2em}A}{}_{}^{}}}$ &  $- 2.2013$ &   $d\indices*{*_{B}^{\tensor*[^{4}]{\hspace{-0.2em}A}{}_{}^{}}_{\bar{B}}^{\tensor*[^{4}]{\hspace{-0.2em}A}{}_{}^{}}}$ &   $6.9692$ &               $d\indices*{*_{B}^{\tensor*[^{4}]{\hspace{-0.2em}A}{}_{}^{}}_{\bar{B}}^{\tensor*[^{5}]{\hspace{-0.2em}A}{}_{}^{}}}$ &  $- 1.3683$ &               $d\indices*{*_{B}^{\tensor*[^{5}]{\hspace{-0.2em}A}{}_{}^{}}_{\bar{B}}^{\tensor*[^{5}]{\hspace{-0.2em}A}{}_{}^{}}}$ &  $12.6662$ \\[0.4em]
                                                                                  $d\indices*{*_{B}^{B_{}^{}}_{\bar{B}}^{B_{}^{}}}$ &  $10.6032$ &                                              $d\indices*{*_{B}^{B_{}^{}}_{\bar{B}}^{\tensor*[^{1}]{\hspace{-0.2em}B}{}_{}^{}}}$ &  $- 2.1960$ &                                    $d\indices*{*_{B}^{B_{}^{}}_{\bar{B}}^{\tensor*[^{2}]{\hspace{-0.2em}B}{}_{}^{}}}$ & $- 0.2653$ &               $d\indices*{*_{B}^{\tensor*[^{1}]{\hspace{-0.2em}B}{}_{}^{}}_{\bar{B}}^{\tensor*[^{1}]{\hspace{-0.2em}B}{}_{}^{}}}$ &    $5.8747$ &               $d\indices*{*_{B}^{\tensor*[^{1}]{\hspace{-0.2em}B}{}_{}^{}}_{\bar{B}}^{\tensor*[^{2}]{\hspace{-0.2em}B}{}_{}^{}}}$ &   $1.3204$ \\[0.4em]
                $d\indices*{*_{B}^{\tensor*[^{2}]{\hspace{-0.2em}B}{}_{}^{}}_{\bar{B}}^{\tensor*[^{2}]{\hspace{-0.2em}B}{}_{}^{}}}$ &  $10.6224$ &                                                                             $d\indices*{*_{A}^{A_{1}^{}}_{\bar{A}}^{A_{1}^{}}}$ &   $31.0538$ &                                  $d\indices*{*_{A}^{A_{1}^{}}_{\bar{A}}^{\tensor*[^{1}]{\hspace{-0.2em}A}{}_{1}^{}}}$ & $- 2.3640$ &                                              $d\indices*{*_{A}^{A_{1}^{}}_{\bar{A}}^{\tensor*[^{2}]{\hspace{-0.2em}A}{}_{1}^{}}}$ &   $10.4027$ &             $d\indices*{*_{A}^{\tensor*[^{1}]{\hspace{-0.2em}A}{}_{1}^{}}_{\bar{A}}^{\tensor*[^{1}]{\hspace{-0.2em}A}{}_{1}^{}}}$ &  $14.6428$ \\[0.4em]
              $d\indices*{*_{A}^{\tensor*[^{1}]{\hspace{-0.2em}A}{}_{1}^{}}_{\bar{A}}^{\tensor*[^{2}]{\hspace{-0.2em}A}{}_{1}^{}}}$ & $- 2.7501$ &           $d\indices*{*_{A}^{\tensor*[^{2}]{\hspace{-0.2em}A}{}_{1}^{}}_{\bar{A}}^{\tensor*[^{2}]{\hspace{-0.2em}A}{}_{1}^{}}}$ &   $16.1843$ &                                                                   $d\indices*{*_{A}^{A_{2}^{}}_{\bar{A}}^{A_{2}^{}}}$ &   $8.5207$ &                                                                               $d\indices*{*_{A}^{B_{1}^{}}_{\bar{A}}^{B_{1}^{}}}$ &    $3.8476$ &                                              $d\indices*{*_{A}^{B_{1}^{}}_{\bar{A}}^{\tensor*[^{1}]{\hspace{-0.2em}B}{}_{1}^{}}}$ & $- 3.1242$ \\[0.4em]
                                               $d\indices*{*_{A}^{B_{1}^{}}_{\bar{A}}^{\tensor*[^{2}]{\hspace{-0.2em}B}{}_{1}^{}}}$ &   $0.6397$ &           $d\indices*{*_{A}^{\tensor*[^{1}]{\hspace{-0.2em}B}{}_{1}^{}}_{\bar{A}}^{\tensor*[^{1}]{\hspace{-0.2em}B}{}_{1}^{}}}$ &   $16.8192$ & $d\indices*{*_{A}^{\tensor*[^{1}]{\hspace{-0.2em}B}{}_{1}^{}}_{\bar{A}}^{\tensor*[^{2}]{\hspace{-0.2em}B}{}_{1}^{}}}$ & $- 1.5602$ &             $d\indices*{*_{A}^{\tensor*[^{2}]{\hspace{-0.2em}B}{}_{1}^{}}_{\bar{A}}^{\tensor*[^{2}]{\hspace{-0.2em}B}{}_{1}^{}}}$ &    $7.8516$ &                                                                               $d\indices*{*_{A}^{B_{2}^{}}_{\bar{A}}^{B_{2}^{}}}$ &   $9.4765$ \\[0.4em]
                                               $d\indices*{*_{A}^{B_{2}^{}}_{\bar{A}}^{\tensor*[^{1}]{\hspace{-0.2em}B}{}_{2}^{}}}$ & $- 3.0272$ &           $d\indices*{*_{A}^{\tensor*[^{1}]{\hspace{-0.2em}B}{}_{2}^{}}_{\bar{A}}^{\tensor*[^{1}]{\hspace{-0.2em}B}{}_{2}^{}}}$ &    $8.2062$ &                                                         $d\indices*{*_{\Delta}^{A_{1}^{}}_{\bar{\Delta}}^{A_{1}^{}}}$ &  $37.9861$ &                                    $d\indices*{*_{\Delta}^{A_{1}^{}}_{\bar{\Delta}}^{\tensor*[^{1}]{\hspace{-0.2em}A}{}_{1}^{}}}$ & $- 16.9061$ &   $d\indices*{*_{\Delta}^{\tensor*[^{1}]{\hspace{-0.2em}A}{}_{1}^{}}_{\bar{\Delta}}^{\tensor*[^{1}]{\hspace{-0.2em}A}{}_{1}^{}}}$ &  $19.0121$ \\[0.4em]
                                                                      $d\indices*{*_{\Delta}^{B_{2}^{}}_{\bar{\Delta}}^{B_{2}^{}}}$ &   $7.8631$ &                                                                     $d\indices*{*_{\Delta}^{E_{}^{}}_{\bar{\Delta}}^{E_{}^{}}}$ &   $10.0583$ &                          $d\indices*{*_{\Delta}^{E_{}^{}}_{\bar{\Delta}}^{\tensor*[^{1}]{\hspace{-0.2em}E}{}_{}^{}}}$ &   $4.1612$ &                                      $d\indices*{*_{\Delta}^{E_{}^{}}_{\bar{\Delta}}^{\tensor*[^{2}]{\hspace{-0.2em}E}{}_{}^{}}}$ &  $- 0.3293$ &     $d\indices*{*_{\Delta}^{\tensor*[^{1}]{\hspace{-0.2em}E}{}_{}^{}}_{\bar{\Delta}}^{\tensor*[^{1}]{\hspace{-0.2em}E}{}_{}^{}}}$ &   $4.2788$ \\[0.4em]
      $d\indices*{*_{\Delta}^{\tensor*[^{1}]{\hspace{-0.2em}E}{}_{}^{}}_{\bar{\Delta}}^{\tensor*[^{2}]{\hspace{-0.2em}E}{}_{}^{}}}$ &   $1.0343$ &   $d\indices*{*_{\Delta}^{\tensor*[^{2}]{\hspace{-0.2em}E}{}_{}^{}}_{\bar{\Delta}}^{\tensor*[^{2}]{\hspace{-0.2em}E}{}_{}^{}}}$ &   $12.0236$ &                                                                   $d\indices*{*_{W}^{A_{1}^{}}_{\bar{W}}^{A_{1}^{}}}$ &  $19.6718$ &                                                                               $d\indices*{*_{W}^{B_{1}^{}}_{\bar{W}}^{B_{1}^{}}}$ &   $12.0778$ &                                              $d\indices*{*_{W}^{B_{1}^{}}_{\bar{W}}^{\tensor*[^{1}]{\hspace{-0.2em}B}{}_{1}^{}}}$ &   $0.1398$ \\[0.4em]
              $d\indices*{*_{W}^{\tensor*[^{1}]{\hspace{-0.2em}B}{}_{1}^{}}_{\bar{W}}^{\tensor*[^{1}]{\hspace{-0.2em}B}{}_{1}^{}}}$ &  $13.3331$ &                                                                             $d\indices*{*_{W}^{A_{2}^{}}_{\bar{W}}^{A_{2}^{}}}$ &    $3.2377$ &                                                                   $d\indices*{*_{W}^{B_{2}^{}}_{\bar{W}}^{B_{2}^{}}}$ &   $3.8226$ &                                                                                 $d\indices*{*_{W}^{E_{}^{}}_{\bar{W}}^{E_{}^{}}}$ &   $24.8946$ &                                                $d\indices*{*_{W}^{E_{}^{}}_{\bar{W}}^{\tensor*[^{1}]{\hspace{-0.2em}E}{}_{}^{}}}$ & $- 1.5154$ \\[0.4em]
                $d\indices*{*_{W}^{\tensor*[^{1}]{\hspace{-0.2em}E}{}_{}^{}}_{\bar{W}}^{\tensor*[^{1}]{\hspace{-0.2em}E}{}_{}^{}}}$ &   $8.5651$ &                                                                                                                                 &             &                                                                                                                       &            &                                                                                                                                   &             &                                                                                                                                   &            \\[0.4em]
%
%

\end{longtable*}

\begin{longtable*}{cccccccccc}
\caption{All third order irreducible derivatives for ThO$_2$. All results were computed using the BID approach, $\hat{\mathbf{S}}_{O}$ supercell, and SCAN functional.  The units are eV/\AA$^3$.\label{table:tho2_3}} \\
\hline\hline 
Derivative & Value & Derivative & Value & Derivative & Value & Derivative & Value & Derivative & Value \\
\hline\\[-0.9em]
\endfirsthead
\hline\hline
Derivative & Value & Derivative & Value & Derivative & Value & Derivative & Value & Derivative & Value \\
\hline\\[-0.9em]
\endhead
\hline\hline
\endfoot
                                                                                                                 $d\indices*{*_{\Gamma}^{T_{2g}^{}}_{\Gamma}^{T_{2g}^{}}_{\Gamma}^{T_{2g}^{}}}$ & $- 69.7696$ &                                                                                                                $d\indices*{*_{\Gamma}^{T_{1u}^{}}_{\Gamma}^{T_{1u}^{}}_{\Gamma}^{T_{2g}^{}}}$ & $- 61.3189$ &                                                                                                                     $d\indices*{*_{X}^{A_{1g}^{}}_{\bar{A}}^{A_{1}^{}}_{\bar{A}}^{A_{1}^{}}}$ &   $27.2398$ &                                                                                                                      $d\indices*{*_{X}^{E_{u}^{}}_{\bar{A}}^{A_{1}^{}}_{\bar{A}}^{A_{1}^{}}}$ &    $1.3535$ &                                                                                     $d\indices*{*_{X}^{\tensor*[^{1}]{\hspace{-0.2em}E}{}_{u}^{}}_{\bar{A}}^{A_{1}^{}}_{\bar{A}}^{A_{1}^{}}}$ & $- 11.5886$ \\[0.4em]
                                                                                     $d\indices*{*_{X}^{A_{1g}^{}}_{\bar{A}}^{A_{1}^{}}_{\bar{A}}^{\tensor*[^{1}]{\hspace{-0.2em}A}{}_{1}^{}}}$ & $- 14.1003$ &                                                                                     $d\indices*{*_{X}^{E_{u}^{}}_{\bar{A}}^{A_{1}^{}}_{\bar{A}}^{\tensor*[^{1}]{\hspace{-0.2em}A}{}_{1}^{}}}$ &  $- 4.6356$ &                                                    $d\indices*{*_{X}^{\tensor*[^{1}]{\hspace{-0.2em}E}{}_{u}^{}}_{\bar{A}}^{A_{1}^{}}_{\bar{A}}^{\tensor*[^{1}]{\hspace{-0.2em}A}{}_{1}^{}}}$ &    $4.2803$ &                                                                                    $d\indices*{*_{X}^{A_{1g}^{}}_{\bar{A}}^{A_{1}^{}}_{\bar{A}}^{\tensor*[^{2}]{\hspace{-0.2em}A}{}_{1}^{}}}$ &  $- 1.1438$ &                                                                                     $d\indices*{*_{X}^{E_{u}^{}}_{\bar{A}}^{A_{1}^{}}_{\bar{A}}^{\tensor*[^{2}]{\hspace{-0.2em}A}{}_{1}^{}}}$ &   $20.1127$ \\[0.4em]
                                                     $d\indices*{*_{X}^{\tensor*[^{1}]{\hspace{-0.2em}E}{}_{u}^{}}_{\bar{A}}^{A_{1}^{}}_{\bar{A}}^{\tensor*[^{2}]{\hspace{-0.2em}A}{}_{1}^{}}}$ & $- 29.6225$ &                                                                                                                      $d\indices*{*_{X}^{E_{g}^{}}_{\bar{A}}^{A_{1}^{}}_{\bar{A}}^{A_{2}^{}}}$ &  $- 2.2750$ &                                                                                                                      $d\indices*{*_{X}^{E_{g}^{}}_{\bar{A}}^{A_{1}^{}}_{\bar{A}}^{B_{1}^{}}}$ &    $3.1350$ &                                                                                                                     $d\indices*{*_{X}^{A_{2u}^{}}_{\bar{A}}^{A_{1}^{}}_{\bar{A}}^{B_{1}^{}}}$ &   $27.7794$ &                                                                                                                     $d\indices*{*_{X}^{B_{1u}^{}}_{\bar{A}}^{A_{1}^{}}_{\bar{A}}^{B_{1}^{}}}$ &   $14.5275$ \\[0.4em]
                                                                                      $d\indices*{*_{X}^{E_{g}^{}}_{\bar{A}}^{A_{1}^{}}_{\bar{A}}^{\tensor*[^{1}]{\hspace{-0.2em}B}{}_{1}^{}}}$ &   $18.7548$ &                                                                                    $d\indices*{*_{X}^{A_{2u}^{}}_{\bar{A}}^{A_{1}^{}}_{\bar{A}}^{\tensor*[^{1}]{\hspace{-0.2em}B}{}_{1}^{}}}$ &   $25.0435$ &                                                                                    $d\indices*{*_{X}^{B_{1u}^{}}_{\bar{A}}^{A_{1}^{}}_{\bar{A}}^{\tensor*[^{1}]{\hspace{-0.2em}B}{}_{1}^{}}}$ &    $1.3933$ &                                                                                     $d\indices*{*_{X}^{E_{g}^{}}_{\bar{A}}^{A_{1}^{}}_{\bar{A}}^{\tensor*[^{2}]{\hspace{-0.2em}B}{}_{1}^{}}}$ &   $20.9367$ &                                                                                    $d\indices*{*_{X}^{A_{2u}^{}}_{\bar{A}}^{A_{1}^{}}_{\bar{A}}^{\tensor*[^{2}]{\hspace{-0.2em}B}{}_{1}^{}}}$ &  $- 1.4935$ \\[0.4em]
                                                                                     $d\indices*{*_{X}^{B_{1u}^{}}_{\bar{A}}^{A_{1}^{}}_{\bar{A}}^{\tensor*[^{2}]{\hspace{-0.2em}B}{}_{1}^{}}}$ &  $- 5.1039$ &                                                                                                                      $d\indices*{*_{X}^{E_{u}^{}}_{\bar{A}}^{A_{1}^{}}_{\bar{A}}^{B_{2}^{}}}$ &    $1.4424$ &                                                                                     $d\indices*{*_{X}^{\tensor*[^{1}]{\hspace{-0.2em}E}{}_{u}^{}}_{\bar{A}}^{A_{1}^{}}_{\bar{A}}^{B_{2}^{}}}$ & $- 12.6584$ &                                                                                     $d\indices*{*_{X}^{E_{u}^{}}_{\bar{A}}^{A_{1}^{}}_{\bar{A}}^{\tensor*[^{1}]{\hspace{-0.2em}B}{}_{2}^{}}}$ &  $- 1.0336$ &                                                    $d\indices*{*_{X}^{\tensor*[^{1}]{\hspace{-0.2em}E}{}_{u}^{}}_{\bar{A}}^{A_{1}^{}}_{\bar{A}}^{\tensor*[^{1}]{\hspace{-0.2em}B}{}_{2}^{}}}$ &    $4.1607$ \\[0.4em]
                                                    $d\indices*{*_{X}^{A_{1g}^{}}_{\bar{A}}^{\tensor*[^{1}]{\hspace{-0.2em}A}{}_{1}^{}}_{\bar{A}}^{\tensor*[^{1}]{\hspace{-0.2em}A}{}_{1}^{}}}$ &  $- 0.9701$ &                                                    $d\indices*{*_{X}^{E_{u}^{}}_{\bar{A}}^{\tensor*[^{1}]{\hspace{-0.2em}A}{}_{1}^{}}_{\bar{A}}^{\tensor*[^{1}]{\hspace{-0.2em}A}{}_{1}^{}}}$ &    $4.5649$ &                   $d\indices*{*_{X}^{\tensor*[^{1}]{\hspace{-0.2em}E}{}_{u}^{}}_{\bar{A}}^{\tensor*[^{1}]{\hspace{-0.2em}A}{}_{1}^{}}_{\bar{A}}^{\tensor*[^{1}]{\hspace{-0.2em}A}{}_{1}^{}}}$ &  $- 5.6848$ &                                                   $d\indices*{*_{X}^{A_{1g}^{}}_{\bar{A}}^{\tensor*[^{1}]{\hspace{-0.2em}A}{}_{1}^{}}_{\bar{A}}^{\tensor*[^{2}]{\hspace{-0.2em}A}{}_{1}^{}}}$ &    $0.7161$ &                                                    $d\indices*{*_{X}^{E_{u}^{}}_{\bar{A}}^{\tensor*[^{1}]{\hspace{-0.2em}A}{}_{1}^{}}_{\bar{A}}^{\tensor*[^{2}]{\hspace{-0.2em}A}{}_{1}^{}}}$ & $- 10.9836$ \\[0.4em]
                    $d\indices*{*_{X}^{\tensor*[^{1}]{\hspace{-0.2em}E}{}_{u}^{}}_{\bar{A}}^{\tensor*[^{1}]{\hspace{-0.2em}A}{}_{1}^{}}_{\bar{A}}^{\tensor*[^{2}]{\hspace{-0.2em}A}{}_{1}^{}}}$ &   $12.2275$ &                                                                                     $d\indices*{*_{X}^{E_{g}^{}}_{\bar{A}}^{\tensor*[^{1}]{\hspace{-0.2em}A}{}_{1}^{}}_{\bar{A}}^{A_{2}^{}}}$ & $- 10.0001$ &                                                                                     $d\indices*{*_{X}^{E_{g}^{}}_{\bar{A}}^{\tensor*[^{1}]{\hspace{-0.2em}A}{}_{1}^{}}_{\bar{A}}^{B_{1}^{}}}$ &  $- 3.2320$ &                                                                                    $d\indices*{*_{X}^{A_{2u}^{}}_{\bar{A}}^{\tensor*[^{1}]{\hspace{-0.2em}A}{}_{1}^{}}_{\bar{A}}^{B_{1}^{}}}$ & $- 22.9783$ &                                                                                    $d\indices*{*_{X}^{B_{1u}^{}}_{\bar{A}}^{\tensor*[^{1}]{\hspace{-0.2em}A}{}_{1}^{}}_{\bar{A}}^{B_{1}^{}}}$ &    $1.0784$ \\[0.4em]
                                                     $d\indices*{*_{X}^{E_{g}^{}}_{\bar{A}}^{\tensor*[^{1}]{\hspace{-0.2em}A}{}_{1}^{}}_{\bar{A}}^{\tensor*[^{1}]{\hspace{-0.2em}B}{}_{1}^{}}}$ &  $- 6.3860$ &                                                   $d\indices*{*_{X}^{A_{2u}^{}}_{\bar{A}}^{\tensor*[^{1}]{\hspace{-0.2em}A}{}_{1}^{}}_{\bar{A}}^{\tensor*[^{1}]{\hspace{-0.2em}B}{}_{1}^{}}}$ &   $13.6274$ &                                                   $d\indices*{*_{X}^{B_{1u}^{}}_{\bar{A}}^{\tensor*[^{1}]{\hspace{-0.2em}A}{}_{1}^{}}_{\bar{A}}^{\tensor*[^{1}]{\hspace{-0.2em}B}{}_{1}^{}}}$ &  $- 8.2127$ &                                                    $d\indices*{*_{X}^{E_{g}^{}}_{\bar{A}}^{\tensor*[^{1}]{\hspace{-0.2em}A}{}_{1}^{}}_{\bar{A}}^{\tensor*[^{2}]{\hspace{-0.2em}B}{}_{1}^{}}}$ & $- 10.0059$ &                                                   $d\indices*{*_{X}^{A_{2u}^{}}_{\bar{A}}^{\tensor*[^{1}]{\hspace{-0.2em}A}{}_{1}^{}}_{\bar{A}}^{\tensor*[^{2}]{\hspace{-0.2em}B}{}_{1}^{}}}$ &    $4.0269$ \\[0.4em]
                                                    $d\indices*{*_{X}^{B_{1u}^{}}_{\bar{A}}^{\tensor*[^{1}]{\hspace{-0.2em}A}{}_{1}^{}}_{\bar{A}}^{\tensor*[^{2}]{\hspace{-0.2em}B}{}_{1}^{}}}$ &    $8.2616$ &                                                                                     $d\indices*{*_{X}^{E_{u}^{}}_{\bar{A}}^{\tensor*[^{1}]{\hspace{-0.2em}A}{}_{1}^{}}_{\bar{A}}^{B_{2}^{}}}$ &   $11.4487$ &                                                    $d\indices*{*_{X}^{\tensor*[^{1}]{\hspace{-0.2em}E}{}_{u}^{}}_{\bar{A}}^{\tensor*[^{1}]{\hspace{-0.2em}A}{}_{1}^{}}_{\bar{A}}^{B_{2}^{}}}$ & $- 14.6822$ &                                                    $d\indices*{*_{X}^{E_{u}^{}}_{\bar{A}}^{\tensor*[^{1}]{\hspace{-0.2em}A}{}_{1}^{}}_{\bar{A}}^{\tensor*[^{1}]{\hspace{-0.2em}B}{}_{2}^{}}}$ &  $- 7.0751$ &                   $d\indices*{*_{X}^{\tensor*[^{1}]{\hspace{-0.2em}E}{}_{u}^{}}_{\bar{A}}^{\tensor*[^{1}]{\hspace{-0.2em}A}{}_{1}^{}}_{\bar{A}}^{\tensor*[^{1}]{\hspace{-0.2em}B}{}_{2}^{}}}$ &   $10.5050$ \\[0.4em]
                                                    $d\indices*{*_{X}^{A_{1g}^{}}_{\bar{A}}^{\tensor*[^{2}]{\hspace{-0.2em}A}{}_{1}^{}}_{\bar{A}}^{\tensor*[^{2}]{\hspace{-0.2em}A}{}_{1}^{}}}$ & $- 21.8122$ &                                                    $d\indices*{*_{X}^{E_{u}^{}}_{\bar{A}}^{\tensor*[^{2}]{\hspace{-0.2em}A}{}_{1}^{}}_{\bar{A}}^{\tensor*[^{2}]{\hspace{-0.2em}A}{}_{1}^{}}}$ &    $1.1935$ &                   $d\indices*{*_{X}^{\tensor*[^{1}]{\hspace{-0.2em}E}{}_{u}^{}}_{\bar{A}}^{\tensor*[^{2}]{\hspace{-0.2em}A}{}_{1}^{}}_{\bar{A}}^{\tensor*[^{2}]{\hspace{-0.2em}A}{}_{1}^{}}}$ &  $- 2.7581$ &                                                                                     $d\indices*{*_{X}^{E_{g}^{}}_{\bar{A}}^{\tensor*[^{2}]{\hspace{-0.2em}A}{}_{1}^{}}_{\bar{A}}^{A_{2}^{}}}$ &    $2.3228$ &                                                                                     $d\indices*{*_{X}^{E_{g}^{}}_{\bar{A}}^{\tensor*[^{2}]{\hspace{-0.2em}A}{}_{1}^{}}_{\bar{A}}^{B_{1}^{}}}$ &   $12.4345$ \\[0.4em]
                                                                                     $d\indices*{*_{X}^{A_{2u}^{}}_{\bar{A}}^{\tensor*[^{2}]{\hspace{-0.2em}A}{}_{1}^{}}_{\bar{A}}^{B_{1}^{}}}$ &    $0.3564$ &                                                                                    $d\indices*{*_{X}^{B_{1u}^{}}_{\bar{A}}^{\tensor*[^{2}]{\hspace{-0.2em}A}{}_{1}^{}}_{\bar{A}}^{B_{1}^{}}}$ &    $3.4445$ &                                                    $d\indices*{*_{X}^{E_{g}^{}}_{\bar{A}}^{\tensor*[^{2}]{\hspace{-0.2em}A}{}_{1}^{}}_{\bar{A}}^{\tensor*[^{1}]{\hspace{-0.2em}B}{}_{1}^{}}}$ &    $4.4445$ &                                                   $d\indices*{*_{X}^{A_{2u}^{}}_{\bar{A}}^{\tensor*[^{2}]{\hspace{-0.2em}A}{}_{1}^{}}_{\bar{A}}^{\tensor*[^{1}]{\hspace{-0.2em}B}{}_{1}^{}}}$ & $- 19.5033$ &                                                   $d\indices*{*_{X}^{B_{1u}^{}}_{\bar{A}}^{\tensor*[^{2}]{\hspace{-0.2em}A}{}_{1}^{}}_{\bar{A}}^{\tensor*[^{1}]{\hspace{-0.2em}B}{}_{1}^{}}}$ & $- 12.3097$ \\[0.4em]
                                                     $d\indices*{*_{X}^{E_{g}^{}}_{\bar{A}}^{\tensor*[^{2}]{\hspace{-0.2em}A}{}_{1}^{}}_{\bar{A}}^{\tensor*[^{2}]{\hspace{-0.2em}B}{}_{1}^{}}}$ &    $3.6263$ &                                                   $d\indices*{*_{X}^{A_{2u}^{}}_{\bar{A}}^{\tensor*[^{2}]{\hspace{-0.2em}A}{}_{1}^{}}_{\bar{A}}^{\tensor*[^{2}]{\hspace{-0.2em}B}{}_{1}^{}}}$ & $- 20.0455$ &                                                   $d\indices*{*_{X}^{B_{1u}^{}}_{\bar{A}}^{\tensor*[^{2}]{\hspace{-0.2em}A}{}_{1}^{}}_{\bar{A}}^{\tensor*[^{2}]{\hspace{-0.2em}B}{}_{1}^{}}}$ & $- 16.7219$ &                                                                                     $d\indices*{*_{X}^{E_{u}^{}}_{\bar{A}}^{\tensor*[^{2}]{\hspace{-0.2em}A}{}_{1}^{}}_{\bar{A}}^{B_{2}^{}}}$ &  $- 4.9510$ &                                                    $d\indices*{*_{X}^{\tensor*[^{1}]{\hspace{-0.2em}E}{}_{u}^{}}_{\bar{A}}^{\tensor*[^{2}]{\hspace{-0.2em}A}{}_{1}^{}}_{\bar{A}}^{B_{2}^{}}}$ &  $- 0.7645$ \\[0.4em]
                                                     $d\indices*{*_{X}^{E_{u}^{}}_{\bar{A}}^{\tensor*[^{2}]{\hspace{-0.2em}A}{}_{1}^{}}_{\bar{A}}^{\tensor*[^{1}]{\hspace{-0.2em}B}{}_{2}^{}}}$ &   $10.0150$ &                   $d\indices*{*_{X}^{\tensor*[^{1}]{\hspace{-0.2em}E}{}_{u}^{}}_{\bar{A}}^{\tensor*[^{2}]{\hspace{-0.2em}A}{}_{1}^{}}_{\bar{A}}^{\tensor*[^{1}]{\hspace{-0.2em}B}{}_{2}^{}}}$ &  $- 2.4191$ &                                                                                                                     $d\indices*{*_{X}^{A_{1g}^{}}_{\bar{A}}^{A_{2}^{}}_{\bar{A}}^{A_{2}^{}}}$ &   $18.4673$ &                                                                                                                      $d\indices*{*_{X}^{E_{u}^{}}_{\bar{A}}^{A_{2}^{}}_{\bar{A}}^{A_{2}^{}}}$ &    $9.7115$ &                                                                                     $d\indices*{*_{X}^{\tensor*[^{1}]{\hspace{-0.2em}E}{}_{u}^{}}_{\bar{A}}^{A_{2}^{}}_{\bar{A}}^{A_{2}^{}}}$ &  $- 0.6209$ \\[0.4em]
                                                                                                                       $d\indices*{*_{X}^{E_{u}^{}}_{\bar{A}}^{A_{2}^{}}_{\bar{A}}^{B_{1}^{}}}$ &    $7.9422$ &                                                                                     $d\indices*{*_{X}^{\tensor*[^{1}]{\hspace{-0.2em}E}{}_{u}^{}}_{\bar{A}}^{A_{2}^{}}_{\bar{A}}^{B_{1}^{}}}$ &  $- 9.2651$ &                                                                                     $d\indices*{*_{X}^{E_{u}^{}}_{\bar{A}}^{A_{2}^{}}_{\bar{A}}^{\tensor*[^{1}]{\hspace{-0.2em}B}{}_{1}^{}}}$ & $- 11.0525$ &                                                    $d\indices*{*_{X}^{\tensor*[^{1}]{\hspace{-0.2em}E}{}_{u}^{}}_{\bar{A}}^{A_{2}^{}}_{\bar{A}}^{\tensor*[^{1}]{\hspace{-0.2em}B}{}_{1}^{}}}$ &   $15.0396$ &                                                                                     $d\indices*{*_{X}^{E_{u}^{}}_{\bar{A}}^{A_{2}^{}}_{\bar{A}}^{\tensor*[^{2}]{\hspace{-0.2em}B}{}_{1}^{}}}$ &  $- 8.7293$ \\[0.4em]
                                                     $d\indices*{*_{X}^{\tensor*[^{1}]{\hspace{-0.2em}E}{}_{u}^{}}_{\bar{A}}^{A_{2}^{}}_{\bar{A}}^{\tensor*[^{2}]{\hspace{-0.2em}B}{}_{1}^{}}}$ &  $- 1.6811$ &                                                                                                                      $d\indices*{*_{X}^{E_{g}^{}}_{\bar{A}}^{A_{2}^{}}_{\bar{A}}^{B_{2}^{}}}$ &    $1.1574$ &                                                                                                                     $d\indices*{*_{X}^{A_{2u}^{}}_{\bar{A}}^{A_{2}^{}}_{\bar{A}}^{B_{2}^{}}}$ &   $29.1685$ &                                                                                                                     $d\indices*{*_{X}^{B_{1u}^{}}_{\bar{A}}^{A_{2}^{}}_{\bar{A}}^{B_{2}^{}}}$ & $- 14.7607$ &                                                                                     $d\indices*{*_{X}^{E_{g}^{}}_{\bar{A}}^{A_{2}^{}}_{\bar{A}}^{\tensor*[^{1}]{\hspace{-0.2em}B}{}_{2}^{}}}$ &    $0.1197$ \\[0.4em]
                                                                                     $d\indices*{*_{X}^{A_{2u}^{}}_{\bar{A}}^{A_{2}^{}}_{\bar{A}}^{\tensor*[^{1}]{\hspace{-0.2em}B}{}_{2}^{}}}$ & $- 15.3408$ &                                                                                    $d\indices*{*_{X}^{B_{1u}^{}}_{\bar{A}}^{A_{2}^{}}_{\bar{A}}^{\tensor*[^{1}]{\hspace{-0.2em}B}{}_{2}^{}}}$ &   $11.2374$ &                                                                                                                     $d\indices*{*_{X}^{A_{1g}^{}}_{\bar{A}}^{B_{1}^{}}_{\bar{A}}^{B_{1}^{}}}$ &    $0.2678$ &                                                                                                                      $d\indices*{*_{X}^{E_{u}^{}}_{\bar{A}}^{B_{1}^{}}_{\bar{A}}^{B_{1}^{}}}$ &  $- 2.0125$ &                                                                                     $d\indices*{*_{X}^{\tensor*[^{1}]{\hspace{-0.2em}E}{}_{u}^{}}_{\bar{A}}^{B_{1}^{}}_{\bar{A}}^{B_{1}^{}}}$ &  $- 0.0455$ \\[0.4em]
                                                                                     $d\indices*{*_{X}^{A_{1g}^{}}_{\bar{A}}^{B_{1}^{}}_{\bar{A}}^{\tensor*[^{1}]{\hspace{-0.2em}B}{}_{1}^{}}}$ &    $9.6101$ &                                                                                     $d\indices*{*_{X}^{E_{u}^{}}_{\bar{A}}^{B_{1}^{}}_{\bar{A}}^{\tensor*[^{1}]{\hspace{-0.2em}B}{}_{1}^{}}}$ &    $6.0941$ &                                                    $d\indices*{*_{X}^{\tensor*[^{1}]{\hspace{-0.2em}E}{}_{u}^{}}_{\bar{A}}^{B_{1}^{}}_{\bar{A}}^{\tensor*[^{1}]{\hspace{-0.2em}B}{}_{1}^{}}}$ &  $- 9.6875$ &                                                                                    $d\indices*{*_{X}^{A_{1g}^{}}_{\bar{A}}^{B_{1}^{}}_{\bar{A}}^{\tensor*[^{2}]{\hspace{-0.2em}B}{}_{1}^{}}}$ &  $- 2.0386$ &                                                                                     $d\indices*{*_{X}^{E_{u}^{}}_{\bar{A}}^{B_{1}^{}}_{\bar{A}}^{\tensor*[^{2}]{\hspace{-0.2em}B}{}_{1}^{}}}$ &    $7.4549$ \\[0.4em]
                                                     $d\indices*{*_{X}^{\tensor*[^{1}]{\hspace{-0.2em}E}{}_{u}^{}}_{\bar{A}}^{B_{1}^{}}_{\bar{A}}^{\tensor*[^{2}]{\hspace{-0.2em}B}{}_{1}^{}}}$ &  $- 8.4798$ &                                                                                                                      $d\indices*{*_{X}^{E_{g}^{}}_{\bar{A}}^{B_{2}^{}}_{\bar{A}}^{B_{1}^{}}}$ & $- 10.0550$ &                                                                                     $d\indices*{*_{X}^{E_{g}^{}}_{\bar{A}}^{\tensor*[^{1}]{\hspace{-0.2em}B}{}_{2}^{}}_{\bar{A}}^{B_{1}^{}}}$ &    $7.2478$ &                                                   $d\indices*{*_{X}^{A_{1g}^{}}_{\bar{A}}^{\tensor*[^{1}]{\hspace{-0.2em}B}{}_{1}^{}}_{\bar{A}}^{\tensor*[^{1}]{\hspace{-0.2em}B}{}_{1}^{}}}$ & $- 30.5331$ &                                                    $d\indices*{*_{X}^{E_{u}^{}}_{\bar{A}}^{\tensor*[^{1}]{\hspace{-0.2em}B}{}_{1}^{}}_{\bar{A}}^{\tensor*[^{1}]{\hspace{-0.2em}B}{}_{1}^{}}}$ &    $3.3593$ \\[0.4em]
                    $d\indices*{*_{X}^{\tensor*[^{1}]{\hspace{-0.2em}E}{}_{u}^{}}_{\bar{A}}^{\tensor*[^{1}]{\hspace{-0.2em}B}{}_{1}^{}}_{\bar{A}}^{\tensor*[^{1}]{\hspace{-0.2em}B}{}_{1}^{}}}$ & $- 23.1531$ &                                                   $d\indices*{*_{X}^{A_{1g}^{}}_{\bar{A}}^{\tensor*[^{1}]{\hspace{-0.2em}B}{}_{1}^{}}_{\bar{A}}^{\tensor*[^{2}]{\hspace{-0.2em}B}{}_{1}^{}}}$ & $- 16.4836$ &                                                    $d\indices*{*_{X}^{E_{u}^{}}_{\bar{A}}^{\tensor*[^{1}]{\hspace{-0.2em}B}{}_{1}^{}}_{\bar{A}}^{\tensor*[^{2}]{\hspace{-0.2em}B}{}_{1}^{}}}$ &  $- 0.2145$ &                   $d\indices*{*_{X}^{\tensor*[^{1}]{\hspace{-0.2em}E}{}_{u}^{}}_{\bar{A}}^{\tensor*[^{1}]{\hspace{-0.2em}B}{}_{1}^{}}_{\bar{A}}^{\tensor*[^{2}]{\hspace{-0.2em}B}{}_{1}^{}}}$ &  $- 0.1635$ &                                                                                     $d\indices*{*_{X}^{E_{g}^{}}_{\bar{A}}^{B_{2}^{}}_{\bar{A}}^{\tensor*[^{1}]{\hspace{-0.2em}B}{}_{1}^{}}}$ &   $16.6913$ \\[0.4em]
                                                     $d\indices*{*_{X}^{E_{g}^{}}_{\bar{A}}^{\tensor*[^{1}]{\hspace{-0.2em}B}{}_{2}^{}}_{\bar{A}}^{\tensor*[^{1}]{\hspace{-0.2em}B}{}_{1}^{}}}$ & $- 11.5214$ &                                                   $d\indices*{*_{X}^{A_{1g}^{}}_{\bar{A}}^{\tensor*[^{2}]{\hspace{-0.2em}B}{}_{1}^{}}_{\bar{A}}^{\tensor*[^{2}]{\hspace{-0.2em}B}{}_{1}^{}}}$ & $- 16.1055$ &                                                    $d\indices*{*_{X}^{E_{u}^{}}_{\bar{A}}^{\tensor*[^{2}]{\hspace{-0.2em}B}{}_{1}^{}}_{\bar{A}}^{\tensor*[^{2}]{\hspace{-0.2em}B}{}_{1}^{}}}$ &    $4.1140$ &                   $d\indices*{*_{X}^{\tensor*[^{1}]{\hspace{-0.2em}E}{}_{u}^{}}_{\bar{A}}^{\tensor*[^{2}]{\hspace{-0.2em}B}{}_{1}^{}}_{\bar{A}}^{\tensor*[^{2}]{\hspace{-0.2em}B}{}_{1}^{}}}$ &    $2.9131$ &                                                                                     $d\indices*{*_{X}^{E_{g}^{}}_{\bar{A}}^{B_{2}^{}}_{\bar{A}}^{\tensor*[^{2}]{\hspace{-0.2em}B}{}_{1}^{}}}$ &  $- 2.2384$ \\[0.4em]
                                                     $d\indices*{*_{X}^{E_{g}^{}}_{\bar{A}}^{\tensor*[^{1}]{\hspace{-0.2em}B}{}_{2}^{}}_{\bar{A}}^{\tensor*[^{2}]{\hspace{-0.2em}B}{}_{1}^{}}}$ &    $1.2708$ &                                                                                                                     $d\indices*{*_{X}^{A_{1g}^{}}_{\bar{A}}^{B_{2}^{}}_{\bar{A}}^{B_{2}^{}}}$ &   $25.5609$ &                                                                                                                      $d\indices*{*_{X}^{E_{u}^{}}_{\bar{A}}^{B_{2}^{}}_{\bar{A}}^{B_{2}^{}}}$ &    $1.4006$ &                                                                                     $d\indices*{*_{X}^{\tensor*[^{1}]{\hspace{-0.2em}E}{}_{u}^{}}_{\bar{A}}^{B_{2}^{}}_{\bar{A}}^{B_{2}^{}}}$ &    $5.3453$ &                                                                                    $d\indices*{*_{X}^{A_{1g}^{}}_{\bar{A}}^{B_{2}^{}}_{\bar{A}}^{\tensor*[^{1}]{\hspace{-0.2em}B}{}_{2}^{}}}$ & $- 20.1300$ \\[0.4em]
                                                                                      $d\indices*{*_{X}^{E_{u}^{}}_{\bar{A}}^{B_{2}^{}}_{\bar{A}}^{\tensor*[^{1}]{\hspace{-0.2em}B}{}_{2}^{}}}$ &    $0.2867$ &                                                    $d\indices*{*_{X}^{\tensor*[^{1}]{\hspace{-0.2em}E}{}_{u}^{}}_{\bar{A}}^{B_{2}^{}}_{\bar{A}}^{\tensor*[^{1}]{\hspace{-0.2em}B}{}_{2}^{}}}$ &    $2.2333$ &                                                   $d\indices*{*_{X}^{A_{1g}^{}}_{\bar{A}}^{\tensor*[^{1}]{\hspace{-0.2em}B}{}_{2}^{}}_{\bar{A}}^{\tensor*[^{1}]{\hspace{-0.2em}B}{}_{2}^{}}}$ &   $14.8389$ &                                                    $d\indices*{*_{X}^{E_{u}^{}}_{\bar{A}}^{\tensor*[^{1}]{\hspace{-0.2em}B}{}_{2}^{}}_{\bar{A}}^{\tensor*[^{1}]{\hspace{-0.2em}B}{}_{2}^{}}}$ & $- 10.0671$ &                   $d\indices*{*_{X}^{\tensor*[^{1}]{\hspace{-0.2em}E}{}_{u}^{}}_{\bar{A}}^{\tensor*[^{1}]{\hspace{-0.2em}B}{}_{2}^{}}_{\bar{A}}^{\tensor*[^{1}]{\hspace{-0.2em}B}{}_{2}^{}}}$ &  $- 0.4299$ \\[0.4em]
                                                                                                                             $d\indices*{*_{\Gamma}^{T_{2g}^{}}_{X}^{E_{g}^{}}_{X}^{E_{g}^{}}}$ &   $26.1873$ &                                                                                                                          $d\indices*{*_{\Gamma}^{T_{2g}^{}}_{X}^{B_{1u}^{}}_{X}^{A_{2u}^{}}}$ &   $27.0755$ &                                                                                                                            $d\indices*{*_{\Gamma}^{T_{2g}^{}}_{X}^{E_{u}^{}}_{X}^{E_{u}^{}}}$ & $- 19.5586$ &                                                                                           $d\indices*{*_{\Gamma}^{T_{2g}^{}}_{X}^{E_{u}^{}}_{X}^{\tensor*[^{1}]{\hspace{-0.2em}E}{}_{u}^{}}}$ &   $11.8258$ &                                                          $d\indices*{*_{\Gamma}^{T_{2g}^{}}_{X}^{\tensor*[^{1}]{\hspace{-0.2em}E}{}_{u}^{}}_{X}^{\tensor*[^{1}]{\hspace{-0.2em}E}{}_{u}^{}}}$ & $- 37.3082$ \\[0.4em]
                                                                                                                            $d\indices*{*_{\Gamma}^{T_{2g}^{}}_{X}^{A_{1g}^{}}_{X}^{E_{g}^{}}}$ & $- 18.5250$ &                                                                                                                           $d\indices*{*_{\Gamma}^{T_{2g}^{}}_{X}^{A_{2u}^{}}_{X}^{E_{u}^{}}}$ &   $12.6099$ &                                                                                          $d\indices*{*_{\Gamma}^{T_{2g}^{}}_{X}^{A_{2u}^{}}_{X}^{\tensor*[^{1}]{\hspace{-0.2em}E}{}_{u}^{}}}$ & $- 28.7389$ &                                                                                                                           $d\indices*{*_{\Gamma}^{T_{2g}^{}}_{X}^{B_{1u}^{}}_{X}^{E_{u}^{}}}$ & $- 11.3804$ &                                                                                          $d\indices*{*_{\Gamma}^{T_{2g}^{}}_{X}^{B_{1u}^{}}_{X}^{\tensor*[^{1}]{\hspace{-0.2em}E}{}_{u}^{}}}$ &   $13.7967$ \\[0.4em]
                                                                                                                            $d\indices*{*_{\Gamma}^{T_{1u}^{}}_{X}^{A_{1g}^{}}_{X}^{E_{u}^{}}}$ & $- 27.4924$ &                                                                                          $d\indices*{*_{\Gamma}^{T_{1u}^{}}_{X}^{A_{1g}^{}}_{X}^{\tensor*[^{1}]{\hspace{-0.2em}E}{}_{u}^{}}}$ &   $30.3782$ &                                                                                                                           $d\indices*{*_{\Gamma}^{T_{1u}^{}}_{X}^{B_{1u}^{}}_{X}^{E_{g}^{}}}$ & $- 18.5015$ &                                                                                                                           $d\indices*{*_{\Gamma}^{T_{1u}^{}}_{X}^{E_{g}^{}}_{X}^{A_{2u}^{}}}$ &   $35.0379$ &                                                                                                                          $d\indices*{*_{\Gamma}^{T_{1u}^{}}_{X}^{A_{1g}^{}}_{X}^{A_{2u}^{}}}$ &   $35.8558$ \\[0.4em]
                                                                                                                             $d\indices*{*_{\Gamma}^{T_{1u}^{}}_{X}^{E_{g}^{}}_{X}^{E_{u}^{}}}$ & $- 17.2141$ &                                                                                           $d\indices*{*_{\Gamma}^{T_{1u}^{}}_{X}^{E_{g}^{}}_{X}^{\tensor*[^{1}]{\hspace{-0.2em}E}{}_{u}^{}}}$ &   $21.1426$ &                                                                                                                      $d\indices*{*_{\Gamma}^{T_{2g}^{}}_{A}^{A_{1}^{}}_{\bar{A}}^{A_{1}^{}}}$ &   $32.3889$ &                                                                                     $d\indices*{*_{\Gamma}^{T_{2g}^{}}_{A}^{A_{1}^{}}_{\bar{A}}^{\tensor*[^{1}]{\hspace{-0.2em}A}{}_{1}^{}}}$ &  $- 7.7883$ &                                                                                     $d\indices*{*_{\Gamma}^{T_{2g}^{}}_{A}^{A_{1}^{}}_{\bar{A}}^{\tensor*[^{2}]{\hspace{-0.2em}A}{}_{1}^{}}}$ &    $6.4936$ \\[0.4em]
                                                     $d\indices*{*_{\Gamma}^{T_{2g}^{}}_{A}^{\tensor*[^{1}]{\hspace{-0.2em}A}{}_{1}^{}}_{\bar{A}}^{\tensor*[^{1}]{\hspace{-0.2em}A}{}_{1}^{}}}$ &  $- 1.8510$ &                                                    $d\indices*{*_{\Gamma}^{T_{2g}^{}}_{A}^{\tensor*[^{1}]{\hspace{-0.2em}A}{}_{1}^{}}_{\bar{A}}^{\tensor*[^{2}]{\hspace{-0.2em}A}{}_{1}^{}}}$ &  $- 4.2579$ &                                                    $d\indices*{*_{\Gamma}^{T_{2g}^{}}_{A}^{\tensor*[^{2}]{\hspace{-0.2em}A}{}_{1}^{}}_{\bar{A}}^{\tensor*[^{2}]{\hspace{-0.2em}A}{}_{1}^{}}}$ &   $19.6028$ &                                                                                                                      $d\indices*{*_{\Gamma}^{T_{2g}^{}}_{A}^{A_{2}^{}}_{\bar{A}}^{A_{2}^{}}}$ & $- 12.5702$ &                                                                                                                      $d\indices*{*_{\Gamma}^{T_{2g}^{}}_{A}^{B_{1}^{}}_{\bar{A}}^{B_{1}^{}}}$ &  $- 0.4749$ \\[0.4em]
                                                                                      $d\indices*{*_{\Gamma}^{T_{2g}^{}}_{A}^{B_{1}^{}}_{\bar{A}}^{\tensor*[^{1}]{\hspace{-0.2em}B}{}_{1}^{}}}$ &   $10.6102$ &                                                                                     $d\indices*{*_{\Gamma}^{T_{2g}^{}}_{A}^{B_{1}^{}}_{\bar{A}}^{\tensor*[^{2}]{\hspace{-0.2em}B}{}_{1}^{}}}$ &  $- 4.8087$ &                                                    $d\indices*{*_{\Gamma}^{T_{2g}^{}}_{A}^{\tensor*[^{1}]{\hspace{-0.2em}B}{}_{1}^{}}_{\bar{A}}^{\tensor*[^{1}]{\hspace{-0.2em}B}{}_{1}^{}}}$ &  $- 3.0863$ &                                                    $d\indices*{*_{\Gamma}^{T_{2g}^{}}_{A}^{\tensor*[^{1}]{\hspace{-0.2em}B}{}_{1}^{}}_{\bar{A}}^{\tensor*[^{2}]{\hspace{-0.2em}B}{}_{1}^{}}}$ &   $10.3969$ &                                                    $d\indices*{*_{\Gamma}^{T_{2g}^{}}_{A}^{\tensor*[^{2}]{\hspace{-0.2em}B}{}_{1}^{}}_{\bar{A}}^{\tensor*[^{2}]{\hspace{-0.2em}B}{}_{1}^{}}}$ &   $11.2986$ \\[0.4em]
                                                                                                                       $d\indices*{*_{\Gamma}^{T_{2g}^{}}_{A}^{B_{2}^{}}_{\bar{A}}^{B_{2}^{}}}$ & $- 19.7557$ &                                                                                     $d\indices*{*_{\Gamma}^{T_{2g}^{}}_{A}^{B_{2}^{}}_{\bar{A}}^{\tensor*[^{1}]{\hspace{-0.2em}B}{}_{2}^{}}}$ &   $13.1319$ &                                                    $d\indices*{*_{\Gamma}^{T_{2g}^{}}_{A}^{\tensor*[^{1}]{\hspace{-0.2em}B}{}_{2}^{}}_{\bar{A}}^{\tensor*[^{1}]{\hspace{-0.2em}B}{}_{2}^{}}}$ & $- 10.1909$ &                                                                                                                      $d\indices*{*_{\Gamma}^{T_{2g}^{}}_{A}^{A_{1}^{}}_{\bar{A}}^{A_{2}^{}}}$ &    $2.7025$ &                                                                                                                      $d\indices*{*_{\Gamma}^{T_{2g}^{}}_{A}^{A_{1}^{}}_{\bar{A}}^{B_{1}^{}}}$ &    $2.3970$ \\[0.4em]
                                                                                      $d\indices*{*_{\Gamma}^{T_{2g}^{}}_{A}^{A_{1}^{}}_{\bar{A}}^{\tensor*[^{1}]{\hspace{-0.2em}B}{}_{1}^{}}}$ & $- 22.1145$ &                                                                                     $d\indices*{*_{\Gamma}^{T_{2g}^{}}_{A}^{A_{1}^{}}_{\bar{A}}^{\tensor*[^{2}]{\hspace{-0.2em}B}{}_{1}^{}}}$ & $- 17.5722$ &                                                                                     $d\indices*{*_{\Gamma}^{T_{2g}^{}}_{A}^{\tensor*[^{1}]{\hspace{-0.2em}A}{}_{1}^{}}_{\bar{A}}^{A_{2}^{}}}$ &    $9.1039$ &                                                                                     $d\indices*{*_{\Gamma}^{T_{2g}^{}}_{A}^{\tensor*[^{1}]{\hspace{-0.2em}A}{}_{1}^{}}_{\bar{A}}^{B_{1}^{}}}$ &  $- 4.2400$ &                                                    $d\indices*{*_{\Gamma}^{T_{2g}^{}}_{A}^{\tensor*[^{1}]{\hspace{-0.2em}A}{}_{1}^{}}_{\bar{A}}^{\tensor*[^{1}]{\hspace{-0.2em}B}{}_{1}^{}}}$ &    $7.4916$ \\[0.4em]
                                                     $d\indices*{*_{\Gamma}^{T_{2g}^{}}_{A}^{\tensor*[^{1}]{\hspace{-0.2em}A}{}_{1}^{}}_{\bar{A}}^{\tensor*[^{2}]{\hspace{-0.2em}B}{}_{1}^{}}}$ &    $9.6127$ &                                                                                     $d\indices*{*_{\Gamma}^{T_{2g}^{}}_{A}^{\tensor*[^{2}]{\hspace{-0.2em}A}{}_{1}^{}}_{\bar{A}}^{A_{2}^{}}}$ & $- 10.0066$ &                                                                                     $d\indices*{*_{\Gamma}^{T_{2g}^{}}_{A}^{\tensor*[^{2}]{\hspace{-0.2em}A}{}_{1}^{}}_{\bar{A}}^{B_{1}^{}}}$ &   $11.8143$ &                                                    $d\indices*{*_{\Gamma}^{T_{2g}^{}}_{A}^{\tensor*[^{2}]{\hspace{-0.2em}A}{}_{1}^{}}_{\bar{A}}^{\tensor*[^{1}]{\hspace{-0.2em}B}{}_{1}^{}}}$ &  $- 4.2533$ &                                                    $d\indices*{*_{\Gamma}^{T_{2g}^{}}_{A}^{\tensor*[^{2}]{\hspace{-0.2em}A}{}_{1}^{}}_{\bar{A}}^{\tensor*[^{2}]{\hspace{-0.2em}B}{}_{1}^{}}}$ &    $0.2704$ \\[0.4em]
                                                                                                                       $d\indices*{*_{\Gamma}^{T_{2g}^{}}_{A}^{A_{2}^{}}_{\bar{A}}^{B_{2}^{}}}$ &    $0.7906$ &                                                                                     $d\indices*{*_{\Gamma}^{T_{2g}^{}}_{A}^{A_{2}^{}}_{\bar{A}}^{\tensor*[^{1}]{\hspace{-0.2em}B}{}_{2}^{}}}$ &  $- 9.2582$ &                                                                                                                      $d\indices*{*_{\Gamma}^{T_{2g}^{}}_{A}^{B_{1}^{}}_{\bar{A}}^{B_{2}^{}}}$ & $- 10.2403$ &                                                                                     $d\indices*{*_{\Gamma}^{T_{2g}^{}}_{A}^{B_{1}^{}}_{\bar{A}}^{\tensor*[^{1}]{\hspace{-0.2em}B}{}_{2}^{}}}$ &    $7.8096$ &                                                                                     $d\indices*{*_{\Gamma}^{T_{2g}^{}}_{A}^{\tensor*[^{1}]{\hspace{-0.2em}B}{}_{1}^{}}_{\bar{A}}^{B_{2}^{}}}$ &   $14.7439$ \\[0.4em]
                                                     $d\indices*{*_{\Gamma}^{T_{2g}^{}}_{A}^{\tensor*[^{1}]{\hspace{-0.2em}B}{}_{1}^{}}_{\bar{A}}^{\tensor*[^{1}]{\hspace{-0.2em}B}{}_{2}^{}}}$ & $- 10.8303$ &                                                                                     $d\indices*{*_{\Gamma}^{T_{2g}^{}}_{A}^{\tensor*[^{2}]{\hspace{-0.2em}B}{}_{1}^{}}_{\bar{A}}^{B_{2}^{}}}$ &  $- 1.7263$ &                                                    $d\indices*{*_{\Gamma}^{T_{2g}^{}}_{A}^{\tensor*[^{2}]{\hspace{-0.2em}B}{}_{1}^{}}_{\bar{A}}^{\tensor*[^{1}]{\hspace{-0.2em}B}{}_{2}^{}}}$ &  $- 8.5922$ &                                                                                     $d\indices*{*_{\Gamma}^{T_{1u}^{}}_{A}^{A_{1}^{}}_{\bar{A}}^{\tensor*[^{1}]{\hspace{-0.2em}A}{}_{1}^{}}}$ &    $1.6975$ &                                                                                     $d\indices*{*_{\Gamma}^{T_{1u}^{}}_{A}^{A_{1}^{}}_{\bar{A}}^{\tensor*[^{2}]{\hspace{-0.2em}A}{}_{1}^{}}}$ &   $38.9377$ \\[0.4em]
                                                                                                                       $d\indices*{*_{\Gamma}^{T_{1u}^{}}_{A}^{A_{1}^{}}_{\bar{A}}^{B_{2}^{}}}$ &  $- 8.9409$ &                                                                                     $d\indices*{*_{\Gamma}^{T_{1u}^{}}_{A}^{A_{1}^{}}_{\bar{A}}^{\tensor*[^{1}]{\hspace{-0.2em}B}{}_{2}^{}}}$ &    $3.5607$ &                                                    $d\indices*{*_{\Gamma}^{T_{1u}^{}}_{A}^{\tensor*[^{1}]{\hspace{-0.2em}A}{}_{1}^{}}_{\bar{A}}^{\tensor*[^{2}]{\hspace{-0.2em}A}{}_{1}^{}}}$ & $- 21.4899$ &                                                                                     $d\indices*{*_{\Gamma}^{T_{1u}^{}}_{A}^{\tensor*[^{1}]{\hspace{-0.2em}A}{}_{1}^{}}_{\bar{A}}^{B_{2}^{}}}$ & $- 18.4898$ &                                                    $d\indices*{*_{\Gamma}^{T_{1u}^{}}_{A}^{\tensor*[^{1}]{\hspace{-0.2em}A}{}_{1}^{}}_{\bar{A}}^{\tensor*[^{1}]{\hspace{-0.2em}B}{}_{2}^{}}}$ &   $13.7639$ \\[0.4em]
                                                                                      $d\indices*{*_{\Gamma}^{T_{1u}^{}}_{A}^{\tensor*[^{2}]{\hspace{-0.2em}A}{}_{1}^{}}_{\bar{A}}^{B_{2}^{}}}$ &    $0.4786$ &                                                    $d\indices*{*_{\Gamma}^{T_{1u}^{}}_{A}^{\tensor*[^{2}]{\hspace{-0.2em}A}{}_{1}^{}}_{\bar{A}}^{\tensor*[^{1}]{\hspace{-0.2em}B}{}_{2}^{}}}$ &  $- 3.7149$ &                                                                                                                      $d\indices*{*_{\Gamma}^{T_{1u}^{}}_{A}^{A_{2}^{}}_{\bar{A}}^{B_{1}^{}}}$ &   $12.8897$ &                                                                                     $d\indices*{*_{\Gamma}^{T_{1u}^{}}_{A}^{A_{2}^{}}_{\bar{A}}^{\tensor*[^{1}]{\hspace{-0.2em}B}{}_{1}^{}}}$ & $- 20.2921$ &                                                                                     $d\indices*{*_{\Gamma}^{T_{1u}^{}}_{A}^{A_{2}^{}}_{\bar{A}}^{\tensor*[^{2}]{\hspace{-0.2em}B}{}_{1}^{}}}$ &    $2.3207$ \\[0.4em]
                                                                                      $d\indices*{*_{\Gamma}^{T_{1u}^{}}_{A}^{B_{1}^{}}_{\bar{A}}^{\tensor*[^{1}]{\hspace{-0.2em}B}{}_{1}^{}}}$ &   $10.7209$ &                                                                                     $d\indices*{*_{\Gamma}^{T_{1u}^{}}_{A}^{B_{1}^{}}_{\bar{A}}^{\tensor*[^{2}]{\hspace{-0.2em}B}{}_{1}^{}}}$ &   $12.3213$ &                                                    $d\indices*{*_{\Gamma}^{T_{1u}^{}}_{A}^{\tensor*[^{1}]{\hspace{-0.2em}B}{}_{1}^{}}_{\bar{A}}^{\tensor*[^{2}]{\hspace{-0.2em}B}{}_{1}^{}}}$ &    $5.5920$ &                                                                                     $d\indices*{*_{\Gamma}^{T_{1u}^{}}_{A}^{B_{2}^{}}_{\bar{A}}^{\tensor*[^{1}]{\hspace{-0.2em}B}{}_{2}^{}}}$ &  $- 0.0420$ &                                                                                                                      $d\indices*{*_{\Gamma}^{T_{1u}^{}}_{A}^{A_{1}^{}}_{\bar{A}}^{B_{1}^{}}}$ &   $24.1775$ \\[0.4em]
                                                                                      $d\indices*{*_{\Gamma}^{T_{1u}^{}}_{A}^{A_{1}^{}}_{\bar{A}}^{\tensor*[^{1}]{\hspace{-0.2em}B}{}_{1}^{}}}$ &    $5.0290$ &                                                                                     $d\indices*{*_{\Gamma}^{T_{1u}^{}}_{A}^{A_{1}^{}}_{\bar{A}}^{\tensor*[^{2}]{\hspace{-0.2em}B}{}_{1}^{}}}$ &    $3.6649$ &                                                                                     $d\indices*{*_{\Gamma}^{T_{1u}^{}}_{A}^{\tensor*[^{1}]{\hspace{-0.2em}A}{}_{1}^{}}_{\bar{A}}^{B_{1}^{}}}$ &    $1.6857$ &                                                    $d\indices*{*_{\Gamma}^{T_{1u}^{}}_{A}^{\tensor*[^{1}]{\hspace{-0.2em}A}{}_{1}^{}}_{\bar{A}}^{\tensor*[^{1}]{\hspace{-0.2em}B}{}_{1}^{}}}$ & $- 21.8555$ &                                                    $d\indices*{*_{\Gamma}^{T_{1u}^{}}_{A}^{\tensor*[^{1}]{\hspace{-0.2em}A}{}_{1}^{}}_{\bar{A}}^{\tensor*[^{2}]{\hspace{-0.2em}B}{}_{1}^{}}}$ &  $- 3.4570$ \\[0.4em]
                                                                                      $d\indices*{*_{\Gamma}^{T_{1u}^{}}_{A}^{\tensor*[^{2}]{\hspace{-0.2em}A}{}_{1}^{}}_{\bar{A}}^{B_{1}^{}}}$ &    $1.1795$ &                                                    $d\indices*{*_{\Gamma}^{T_{1u}^{}}_{A}^{\tensor*[^{2}]{\hspace{-0.2em}A}{}_{1}^{}}_{\bar{A}}^{\tensor*[^{1}]{\hspace{-0.2em}B}{}_{1}^{}}}$ &   $19.8976$ &                                                    $d\indices*{*_{\Gamma}^{T_{1u}^{}}_{A}^{\tensor*[^{2}]{\hspace{-0.2em}A}{}_{1}^{}}_{\bar{A}}^{\tensor*[^{2}]{\hspace{-0.2em}B}{}_{1}^{}}}$ &   $26.5325$ &                                                                                                                      $d\indices*{*_{\Gamma}^{T_{1u}^{}}_{A}^{A_{2}^{}}_{\bar{A}}^{B_{2}^{}}}$ &   $22.8275$ &                                                                                     $d\indices*{*_{\Gamma}^{T_{1u}^{}}_{A}^{A_{2}^{}}_{\bar{A}}^{\tensor*[^{1}]{\hspace{-0.2em}B}{}_{2}^{}}}$ & $- 16.4968$ \\[0.4em]
                                                                                                                $d\indices*{*_{A_{3}}^{A_{1}^{}}_{X_{2}}^{A_{1g}^{}}_{\bar{A}_{1}}^{A_{1}^{}}}$ &    $1.3381$ &                                                                                                               $d\indices*{*_{A_{3}}^{A_{1}^{}}_{X_{2}}^{A_{2u}^{}}_{\bar{A}_{1}}^{A_{1}^{}}}$ &  $- 8.5294$ &                                                                              $d\indices*{*_{A_{3}}^{A_{1}^{}}_{X_{2}}^{A_{1g}^{}}_{\bar{A}_{1}}^{\tensor*[^{1}]{\hspace{-0.2em}A}{}_{1}^{}}}$ &    $0.7699$ &                                                                               $d\indices*{*_{A_{3}}^{A_{1}^{}}_{X_{2}}^{E_{g}^{}}_{\bar{A}_{1}}^{\tensor*[^{1}]{\hspace{-0.2em}A}{}_{1}^{}}}$ &  $- 1.1880$ &                                                                              $d\indices*{*_{A_{3}}^{A_{1}^{}}_{X_{2}}^{A_{2u}^{}}_{\bar{A}_{1}}^{\tensor*[^{1}]{\hspace{-0.2em}A}{}_{1}^{}}}$ & $- 21.6344$ \\[0.4em]
                                                                                $d\indices*{*_{A_{3}}^{A_{1}^{}}_{X_{2}}^{E_{u}^{}}_{\bar{A}_{1}}^{\tensor*[^{1}]{\hspace{-0.2em}A}{}_{1}^{}}}$ &  $- 9.8014$ &                                              $d\indices*{*_{A_{3}}^{A_{1}^{}}_{X_{2}}^{\tensor*[^{1}]{\hspace{-0.2em}E}{}_{u}^{}}_{\bar{A}_{1}}^{\tensor*[^{1}]{\hspace{-0.2em}A}{}_{1}^{}}}$ &   $18.3975$ &                                                                              $d\indices*{*_{A_{3}}^{A_{1}^{}}_{X_{2}}^{A_{1g}^{}}_{\bar{A}_{1}}^{\tensor*[^{2}]{\hspace{-0.2em}A}{}_{1}^{}}}$ &  $- 1.0629$ &                                                                               $d\indices*{*_{A_{3}}^{A_{1}^{}}_{X_{2}}^{E_{g}^{}}_{\bar{A}_{1}}^{\tensor*[^{2}]{\hspace{-0.2em}A}{}_{1}^{}}}$ &    $0.0550$ &                                                                              $d\indices*{*_{A_{3}}^{A_{1}^{}}_{X_{2}}^{A_{2u}^{}}_{\bar{A}_{1}}^{\tensor*[^{2}]{\hspace{-0.2em}A}{}_{1}^{}}}$ &  $- 6.2338$ \\[0.4em]
                                                                                $d\indices*{*_{A_{3}}^{A_{1}^{}}_{X_{2}}^{E_{u}^{}}_{\bar{A}_{1}}^{\tensor*[^{2}]{\hspace{-0.2em}A}{}_{1}^{}}}$ &    $2.3496$ &                                              $d\indices*{*_{A_{3}}^{A_{1}^{}}_{X_{2}}^{\tensor*[^{1}]{\hspace{-0.2em}E}{}_{u}^{}}_{\bar{A}_{1}}^{\tensor*[^{2}]{\hspace{-0.2em}A}{}_{1}^{}}}$ &    $2.2521$ &                                                                                                                $d\indices*{*_{A_{3}}^{A_{1}^{}}_{X_{2}}^{E_{g}^{}}_{\bar{A}_{1}}^{A_{2}^{}}}$ &    $3.9903$ &                                                                                                               $d\indices*{*_{A_{3}}^{A_{1}^{}}_{X_{2}}^{B_{1u}^{}}_{\bar{A}_{1}}^{A_{2}^{}}}$ & $- 21.3163$ &                                                                                                                $d\indices*{*_{A_{3}}^{A_{1}^{}}_{X_{2}}^{E_{u}^{}}_{\bar{A}_{1}}^{A_{2}^{}}}$ &   $17.7577$ \\[0.4em]
                                                                                $d\indices*{*_{A_{3}}^{A_{1}^{}}_{X_{2}}^{\tensor*[^{1}]{\hspace{-0.2em}E}{}_{u}^{}}_{\bar{A}_{1}}^{A_{2}^{}}}$ & $- 28.4127$ &                                                                                                                $d\indices*{*_{A_{3}}^{A_{1}^{}}_{X_{2}}^{E_{g}^{}}_{\bar{A}_{1}}^{B_{1}^{}}}$ & $- 14.7681$ &                                                                                                               $d\indices*{*_{A_{3}}^{A_{1}^{}}_{X_{2}}^{B_{1u}^{}}_{\bar{A}_{1}}^{B_{1}^{}}}$ &  $- 2.2489$ &                                                                                                                $d\indices*{*_{A_{3}}^{A_{1}^{}}_{X_{2}}^{E_{u}^{}}_{\bar{A}_{1}}^{B_{1}^{}}}$ &    $2.8249$ &                                                                               $d\indices*{*_{A_{3}}^{A_{1}^{}}_{X_{2}}^{\tensor*[^{1}]{\hspace{-0.2em}E}{}_{u}^{}}_{\bar{A}_{1}}^{B_{1}^{}}}$ &  $- 0.6668$ \\[0.4em]
                                                                                $d\indices*{*_{A_{3}}^{A_{1}^{}}_{X_{2}}^{E_{g}^{}}_{\bar{A}_{1}}^{\tensor*[^{1}]{\hspace{-0.2em}B}{}_{1}^{}}}$ &   $22.1748$ &                                                                              $d\indices*{*_{A_{3}}^{A_{1}^{}}_{X_{2}}^{B_{1u}^{}}_{\bar{A}_{1}}^{\tensor*[^{1}]{\hspace{-0.2em}B}{}_{1}^{}}}$ &    $4.3324$ &                                                                               $d\indices*{*_{A_{3}}^{A_{1}^{}}_{X_{2}}^{E_{u}^{}}_{\bar{A}_{1}}^{\tensor*[^{1}]{\hspace{-0.2em}B}{}_{1}^{}}}$ &  $- 0.6906$ &                                              $d\indices*{*_{A_{3}}^{A_{1}^{}}_{X_{2}}^{\tensor*[^{1}]{\hspace{-0.2em}E}{}_{u}^{}}_{\bar{A}_{1}}^{\tensor*[^{1}]{\hspace{-0.2em}B}{}_{1}^{}}}$ & $- 11.9815$ &                                                                               $d\indices*{*_{A_{3}}^{A_{1}^{}}_{X_{2}}^{E_{g}^{}}_{\bar{A}_{1}}^{\tensor*[^{2}]{\hspace{-0.2em}B}{}_{1}^{}}}$ &  $- 2.4482$ \\[0.4em]
                                                                               $d\indices*{*_{A_{3}}^{A_{1}^{}}_{X_{2}}^{B_{1u}^{}}_{\bar{A}_{1}}^{\tensor*[^{2}]{\hspace{-0.2em}B}{}_{1}^{}}}$ &    $2.5775$ &                                                                               $d\indices*{*_{A_{3}}^{A_{1}^{}}_{X_{2}}^{E_{u}^{}}_{\bar{A}_{1}}^{\tensor*[^{2}]{\hspace{-0.2em}B}{}_{1}^{}}}$ &    $1.8698$ &                                              $d\indices*{*_{A_{3}}^{A_{1}^{}}_{X_{2}}^{\tensor*[^{1}]{\hspace{-0.2em}E}{}_{u}^{}}_{\bar{A}_{1}}^{\tensor*[^{2}]{\hspace{-0.2em}B}{}_{1}^{}}}$ &  $- 2.4141$ &                                                                                                               $d\indices*{*_{A_{3}}^{A_{1}^{}}_{X_{2}}^{A_{1g}^{}}_{\bar{A}_{1}}^{B_{2}^{}}}$ & $- 28.2045$ &                                                                                                                $d\indices*{*_{A_{3}}^{A_{1}^{}}_{X_{2}}^{E_{g}^{}}_{\bar{A}_{1}}^{B_{2}^{}}}$ & $- 27.7138$ \\[0.4em]
                                                                                                                $d\indices*{*_{A_{3}}^{A_{1}^{}}_{X_{2}}^{A_{2u}^{}}_{\bar{A}_{1}}^{B_{2}^{}}}$ &    $1.2013$ &                                                                                                                $d\indices*{*_{A_{3}}^{A_{1}^{}}_{X_{2}}^{E_{u}^{}}_{\bar{A}_{1}}^{B_{2}^{}}}$ &    $4.7359$ &                                                                               $d\indices*{*_{A_{3}}^{A_{1}^{}}_{X_{2}}^{\tensor*[^{1}]{\hspace{-0.2em}E}{}_{u}^{}}_{\bar{A}_{1}}^{B_{2}^{}}}$ & $- 12.1849$ &                                                                              $d\indices*{*_{A_{3}}^{A_{1}^{}}_{X_{2}}^{A_{1g}^{}}_{\bar{A}_{1}}^{\tensor*[^{1}]{\hspace{-0.2em}B}{}_{2}^{}}}$ &   $21.7650$ &                                                                               $d\indices*{*_{A_{3}}^{A_{1}^{}}_{X_{2}}^{E_{g}^{}}_{\bar{A}_{1}}^{\tensor*[^{1}]{\hspace{-0.2em}B}{}_{2}^{}}}$ &   $18.5000$ \\[0.4em]
                                                                               $d\indices*{*_{A_{3}}^{A_{1}^{}}_{X_{2}}^{A_{2u}^{}}_{\bar{A}_{1}}^{\tensor*[^{1}]{\hspace{-0.2em}B}{}_{2}^{}}}$ &    $2.9821$ &                                                                               $d\indices*{*_{A_{3}}^{A_{1}^{}}_{X_{2}}^{E_{u}^{}}_{\bar{A}_{1}}^{\tensor*[^{1}]{\hspace{-0.2em}B}{}_{2}^{}}}$ &  $- 2.6410$ &                                              $d\indices*{*_{A_{3}}^{A_{1}^{}}_{X_{2}}^{\tensor*[^{1}]{\hspace{-0.2em}E}{}_{u}^{}}_{\bar{A}_{1}}^{\tensor*[^{1}]{\hspace{-0.2em}B}{}_{2}^{}}}$ &    $2.5251$ &                                             $d\indices*{*_{A_{3}}^{\tensor*[^{1}]{\hspace{-0.2em}A}{}_{1}^{}}_{X_{2}}^{A_{1g}^{}}_{\bar{A}_{1}}^{\tensor*[^{1}]{\hspace{-0.2em}A}{}_{1}^{}}}$ &  $- 3.3896$ &                                             $d\indices*{*_{A_{3}}^{\tensor*[^{1}]{\hspace{-0.2em}A}{}_{1}^{}}_{X_{2}}^{A_{2u}^{}}_{\bar{A}_{1}}^{\tensor*[^{1}]{\hspace{-0.2em}A}{}_{1}^{}}}$ &   $15.8344$ \\[0.4em]
                                              $d\indices*{*_{A_{3}}^{\tensor*[^{1}]{\hspace{-0.2em}A}{}_{1}^{}}_{X_{2}}^{A_{1g}^{}}_{\bar{A}_{1}}^{\tensor*[^{2}]{\hspace{-0.2em}A}{}_{1}^{}}}$ &   $13.6097$ &                                              $d\indices*{*_{A_{3}}^{\tensor*[^{1}]{\hspace{-0.2em}A}{}_{1}^{}}_{X_{2}}^{E_{g}^{}}_{\bar{A}_{1}}^{\tensor*[^{2}]{\hspace{-0.2em}A}{}_{1}^{}}}$ &   $13.6179$ &                                             $d\indices*{*_{A_{3}}^{\tensor*[^{1}]{\hspace{-0.2em}A}{}_{1}^{}}_{X_{2}}^{A_{2u}^{}}_{\bar{A}_{1}}^{\tensor*[^{2}]{\hspace{-0.2em}A}{}_{1}^{}}}$ &  $- 3.5418$ &                                              $d\indices*{*_{A_{3}}^{\tensor*[^{1}]{\hspace{-0.2em}A}{}_{1}^{}}_{X_{2}}^{E_{u}^{}}_{\bar{A}_{1}}^{\tensor*[^{2}]{\hspace{-0.2em}A}{}_{1}^{}}}$ &    $1.7504$ &             $d\indices*{*_{A_{3}}^{\tensor*[^{1}]{\hspace{-0.2em}A}{}_{1}^{}}_{X_{2}}^{\tensor*[^{1}]{\hspace{-0.2em}E}{}_{u}^{}}_{\bar{A}_{1}}^{\tensor*[^{2}]{\hspace{-0.2em}A}{}_{1}^{}}}$ &  $- 0.1148$ \\[0.4em]
                                                                                $d\indices*{*_{A_{3}}^{\tensor*[^{1}]{\hspace{-0.2em}A}{}_{1}^{}}_{X_{2}}^{E_{g}^{}}_{\bar{A}_{1}}^{A_{2}^{}}}$ &  $- 5.2409$ &                                                                              $d\indices*{*_{A_{3}}^{\tensor*[^{1}]{\hspace{-0.2em}A}{}_{1}^{}}_{X_{2}}^{B_{1u}^{}}_{\bar{A}_{1}}^{A_{2}^{}}}$ &   $10.1167$ &                                                                               $d\indices*{*_{A_{3}}^{\tensor*[^{1}]{\hspace{-0.2em}A}{}_{1}^{}}_{X_{2}}^{E_{u}^{}}_{\bar{A}_{1}}^{A_{2}^{}}}$ &  $- 9.5060$ &                                              $d\indices*{*_{A_{3}}^{\tensor*[^{1}]{\hspace{-0.2em}A}{}_{1}^{}}_{X_{2}}^{\tensor*[^{1}]{\hspace{-0.2em}E}{}_{u}^{}}_{\bar{A}_{1}}^{A_{2}^{}}}$ &    $9.5768$ &                                                                               $d\indices*{*_{A_{3}}^{\tensor*[^{1}]{\hspace{-0.2em}A}{}_{1}^{}}_{X_{2}}^{E_{g}^{}}_{\bar{A}_{1}}^{B_{1}^{}}}$ &  $- 0.7376$ \\[0.4em]
                                                                               $d\indices*{*_{A_{3}}^{\tensor*[^{1}]{\hspace{-0.2em}A}{}_{1}^{}}_{X_{2}}^{B_{1u}^{}}_{\bar{A}_{1}}^{B_{1}^{}}}$ &    $2.8684$ &                                                                               $d\indices*{*_{A_{3}}^{\tensor*[^{1}]{\hspace{-0.2em}A}{}_{1}^{}}_{X_{2}}^{E_{u}^{}}_{\bar{A}_{1}}^{B_{1}^{}}}$ &  $- 3.8491$ &                                              $d\indices*{*_{A_{3}}^{\tensor*[^{1}]{\hspace{-0.2em}A}{}_{1}^{}}_{X_{2}}^{\tensor*[^{1}]{\hspace{-0.2em}E}{}_{u}^{}}_{\bar{A}_{1}}^{B_{1}^{}}}$ &    $2.2558$ &                                              $d\indices*{*_{A_{3}}^{\tensor*[^{1}]{\hspace{-0.2em}A}{}_{1}^{}}_{X_{2}}^{E_{g}^{}}_{\bar{A}_{1}}^{\tensor*[^{1}]{\hspace{-0.2em}B}{}_{1}^{}}}$ &  $- 9.0614$ &                                             $d\indices*{*_{A_{3}}^{\tensor*[^{1}]{\hspace{-0.2em}A}{}_{1}^{}}_{X_{2}}^{B_{1u}^{}}_{\bar{A}_{1}}^{\tensor*[^{1}]{\hspace{-0.2em}B}{}_{1}^{}}}$ &    $6.6815$ \\[0.4em]
                                               $d\indices*{*_{A_{3}}^{\tensor*[^{1}]{\hspace{-0.2em}A}{}_{1}^{}}_{X_{2}}^{E_{u}^{}}_{\bar{A}_{1}}^{\tensor*[^{1}]{\hspace{-0.2em}B}{}_{1}^{}}}$ &   $11.0263$ &             $d\indices*{*_{A_{3}}^{\tensor*[^{1}]{\hspace{-0.2em}A}{}_{1}^{}}_{X_{2}}^{\tensor*[^{1}]{\hspace{-0.2em}E}{}_{u}^{}}_{\bar{A}_{1}}^{\tensor*[^{1}]{\hspace{-0.2em}B}{}_{1}^{}}}$ & $- 14.3866$ &                                              $d\indices*{*_{A_{3}}^{\tensor*[^{1}]{\hspace{-0.2em}A}{}_{1}^{}}_{X_{2}}^{E_{g}^{}}_{\bar{A}_{1}}^{\tensor*[^{2}]{\hspace{-0.2em}B}{}_{1}^{}}}$ &    $3.0739$ &                                             $d\indices*{*_{A_{3}}^{\tensor*[^{1}]{\hspace{-0.2em}A}{}_{1}^{}}_{X_{2}}^{B_{1u}^{}}_{\bar{A}_{1}}^{\tensor*[^{2}]{\hspace{-0.2em}B}{}_{1}^{}}}$ &    $9.5749$ &                                              $d\indices*{*_{A_{3}}^{\tensor*[^{1}]{\hspace{-0.2em}A}{}_{1}^{}}_{X_{2}}^{E_{u}^{}}_{\bar{A}_{1}}^{\tensor*[^{2}]{\hspace{-0.2em}B}{}_{1}^{}}}$ &    $7.5752$ \\[0.4em]
              $d\indices*{*_{A_{3}}^{\tensor*[^{1}]{\hspace{-0.2em}A}{}_{1}^{}}_{X_{2}}^{\tensor*[^{1}]{\hspace{-0.2em}E}{}_{u}^{}}_{\bar{A}_{1}}^{\tensor*[^{2}]{\hspace{-0.2em}B}{}_{1}^{}}}$ & $- 12.1305$ &                                                                              $d\indices*{*_{A_{3}}^{\tensor*[^{1}]{\hspace{-0.2em}A}{}_{1}^{}}_{X_{2}}^{A_{1g}^{}}_{\bar{A}_{1}}^{B_{2}^{}}}$ &   $16.5197$ &                                                                               $d\indices*{*_{A_{3}}^{\tensor*[^{1}]{\hspace{-0.2em}A}{}_{1}^{}}_{X_{2}}^{E_{g}^{}}_{\bar{A}_{1}}^{B_{2}^{}}}$ &   $11.1167$ &                                                                              $d\indices*{*_{A_{3}}^{\tensor*[^{1}]{\hspace{-0.2em}A}{}_{1}^{}}_{X_{2}}^{A_{2u}^{}}_{\bar{A}_{1}}^{B_{2}^{}}}$ &  $- 0.1260$ &                                                                               $d\indices*{*_{A_{3}}^{\tensor*[^{1}]{\hspace{-0.2em}A}{}_{1}^{}}_{X_{2}}^{E_{u}^{}}_{\bar{A}_{1}}^{B_{2}^{}}}$ &  $- 4.6401$ \\[0.4em]
                                               $d\indices*{*_{A_{3}}^{\tensor*[^{1}]{\hspace{-0.2em}A}{}_{1}^{}}_{X_{2}}^{\tensor*[^{1}]{\hspace{-0.2em}E}{}_{u}^{}}_{\bar{A}_{1}}^{B_{2}^{}}}$ &    $4.3049$ &                                             $d\indices*{*_{A_{3}}^{\tensor*[^{1}]{\hspace{-0.2em}A}{}_{1}^{}}_{X_{2}}^{A_{1g}^{}}_{\bar{A}_{1}}^{\tensor*[^{1}]{\hspace{-0.2em}B}{}_{2}^{}}}$ & $- 11.6075$ &                                              $d\indices*{*_{A_{3}}^{\tensor*[^{1}]{\hspace{-0.2em}A}{}_{1}^{}}_{X_{2}}^{E_{g}^{}}_{\bar{A}_{1}}^{\tensor*[^{1}]{\hspace{-0.2em}B}{}_{2}^{}}}$ &  $- 9.1661$ &                                             $d\indices*{*_{A_{3}}^{\tensor*[^{1}]{\hspace{-0.2em}A}{}_{1}^{}}_{X_{2}}^{A_{2u}^{}}_{\bar{A}_{1}}^{\tensor*[^{1}]{\hspace{-0.2em}B}{}_{2}^{}}}$ &    $2.4046$ &                                              $d\indices*{*_{A_{3}}^{\tensor*[^{1}]{\hspace{-0.2em}A}{}_{1}^{}}_{X_{2}}^{E_{u}^{}}_{\bar{A}_{1}}^{\tensor*[^{1}]{\hspace{-0.2em}B}{}_{2}^{}}}$ &    $4.4657$ \\[0.4em]
              $d\indices*{*_{A_{3}}^{\tensor*[^{1}]{\hspace{-0.2em}A}{}_{1}^{}}_{X_{2}}^{\tensor*[^{1}]{\hspace{-0.2em}E}{}_{u}^{}}_{\bar{A}_{1}}^{\tensor*[^{1}]{\hspace{-0.2em}B}{}_{2}^{}}}$ &  $- 3.1804$ &                                             $d\indices*{*_{A_{3}}^{\tensor*[^{2}]{\hspace{-0.2em}A}{}_{1}^{}}_{X_{2}}^{A_{1g}^{}}_{\bar{A}_{1}}^{\tensor*[^{2}]{\hspace{-0.2em}A}{}_{1}^{}}}$ &  $- 7.9327$ &                                             $d\indices*{*_{A_{3}}^{\tensor*[^{2}]{\hspace{-0.2em}A}{}_{1}^{}}_{X_{2}}^{A_{2u}^{}}_{\bar{A}_{1}}^{\tensor*[^{2}]{\hspace{-0.2em}A}{}_{1}^{}}}$ &  $- 2.6338$ &                                                                               $d\indices*{*_{A_{3}}^{\tensor*[^{2}]{\hspace{-0.2em}A}{}_{1}^{}}_{X_{2}}^{E_{g}^{}}_{\bar{A}_{1}}^{A_{2}^{}}}$ &   $17.9252$ &                                                                              $d\indices*{*_{A_{3}}^{\tensor*[^{2}]{\hspace{-0.2em}A}{}_{1}^{}}_{X_{2}}^{B_{1u}^{}}_{\bar{A}_{1}}^{A_{2}^{}}}$ &  $- 6.4501$ \\[0.4em]
                                                                                $d\indices*{*_{A_{3}}^{\tensor*[^{2}]{\hspace{-0.2em}A}{}_{1}^{}}_{X_{2}}^{E_{u}^{}}_{\bar{A}_{1}}^{A_{2}^{}}}$ &  $- 3.1372$ &                                              $d\indices*{*_{A_{3}}^{\tensor*[^{2}]{\hspace{-0.2em}A}{}_{1}^{}}_{X_{2}}^{\tensor*[^{1}]{\hspace{-0.2em}E}{}_{u}^{}}_{\bar{A}_{1}}^{A_{2}^{}}}$ &  $- 5.1078$ &                                                                               $d\indices*{*_{A_{3}}^{\tensor*[^{2}]{\hspace{-0.2em}A}{}_{1}^{}}_{X_{2}}^{E_{g}^{}}_{\bar{A}_{1}}^{B_{1}^{}}}$ &  $- 1.8497$ &                                                                              $d\indices*{*_{A_{3}}^{\tensor*[^{2}]{\hspace{-0.2em}A}{}_{1}^{}}_{X_{2}}^{B_{1u}^{}}_{\bar{A}_{1}}^{B_{1}^{}}}$ & $- 12.1283$ &                                                                               $d\indices*{*_{A_{3}}^{\tensor*[^{2}]{\hspace{-0.2em}A}{}_{1}^{}}_{X_{2}}^{E_{u}^{}}_{\bar{A}_{1}}^{B_{1}^{}}}$ &   $11.0476$ \\[0.4em]
                                               $d\indices*{*_{A_{3}}^{\tensor*[^{2}]{\hspace{-0.2em}A}{}_{1}^{}}_{X_{2}}^{\tensor*[^{1}]{\hspace{-0.2em}E}{}_{u}^{}}_{\bar{A}_{1}}^{B_{1}^{}}}$ & $- 11.8520$ &                                              $d\indices*{*_{A_{3}}^{\tensor*[^{2}]{\hspace{-0.2em}A}{}_{1}^{}}_{X_{2}}^{E_{g}^{}}_{\bar{A}_{1}}^{\tensor*[^{1}]{\hspace{-0.2em}B}{}_{1}^{}}}$ &    $0.2404$ &                                             $d\indices*{*_{A_{3}}^{\tensor*[^{2}]{\hspace{-0.2em}A}{}_{1}^{}}_{X_{2}}^{B_{1u}^{}}_{\bar{A}_{1}}^{\tensor*[^{1}]{\hspace{-0.2em}B}{}_{1}^{}}}$ &   $13.9103$ &                                              $d\indices*{*_{A_{3}}^{\tensor*[^{2}]{\hspace{-0.2em}A}{}_{1}^{}}_{X_{2}}^{E_{u}^{}}_{\bar{A}_{1}}^{\tensor*[^{1}]{\hspace{-0.2em}B}{}_{1}^{}}}$ & $- 16.4198$ &             $d\indices*{*_{A_{3}}^{\tensor*[^{2}]{\hspace{-0.2em}A}{}_{1}^{}}_{X_{2}}^{\tensor*[^{1}]{\hspace{-0.2em}E}{}_{u}^{}}_{\bar{A}_{1}}^{\tensor*[^{1}]{\hspace{-0.2em}B}{}_{1}^{}}}$ &   $17.9355$ \\[0.4em]
                                               $d\indices*{*_{A_{3}}^{\tensor*[^{2}]{\hspace{-0.2em}A}{}_{1}^{}}_{X_{2}}^{E_{g}^{}}_{\bar{A}_{1}}^{\tensor*[^{2}]{\hspace{-0.2em}B}{}_{1}^{}}}$ &    $0.0747$ &                                             $d\indices*{*_{A_{3}}^{\tensor*[^{2}]{\hspace{-0.2em}A}{}_{1}^{}}_{X_{2}}^{B_{1u}^{}}_{\bar{A}_{1}}^{\tensor*[^{2}]{\hspace{-0.2em}B}{}_{1}^{}}}$ &  $- 3.3542$ &                                              $d\indices*{*_{A_{3}}^{\tensor*[^{2}]{\hspace{-0.2em}A}{}_{1}^{}}_{X_{2}}^{E_{u}^{}}_{\bar{A}_{1}}^{\tensor*[^{2}]{\hspace{-0.2em}B}{}_{1}^{}}}$ &  $- 8.6171$ &             $d\indices*{*_{A_{3}}^{\tensor*[^{2}]{\hspace{-0.2em}A}{}_{1}^{}}_{X_{2}}^{\tensor*[^{1}]{\hspace{-0.2em}E}{}_{u}^{}}_{\bar{A}_{1}}^{\tensor*[^{2}]{\hspace{-0.2em}B}{}_{1}^{}}}$ &    $1.5470$ &                                                                              $d\indices*{*_{A_{3}}^{\tensor*[^{2}]{\hspace{-0.2em}A}{}_{1}^{}}_{X_{2}}^{A_{1g}^{}}_{\bar{A}_{1}}^{B_{2}^{}}}$ &    $1.5044$ \\[0.4em]
                                                                                $d\indices*{*_{A_{3}}^{\tensor*[^{2}]{\hspace{-0.2em}A}{}_{1}^{}}_{X_{2}}^{E_{g}^{}}_{\bar{A}_{1}}^{B_{2}^{}}}$ &  $- 5.4534$ &                                                                              $d\indices*{*_{A_{3}}^{\tensor*[^{2}]{\hspace{-0.2em}A}{}_{1}^{}}_{X_{2}}^{A_{2u}^{}}_{\bar{A}_{1}}^{B_{2}^{}}}$ &   $32.5595$ &                                                                               $d\indices*{*_{A_{3}}^{\tensor*[^{2}]{\hspace{-0.2em}A}{}_{1}^{}}_{X_{2}}^{E_{u}^{}}_{\bar{A}_{1}}^{B_{2}^{}}}$ &   $20.1477$ &                                              $d\indices*{*_{A_{3}}^{\tensor*[^{2}]{\hspace{-0.2em}A}{}_{1}^{}}_{X_{2}}^{\tensor*[^{1}]{\hspace{-0.2em}E}{}_{u}^{}}_{\bar{A}_{1}}^{B_{2}^{}}}$ & $- 28.2765$ &                                             $d\indices*{*_{A_{3}}^{\tensor*[^{2}]{\hspace{-0.2em}A}{}_{1}^{}}_{X_{2}}^{A_{1g}^{}}_{\bar{A}_{1}}^{\tensor*[^{1}]{\hspace{-0.2em}B}{}_{2}^{}}}$ &  $- 3.1687$ \\[0.4em]
                                               $d\indices*{*_{A_{3}}^{\tensor*[^{2}]{\hspace{-0.2em}A}{}_{1}^{}}_{X_{2}}^{E_{g}^{}}_{\bar{A}_{1}}^{\tensor*[^{1}]{\hspace{-0.2em}B}{}_{2}^{}}}$ &    $6.1814$ &                                             $d\indices*{*_{A_{3}}^{\tensor*[^{2}]{\hspace{-0.2em}A}{}_{1}^{}}_{X_{2}}^{A_{2u}^{}}_{\bar{A}_{1}}^{\tensor*[^{1}]{\hspace{-0.2em}B}{}_{2}^{}}}$ & $- 20.1316$ &                                              $d\indices*{*_{A_{3}}^{\tensor*[^{2}]{\hspace{-0.2em}A}{}_{1}^{}}_{X_{2}}^{E_{u}^{}}_{\bar{A}_{1}}^{\tensor*[^{1}]{\hspace{-0.2em}B}{}_{2}^{}}}$ & $- 13.6920$ &             $d\indices*{*_{A_{3}}^{\tensor*[^{2}]{\hspace{-0.2em}A}{}_{1}^{}}_{X_{2}}^{\tensor*[^{1}]{\hspace{-0.2em}E}{}_{u}^{}}_{\bar{A}_{1}}^{\tensor*[^{1}]{\hspace{-0.2em}B}{}_{2}^{}}}$ &   $18.0362$ &                                                                                                               $d\indices*{*_{A_{3}}^{A_{2}^{}}_{X_{2}}^{A_{1g}^{}}_{\bar{A}_{1}}^{A_{2}^{}}}$ &    $7.7030$ \\[0.4em]
                                                                                                                $d\indices*{*_{A_{3}}^{A_{2}^{}}_{X_{2}}^{A_{2u}^{}}_{\bar{A}_{1}}^{A_{2}^{}}}$ &    $2.5615$ &                                                                                                               $d\indices*{*_{A_{3}}^{A_{2}^{}}_{X_{2}}^{A_{1g}^{}}_{\bar{A}_{1}}^{B_{1}^{}}}$ & $- 13.3270$ &                                                                                                                $d\indices*{*_{A_{3}}^{A_{2}^{}}_{X_{2}}^{E_{g}^{}}_{\bar{A}_{1}}^{B_{1}^{}}}$ &  $- 8.7783$ &                                                                                                               $d\indices*{*_{A_{3}}^{A_{2}^{}}_{X_{2}}^{A_{2u}^{}}_{\bar{A}_{1}}^{B_{1}^{}}}$ &    $0.0167$ &                                                                                                                $d\indices*{*_{A_{3}}^{A_{2}^{}}_{X_{2}}^{E_{u}^{}}_{\bar{A}_{1}}^{B_{1}^{}}}$ &  $- 1.3946$ \\[0.4em]
                                                                                $d\indices*{*_{A_{3}}^{A_{2}^{}}_{X_{2}}^{\tensor*[^{1}]{\hspace{-0.2em}E}{}_{u}^{}}_{\bar{A}_{1}}^{B_{1}^{}}}$ &  $- 0.6420$ &                                                                              $d\indices*{*_{A_{3}}^{A_{2}^{}}_{X_{2}}^{A_{1g}^{}}_{\bar{A}_{1}}^{\tensor*[^{1}]{\hspace{-0.2em}B}{}_{1}^{}}}$ &    $0.7637$ &                                                                               $d\indices*{*_{A_{3}}^{A_{2}^{}}_{X_{2}}^{E_{g}^{}}_{\bar{A}_{1}}^{\tensor*[^{1}]{\hspace{-0.2em}B}{}_{1}^{}}}$ &  $- 2.2108$ &                                                                              $d\indices*{*_{A_{3}}^{A_{2}^{}}_{X_{2}}^{A_{2u}^{}}_{\bar{A}_{1}}^{\tensor*[^{1}]{\hspace{-0.2em}B}{}_{1}^{}}}$ & $- 24.8679$ &                                                                               $d\indices*{*_{A_{3}}^{A_{2}^{}}_{X_{2}}^{E_{u}^{}}_{\bar{A}_{1}}^{\tensor*[^{1}]{\hspace{-0.2em}B}{}_{1}^{}}}$ &    $8.5779$ \\[0.4em]
                                               $d\indices*{*_{A_{3}}^{A_{2}^{}}_{X_{2}}^{\tensor*[^{1}]{\hspace{-0.2em}E}{}_{u}^{}}_{\bar{A}_{1}}^{\tensor*[^{1}]{\hspace{-0.2em}B}{}_{1}^{}}}$ & $- 13.9623$ &                                                                              $d\indices*{*_{A_{3}}^{A_{2}^{}}_{X_{2}}^{A_{1g}^{}}_{\bar{A}_{1}}^{\tensor*[^{2}]{\hspace{-0.2em}B}{}_{1}^{}}}$ &  $- 5.7667$ &                                                                               $d\indices*{*_{A_{3}}^{A_{2}^{}}_{X_{2}}^{E_{g}^{}}_{\bar{A}_{1}}^{\tensor*[^{2}]{\hspace{-0.2em}B}{}_{1}^{}}}$ &    $3.2307$ &                                                                              $d\indices*{*_{A_{3}}^{A_{2}^{}}_{X_{2}}^{A_{2u}^{}}_{\bar{A}_{1}}^{\tensor*[^{2}]{\hspace{-0.2em}B}{}_{1}^{}}}$ & $- 20.7536$ &                                                                               $d\indices*{*_{A_{3}}^{A_{2}^{}}_{X_{2}}^{E_{u}^{}}_{\bar{A}_{1}}^{\tensor*[^{2}]{\hspace{-0.2em}B}{}_{1}^{}}}$ &   $11.0365$ \\[0.4em]
                                               $d\indices*{*_{A_{3}}^{A_{2}^{}}_{X_{2}}^{\tensor*[^{1}]{\hspace{-0.2em}E}{}_{u}^{}}_{\bar{A}_{1}}^{\tensor*[^{2}]{\hspace{-0.2em}B}{}_{1}^{}}}$ & $- 12.7066$ &                                                                                                                $d\indices*{*_{A_{3}}^{B_{2}^{}}_{X_{2}}^{E_{g}^{}}_{\bar{A}_{1}}^{A_{2}^{}}}$ &    $0.0429$ &                                                                                                               $d\indices*{*_{A_{3}}^{B_{2}^{}}_{X_{2}}^{B_{1u}^{}}_{\bar{A}_{1}}^{A_{2}^{}}}$ &    $0.4704$ &                                                                                                                $d\indices*{*_{A_{3}}^{B_{2}^{}}_{X_{2}}^{E_{u}^{}}_{\bar{A}_{1}}^{A_{2}^{}}}$ &  $- 0.6953$ &                                                                               $d\indices*{*_{A_{3}}^{B_{2}^{}}_{X_{2}}^{\tensor*[^{1}]{\hspace{-0.2em}E}{}_{u}^{}}_{\bar{A}_{1}}^{A_{2}^{}}}$ &  $- 1.5514$ \\[0.4em]
                                                                                $d\indices*{*_{A_{3}}^{\tensor*[^{1}]{\hspace{-0.2em}B}{}_{2}^{}}_{X_{2}}^{E_{g}^{}}_{\bar{A}_{1}}^{A_{2}^{}}}$ &  $- 0.0283$ &                                                                              $d\indices*{*_{A_{3}}^{\tensor*[^{1}]{\hspace{-0.2em}B}{}_{2}^{}}_{X_{2}}^{B_{1u}^{}}_{\bar{A}_{1}}^{A_{2}^{}}}$ &    $0.1489$ &                                                                               $d\indices*{*_{A_{3}}^{\tensor*[^{1}]{\hspace{-0.2em}B}{}_{2}^{}}_{X_{2}}^{E_{u}^{}}_{\bar{A}_{1}}^{A_{2}^{}}}$ &    $9.7812$ &                                              $d\indices*{*_{A_{3}}^{\tensor*[^{1}]{\hspace{-0.2em}B}{}_{2}^{}}_{X_{2}}^{\tensor*[^{1}]{\hspace{-0.2em}E}{}_{u}^{}}_{\bar{A}_{1}}^{A_{2}^{}}}$ &  $- 1.0332$ &                                                                                                               $d\indices*{*_{A_{3}}^{B_{1}^{}}_{X_{2}}^{A_{1g}^{}}_{\bar{A}_{1}}^{B_{1}^{}}}$ &    $1.2489$ \\[0.4em]
                                                                                                                $d\indices*{*_{A_{3}}^{B_{1}^{}}_{X_{2}}^{A_{2u}^{}}_{\bar{A}_{1}}^{B_{1}^{}}}$ &   $22.2988$ &                                                                              $d\indices*{*_{A_{3}}^{B_{1}^{}}_{X_{2}}^{A_{1g}^{}}_{\bar{A}_{1}}^{\tensor*[^{1}]{\hspace{-0.2em}B}{}_{1}^{}}}$ &  $- 9.5017$ &                                                                               $d\indices*{*_{A_{3}}^{B_{1}^{}}_{X_{2}}^{E_{g}^{}}_{\bar{A}_{1}}^{\tensor*[^{1}]{\hspace{-0.2em}B}{}_{1}^{}}}$ &  $- 6.5135$ &                                                                              $d\indices*{*_{A_{3}}^{B_{1}^{}}_{X_{2}}^{A_{2u}^{}}_{\bar{A}_{1}}^{\tensor*[^{1}]{\hspace{-0.2em}B}{}_{1}^{}}}$ & $- 14.6738$ &                                                                               $d\indices*{*_{A_{3}}^{B_{1}^{}}_{X_{2}}^{E_{u}^{}}_{\bar{A}_{1}}^{\tensor*[^{1}]{\hspace{-0.2em}B}{}_{1}^{}}}$ &  $- 8.9814$ \\[0.4em]
                                               $d\indices*{*_{A_{3}}^{B_{1}^{}}_{X_{2}}^{\tensor*[^{1}]{\hspace{-0.2em}E}{}_{u}^{}}_{\bar{A}_{1}}^{\tensor*[^{1}]{\hspace{-0.2em}B}{}_{1}^{}}}$ &   $16.1559$ &                                                                              $d\indices*{*_{A_{3}}^{B_{1}^{}}_{X_{2}}^{A_{1g}^{}}_{\bar{A}_{1}}^{\tensor*[^{2}]{\hspace{-0.2em}B}{}_{1}^{}}}$ & $- 12.4485$ &                                                                               $d\indices*{*_{A_{3}}^{B_{1}^{}}_{X_{2}}^{E_{g}^{}}_{\bar{A}_{1}}^{\tensor*[^{2}]{\hspace{-0.2em}B}{}_{1}^{}}}$ &  $- 8.3388$ &                                                                              $d\indices*{*_{A_{3}}^{B_{1}^{}}_{X_{2}}^{A_{2u}^{}}_{\bar{A}_{1}}^{\tensor*[^{2}]{\hspace{-0.2em}B}{}_{1}^{}}}$ &    $0.4714$ &                                                                               $d\indices*{*_{A_{3}}^{B_{1}^{}}_{X_{2}}^{E_{u}^{}}_{\bar{A}_{1}}^{\tensor*[^{2}]{\hspace{-0.2em}B}{}_{1}^{}}}$ &    $3.2574$ \\[0.4em]
                                               $d\indices*{*_{A_{3}}^{B_{1}^{}}_{X_{2}}^{\tensor*[^{1}]{\hspace{-0.2em}E}{}_{u}^{}}_{\bar{A}_{1}}^{\tensor*[^{2}]{\hspace{-0.2em}B}{}_{1}^{}}}$ &  $- 3.2178$ &                                                                                                                $d\indices*{*_{A_{3}}^{B_{2}^{}}_{X_{2}}^{E_{g}^{}}_{\bar{A}_{1}}^{B_{1}^{}}}$ &    $0.2826$ &                                                                                                               $d\indices*{*_{A_{3}}^{B_{2}^{}}_{X_{2}}^{B_{1u}^{}}_{\bar{A}_{1}}^{B_{1}^{}}}$ &   $10.0716$ &                                                                                                                $d\indices*{*_{A_{3}}^{B_{2}^{}}_{X_{2}}^{E_{u}^{}}_{\bar{A}_{1}}^{B_{1}^{}}}$ &    $8.8323$ &                                                                               $d\indices*{*_{A_{3}}^{B_{2}^{}}_{X_{2}}^{\tensor*[^{1}]{\hspace{-0.2em}E}{}_{u}^{}}_{\bar{A}_{1}}^{B_{1}^{}}}$ & $- 12.3931$ \\[0.4em]
                                                                                $d\indices*{*_{A_{3}}^{\tensor*[^{1}]{\hspace{-0.2em}B}{}_{2}^{}}_{X_{2}}^{E_{g}^{}}_{\bar{A}_{1}}^{B_{1}^{}}}$ &    $1.4901$ &                                                                              $d\indices*{*_{A_{3}}^{\tensor*[^{1}]{\hspace{-0.2em}B}{}_{2}^{}}_{X_{2}}^{B_{1u}^{}}_{\bar{A}_{1}}^{B_{1}^{}}}$ &  $- 7.5542$ &                                                                               $d\indices*{*_{A_{3}}^{\tensor*[^{1}]{\hspace{-0.2em}B}{}_{2}^{}}_{X_{2}}^{E_{u}^{}}_{\bar{A}_{1}}^{B_{1}^{}}}$ &  $- 6.6207$ &                                              $d\indices*{*_{A_{3}}^{\tensor*[^{1}]{\hspace{-0.2em}B}{}_{2}^{}}_{X_{2}}^{\tensor*[^{1}]{\hspace{-0.2em}E}{}_{u}^{}}_{\bar{A}_{1}}^{B_{1}^{}}}$ &    $8.5416$ &                                             $d\indices*{*_{A_{3}}^{\tensor*[^{1}]{\hspace{-0.2em}B}{}_{1}^{}}_{X_{2}}^{A_{1g}^{}}_{\bar{A}_{1}}^{\tensor*[^{1}]{\hspace{-0.2em}B}{}_{1}^{}}}$ &   $24.7133$ \\[0.4em]
                                              $d\indices*{*_{A_{3}}^{\tensor*[^{1}]{\hspace{-0.2em}B}{}_{1}^{}}_{X_{2}}^{A_{2u}^{}}_{\bar{A}_{1}}^{\tensor*[^{1}]{\hspace{-0.2em}B}{}_{1}^{}}}$ & $- 17.7853$ &                                             $d\indices*{*_{A_{3}}^{\tensor*[^{1}]{\hspace{-0.2em}B}{}_{1}^{}}_{X_{2}}^{A_{1g}^{}}_{\bar{A}_{1}}^{\tensor*[^{2}]{\hspace{-0.2em}B}{}_{1}^{}}}$ &   $15.0099$ &                                              $d\indices*{*_{A_{3}}^{\tensor*[^{1}]{\hspace{-0.2em}B}{}_{1}^{}}_{X_{2}}^{E_{g}^{}}_{\bar{A}_{1}}^{\tensor*[^{2}]{\hspace{-0.2em}B}{}_{1}^{}}}$ &   $13.9646$ &                                             $d\indices*{*_{A_{3}}^{\tensor*[^{1}]{\hspace{-0.2em}B}{}_{1}^{}}_{X_{2}}^{A_{2u}^{}}_{\bar{A}_{1}}^{\tensor*[^{2}]{\hspace{-0.2em}B}{}_{1}^{}}}$ &    $3.6305$ &                                              $d\indices*{*_{A_{3}}^{\tensor*[^{1}]{\hspace{-0.2em}B}{}_{1}^{}}_{X_{2}}^{E_{u}^{}}_{\bar{A}_{1}}^{\tensor*[^{2}]{\hspace{-0.2em}B}{}_{1}^{}}}$ &  $- 0.9625$ \\[0.4em]
              $d\indices*{*_{A_{3}}^{\tensor*[^{1}]{\hspace{-0.2em}B}{}_{1}^{}}_{X_{2}}^{\tensor*[^{1}]{\hspace{-0.2em}E}{}_{u}^{}}_{\bar{A}_{1}}^{\tensor*[^{2}]{\hspace{-0.2em}B}{}_{1}^{}}}$ &    $2.6140$ &                                                                               $d\indices*{*_{A_{3}}^{B_{2}^{}}_{X_{2}}^{E_{g}^{}}_{\bar{A}_{1}}^{\tensor*[^{1}]{\hspace{-0.2em}B}{}_{1}^{}}}$ &   $13.3116$ &                                                                              $d\indices*{*_{A_{3}}^{B_{2}^{}}_{X_{2}}^{B_{1u}^{}}_{\bar{A}_{1}}^{\tensor*[^{1}]{\hspace{-0.2em}B}{}_{1}^{}}}$ &    $3.8953$ &                                                                               $d\indices*{*_{A_{3}}^{B_{2}^{}}_{X_{2}}^{E_{u}^{}}_{\bar{A}_{1}}^{\tensor*[^{1}]{\hspace{-0.2em}B}{}_{1}^{}}}$ &    $1.3759$ &                                              $d\indices*{*_{A_{3}}^{B_{2}^{}}_{X_{2}}^{\tensor*[^{1}]{\hspace{-0.2em}E}{}_{u}^{}}_{\bar{A}_{1}}^{\tensor*[^{1}]{\hspace{-0.2em}B}{}_{1}^{}}}$ & $- 16.5303$ \\[0.4em]
                                               $d\indices*{*_{A_{3}}^{\tensor*[^{1}]{\hspace{-0.2em}B}{}_{2}^{}}_{X_{2}}^{E_{g}^{}}_{\bar{A}_{1}}^{\tensor*[^{1}]{\hspace{-0.2em}B}{}_{1}^{}}}$ &  $- 9.0557$ &                                             $d\indices*{*_{A_{3}}^{\tensor*[^{1}]{\hspace{-0.2em}B}{}_{2}^{}}_{X_{2}}^{B_{1u}^{}}_{\bar{A}_{1}}^{\tensor*[^{1}]{\hspace{-0.2em}B}{}_{1}^{}}}$ &  $- 2.0548$ &                                              $d\indices*{*_{A_{3}}^{\tensor*[^{1}]{\hspace{-0.2em}B}{}_{2}^{}}_{X_{2}}^{E_{u}^{}}_{\bar{A}_{1}}^{\tensor*[^{1}]{\hspace{-0.2em}B}{}_{1}^{}}}$ &  $- 1.2641$ &             $d\indices*{*_{A_{3}}^{\tensor*[^{1}]{\hspace{-0.2em}B}{}_{2}^{}}_{X_{2}}^{\tensor*[^{1}]{\hspace{-0.2em}E}{}_{u}^{}}_{\bar{A}_{1}}^{\tensor*[^{1}]{\hspace{-0.2em}B}{}_{1}^{}}}$ &    $2.3688$ &                                             $d\indices*{*_{A_{3}}^{\tensor*[^{2}]{\hspace{-0.2em}B}{}_{1}^{}}_{X_{2}}^{A_{1g}^{}}_{\bar{A}_{1}}^{\tensor*[^{2}]{\hspace{-0.2em}B}{}_{1}^{}}}$ &    $5.9379$ \\[0.4em]
                                              $d\indices*{*_{A_{3}}^{\tensor*[^{2}]{\hspace{-0.2em}B}{}_{1}^{}}_{X_{2}}^{A_{2u}^{}}_{\bar{A}_{1}}^{\tensor*[^{2}]{\hspace{-0.2em}B}{}_{1}^{}}}$ &    $5.0361$ &                                                                               $d\indices*{*_{A_{3}}^{B_{2}^{}}_{X_{2}}^{E_{g}^{}}_{\bar{A}_{1}}^{\tensor*[^{2}]{\hspace{-0.2em}B}{}_{1}^{}}}$ &   $13.3411$ &                                                                              $d\indices*{*_{A_{3}}^{B_{2}^{}}_{X_{2}}^{B_{1u}^{}}_{\bar{A}_{1}}^{\tensor*[^{2}]{\hspace{-0.2em}B}{}_{1}^{}}}$ &  $- 1.3586$ &                                                                               $d\indices*{*_{A_{3}}^{B_{2}^{}}_{X_{2}}^{E_{u}^{}}_{\bar{A}_{1}}^{\tensor*[^{2}]{\hspace{-0.2em}B}{}_{1}^{}}}$ &  $- 1.0007$ &                                              $d\indices*{*_{A_{3}}^{B_{2}^{}}_{X_{2}}^{\tensor*[^{1}]{\hspace{-0.2em}E}{}_{u}^{}}_{\bar{A}_{1}}^{\tensor*[^{2}]{\hspace{-0.2em}B}{}_{1}^{}}}$ &    $4.8322$ \\[0.4em]
                                               $d\indices*{*_{A_{3}}^{\tensor*[^{1}]{\hspace{-0.2em}B}{}_{2}^{}}_{X_{2}}^{E_{g}^{}}_{\bar{A}_{1}}^{\tensor*[^{2}]{\hspace{-0.2em}B}{}_{1}^{}}}$ & $- 10.5832$ &                                             $d\indices*{*_{A_{3}}^{\tensor*[^{1}]{\hspace{-0.2em}B}{}_{2}^{}}_{X_{2}}^{B_{1u}^{}}_{\bar{A}_{1}}^{\tensor*[^{2}]{\hspace{-0.2em}B}{}_{1}^{}}}$ &    $3.5330$ &                                              $d\indices*{*_{A_{3}}^{\tensor*[^{1}]{\hspace{-0.2em}B}{}_{2}^{}}_{X_{2}}^{E_{u}^{}}_{\bar{A}_{1}}^{\tensor*[^{2}]{\hspace{-0.2em}B}{}_{1}^{}}}$ &  $- 7.0915$ &             $d\indices*{*_{A_{3}}^{\tensor*[^{1}]{\hspace{-0.2em}B}{}_{2}^{}}_{X_{2}}^{\tensor*[^{1}]{\hspace{-0.2em}E}{}_{u}^{}}_{\bar{A}_{1}}^{\tensor*[^{2}]{\hspace{-0.2em}B}{}_{1}^{}}}$ &  $- 1.0847$ &                                                                                                               $d\indices*{*_{A_{3}}^{B_{2}^{}}_{X_{2}}^{A_{1g}^{}}_{\bar{A}_{1}}^{B_{2}^{}}}$ &    $3.2768$ \\[0.4em]
                                                                                                                $d\indices*{*_{A_{3}}^{B_{2}^{}}_{X_{2}}^{A_{2u}^{}}_{\bar{A}_{1}}^{B_{2}^{}}}$ &  $- 5.5258$ &                                                                              $d\indices*{*_{A_{3}}^{B_{2}^{}}_{X_{2}}^{A_{1g}^{}}_{\bar{A}_{1}}^{\tensor*[^{1}]{\hspace{-0.2em}B}{}_{2}^{}}}$ &  $- 0.0927$ &                                                                               $d\indices*{*_{A_{3}}^{B_{2}^{}}_{X_{2}}^{E_{g}^{}}_{\bar{A}_{1}}^{\tensor*[^{1}]{\hspace{-0.2em}B}{}_{2}^{}}}$ &    $0.3655$ &                                                                              $d\indices*{*_{A_{3}}^{B_{2}^{}}_{X_{2}}^{A_{2u}^{}}_{\bar{A}_{1}}^{\tensor*[^{1}]{\hspace{-0.2em}B}{}_{2}^{}}}$ &  $- 1.6295$ &                                                                               $d\indices*{*_{A_{3}}^{B_{2}^{}}_{X_{2}}^{E_{u}^{}}_{\bar{A}_{1}}^{\tensor*[^{1}]{\hspace{-0.2em}B}{}_{2}^{}}}$ &  $- 0.4028$ \\[0.4em]
                                               $d\indices*{*_{A_{3}}^{B_{2}^{}}_{X_{2}}^{\tensor*[^{1}]{\hspace{-0.2em}E}{}_{u}^{}}_{\bar{A}_{1}}^{\tensor*[^{1}]{\hspace{-0.2em}B}{}_{2}^{}}}$ &    $0.5993$ &                                             $d\indices*{*_{A_{3}}^{\tensor*[^{1}]{\hspace{-0.2em}B}{}_{2}^{}}_{X_{2}}^{A_{1g}^{}}_{\bar{A}_{1}}^{\tensor*[^{1}]{\hspace{-0.2em}B}{}_{2}^{}}}$ &  $- 8.2815$ &                                             $d\indices*{*_{A_{3}}^{\tensor*[^{1}]{\hspace{-0.2em}B}{}_{2}^{}}_{X_{2}}^{A_{2u}^{}}_{\bar{A}_{1}}^{\tensor*[^{1}]{\hspace{-0.2em}B}{}_{2}^{}}}$ &    $5.5878$ &                                                                                                          $d\indices*{*_{A_{5}}^{A_{1}^{}}_{\bar{A}_{4}}^{A_{1}^{}}_{\bar{A}_{1}}^{A_{1}^{}}}$ &  $- 8.1272$ &                                                                         $d\indices*{*_{A_{5}}^{A_{1}^{}}_{\bar{A}_{4}}^{A_{1}^{}}_{\bar{A}_{1}}^{\tensor*[^{1}]{\hspace{-0.2em}A}{}_{1}^{}}}$ &    $0.9005$ \\[0.4em]
                                                                          $d\indices*{*_{A_{5}}^{A_{1}^{}}_{\bar{A}_{4}}^{A_{1}^{}}_{\bar{A}_{1}}^{\tensor*[^{2}]{\hspace{-0.2em}A}{}_{1}^{}}}$ &   $21.9342$ &                                                                                                          $d\indices*{*_{A_{5}}^{A_{1}^{}}_{\bar{A}_{4}}^{A_{1}^{}}_{\bar{A}_{1}}^{B_{1}^{}}}$ &   $13.0670$ &                                                                         $d\indices*{*_{A_{5}}^{A_{1}^{}}_{\bar{A}_{4}}^{A_{1}^{}}_{\bar{A}_{1}}^{\tensor*[^{1}]{\hspace{-0.2em}B}{}_{1}^{}}}$ &    $9.3672$ &                                                                         $d\indices*{*_{A_{5}}^{A_{1}^{}}_{\bar{A}_{4}}^{A_{1}^{}}_{\bar{A}_{1}}^{\tensor*[^{2}]{\hspace{-0.2em}B}{}_{1}^{}}}$ &  $- 1.4900$ &                                        $d\indices*{*_{A_{5}}^{A_{1}^{}}_{\bar{A}_{4}}^{\tensor*[^{1}]{\hspace{-0.2em}A}{}_{1}^{}}_{\bar{A}_{1}}^{\tensor*[^{1}]{\hspace{-0.2em}A}{}_{1}^{}}}$ &    $2.0059$ \\[0.4em]
                                         $d\indices*{*_{A_{5}}^{A_{1}^{}}_{\bar{A}_{4}}^{\tensor*[^{1}]{\hspace{-0.2em}A}{}_{1}^{}}_{\bar{A}_{1}}^{\tensor*[^{2}]{\hspace{-0.2em}A}{}_{1}^{}}}$ &  $- 9.7489$ &                                                                         $d\indices*{*_{A_{5}}^{A_{1}^{}}_{\bar{A}_{4}}^{\tensor*[^{1}]{\hspace{-0.2em}A}{}_{1}^{}}_{\bar{A}_{1}}^{A_{2}^{}}}$ &    $0.8688$ &                                                                         $d\indices*{*_{A_{5}}^{A_{1}^{}}_{\bar{A}_{4}}^{\tensor*[^{1}]{\hspace{-0.2em}A}{}_{1}^{}}_{\bar{A}_{1}}^{B_{1}^{}}}$ & $- 13.7873$ &                                        $d\indices*{*_{A_{5}}^{A_{1}^{}}_{\bar{A}_{4}}^{\tensor*[^{1}]{\hspace{-0.2em}A}{}_{1}^{}}_{\bar{A}_{1}}^{\tensor*[^{1}]{\hspace{-0.2em}B}{}_{1}^{}}}$ &   $10.2185$ &                                        $d\indices*{*_{A_{5}}^{A_{1}^{}}_{\bar{A}_{4}}^{\tensor*[^{1}]{\hspace{-0.2em}A}{}_{1}^{}}_{\bar{A}_{1}}^{\tensor*[^{2}]{\hspace{-0.2em}B}{}_{1}^{}}}$ &    $3.3671$ \\[0.4em]
                                                                          $d\indices*{*_{A_{5}}^{A_{1}^{}}_{\bar{A}_{4}}^{\tensor*[^{1}]{\hspace{-0.2em}A}{}_{1}^{}}_{\bar{A}_{1}}^{B_{2}^{}}}$ & $- 13.6593$ &                                        $d\indices*{*_{A_{5}}^{A_{1}^{}}_{\bar{A}_{4}}^{\tensor*[^{1}]{\hspace{-0.2em}A}{}_{1}^{}}_{\bar{A}_{1}}^{\tensor*[^{1}]{\hspace{-0.2em}B}{}_{2}^{}}}$ &   $10.6549$ &                                        $d\indices*{*_{A_{5}}^{A_{1}^{}}_{\bar{A}_{4}}^{\tensor*[^{2}]{\hspace{-0.2em}A}{}_{1}^{}}_{\bar{A}_{1}}^{\tensor*[^{2}]{\hspace{-0.2em}A}{}_{1}^{}}}$ &    $0.6873$ &                                                                         $d\indices*{*_{A_{5}}^{A_{1}^{}}_{\bar{A}_{4}}^{\tensor*[^{2}]{\hspace{-0.2em}A}{}_{1}^{}}_{\bar{A}_{1}}^{A_{2}^{}}}$ &    $0.8201$ &                                                                         $d\indices*{*_{A_{5}}^{A_{1}^{}}_{\bar{A}_{4}}^{\tensor*[^{2}]{\hspace{-0.2em}A}{}_{1}^{}}_{\bar{A}_{1}}^{B_{1}^{}}}$ &    $1.1004$ \\[0.4em]
                                         $d\indices*{*_{A_{5}}^{A_{1}^{}}_{\bar{A}_{4}}^{\tensor*[^{2}]{\hspace{-0.2em}A}{}_{1}^{}}_{\bar{A}_{1}}^{\tensor*[^{1}]{\hspace{-0.2em}B}{}_{1}^{}}}$ & $- 12.6588$ &                                        $d\indices*{*_{A_{5}}^{A_{1}^{}}_{\bar{A}_{4}}^{\tensor*[^{2}]{\hspace{-0.2em}A}{}_{1}^{}}_{\bar{A}_{1}}^{\tensor*[^{2}]{\hspace{-0.2em}B}{}_{1}^{}}}$ & $- 13.5239$ &                                                                         $d\indices*{*_{A_{5}}^{A_{1}^{}}_{\bar{A}_{4}}^{\tensor*[^{2}]{\hspace{-0.2em}A}{}_{1}^{}}_{\bar{A}_{1}}^{B_{2}^{}}}$ &  $- 2.8547$ &                                        $d\indices*{*_{A_{5}}^{A_{1}^{}}_{\bar{A}_{4}}^{\tensor*[^{2}]{\hspace{-0.2em}A}{}_{1}^{}}_{\bar{A}_{1}}^{\tensor*[^{1}]{\hspace{-0.2em}B}{}_{2}^{}}}$ &    $0.3392$ &                                                                                                          $d\indices*{*_{A_{5}}^{A_{1}^{}}_{\bar{A}_{4}}^{A_{2}^{}}_{\bar{A}_{1}}^{A_{2}^{}}}$ &  $- 3.0583$ \\[0.4em]
                                                                                                           $d\indices*{*_{A_{5}}^{A_{1}^{}}_{\bar{A}_{4}}^{A_{2}^{}}_{\bar{A}_{1}}^{B_{1}^{}}}$ &   $11.1191$ &                                                                         $d\indices*{*_{A_{5}}^{A_{1}^{}}_{\bar{A}_{4}}^{A_{2}^{}}_{\bar{A}_{1}}^{\tensor*[^{1}]{\hspace{-0.2em}B}{}_{1}^{}}}$ & $- 14.0189$ &                                                                         $d\indices*{*_{A_{5}}^{A_{1}^{}}_{\bar{A}_{4}}^{A_{2}^{}}_{\bar{A}_{1}}^{\tensor*[^{2}]{\hspace{-0.2em}B}{}_{1}^{}}}$ &    $1.6498$ &                                                                                                          $d\indices*{*_{A_{5}}^{A_{1}^{}}_{\bar{A}_{4}}^{A_{2}^{}}_{\bar{A}_{1}}^{B_{2}^{}}}$ & $- 18.3532$ &                                                                         $d\indices*{*_{A_{5}}^{A_{1}^{}}_{\bar{A}_{4}}^{A_{2}^{}}_{\bar{A}_{1}}^{\tensor*[^{1}]{\hspace{-0.2em}B}{}_{2}^{}}}$ &   $11.1399$ \\[0.4em]
                                                                                                           $d\indices*{*_{A_{5}}^{A_{1}^{}}_{\bar{A}_{4}}^{B_{1}^{}}_{\bar{A}_{1}}^{B_{1}^{}}}$ &  $- 0.1756$ &                                                                         $d\indices*{*_{A_{5}}^{A_{1}^{}}_{\bar{A}_{4}}^{B_{1}^{}}_{\bar{A}_{1}}^{\tensor*[^{1}]{\hspace{-0.2em}B}{}_{1}^{}}}$ &   $11.9351$ &                                                                         $d\indices*{*_{A_{5}}^{A_{1}^{}}_{\bar{A}_{4}}^{B_{1}^{}}_{\bar{A}_{1}}^{\tensor*[^{2}]{\hspace{-0.2em}B}{}_{1}^{}}}$ &   $11.9270$ &                                                                                                          $d\indices*{*_{A_{5}}^{A_{1}^{}}_{\bar{A}_{4}}^{B_{2}^{}}_{\bar{A}_{1}}^{B_{1}^{}}}$ &  $- 1.0933$ &                                                                         $d\indices*{*_{A_{5}}^{A_{1}^{}}_{\bar{A}_{4}}^{\tensor*[^{1}]{\hspace{-0.2em}B}{}_{2}^{}}_{\bar{A}_{1}}^{B_{1}^{}}}$ &    $2.1282$ \\[0.4em]
                                         $d\indices*{*_{A_{5}}^{A_{1}^{}}_{\bar{A}_{4}}^{\tensor*[^{1}]{\hspace{-0.2em}B}{}_{1}^{}}_{\bar{A}_{1}}^{\tensor*[^{1}]{\hspace{-0.2em}B}{}_{1}^{}}}$ &    $2.7159$ &                                        $d\indices*{*_{A_{5}}^{A_{1}^{}}_{\bar{A}_{4}}^{\tensor*[^{1}]{\hspace{-0.2em}B}{}_{1}^{}}_{\bar{A}_{1}}^{\tensor*[^{2}]{\hspace{-0.2em}B}{}_{1}^{}}}$ &  $- 3.2220$ &                                                                         $d\indices*{*_{A_{5}}^{A_{1}^{}}_{\bar{A}_{4}}^{B_{2}^{}}_{\bar{A}_{1}}^{\tensor*[^{1}]{\hspace{-0.2em}B}{}_{1}^{}}}$ &  $- 4.0556$ &                                        $d\indices*{*_{A_{5}}^{A_{1}^{}}_{\bar{A}_{4}}^{\tensor*[^{1}]{\hspace{-0.2em}B}{}_{2}^{}}_{\bar{A}_{1}}^{\tensor*[^{1}]{\hspace{-0.2em}B}{}_{1}^{}}}$ &    $0.7848$ &                                        $d\indices*{*_{A_{5}}^{A_{1}^{}}_{\bar{A}_{4}}^{\tensor*[^{2}]{\hspace{-0.2em}B}{}_{1}^{}}_{\bar{A}_{1}}^{\tensor*[^{2}]{\hspace{-0.2em}B}{}_{1}^{}}}$ &  $- 4.5800$ \\[0.4em]
                                                                          $d\indices*{*_{A_{5}}^{A_{1}^{}}_{\bar{A}_{4}}^{B_{2}^{}}_{\bar{A}_{1}}^{\tensor*[^{2}]{\hspace{-0.2em}B}{}_{1}^{}}}$ &  $- 0.3555$ &                                        $d\indices*{*_{A_{5}}^{A_{1}^{}}_{\bar{A}_{4}}^{\tensor*[^{1}]{\hspace{-0.2em}B}{}_{2}^{}}_{\bar{A}_{1}}^{\tensor*[^{2}]{\hspace{-0.2em}B}{}_{1}^{}}}$ &  $- 1.5955$ &                                                                                                          $d\indices*{*_{A_{5}}^{A_{1}^{}}_{\bar{A}_{4}}^{B_{2}^{}}_{\bar{A}_{1}}^{B_{2}^{}}}$ &  $- 1.8437$ &                                                                         $d\indices*{*_{A_{5}}^{A_{1}^{}}_{\bar{A}_{4}}^{B_{2}^{}}_{\bar{A}_{1}}^{\tensor*[^{1}]{\hspace{-0.2em}B}{}_{2}^{}}}$ &    $3.7120$ &                                        $d\indices*{*_{A_{5}}^{A_{1}^{}}_{\bar{A}_{4}}^{\tensor*[^{1}]{\hspace{-0.2em}B}{}_{2}^{}}_{\bar{A}_{1}}^{\tensor*[^{1}]{\hspace{-0.2em}B}{}_{2}^{}}}$ &  $- 2.4520$ \\[0.4em]
        $d\indices*{*_{A_{5}}^{\tensor*[^{1}]{\hspace{-0.2em}A}{}_{1}^{}}_{\bar{A}_{4}}^{\tensor*[^{1}]{\hspace{-0.2em}A}{}_{1}^{}}_{\bar{A}_{1}}^{\tensor*[^{1}]{\hspace{-0.2em}A}{}_{1}^{}}}$ &    $1.2379$ &       $d\indices*{*_{A_{5}}^{\tensor*[^{1}]{\hspace{-0.2em}A}{}_{1}^{}}_{\bar{A}_{4}}^{\tensor*[^{1}]{\hspace{-0.2em}A}{}_{1}^{}}_{\bar{A}_{1}}^{\tensor*[^{2}]{\hspace{-0.2em}A}{}_{1}^{}}}$ &    $0.9838$ &                                        $d\indices*{*_{A_{5}}^{\tensor*[^{1}]{\hspace{-0.2em}A}{}_{1}^{}}_{\bar{A}_{4}}^{\tensor*[^{1}]{\hspace{-0.2em}A}{}_{1}^{}}_{\bar{A}_{1}}^{B_{1}^{}}}$ &   $15.1149$ &       $d\indices*{*_{A_{5}}^{\tensor*[^{1}]{\hspace{-0.2em}A}{}_{1}^{}}_{\bar{A}_{4}}^{\tensor*[^{1}]{\hspace{-0.2em}A}{}_{1}^{}}_{\bar{A}_{1}}^{\tensor*[^{1}]{\hspace{-0.2em}B}{}_{1}^{}}}$ & $- 12.2046$ &       $d\indices*{*_{A_{5}}^{\tensor*[^{1}]{\hspace{-0.2em}A}{}_{1}^{}}_{\bar{A}_{4}}^{\tensor*[^{1}]{\hspace{-0.2em}A}{}_{1}^{}}_{\bar{A}_{1}}^{\tensor*[^{2}]{\hspace{-0.2em}B}{}_{1}^{}}}$ &  $- 4.4258$ \\[0.4em]
        $d\indices*{*_{A_{5}}^{\tensor*[^{1}]{\hspace{-0.2em}A}{}_{1}^{}}_{\bar{A}_{4}}^{\tensor*[^{2}]{\hspace{-0.2em}A}{}_{1}^{}}_{\bar{A}_{1}}^{\tensor*[^{2}]{\hspace{-0.2em}A}{}_{1}^{}}}$ &    $5.9835$ &                                        $d\indices*{*_{A_{5}}^{\tensor*[^{1}]{\hspace{-0.2em}A}{}_{1}^{}}_{\bar{A}_{4}}^{\tensor*[^{2}]{\hspace{-0.2em}A}{}_{1}^{}}_{\bar{A}_{1}}^{A_{2}^{}}}$ &  $- 9.5796$ &                                        $d\indices*{*_{A_{5}}^{\tensor*[^{1}]{\hspace{-0.2em}A}{}_{1}^{}}_{\bar{A}_{4}}^{\tensor*[^{2}]{\hspace{-0.2em}A}{}_{1}^{}}_{\bar{A}_{1}}^{B_{1}^{}}}$ &  $- 3.7091$ &       $d\indices*{*_{A_{5}}^{\tensor*[^{1}]{\hspace{-0.2em}A}{}_{1}^{}}_{\bar{A}_{4}}^{\tensor*[^{2}]{\hspace{-0.2em}A}{}_{1}^{}}_{\bar{A}_{1}}^{\tensor*[^{1}]{\hspace{-0.2em}B}{}_{1}^{}}}$ &    $8.2027$ &       $d\indices*{*_{A_{5}}^{\tensor*[^{1}]{\hspace{-0.2em}A}{}_{1}^{}}_{\bar{A}_{4}}^{\tensor*[^{2}]{\hspace{-0.2em}A}{}_{1}^{}}_{\bar{A}_{1}}^{\tensor*[^{2}]{\hspace{-0.2em}B}{}_{1}^{}}}$ &    $7.5431$ \\[0.4em]
                                         $d\indices*{*_{A_{5}}^{\tensor*[^{1}]{\hspace{-0.2em}A}{}_{1}^{}}_{\bar{A}_{4}}^{\tensor*[^{2}]{\hspace{-0.2em}A}{}_{1}^{}}_{\bar{A}_{1}}^{B_{2}^{}}}$ &  $- 1.5619$ &       $d\indices*{*_{A_{5}}^{\tensor*[^{1}]{\hspace{-0.2em}A}{}_{1}^{}}_{\bar{A}_{4}}^{\tensor*[^{2}]{\hspace{-0.2em}A}{}_{1}^{}}_{\bar{A}_{1}}^{\tensor*[^{1}]{\hspace{-0.2em}B}{}_{2}^{}}}$ &    $6.1174$ &                                                                         $d\indices*{*_{A_{5}}^{\tensor*[^{1}]{\hspace{-0.2em}A}{}_{1}^{}}_{\bar{A}_{4}}^{A_{2}^{}}_{\bar{A}_{1}}^{A_{2}^{}}}$ &   $11.8812$ &                                                                         $d\indices*{*_{A_{5}}^{\tensor*[^{1}]{\hspace{-0.2em}A}{}_{1}^{}}_{\bar{A}_{4}}^{A_{2}^{}}_{\bar{A}_{1}}^{B_{1}^{}}}$ &  $- 0.2902$ &                                        $d\indices*{*_{A_{5}}^{\tensor*[^{1}]{\hspace{-0.2em}A}{}_{1}^{}}_{\bar{A}_{4}}^{A_{2}^{}}_{\bar{A}_{1}}^{\tensor*[^{1}]{\hspace{-0.2em}B}{}_{1}^{}}}$ &    $6.0715$ \\[0.4em]
                                         $d\indices*{*_{A_{5}}^{\tensor*[^{1}]{\hspace{-0.2em}A}{}_{1}^{}}_{\bar{A}_{4}}^{A_{2}^{}}_{\bar{A}_{1}}^{\tensor*[^{2}]{\hspace{-0.2em}B}{}_{1}^{}}}$ &    $7.0789$ &                                                                         $d\indices*{*_{A_{5}}^{\tensor*[^{1}]{\hspace{-0.2em}A}{}_{1}^{}}_{\bar{A}_{4}}^{A_{2}^{}}_{\bar{A}_{1}}^{B_{2}^{}}}$ &    $7.6367$ &                                        $d\indices*{*_{A_{5}}^{\tensor*[^{1}]{\hspace{-0.2em}A}{}_{1}^{}}_{\bar{A}_{4}}^{A_{2}^{}}_{\bar{A}_{1}}^{\tensor*[^{1}]{\hspace{-0.2em}B}{}_{2}^{}}}$ &  $- 5.7200$ &                                                                         $d\indices*{*_{A_{5}}^{\tensor*[^{1}]{\hspace{-0.2em}A}{}_{1}^{}}_{\bar{A}_{4}}^{B_{1}^{}}_{\bar{A}_{1}}^{B_{1}^{}}}$ &  $- 0.1644$ &                                        $d\indices*{*_{A_{5}}^{\tensor*[^{1}]{\hspace{-0.2em}A}{}_{1}^{}}_{\bar{A}_{4}}^{B_{1}^{}}_{\bar{A}_{1}}^{\tensor*[^{1}]{\hspace{-0.2em}B}{}_{1}^{}}}$ &  $- 0.0132$ \\[0.4em]
                                         $d\indices*{*_{A_{5}}^{\tensor*[^{1}]{\hspace{-0.2em}A}{}_{1}^{}}_{\bar{A}_{4}}^{B_{1}^{}}_{\bar{A}_{1}}^{\tensor*[^{2}]{\hspace{-0.2em}B}{}_{1}^{}}}$ &  $- 0.2503$ &                                                                         $d\indices*{*_{A_{5}}^{\tensor*[^{1}]{\hspace{-0.2em}A}{}_{1}^{}}_{\bar{A}_{4}}^{B_{2}^{}}_{\bar{A}_{1}}^{B_{1}^{}}}$ &    $1.5807$ &                                        $d\indices*{*_{A_{5}}^{\tensor*[^{1}]{\hspace{-0.2em}A}{}_{1}^{}}_{\bar{A}_{4}}^{\tensor*[^{1}]{\hspace{-0.2em}B}{}_{2}^{}}_{\bar{A}_{1}}^{B_{1}^{}}}$ &  $- 3.0652$ &       $d\indices*{*_{A_{5}}^{\tensor*[^{1}]{\hspace{-0.2em}A}{}_{1}^{}}_{\bar{A}_{4}}^{\tensor*[^{1}]{\hspace{-0.2em}B}{}_{1}^{}}_{\bar{A}_{1}}^{\tensor*[^{1}]{\hspace{-0.2em}B}{}_{1}^{}}}$ & $- 10.3287$ &       $d\indices*{*_{A_{5}}^{\tensor*[^{1}]{\hspace{-0.2em}A}{}_{1}^{}}_{\bar{A}_{4}}^{\tensor*[^{1}]{\hspace{-0.2em}B}{}_{1}^{}}_{\bar{A}_{1}}^{\tensor*[^{2}]{\hspace{-0.2em}B}{}_{1}^{}}}$ &  $- 3.2027$ \\[0.4em]
                                         $d\indices*{*_{A_{5}}^{\tensor*[^{1}]{\hspace{-0.2em}A}{}_{1}^{}}_{\bar{A}_{4}}^{B_{2}^{}}_{\bar{A}_{1}}^{\tensor*[^{1}]{\hspace{-0.2em}B}{}_{1}^{}}}$ &  $- 9.4692$ &       $d\indices*{*_{A_{5}}^{\tensor*[^{1}]{\hspace{-0.2em}A}{}_{1}^{}}_{\bar{A}_{4}}^{\tensor*[^{1}]{\hspace{-0.2em}B}{}_{2}^{}}_{\bar{A}_{1}}^{\tensor*[^{1}]{\hspace{-0.2em}B}{}_{1}^{}}}$ &    $6.0951$ &       $d\indices*{*_{A_{5}}^{\tensor*[^{1}]{\hspace{-0.2em}A}{}_{1}^{}}_{\bar{A}_{4}}^{\tensor*[^{2}]{\hspace{-0.2em}B}{}_{1}^{}}_{\bar{A}_{1}}^{\tensor*[^{2}]{\hspace{-0.2em}B}{}_{1}^{}}}$ &  $- 2.2336$ &                                        $d\indices*{*_{A_{5}}^{\tensor*[^{1}]{\hspace{-0.2em}A}{}_{1}^{}}_{\bar{A}_{4}}^{B_{2}^{}}_{\bar{A}_{1}}^{\tensor*[^{2}]{\hspace{-0.2em}B}{}_{1}^{}}}$ &  $- 8.9638$ &       $d\indices*{*_{A_{5}}^{\tensor*[^{1}]{\hspace{-0.2em}A}{}_{1}^{}}_{\bar{A}_{4}}^{\tensor*[^{1}]{\hspace{-0.2em}B}{}_{2}^{}}_{\bar{A}_{1}}^{\tensor*[^{2}]{\hspace{-0.2em}B}{}_{1}^{}}}$ &    $6.1460$ \\[0.4em]
                                                                          $d\indices*{*_{A_{5}}^{\tensor*[^{1}]{\hspace{-0.2em}A}{}_{1}^{}}_{\bar{A}_{4}}^{B_{2}^{}}_{\bar{A}_{1}}^{B_{2}^{}}}$ &    $5.4699$ &                                        $d\indices*{*_{A_{5}}^{\tensor*[^{1}]{\hspace{-0.2em}A}{}_{1}^{}}_{\bar{A}_{4}}^{B_{2}^{}}_{\bar{A}_{1}}^{\tensor*[^{1}]{\hspace{-0.2em}B}{}_{2}^{}}}$ &  $- 2.0489$ &       $d\indices*{*_{A_{5}}^{\tensor*[^{1}]{\hspace{-0.2em}A}{}_{1}^{}}_{\bar{A}_{4}}^{\tensor*[^{1}]{\hspace{-0.2em}B}{}_{2}^{}}_{\bar{A}_{1}}^{\tensor*[^{1}]{\hspace{-0.2em}B}{}_{2}^{}}}$ &  $- 6.2816$ &       $d\indices*{*_{A_{5}}^{\tensor*[^{2}]{\hspace{-0.2em}A}{}_{1}^{}}_{\bar{A}_{4}}^{\tensor*[^{2}]{\hspace{-0.2em}A}{}_{1}^{}}_{\bar{A}_{1}}^{\tensor*[^{2}]{\hspace{-0.2em}A}{}_{1}^{}}}$ & $- 14.9144$ &                                        $d\indices*{*_{A_{5}}^{\tensor*[^{2}]{\hspace{-0.2em}A}{}_{1}^{}}_{\bar{A}_{4}}^{\tensor*[^{2}]{\hspace{-0.2em}A}{}_{1}^{}}_{\bar{A}_{1}}^{B_{1}^{}}}$ & $- 10.9129$ \\[0.4em]
        $d\indices*{*_{A_{5}}^{\tensor*[^{2}]{\hspace{-0.2em}A}{}_{1}^{}}_{\bar{A}_{4}}^{\tensor*[^{2}]{\hspace{-0.2em}A}{}_{1}^{}}_{\bar{A}_{1}}^{\tensor*[^{1}]{\hspace{-0.2em}B}{}_{1}^{}}}$ &    $1.3741$ &       $d\indices*{*_{A_{5}}^{\tensor*[^{2}]{\hspace{-0.2em}A}{}_{1}^{}}_{\bar{A}_{4}}^{\tensor*[^{2}]{\hspace{-0.2em}A}{}_{1}^{}}_{\bar{A}_{1}}^{\tensor*[^{2}]{\hspace{-0.2em}B}{}_{1}^{}}}$ &  $- 4.0114$ &                                                                         $d\indices*{*_{A_{5}}^{\tensor*[^{2}]{\hspace{-0.2em}A}{}_{1}^{}}_{\bar{A}_{4}}^{A_{2}^{}}_{\bar{A}_{1}}^{A_{2}^{}}}$ & $- 12.2067$ &                                                                         $d\indices*{*_{A_{5}}^{\tensor*[^{2}]{\hspace{-0.2em}A}{}_{1}^{}}_{\bar{A}_{4}}^{A_{2}^{}}_{\bar{A}_{1}}^{B_{1}^{}}}$ &    $1.6463$ &                                        $d\indices*{*_{A_{5}}^{\tensor*[^{2}]{\hspace{-0.2em}A}{}_{1}^{}}_{\bar{A}_{4}}^{A_{2}^{}}_{\bar{A}_{1}}^{\tensor*[^{1}]{\hspace{-0.2em}B}{}_{1}^{}}}$ &  $- 1.3176$ \\[0.4em]
                                         $d\indices*{*_{A_{5}}^{\tensor*[^{2}]{\hspace{-0.2em}A}{}_{1}^{}}_{\bar{A}_{4}}^{A_{2}^{}}_{\bar{A}_{1}}^{\tensor*[^{2}]{\hspace{-0.2em}B}{}_{1}^{}}}$ &    $0.5322$ &                                                                         $d\indices*{*_{A_{5}}^{\tensor*[^{2}]{\hspace{-0.2em}A}{}_{1}^{}}_{\bar{A}_{4}}^{A_{2}^{}}_{\bar{A}_{1}}^{B_{2}^{}}}$ &  $- 3.0342$ &                                        $d\indices*{*_{A_{5}}^{\tensor*[^{2}]{\hspace{-0.2em}A}{}_{1}^{}}_{\bar{A}_{4}}^{A_{2}^{}}_{\bar{A}_{1}}^{\tensor*[^{1}]{\hspace{-0.2em}B}{}_{2}^{}}}$ &    $3.4424$ &                                                                         $d\indices*{*_{A_{5}}^{\tensor*[^{2}]{\hspace{-0.2em}A}{}_{1}^{}}_{\bar{A}_{4}}^{B_{1}^{}}_{\bar{A}_{1}}^{B_{1}^{}}}$ &    $0.0679$ &                                        $d\indices*{*_{A_{5}}^{\tensor*[^{2}]{\hspace{-0.2em}A}{}_{1}^{}}_{\bar{A}_{4}}^{B_{1}^{}}_{\bar{A}_{1}}^{\tensor*[^{1}]{\hspace{-0.2em}B}{}_{1}^{}}}$ &    $9.1338$ \\[0.4em]
                                         $d\indices*{*_{A_{5}}^{\tensor*[^{2}]{\hspace{-0.2em}A}{}_{1}^{}}_{\bar{A}_{4}}^{B_{1}^{}}_{\bar{A}_{1}}^{\tensor*[^{2}]{\hspace{-0.2em}B}{}_{1}^{}}}$ &    $1.0972$ &                                                                         $d\indices*{*_{A_{5}}^{\tensor*[^{2}]{\hspace{-0.2em}A}{}_{1}^{}}_{\bar{A}_{4}}^{B_{2}^{}}_{\bar{A}_{1}}^{B_{1}^{}}}$ &  $- 9.7310$ &                                        $d\indices*{*_{A_{5}}^{\tensor*[^{2}]{\hspace{-0.2em}A}{}_{1}^{}}_{\bar{A}_{4}}^{\tensor*[^{1}]{\hspace{-0.2em}B}{}_{2}^{}}_{\bar{A}_{1}}^{B_{1}^{}}}$ &    $7.8088$ &       $d\indices*{*_{A_{5}}^{\tensor*[^{2}]{\hspace{-0.2em}A}{}_{1}^{}}_{\bar{A}_{4}}^{\tensor*[^{1}]{\hspace{-0.2em}B}{}_{1}^{}}_{\bar{A}_{1}}^{\tensor*[^{1}]{\hspace{-0.2em}B}{}_{1}^{}}}$ & $- 23.2440$ &       $d\indices*{*_{A_{5}}^{\tensor*[^{2}]{\hspace{-0.2em}A}{}_{1}^{}}_{\bar{A}_{4}}^{\tensor*[^{1}]{\hspace{-0.2em}B}{}_{1}^{}}_{\bar{A}_{1}}^{\tensor*[^{2}]{\hspace{-0.2em}B}{}_{1}^{}}}$ &  $- 7.4504$ \\[0.4em]
                                         $d\indices*{*_{A_{5}}^{\tensor*[^{2}]{\hspace{-0.2em}A}{}_{1}^{}}_{\bar{A}_{4}}^{B_{2}^{}}_{\bar{A}_{1}}^{\tensor*[^{1}]{\hspace{-0.2em}B}{}_{1}^{}}}$ &   $15.0871$ &       $d\indices*{*_{A_{5}}^{\tensor*[^{2}]{\hspace{-0.2em}A}{}_{1}^{}}_{\bar{A}_{4}}^{\tensor*[^{1}]{\hspace{-0.2em}B}{}_{2}^{}}_{\bar{A}_{1}}^{\tensor*[^{1}]{\hspace{-0.2em}B}{}_{1}^{}}}$ & $- 10.8928$ &       $d\indices*{*_{A_{5}}^{\tensor*[^{2}]{\hspace{-0.2em}A}{}_{1}^{}}_{\bar{A}_{4}}^{\tensor*[^{2}]{\hspace{-0.2em}B}{}_{1}^{}}_{\bar{A}_{1}}^{\tensor*[^{2}]{\hspace{-0.2em}B}{}_{1}^{}}}$ & $- 11.7027$ &                                        $d\indices*{*_{A_{5}}^{\tensor*[^{2}]{\hspace{-0.2em}A}{}_{1}^{}}_{\bar{A}_{4}}^{B_{2}^{}}_{\bar{A}_{1}}^{\tensor*[^{2}]{\hspace{-0.2em}B}{}_{1}^{}}}$ &    $1.9293$ &       $d\indices*{*_{A_{5}}^{\tensor*[^{2}]{\hspace{-0.2em}A}{}_{1}^{}}_{\bar{A}_{4}}^{\tensor*[^{1}]{\hspace{-0.2em}B}{}_{2}^{}}_{\bar{A}_{1}}^{\tensor*[^{2}]{\hspace{-0.2em}B}{}_{1}^{}}}$ &  $- 1.0617$ \\[0.4em]
                                                                          $d\indices*{*_{A_{5}}^{\tensor*[^{2}]{\hspace{-0.2em}A}{}_{1}^{}}_{\bar{A}_{4}}^{B_{2}^{}}_{\bar{A}_{1}}^{B_{2}^{}}}$ & $- 18.1152$ &                                        $d\indices*{*_{A_{5}}^{\tensor*[^{2}]{\hspace{-0.2em}A}{}_{1}^{}}_{\bar{A}_{4}}^{B_{2}^{}}_{\bar{A}_{1}}^{\tensor*[^{1}]{\hspace{-0.2em}B}{}_{2}^{}}}$ &   $12.7893$ &       $d\indices*{*_{A_{5}}^{\tensor*[^{2}]{\hspace{-0.2em}A}{}_{1}^{}}_{\bar{A}_{4}}^{\tensor*[^{1}]{\hspace{-0.2em}B}{}_{2}^{}}_{\bar{A}_{1}}^{\tensor*[^{1}]{\hspace{-0.2em}B}{}_{2}^{}}}$ & $- 10.1903$ &                                                                                                          $d\indices*{*_{A_{5}}^{A_{2}^{}}_{\bar{A}_{4}}^{A_{2}^{}}_{\bar{A}_{1}}^{B_{1}^{}}}$ &    $5.7923$ &                                                                         $d\indices*{*_{A_{5}}^{A_{2}^{}}_{\bar{A}_{4}}^{A_{2}^{}}_{\bar{A}_{1}}^{\tensor*[^{1}]{\hspace{-0.2em}B}{}_{1}^{}}}$ &    $1.0691$ \\[0.4em]
                                                                          $d\indices*{*_{A_{5}}^{A_{2}^{}}_{\bar{A}_{4}}^{A_{2}^{}}_{\bar{A}_{1}}^{\tensor*[^{2}]{\hspace{-0.2em}B}{}_{1}^{}}}$ &  $- 2.1920$ &                                                                         $d\indices*{*_{A_{5}}^{A_{2}^{}}_{\bar{A}_{4}}^{B_{1}^{}}_{\bar{A}_{1}}^{\tensor*[^{1}]{\hspace{-0.2em}B}{}_{1}^{}}}$ &    $4.6695$ &                                                                         $d\indices*{*_{A_{5}}^{A_{2}^{}}_{\bar{A}_{4}}^{B_{1}^{}}_{\bar{A}_{1}}^{\tensor*[^{2}]{\hspace{-0.2em}B}{}_{1}^{}}}$ &    $6.1293$ &                                                                                                          $d\indices*{*_{A_{5}}^{A_{2}^{}}_{\bar{A}_{4}}^{B_{2}^{}}_{\bar{A}_{1}}^{B_{1}^{}}}$ &  $- 0.2431$ &                                                                         $d\indices*{*_{A_{5}}^{A_{2}^{}}_{\bar{A}_{4}}^{\tensor*[^{1}]{\hspace{-0.2em}B}{}_{2}^{}}_{\bar{A}_{1}}^{B_{1}^{}}}$ &    $0.7799$ \\[0.4em]
                                         $d\indices*{*_{A_{5}}^{A_{2}^{}}_{\bar{A}_{4}}^{\tensor*[^{1}]{\hspace{-0.2em}B}{}_{1}^{}}_{\bar{A}_{1}}^{\tensor*[^{2}]{\hspace{-0.2em}B}{}_{1}^{}}}$ & $- 10.2751$ &                                                                         $d\indices*{*_{A_{5}}^{A_{2}^{}}_{\bar{A}_{4}}^{B_{2}^{}}_{\bar{A}_{1}}^{\tensor*[^{1}]{\hspace{-0.2em}B}{}_{1}^{}}}$ &   $10.0571$ &                                        $d\indices*{*_{A_{5}}^{A_{2}^{}}_{\bar{A}_{4}}^{\tensor*[^{1}]{\hspace{-0.2em}B}{}_{2}^{}}_{\bar{A}_{1}}^{\tensor*[^{1}]{\hspace{-0.2em}B}{}_{1}^{}}}$ &  $- 6.1371$ &                                                                         $d\indices*{*_{A_{5}}^{A_{2}^{}}_{\bar{A}_{4}}^{B_{2}^{}}_{\bar{A}_{1}}^{\tensor*[^{2}]{\hspace{-0.2em}B}{}_{1}^{}}}$ &    $9.7152$ &                                        $d\indices*{*_{A_{5}}^{A_{2}^{}}_{\bar{A}_{4}}^{\tensor*[^{1}]{\hspace{-0.2em}B}{}_{2}^{}}_{\bar{A}_{1}}^{\tensor*[^{2}]{\hspace{-0.2em}B}{}_{1}^{}}}$ &  $- 7.5989$ \\[0.4em]
                                                                          $d\indices*{*_{A_{5}}^{A_{2}^{}}_{\bar{A}_{4}}^{B_{2}^{}}_{\bar{A}_{1}}^{\tensor*[^{1}]{\hspace{-0.2em}B}{}_{2}^{}}}$ &    $0.0418$ &                                                                                                          $d\indices*{*_{A_{5}}^{B_{1}^{}}_{\bar{A}_{4}}^{B_{1}^{}}_{\bar{A}_{1}}^{B_{1}^{}}}$ &   $11.0952$ &                                                                         $d\indices*{*_{A_{5}}^{B_{1}^{}}_{\bar{A}_{4}}^{B_{1}^{}}_{\bar{A}_{1}}^{\tensor*[^{1}]{\hspace{-0.2em}B}{}_{1}^{}}}$ &  $- 7.6856$ &                                                                         $d\indices*{*_{A_{5}}^{B_{1}^{}}_{\bar{A}_{4}}^{B_{1}^{}}_{\bar{A}_{1}}^{\tensor*[^{2}]{\hspace{-0.2em}B}{}_{1}^{}}}$ &    $0.1498$ &                                        $d\indices*{*_{A_{5}}^{B_{1}^{}}_{\bar{A}_{4}}^{\tensor*[^{1}]{\hspace{-0.2em}B}{}_{1}^{}}_{\bar{A}_{1}}^{\tensor*[^{1}]{\hspace{-0.2em}B}{}_{1}^{}}}$ &  $- 0.2613$ \\[0.4em]
                                         $d\indices*{*_{A_{5}}^{B_{1}^{}}_{\bar{A}_{4}}^{\tensor*[^{1}]{\hspace{-0.2em}B}{}_{1}^{}}_{\bar{A}_{1}}^{\tensor*[^{2}]{\hspace{-0.2em}B}{}_{1}^{}}}$ &  $- 5.0128$ &                                                                         $d\indices*{*_{A_{5}}^{B_{2}^{}}_{\bar{A}_{4}}^{B_{1}^{}}_{\bar{A}_{1}}^{\tensor*[^{1}]{\hspace{-0.2em}B}{}_{1}^{}}}$ & $- 10.2015$ &                                        $d\indices*{*_{A_{5}}^{\tensor*[^{1}]{\hspace{-0.2em}B}{}_{2}^{}}_{\bar{A}_{4}}^{B_{1}^{}}_{\bar{A}_{1}}^{\tensor*[^{1}]{\hspace{-0.2em}B}{}_{1}^{}}}$ &    $6.2490$ &                                        $d\indices*{*_{A_{5}}^{B_{1}^{}}_{\bar{A}_{4}}^{\tensor*[^{2}]{\hspace{-0.2em}B}{}_{1}^{}}_{\bar{A}_{1}}^{\tensor*[^{2}]{\hspace{-0.2em}B}{}_{1}^{}}}$ &  $- 5.9144$ &                                                                         $d\indices*{*_{A_{5}}^{B_{2}^{}}_{\bar{A}_{4}}^{B_{1}^{}}_{\bar{A}_{1}}^{\tensor*[^{2}]{\hspace{-0.2em}B}{}_{1}^{}}}$ &    $1.9883$ \\[0.4em]
                                         $d\indices*{*_{A_{5}}^{\tensor*[^{1}]{\hspace{-0.2em}B}{}_{2}^{}}_{\bar{A}_{4}}^{B_{1}^{}}_{\bar{A}_{1}}^{\tensor*[^{2}]{\hspace{-0.2em}B}{}_{1}^{}}}$ &  $- 1.8833$ &                                                                                                          $d\indices*{*_{A_{5}}^{B_{2}^{}}_{\bar{A}_{4}}^{B_{2}^{}}_{\bar{A}_{1}}^{B_{1}^{}}}$ &    $9.5248$ &                                                                         $d\indices*{*_{A_{5}}^{B_{2}^{}}_{\bar{A}_{4}}^{\tensor*[^{1}]{\hspace{-0.2em}B}{}_{2}^{}}_{\bar{A}_{1}}^{B_{1}^{}}}$ &  $- 5.9684$ &                                        $d\indices*{*_{A_{5}}^{\tensor*[^{1}]{\hspace{-0.2em}B}{}_{2}^{}}_{\bar{A}_{4}}^{\tensor*[^{1}]{\hspace{-0.2em}B}{}_{2}^{}}_{\bar{A}_{1}}^{B_{1}^{}}}$ &    $4.4927$ &       $d\indices*{*_{A_{5}}^{\tensor*[^{1}]{\hspace{-0.2em}B}{}_{1}^{}}_{\bar{A}_{4}}^{\tensor*[^{1}]{\hspace{-0.2em}B}{}_{1}^{}}_{\bar{A}_{1}}^{\tensor*[^{1}]{\hspace{-0.2em}B}{}_{1}^{}}}$ &  $- 3.5156$ \\[0.4em]
        $d\indices*{*_{A_{5}}^{\tensor*[^{1}]{\hspace{-0.2em}B}{}_{1}^{}}_{\bar{A}_{4}}^{\tensor*[^{1}]{\hspace{-0.2em}B}{}_{1}^{}}_{\bar{A}_{1}}^{\tensor*[^{2}]{\hspace{-0.2em}B}{}_{1}^{}}}$ &  $- 4.9464$ &       $d\indices*{*_{A_{5}}^{\tensor*[^{1}]{\hspace{-0.2em}B}{}_{1}^{}}_{\bar{A}_{4}}^{\tensor*[^{2}]{\hspace{-0.2em}B}{}_{1}^{}}_{\bar{A}_{1}}^{\tensor*[^{2}]{\hspace{-0.2em}B}{}_{1}^{}}}$ &    $0.5672$ &                                        $d\indices*{*_{A_{5}}^{B_{2}^{}}_{\bar{A}_{4}}^{\tensor*[^{1}]{\hspace{-0.2em}B}{}_{1}^{}}_{\bar{A}_{1}}^{\tensor*[^{2}]{\hspace{-0.2em}B}{}_{1}^{}}}$ &  $- 1.6826$ &       $d\indices*{*_{A_{5}}^{\tensor*[^{1}]{\hspace{-0.2em}B}{}_{2}^{}}_{\bar{A}_{4}}^{\tensor*[^{1}]{\hspace{-0.2em}B}{}_{1}^{}}_{\bar{A}_{1}}^{\tensor*[^{2}]{\hspace{-0.2em}B}{}_{1}^{}}}$ &    $0.2761$ &                                                                         $d\indices*{*_{A_{5}}^{B_{2}^{}}_{\bar{A}_{4}}^{B_{2}^{}}_{\bar{A}_{1}}^{\tensor*[^{1}]{\hspace{-0.2em}B}{}_{1}^{}}}$ &    $5.9124$ \\[0.4em]
                                         $d\indices*{*_{A_{5}}^{B_{2}^{}}_{\bar{A}_{4}}^{\tensor*[^{1}]{\hspace{-0.2em}B}{}_{2}^{}}_{\bar{A}_{1}}^{\tensor*[^{1}]{\hspace{-0.2em}B}{}_{1}^{}}}$ &  $- 2.7206$ &       $d\indices*{*_{A_{5}}^{\tensor*[^{1}]{\hspace{-0.2em}B}{}_{2}^{}}_{\bar{A}_{4}}^{\tensor*[^{1}]{\hspace{-0.2em}B}{}_{2}^{}}_{\bar{A}_{1}}^{\tensor*[^{1}]{\hspace{-0.2em}B}{}_{1}^{}}}$ &    $2.5275$ &       $d\indices*{*_{A_{5}}^{\tensor*[^{2}]{\hspace{-0.2em}B}{}_{1}^{}}_{\bar{A}_{4}}^{\tensor*[^{2}]{\hspace{-0.2em}B}{}_{1}^{}}_{\bar{A}_{1}}^{\tensor*[^{2}]{\hspace{-0.2em}B}{}_{1}^{}}}$ &    $6.8376$ &                                                                         $d\indices*{*_{A_{5}}^{B_{2}^{}}_{\bar{A}_{4}}^{B_{2}^{}}_{\bar{A}_{1}}^{\tensor*[^{2}]{\hspace{-0.2em}B}{}_{1}^{}}}$ &  $- 2.2342$ &                                        $d\indices*{*_{A_{5}}^{B_{2}^{}}_{\bar{A}_{4}}^{\tensor*[^{1}]{\hspace{-0.2em}B}{}_{2}^{}}_{\bar{A}_{1}}^{\tensor*[^{2}]{\hspace{-0.2em}B}{}_{1}^{}}}$ &    $1.0182$ \\[0.4em]
        $d\indices*{*_{A_{5}}^{\tensor*[^{1}]{\hspace{-0.2em}B}{}_{2}^{}}_{\bar{A}_{4}}^{\tensor*[^{1}]{\hspace{-0.2em}B}{}_{2}^{}}_{\bar{A}_{1}}^{\tensor*[^{2}]{\hspace{-0.2em}B}{}_{1}^{}}}$ &  $- 2.2657$ &                                                                                                                       $d\indices*{*_{X_{1}}^{A_{1g}^{}}_{X_{2}}^{A_{1g}^{}}_{X}^{A_{1g}^{}}}$ & $- 21.0240$ &                                                                                                                         $d\indices*{*_{X_{1}}^{A_{1g}^{}}_{X_{2}}^{E_{g}^{}}_{X}^{E_{g}^{}}}$ &   $20.0262$ &                                                                                                                        $d\indices*{*_{X_{1}}^{A_{1g}^{}}_{X_{2}}^{A_{2u}^{}}_{X}^{E_{u}^{}}}$ &   $18.6559$ &                                                                                       $d\indices*{*_{X_{1}}^{A_{1g}^{}}_{X_{2}}^{A_{2u}^{}}_{X}^{\tensor*[^{1}]{\hspace{-0.2em}E}{}_{u}^{}}}$ & $- 25.4910$ \\[0.4em]
                                                                                                                        $d\indices*{*_{X_{1}}^{A_{1g}^{}}_{X_{2}}^{B_{1u}^{}}_{X}^{B_{1u}^{}}}$ & $- 18.9199$ &                                                                                                                         $d\indices*{*_{X_{1}}^{A_{1g}^{}}_{X_{2}}^{E_{u}^{}}_{X}^{E_{u}^{}}}$ & $- 13.7856$ &                                                                                        $d\indices*{*_{X_{1}}^{A_{1g}^{}}_{X_{2}}^{E_{u}^{}}_{X}^{\tensor*[^{1}]{\hspace{-0.2em}E}{}_{u}^{}}}$ &   $22.1526$ &                                                       $d\indices*{*_{X_{1}}^{A_{1g}^{}}_{X_{2}}^{\tensor*[^{1}]{\hspace{-0.2em}E}{}_{u}^{}}_{X}^{\tensor*[^{1}]{\hspace{-0.2em}E}{}_{u}^{}}}$ & $- 30.8607$ &                                                                                                                          $d\indices*{*_{X_{1}}^{E_{g}^{}}_{X_{2}}^{E_{g}^{}}_{X}^{E_{g}^{}}}$ & $- 11.7446$ \\[0.4em]
                                                                                                                          $d\indices*{*_{X_{1}}^{A_{2u}^{}}_{X_{2}}^{E_{g}^{}}_{X}^{E_{u}^{}}}$ &   $22.0896$ &                                                                                        $d\indices*{*_{X_{1}}^{A_{2u}^{}}_{X_{2}}^{E_{g}^{}}_{X}^{\tensor*[^{1}]{\hspace{-0.2em}E}{}_{u}^{}}}$ & $- 37.9132$ &                                                                                                                        $d\indices*{*_{X_{1}}^{B_{1u}^{}}_{X_{2}}^{A_{2u}^{}}_{X}^{E_{g}^{}}}$ &   $23.5111$ &                                                                                                                         $d\indices*{*_{X_{1}}^{B_{1u}^{}}_{X_{2}}^{E_{g}^{}}_{X}^{E_{u}^{}}}$ &   $11.5661$ &                                                                                                                          $d\indices*{*_{X_{1}}^{E_{g}^{}}_{X_{2}}^{E_{u}^{}}_{X}^{E_{u}^{}}}$ & $- 10.5328$ \\[0.4em]
                                                                        $\tensor*[^{1}]{d}{}\indices*{*_{X_{1}}^{E_{g}^{}}_{X_{2}}^{E_{u}^{}}_{X}^{\tensor*[^{1}]{\hspace{-0.2em}E}{}_{u}^{}}}$ &   $12.7812$ &                                                                                        $d\indices*{*_{X_{1}}^{B_{1u}^{}}_{X_{2}}^{E_{g}^{}}_{X}^{\tensor*[^{1}]{\hspace{-0.2em}E}{}_{u}^{}}}$ & $- 13.4626$ &                                                                       $\tensor*[^{0}]{d}{}\indices*{*_{X_{1}}^{E_{g}^{}}_{X_{2}}^{E_{u}^{}}_{X}^{\tensor*[^{1}]{\hspace{-0.2em}E}{}_{u}^{}}}$ &   $11.3955$ &                                                        $d\indices*{*_{X_{1}}^{E_{g}^{}}_{X_{2}}^{\tensor*[^{1}]{\hspace{-0.2em}E}{}_{u}^{}}_{X}^{\tensor*[^{1}]{\hspace{-0.2em}E}{}_{u}^{}}}$ & $- 19.1670$ &                                                                                                    $d\indices*{*_{\bar{A}_{1}}^{A_{1}^{}}_{\bar{A}_{2}}^{A_{1}^{}}_{\bar{A}_{5}}^{A_{1}^{}}}$ &   $29.8536$ \\[0.4em]
                                                                    $d\indices*{*_{\bar{A}_{1}}^{A_{1}^{}}_{\bar{A}_{2}}^{A_{1}^{}}_{\bar{A}_{5}}^{\tensor*[^{1}]{\hspace{-0.2em}A}{}_{1}^{}}}$ &   $13.3563$ &                                                                   $d\indices*{*_{\bar{A}_{1}}^{A_{1}^{}}_{\bar{A}_{2}}^{A_{1}^{}}_{\bar{A}_{5}}^{\tensor*[^{2}]{\hspace{-0.2em}A}{}_{1}^{}}}$ &    $2.8011$ &                                                                                                    $d\indices*{*_{\bar{A}_{1}}^{A_{1}^{}}_{\bar{A}_{2}}^{A_{1}^{}}_{\bar{A}_{5}}^{A_{2}^{}}}$ &   $25.5682$ &                                  $d\indices*{*_{\bar{A}_{1}}^{A_{1}^{}}_{\bar{A}_{2}}^{\tensor*[^{1}]{\hspace{-0.2em}A}{}_{1}^{}}_{\bar{A}_{5}}^{\tensor*[^{1}]{\hspace{-0.2em}A}{}_{1}^{}}}$ & $- 12.6505$ &                                  $d\indices*{*_{\bar{A}_{1}}^{A_{1}^{}}_{\bar{A}_{2}}^{\tensor*[^{1}]{\hspace{-0.2em}A}{}_{1}^{}}_{\bar{A}_{5}}^{\tensor*[^{2}]{\hspace{-0.2em}A}{}_{1}^{}}}$ &    $3.8005$ \\[0.4em]
                                                                    $d\indices*{*_{\bar{A}_{1}}^{A_{1}^{}}_{\bar{A}_{2}}^{\tensor*[^{1}]{\hspace{-0.2em}A}{}_{1}^{}}_{\bar{A}_{5}}^{A_{2}^{}}}$ &  $- 8.7706$ &                                                                   $d\indices*{*_{\bar{A}_{1}}^{A_{1}^{}}_{\bar{A}_{2}}^{\tensor*[^{1}]{\hspace{-0.2em}A}{}_{1}^{}}_{\bar{A}_{5}}^{B_{1}^{}}}$ &    $0.1936$ &                                  $d\indices*{*_{\bar{A}_{1}}^{A_{1}^{}}_{\bar{A}_{2}}^{\tensor*[^{1}]{\hspace{-0.2em}A}{}_{1}^{}}_{\bar{A}_{5}}^{\tensor*[^{1}]{\hspace{-0.2em}B}{}_{1}^{}}}$ & $- 10.2936$ &                                  $d\indices*{*_{\bar{A}_{1}}^{A_{1}^{}}_{\bar{A}_{2}}^{\tensor*[^{1}]{\hspace{-0.2em}A}{}_{1}^{}}_{\bar{A}_{5}}^{\tensor*[^{2}]{\hspace{-0.2em}B}{}_{1}^{}}}$ & $- 11.0956$ &                                                                   $d\indices*{*_{\bar{A}_{1}}^{A_{1}^{}}_{\bar{A}_{2}}^{\tensor*[^{1}]{\hspace{-0.2em}A}{}_{1}^{}}_{\bar{A}_{5}}^{B_{2}^{}}}$ &    $0.0941$ \\[0.4em]
                                   $d\indices*{*_{\bar{A}_{1}}^{A_{1}^{}}_{\bar{A}_{2}}^{\tensor*[^{1}]{\hspace{-0.2em}A}{}_{1}^{}}_{\bar{A}_{5}}^{\tensor*[^{1}]{\hspace{-0.2em}B}{}_{2}^{}}}$ &    $1.3322$ &                                  $d\indices*{*_{\bar{A}_{1}}^{A_{1}^{}}_{\bar{A}_{2}}^{\tensor*[^{2}]{\hspace{-0.2em}A}{}_{1}^{}}_{\bar{A}_{5}}^{\tensor*[^{2}]{\hspace{-0.2em}A}{}_{1}^{}}}$ &    $1.3886$ &                                                                   $d\indices*{*_{\bar{A}_{1}}^{A_{1}^{}}_{\bar{A}_{2}}^{\tensor*[^{2}]{\hspace{-0.2em}A}{}_{1}^{}}_{\bar{A}_{5}}^{A_{2}^{}}}$ &    $4.5332$ &                                                                   $d\indices*{*_{\bar{A}_{1}}^{A_{1}^{}}_{\bar{A}_{2}}^{\tensor*[^{2}]{\hspace{-0.2em}A}{}_{1}^{}}_{\bar{A}_{5}}^{B_{1}^{}}}$ & $- 11.5899$ &                                  $d\indices*{*_{\bar{A}_{1}}^{A_{1}^{}}_{\bar{A}_{2}}^{\tensor*[^{2}]{\hspace{-0.2em}A}{}_{1}^{}}_{\bar{A}_{5}}^{\tensor*[^{1}]{\hspace{-0.2em}B}{}_{1}^{}}}$ &   $14.5342$ \\[0.4em]
                                   $d\indices*{*_{\bar{A}_{1}}^{A_{1}^{}}_{\bar{A}_{2}}^{\tensor*[^{2}]{\hspace{-0.2em}A}{}_{1}^{}}_{\bar{A}_{5}}^{\tensor*[^{2}]{\hspace{-0.2em}B}{}_{1}^{}}}$ &    $0.2322$ &                                                                   $d\indices*{*_{\bar{A}_{1}}^{A_{1}^{}}_{\bar{A}_{2}}^{\tensor*[^{2}]{\hspace{-0.2em}A}{}_{1}^{}}_{\bar{A}_{5}}^{B_{2}^{}}}$ &   $22.0956$ &                                  $d\indices*{*_{\bar{A}_{1}}^{A_{1}^{}}_{\bar{A}_{2}}^{\tensor*[^{2}]{\hspace{-0.2em}A}{}_{1}^{}}_{\bar{A}_{5}}^{\tensor*[^{1}]{\hspace{-0.2em}B}{}_{2}^{}}}$ & $- 14.7084$ &                                                                                                    $d\indices*{*_{\bar{A}_{1}}^{A_{1}^{}}_{\bar{A}_{2}}^{A_{2}^{}}_{\bar{A}_{5}}^{A_{2}^{}}}$ &  $- 0.4115$ &                                                                                                    $d\indices*{*_{\bar{A}_{1}}^{A_{1}^{}}_{\bar{A}_{2}}^{A_{2}^{}}_{\bar{A}_{5}}^{B_{1}^{}}}$ &  $- 2.0569$ \\[0.4em]
                                                                    $d\indices*{*_{\bar{A}_{1}}^{A_{1}^{}}_{\bar{A}_{2}}^{A_{2}^{}}_{\bar{A}_{5}}^{\tensor*[^{1}]{\hspace{-0.2em}B}{}_{1}^{}}}$ &   $15.9917$ &                                                                   $d\indices*{*_{\bar{A}_{1}}^{A_{1}^{}}_{\bar{A}_{2}}^{A_{2}^{}}_{\bar{A}_{5}}^{\tensor*[^{2}]{\hspace{-0.2em}B}{}_{1}^{}}}$ &   $10.7662$ &                                                                                                    $d\indices*{*_{\bar{A}_{1}}^{A_{1}^{}}_{\bar{A}_{2}}^{A_{2}^{}}_{\bar{A}_{5}}^{B_{2}^{}}}$ &    $1.2190$ &                                                                   $d\indices*{*_{\bar{A}_{1}}^{A_{1}^{}}_{\bar{A}_{2}}^{A_{2}^{}}_{\bar{A}_{5}}^{\tensor*[^{1}]{\hspace{-0.2em}B}{}_{2}^{}}}$ &    $0.0476$ &                                                                                                    $d\indices*{*_{\bar{A}_{1}}^{A_{1}^{}}_{\bar{A}_{2}}^{B_{1}^{}}_{\bar{A}_{5}}^{B_{1}^{}}}$ & $- 12.7181$ \\[0.4em]
                                                                    $d\indices*{*_{\bar{A}_{1}}^{A_{1}^{}}_{\bar{A}_{2}}^{B_{1}^{}}_{\bar{A}_{5}}^{\tensor*[^{1}]{\hspace{-0.2em}B}{}_{1}^{}}}$ &    $6.5848$ &                                                                   $d\indices*{*_{\bar{A}_{1}}^{A_{1}^{}}_{\bar{A}_{2}}^{B_{1}^{}}_{\bar{A}_{5}}^{\tensor*[^{2}]{\hspace{-0.2em}B}{}_{1}^{}}}$ &  $- 1.9160$ &                                                                                                    $d\indices*{*_{\bar{A}_{1}}^{A_{1}^{}}_{\bar{A}_{2}}^{B_{2}^{}}_{\bar{A}_{5}}^{B_{1}^{}}}$ & $- 16.0565$ &                                                                   $d\indices*{*_{\bar{A}_{1}}^{A_{1}^{}}_{\bar{A}_{2}}^{\tensor*[^{1}]{\hspace{-0.2em}B}{}_{2}^{}}_{\bar{A}_{5}}^{B_{1}^{}}}$ &   $10.7492$ &                                  $d\indices*{*_{\bar{A}_{1}}^{A_{1}^{}}_{\bar{A}_{2}}^{\tensor*[^{1}]{\hspace{-0.2em}B}{}_{1}^{}}_{\bar{A}_{5}}^{\tensor*[^{1}]{\hspace{-0.2em}B}{}_{1}^{}}}$ &   $24.7733$ \\[0.4em]
                                   $d\indices*{*_{\bar{A}_{1}}^{A_{1}^{}}_{\bar{A}_{2}}^{\tensor*[^{1}]{\hspace{-0.2em}B}{}_{1}^{}}_{\bar{A}_{5}}^{\tensor*[^{2}]{\hspace{-0.2em}B}{}_{1}^{}}}$ &    $1.4014$ &                                                                   $d\indices*{*_{\bar{A}_{1}}^{A_{1}^{}}_{\bar{A}_{2}}^{B_{2}^{}}_{\bar{A}_{5}}^{\tensor*[^{1}]{\hspace{-0.2em}B}{}_{1}^{}}}$ & $- 13.5599$ &                                  $d\indices*{*_{\bar{A}_{1}}^{A_{1}^{}}_{\bar{A}_{2}}^{\tensor*[^{1}]{\hspace{-0.2em}B}{}_{2}^{}}_{\bar{A}_{5}}^{\tensor*[^{1}]{\hspace{-0.2em}B}{}_{1}^{}}}$ &    $1.4233$ &                                  $d\indices*{*_{\bar{A}_{1}}^{A_{1}^{}}_{\bar{A}_{2}}^{\tensor*[^{2}]{\hspace{-0.2em}B}{}_{1}^{}}_{\bar{A}_{5}}^{\tensor*[^{2}]{\hspace{-0.2em}B}{}_{1}^{}}}$ &  $- 0.4443$ &                                                                   $d\indices*{*_{\bar{A}_{1}}^{A_{1}^{}}_{\bar{A}_{2}}^{B_{2}^{}}_{\bar{A}_{5}}^{\tensor*[^{2}]{\hspace{-0.2em}B}{}_{1}^{}}}$ &    $0.0979$ \\[0.4em]
                                   $d\indices*{*_{\bar{A}_{1}}^{A_{1}^{}}_{\bar{A}_{2}}^{\tensor*[^{1}]{\hspace{-0.2em}B}{}_{2}^{}}_{\bar{A}_{5}}^{\tensor*[^{2}]{\hspace{-0.2em}B}{}_{1}^{}}}$ &    $0.7698$ &                                                                                                    $d\indices*{*_{\bar{A}_{1}}^{A_{1}^{}}_{\bar{A}_{2}}^{B_{2}^{}}_{\bar{A}_{5}}^{B_{2}^{}}}$ & $- 12.9525$ &                                                                   $d\indices*{*_{\bar{A}_{1}}^{A_{1}^{}}_{\bar{A}_{2}}^{B_{2}^{}}_{\bar{A}_{5}}^{\tensor*[^{1}]{\hspace{-0.2em}B}{}_{2}^{}}}$ &    $2.0825$ &                                  $d\indices*{*_{\bar{A}_{1}}^{A_{1}^{}}_{\bar{A}_{2}}^{\tensor*[^{1}]{\hspace{-0.2em}B}{}_{2}^{}}_{\bar{A}_{5}}^{\tensor*[^{1}]{\hspace{-0.2em}B}{}_{2}^{}}}$ &    $0.8794$ & $d\indices*{*_{\bar{A}_{1}}^{\tensor*[^{1}]{\hspace{-0.2em}A}{}_{1}^{}}_{\bar{A}_{2}}^{\tensor*[^{1}]{\hspace{-0.2em}A}{}_{1}^{}}_{\bar{A}_{5}}^{\tensor*[^{1}]{\hspace{-0.2em}A}{}_{1}^{}}}$ &   $14.0929$ \\[0.4em]
  $d\indices*{*_{\bar{A}_{1}}^{\tensor*[^{1}]{\hspace{-0.2em}A}{}_{1}^{}}_{\bar{A}_{2}}^{\tensor*[^{1}]{\hspace{-0.2em}A}{}_{1}^{}}_{\bar{A}_{5}}^{\tensor*[^{2}]{\hspace{-0.2em}A}{}_{1}^{}}}$ &  $- 6.3324$ &                                  $d\indices*{*_{\bar{A}_{1}}^{\tensor*[^{1}]{\hspace{-0.2em}A}{}_{1}^{}}_{\bar{A}_{2}}^{\tensor*[^{1}]{\hspace{-0.2em}A}{}_{1}^{}}_{\bar{A}_{5}}^{A_{2}^{}}}$ &    $0.2835$ & $d\indices*{*_{\bar{A}_{1}}^{\tensor*[^{1}]{\hspace{-0.2em}A}{}_{1}^{}}_{\bar{A}_{2}}^{\tensor*[^{2}]{\hspace{-0.2em}A}{}_{1}^{}}_{\bar{A}_{5}}^{\tensor*[^{2}]{\hspace{-0.2em}A}{}_{1}^{}}}$ &    $9.3205$ &                                  $d\indices*{*_{\bar{A}_{1}}^{\tensor*[^{1}]{\hspace{-0.2em}A}{}_{1}^{}}_{\bar{A}_{2}}^{\tensor*[^{2}]{\hspace{-0.2em}A}{}_{1}^{}}_{\bar{A}_{5}}^{A_{2}^{}}}$ &    $4.0823$ &                                  $d\indices*{*_{\bar{A}_{1}}^{\tensor*[^{1}]{\hspace{-0.2em}A}{}_{1}^{}}_{\bar{A}_{2}}^{\tensor*[^{2}]{\hspace{-0.2em}A}{}_{1}^{}}_{\bar{A}_{5}}^{B_{1}^{}}}$ &    $0.1717$ \\[0.4em]
  $d\indices*{*_{\bar{A}_{1}}^{\tensor*[^{1}]{\hspace{-0.2em}A}{}_{1}^{}}_{\bar{A}_{2}}^{\tensor*[^{2}]{\hspace{-0.2em}A}{}_{1}^{}}_{\bar{A}_{5}}^{\tensor*[^{1}]{\hspace{-0.2em}B}{}_{1}^{}}}$ &  $- 7.7174$ & $d\indices*{*_{\bar{A}_{1}}^{\tensor*[^{1}]{\hspace{-0.2em}A}{}_{1}^{}}_{\bar{A}_{2}}^{\tensor*[^{2}]{\hspace{-0.2em}A}{}_{1}^{}}_{\bar{A}_{5}}^{\tensor*[^{2}]{\hspace{-0.2em}B}{}_{1}^{}}}$ &  $- 4.1124$ &                                  $d\indices*{*_{\bar{A}_{1}}^{\tensor*[^{1}]{\hspace{-0.2em}A}{}_{1}^{}}_{\bar{A}_{2}}^{\tensor*[^{2}]{\hspace{-0.2em}A}{}_{1}^{}}_{\bar{A}_{5}}^{B_{2}^{}}}$ & $- 11.5820$ & $d\indices*{*_{\bar{A}_{1}}^{\tensor*[^{1}]{\hspace{-0.2em}A}{}_{1}^{}}_{\bar{A}_{2}}^{\tensor*[^{2}]{\hspace{-0.2em}A}{}_{1}^{}}_{\bar{A}_{5}}^{\tensor*[^{1}]{\hspace{-0.2em}B}{}_{2}^{}}}$ &    $7.8204$ &                                                                   $d\indices*{*_{\bar{A}_{1}}^{\tensor*[^{1}]{\hspace{-0.2em}A}{}_{1}^{}}_{\bar{A}_{2}}^{A_{2}^{}}_{\bar{A}_{5}}^{A_{2}^{}}}$ &  $- 7.3294$ \\[0.4em]
                                                                    $d\indices*{*_{\bar{A}_{1}}^{\tensor*[^{1}]{\hspace{-0.2em}A}{}_{1}^{}}_{\bar{A}_{2}}^{A_{2}^{}}_{\bar{A}_{5}}^{B_{1}^{}}}$ &    $2.8152$ &                                  $d\indices*{*_{\bar{A}_{1}}^{\tensor*[^{1}]{\hspace{-0.2em}A}{}_{1}^{}}_{\bar{A}_{2}}^{A_{2}^{}}_{\bar{A}_{5}}^{\tensor*[^{1}]{\hspace{-0.2em}B}{}_{1}^{}}}$ &  $- 5.9271$ &                                  $d\indices*{*_{\bar{A}_{1}}^{\tensor*[^{1}]{\hspace{-0.2em}A}{}_{1}^{}}_{\bar{A}_{2}}^{A_{2}^{}}_{\bar{A}_{5}}^{\tensor*[^{2}]{\hspace{-0.2em}B}{}_{1}^{}}}$ &  $- 6.1516$ &                                                                   $d\indices*{*_{\bar{A}_{1}}^{\tensor*[^{1}]{\hspace{-0.2em}A}{}_{1}^{}}_{\bar{A}_{2}}^{A_{2}^{}}_{\bar{A}_{5}}^{B_{2}^{}}}$ &    $0.8628$ &                                  $d\indices*{*_{\bar{A}_{1}}^{\tensor*[^{1}]{\hspace{-0.2em}A}{}_{1}^{}}_{\bar{A}_{2}}^{A_{2}^{}}_{\bar{A}_{5}}^{\tensor*[^{1}]{\hspace{-0.2em}B}{}_{2}^{}}}$ &  $- 9.1871$ \\[0.4em]
                                                                    $d\indices*{*_{\bar{A}_{1}}^{\tensor*[^{1}]{\hspace{-0.2em}A}{}_{1}^{}}_{\bar{A}_{2}}^{B_{1}^{}}_{\bar{A}_{5}}^{B_{1}^{}}}$ &   $13.7698$ &                                  $d\indices*{*_{\bar{A}_{1}}^{\tensor*[^{1}]{\hspace{-0.2em}A}{}_{1}^{}}_{\bar{A}_{2}}^{B_{1}^{}}_{\bar{A}_{5}}^{\tensor*[^{1}]{\hspace{-0.2em}B}{}_{1}^{}}}$ &  $- 6.8728$ &                                  $d\indices*{*_{\bar{A}_{1}}^{\tensor*[^{1}]{\hspace{-0.2em}A}{}_{1}^{}}_{\bar{A}_{2}}^{B_{1}^{}}_{\bar{A}_{5}}^{\tensor*[^{2}]{\hspace{-0.2em}B}{}_{1}^{}}}$ &    $0.4946$ &                                                                   $d\indices*{*_{\bar{A}_{1}}^{\tensor*[^{1}]{\hspace{-0.2em}A}{}_{1}^{}}_{\bar{A}_{2}}^{B_{2}^{}}_{\bar{A}_{5}}^{B_{1}^{}}}$ &  $- 0.4228$ &                                  $d\indices*{*_{\bar{A}_{1}}^{\tensor*[^{1}]{\hspace{-0.2em}A}{}_{1}^{}}_{\bar{A}_{2}}^{\tensor*[^{1}]{\hspace{-0.2em}B}{}_{2}^{}}_{\bar{A}_{5}}^{B_{1}^{}}}$ &    $0.2139$ \\[0.4em]
  $d\indices*{*_{\bar{A}_{1}}^{\tensor*[^{1}]{\hspace{-0.2em}A}{}_{1}^{}}_{\bar{A}_{2}}^{\tensor*[^{1}]{\hspace{-0.2em}B}{}_{1}^{}}_{\bar{A}_{5}}^{\tensor*[^{1}]{\hspace{-0.2em}B}{}_{1}^{}}}$ &    $8.9868$ & $d\indices*{*_{\bar{A}_{1}}^{\tensor*[^{1}]{\hspace{-0.2em}A}{}_{1}^{}}_{\bar{A}_{2}}^{\tensor*[^{1}]{\hspace{-0.2em}B}{}_{1}^{}}_{\bar{A}_{5}}^{\tensor*[^{2}]{\hspace{-0.2em}B}{}_{1}^{}}}$ &    $3.7020$ &                                  $d\indices*{*_{\bar{A}_{1}}^{\tensor*[^{1}]{\hspace{-0.2em}A}{}_{1}^{}}_{\bar{A}_{2}}^{B_{2}^{}}_{\bar{A}_{5}}^{\tensor*[^{1}]{\hspace{-0.2em}B}{}_{1}^{}}}$ &   $10.4627$ & $d\indices*{*_{\bar{A}_{1}}^{\tensor*[^{1}]{\hspace{-0.2em}A}{}_{1}^{}}_{\bar{A}_{2}}^{\tensor*[^{1}]{\hspace{-0.2em}B}{}_{2}^{}}_{\bar{A}_{5}}^{\tensor*[^{1}]{\hspace{-0.2em}B}{}_{1}^{}}}$ &  $- 8.1820$ & $d\indices*{*_{\bar{A}_{1}}^{\tensor*[^{1}]{\hspace{-0.2em}A}{}_{1}^{}}_{\bar{A}_{2}}^{\tensor*[^{2}]{\hspace{-0.2em}B}{}_{1}^{}}_{\bar{A}_{5}}^{\tensor*[^{2}]{\hspace{-0.2em}B}{}_{1}^{}}}$ &    $6.5944$ \\[0.4em]
                                   $d\indices*{*_{\bar{A}_{1}}^{\tensor*[^{1}]{\hspace{-0.2em}A}{}_{1}^{}}_{\bar{A}_{2}}^{B_{2}^{}}_{\bar{A}_{5}}^{\tensor*[^{2}]{\hspace{-0.2em}B}{}_{1}^{}}}$ &    $2.0126$ & $d\indices*{*_{\bar{A}_{1}}^{\tensor*[^{1}]{\hspace{-0.2em}A}{}_{1}^{}}_{\bar{A}_{2}}^{\tensor*[^{1}]{\hspace{-0.2em}B}{}_{2}^{}}_{\bar{A}_{5}}^{\tensor*[^{2}]{\hspace{-0.2em}B}{}_{1}^{}}}$ & $- 10.8197$ &                                                                   $d\indices*{*_{\bar{A}_{1}}^{\tensor*[^{1}]{\hspace{-0.2em}A}{}_{1}^{}}_{\bar{A}_{2}}^{B_{2}^{}}_{\bar{A}_{5}}^{B_{2}^{}}}$ & $- 11.2852$ &                                  $d\indices*{*_{\bar{A}_{1}}^{\tensor*[^{1}]{\hspace{-0.2em}A}{}_{1}^{}}_{\bar{A}_{2}}^{B_{2}^{}}_{\bar{A}_{5}}^{\tensor*[^{1}]{\hspace{-0.2em}B}{}_{2}^{}}}$ &    $8.2919$ & $d\indices*{*_{\bar{A}_{1}}^{\tensor*[^{1}]{\hspace{-0.2em}A}{}_{1}^{}}_{\bar{A}_{2}}^{\tensor*[^{1}]{\hspace{-0.2em}B}{}_{2}^{}}_{\bar{A}_{5}}^{\tensor*[^{1}]{\hspace{-0.2em}B}{}_{2}^{}}}$ &  $- 6.1429$ \\[0.4em]
  $d\indices*{*_{\bar{A}_{1}}^{\tensor*[^{2}]{\hspace{-0.2em}A}{}_{1}^{}}_{\bar{A}_{2}}^{\tensor*[^{2}]{\hspace{-0.2em}A}{}_{1}^{}}_{\bar{A}_{5}}^{\tensor*[^{2}]{\hspace{-0.2em}A}{}_{1}^{}}}$ &  $- 3.6258$ &                                  $d\indices*{*_{\bar{A}_{1}}^{\tensor*[^{2}]{\hspace{-0.2em}A}{}_{1}^{}}_{\bar{A}_{2}}^{\tensor*[^{2}]{\hspace{-0.2em}A}{}_{1}^{}}_{\bar{A}_{5}}^{A_{2}^{}}}$ &   $14.8337$ &                                                                   $d\indices*{*_{\bar{A}_{1}}^{\tensor*[^{2}]{\hspace{-0.2em}A}{}_{1}^{}}_{\bar{A}_{2}}^{A_{2}^{}}_{\bar{A}_{5}}^{A_{2}^{}}}$ &    $1.8444$ &                                                                   $d\indices*{*_{\bar{A}_{1}}^{\tensor*[^{2}]{\hspace{-0.2em}A}{}_{1}^{}}_{\bar{A}_{2}}^{A_{2}^{}}_{\bar{A}_{5}}^{B_{1}^{}}}$ &  $- 8.7378$ &                                  $d\indices*{*_{\bar{A}_{1}}^{\tensor*[^{2}]{\hspace{-0.2em}A}{}_{1}^{}}_{\bar{A}_{2}}^{A_{2}^{}}_{\bar{A}_{5}}^{\tensor*[^{1}]{\hspace{-0.2em}B}{}_{1}^{}}}$ &    $2.8579$ \\[0.4em]
                                   $d\indices*{*_{\bar{A}_{1}}^{\tensor*[^{2}]{\hspace{-0.2em}A}{}_{1}^{}}_{\bar{A}_{2}}^{A_{2}^{}}_{\bar{A}_{5}}^{\tensor*[^{2}]{\hspace{-0.2em}B}{}_{1}^{}}}$ &    $5.4066$ &                                                                   $d\indices*{*_{\bar{A}_{1}}^{\tensor*[^{2}]{\hspace{-0.2em}A}{}_{1}^{}}_{\bar{A}_{2}}^{A_{2}^{}}_{\bar{A}_{5}}^{B_{2}^{}}}$ &    $0.0948$ &                                  $d\indices*{*_{\bar{A}_{1}}^{\tensor*[^{2}]{\hspace{-0.2em}A}{}_{1}^{}}_{\bar{A}_{2}}^{A_{2}^{}}_{\bar{A}_{5}}^{\tensor*[^{1}]{\hspace{-0.2em}B}{}_{2}^{}}}$ &  $- 0.7093$ &                                                                   $d\indices*{*_{\bar{A}_{1}}^{\tensor*[^{2}]{\hspace{-0.2em}A}{}_{1}^{}}_{\bar{A}_{2}}^{B_{1}^{}}_{\bar{A}_{5}}^{B_{1}^{}}}$ &  $- 5.1401$ &                                  $d\indices*{*_{\bar{A}_{1}}^{\tensor*[^{2}]{\hspace{-0.2em}A}{}_{1}^{}}_{\bar{A}_{2}}^{B_{1}^{}}_{\bar{A}_{5}}^{\tensor*[^{1}]{\hspace{-0.2em}B}{}_{1}^{}}}$ &  $- 4.5771$ \\[0.4em]
                                   $d\indices*{*_{\bar{A}_{1}}^{\tensor*[^{2}]{\hspace{-0.2em}A}{}_{1}^{}}_{\bar{A}_{2}}^{B_{1}^{}}_{\bar{A}_{5}}^{\tensor*[^{2}]{\hspace{-0.2em}B}{}_{1}^{}}}$ &  $- 8.5460$ &                                                                   $d\indices*{*_{\bar{A}_{1}}^{\tensor*[^{2}]{\hspace{-0.2em}A}{}_{1}^{}}_{\bar{A}_{2}}^{B_{2}^{}}_{\bar{A}_{5}}^{B_{1}^{}}}$ &  $- 2.6518$ &                                  $d\indices*{*_{\bar{A}_{1}}^{\tensor*[^{2}]{\hspace{-0.2em}A}{}_{1}^{}}_{\bar{A}_{2}}^{\tensor*[^{1}]{\hspace{-0.2em}B}{}_{2}^{}}_{\bar{A}_{5}}^{B_{1}^{}}}$ &    $1.3990$ & $d\indices*{*_{\bar{A}_{1}}^{\tensor*[^{2}]{\hspace{-0.2em}A}{}_{1}^{}}_{\bar{A}_{2}}^{\tensor*[^{1}]{\hspace{-0.2em}B}{}_{1}^{}}_{\bar{A}_{5}}^{\tensor*[^{1}]{\hspace{-0.2em}B}{}_{1}^{}}}$ &   $21.2876$ & $d\indices*{*_{\bar{A}_{1}}^{\tensor*[^{2}]{\hspace{-0.2em}A}{}_{1}^{}}_{\bar{A}_{2}}^{\tensor*[^{1}]{\hspace{-0.2em}B}{}_{1}^{}}_{\bar{A}_{5}}^{\tensor*[^{2}]{\hspace{-0.2em}B}{}_{1}^{}}}$ &    $8.5982$ \\[0.4em]
                                   $d\indices*{*_{\bar{A}_{1}}^{\tensor*[^{2}]{\hspace{-0.2em}A}{}_{1}^{}}_{\bar{A}_{2}}^{B_{2}^{}}_{\bar{A}_{5}}^{\tensor*[^{1}]{\hspace{-0.2em}B}{}_{1}^{}}}$ & $- 12.9720$ & $d\indices*{*_{\bar{A}_{1}}^{\tensor*[^{2}]{\hspace{-0.2em}A}{}_{1}^{}}_{\bar{A}_{2}}^{\tensor*[^{1}]{\hspace{-0.2em}B}{}_{2}^{}}_{\bar{A}_{5}}^{\tensor*[^{1}]{\hspace{-0.2em}B}{}_{1}^{}}}$ &   $10.4261$ & $d\indices*{*_{\bar{A}_{1}}^{\tensor*[^{2}]{\hspace{-0.2em}A}{}_{1}^{}}_{\bar{A}_{2}}^{\tensor*[^{2}]{\hspace{-0.2em}B}{}_{1}^{}}_{\bar{A}_{5}}^{\tensor*[^{2}]{\hspace{-0.2em}B}{}_{1}^{}}}$ &  $- 3.3932$ &                                  $d\indices*{*_{\bar{A}_{1}}^{\tensor*[^{2}]{\hspace{-0.2em}A}{}_{1}^{}}_{\bar{A}_{2}}^{B_{2}^{}}_{\bar{A}_{5}}^{\tensor*[^{2}]{\hspace{-0.2em}B}{}_{1}^{}}}$ & $- 15.2531$ & $d\indices*{*_{\bar{A}_{1}}^{\tensor*[^{2}]{\hspace{-0.2em}A}{}_{1}^{}}_{\bar{A}_{2}}^{\tensor*[^{1}]{\hspace{-0.2em}B}{}_{2}^{}}_{\bar{A}_{5}}^{\tensor*[^{2}]{\hspace{-0.2em}B}{}_{1}^{}}}$ &    $9.9593$ \\[0.4em]
                                                                    $d\indices*{*_{\bar{A}_{1}}^{\tensor*[^{2}]{\hspace{-0.2em}A}{}_{1}^{}}_{\bar{A}_{2}}^{B_{2}^{}}_{\bar{A}_{5}}^{B_{2}^{}}}$ &  $- 0.5040$ &                                  $d\indices*{*_{\bar{A}_{1}}^{\tensor*[^{2}]{\hspace{-0.2em}A}{}_{1}^{}}_{\bar{A}_{2}}^{B_{2}^{}}_{\bar{A}_{5}}^{\tensor*[^{1}]{\hspace{-0.2em}B}{}_{2}^{}}}$ &  $- 1.3905$ & $d\indices*{*_{\bar{A}_{1}}^{\tensor*[^{2}]{\hspace{-0.2em}A}{}_{1}^{}}_{\bar{A}_{2}}^{\tensor*[^{1}]{\hspace{-0.2em}B}{}_{2}^{}}_{\bar{A}_{5}}^{\tensor*[^{1}]{\hspace{-0.2em}B}{}_{2}^{}}}$ &    $1.9797$ &                                                                                                    $d\indices*{*_{\bar{A}_{1}}^{A_{2}^{}}_{\bar{A}_{2}}^{A_{2}^{}}_{\bar{A}_{5}}^{A_{2}^{}}}$ &    $9.6029$ &                                                                                                    $d\indices*{*_{\bar{A}_{1}}^{A_{2}^{}}_{\bar{A}_{2}}^{B_{1}^{}}_{\bar{A}_{5}}^{B_{1}^{}}}$ &    $0.3898$ \\[0.4em]
                                                                    $d\indices*{*_{\bar{A}_{1}}^{A_{2}^{}}_{\bar{A}_{2}}^{B_{1}^{}}_{\bar{A}_{5}}^{\tensor*[^{1}]{\hspace{-0.2em}B}{}_{1}^{}}}$ &    $7.3313$ &                                                                   $d\indices*{*_{\bar{A}_{1}}^{A_{2}^{}}_{\bar{A}_{2}}^{B_{1}^{}}_{\bar{A}_{5}}^{\tensor*[^{2}]{\hspace{-0.2em}B}{}_{1}^{}}}$ &  $- 3.2667$ &                                                                                                    $d\indices*{*_{\bar{A}_{1}}^{A_{2}^{}}_{\bar{A}_{2}}^{B_{2}^{}}_{\bar{A}_{5}}^{B_{1}^{}}}$ &  $- 6.9992$ &                                                                   $d\indices*{*_{\bar{A}_{1}}^{A_{2}^{}}_{\bar{A}_{2}}^{\tensor*[^{1}]{\hspace{-0.2em}B}{}_{2}^{}}_{\bar{A}_{5}}^{B_{1}^{}}}$ &    $5.6354$ &                                  $d\indices*{*_{\bar{A}_{1}}^{A_{2}^{}}_{\bar{A}_{2}}^{\tensor*[^{1}]{\hspace{-0.2em}B}{}_{1}^{}}_{\bar{A}_{5}}^{\tensor*[^{1}]{\hspace{-0.2em}B}{}_{1}^{}}}$ &  $- 2.1379$ \\[0.4em]
                                   $d\indices*{*_{\bar{A}_{1}}^{A_{2}^{}}_{\bar{A}_{2}}^{\tensor*[^{1}]{\hspace{-0.2em}B}{}_{1}^{}}_{\bar{A}_{5}}^{\tensor*[^{2}]{\hspace{-0.2em}B}{}_{1}^{}}}$ &    $7.2924$ &                                                                   $d\indices*{*_{\bar{A}_{1}}^{A_{2}^{}}_{\bar{A}_{2}}^{B_{2}^{}}_{\bar{A}_{5}}^{\tensor*[^{1}]{\hspace{-0.2em}B}{}_{1}^{}}}$ &   $11.9822$ &                                  $d\indices*{*_{\bar{A}_{1}}^{A_{2}^{}}_{\bar{A}_{2}}^{\tensor*[^{1}]{\hspace{-0.2em}B}{}_{2}^{}}_{\bar{A}_{5}}^{\tensor*[^{1}]{\hspace{-0.2em}B}{}_{1}^{}}}$ &  $- 7.8674$ &                                  $d\indices*{*_{\bar{A}_{1}}^{A_{2}^{}}_{\bar{A}_{2}}^{\tensor*[^{2}]{\hspace{-0.2em}B}{}_{1}^{}}_{\bar{A}_{5}}^{\tensor*[^{2}]{\hspace{-0.2em}B}{}_{1}^{}}}$ &    $8.4074$ &                                                                   $d\indices*{*_{\bar{A}_{1}}^{A_{2}^{}}_{\bar{A}_{2}}^{B_{2}^{}}_{\bar{A}_{5}}^{\tensor*[^{2}]{\hspace{-0.2em}B}{}_{1}^{}}}$ &  $- 1.8056$ \\[0.4em]
                                   $d\indices*{*_{\bar{A}_{1}}^{A_{2}^{}}_{\bar{A}_{2}}^{\tensor*[^{1}]{\hspace{-0.2em}B}{}_{2}^{}}_{\bar{A}_{5}}^{\tensor*[^{2}]{\hspace{-0.2em}B}{}_{1}^{}}}$ &    $0.4871$ &                                                                                                    $d\indices*{*_{\bar{A}_{1}}^{A_{2}^{}}_{\bar{A}_{2}}^{B_{2}^{}}_{\bar{A}_{5}}^{B_{2}^{}}}$ &   $13.3607$ &                                                                   $d\indices*{*_{\bar{A}_{1}}^{A_{2}^{}}_{\bar{A}_{2}}^{B_{2}^{}}_{\bar{A}_{5}}^{\tensor*[^{1}]{\hspace{-0.2em}B}{}_{2}^{}}}$ &  $- 9.0880$ &                                  $d\indices*{*_{\bar{A}_{1}}^{A_{2}^{}}_{\bar{A}_{2}}^{\tensor*[^{1}]{\hspace{-0.2em}B}{}_{2}^{}}_{\bar{A}_{5}}^{\tensor*[^{1}]{\hspace{-0.2em}B}{}_{2}^{}}}$ &    $6.7828$ &                                  $d\indices*{*_{\bar{A}_{1}}^{B_{1}^{}}_{\bar{A}_{2}}^{\tensor*[^{1}]{\hspace{-0.2em}B}{}_{1}^{}}_{\bar{A}_{5}}^{\tensor*[^{2}]{\hspace{-0.2em}B}{}_{1}^{}}}$ &  $- 8.4523$ \\[0.4em]
                                                                    $d\indices*{*_{\bar{A}_{1}}^{B_{2}^{}}_{\bar{A}_{2}}^{B_{1}^{}}_{\bar{A}_{5}}^{\tensor*[^{1}]{\hspace{-0.2em}B}{}_{1}^{}}}$ &    $6.5073$ &                                  $d\indices*{*_{\bar{A}_{1}}^{\tensor*[^{1}]{\hspace{-0.2em}B}{}_{2}^{}}_{\bar{A}_{2}}^{B_{1}^{}}_{\bar{A}_{5}}^{\tensor*[^{1}]{\hspace{-0.2em}B}{}_{1}^{}}}$ &  $- 4.0682$ &                                                                   $d\indices*{*_{\bar{A}_{1}}^{B_{2}^{}}_{\bar{A}_{2}}^{B_{1}^{}}_{\bar{A}_{5}}^{\tensor*[^{2}]{\hspace{-0.2em}B}{}_{1}^{}}}$ &    $6.7099$ &                                  $d\indices*{*_{\bar{A}_{1}}^{\tensor*[^{1}]{\hspace{-0.2em}B}{}_{2}^{}}_{\bar{A}_{2}}^{B_{1}^{}}_{\bar{A}_{5}}^{\tensor*[^{2}]{\hspace{-0.2em}B}{}_{1}^{}}}$ &  $- 4.8655$ &                                                                   $d\indices*{*_{\bar{A}_{1}}^{B_{2}^{}}_{\bar{A}_{2}}^{\tensor*[^{1}]{\hspace{-0.2em}B}{}_{2}^{}}_{\bar{A}_{5}}^{B_{1}^{}}}$ &    $0.0082$ \\[0.4em]
                                   $d\indices*{*_{\bar{A}_{1}}^{B_{2}^{}}_{\bar{A}_{2}}^{\tensor*[^{1}]{\hspace{-0.2em}B}{}_{1}^{}}_{\bar{A}_{5}}^{\tensor*[^{2}]{\hspace{-0.2em}B}{}_{1}^{}}}$ &    $3.8105$ & $d\indices*{*_{\bar{A}_{1}}^{\tensor*[^{1}]{\hspace{-0.2em}B}{}_{2}^{}}_{\bar{A}_{2}}^{\tensor*[^{1}]{\hspace{-0.2em}B}{}_{1}^{}}_{\bar{A}_{5}}^{\tensor*[^{2}]{\hspace{-0.2em}B}{}_{1}^{}}}$ &  $- 2.1064$ &                                  $d\indices*{*_{\bar{A}_{1}}^{B_{2}^{}}_{\bar{A}_{2}}^{\tensor*[^{1}]{\hspace{-0.2em}B}{}_{2}^{}}_{\bar{A}_{5}}^{\tensor*[^{1}]{\hspace{-0.2em}B}{}_{1}^{}}}$ &    $0.4502$ &                                  $d\indices*{*_{\bar{A}_{1}}^{B_{2}^{}}_{\bar{A}_{2}}^{\tensor*[^{1}]{\hspace{-0.2em}B}{}_{2}^{}}_{\bar{A}_{5}}^{\tensor*[^{2}]{\hspace{-0.2em}B}{}_{1}^{}}}$ &    $0.7519$ &                                                                                                                                                                                               &              \\[0.4em]
%
%
%

\end{longtable*}

\begin{longtable*}{cccccccccc}
\caption{All the second order irreducible derivatives for CaF$_2$ at $V_{300K}$. All results were computed using the LID approach, $4\hat{\mathbf1}$ supercell, and LDA functional. The units are eV/\AA$^2$.\label{table:caf2_2_300_lda}} \\
\hline\hline 
Derivative & Value & Derivative & Value & Derivative & Value & Derivative & Value & Derivative & Value \\
\hline\\[-0.9em]
\endfirsthead
\hline\hline
Derivative & Value & Derivative & Value & Derivative & Value & Derivative & Value & Derivative & Value  \\
\hline\\[-0.9em]
\endhead
\hline\hline
\endfoot
                                                                          $d\indices*{*_{\Gamma}^{T_{2g}^{}}_{\Gamma}^{T_{2g}^{}}}$ &   $7.6279$ &                                                                         $d\indices*{*_{\Gamma}^{T_{1u}^{}}_{\Gamma}^{T_{1u}^{}}}$ &   $7.8616$ &                                                                                   $d\indices*{*_{L}^{A_{1g}^{}}_{L}^{A_{1g}^{}}}$ &   $8.1961$ &                                                                                   $d\indices*{*_{L}^{E_{g}^{}}_{L}^{E_{g}^{}}}$ &   $5.7311$ &                                                                               $d\indices*{*_{L}^{A_{2u}^{}}_{L}^{A_{2u}^{}}}$ &  $15.0610$ \\[0.4em]
                                                   $d\indices*{*_{L}^{A_{2u}^{}}_{L}^{\tensor*[^{1}]{\hspace{-0.2em}A}{}_{2u}^{}}}$ &   $2.5723$ &                 $d\indices*{*_{L}^{\tensor*[^{1}]{\hspace{-0.2em}A}{}_{2u}^{}}_{L}^{\tensor*[^{1}]{\hspace{-0.2em}A}{}_{2u}^{}}}$ &   $6.4142$ &                                                                                     $d\indices*{*_{L}^{E_{u}^{}}_{L}^{E_{u}^{}}}$ &   $5.5838$ &                                                  $d\indices*{*_{L}^{E_{u}^{}}_{L}^{\tensor*[^{1}]{\hspace{-0.2em}E}{}_{u}^{}}}$ & $- 2.9041$ &               $d\indices*{*_{L}^{\tensor*[^{1}]{\hspace{-0.2em}E}{}_{u}^{}}_{L}^{\tensor*[^{1}]{\hspace{-0.2em}E}{}_{u}^{}}}$ &   $4.8096$ \\[0.4em]
                                                                                    $d\indices*{*_{X}^{A_{1g}^{}}_{X}^{A_{1g}^{}}}$ &  $11.0201$ &                                                                                     $d\indices*{*_{X}^{E_{g}^{}}_{X}^{E_{g}^{}}}$ &   $2.6811$ &                                                                                   $d\indices*{*_{X}^{A_{2u}^{}}_{X}^{A_{2u}^{}}}$ &  $15.9552$ &                                                                                 $d\indices*{*_{X}^{B_{1u}^{}}_{X}^{B_{1u}^{}}}$ &   $1.6483$ &                                                                                 $d\indices*{*_{X}^{E_{u}^{}}_{X}^{E_{u}^{}}}$ &   $7.7258$ \\[0.4em]
                                                     $d\indices*{*_{X}^{E_{u}^{}}_{X}^{\tensor*[^{1}]{\hspace{-0.2em}E}{}_{u}^{}}}$ & $- 2.0960$ &                   $d\indices*{*_{X}^{\tensor*[^{1}]{\hspace{-0.2em}E}{}_{u}^{}}_{X}^{\tensor*[^{1}]{\hspace{-0.2em}E}{}_{u}^{}}}$ &   $6.4638$ &                                                                               $d\indices*{*_{A}^{A_{1}^{}}_{\bar{A}}^{A_{1}^{}}}$ &  $13.5046$ &                                            $d\indices*{*_{A}^{A_{1}^{}}_{\bar{A}}^{\tensor*[^{1}]{\hspace{-0.2em}A}{}_{1}^{}}}$ & $- 1.9639$ &                                          $d\indices*{*_{A}^{A_{1}^{}}_{\bar{A}}^{\tensor*[^{2}]{\hspace{-0.2em}A}{}_{1}^{}}}$ &   $4.2085$ \\[0.4em]
              $d\indices*{*_{A}^{\tensor*[^{1}]{\hspace{-0.2em}A}{}_{1}^{}}_{\bar{A}}^{\tensor*[^{1}]{\hspace{-0.2em}A}{}_{1}^{}}}$ &   $8.6242$ &             $d\indices*{*_{A}^{\tensor*[^{1}]{\hspace{-0.2em}A}{}_{1}^{}}_{\bar{A}}^{\tensor*[^{2}]{\hspace{-0.2em}A}{}_{1}^{}}}$ & $- 0.6038$ &             $d\indices*{*_{A}^{\tensor*[^{2}]{\hspace{-0.2em}A}{}_{1}^{}}_{\bar{A}}^{\tensor*[^{2}]{\hspace{-0.2em}A}{}_{1}^{}}}$ &   $8.4374$ &                                                                             $d\indices*{*_{A}^{A_{2}^{}}_{\bar{A}}^{A_{2}^{}}}$ &   $5.2423$ &                                                                           $d\indices*{*_{A}^{B_{1}^{}}_{\bar{A}}^{B_{1}^{}}}$ &   $2.4348$ \\[0.4em]
                                               $d\indices*{*_{A}^{B_{1}^{}}_{\bar{A}}^{\tensor*[^{1}]{\hspace{-0.2em}B}{}_{1}^{}}}$ & $- 1.9582$ &                                              $d\indices*{*_{A}^{B_{1}^{}}_{\bar{A}}^{\tensor*[^{2}]{\hspace{-0.2em}B}{}_{1}^{}}}$ &   $0.2431$ &             $d\indices*{*_{A}^{\tensor*[^{1}]{\hspace{-0.2em}B}{}_{1}^{}}_{\bar{A}}^{\tensor*[^{1}]{\hspace{-0.2em}B}{}_{1}^{}}}$ &   $8.0376$ &           $d\indices*{*_{A}^{\tensor*[^{1}]{\hspace{-0.2em}B}{}_{1}^{}}_{\bar{A}}^{\tensor*[^{2}]{\hspace{-0.2em}B}{}_{1}^{}}}$ &   $0.7126$ &         $d\indices*{*_{A}^{\tensor*[^{2}]{\hspace{-0.2em}B}{}_{1}^{}}_{\bar{A}}^{\tensor*[^{2}]{\hspace{-0.2em}B}{}_{1}^{}}}$ &   $5.1371$ \\[0.4em]
                                                                                $d\indices*{*_{A}^{B_{2}^{}}_{\bar{A}}^{B_{2}^{}}}$ &   $5.5873$ &                                              $d\indices*{*_{A}^{B_{2}^{}}_{\bar{A}}^{\tensor*[^{1}]{\hspace{-0.2em}B}{}_{2}^{}}}$ & $- 2.8799$ &             $d\indices*{*_{A}^{\tensor*[^{1}]{\hspace{-0.2em}B}{}_{2}^{}}_{\bar{A}}^{\tensor*[^{1}]{\hspace{-0.2em}B}{}_{2}^{}}}$ &   $5.3284$ &                                                                             $d\indices*{*_{W}^{A_{1}^{}}_{\bar{W}}^{A_{1}^{}}}$ &  $10.1433$ &                                                                           $d\indices*{*_{W}^{B_{1}^{}}_{\bar{W}}^{B_{1}^{}}}$ &   $6.6529$ \\[0.4em]
                                               $d\indices*{*_{W}^{B_{1}^{}}_{\bar{W}}^{\tensor*[^{1}]{\hspace{-0.2em}B}{}_{1}^{}}}$ & $- 1.7657$ &             $d\indices*{*_{W}^{\tensor*[^{1}]{\hspace{-0.2em}B}{}_{1}^{}}_{\bar{W}}^{\tensor*[^{1}]{\hspace{-0.2em}B}{}_{1}^{}}}$ &   $8.5818$ &                                                                               $d\indices*{*_{W}^{A_{2}^{}}_{\bar{W}}^{A_{2}^{}}}$ &   $2.0020$ &                                                                             $d\indices*{*_{W}^{B_{2}^{}}_{\bar{W}}^{B_{2}^{}}}$ &   $2.2970$ &                                                                             $d\indices*{*_{W}^{E_{}^{}}_{\bar{W}}^{E_{}^{}}}$ &  $10.6648$ \\[0.4em]
                                                 $d\indices*{*_{W}^{E_{}^{}}_{\bar{W}}^{\tensor*[^{1}]{\hspace{-0.2em}E}{}_{}^{}}}$ &   $0.6835$ &               $d\indices*{*_{W}^{\tensor*[^{1}]{\hspace{-0.2em}E}{}_{}^{}}_{\bar{W}}^{\tensor*[^{1}]{\hspace{-0.2em}E}{}_{}^{}}}$ &   $5.4178$ &                                                                   $d\indices*{*_{\Lambda}^{A_{1}^{}}_{\bar{\Lambda}}^{A_{1}^{}}}$ &  $15.8166$ &                                $d\indices*{*_{\Lambda}^{A_{1}^{}}_{\bar{\Lambda}}^{\tensor*[^{1}]{\hspace{-0.2em}A}{}_{1}^{}}}$ & $- 8.2477$ &                              $d\indices*{*_{\Lambda}^{A_{1}^{}}_{\bar{\Lambda}}^{\tensor*[^{2}]{\hspace{-0.2em}A}{}_{1}^{}}}$ &   $1.6258$ \\[0.4em]
  $d\indices*{*_{\Lambda}^{\tensor*[^{1}]{\hspace{-0.2em}A}{}_{1}^{}}_{\bar{\Lambda}}^{\tensor*[^{1}]{\hspace{-0.2em}A}{}_{1}^{}}}$ &   $8.1807$ & $d\indices*{*_{\Lambda}^{\tensor*[^{1}]{\hspace{-0.2em}A}{}_{1}^{}}_{\bar{\Lambda}}^{\tensor*[^{2}]{\hspace{-0.2em}A}{}_{1}^{}}}$ & $- 0.0063$ & $d\indices*{*_{\Lambda}^{\tensor*[^{2}]{\hspace{-0.2em}A}{}_{1}^{}}_{\bar{\Lambda}}^{\tensor*[^{2}]{\hspace{-0.2em}A}{}_{1}^{}}}$ &   $6.9576$ &                                                                   $d\indices*{*_{\Lambda}^{E_{}^{}}_{\bar{\Lambda}}^{E_{}^{}}}$ &   $5.4174$ &                                $d\indices*{*_{\Lambda}^{E_{}^{}}_{\bar{\Lambda}}^{\tensor*[^{1}]{\hspace{-0.2em}E}{}_{}^{}}}$ & $- 2.6111$ \\[0.4em]
                                     $d\indices*{*_{\Lambda}^{E_{}^{}}_{\bar{\Lambda}}^{\tensor*[^{2}]{\hspace{-0.2em}E}{}_{}^{}}}$ & $- 2.0639$ &   $d\indices*{*_{\Lambda}^{\tensor*[^{1}]{\hspace{-0.2em}E}{}_{}^{}}_{\bar{\Lambda}}^{\tensor*[^{1}]{\hspace{-0.2em}E}{}_{}^{}}}$ &   $4.1413$ &   $d\indices*{*_{\Lambda}^{\tensor*[^{1}]{\hspace{-0.2em}E}{}_{}^{}}_{\bar{\Lambda}}^{\tensor*[^{2}]{\hspace{-0.2em}E}{}_{}^{}}}$ & $- 1.4628$ & $d\indices*{*_{\Lambda}^{\tensor*[^{2}]{\hspace{-0.2em}E}{}_{}^{}}_{\bar{\Lambda}}^{\tensor*[^{2}]{\hspace{-0.2em}E}{}_{}^{}}}$ &   $6.1915$ &                                                                             $d\indices*{*_{B}^{A_{}^{}}_{\bar{B}}^{A_{}^{}}}$ &  $12.7196$ \\[0.4em]
                                                 $d\indices*{*_{B}^{A_{}^{}}_{\bar{B}}^{\tensor*[^{1}]{\hspace{-0.2em}A}{}_{}^{}}}$ & $- 3.0354$ &                                                $d\indices*{*_{B}^{A_{}^{}}_{\bar{B}}^{\tensor*[^{2}]{\hspace{-0.2em}A}{}_{}^{}}}$ & $- 2.3591$ &                                                $d\indices*{*_{B}^{A_{}^{}}_{\bar{B}}^{\tensor*[^{3}]{\hspace{-0.2em}A}{}_{}^{}}}$ & $- 0.1606$ &                                              $d\indices*{*_{B}^{A_{}^{}}_{\bar{B}}^{\tensor*[^{4}]{\hspace{-0.2em}A}{}_{}^{}}}$ & $- 0.7234$ &                                            $d\indices*{*_{B}^{A_{}^{}}_{\bar{B}}^{\tensor*[^{5}]{\hspace{-0.2em}A}{}_{}^{}}}$ &   $0.6057$ \\[0.4em]
                $d\indices*{*_{B}^{\tensor*[^{1}]{\hspace{-0.2em}A}{}_{}^{}}_{\bar{B}}^{\tensor*[^{1}]{\hspace{-0.2em}A}{}_{}^{}}}$ &   $8.4461$ &               $d\indices*{*_{B}^{\tensor*[^{1}]{\hspace{-0.2em}A}{}_{}^{}}_{\bar{B}}^{\tensor*[^{2}]{\hspace{-0.2em}A}{}_{}^{}}}$ & $- 0.3488$ &               $d\indices*{*_{B}^{\tensor*[^{1}]{\hspace{-0.2em}A}{}_{}^{}}_{\bar{B}}^{\tensor*[^{3}]{\hspace{-0.2em}A}{}_{}^{}}}$ &   $0.5170$ &             $d\indices*{*_{B}^{\tensor*[^{1}]{\hspace{-0.2em}A}{}_{}^{}}_{\bar{B}}^{\tensor*[^{4}]{\hspace{-0.2em}A}{}_{}^{}}}$ &   $0.7844$ &           $d\indices*{*_{B}^{\tensor*[^{1}]{\hspace{-0.2em}A}{}_{}^{}}_{\bar{B}}^{\tensor*[^{5}]{\hspace{-0.2em}A}{}_{}^{}}}$ & $- 2.4745$ \\[0.4em]
                $d\indices*{*_{B}^{\tensor*[^{2}]{\hspace{-0.2em}A}{}_{}^{}}_{\bar{B}}^{\tensor*[^{2}]{\hspace{-0.2em}A}{}_{}^{}}}$ &   $8.9225$ &               $d\indices*{*_{B}^{\tensor*[^{2}]{\hspace{-0.2em}A}{}_{}^{}}_{\bar{B}}^{\tensor*[^{3}]{\hspace{-0.2em}A}{}_{}^{}}}$ &   $1.2127$ &               $d\indices*{*_{B}^{\tensor*[^{2}]{\hspace{-0.2em}A}{}_{}^{}}_{\bar{B}}^{\tensor*[^{4}]{\hspace{-0.2em}A}{}_{}^{}}}$ &   $1.0564$ &             $d\indices*{*_{B}^{\tensor*[^{2}]{\hspace{-0.2em}A}{}_{}^{}}_{\bar{B}}^{\tensor*[^{5}]{\hspace{-0.2em}A}{}_{}^{}}}$ &   $0.1904$ &           $d\indices*{*_{B}^{\tensor*[^{3}]{\hspace{-0.2em}A}{}_{}^{}}_{\bar{B}}^{\tensor*[^{3}]{\hspace{-0.2em}A}{}_{}^{}}}$ &   $4.2345$ \\[0.4em]
                $d\indices*{*_{B}^{\tensor*[^{3}]{\hspace{-0.2em}A}{}_{}^{}}_{\bar{B}}^{\tensor*[^{4}]{\hspace{-0.2em}A}{}_{}^{}}}$ & $- 1.4234$ &               $d\indices*{*_{B}^{\tensor*[^{3}]{\hspace{-0.2em}A}{}_{}^{}}_{\bar{B}}^{\tensor*[^{5}]{\hspace{-0.2em}A}{}_{}^{}}}$ & $- 0.1686$ &               $d\indices*{*_{B}^{\tensor*[^{4}]{\hspace{-0.2em}A}{}_{}^{}}_{\bar{B}}^{\tensor*[^{4}]{\hspace{-0.2em}A}{}_{}^{}}}$ &   $7.3851$ &             $d\indices*{*_{B}^{\tensor*[^{4}]{\hspace{-0.2em}A}{}_{}^{}}_{\bar{B}}^{\tensor*[^{5}]{\hspace{-0.2em}A}{}_{}^{}}}$ & $- 0.4370$ &           $d\indices*{*_{B}^{\tensor*[^{5}]{\hspace{-0.2em}A}{}_{}^{}}_{\bar{B}}^{\tensor*[^{5}]{\hspace{-0.2em}A}{}_{}^{}}}$ &   $4.0379$ \\[0.4em]
                                                                                  $d\indices*{*_{B}^{B_{}^{}}_{\bar{B}}^{B_{}^{}}}$ &   $6.0225$ &                                                $d\indices*{*_{B}^{B_{}^{}}_{\bar{B}}^{\tensor*[^{1}]{\hspace{-0.2em}B}{}_{}^{}}}$ & $- 2.0551$ &                                                $d\indices*{*_{B}^{B_{}^{}}_{\bar{B}}^{\tensor*[^{2}]{\hspace{-0.2em}B}{}_{}^{}}}$ & $- 1.4781$ &             $d\indices*{*_{B}^{\tensor*[^{1}]{\hspace{-0.2em}B}{}_{}^{}}_{\bar{B}}^{\tensor*[^{1}]{\hspace{-0.2em}B}{}_{}^{}}}$ &   $3.7107$ &           $d\indices*{*_{B}^{\tensor*[^{1}]{\hspace{-0.2em}B}{}_{}^{}}_{\bar{B}}^{\tensor*[^{2}]{\hspace{-0.2em}B}{}_{}^{}}}$ &   $1.0508$ \\[0.4em]
                $d\indices*{*_{B}^{\tensor*[^{2}]{\hspace{-0.2em}B}{}_{}^{}}_{\bar{B}}^{\tensor*[^{2}]{\hspace{-0.2em}B}{}_{}^{}}}$ &   $6.6858$ &                                                                     $d\indices*{*_{\Delta}^{A_{1}^{}}_{\bar{\Delta}}^{A_{1}^{}}}$ &  $16.2899$ &                                    $d\indices*{*_{\Delta}^{A_{1}^{}}_{\bar{\Delta}}^{\tensor*[^{1}]{\hspace{-0.2em}A}{}_{1}^{}}}$ & $- 8.0296$ & $d\indices*{*_{\Delta}^{\tensor*[^{1}]{\hspace{-0.2em}A}{}_{1}^{}}_{\bar{\Delta}}^{\tensor*[^{1}]{\hspace{-0.2em}A}{}_{1}^{}}}$ &   $9.6203$ &                                                                 $d\indices*{*_{\Delta}^{B_{2}^{}}_{\bar{\Delta}}^{B_{2}^{}}}$ &   $4.8891$ \\[0.4em]
                                                                        $d\indices*{*_{\Delta}^{E_{}^{}}_{\bar{\Delta}}^{E_{}^{}}}$ &   $5.8485$ &                                      $d\indices*{*_{\Delta}^{E_{}^{}}_{\bar{\Delta}}^{\tensor*[^{1}]{\hspace{-0.2em}E}{}_{}^{}}}$ &   $2.6267$ &                                      $d\indices*{*_{\Delta}^{E_{}^{}}_{\bar{\Delta}}^{\tensor*[^{2}]{\hspace{-0.2em}E}{}_{}^{}}}$ & $- 1.5037$ &   $d\indices*{*_{\Delta}^{\tensor*[^{1}]{\hspace{-0.2em}E}{}_{}^{}}_{\bar{\Delta}}^{\tensor*[^{1}]{\hspace{-0.2em}E}{}_{}^{}}}$ &   $2.6247$ & $d\indices*{*_{\Delta}^{\tensor*[^{1}]{\hspace{-0.2em}E}{}_{}^{}}_{\bar{\Delta}}^{\tensor*[^{2}]{\hspace{-0.2em}E}{}_{}^{}}}$ &   $0.4169$ \\[0.4em]
      $d\indices*{*_{\Delta}^{\tensor*[^{2}]{\hspace{-0.2em}E}{}_{}^{}}_{\bar{\Delta}}^{\tensor*[^{2}]{\hspace{-0.2em}E}{}_{}^{}}}$ &   $7.6603$ &                                                                                                                                   &            &                                                                                                                                   &            &                                                                                                                                 &            &                                                                                                                               &             \\
%
%
%

\end{longtable*}

\begin{longtable*}{cccccccccc}
\caption{All the third order space group irreducible derivatives for CaF$_2$ computed at  $V_{300K}$. All results were computed using the BID approach, $\hat{\mathbf{S}}_{O}$ supercell, and LDA functional.  The units are eV/\AA$^3$.\label{table:caf2_3_300_lda}} \\
\hline\hline 
Derivative & Value & Derivative & Value & Derivative & Value & Derivative & Value & Derivative & Value\\
\hline\\[-0.9em]
\endfirsthead
\hline\hline
Derivative & Value & Derivative & Value & Derivative & Value & Derivative & Value  & Derivative & Value\\
\hline\\[-0.9em]
\endhead
\hline\hline
\endfoot
                                                                                                                 $d\indices*{*_{\Gamma}^{T_{2g}^{}}_{\Gamma}^{T_{2g}^{}}_{\Gamma}^{T_{2g}^{}}}$ & $- 62.3936$ &                                                                                                                $d\indices*{*_{\Gamma}^{T_{1u}^{}}_{\Gamma}^{T_{1u}^{}}_{\Gamma}^{T_{2g}^{}}}$ & $- 56.2518$ &                                                                                                                     $d\indices*{*_{X}^{A_{1g}^{}}_{\bar{A}}^{A_{1}^{}}_{\bar{A}}^{A_{1}^{}}}$ &   $17.6441$ &                                                                                                                      $d\indices*{*_{X}^{E_{u}^{}}_{\bar{A}}^{A_{1}^{}}_{\bar{A}}^{A_{1}^{}}}$ &  $- 2.3114$ &                                                                                     $d\indices*{*_{X}^{\tensor*[^{1}]{\hspace{-0.2em}E}{}_{u}^{}}_{\bar{A}}^{A_{1}^{}}_{\bar{A}}^{A_{1}^{}}}$ &  $- 3.5055$ \\[0.4em]
                                                                                     $d\indices*{*_{X}^{A_{1g}^{}}_{\bar{A}}^{A_{1}^{}}_{\bar{A}}^{\tensor*[^{1}]{\hspace{-0.2em}A}{}_{1}^{}}}$ &  $- 8.6560$ &                                                                                     $d\indices*{*_{X}^{E_{u}^{}}_{\bar{A}}^{A_{1}^{}}_{\bar{A}}^{\tensor*[^{1}]{\hspace{-0.2em}A}{}_{1}^{}}}$ &  $- 1.7773$ &                                                    $d\indices*{*_{X}^{\tensor*[^{1}]{\hspace{-0.2em}E}{}_{u}^{}}_{\bar{A}}^{A_{1}^{}}_{\bar{A}}^{\tensor*[^{1}]{\hspace{-0.2em}A}{}_{1}^{}}}$ &  $- 0.1030$ &                                                                                    $d\indices*{*_{X}^{A_{1g}^{}}_{\bar{A}}^{A_{1}^{}}_{\bar{A}}^{\tensor*[^{2}]{\hspace{-0.2em}A}{}_{1}^{}}}$ &  $- 1.2883$ &                                                                                     $d\indices*{*_{X}^{E_{u}^{}}_{\bar{A}}^{A_{1}^{}}_{\bar{A}}^{\tensor*[^{2}]{\hspace{-0.2em}A}{}_{1}^{}}}$ &   $13.1894$ \\[0.4em]
                                                     $d\indices*{*_{X}^{\tensor*[^{1}]{\hspace{-0.2em}E}{}_{u}^{}}_{\bar{A}}^{A_{1}^{}}_{\bar{A}}^{\tensor*[^{2}]{\hspace{-0.2em}A}{}_{1}^{}}}$ & $- 18.1721$ &                                                                                                                      $d\indices*{*_{X}^{E_{g}^{}}_{\bar{A}}^{A_{1}^{}}_{\bar{A}}^{A_{2}^{}}}$ &  $- 0.8078$ &                                                                                                                      $d\indices*{*_{X}^{E_{g}^{}}_{\bar{A}}^{A_{1}^{}}_{\bar{A}}^{B_{1}^{}}}$ &    $2.4265$ &                                                                                                                     $d\indices*{*_{X}^{A_{2u}^{}}_{\bar{A}}^{A_{1}^{}}_{\bar{A}}^{B_{1}^{}}}$ &   $14.0671$ &                                                                                                                     $d\indices*{*_{X}^{B_{1u}^{}}_{\bar{A}}^{A_{1}^{}}_{\bar{A}}^{B_{1}^{}}}$ &   $10.0372$ \\[0.4em]
                                                                                      $d\indices*{*_{X}^{E_{g}^{}}_{\bar{A}}^{A_{1}^{}}_{\bar{A}}^{\tensor*[^{1}]{\hspace{-0.2em}B}{}_{1}^{}}}$ &   $13.7311$ &                                                                                    $d\indices*{*_{X}^{A_{2u}^{}}_{\bar{A}}^{A_{1}^{}}_{\bar{A}}^{\tensor*[^{1}]{\hspace{-0.2em}B}{}_{1}^{}}}$ &    $5.9836$ &                                                                                    $d\indices*{*_{X}^{B_{1u}^{}}_{\bar{A}}^{A_{1}^{}}_{\bar{A}}^{\tensor*[^{1}]{\hspace{-0.2em}B}{}_{1}^{}}}$ &    $0.1755$ &                                                                                     $d\indices*{*_{X}^{E_{g}^{}}_{\bar{A}}^{A_{1}^{}}_{\bar{A}}^{\tensor*[^{2}]{\hspace{-0.2em}B}{}_{1}^{}}}$ &   $12.6400$ &                                                                                    $d\indices*{*_{X}^{A_{2u}^{}}_{\bar{A}}^{A_{1}^{}}_{\bar{A}}^{\tensor*[^{2}]{\hspace{-0.2em}B}{}_{1}^{}}}$ &    $1.1096$ \\[0.4em]
                                                                                     $d\indices*{*_{X}^{B_{1u}^{}}_{\bar{A}}^{A_{1}^{}}_{\bar{A}}^{\tensor*[^{2}]{\hspace{-0.2em}B}{}_{1}^{}}}$ &  $- 0.7918$ &                                                                                                                      $d\indices*{*_{X}^{E_{u}^{}}_{\bar{A}}^{A_{1}^{}}_{\bar{A}}^{B_{2}^{}}}$ &    $0.5858$ &                                                                                     $d\indices*{*_{X}^{\tensor*[^{1}]{\hspace{-0.2em}E}{}_{u}^{}}_{\bar{A}}^{A_{1}^{}}_{\bar{A}}^{B_{2}^{}}}$ &  $- 3.4183$ &                                                                                     $d\indices*{*_{X}^{E_{u}^{}}_{\bar{A}}^{A_{1}^{}}_{\bar{A}}^{\tensor*[^{1}]{\hspace{-0.2em}B}{}_{2}^{}}}$ &  $- 1.2806$ &                                                    $d\indices*{*_{X}^{\tensor*[^{1}]{\hspace{-0.2em}E}{}_{u}^{}}_{\bar{A}}^{A_{1}^{}}_{\bar{A}}^{\tensor*[^{1}]{\hspace{-0.2em}B}{}_{2}^{}}}$ &    $1.3382$ \\[0.4em]
                                                    $d\indices*{*_{X}^{A_{1g}^{}}_{\bar{A}}^{\tensor*[^{1}]{\hspace{-0.2em}A}{}_{1}^{}}_{\bar{A}}^{\tensor*[^{1}]{\hspace{-0.2em}A}{}_{1}^{}}}$ &  $- 0.3728$ &                                                    $d\indices*{*_{X}^{E_{u}^{}}_{\bar{A}}^{\tensor*[^{1}]{\hspace{-0.2em}A}{}_{1}^{}}_{\bar{A}}^{\tensor*[^{1}]{\hspace{-0.2em}A}{}_{1}^{}}}$ &    $1.1072$ &                   $d\indices*{*_{X}^{\tensor*[^{1}]{\hspace{-0.2em}E}{}_{u}^{}}_{\bar{A}}^{\tensor*[^{1}]{\hspace{-0.2em}A}{}_{1}^{}}_{\bar{A}}^{\tensor*[^{1}]{\hspace{-0.2em}A}{}_{1}^{}}}$ &  $- 1.0283$ &                                                   $d\indices*{*_{X}^{A_{1g}^{}}_{\bar{A}}^{\tensor*[^{1}]{\hspace{-0.2em}A}{}_{1}^{}}_{\bar{A}}^{\tensor*[^{2}]{\hspace{-0.2em}A}{}_{1}^{}}}$ &    $0.7059$ &                                                    $d\indices*{*_{X}^{E_{u}^{}}_{\bar{A}}^{\tensor*[^{1}]{\hspace{-0.2em}A}{}_{1}^{}}_{\bar{A}}^{\tensor*[^{2}]{\hspace{-0.2em}A}{}_{1}^{}}}$ &  $- 6.6312$ \\[0.4em]
                    $d\indices*{*_{X}^{\tensor*[^{1}]{\hspace{-0.2em}E}{}_{u}^{}}_{\bar{A}}^{\tensor*[^{1}]{\hspace{-0.2em}A}{}_{1}^{}}_{\bar{A}}^{\tensor*[^{2}]{\hspace{-0.2em}A}{}_{1}^{}}}$ &    $8.5625$ &                                                                                     $d\indices*{*_{X}^{E_{g}^{}}_{\bar{A}}^{\tensor*[^{1}]{\hspace{-0.2em}A}{}_{1}^{}}_{\bar{A}}^{A_{2}^{}}}$ &  $- 8.3433$ &                                                                                     $d\indices*{*_{X}^{E_{g}^{}}_{\bar{A}}^{\tensor*[^{1}]{\hspace{-0.2em}A}{}_{1}^{}}_{\bar{A}}^{B_{1}^{}}}$ &  $- 1.4233$ &                                                                                    $d\indices*{*_{X}^{A_{2u}^{}}_{\bar{A}}^{\tensor*[^{1}]{\hspace{-0.2em}A}{}_{1}^{}}_{\bar{A}}^{B_{1}^{}}}$ & $- 11.1780$ &                                                                                    $d\indices*{*_{X}^{B_{1u}^{}}_{\bar{A}}^{\tensor*[^{1}]{\hspace{-0.2em}A}{}_{1}^{}}_{\bar{A}}^{B_{1}^{}}}$ &  $- 0.5686$ \\[0.4em]
                                                     $d\indices*{*_{X}^{E_{g}^{}}_{\bar{A}}^{\tensor*[^{1}]{\hspace{-0.2em}A}{}_{1}^{}}_{\bar{A}}^{\tensor*[^{1}]{\hspace{-0.2em}B}{}_{1}^{}}}$ &  $- 4.6998$ &                                                   $d\indices*{*_{X}^{A_{2u}^{}}_{\bar{A}}^{\tensor*[^{1}]{\hspace{-0.2em}A}{}_{1}^{}}_{\bar{A}}^{\tensor*[^{1}]{\hspace{-0.2em}B}{}_{1}^{}}}$ &    $6.1048$ &                                                   $d\indices*{*_{X}^{B_{1u}^{}}_{\bar{A}}^{\tensor*[^{1}]{\hspace{-0.2em}A}{}_{1}^{}}_{\bar{A}}^{\tensor*[^{1}]{\hspace{-0.2em}B}{}_{1}^{}}}$ &  $- 3.7742$ &                                                    $d\indices*{*_{X}^{E_{g}^{}}_{\bar{A}}^{\tensor*[^{1}]{\hspace{-0.2em}A}{}_{1}^{}}_{\bar{A}}^{\tensor*[^{2}]{\hspace{-0.2em}B}{}_{1}^{}}}$ &  $- 6.0817$ &                                                   $d\indices*{*_{X}^{A_{2u}^{}}_{\bar{A}}^{\tensor*[^{1}]{\hspace{-0.2em}A}{}_{1}^{}}_{\bar{A}}^{\tensor*[^{2}]{\hspace{-0.2em}B}{}_{1}^{}}}$ &    $2.1196$ \\[0.4em]
                                                    $d\indices*{*_{X}^{B_{1u}^{}}_{\bar{A}}^{\tensor*[^{1}]{\hspace{-0.2em}A}{}_{1}^{}}_{\bar{A}}^{\tensor*[^{2}]{\hspace{-0.2em}B}{}_{1}^{}}}$ &    $4.1474$ &                                                                                     $d\indices*{*_{X}^{E_{u}^{}}_{\bar{A}}^{\tensor*[^{1}]{\hspace{-0.2em}A}{}_{1}^{}}_{\bar{A}}^{B_{2}^{}}}$ &    $9.4202$ &                                                    $d\indices*{*_{X}^{\tensor*[^{1}]{\hspace{-0.2em}E}{}_{u}^{}}_{\bar{A}}^{\tensor*[^{1}]{\hspace{-0.2em}A}{}_{1}^{}}_{\bar{A}}^{B_{2}^{}}}$ & $- 11.1107$ &                                                    $d\indices*{*_{X}^{E_{u}^{}}_{\bar{A}}^{\tensor*[^{1}]{\hspace{-0.2em}A}{}_{1}^{}}_{\bar{A}}^{\tensor*[^{1}]{\hspace{-0.2em}B}{}_{2}^{}}}$ &  $- 6.7037$ &                   $d\indices*{*_{X}^{\tensor*[^{1}]{\hspace{-0.2em}E}{}_{u}^{}}_{\bar{A}}^{\tensor*[^{1}]{\hspace{-0.2em}A}{}_{1}^{}}_{\bar{A}}^{\tensor*[^{1}]{\hspace{-0.2em}B}{}_{2}^{}}}$ &    $7.6983$ \\[0.4em]
                                                    $d\indices*{*_{X}^{A_{1g}^{}}_{\bar{A}}^{\tensor*[^{2}]{\hspace{-0.2em}A}{}_{1}^{}}_{\bar{A}}^{\tensor*[^{2}]{\hspace{-0.2em}A}{}_{1}^{}}}$ & $- 14.6438$ &                                                    $d\indices*{*_{X}^{E_{u}^{}}_{\bar{A}}^{\tensor*[^{2}]{\hspace{-0.2em}A}{}_{1}^{}}_{\bar{A}}^{\tensor*[^{2}]{\hspace{-0.2em}A}{}_{1}^{}}}$ &  $- 0.5352$ &                   $d\indices*{*_{X}^{\tensor*[^{1}]{\hspace{-0.2em}E}{}_{u}^{}}_{\bar{A}}^{\tensor*[^{2}]{\hspace{-0.2em}A}{}_{1}^{}}_{\bar{A}}^{\tensor*[^{2}]{\hspace{-0.2em}A}{}_{1}^{}}}$ &    $0.4225$ &                                                                                     $d\indices*{*_{X}^{E_{g}^{}}_{\bar{A}}^{\tensor*[^{2}]{\hspace{-0.2em}A}{}_{1}^{}}_{\bar{A}}^{A_{2}^{}}}$ &    $2.0457$ &                                                                                     $d\indices*{*_{X}^{E_{g}^{}}_{\bar{A}}^{\tensor*[^{2}]{\hspace{-0.2em}A}{}_{1}^{}}_{\bar{A}}^{B_{1}^{}}}$ &    $7.3374$ \\[0.4em]
                                                                                     $d\indices*{*_{X}^{A_{2u}^{}}_{\bar{A}}^{\tensor*[^{2}]{\hspace{-0.2em}A}{}_{1}^{}}_{\bar{A}}^{B_{1}^{}}}$ &  $- 3.5354$ &                                                                                    $d\indices*{*_{X}^{B_{1u}^{}}_{\bar{A}}^{\tensor*[^{2}]{\hspace{-0.2em}A}{}_{1}^{}}_{\bar{A}}^{B_{1}^{}}}$ &    $2.8315$ &                                                    $d\indices*{*_{X}^{E_{g}^{}}_{\bar{A}}^{\tensor*[^{2}]{\hspace{-0.2em}A}{}_{1}^{}}_{\bar{A}}^{\tensor*[^{1}]{\hspace{-0.2em}B}{}_{1}^{}}}$ &    $2.6255$ &                                                   $d\indices*{*_{X}^{A_{2u}^{}}_{\bar{A}}^{\tensor*[^{2}]{\hspace{-0.2em}A}{}_{1}^{}}_{\bar{A}}^{\tensor*[^{1}]{\hspace{-0.2em}B}{}_{1}^{}}}$ & $- 12.2657$ &                                                   $d\indices*{*_{X}^{B_{1u}^{}}_{\bar{A}}^{\tensor*[^{2}]{\hspace{-0.2em}A}{}_{1}^{}}_{\bar{A}}^{\tensor*[^{1}]{\hspace{-0.2em}B}{}_{1}^{}}}$ &  $- 7.7845$ \\[0.4em]
                                                     $d\indices*{*_{X}^{E_{g}^{}}_{\bar{A}}^{\tensor*[^{2}]{\hspace{-0.2em}A}{}_{1}^{}}_{\bar{A}}^{\tensor*[^{2}]{\hspace{-0.2em}B}{}_{1}^{}}}$ &    $1.7696$ &                                                   $d\indices*{*_{X}^{A_{2u}^{}}_{\bar{A}}^{\tensor*[^{2}]{\hspace{-0.2em}A}{}_{1}^{}}_{\bar{A}}^{\tensor*[^{2}]{\hspace{-0.2em}B}{}_{1}^{}}}$ & $- 12.1602$ &                                                   $d\indices*{*_{X}^{B_{1u}^{}}_{\bar{A}}^{\tensor*[^{2}]{\hspace{-0.2em}A}{}_{1}^{}}_{\bar{A}}^{\tensor*[^{2}]{\hspace{-0.2em}B}{}_{1}^{}}}$ & $- 10.3847$ &                                                                                     $d\indices*{*_{X}^{E_{u}^{}}_{\bar{A}}^{\tensor*[^{2}]{\hspace{-0.2em}A}{}_{1}^{}}_{\bar{A}}^{B_{2}^{}}}$ &  $- 3.0913$ &                                                    $d\indices*{*_{X}^{\tensor*[^{1}]{\hspace{-0.2em}E}{}_{u}^{}}_{\bar{A}}^{\tensor*[^{2}]{\hspace{-0.2em}A}{}_{1}^{}}_{\bar{A}}^{B_{2}^{}}}$ &  $- 1.8992$ \\[0.4em]
                                                     $d\indices*{*_{X}^{E_{u}^{}}_{\bar{A}}^{\tensor*[^{2}]{\hspace{-0.2em}A}{}_{1}^{}}_{\bar{A}}^{\tensor*[^{1}]{\hspace{-0.2em}B}{}_{2}^{}}}$ &    $8.6372$ &                   $d\indices*{*_{X}^{\tensor*[^{1}]{\hspace{-0.2em}E}{}_{u}^{}}_{\bar{A}}^{\tensor*[^{2}]{\hspace{-0.2em}A}{}_{1}^{}}_{\bar{A}}^{\tensor*[^{1}]{\hspace{-0.2em}B}{}_{2}^{}}}$ &  $- 0.4688$ &                                                                                                                     $d\indices*{*_{X}^{A_{1g}^{}}_{\bar{A}}^{A_{2}^{}}_{\bar{A}}^{A_{2}^{}}}$ &   $14.6610$ &                                                                                                                      $d\indices*{*_{X}^{E_{u}^{}}_{\bar{A}}^{A_{2}^{}}_{\bar{A}}^{A_{2}^{}}}$ &    $6.6638$ &                                                                                     $d\indices*{*_{X}^{\tensor*[^{1}]{\hspace{-0.2em}E}{}_{u}^{}}_{\bar{A}}^{A_{2}^{}}_{\bar{A}}^{A_{2}^{}}}$ &  $- 1.1567$ \\[0.4em]
                                                                                                                       $d\indices*{*_{X}^{E_{u}^{}}_{\bar{A}}^{A_{2}^{}}_{\bar{A}}^{B_{1}^{}}}$ &    $6.8811$ &                                                                                     $d\indices*{*_{X}^{\tensor*[^{1}]{\hspace{-0.2em}E}{}_{u}^{}}_{\bar{A}}^{A_{2}^{}}_{\bar{A}}^{B_{1}^{}}}$ &  $- 7.5253$ &                                                                                     $d\indices*{*_{X}^{E_{u}^{}}_{\bar{A}}^{A_{2}^{}}_{\bar{A}}^{\tensor*[^{1}]{\hspace{-0.2em}B}{}_{1}^{}}}$ &  $- 9.9476$ &                                                    $d\indices*{*_{X}^{\tensor*[^{1}]{\hspace{-0.2em}E}{}_{u}^{}}_{\bar{A}}^{A_{2}^{}}_{\bar{A}}^{\tensor*[^{1}]{\hspace{-0.2em}B}{}_{1}^{}}}$ &   $12.1428$ &                                                                                     $d\indices*{*_{X}^{E_{u}^{}}_{\bar{A}}^{A_{2}^{}}_{\bar{A}}^{\tensor*[^{2}]{\hspace{-0.2em}B}{}_{1}^{}}}$ &  $- 7.4255$ \\[0.4em]
                                                     $d\indices*{*_{X}^{\tensor*[^{1}]{\hspace{-0.2em}E}{}_{u}^{}}_{\bar{A}}^{A_{2}^{}}_{\bar{A}}^{\tensor*[^{2}]{\hspace{-0.2em}B}{}_{1}^{}}}$ &  $- 0.9101$ &                                                                                                                      $d\indices*{*_{X}^{E_{g}^{}}_{\bar{A}}^{A_{2}^{}}_{\bar{A}}^{B_{2}^{}}}$ &    $0.6681$ &                                                                                                                     $d\indices*{*_{X}^{A_{2u}^{}}_{\bar{A}}^{A_{2}^{}}_{\bar{A}}^{B_{2}^{}}}$ &   $20.0521$ &                                                                                                                     $d\indices*{*_{X}^{B_{1u}^{}}_{\bar{A}}^{A_{2}^{}}_{\bar{A}}^{B_{2}^{}}}$ & $- 12.1239$ &                                                                                     $d\indices*{*_{X}^{E_{g}^{}}_{\bar{A}}^{A_{2}^{}}_{\bar{A}}^{\tensor*[^{1}]{\hspace{-0.2em}B}{}_{2}^{}}}$ &    $0.1875$ \\[0.4em]
                                                                                     $d\indices*{*_{X}^{A_{2u}^{}}_{\bar{A}}^{A_{2}^{}}_{\bar{A}}^{\tensor*[^{1}]{\hspace{-0.2em}B}{}_{2}^{}}}$ & $- 10.0653$ &                                                                                    $d\indices*{*_{X}^{B_{1u}^{}}_{\bar{A}}^{A_{2}^{}}_{\bar{A}}^{\tensor*[^{1}]{\hspace{-0.2em}B}{}_{2}^{}}}$ &    $8.8409$ &                                                                                                                     $d\indices*{*_{X}^{A_{1g}^{}}_{\bar{A}}^{B_{1}^{}}_{\bar{A}}^{B_{1}^{}}}$ &    $2.5214$ &                                                                                                                      $d\indices*{*_{X}^{E_{u}^{}}_{\bar{A}}^{B_{1}^{}}_{\bar{A}}^{B_{1}^{}}}$ &  $- 1.6457$ &                                                                                     $d\indices*{*_{X}^{\tensor*[^{1}]{\hspace{-0.2em}E}{}_{u}^{}}_{\bar{A}}^{B_{1}^{}}_{\bar{A}}^{B_{1}^{}}}$ &  $- 0.8505$ \\[0.4em]
                                                                                     $d\indices*{*_{X}^{A_{1g}^{}}_{\bar{A}}^{B_{1}^{}}_{\bar{A}}^{\tensor*[^{1}]{\hspace{-0.2em}B}{}_{1}^{}}}$ &    $4.3450$ &                                                                                     $d\indices*{*_{X}^{E_{u}^{}}_{\bar{A}}^{B_{1}^{}}_{\bar{A}}^{\tensor*[^{1}]{\hspace{-0.2em}B}{}_{1}^{}}}$ &    $5.2678$ &                                                    $d\indices*{*_{X}^{\tensor*[^{1}]{\hspace{-0.2em}E}{}_{u}^{}}_{\bar{A}}^{B_{1}^{}}_{\bar{A}}^{\tensor*[^{1}]{\hspace{-0.2em}B}{}_{1}^{}}}$ &  $- 8.7261$ &                                                                                    $d\indices*{*_{X}^{A_{1g}^{}}_{\bar{A}}^{B_{1}^{}}_{\bar{A}}^{\tensor*[^{2}]{\hspace{-0.2em}B}{}_{1}^{}}}$ &    $1.0047$ &                                                                                     $d\indices*{*_{X}^{E_{u}^{}}_{\bar{A}}^{B_{1}^{}}_{\bar{A}}^{\tensor*[^{2}]{\hspace{-0.2em}B}{}_{1}^{}}}$ &    $7.5603$ \\[0.4em]
                                                     $d\indices*{*_{X}^{\tensor*[^{1}]{\hspace{-0.2em}E}{}_{u}^{}}_{\bar{A}}^{B_{1}^{}}_{\bar{A}}^{\tensor*[^{2}]{\hspace{-0.2em}B}{}_{1}^{}}}$ &  $- 7.4427$ &                                                                                                                      $d\indices*{*_{X}^{E_{g}^{}}_{\bar{A}}^{B_{2}^{}}_{\bar{A}}^{B_{1}^{}}}$ &  $- 9.6188$ &                                                                                     $d\indices*{*_{X}^{E_{g}^{}}_{\bar{A}}^{\tensor*[^{1}]{\hspace{-0.2em}B}{}_{2}^{}}_{\bar{A}}^{B_{1}^{}}}$ &    $5.9357$ &                                                   $d\indices*{*_{X}^{A_{1g}^{}}_{\bar{A}}^{\tensor*[^{1}]{\hspace{-0.2em}B}{}_{1}^{}}_{\bar{A}}^{\tensor*[^{1}]{\hspace{-0.2em}B}{}_{1}^{}}}$ & $- 17.3460$ &                                                    $d\indices*{*_{X}^{E_{u}^{}}_{\bar{A}}^{\tensor*[^{1}]{\hspace{-0.2em}B}{}_{1}^{}}_{\bar{A}}^{\tensor*[^{1}]{\hspace{-0.2em}B}{}_{1}^{}}}$ &    $1.1130$ \\[0.4em]
                    $d\indices*{*_{X}^{\tensor*[^{1}]{\hspace{-0.2em}E}{}_{u}^{}}_{\bar{A}}^{\tensor*[^{1}]{\hspace{-0.2em}B}{}_{1}^{}}_{\bar{A}}^{\tensor*[^{1}]{\hspace{-0.2em}B}{}_{1}^{}}}$ &  $- 3.9095$ &                                                   $d\indices*{*_{X}^{A_{1g}^{}}_{\bar{A}}^{\tensor*[^{1}]{\hspace{-0.2em}B}{}_{1}^{}}_{\bar{A}}^{\tensor*[^{2}]{\hspace{-0.2em}B}{}_{1}^{}}}$ & $- 10.8822$ &                                                    $d\indices*{*_{X}^{E_{u}^{}}_{\bar{A}}^{\tensor*[^{1}]{\hspace{-0.2em}B}{}_{1}^{}}_{\bar{A}}^{\tensor*[^{2}]{\hspace{-0.2em}B}{}_{1}^{}}}$ &    $1.6188$ &                   $d\indices*{*_{X}^{\tensor*[^{1}]{\hspace{-0.2em}E}{}_{u}^{}}_{\bar{A}}^{\tensor*[^{1}]{\hspace{-0.2em}B}{}_{1}^{}}_{\bar{A}}^{\tensor*[^{2}]{\hspace{-0.2em}B}{}_{1}^{}}}$ &    $0.6494$ &                                                                                     $d\indices*{*_{X}^{E_{g}^{}}_{\bar{A}}^{B_{2}^{}}_{\bar{A}}^{\tensor*[^{1}]{\hspace{-0.2em}B}{}_{1}^{}}}$ &   $13.2653$ \\[0.4em]
                                                     $d\indices*{*_{X}^{E_{g}^{}}_{\bar{A}}^{\tensor*[^{1}]{\hspace{-0.2em}B}{}_{2}^{}}_{\bar{A}}^{\tensor*[^{1}]{\hspace{-0.2em}B}{}_{1}^{}}}$ &  $- 9.7467$ &                                                   $d\indices*{*_{X}^{A_{1g}^{}}_{\bar{A}}^{\tensor*[^{2}]{\hspace{-0.2em}B}{}_{1}^{}}_{\bar{A}}^{\tensor*[^{2}]{\hspace{-0.2em}B}{}_{1}^{}}}$ & $- 10.1338$ &                                                    $d\indices*{*_{X}^{E_{u}^{}}_{\bar{A}}^{\tensor*[^{2}]{\hspace{-0.2em}B}{}_{1}^{}}_{\bar{A}}^{\tensor*[^{2}]{\hspace{-0.2em}B}{}_{1}^{}}}$ &    $3.5660$ &                   $d\indices*{*_{X}^{\tensor*[^{1}]{\hspace{-0.2em}E}{}_{u}^{}}_{\bar{A}}^{\tensor*[^{2}]{\hspace{-0.2em}B}{}_{1}^{}}_{\bar{A}}^{\tensor*[^{2}]{\hspace{-0.2em}B}{}_{1}^{}}}$ &    $0.9737$ &                                                                                     $d\indices*{*_{X}^{E_{g}^{}}_{\bar{A}}^{B_{2}^{}}_{\bar{A}}^{\tensor*[^{2}]{\hspace{-0.2em}B}{}_{1}^{}}}$ &  $- 1.2145$ \\[0.4em]
                                                     $d\indices*{*_{X}^{E_{g}^{}}_{\bar{A}}^{\tensor*[^{1}]{\hspace{-0.2em}B}{}_{2}^{}}_{\bar{A}}^{\tensor*[^{2}]{\hspace{-0.2em}B}{}_{1}^{}}}$ &    $0.1467$ &                                                                                                                     $d\indices*{*_{X}^{A_{1g}^{}}_{\bar{A}}^{B_{2}^{}}_{\bar{A}}^{B_{2}^{}}}$ &   $15.5141$ &                                                                                                                      $d\indices*{*_{X}^{E_{u}^{}}_{\bar{A}}^{B_{2}^{}}_{\bar{A}}^{B_{2}^{}}}$ &  $- 2.7771$ &                                                                                     $d\indices*{*_{X}^{\tensor*[^{1}]{\hspace{-0.2em}E}{}_{u}^{}}_{\bar{A}}^{B_{2}^{}}_{\bar{A}}^{B_{2}^{}}}$ &    $0.2873$ &                                                                                    $d\indices*{*_{X}^{A_{1g}^{}}_{\bar{A}}^{B_{2}^{}}_{\bar{A}}^{\tensor*[^{1}]{\hspace{-0.2em}B}{}_{2}^{}}}$ & $- 10.0802$ \\[0.4em]
                                                                                      $d\indices*{*_{X}^{E_{u}^{}}_{\bar{A}}^{B_{2}^{}}_{\bar{A}}^{\tensor*[^{1}]{\hspace{-0.2em}B}{}_{2}^{}}}$ &    $0.2751$ &                                                    $d\indices*{*_{X}^{\tensor*[^{1}]{\hspace{-0.2em}E}{}_{u}^{}}_{\bar{A}}^{B_{2}^{}}_{\bar{A}}^{\tensor*[^{1}]{\hspace{-0.2em}B}{}_{2}^{}}}$ &  $- 0.9627$ &                                                   $d\indices*{*_{X}^{A_{1g}^{}}_{\bar{A}}^{\tensor*[^{1}]{\hspace{-0.2em}B}{}_{2}^{}}_{\bar{A}}^{\tensor*[^{1}]{\hspace{-0.2em}B}{}_{2}^{}}}$ &   $10.1312$ &                                                    $d\indices*{*_{X}^{E_{u}^{}}_{\bar{A}}^{\tensor*[^{1}]{\hspace{-0.2em}B}{}_{2}^{}}_{\bar{A}}^{\tensor*[^{1}]{\hspace{-0.2em}B}{}_{2}^{}}}$ &  $- 6.9644$ &                   $d\indices*{*_{X}^{\tensor*[^{1}]{\hspace{-0.2em}E}{}_{u}^{}}_{\bar{A}}^{\tensor*[^{1}]{\hspace{-0.2em}B}{}_{2}^{}}_{\bar{A}}^{\tensor*[^{1}]{\hspace{-0.2em}B}{}_{2}^{}}}$ &    $1.0266$ \\[0.4em]
                                                                                                                             $d\indices*{*_{\Gamma}^{T_{2g}^{}}_{X}^{E_{g}^{}}_{X}^{E_{g}^{}}}$ &   $22.5282$ &                                                                                                                          $d\indices*{*_{\Gamma}^{T_{2g}^{}}_{X}^{B_{1u}^{}}_{X}^{A_{2u}^{}}}$ &   $13.1707$ &                                                                                                                            $d\indices*{*_{\Gamma}^{T_{2g}^{}}_{X}^{E_{u}^{}}_{X}^{E_{u}^{}}}$ & $- 16.2149$ &                                                                                           $d\indices*{*_{\Gamma}^{T_{2g}^{}}_{X}^{E_{u}^{}}_{X}^{\tensor*[^{1}]{\hspace{-0.2em}E}{}_{u}^{}}}$ &    $9.2043$ &                                                          $d\indices*{*_{\Gamma}^{T_{2g}^{}}_{X}^{\tensor*[^{1}]{\hspace{-0.2em}E}{}_{u}^{}}_{X}^{\tensor*[^{1}]{\hspace{-0.2em}E}{}_{u}^{}}}$ & $- 34.0148$ \\[0.4em]
                                                                                                                            $d\indices*{*_{\Gamma}^{T_{2g}^{}}_{X}^{A_{1g}^{}}_{X}^{E_{g}^{}}}$ & $- 10.2044$ &                                                                                                                           $d\indices*{*_{\Gamma}^{T_{2g}^{}}_{X}^{A_{2u}^{}}_{X}^{E_{u}^{}}}$ &   $11.2352$ &                                                                                          $d\indices*{*_{\Gamma}^{T_{2g}^{}}_{X}^{A_{2u}^{}}_{X}^{\tensor*[^{1}]{\hspace{-0.2em}E}{}_{u}^{}}}$ & $- 19.3272$ &                                                                                                                           $d\indices*{*_{\Gamma}^{T_{2g}^{}}_{X}^{B_{1u}^{}}_{X}^{E_{u}^{}}}$ &  $- 9.5933$ &                                                                                          $d\indices*{*_{\Gamma}^{T_{2g}^{}}_{X}^{B_{1u}^{}}_{X}^{\tensor*[^{1}]{\hspace{-0.2em}E}{}_{u}^{}}}$ &   $11.9337$ \\[0.4em]
                                                                                                                            $d\indices*{*_{\Gamma}^{T_{1u}^{}}_{X}^{A_{1g}^{}}_{X}^{E_{u}^{}}}$ & $- 16.1265$ &                                                                                          $d\indices*{*_{\Gamma}^{T_{1u}^{}}_{X}^{A_{1g}^{}}_{X}^{\tensor*[^{1}]{\hspace{-0.2em}E}{}_{u}^{}}}$ &   $18.9041$ &                                                                                                                           $d\indices*{*_{\Gamma}^{T_{1u}^{}}_{X}^{B_{1u}^{}}_{X}^{E_{g}^{}}}$ & $- 16.4616$ &                                                                                                                           $d\indices*{*_{\Gamma}^{T_{1u}^{}}_{X}^{E_{g}^{}}_{X}^{A_{2u}^{}}}$ &   $23.6967$ &                                                                                                                          $d\indices*{*_{\Gamma}^{T_{1u}^{}}_{X}^{A_{1g}^{}}_{X}^{A_{2u}^{}}}$ &   $19.8390$ \\[0.4em]
                                                                                                                             $d\indices*{*_{\Gamma}^{T_{1u}^{}}_{X}^{E_{g}^{}}_{X}^{E_{u}^{}}}$ & $- 16.1182$ &                                                                                           $d\indices*{*_{\Gamma}^{T_{1u}^{}}_{X}^{E_{g}^{}}_{X}^{\tensor*[^{1}]{\hspace{-0.2em}E}{}_{u}^{}}}$ &   $20.3291$ &                                                                                                                      $d\indices*{*_{\Gamma}^{T_{2g}^{}}_{A}^{A_{1}^{}}_{\bar{A}}^{A_{1}^{}}}$ &   $20.5198$ &                                                                                     $d\indices*{*_{\Gamma}^{T_{2g}^{}}_{A}^{A_{1}^{}}_{\bar{A}}^{\tensor*[^{1}]{\hspace{-0.2em}A}{}_{1}^{}}}$ &  $- 6.9758$ &                                                                                     $d\indices*{*_{\Gamma}^{T_{2g}^{}}_{A}^{A_{1}^{}}_{\bar{A}}^{\tensor*[^{2}]{\hspace{-0.2em}A}{}_{1}^{}}}$ &    $1.9380$ \\[0.4em]
                                                     $d\indices*{*_{\Gamma}^{T_{2g}^{}}_{A}^{\tensor*[^{1}]{\hspace{-0.2em}A}{}_{1}^{}}_{\bar{A}}^{\tensor*[^{1}]{\hspace{-0.2em}A}{}_{1}^{}}}$ &  $- 1.6417$ &                                                    $d\indices*{*_{\Gamma}^{T_{2g}^{}}_{A}^{\tensor*[^{1}]{\hspace{-0.2em}A}{}_{1}^{}}_{\bar{A}}^{\tensor*[^{2}]{\hspace{-0.2em}A}{}_{1}^{}}}$ &  $- 2.2069$ &                                                    $d\indices*{*_{\Gamma}^{T_{2g}^{}}_{A}^{\tensor*[^{2}]{\hspace{-0.2em}A}{}_{1}^{}}_{\bar{A}}^{\tensor*[^{2}]{\hspace{-0.2em}A}{}_{1}^{}}}$ &   $10.8195$ &                                                                                                                      $d\indices*{*_{\Gamma}^{T_{2g}^{}}_{A}^{A_{2}^{}}_{\bar{A}}^{A_{2}^{}}}$ & $- 10.0224$ &                                                                                                                      $d\indices*{*_{\Gamma}^{T_{2g}^{}}_{A}^{B_{1}^{}}_{\bar{A}}^{B_{1}^{}}}$ &  $- 1.8240$ \\[0.4em]
                                                                                      $d\indices*{*_{\Gamma}^{T_{2g}^{}}_{A}^{B_{1}^{}}_{\bar{A}}^{\tensor*[^{1}]{\hspace{-0.2em}B}{}_{1}^{}}}$ &    $5.9008$ &                                                                                     $d\indices*{*_{\Gamma}^{T_{2g}^{}}_{A}^{B_{1}^{}}_{\bar{A}}^{\tensor*[^{2}]{\hspace{-0.2em}B}{}_{1}^{}}}$ &  $- 2.1900$ &                                                    $d\indices*{*_{\Gamma}^{T_{2g}^{}}_{A}^{\tensor*[^{1}]{\hspace{-0.2em}B}{}_{1}^{}}_{\bar{A}}^{\tensor*[^{1}]{\hspace{-0.2em}B}{}_{1}^{}}}$ &  $- 0.0679$ &                                                    $d\indices*{*_{\Gamma}^{T_{2g}^{}}_{A}^{\tensor*[^{1}]{\hspace{-0.2em}B}{}_{1}^{}}_{\bar{A}}^{\tensor*[^{2}]{\hspace{-0.2em}B}{}_{1}^{}}}$ &   $10.1283$ &                                                    $d\indices*{*_{\Gamma}^{T_{2g}^{}}_{A}^{\tensor*[^{2}]{\hspace{-0.2em}B}{}_{1}^{}}_{\bar{A}}^{\tensor*[^{2}]{\hspace{-0.2em}B}{}_{1}^{}}}$ &   $10.1715$ \\[0.4em]
                                                                                                                       $d\indices*{*_{\Gamma}^{T_{2g}^{}}_{A}^{B_{2}^{}}_{\bar{A}}^{B_{2}^{}}}$ & $- 14.4232$ &                                                                                     $d\indices*{*_{\Gamma}^{T_{2g}^{}}_{A}^{B_{2}^{}}_{\bar{A}}^{\tensor*[^{1}]{\hspace{-0.2em}B}{}_{2}^{}}}$ &   $10.6408$ &                                                    $d\indices*{*_{\Gamma}^{T_{2g}^{}}_{A}^{\tensor*[^{1}]{\hspace{-0.2em}B}{}_{2}^{}}_{\bar{A}}^{\tensor*[^{1}]{\hspace{-0.2em}B}{}_{2}^{}}}$ &  $- 8.1975$ &                                                                                                                      $d\indices*{*_{\Gamma}^{T_{2g}^{}}_{A}^{A_{1}^{}}_{\bar{A}}^{A_{2}^{}}}$ &  $- 0.4895$ &                                                                                                                      $d\indices*{*_{\Gamma}^{T_{2g}^{}}_{A}^{A_{1}^{}}_{\bar{A}}^{B_{1}^{}}}$ &    $0.0234$ \\[0.4em]
                                                                                      $d\indices*{*_{\Gamma}^{T_{2g}^{}}_{A}^{A_{1}^{}}_{\bar{A}}^{\tensor*[^{1}]{\hspace{-0.2em}B}{}_{1}^{}}}$ & $- 13.5409$ &                                                                                     $d\indices*{*_{\Gamma}^{T_{2g}^{}}_{A}^{A_{1}^{}}_{\bar{A}}^{\tensor*[^{2}]{\hspace{-0.2em}B}{}_{1}^{}}}$ & $- 13.0562$ &                                                                                     $d\indices*{*_{\Gamma}^{T_{2g}^{}}_{A}^{\tensor*[^{1}]{\hspace{-0.2em}A}{}_{1}^{}}_{\bar{A}}^{A_{2}^{}}}$ &    $7.4200$ &                                                                                     $d\indices*{*_{\Gamma}^{T_{2g}^{}}_{A}^{\tensor*[^{1}]{\hspace{-0.2em}A}{}_{1}^{}}_{\bar{A}}^{B_{1}^{}}}$ &  $- 3.7740$ &                                                    $d\indices*{*_{\Gamma}^{T_{2g}^{}}_{A}^{\tensor*[^{1}]{\hspace{-0.2em}A}{}_{1}^{}}_{\bar{A}}^{\tensor*[^{1}]{\hspace{-0.2em}B}{}_{1}^{}}}$ &    $5.3085$ \\[0.4em]
                                                     $d\indices*{*_{\Gamma}^{T_{2g}^{}}_{A}^{\tensor*[^{1}]{\hspace{-0.2em}A}{}_{1}^{}}_{\bar{A}}^{\tensor*[^{2}]{\hspace{-0.2em}B}{}_{1}^{}}}$ &    $6.9066$ &                                                                                     $d\indices*{*_{\Gamma}^{T_{2g}^{}}_{A}^{\tensor*[^{2}]{\hspace{-0.2em}A}{}_{1}^{}}_{\bar{A}}^{A_{2}^{}}}$ &  $- 5.0512$ &                                                                                     $d\indices*{*_{\Gamma}^{T_{2g}^{}}_{A}^{\tensor*[^{2}]{\hspace{-0.2em}A}{}_{1}^{}}_{\bar{A}}^{B_{1}^{}}}$ &    $9.7995$ &                                                    $d\indices*{*_{\Gamma}^{T_{2g}^{}}_{A}^{\tensor*[^{2}]{\hspace{-0.2em}A}{}_{1}^{}}_{\bar{A}}^{\tensor*[^{1}]{\hspace{-0.2em}B}{}_{1}^{}}}$ &  $- 0.1744$ &                                                    $d\indices*{*_{\Gamma}^{T_{2g}^{}}_{A}^{\tensor*[^{2}]{\hspace{-0.2em}A}{}_{1}^{}}_{\bar{A}}^{\tensor*[^{2}]{\hspace{-0.2em}B}{}_{1}^{}}}$ &    $3.2244$ \\[0.4em]
                                                                                                                       $d\indices*{*_{\Gamma}^{T_{2g}^{}}_{A}^{A_{2}^{}}_{\bar{A}}^{B_{2}^{}}}$ &    $2.3193$ &                                                                                     $d\indices*{*_{\Gamma}^{T_{2g}^{}}_{A}^{A_{2}^{}}_{\bar{A}}^{\tensor*[^{1}]{\hspace{-0.2em}B}{}_{2}^{}}}$ &  $- 4.7277$ &                                                                                                                      $d\indices*{*_{\Gamma}^{T_{2g}^{}}_{A}^{B_{1}^{}}_{\bar{A}}^{B_{2}^{}}}$ &  $- 7.6359$ &                                                                                     $d\indices*{*_{\Gamma}^{T_{2g}^{}}_{A}^{B_{1}^{}}_{\bar{A}}^{\tensor*[^{1}]{\hspace{-0.2em}B}{}_{2}^{}}}$ &    $6.9321$ &                                                                                     $d\indices*{*_{\Gamma}^{T_{2g}^{}}_{A}^{\tensor*[^{1}]{\hspace{-0.2em}B}{}_{1}^{}}_{\bar{A}}^{B_{2}^{}}}$ &   $12.2361$ \\[0.4em]
                                                     $d\indices*{*_{\Gamma}^{T_{2g}^{}}_{A}^{\tensor*[^{1}]{\hspace{-0.2em}B}{}_{1}^{}}_{\bar{A}}^{\tensor*[^{1}]{\hspace{-0.2em}B}{}_{2}^{}}}$ &  $- 8.1045$ &                                                                                     $d\indices*{*_{\Gamma}^{T_{2g}^{}}_{A}^{\tensor*[^{2}]{\hspace{-0.2em}B}{}_{1}^{}}_{\bar{A}}^{B_{2}^{}}}$ &    $0.1424$ &                                                    $d\indices*{*_{\Gamma}^{T_{2g}^{}}_{A}^{\tensor*[^{2}]{\hspace{-0.2em}B}{}_{1}^{}}_{\bar{A}}^{\tensor*[^{1}]{\hspace{-0.2em}B}{}_{2}^{}}}$ &  $- 5.9884$ &                                                                                     $d\indices*{*_{\Gamma}^{T_{1u}^{}}_{A}^{A_{1}^{}}_{\bar{A}}^{\tensor*[^{1}]{\hspace{-0.2em}A}{}_{1}^{}}}$ &    $2.0312$ &                                                                                     $d\indices*{*_{\Gamma}^{T_{1u}^{}}_{A}^{A_{1}^{}}_{\bar{A}}^{\tensor*[^{2}]{\hspace{-0.2em}A}{}_{1}^{}}}$ &   $23.1879$ \\[0.4em]
                                                                                                                       $d\indices*{*_{\Gamma}^{T_{1u}^{}}_{A}^{A_{1}^{}}_{\bar{A}}^{B_{2}^{}}}$ &  $- 2.5818$ &                                                                                     $d\indices*{*_{\Gamma}^{T_{1u}^{}}_{A}^{A_{1}^{}}_{\bar{A}}^{\tensor*[^{1}]{\hspace{-0.2em}B}{}_{2}^{}}}$ &    $0.2349$ &                                                    $d\indices*{*_{\Gamma}^{T_{1u}^{}}_{A}^{\tensor*[^{1}]{\hspace{-0.2em}A}{}_{1}^{}}_{\bar{A}}^{\tensor*[^{2}]{\hspace{-0.2em}A}{}_{1}^{}}}$ & $- 13.2554$ &                                                                                     $d\indices*{*_{\Gamma}^{T_{1u}^{}}_{A}^{\tensor*[^{1}]{\hspace{-0.2em}A}{}_{1}^{}}_{\bar{A}}^{B_{2}^{}}}$ & $- 14.4460$ &                                                    $d\indices*{*_{\Gamma}^{T_{1u}^{}}_{A}^{\tensor*[^{1}]{\hspace{-0.2em}A}{}_{1}^{}}_{\bar{A}}^{\tensor*[^{1}]{\hspace{-0.2em}B}{}_{2}^{}}}$ &   $11.6670$ \\[0.4em]
                                                                                      $d\indices*{*_{\Gamma}^{T_{1u}^{}}_{A}^{\tensor*[^{2}]{\hspace{-0.2em}A}{}_{1}^{}}_{\bar{A}}^{B_{2}^{}}}$ &  $- 1.2601$ &                                                    $d\indices*{*_{\Gamma}^{T_{1u}^{}}_{A}^{\tensor*[^{2}]{\hspace{-0.2em}A}{}_{1}^{}}_{\bar{A}}^{\tensor*[^{1}]{\hspace{-0.2em}B}{}_{2}^{}}}$ &  $- 1.2989$ &                                                                                                                      $d\indices*{*_{\Gamma}^{T_{1u}^{}}_{A}^{A_{2}^{}}_{\bar{A}}^{B_{1}^{}}}$ &   $11.5190$ &                                                                                     $d\indices*{*_{\Gamma}^{T_{1u}^{}}_{A}^{A_{2}^{}}_{\bar{A}}^{\tensor*[^{1}]{\hspace{-0.2em}B}{}_{1}^{}}}$ & $- 17.4191$ &                                                                                     $d\indices*{*_{\Gamma}^{T_{1u}^{}}_{A}^{A_{2}^{}}_{\bar{A}}^{\tensor*[^{2}]{\hspace{-0.2em}B}{}_{1}^{}}}$ &  $- 0.1463$ \\[0.4em]
                                                                                      $d\indices*{*_{\Gamma}^{T_{1u}^{}}_{A}^{B_{1}^{}}_{\bar{A}}^{\tensor*[^{1}]{\hspace{-0.2em}B}{}_{1}^{}}}$ &   $10.9394$ &                                                                                     $d\indices*{*_{\Gamma}^{T_{1u}^{}}_{A}^{B_{1}^{}}_{\bar{A}}^{\tensor*[^{2}]{\hspace{-0.2em}B}{}_{1}^{}}}$ &    $9.5904$ &                                                    $d\indices*{*_{\Gamma}^{T_{1u}^{}}_{A}^{\tensor*[^{1}]{\hspace{-0.2em}B}{}_{1}^{}}_{\bar{A}}^{\tensor*[^{2}]{\hspace{-0.2em}B}{}_{1}^{}}}$ &    $2.2727$ &                                                                                     $d\indices*{*_{\Gamma}^{T_{1u}^{}}_{A}^{B_{2}^{}}_{\bar{A}}^{\tensor*[^{1}]{\hspace{-0.2em}B}{}_{2}^{}}}$ &    $1.7326$ &                                                                                                                      $d\indices*{*_{\Gamma}^{T_{1u}^{}}_{A}^{A_{1}^{}}_{\bar{A}}^{B_{1}^{}}}$ &   $13.2330$ \\[0.4em]
                                                                                      $d\indices*{*_{\Gamma}^{T_{1u}^{}}_{A}^{A_{1}^{}}_{\bar{A}}^{\tensor*[^{1}]{\hspace{-0.2em}B}{}_{1}^{}}}$ &    $1.1383$ &                                                                                     $d\indices*{*_{\Gamma}^{T_{1u}^{}}_{A}^{A_{1}^{}}_{\bar{A}}^{\tensor*[^{2}]{\hspace{-0.2em}B}{}_{1}^{}}}$ &    $2.6949$ &                                                                                     $d\indices*{*_{\Gamma}^{T_{1u}^{}}_{A}^{\tensor*[^{1}]{\hspace{-0.2em}A}{}_{1}^{}}_{\bar{A}}^{B_{1}^{}}}$ &  $- 0.6572$ &                                                    $d\indices*{*_{\Gamma}^{T_{1u}^{}}_{A}^{\tensor*[^{1}]{\hspace{-0.2em}A}{}_{1}^{}}_{\bar{A}}^{\tensor*[^{1}]{\hspace{-0.2em}B}{}_{1}^{}}}$ & $- 10.4407$ &                                                    $d\indices*{*_{\Gamma}^{T_{1u}^{}}_{A}^{\tensor*[^{1}]{\hspace{-0.2em}A}{}_{1}^{}}_{\bar{A}}^{\tensor*[^{2}]{\hspace{-0.2em}B}{}_{1}^{}}}$ &    $0.4882$ \\[0.4em]
                                                                                      $d\indices*{*_{\Gamma}^{T_{1u}^{}}_{A}^{\tensor*[^{2}]{\hspace{-0.2em}A}{}_{1}^{}}_{\bar{A}}^{B_{1}^{}}}$ &    $2.3729$ &                                                    $d\indices*{*_{\Gamma}^{T_{1u}^{}}_{A}^{\tensor*[^{2}]{\hspace{-0.2em}A}{}_{1}^{}}_{\bar{A}}^{\tensor*[^{1}]{\hspace{-0.2em}B}{}_{1}^{}}}$ &   $10.3634$ &                                                    $d\indices*{*_{\Gamma}^{T_{1u}^{}}_{A}^{\tensor*[^{2}]{\hspace{-0.2em}A}{}_{1}^{}}_{\bar{A}}^{\tensor*[^{2}]{\hspace{-0.2em}B}{}_{1}^{}}}$ &   $14.5553$ &                                                                                                                      $d\indices*{*_{\Gamma}^{T_{1u}^{}}_{A}^{A_{2}^{}}_{\bar{A}}^{B_{2}^{}}}$ &   $20.0296$ &                                                                                     $d\indices*{*_{\Gamma}^{T_{1u}^{}}_{A}^{A_{2}^{}}_{\bar{A}}^{\tensor*[^{1}]{\hspace{-0.2em}B}{}_{2}^{}}}$ & $- 13.2082$ \\[0.4em]
                                                                                                                $d\indices*{*_{A_{3}}^{A_{1}^{}}_{X_{2}}^{A_{1g}^{}}_{\bar{A}_{1}}^{A_{1}^{}}}$ &    $2.7508$ &                                                                                                               $d\indices*{*_{A_{3}}^{A_{1}^{}}_{X_{2}}^{A_{2u}^{}}_{\bar{A}_{1}}^{A_{1}^{}}}$ &  $- 1.6811$ &                                                                              $d\indices*{*_{A_{3}}^{A_{1}^{}}_{X_{2}}^{A_{1g}^{}}_{\bar{A}_{1}}^{\tensor*[^{1}]{\hspace{-0.2em}A}{}_{1}^{}}}$ &    $0.9743$ &                                                                               $d\indices*{*_{A_{3}}^{A_{1}^{}}_{X_{2}}^{E_{g}^{}}_{\bar{A}_{1}}^{\tensor*[^{1}]{\hspace{-0.2em}A}{}_{1}^{}}}$ &    $2.6143$ &                                                                              $d\indices*{*_{A_{3}}^{A_{1}^{}}_{X_{2}}^{A_{2u}^{}}_{\bar{A}_{1}}^{\tensor*[^{1}]{\hspace{-0.2em}A}{}_{1}^{}}}$ & $- 13.9042$ \\[0.4em]
                                                                                $d\indices*{*_{A_{3}}^{A_{1}^{}}_{X_{2}}^{E_{u}^{}}_{\bar{A}_{1}}^{\tensor*[^{1}]{\hspace{-0.2em}A}{}_{1}^{}}}$ &  $- 5.7154$ &                                              $d\indices*{*_{A_{3}}^{A_{1}^{}}_{X_{2}}^{\tensor*[^{1}]{\hspace{-0.2em}E}{}_{u}^{}}_{\bar{A}_{1}}^{\tensor*[^{1}]{\hspace{-0.2em}A}{}_{1}^{}}}$ &   $11.5538$ &                                                                              $d\indices*{*_{A_{3}}^{A_{1}^{}}_{X_{2}}^{A_{1g}^{}}_{\bar{A}_{1}}^{\tensor*[^{2}]{\hspace{-0.2em}A}{}_{1}^{}}}$ &    $1.2335$ &                                                                               $d\indices*{*_{A_{3}}^{A_{1}^{}}_{X_{2}}^{E_{g}^{}}_{\bar{A}_{1}}^{\tensor*[^{2}]{\hspace{-0.2em}A}{}_{1}^{}}}$ &  $- 1.2578$ &                                                                              $d\indices*{*_{A_{3}}^{A_{1}^{}}_{X_{2}}^{A_{2u}^{}}_{\bar{A}_{1}}^{\tensor*[^{2}]{\hspace{-0.2em}A}{}_{1}^{}}}$ &  $- 0.7534$ \\[0.4em]
                                                                                $d\indices*{*_{A_{3}}^{A_{1}^{}}_{X_{2}}^{E_{u}^{}}_{\bar{A}_{1}}^{\tensor*[^{2}]{\hspace{-0.2em}A}{}_{1}^{}}}$ &  $- 0.8884$ &                                              $d\indices*{*_{A_{3}}^{A_{1}^{}}_{X_{2}}^{\tensor*[^{1}]{\hspace{-0.2em}E}{}_{u}^{}}_{\bar{A}_{1}}^{\tensor*[^{2}]{\hspace{-0.2em}A}{}_{1}^{}}}$ &    $4.4400$ &                                                                                                                $d\indices*{*_{A_{3}}^{A_{1}^{}}_{X_{2}}^{E_{g}^{}}_{\bar{A}_{1}}^{A_{2}^{}}}$ &    $0.9123$ &                                                                                                               $d\indices*{*_{A_{3}}^{A_{1}^{}}_{X_{2}}^{B_{1u}^{}}_{\bar{A}_{1}}^{A_{2}^{}}}$ & $- 13.8371$ &                                                                                                                $d\indices*{*_{A_{3}}^{A_{1}^{}}_{X_{2}}^{E_{u}^{}}_{\bar{A}_{1}}^{A_{2}^{}}}$ &   $15.1723$ \\[0.4em]
                                                                                $d\indices*{*_{A_{3}}^{A_{1}^{}}_{X_{2}}^{\tensor*[^{1}]{\hspace{-0.2em}E}{}_{u}^{}}_{\bar{A}_{1}}^{A_{2}^{}}}$ & $- 16.3786$ &                                                                                                                $d\indices*{*_{A_{3}}^{A_{1}^{}}_{X_{2}}^{E_{g}^{}}_{\bar{A}_{1}}^{B_{1}^{}}}$ & $- 10.6868$ &                                                                                                               $d\indices*{*_{A_{3}}^{A_{1}^{}}_{X_{2}}^{B_{1u}^{}}_{\bar{A}_{1}}^{B_{1}^{}}}$ &  $- 0.9985$ &                                                                                                                $d\indices*{*_{A_{3}}^{A_{1}^{}}_{X_{2}}^{E_{u}^{}}_{\bar{A}_{1}}^{B_{1}^{}}}$ &  $- 2.3912$ &                                                                               $d\indices*{*_{A_{3}}^{A_{1}^{}}_{X_{2}}^{\tensor*[^{1}]{\hspace{-0.2em}E}{}_{u}^{}}_{\bar{A}_{1}}^{B_{1}^{}}}$ &  $- 4.1996$ \\[0.4em]
                                                                                $d\indices*{*_{A_{3}}^{A_{1}^{}}_{X_{2}}^{E_{g}^{}}_{\bar{A}_{1}}^{\tensor*[^{1}]{\hspace{-0.2em}B}{}_{1}^{}}}$ &   $10.6464$ &                                                                              $d\indices*{*_{A_{3}}^{A_{1}^{}}_{X_{2}}^{B_{1u}^{}}_{\bar{A}_{1}}^{\tensor*[^{1}]{\hspace{-0.2em}B}{}_{1}^{}}}$ &    $0.3264$ &                                                                               $d\indices*{*_{A_{3}}^{A_{1}^{}}_{X_{2}}^{E_{u}^{}}_{\bar{A}_{1}}^{\tensor*[^{1}]{\hspace{-0.2em}B}{}_{1}^{}}}$ &  $- 2.5984$ &                                              $d\indices*{*_{A_{3}}^{A_{1}^{}}_{X_{2}}^{\tensor*[^{1}]{\hspace{-0.2em}E}{}_{u}^{}}_{\bar{A}_{1}}^{\tensor*[^{1}]{\hspace{-0.2em}B}{}_{1}^{}}}$ &  $- 4.3111$ &                                                                               $d\indices*{*_{A_{3}}^{A_{1}^{}}_{X_{2}}^{E_{g}^{}}_{\bar{A}_{1}}^{\tensor*[^{2}]{\hspace{-0.2em}B}{}_{1}^{}}}$ &  $- 0.3668$ \\[0.4em]
                                                                               $d\indices*{*_{A_{3}}^{A_{1}^{}}_{X_{2}}^{B_{1u}^{}}_{\bar{A}_{1}}^{\tensor*[^{2}]{\hspace{-0.2em}B}{}_{1}^{}}}$ &  $- 0.3924$ &                                                                               $d\indices*{*_{A_{3}}^{A_{1}^{}}_{X_{2}}^{E_{u}^{}}_{\bar{A}_{1}}^{\tensor*[^{2}]{\hspace{-0.2em}B}{}_{1}^{}}}$ &    $3.5661$ &                                              $d\indices*{*_{A_{3}}^{A_{1}^{}}_{X_{2}}^{\tensor*[^{1}]{\hspace{-0.2em}E}{}_{u}^{}}_{\bar{A}_{1}}^{\tensor*[^{2}]{\hspace{-0.2em}B}{}_{1}^{}}}$ &    $1.4956$ &                                                                                                               $d\indices*{*_{A_{3}}^{A_{1}^{}}_{X_{2}}^{A_{1g}^{}}_{\bar{A}_{1}}^{B_{2}^{}}}$ & $- 17.4761$ &                                                                                                                $d\indices*{*_{A_{3}}^{A_{1}^{}}_{X_{2}}^{E_{g}^{}}_{\bar{A}_{1}}^{B_{2}^{}}}$ & $- 20.7389$ \\[0.4em]
                                                                                                                $d\indices*{*_{A_{3}}^{A_{1}^{}}_{X_{2}}^{A_{2u}^{}}_{\bar{A}_{1}}^{B_{2}^{}}}$ &    $1.1376$ &                                                                                                                $d\indices*{*_{A_{3}}^{A_{1}^{}}_{X_{2}}^{E_{u}^{}}_{\bar{A}_{1}}^{B_{2}^{}}}$ &    $2.5672$ &                                                                               $d\indices*{*_{A_{3}}^{A_{1}^{}}_{X_{2}}^{\tensor*[^{1}]{\hspace{-0.2em}E}{}_{u}^{}}_{\bar{A}_{1}}^{B_{2}^{}}}$ &  $- 2.7677$ &                                                                              $d\indices*{*_{A_{3}}^{A_{1}^{}}_{X_{2}}^{A_{1g}^{}}_{\bar{A}_{1}}^{\tensor*[^{1}]{\hspace{-0.2em}B}{}_{2}^{}}}$ &   $11.2590$ &                                                                               $d\indices*{*_{A_{3}}^{A_{1}^{}}_{X_{2}}^{E_{g}^{}}_{\bar{A}_{1}}^{\tensor*[^{1}]{\hspace{-0.2em}B}{}_{2}^{}}}$ &    $8.1256$ \\[0.4em]
                                                                               $d\indices*{*_{A_{3}}^{A_{1}^{}}_{X_{2}}^{A_{2u}^{}}_{\bar{A}_{1}}^{\tensor*[^{1}]{\hspace{-0.2em}B}{}_{2}^{}}}$ &    $2.2920$ &                                                                               $d\indices*{*_{A_{3}}^{A_{1}^{}}_{X_{2}}^{E_{u}^{}}_{\bar{A}_{1}}^{\tensor*[^{1}]{\hspace{-0.2em}B}{}_{2}^{}}}$ &    $0.8879$ &                                              $d\indices*{*_{A_{3}}^{A_{1}^{}}_{X_{2}}^{\tensor*[^{1}]{\hspace{-0.2em}E}{}_{u}^{}}_{\bar{A}_{1}}^{\tensor*[^{1}]{\hspace{-0.2em}B}{}_{2}^{}}}$ &    $2.3181$ &                                             $d\indices*{*_{A_{3}}^{\tensor*[^{1}]{\hspace{-0.2em}A}{}_{1}^{}}_{X_{2}}^{A_{1g}^{}}_{\bar{A}_{1}}^{\tensor*[^{1}]{\hspace{-0.2em}A}{}_{1}^{}}}$ &  $- 3.1741$ &                                             $d\indices*{*_{A_{3}}^{\tensor*[^{1}]{\hspace{-0.2em}A}{}_{1}^{}}_{X_{2}}^{A_{2u}^{}}_{\bar{A}_{1}}^{\tensor*[^{1}]{\hspace{-0.2em}A}{}_{1}^{}}}$ &    $1.9495$ \\[0.4em]
                                              $d\indices*{*_{A_{3}}^{\tensor*[^{1}]{\hspace{-0.2em}A}{}_{1}^{}}_{X_{2}}^{A_{1g}^{}}_{\bar{A}_{1}}^{\tensor*[^{2}]{\hspace{-0.2em}A}{}_{1}^{}}}$ &    $8.8771$ &                                              $d\indices*{*_{A_{3}}^{\tensor*[^{1}]{\hspace{-0.2em}A}{}_{1}^{}}_{X_{2}}^{E_{g}^{}}_{\bar{A}_{1}}^{\tensor*[^{2}]{\hspace{-0.2em}A}{}_{1}^{}}}$ &    $9.2195$ &                                             $d\indices*{*_{A_{3}}^{\tensor*[^{1}]{\hspace{-0.2em}A}{}_{1}^{}}_{X_{2}}^{A_{2u}^{}}_{\bar{A}_{1}}^{\tensor*[^{2}]{\hspace{-0.2em}A}{}_{1}^{}}}$ &  $- 0.3277$ &                                              $d\indices*{*_{A_{3}}^{\tensor*[^{1}]{\hspace{-0.2em}A}{}_{1}^{}}_{X_{2}}^{E_{u}^{}}_{\bar{A}_{1}}^{\tensor*[^{2}]{\hspace{-0.2em}A}{}_{1}^{}}}$ &  $- 0.7220$ &             $d\indices*{*_{A_{3}}^{\tensor*[^{1}]{\hspace{-0.2em}A}{}_{1}^{}}_{X_{2}}^{\tensor*[^{1}]{\hspace{-0.2em}E}{}_{u}^{}}_{\bar{A}_{1}}^{\tensor*[^{2}]{\hspace{-0.2em}A}{}_{1}^{}}}$ &  $- 0.8950$ \\[0.4em]
                                                                                $d\indices*{*_{A_{3}}^{\tensor*[^{1}]{\hspace{-0.2em}A}{}_{1}^{}}_{X_{2}}^{E_{g}^{}}_{\bar{A}_{1}}^{A_{2}^{}}}$ &  $- 1.6862$ &                                                                              $d\indices*{*_{A_{3}}^{\tensor*[^{1}]{\hspace{-0.2em}A}{}_{1}^{}}_{X_{2}}^{B_{1u}^{}}_{\bar{A}_{1}}^{A_{2}^{}}}$ &    $9.1966$ &                                                                               $d\indices*{*_{A_{3}}^{\tensor*[^{1}]{\hspace{-0.2em}A}{}_{1}^{}}_{X_{2}}^{E_{u}^{}}_{\bar{A}_{1}}^{A_{2}^{}}}$ &  $- 7.8365$ &                                              $d\indices*{*_{A_{3}}^{\tensor*[^{1}]{\hspace{-0.2em}A}{}_{1}^{}}_{X_{2}}^{\tensor*[^{1}]{\hspace{-0.2em}E}{}_{u}^{}}_{\bar{A}_{1}}^{A_{2}^{}}}$ &    $6.0947$ &                                                                               $d\indices*{*_{A_{3}}^{\tensor*[^{1}]{\hspace{-0.2em}A}{}_{1}^{}}_{X_{2}}^{E_{g}^{}}_{\bar{A}_{1}}^{B_{1}^{}}}$ &  $- 2.0130$ \\[0.4em]
                                                                               $d\indices*{*_{A_{3}}^{\tensor*[^{1}]{\hspace{-0.2em}A}{}_{1}^{}}_{X_{2}}^{B_{1u}^{}}_{\bar{A}_{1}}^{B_{1}^{}}}$ &    $3.1088$ &                                                                               $d\indices*{*_{A_{3}}^{\tensor*[^{1}]{\hspace{-0.2em}A}{}_{1}^{}}_{X_{2}}^{E_{u}^{}}_{\bar{A}_{1}}^{B_{1}^{}}}$ &  $- 1.2316$ &                                              $d\indices*{*_{A_{3}}^{\tensor*[^{1}]{\hspace{-0.2em}A}{}_{1}^{}}_{X_{2}}^{\tensor*[^{1}]{\hspace{-0.2em}E}{}_{u}^{}}_{\bar{A}_{1}}^{B_{1}^{}}}$ &    $2.1896$ &                                              $d\indices*{*_{A_{3}}^{\tensor*[^{1}]{\hspace{-0.2em}A}{}_{1}^{}}_{X_{2}}^{E_{g}^{}}_{\bar{A}_{1}}^{\tensor*[^{1}]{\hspace{-0.2em}B}{}_{1}^{}}}$ &  $- 4.0328$ &                                             $d\indices*{*_{A_{3}}^{\tensor*[^{1}]{\hspace{-0.2em}A}{}_{1}^{}}_{X_{2}}^{B_{1u}^{}}_{\bar{A}_{1}}^{\tensor*[^{1}]{\hspace{-0.2em}B}{}_{1}^{}}}$ &    $6.4874$ \\[0.4em]
                                               $d\indices*{*_{A_{3}}^{\tensor*[^{1}]{\hspace{-0.2em}A}{}_{1}^{}}_{X_{2}}^{E_{u}^{}}_{\bar{A}_{1}}^{\tensor*[^{1}]{\hspace{-0.2em}B}{}_{1}^{}}}$ &    $1.5782$ &             $d\indices*{*_{A_{3}}^{\tensor*[^{1}]{\hspace{-0.2em}A}{}_{1}^{}}_{X_{2}}^{\tensor*[^{1}]{\hspace{-0.2em}E}{}_{u}^{}}_{\bar{A}_{1}}^{\tensor*[^{1}]{\hspace{-0.2em}B}{}_{1}^{}}}$ & $- 11.1375$ &                                              $d\indices*{*_{A_{3}}^{\tensor*[^{1}]{\hspace{-0.2em}A}{}_{1}^{}}_{X_{2}}^{E_{g}^{}}_{\bar{A}_{1}}^{\tensor*[^{2}]{\hspace{-0.2em}B}{}_{1}^{}}}$ &    $2.5315$ &                                             $d\indices*{*_{A_{3}}^{\tensor*[^{1}]{\hspace{-0.2em}A}{}_{1}^{}}_{X_{2}}^{B_{1u}^{}}_{\bar{A}_{1}}^{\tensor*[^{2}]{\hspace{-0.2em}B}{}_{1}^{}}}$ &    $7.0024$ &                                              $d\indices*{*_{A_{3}}^{\tensor*[^{1}]{\hspace{-0.2em}A}{}_{1}^{}}_{X_{2}}^{E_{u}^{}}_{\bar{A}_{1}}^{\tensor*[^{2}]{\hspace{-0.2em}B}{}_{1}^{}}}$ &    $3.9802$ \\[0.4em]
              $d\indices*{*_{A_{3}}^{\tensor*[^{1}]{\hspace{-0.2em}A}{}_{1}^{}}_{X_{2}}^{\tensor*[^{1}]{\hspace{-0.2em}E}{}_{u}^{}}_{\bar{A}_{1}}^{\tensor*[^{2}]{\hspace{-0.2em}B}{}_{1}^{}}}$ & $- 11.7183$ &                                                                              $d\indices*{*_{A_{3}}^{\tensor*[^{1}]{\hspace{-0.2em}A}{}_{1}^{}}_{X_{2}}^{A_{1g}^{}}_{\bar{A}_{1}}^{B_{2}^{}}}$ &    $9.6090$ &                                                                               $d\indices*{*_{A_{3}}^{\tensor*[^{1}]{\hspace{-0.2em}A}{}_{1}^{}}_{X_{2}}^{E_{g}^{}}_{\bar{A}_{1}}^{B_{2}^{}}}$ &   $12.7127$ &                                                                              $d\indices*{*_{A_{3}}^{\tensor*[^{1}]{\hspace{-0.2em}A}{}_{1}^{}}_{X_{2}}^{A_{2u}^{}}_{\bar{A}_{1}}^{B_{2}^{}}}$ &    $0.1549$ &                                                                               $d\indices*{*_{A_{3}}^{\tensor*[^{1}]{\hspace{-0.2em}A}{}_{1}^{}}_{X_{2}}^{E_{u}^{}}_{\bar{A}_{1}}^{B_{2}^{}}}$ &  $- 3.3905$ \\[0.4em]
                                               $d\indices*{*_{A_{3}}^{\tensor*[^{1}]{\hspace{-0.2em}A}{}_{1}^{}}_{X_{2}}^{\tensor*[^{1}]{\hspace{-0.2em}E}{}_{u}^{}}_{\bar{A}_{1}}^{B_{2}^{}}}$ &    $2.2080$ &                                             $d\indices*{*_{A_{3}}^{\tensor*[^{1}]{\hspace{-0.2em}A}{}_{1}^{}}_{X_{2}}^{A_{1g}^{}}_{\bar{A}_{1}}^{\tensor*[^{1}]{\hspace{-0.2em}B}{}_{2}^{}}}$ &  $- 6.4858$ &                                              $d\indices*{*_{A_{3}}^{\tensor*[^{1}]{\hspace{-0.2em}A}{}_{1}^{}}_{X_{2}}^{E_{g}^{}}_{\bar{A}_{1}}^{\tensor*[^{1}]{\hspace{-0.2em}B}{}_{2}^{}}}$ &  $- 5.3861$ &                                             $d\indices*{*_{A_{3}}^{\tensor*[^{1}]{\hspace{-0.2em}A}{}_{1}^{}}_{X_{2}}^{A_{2u}^{}}_{\bar{A}_{1}}^{\tensor*[^{1}]{\hspace{-0.2em}B}{}_{2}^{}}}$ &    $2.7035$ &                                              $d\indices*{*_{A_{3}}^{\tensor*[^{1}]{\hspace{-0.2em}A}{}_{1}^{}}_{X_{2}}^{E_{u}^{}}_{\bar{A}_{1}}^{\tensor*[^{1}]{\hspace{-0.2em}B}{}_{2}^{}}}$ &    $4.7150$ \\[0.4em]
              $d\indices*{*_{A_{3}}^{\tensor*[^{1}]{\hspace{-0.2em}A}{}_{1}^{}}_{X_{2}}^{\tensor*[^{1}]{\hspace{-0.2em}E}{}_{u}^{}}_{\bar{A}_{1}}^{\tensor*[^{1}]{\hspace{-0.2em}B}{}_{2}^{}}}$ &  $- 2.1375$ &                                             $d\indices*{*_{A_{3}}^{\tensor*[^{2}]{\hspace{-0.2em}A}{}_{1}^{}}_{X_{2}}^{A_{1g}^{}}_{\bar{A}_{1}}^{\tensor*[^{2}]{\hspace{-0.2em}A}{}_{1}^{}}}$ &  $- 2.1838$ &                                             $d\indices*{*_{A_{3}}^{\tensor*[^{2}]{\hspace{-0.2em}A}{}_{1}^{}}_{X_{2}}^{A_{2u}^{}}_{\bar{A}_{1}}^{\tensor*[^{2}]{\hspace{-0.2em}A}{}_{1}^{}}}$ &  $- 0.5690$ &                                                                               $d\indices*{*_{A_{3}}^{\tensor*[^{2}]{\hspace{-0.2em}A}{}_{1}^{}}_{X_{2}}^{E_{g}^{}}_{\bar{A}_{1}}^{A_{2}^{}}}$ &   $11.9568$ &                                                                              $d\indices*{*_{A_{3}}^{\tensor*[^{2}]{\hspace{-0.2em}A}{}_{1}^{}}_{X_{2}}^{B_{1u}^{}}_{\bar{A}_{1}}^{A_{2}^{}}}$ &  $- 4.7445$ \\[0.4em]
                                                                                $d\indices*{*_{A_{3}}^{\tensor*[^{2}]{\hspace{-0.2em}A}{}_{1}^{}}_{X_{2}}^{E_{u}^{}}_{\bar{A}_{1}}^{A_{2}^{}}}$ &    $3.5129$ &                                              $d\indices*{*_{A_{3}}^{\tensor*[^{2}]{\hspace{-0.2em}A}{}_{1}^{}}_{X_{2}}^{\tensor*[^{1}]{\hspace{-0.2em}E}{}_{u}^{}}_{\bar{A}_{1}}^{A_{2}^{}}}$ &    $2.5731$ &                                                                               $d\indices*{*_{A_{3}}^{\tensor*[^{2}]{\hspace{-0.2em}A}{}_{1}^{}}_{X_{2}}^{E_{g}^{}}_{\bar{A}_{1}}^{B_{1}^{}}}$ &    $1.7995$ &                                                                              $d\indices*{*_{A_{3}}^{\tensor*[^{2}]{\hspace{-0.2em}A}{}_{1}^{}}_{X_{2}}^{B_{1u}^{}}_{\bar{A}_{1}}^{B_{1}^{}}}$ &  $- 9.6596$ &                                                                               $d\indices*{*_{A_{3}}^{\tensor*[^{2}]{\hspace{-0.2em}A}{}_{1}^{}}_{X_{2}}^{E_{u}^{}}_{\bar{A}_{1}}^{B_{1}^{}}}$ &    $5.6571$ \\[0.4em]
                                               $d\indices*{*_{A_{3}}^{\tensor*[^{2}]{\hspace{-0.2em}A}{}_{1}^{}}_{X_{2}}^{\tensor*[^{1}]{\hspace{-0.2em}E}{}_{u}^{}}_{\bar{A}_{1}}^{B_{1}^{}}}$ & $- 13.1954$ &                                              $d\indices*{*_{A_{3}}^{\tensor*[^{2}]{\hspace{-0.2em}A}{}_{1}^{}}_{X_{2}}^{E_{g}^{}}_{\bar{A}_{1}}^{\tensor*[^{1}]{\hspace{-0.2em}B}{}_{1}^{}}}$ &    $1.5687$ &                                             $d\indices*{*_{A_{3}}^{\tensor*[^{2}]{\hspace{-0.2em}A}{}_{1}^{}}_{X_{2}}^{B_{1u}^{}}_{\bar{A}_{1}}^{\tensor*[^{1}]{\hspace{-0.2em}B}{}_{1}^{}}}$ &    $9.7615$ &                                              $d\indices*{*_{A_{3}}^{\tensor*[^{2}]{\hspace{-0.2em}A}{}_{1}^{}}_{X_{2}}^{E_{u}^{}}_{\bar{A}_{1}}^{\tensor*[^{1}]{\hspace{-0.2em}B}{}_{1}^{}}}$ &  $- 9.6089$ &             $d\indices*{*_{A_{3}}^{\tensor*[^{2}]{\hspace{-0.2em}A}{}_{1}^{}}_{X_{2}}^{\tensor*[^{1}]{\hspace{-0.2em}E}{}_{u}^{}}_{\bar{A}_{1}}^{\tensor*[^{1}]{\hspace{-0.2em}B}{}_{1}^{}}}$ &    $9.5832$ \\[0.4em]
                                               $d\indices*{*_{A_{3}}^{\tensor*[^{2}]{\hspace{-0.2em}A}{}_{1}^{}}_{X_{2}}^{E_{g}^{}}_{\bar{A}_{1}}^{\tensor*[^{2}]{\hspace{-0.2em}B}{}_{1}^{}}}$ &  $- 2.6857$ &                                             $d\indices*{*_{A_{3}}^{\tensor*[^{2}]{\hspace{-0.2em}A}{}_{1}^{}}_{X_{2}}^{B_{1u}^{}}_{\bar{A}_{1}}^{\tensor*[^{2}]{\hspace{-0.2em}B}{}_{1}^{}}}$ &  $- 0.2901$ &                                              $d\indices*{*_{A_{3}}^{\tensor*[^{2}]{\hspace{-0.2em}A}{}_{1}^{}}_{X_{2}}^{E_{u}^{}}_{\bar{A}_{1}}^{\tensor*[^{2}]{\hspace{-0.2em}B}{}_{1}^{}}}$ &  $- 6.0819$ &             $d\indices*{*_{A_{3}}^{\tensor*[^{2}]{\hspace{-0.2em}A}{}_{1}^{}}_{X_{2}}^{\tensor*[^{1}]{\hspace{-0.2em}E}{}_{u}^{}}_{\bar{A}_{1}}^{\tensor*[^{2}]{\hspace{-0.2em}B}{}_{1}^{}}}$ &    $2.0429$ &                                                                              $d\indices*{*_{A_{3}}^{\tensor*[^{2}]{\hspace{-0.2em}A}{}_{1}^{}}_{X_{2}}^{A_{1g}^{}}_{\bar{A}_{1}}^{B_{2}^{}}}$ &    $2.8177$ \\[0.4em]
                                                                                $d\indices*{*_{A_{3}}^{\tensor*[^{2}]{\hspace{-0.2em}A}{}_{1}^{}}_{X_{2}}^{E_{g}^{}}_{\bar{A}_{1}}^{B_{2}^{}}}$ &  $- 4.9247$ &                                                                              $d\indices*{*_{A_{3}}^{\tensor*[^{2}]{\hspace{-0.2em}A}{}_{1}^{}}_{X_{2}}^{A_{2u}^{}}_{\bar{A}_{1}}^{B_{2}^{}}}$ &   $14.7202$ &                                                                               $d\indices*{*_{A_{3}}^{\tensor*[^{2}]{\hspace{-0.2em}A}{}_{1}^{}}_{X_{2}}^{E_{u}^{}}_{\bar{A}_{1}}^{B_{2}^{}}}$ &   $13.9551$ &                                              $d\indices*{*_{A_{3}}^{\tensor*[^{2}]{\hspace{-0.2em}A}{}_{1}^{}}_{X_{2}}^{\tensor*[^{1}]{\hspace{-0.2em}E}{}_{u}^{}}_{\bar{A}_{1}}^{B_{2}^{}}}$ & $- 16.9968$ &                                             $d\indices*{*_{A_{3}}^{\tensor*[^{2}]{\hspace{-0.2em}A}{}_{1}^{}}_{X_{2}}^{A_{1g}^{}}_{\bar{A}_{1}}^{\tensor*[^{1}]{\hspace{-0.2em}B}{}_{2}^{}}}$ &  $- 3.9131$ \\[0.4em]
                                               $d\indices*{*_{A_{3}}^{\tensor*[^{2}]{\hspace{-0.2em}A}{}_{1}^{}}_{X_{2}}^{E_{g}^{}}_{\bar{A}_{1}}^{\tensor*[^{1}]{\hspace{-0.2em}B}{}_{2}^{}}}$ &    $2.0046$ &                                             $d\indices*{*_{A_{3}}^{\tensor*[^{2}]{\hspace{-0.2em}A}{}_{1}^{}}_{X_{2}}^{A_{2u}^{}}_{\bar{A}_{1}}^{\tensor*[^{1}]{\hspace{-0.2em}B}{}_{2}^{}}}$ & $- 11.4815$ &                                              $d\indices*{*_{A_{3}}^{\tensor*[^{2}]{\hspace{-0.2em}A}{}_{1}^{}}_{X_{2}}^{E_{u}^{}}_{\bar{A}_{1}}^{\tensor*[^{1}]{\hspace{-0.2em}B}{}_{2}^{}}}$ & $- 12.2344$ &             $d\indices*{*_{A_{3}}^{\tensor*[^{2}]{\hspace{-0.2em}A}{}_{1}^{}}_{X_{2}}^{\tensor*[^{1}]{\hspace{-0.2em}E}{}_{u}^{}}_{\bar{A}_{1}}^{\tensor*[^{1}]{\hspace{-0.2em}B}{}_{2}^{}}}$ &   $11.0185$ &                                                                                                               $d\indices*{*_{A_{3}}^{A_{2}^{}}_{X_{2}}^{A_{1g}^{}}_{\bar{A}_{1}}^{A_{2}^{}}}$ &    $5.5555$ \\[0.4em]
                                                                                                                $d\indices*{*_{A_{3}}^{A_{2}^{}}_{X_{2}}^{A_{2u}^{}}_{\bar{A}_{1}}^{A_{2}^{}}}$ &  $- 3.1657$ &                                                                                                               $d\indices*{*_{A_{3}}^{A_{2}^{}}_{X_{2}}^{A_{1g}^{}}_{\bar{A}_{1}}^{B_{1}^{}}}$ & $- 10.1624$ &                                                                                                                $d\indices*{*_{A_{3}}^{A_{2}^{}}_{X_{2}}^{E_{g}^{}}_{\bar{A}_{1}}^{B_{1}^{}}}$ &  $- 1.3771$ &                                                                                                               $d\indices*{*_{A_{3}}^{A_{2}^{}}_{X_{2}}^{A_{2u}^{}}_{\bar{A}_{1}}^{B_{1}^{}}}$ &  $- 2.6473$ &                                                                                                                $d\indices*{*_{A_{3}}^{A_{2}^{}}_{X_{2}}^{E_{u}^{}}_{\bar{A}_{1}}^{B_{1}^{}}}$ &  $- 0.9663$ \\[0.4em]
                                                                                $d\indices*{*_{A_{3}}^{A_{2}^{}}_{X_{2}}^{\tensor*[^{1}]{\hspace{-0.2em}E}{}_{u}^{}}_{\bar{A}_{1}}^{B_{1}^{}}}$ &  $- 1.2810$ &                                                                              $d\indices*{*_{A_{3}}^{A_{2}^{}}_{X_{2}}^{A_{1g}^{}}_{\bar{A}_{1}}^{\tensor*[^{1}]{\hspace{-0.2em}B}{}_{1}^{}}}$ &    $1.9099$ &                                                                               $d\indices*{*_{A_{3}}^{A_{2}^{}}_{X_{2}}^{E_{g}^{}}_{\bar{A}_{1}}^{\tensor*[^{1}]{\hspace{-0.2em}B}{}_{1}^{}}}$ &  $- 2.2603$ &                                                                              $d\indices*{*_{A_{3}}^{A_{2}^{}}_{X_{2}}^{A_{2u}^{}}_{\bar{A}_{1}}^{\tensor*[^{1}]{\hspace{-0.2em}B}{}_{1}^{}}}$ & $- 14.8070$ &                                                                               $d\indices*{*_{A_{3}}^{A_{2}^{}}_{X_{2}}^{E_{u}^{}}_{\bar{A}_{1}}^{\tensor*[^{1}]{\hspace{-0.2em}B}{}_{1}^{}}}$ &   $10.4280$ \\[0.4em]
                                               $d\indices*{*_{A_{3}}^{A_{2}^{}}_{X_{2}}^{\tensor*[^{1}]{\hspace{-0.2em}E}{}_{u}^{}}_{\bar{A}_{1}}^{\tensor*[^{1}]{\hspace{-0.2em}B}{}_{1}^{}}}$ & $- 11.8854$ &                                                                              $d\indices*{*_{A_{3}}^{A_{2}^{}}_{X_{2}}^{A_{1g}^{}}_{\bar{A}_{1}}^{\tensor*[^{2}]{\hspace{-0.2em}B}{}_{1}^{}}}$ &  $- 2.3463$ &                                                                               $d\indices*{*_{A_{3}}^{A_{2}^{}}_{X_{2}}^{E_{g}^{}}_{\bar{A}_{1}}^{\tensor*[^{2}]{\hspace{-0.2em}B}{}_{1}^{}}}$ &  $- 3.2727$ &                                                                              $d\indices*{*_{A_{3}}^{A_{2}^{}}_{X_{2}}^{A_{2u}^{}}_{\bar{A}_{1}}^{\tensor*[^{2}]{\hspace{-0.2em}B}{}_{1}^{}}}$ & $- 11.2370$ &                                                                               $d\indices*{*_{A_{3}}^{A_{2}^{}}_{X_{2}}^{E_{u}^{}}_{\bar{A}_{1}}^{\tensor*[^{2}]{\hspace{-0.2em}B}{}_{1}^{}}}$ &    $9.8421$ \\[0.4em]
                                               $d\indices*{*_{A_{3}}^{A_{2}^{}}_{X_{2}}^{\tensor*[^{1}]{\hspace{-0.2em}E}{}_{u}^{}}_{\bar{A}_{1}}^{\tensor*[^{2}]{\hspace{-0.2em}B}{}_{1}^{}}}$ & $- 11.4649$ &                                                                                                                $d\indices*{*_{A_{3}}^{B_{2}^{}}_{X_{2}}^{E_{g}^{}}_{\bar{A}_{1}}^{A_{2}^{}}}$ &    $1.5966$ &                                                                                                               $d\indices*{*_{A_{3}}^{B_{2}^{}}_{X_{2}}^{B_{1u}^{}}_{\bar{A}_{1}}^{A_{2}^{}}}$ &    $2.1227$ &                                                                                                                $d\indices*{*_{A_{3}}^{B_{2}^{}}_{X_{2}}^{E_{u}^{}}_{\bar{A}_{1}}^{A_{2}^{}}}$ &  $- 0.0529$ &                                                                               $d\indices*{*_{A_{3}}^{B_{2}^{}}_{X_{2}}^{\tensor*[^{1}]{\hspace{-0.2em}E}{}_{u}^{}}_{\bar{A}_{1}}^{A_{2}^{}}}$ &  $- 1.9433$ \\[0.4em]
                                                                                $d\indices*{*_{A_{3}}^{\tensor*[^{1}]{\hspace{-0.2em}B}{}_{2}^{}}_{X_{2}}^{E_{g}^{}}_{\bar{A}_{1}}^{A_{2}^{}}}$ &    $3.4319$ &                                                                              $d\indices*{*_{A_{3}}^{\tensor*[^{1}]{\hspace{-0.2em}B}{}_{2}^{}}_{X_{2}}^{B_{1u}^{}}_{\bar{A}_{1}}^{A_{2}^{}}}$ &    $2.1861$ &                                                                               $d\indices*{*_{A_{3}}^{\tensor*[^{1}]{\hspace{-0.2em}B}{}_{2}^{}}_{X_{2}}^{E_{u}^{}}_{\bar{A}_{1}}^{A_{2}^{}}}$ &    $6.0687$ &                                              $d\indices*{*_{A_{3}}^{\tensor*[^{1}]{\hspace{-0.2em}B}{}_{2}^{}}_{X_{2}}^{\tensor*[^{1}]{\hspace{-0.2em}E}{}_{u}^{}}_{\bar{A}_{1}}^{A_{2}^{}}}$ &  $- 3.7701$ &                                                                                                               $d\indices*{*_{A_{3}}^{B_{1}^{}}_{X_{2}}^{A_{1g}^{}}_{\bar{A}_{1}}^{B_{1}^{}}}$ &    $3.6441$ \\[0.4em]
                                                                                                                $d\indices*{*_{A_{3}}^{B_{1}^{}}_{X_{2}}^{A_{2u}^{}}_{\bar{A}_{1}}^{B_{1}^{}}}$ &   $11.5258$ &                                                                              $d\indices*{*_{A_{3}}^{B_{1}^{}}_{X_{2}}^{A_{1g}^{}}_{\bar{A}_{1}}^{\tensor*[^{1}]{\hspace{-0.2em}B}{}_{1}^{}}}$ &  $- 9.6248$ &                                                                               $d\indices*{*_{A_{3}}^{B_{1}^{}}_{X_{2}}^{E_{g}^{}}_{\bar{A}_{1}}^{\tensor*[^{1}]{\hspace{-0.2em}B}{}_{1}^{}}}$ &  $- 2.0780$ &                                                                              $d\indices*{*_{A_{3}}^{B_{1}^{}}_{X_{2}}^{A_{2u}^{}}_{\bar{A}_{1}}^{\tensor*[^{1}]{\hspace{-0.2em}B}{}_{1}^{}}}$ &  $- 6.6320$ &                                                                               $d\indices*{*_{A_{3}}^{B_{1}^{}}_{X_{2}}^{E_{u}^{}}_{\bar{A}_{1}}^{\tensor*[^{1}]{\hspace{-0.2em}B}{}_{1}^{}}}$ &  $- 0.7526$ \\[0.4em]
                                               $d\indices*{*_{A_{3}}^{B_{1}^{}}_{X_{2}}^{\tensor*[^{1}]{\hspace{-0.2em}E}{}_{u}^{}}_{\bar{A}_{1}}^{\tensor*[^{1}]{\hspace{-0.2em}B}{}_{1}^{}}}$ &    $7.1412$ &                                                                              $d\indices*{*_{A_{3}}^{B_{1}^{}}_{X_{2}}^{A_{1g}^{}}_{\bar{A}_{1}}^{\tensor*[^{2}]{\hspace{-0.2em}B}{}_{1}^{}}}$ &  $- 8.3714$ &                                                                               $d\indices*{*_{A_{3}}^{B_{1}^{}}_{X_{2}}^{E_{g}^{}}_{\bar{A}_{1}}^{\tensor*[^{2}]{\hspace{-0.2em}B}{}_{1}^{}}}$ & $- 10.2737$ &                                                                              $d\indices*{*_{A_{3}}^{B_{1}^{}}_{X_{2}}^{A_{2u}^{}}_{\bar{A}_{1}}^{\tensor*[^{2}]{\hspace{-0.2em}B}{}_{1}^{}}}$ &    $3.3542$ &                                                                               $d\indices*{*_{A_{3}}^{B_{1}^{}}_{X_{2}}^{E_{u}^{}}_{\bar{A}_{1}}^{\tensor*[^{2}]{\hspace{-0.2em}B}{}_{1}^{}}}$ &    $4.2166$ \\[0.4em]
                                               $d\indices*{*_{A_{3}}^{B_{1}^{}}_{X_{2}}^{\tensor*[^{1}]{\hspace{-0.2em}E}{}_{u}^{}}_{\bar{A}_{1}}^{\tensor*[^{2}]{\hspace{-0.2em}B}{}_{1}^{}}}$ &  $- 3.2806$ &                                                                                                                $d\indices*{*_{A_{3}}^{B_{2}^{}}_{X_{2}}^{E_{g}^{}}_{\bar{A}_{1}}^{B_{1}^{}}}$ &  $- 1.2006$ &                                                                                                               $d\indices*{*_{A_{3}}^{B_{2}^{}}_{X_{2}}^{B_{1u}^{}}_{\bar{A}_{1}}^{B_{1}^{}}}$ &    $9.9445$ &                                                                                                                $d\indices*{*_{A_{3}}^{B_{2}^{}}_{X_{2}}^{E_{u}^{}}_{\bar{A}_{1}}^{B_{1}^{}}}$ &    $9.1266$ &                                                                               $d\indices*{*_{A_{3}}^{B_{2}^{}}_{X_{2}}^{\tensor*[^{1}]{\hspace{-0.2em}E}{}_{u}^{}}_{\bar{A}_{1}}^{B_{1}^{}}}$ & $- 11.2070$ \\[0.4em]
                                                                                $d\indices*{*_{A_{3}}^{\tensor*[^{1}]{\hspace{-0.2em}B}{}_{2}^{}}_{X_{2}}^{E_{g}^{}}_{\bar{A}_{1}}^{B_{1}^{}}}$ &  $- 2.3371$ &                                                                              $d\indices*{*_{A_{3}}^{\tensor*[^{1}]{\hspace{-0.2em}B}{}_{2}^{}}_{X_{2}}^{B_{1u}^{}}_{\bar{A}_{1}}^{B_{1}^{}}}$ &  $- 7.5731$ &                                                                               $d\indices*{*_{A_{3}}^{\tensor*[^{1}]{\hspace{-0.2em}B}{}_{2}^{}}_{X_{2}}^{E_{u}^{}}_{\bar{A}_{1}}^{B_{1}^{}}}$ &  $- 4.3392$ &                                              $d\indices*{*_{A_{3}}^{\tensor*[^{1}]{\hspace{-0.2em}B}{}_{2}^{}}_{X_{2}}^{\tensor*[^{1}]{\hspace{-0.2em}E}{}_{u}^{}}_{\bar{A}_{1}}^{B_{1}^{}}}$ &    $7.3662$ &                                             $d\indices*{*_{A_{3}}^{\tensor*[^{1}]{\hspace{-0.2em}B}{}_{1}^{}}_{X_{2}}^{A_{1g}^{}}_{\bar{A}_{1}}^{\tensor*[^{1}]{\hspace{-0.2em}B}{}_{1}^{}}}$ &   $15.6218$ \\[0.4em]
                                              $d\indices*{*_{A_{3}}^{\tensor*[^{1}]{\hspace{-0.2em}B}{}_{1}^{}}_{X_{2}}^{A_{2u}^{}}_{\bar{A}_{1}}^{\tensor*[^{1}]{\hspace{-0.2em}B}{}_{1}^{}}}$ &    $1.7639$ &                                             $d\indices*{*_{A_{3}}^{\tensor*[^{1}]{\hspace{-0.2em}B}{}_{1}^{}}_{X_{2}}^{A_{1g}^{}}_{\bar{A}_{1}}^{\tensor*[^{2}]{\hspace{-0.2em}B}{}_{1}^{}}}$ &   $13.3198$ &                                              $d\indices*{*_{A_{3}}^{\tensor*[^{1}]{\hspace{-0.2em}B}{}_{1}^{}}_{X_{2}}^{E_{g}^{}}_{\bar{A}_{1}}^{\tensor*[^{2}]{\hspace{-0.2em}B}{}_{1}^{}}}$ &    $9.4924$ &                                             $d\indices*{*_{A_{3}}^{\tensor*[^{1}]{\hspace{-0.2em}B}{}_{1}^{}}_{X_{2}}^{A_{2u}^{}}_{\bar{A}_{1}}^{\tensor*[^{2}]{\hspace{-0.2em}B}{}_{1}^{}}}$ &    $0.4201$ &                                              $d\indices*{*_{A_{3}}^{\tensor*[^{1}]{\hspace{-0.2em}B}{}_{1}^{}}_{X_{2}}^{E_{u}^{}}_{\bar{A}_{1}}^{\tensor*[^{2}]{\hspace{-0.2em}B}{}_{1}^{}}}$ &    $3.2283$ \\[0.4em]
              $d\indices*{*_{A_{3}}^{\tensor*[^{1}]{\hspace{-0.2em}B}{}_{1}^{}}_{X_{2}}^{\tensor*[^{1}]{\hspace{-0.2em}E}{}_{u}^{}}_{\bar{A}_{1}}^{\tensor*[^{2}]{\hspace{-0.2em}B}{}_{1}^{}}}$ &  $- 1.2219$ &                                                                               $d\indices*{*_{A_{3}}^{B_{2}^{}}_{X_{2}}^{E_{g}^{}}_{\bar{A}_{1}}^{\tensor*[^{1}]{\hspace{-0.2em}B}{}_{1}^{}}}$ &    $9.1216$ &                                                                              $d\indices*{*_{A_{3}}^{B_{2}^{}}_{X_{2}}^{B_{1u}^{}}_{\bar{A}_{1}}^{\tensor*[^{1}]{\hspace{-0.2em}B}{}_{1}^{}}}$ &    $1.9143$ &                                                                               $d\indices*{*_{A_{3}}^{B_{2}^{}}_{X_{2}}^{E_{u}^{}}_{\bar{A}_{1}}^{\tensor*[^{1}]{\hspace{-0.2em}B}{}_{1}^{}}}$ &  $- 3.5309$ &                                              $d\indices*{*_{A_{3}}^{B_{2}^{}}_{X_{2}}^{\tensor*[^{1}]{\hspace{-0.2em}E}{}_{u}^{}}_{\bar{A}_{1}}^{\tensor*[^{1}]{\hspace{-0.2em}B}{}_{1}^{}}}$ &  $- 5.7101$ \\[0.4em]
                                               $d\indices*{*_{A_{3}}^{\tensor*[^{1}]{\hspace{-0.2em}B}{}_{2}^{}}_{X_{2}}^{E_{g}^{}}_{\bar{A}_{1}}^{\tensor*[^{1}]{\hspace{-0.2em}B}{}_{1}^{}}}$ & $- 11.4799$ &                                             $d\indices*{*_{A_{3}}^{\tensor*[^{1}]{\hspace{-0.2em}B}{}_{2}^{}}_{X_{2}}^{B_{1u}^{}}_{\bar{A}_{1}}^{\tensor*[^{1}]{\hspace{-0.2em}B}{}_{1}^{}}}$ &  $- 0.3070$ &                                              $d\indices*{*_{A_{3}}^{\tensor*[^{1}]{\hspace{-0.2em}B}{}_{2}^{}}_{X_{2}}^{E_{u}^{}}_{\bar{A}_{1}}^{\tensor*[^{1}]{\hspace{-0.2em}B}{}_{1}^{}}}$ &  $- 2.7652$ &             $d\indices*{*_{A_{3}}^{\tensor*[^{1}]{\hspace{-0.2em}B}{}_{2}^{}}_{X_{2}}^{\tensor*[^{1}]{\hspace{-0.2em}E}{}_{u}^{}}_{\bar{A}_{1}}^{\tensor*[^{1}]{\hspace{-0.2em}B}{}_{1}^{}}}$ &  $- 2.4531$ &                                             $d\indices*{*_{A_{3}}^{\tensor*[^{2}]{\hspace{-0.2em}B}{}_{1}^{}}_{X_{2}}^{A_{1g}^{}}_{\bar{A}_{1}}^{\tensor*[^{2}]{\hspace{-0.2em}B}{}_{1}^{}}}$ &   $10.6576$ \\[0.4em]
                                              $d\indices*{*_{A_{3}}^{\tensor*[^{2}]{\hspace{-0.2em}B}{}_{1}^{}}_{X_{2}}^{A_{2u}^{}}_{\bar{A}_{1}}^{\tensor*[^{2}]{\hspace{-0.2em}B}{}_{1}^{}}}$ &  $- 4.0928$ &                                                                               $d\indices*{*_{A_{3}}^{B_{2}^{}}_{X_{2}}^{E_{g}^{}}_{\bar{A}_{1}}^{\tensor*[^{2}]{\hspace{-0.2em}B}{}_{1}^{}}}$ &    $9.7267$ &                                                                              $d\indices*{*_{A_{3}}^{B_{2}^{}}_{X_{2}}^{B_{1u}^{}}_{\bar{A}_{1}}^{\tensor*[^{2}]{\hspace{-0.2em}B}{}_{1}^{}}}$ &  $- 1.4560$ &                                                                               $d\indices*{*_{A_{3}}^{B_{2}^{}}_{X_{2}}^{E_{u}^{}}_{\bar{A}_{1}}^{\tensor*[^{2}]{\hspace{-0.2em}B}{}_{1}^{}}}$ &  $- 4.5567$ &                                              $d\indices*{*_{A_{3}}^{B_{2}^{}}_{X_{2}}^{\tensor*[^{1}]{\hspace{-0.2em}E}{}_{u}^{}}_{\bar{A}_{1}}^{\tensor*[^{2}]{\hspace{-0.2em}B}{}_{1}^{}}}$ &  $- 3.0276$ \\[0.4em]
                                               $d\indices*{*_{A_{3}}^{\tensor*[^{1}]{\hspace{-0.2em}B}{}_{2}^{}}_{X_{2}}^{E_{g}^{}}_{\bar{A}_{1}}^{\tensor*[^{2}]{\hspace{-0.2em}B}{}_{1}^{}}}$ & $- 11.4646$ &                                             $d\indices*{*_{A_{3}}^{\tensor*[^{1}]{\hspace{-0.2em}B}{}_{2}^{}}_{X_{2}}^{B_{1u}^{}}_{\bar{A}_{1}}^{\tensor*[^{2}]{\hspace{-0.2em}B}{}_{1}^{}}}$ &    $3.0608$ &                                              $d\indices*{*_{A_{3}}^{\tensor*[^{1}]{\hspace{-0.2em}B}{}_{2}^{}}_{X_{2}}^{E_{u}^{}}_{\bar{A}_{1}}^{\tensor*[^{2}]{\hspace{-0.2em}B}{}_{1}^{}}}$ & $- 11.1364$ &             $d\indices*{*_{A_{3}}^{\tensor*[^{1}]{\hspace{-0.2em}B}{}_{2}^{}}_{X_{2}}^{\tensor*[^{1}]{\hspace{-0.2em}E}{}_{u}^{}}_{\bar{A}_{1}}^{\tensor*[^{2}]{\hspace{-0.2em}B}{}_{1}^{}}}$ &  $- 7.3341$ &                                                                                                               $d\indices*{*_{A_{3}}^{B_{2}^{}}_{X_{2}}^{A_{1g}^{}}_{\bar{A}_{1}}^{B_{2}^{}}}$ &    $0.3566$ \\[0.4em]
                                                                                                                $d\indices*{*_{A_{3}}^{B_{2}^{}}_{X_{2}}^{A_{2u}^{}}_{\bar{A}_{1}}^{B_{2}^{}}}$ &    $3.2249$ &                                                                              $d\indices*{*_{A_{3}}^{B_{2}^{}}_{X_{2}}^{A_{1g}^{}}_{\bar{A}_{1}}^{\tensor*[^{1}]{\hspace{-0.2em}B}{}_{2}^{}}}$ &  $- 1.2821$ &                                                                               $d\indices*{*_{A_{3}}^{B_{2}^{}}_{X_{2}}^{E_{g}^{}}_{\bar{A}_{1}}^{\tensor*[^{1}]{\hspace{-0.2em}B}{}_{2}^{}}}$ &  $- 1.3146$ &                                                                              $d\indices*{*_{A_{3}}^{B_{2}^{}}_{X_{2}}^{A_{2u}^{}}_{\bar{A}_{1}}^{\tensor*[^{1}]{\hspace{-0.2em}B}{}_{2}^{}}}$ &  $- 1.5301$ &                                                                               $d\indices*{*_{A_{3}}^{B_{2}^{}}_{X_{2}}^{E_{u}^{}}_{\bar{A}_{1}}^{\tensor*[^{1}]{\hspace{-0.2em}B}{}_{2}^{}}}$ &    $0.6785$ \\[0.4em]
                                               $d\indices*{*_{A_{3}}^{B_{2}^{}}_{X_{2}}^{\tensor*[^{1}]{\hspace{-0.2em}E}{}_{u}^{}}_{\bar{A}_{1}}^{\tensor*[^{1}]{\hspace{-0.2em}B}{}_{2}^{}}}$ &  $- 0.2734$ &                                             $d\indices*{*_{A_{3}}^{\tensor*[^{1}]{\hspace{-0.2em}B}{}_{2}^{}}_{X_{2}}^{A_{1g}^{}}_{\bar{A}_{1}}^{\tensor*[^{1}]{\hspace{-0.2em}B}{}_{2}^{}}}$ &  $- 4.9859$ &                                             $d\indices*{*_{A_{3}}^{\tensor*[^{1}]{\hspace{-0.2em}B}{}_{2}^{}}_{X_{2}}^{A_{2u}^{}}_{\bar{A}_{1}}^{\tensor*[^{1}]{\hspace{-0.2em}B}{}_{2}^{}}}$ &    $1.4152$ &                                                                                                          $d\indices*{*_{A_{5}}^{A_{1}^{}}_{\bar{A}_{4}}^{A_{1}^{}}_{\bar{A}_{1}}^{A_{1}^{}}}$ &    $9.4988$ &                                                                         $d\indices*{*_{A_{5}}^{A_{1}^{}}_{\bar{A}_{4}}^{A_{1}^{}}_{\bar{A}_{1}}^{\tensor*[^{1}]{\hspace{-0.2em}A}{}_{1}^{}}}$ &    $0.3998$ \\[0.4em]
                                                                          $d\indices*{*_{A_{5}}^{A_{1}^{}}_{\bar{A}_{4}}^{A_{1}^{}}_{\bar{A}_{1}}^{\tensor*[^{2}]{\hspace{-0.2em}A}{}_{1}^{}}}$ &   $11.4948$ &                                                                                                          $d\indices*{*_{A_{5}}^{A_{1}^{}}_{\bar{A}_{4}}^{A_{1}^{}}_{\bar{A}_{1}}^{B_{1}^{}}}$ &    $6.4200$ &                                                                         $d\indices*{*_{A_{5}}^{A_{1}^{}}_{\bar{A}_{4}}^{A_{1}^{}}_{\bar{A}_{1}}^{\tensor*[^{1}]{\hspace{-0.2em}B}{}_{1}^{}}}$ &    $3.9299$ &                                                                         $d\indices*{*_{A_{5}}^{A_{1}^{}}_{\bar{A}_{4}}^{A_{1}^{}}_{\bar{A}_{1}}^{\tensor*[^{2}]{\hspace{-0.2em}B}{}_{1}^{}}}$ &  $- 3.7716$ &                                        $d\indices*{*_{A_{5}}^{A_{1}^{}}_{\bar{A}_{4}}^{\tensor*[^{1}]{\hspace{-0.2em}A}{}_{1}^{}}_{\bar{A}_{1}}^{\tensor*[^{1}]{\hspace{-0.2em}A}{}_{1}^{}}}$ &  $- 1.3255$ \\[0.4em]
                                         $d\indices*{*_{A_{5}}^{A_{1}^{}}_{\bar{A}_{4}}^{\tensor*[^{1}]{\hspace{-0.2em}A}{}_{1}^{}}_{\bar{A}_{1}}^{\tensor*[^{2}]{\hspace{-0.2em}A}{}_{1}^{}}}$ &  $- 6.5481$ &                                                                         $d\indices*{*_{A_{5}}^{A_{1}^{}}_{\bar{A}_{4}}^{\tensor*[^{1}]{\hspace{-0.2em}A}{}_{1}^{}}_{\bar{A}_{1}}^{A_{2}^{}}}$ &    $1.5649$ &                                                                         $d\indices*{*_{A_{5}}^{A_{1}^{}}_{\bar{A}_{4}}^{\tensor*[^{1}]{\hspace{-0.2em}A}{}_{1}^{}}_{\bar{A}_{1}}^{B_{1}^{}}}$ &  $- 5.9120$ &                                        $d\indices*{*_{A_{5}}^{A_{1}^{}}_{\bar{A}_{4}}^{\tensor*[^{1}]{\hspace{-0.2em}A}{}_{1}^{}}_{\bar{A}_{1}}^{\tensor*[^{1}]{\hspace{-0.2em}B}{}_{1}^{}}}$ &    $6.4060$ &                                        $d\indices*{*_{A_{5}}^{A_{1}^{}}_{\bar{A}_{4}}^{\tensor*[^{1}]{\hspace{-0.2em}A}{}_{1}^{}}_{\bar{A}_{1}}^{\tensor*[^{2}]{\hspace{-0.2em}B}{}_{1}^{}}}$ &  $- 1.4371$ \\[0.4em]
                                                                          $d\indices*{*_{A_{5}}^{A_{1}^{}}_{\bar{A}_{4}}^{\tensor*[^{1}]{\hspace{-0.2em}A}{}_{1}^{}}_{\bar{A}_{1}}^{B_{2}^{}}}$ & $- 12.0852$ &                                        $d\indices*{*_{A_{5}}^{A_{1}^{}}_{\bar{A}_{4}}^{\tensor*[^{1}]{\hspace{-0.2em}A}{}_{1}^{}}_{\bar{A}_{1}}^{\tensor*[^{1}]{\hspace{-0.2em}B}{}_{2}^{}}}$ &    $5.0162$ &                                        $d\indices*{*_{A_{5}}^{A_{1}^{}}_{\bar{A}_{4}}^{\tensor*[^{2}]{\hspace{-0.2em}A}{}_{1}^{}}_{\bar{A}_{1}}^{\tensor*[^{2}]{\hspace{-0.2em}A}{}_{1}^{}}}$ &    $0.4440$ &                                                                         $d\indices*{*_{A_{5}}^{A_{1}^{}}_{\bar{A}_{4}}^{\tensor*[^{2}]{\hspace{-0.2em}A}{}_{1}^{}}_{\bar{A}_{1}}^{A_{2}^{}}}$ &  $- 0.8222$ &                                                                         $d\indices*{*_{A_{5}}^{A_{1}^{}}_{\bar{A}_{4}}^{\tensor*[^{2}]{\hspace{-0.2em}A}{}_{1}^{}}_{\bar{A}_{1}}^{B_{1}^{}}}$ &  $- 0.2272$ \\[0.4em]
                                         $d\indices*{*_{A_{5}}^{A_{1}^{}}_{\bar{A}_{4}}^{\tensor*[^{2}]{\hspace{-0.2em}A}{}_{1}^{}}_{\bar{A}_{1}}^{\tensor*[^{1}]{\hspace{-0.2em}B}{}_{1}^{}}}$ &  $- 9.3259$ &                                        $d\indices*{*_{A_{5}}^{A_{1}^{}}_{\bar{A}_{4}}^{\tensor*[^{2}]{\hspace{-0.2em}A}{}_{1}^{}}_{\bar{A}_{1}}^{\tensor*[^{2}]{\hspace{-0.2em}B}{}_{1}^{}}}$ &  $- 9.8257$ &                                                                         $d\indices*{*_{A_{5}}^{A_{1}^{}}_{\bar{A}_{4}}^{\tensor*[^{2}]{\hspace{-0.2em}A}{}_{1}^{}}_{\bar{A}_{1}}^{B_{2}^{}}}$ &  $- 3.0041$ &                                        $d\indices*{*_{A_{5}}^{A_{1}^{}}_{\bar{A}_{4}}^{\tensor*[^{2}]{\hspace{-0.2em}A}{}_{1}^{}}_{\bar{A}_{1}}^{\tensor*[^{1}]{\hspace{-0.2em}B}{}_{2}^{}}}$ &    $1.4916$ &                                                                                                          $d\indices*{*_{A_{5}}^{A_{1}^{}}_{\bar{A}_{4}}^{A_{2}^{}}_{\bar{A}_{1}}^{A_{2}^{}}}$ &    $0.4772$ \\[0.4em]
                                                                                                           $d\indices*{*_{A_{5}}^{A_{1}^{}}_{\bar{A}_{4}}^{A_{2}^{}}_{\bar{A}_{1}}^{B_{1}^{}}}$ &    $4.8772$ &                                                                         $d\indices*{*_{A_{5}}^{A_{1}^{}}_{\bar{A}_{4}}^{A_{2}^{}}_{\bar{A}_{1}}^{\tensor*[^{1}]{\hspace{-0.2em}B}{}_{1}^{}}}$ &  $- 7.7028$ &                                                                         $d\indices*{*_{A_{5}}^{A_{1}^{}}_{\bar{A}_{4}}^{A_{2}^{}}_{\bar{A}_{1}}^{\tensor*[^{2}]{\hspace{-0.2em}B}{}_{1}^{}}}$ &    $1.7204$ &                                                                                                          $d\indices*{*_{A_{5}}^{A_{1}^{}}_{\bar{A}_{4}}^{A_{2}^{}}_{\bar{A}_{1}}^{B_{2}^{}}}$ & $- 11.5842$ &                                                                         $d\indices*{*_{A_{5}}^{A_{1}^{}}_{\bar{A}_{4}}^{A_{2}^{}}_{\bar{A}_{1}}^{\tensor*[^{1}]{\hspace{-0.2em}B}{}_{2}^{}}}$ &    $5.7927$ \\[0.4em]
                                                                                                           $d\indices*{*_{A_{5}}^{A_{1}^{}}_{\bar{A}_{4}}^{B_{1}^{}}_{\bar{A}_{1}}^{B_{1}^{}}}$ &    $0.8455$ &                                                                         $d\indices*{*_{A_{5}}^{A_{1}^{}}_{\bar{A}_{4}}^{B_{1}^{}}_{\bar{A}_{1}}^{\tensor*[^{1}]{\hspace{-0.2em}B}{}_{1}^{}}}$ &    $6.5419$ &                                                                         $d\indices*{*_{A_{5}}^{A_{1}^{}}_{\bar{A}_{4}}^{B_{1}^{}}_{\bar{A}_{1}}^{\tensor*[^{2}]{\hspace{-0.2em}B}{}_{1}^{}}}$ &    $5.9103$ &                                                                                                          $d\indices*{*_{A_{5}}^{A_{1}^{}}_{\bar{A}_{4}}^{B_{2}^{}}_{\bar{A}_{1}}^{B_{1}^{}}}$ &  $- 0.9508$ &                                                                         $d\indices*{*_{A_{5}}^{A_{1}^{}}_{\bar{A}_{4}}^{\tensor*[^{1}]{\hspace{-0.2em}B}{}_{2}^{}}_{\bar{A}_{1}}^{B_{1}^{}}}$ &  $- 2.2467$ \\[0.4em]
                                         $d\indices*{*_{A_{5}}^{A_{1}^{}}_{\bar{A}_{4}}^{\tensor*[^{1}]{\hspace{-0.2em}B}{}_{1}^{}}_{\bar{A}_{1}}^{\tensor*[^{1}]{\hspace{-0.2em}B}{}_{1}^{}}}$ &    $3.1529$ &                                        $d\indices*{*_{A_{5}}^{A_{1}^{}}_{\bar{A}_{4}}^{\tensor*[^{1}]{\hspace{-0.2em}B}{}_{1}^{}}_{\bar{A}_{1}}^{\tensor*[^{2}]{\hspace{-0.2em}B}{}_{1}^{}}}$ &    $0.2895$ &                                                                         $d\indices*{*_{A_{5}}^{A_{1}^{}}_{\bar{A}_{4}}^{B_{2}^{}}_{\bar{A}_{1}}^{\tensor*[^{1}]{\hspace{-0.2em}B}{}_{1}^{}}}$ &    $0.8951$ &                                        $d\indices*{*_{A_{5}}^{A_{1}^{}}_{\bar{A}_{4}}^{\tensor*[^{1}]{\hspace{-0.2em}B}{}_{2}^{}}_{\bar{A}_{1}}^{\tensor*[^{1}]{\hspace{-0.2em}B}{}_{1}^{}}}$ &  $- 0.5964$ &                                        $d\indices*{*_{A_{5}}^{A_{1}^{}}_{\bar{A}_{4}}^{\tensor*[^{2}]{\hspace{-0.2em}B}{}_{1}^{}}_{\bar{A}_{1}}^{\tensor*[^{2}]{\hspace{-0.2em}B}{}_{1}^{}}}$ &  $- 1.8506$ \\[0.4em]
                                                                          $d\indices*{*_{A_{5}}^{A_{1}^{}}_{\bar{A}_{4}}^{B_{2}^{}}_{\bar{A}_{1}}^{\tensor*[^{2}]{\hspace{-0.2em}B}{}_{1}^{}}}$ &    $2.4286$ &                                        $d\indices*{*_{A_{5}}^{A_{1}^{}}_{\bar{A}_{4}}^{\tensor*[^{1}]{\hspace{-0.2em}B}{}_{2}^{}}_{\bar{A}_{1}}^{\tensor*[^{2}]{\hspace{-0.2em}B}{}_{1}^{}}}$ &  $- 1.6919$ &                                                                                                          $d\indices*{*_{A_{5}}^{A_{1}^{}}_{\bar{A}_{4}}^{B_{2}^{}}_{\bar{A}_{1}}^{B_{2}^{}}}$ &    $0.5337$ &                                                                         $d\indices*{*_{A_{5}}^{A_{1}^{}}_{\bar{A}_{4}}^{B_{2}^{}}_{\bar{A}_{1}}^{\tensor*[^{1}]{\hspace{-0.2em}B}{}_{2}^{}}}$ &    $0.5510$ &                                        $d\indices*{*_{A_{5}}^{A_{1}^{}}_{\bar{A}_{4}}^{\tensor*[^{1}]{\hspace{-0.2em}B}{}_{2}^{}}_{\bar{A}_{1}}^{\tensor*[^{1}]{\hspace{-0.2em}B}{}_{2}^{}}}$ &    $1.2462$ \\[0.4em]
        $d\indices*{*_{A_{5}}^{\tensor*[^{1}]{\hspace{-0.2em}A}{}_{1}^{}}_{\bar{A}_{4}}^{\tensor*[^{1}]{\hspace{-0.2em}A}{}_{1}^{}}_{\bar{A}_{1}}^{\tensor*[^{1}]{\hspace{-0.2em}A}{}_{1}^{}}}$ &    $1.5355$ &       $d\indices*{*_{A_{5}}^{\tensor*[^{1}]{\hspace{-0.2em}A}{}_{1}^{}}_{\bar{A}_{4}}^{\tensor*[^{1}]{\hspace{-0.2em}A}{}_{1}^{}}_{\bar{A}_{1}}^{\tensor*[^{2}]{\hspace{-0.2em}A}{}_{1}^{}}}$ &    $8.3374$ &                                        $d\indices*{*_{A_{5}}^{\tensor*[^{1}]{\hspace{-0.2em}A}{}_{1}^{}}_{\bar{A}_{4}}^{\tensor*[^{1}]{\hspace{-0.2em}A}{}_{1}^{}}_{\bar{A}_{1}}^{B_{1}^{}}}$ &    $9.7988$ &       $d\indices*{*_{A_{5}}^{\tensor*[^{1}]{\hspace{-0.2em}A}{}_{1}^{}}_{\bar{A}_{4}}^{\tensor*[^{1}]{\hspace{-0.2em}A}{}_{1}^{}}_{\bar{A}_{1}}^{\tensor*[^{1}]{\hspace{-0.2em}B}{}_{1}^{}}}$ & $- 12.8345$ &       $d\indices*{*_{A_{5}}^{\tensor*[^{1}]{\hspace{-0.2em}A}{}_{1}^{}}_{\bar{A}_{4}}^{\tensor*[^{1}]{\hspace{-0.2em}A}{}_{1}^{}}_{\bar{A}_{1}}^{\tensor*[^{2}]{\hspace{-0.2em}B}{}_{1}^{}}}$ &  $- 6.6468$ \\[0.4em]
        $d\indices*{*_{A_{5}}^{\tensor*[^{1}]{\hspace{-0.2em}A}{}_{1}^{}}_{\bar{A}_{4}}^{\tensor*[^{2}]{\hspace{-0.2em}A}{}_{1}^{}}_{\bar{A}_{1}}^{\tensor*[^{2}]{\hspace{-0.2em}A}{}_{1}^{}}}$ &    $7.9316$ &                                        $d\indices*{*_{A_{5}}^{\tensor*[^{1}]{\hspace{-0.2em}A}{}_{1}^{}}_{\bar{A}_{4}}^{\tensor*[^{2}]{\hspace{-0.2em}A}{}_{1}^{}}_{\bar{A}_{1}}^{A_{2}^{}}}$ &  $- 4.5016$ &                                        $d\indices*{*_{A_{5}}^{\tensor*[^{1}]{\hspace{-0.2em}A}{}_{1}^{}}_{\bar{A}_{4}}^{\tensor*[^{2}]{\hspace{-0.2em}A}{}_{1}^{}}_{\bar{A}_{1}}^{B_{1}^{}}}$ &  $- 3.0258$ &       $d\indices*{*_{A_{5}}^{\tensor*[^{1}]{\hspace{-0.2em}A}{}_{1}^{}}_{\bar{A}_{4}}^{\tensor*[^{2}]{\hspace{-0.2em}A}{}_{1}^{}}_{\bar{A}_{1}}^{\tensor*[^{1}]{\hspace{-0.2em}B}{}_{1}^{}}}$ &    $2.5669$ &       $d\indices*{*_{A_{5}}^{\tensor*[^{1}]{\hspace{-0.2em}A}{}_{1}^{}}_{\bar{A}_{4}}^{\tensor*[^{2}]{\hspace{-0.2em}A}{}_{1}^{}}_{\bar{A}_{1}}^{\tensor*[^{2}]{\hspace{-0.2em}B}{}_{1}^{}}}$ &    $5.4105$ \\[0.4em]
                                         $d\indices*{*_{A_{5}}^{\tensor*[^{1}]{\hspace{-0.2em}A}{}_{1}^{}}_{\bar{A}_{4}}^{\tensor*[^{2}]{\hspace{-0.2em}A}{}_{1}^{}}_{\bar{A}_{1}}^{B_{2}^{}}}$ &  $- 2.7422$ &       $d\indices*{*_{A_{5}}^{\tensor*[^{1}]{\hspace{-0.2em}A}{}_{1}^{}}_{\bar{A}_{4}}^{\tensor*[^{2}]{\hspace{-0.2em}A}{}_{1}^{}}_{\bar{A}_{1}}^{\tensor*[^{1}]{\hspace{-0.2em}B}{}_{2}^{}}}$ &    $4.3968$ &                                                                         $d\indices*{*_{A_{5}}^{\tensor*[^{1}]{\hspace{-0.2em}A}{}_{1}^{}}_{\bar{A}_{4}}^{A_{2}^{}}_{\bar{A}_{1}}^{A_{2}^{}}}$ &    $7.5277$ &                                                                         $d\indices*{*_{A_{5}}^{\tensor*[^{1}]{\hspace{-0.2em}A}{}_{1}^{}}_{\bar{A}_{4}}^{A_{2}^{}}_{\bar{A}_{1}}^{B_{1}^{}}}$ &    $5.4111$ &                                        $d\indices*{*_{A_{5}}^{\tensor*[^{1}]{\hspace{-0.2em}A}{}_{1}^{}}_{\bar{A}_{4}}^{A_{2}^{}}_{\bar{A}_{1}}^{\tensor*[^{1}]{\hspace{-0.2em}B}{}_{1}^{}}}$ &    $2.3276$ \\[0.4em]
                                         $d\indices*{*_{A_{5}}^{\tensor*[^{1}]{\hspace{-0.2em}A}{}_{1}^{}}_{\bar{A}_{4}}^{A_{2}^{}}_{\bar{A}_{1}}^{\tensor*[^{2}]{\hspace{-0.2em}B}{}_{1}^{}}}$ &    $2.5683$ &                                                                         $d\indices*{*_{A_{5}}^{\tensor*[^{1}]{\hspace{-0.2em}A}{}_{1}^{}}_{\bar{A}_{4}}^{A_{2}^{}}_{\bar{A}_{1}}^{B_{2}^{}}}$ &    $6.5325$ &                                        $d\indices*{*_{A_{5}}^{\tensor*[^{1}]{\hspace{-0.2em}A}{}_{1}^{}}_{\bar{A}_{4}}^{A_{2}^{}}_{\bar{A}_{1}}^{\tensor*[^{1}]{\hspace{-0.2em}B}{}_{2}^{}}}$ &  $- 6.5096$ &                                                                         $d\indices*{*_{A_{5}}^{\tensor*[^{1}]{\hspace{-0.2em}A}{}_{1}^{}}_{\bar{A}_{4}}^{B_{1}^{}}_{\bar{A}_{1}}^{B_{1}^{}}}$ &  $- 0.5109$ &                                        $d\indices*{*_{A_{5}}^{\tensor*[^{1}]{\hspace{-0.2em}A}{}_{1}^{}}_{\bar{A}_{4}}^{B_{1}^{}}_{\bar{A}_{1}}^{\tensor*[^{1}]{\hspace{-0.2em}B}{}_{1}^{}}}$ &  $- 2.7275$ \\[0.4em]
                                         $d\indices*{*_{A_{5}}^{\tensor*[^{1}]{\hspace{-0.2em}A}{}_{1}^{}}_{\bar{A}_{4}}^{B_{1}^{}}_{\bar{A}_{1}}^{\tensor*[^{2}]{\hspace{-0.2em}B}{}_{1}^{}}}$ &  $- 4.0183$ &                                                                         $d\indices*{*_{A_{5}}^{\tensor*[^{1}]{\hspace{-0.2em}A}{}_{1}^{}}_{\bar{A}_{4}}^{B_{2}^{}}_{\bar{A}_{1}}^{B_{1}^{}}}$ &  $- 1.7191$ &                                        $d\indices*{*_{A_{5}}^{\tensor*[^{1}]{\hspace{-0.2em}A}{}_{1}^{}}_{\bar{A}_{4}}^{\tensor*[^{1}]{\hspace{-0.2em}B}{}_{2}^{}}_{\bar{A}_{1}}^{B_{1}^{}}}$ &  $- 5.5367$ &       $d\indices*{*_{A_{5}}^{\tensor*[^{1}]{\hspace{-0.2em}A}{}_{1}^{}}_{\bar{A}_{4}}^{\tensor*[^{1}]{\hspace{-0.2em}B}{}_{1}^{}}_{\bar{A}_{1}}^{\tensor*[^{1}]{\hspace{-0.2em}B}{}_{1}^{}}}$ & $- 10.5789$ &       $d\indices*{*_{A_{5}}^{\tensor*[^{1}]{\hspace{-0.2em}A}{}_{1}^{}}_{\bar{A}_{4}}^{\tensor*[^{1}]{\hspace{-0.2em}B}{}_{1}^{}}_{\bar{A}_{1}}^{\tensor*[^{2}]{\hspace{-0.2em}B}{}_{1}^{}}}$ &  $- 8.5366$ \\[0.4em]
                                         $d\indices*{*_{A_{5}}^{\tensor*[^{1}]{\hspace{-0.2em}A}{}_{1}^{}}_{\bar{A}_{4}}^{B_{2}^{}}_{\bar{A}_{1}}^{\tensor*[^{1}]{\hspace{-0.2em}B}{}_{1}^{}}}$ &  $- 5.4477$ &       $d\indices*{*_{A_{5}}^{\tensor*[^{1}]{\hspace{-0.2em}A}{}_{1}^{}}_{\bar{A}_{4}}^{\tensor*[^{1}]{\hspace{-0.2em}B}{}_{2}^{}}_{\bar{A}_{1}}^{\tensor*[^{1}]{\hspace{-0.2em}B}{}_{1}^{}}}$ &    $5.3151$ &       $d\indices*{*_{A_{5}}^{\tensor*[^{1}]{\hspace{-0.2em}A}{}_{1}^{}}_{\bar{A}_{4}}^{\tensor*[^{2}]{\hspace{-0.2em}B}{}_{1}^{}}_{\bar{A}_{1}}^{\tensor*[^{2}]{\hspace{-0.2em}B}{}_{1}^{}}}$ &    $2.2358$ &                                        $d\indices*{*_{A_{5}}^{\tensor*[^{1}]{\hspace{-0.2em}A}{}_{1}^{}}_{\bar{A}_{4}}^{B_{2}^{}}_{\bar{A}_{1}}^{\tensor*[^{2}]{\hspace{-0.2em}B}{}_{1}^{}}}$ &  $- 7.1602$ &       $d\indices*{*_{A_{5}}^{\tensor*[^{1}]{\hspace{-0.2em}A}{}_{1}^{}}_{\bar{A}_{4}}^{\tensor*[^{1}]{\hspace{-0.2em}B}{}_{2}^{}}_{\bar{A}_{1}}^{\tensor*[^{2}]{\hspace{-0.2em}B}{}_{1}^{}}}$ &    $6.9029$ \\[0.4em]
                                                                          $d\indices*{*_{A_{5}}^{\tensor*[^{1}]{\hspace{-0.2em}A}{}_{1}^{}}_{\bar{A}_{4}}^{B_{2}^{}}_{\bar{A}_{1}}^{B_{2}^{}}}$ &    $4.6207$ &                                        $d\indices*{*_{A_{5}}^{\tensor*[^{1}]{\hspace{-0.2em}A}{}_{1}^{}}_{\bar{A}_{4}}^{B_{2}^{}}_{\bar{A}_{1}}^{\tensor*[^{1}]{\hspace{-0.2em}B}{}_{2}^{}}}$ &  $- 0.1431$ &       $d\indices*{*_{A_{5}}^{\tensor*[^{1}]{\hspace{-0.2em}A}{}_{1}^{}}_{\bar{A}_{4}}^{\tensor*[^{1}]{\hspace{-0.2em}B}{}_{2}^{}}_{\bar{A}_{1}}^{\tensor*[^{1}]{\hspace{-0.2em}B}{}_{2}^{}}}$ &  $- 5.0458$ &       $d\indices*{*_{A_{5}}^{\tensor*[^{2}]{\hspace{-0.2em}A}{}_{1}^{}}_{\bar{A}_{4}}^{\tensor*[^{2}]{\hspace{-0.2em}A}{}_{1}^{}}_{\bar{A}_{1}}^{\tensor*[^{2}]{\hspace{-0.2em}A}{}_{1}^{}}}$ & $- 10.1328$ &                                        $d\indices*{*_{A_{5}}^{\tensor*[^{2}]{\hspace{-0.2em}A}{}_{1}^{}}_{\bar{A}_{4}}^{\tensor*[^{2}]{\hspace{-0.2em}A}{}_{1}^{}}_{\bar{A}_{1}}^{B_{1}^{}}}$ &  $- 3.5291$ \\[0.4em]
        $d\indices*{*_{A_{5}}^{\tensor*[^{2}]{\hspace{-0.2em}A}{}_{1}^{}}_{\bar{A}_{4}}^{\tensor*[^{2}]{\hspace{-0.2em}A}{}_{1}^{}}_{\bar{A}_{1}}^{\tensor*[^{1}]{\hspace{-0.2em}B}{}_{1}^{}}}$ &    $3.9052$ &       $d\indices*{*_{A_{5}}^{\tensor*[^{2}]{\hspace{-0.2em}A}{}_{1}^{}}_{\bar{A}_{4}}^{\tensor*[^{2}]{\hspace{-0.2em}A}{}_{1}^{}}_{\bar{A}_{1}}^{\tensor*[^{2}]{\hspace{-0.2em}B}{}_{1}^{}}}$ &  $- 4.5392$ &                                                                         $d\indices*{*_{A_{5}}^{\tensor*[^{2}]{\hspace{-0.2em}A}{}_{1}^{}}_{\bar{A}_{4}}^{A_{2}^{}}_{\bar{A}_{1}}^{A_{2}^{}}}$ & $- 11.2705$ &                                                                         $d\indices*{*_{A_{5}}^{\tensor*[^{2}]{\hspace{-0.2em}A}{}_{1}^{}}_{\bar{A}_{4}}^{A_{2}^{}}_{\bar{A}_{1}}^{B_{1}^{}}}$ &    $3.5340$ &                                        $d\indices*{*_{A_{5}}^{\tensor*[^{2}]{\hspace{-0.2em}A}{}_{1}^{}}_{\bar{A}_{4}}^{A_{2}^{}}_{\bar{A}_{1}}^{\tensor*[^{1}]{\hspace{-0.2em}B}{}_{1}^{}}}$ &  $- 0.4071$ \\[0.4em]
                                         $d\indices*{*_{A_{5}}^{\tensor*[^{2}]{\hspace{-0.2em}A}{}_{1}^{}}_{\bar{A}_{4}}^{A_{2}^{}}_{\bar{A}_{1}}^{\tensor*[^{2}]{\hspace{-0.2em}B}{}_{1}^{}}}$ &  $- 2.3705$ &                                                                         $d\indices*{*_{A_{5}}^{\tensor*[^{2}]{\hspace{-0.2em}A}{}_{1}^{}}_{\bar{A}_{4}}^{A_{2}^{}}_{\bar{A}_{1}}^{B_{2}^{}}}$ &    $0.9538$ &                                        $d\indices*{*_{A_{5}}^{\tensor*[^{2}]{\hspace{-0.2em}A}{}_{1}^{}}_{\bar{A}_{4}}^{A_{2}^{}}_{\bar{A}_{1}}^{\tensor*[^{1}]{\hspace{-0.2em}B}{}_{2}^{}}}$ &  $- 0.0429$ &                                                                         $d\indices*{*_{A_{5}}^{\tensor*[^{2}]{\hspace{-0.2em}A}{}_{1}^{}}_{\bar{A}_{4}}^{B_{1}^{}}_{\bar{A}_{1}}^{B_{1}^{}}}$ &  $- 0.5155$ &                                        $d\indices*{*_{A_{5}}^{\tensor*[^{2}]{\hspace{-0.2em}A}{}_{1}^{}}_{\bar{A}_{4}}^{B_{1}^{}}_{\bar{A}_{1}}^{\tensor*[^{1}]{\hspace{-0.2em}B}{}_{1}^{}}}$ &    $5.8215$ \\[0.4em]
                                         $d\indices*{*_{A_{5}}^{\tensor*[^{2}]{\hspace{-0.2em}A}{}_{1}^{}}_{\bar{A}_{4}}^{B_{1}^{}}_{\bar{A}_{1}}^{\tensor*[^{2}]{\hspace{-0.2em}B}{}_{1}^{}}}$ &  $- 0.4895$ &                                                                         $d\indices*{*_{A_{5}}^{\tensor*[^{2}]{\hspace{-0.2em}A}{}_{1}^{}}_{\bar{A}_{4}}^{B_{2}^{}}_{\bar{A}_{1}}^{B_{1}^{}}}$ &  $- 6.2263$ &                                        $d\indices*{*_{A_{5}}^{\tensor*[^{2}]{\hspace{-0.2em}A}{}_{1}^{}}_{\bar{A}_{4}}^{\tensor*[^{1}]{\hspace{-0.2em}B}{}_{2}^{}}_{\bar{A}_{1}}^{B_{1}^{}}}$ &    $4.7172$ &       $d\indices*{*_{A_{5}}^{\tensor*[^{2}]{\hspace{-0.2em}A}{}_{1}^{}}_{\bar{A}_{4}}^{\tensor*[^{1}]{\hspace{-0.2em}B}{}_{1}^{}}_{\bar{A}_{1}}^{\tensor*[^{1}]{\hspace{-0.2em}B}{}_{1}^{}}}$ & $- 13.2548$ &       $d\indices*{*_{A_{5}}^{\tensor*[^{2}]{\hspace{-0.2em}A}{}_{1}^{}}_{\bar{A}_{4}}^{\tensor*[^{1}]{\hspace{-0.2em}B}{}_{1}^{}}_{\bar{A}_{1}}^{\tensor*[^{2}]{\hspace{-0.2em}B}{}_{1}^{}}}$ &  $- 8.0324$ \\[0.4em]
                                         $d\indices*{*_{A_{5}}^{\tensor*[^{2}]{\hspace{-0.2em}A}{}_{1}^{}}_{\bar{A}_{4}}^{B_{2}^{}}_{\bar{A}_{1}}^{\tensor*[^{1}]{\hspace{-0.2em}B}{}_{1}^{}}}$ &    $8.3425$ &       $d\indices*{*_{A_{5}}^{\tensor*[^{2}]{\hspace{-0.2em}A}{}_{1}^{}}_{\bar{A}_{4}}^{\tensor*[^{1}]{\hspace{-0.2em}B}{}_{2}^{}}_{\bar{A}_{1}}^{\tensor*[^{1}]{\hspace{-0.2em}B}{}_{1}^{}}}$ &  $- 5.5804$ &       $d\indices*{*_{A_{5}}^{\tensor*[^{2}]{\hspace{-0.2em}A}{}_{1}^{}}_{\bar{A}_{4}}^{\tensor*[^{2}]{\hspace{-0.2em}B}{}_{1}^{}}_{\bar{A}_{1}}^{\tensor*[^{2}]{\hspace{-0.2em}B}{}_{1}^{}}}$ &  $- 5.8186$ &                                        $d\indices*{*_{A_{5}}^{\tensor*[^{2}]{\hspace{-0.2em}A}{}_{1}^{}}_{\bar{A}_{4}}^{B_{2}^{}}_{\bar{A}_{1}}^{\tensor*[^{2}]{\hspace{-0.2em}B}{}_{1}^{}}}$ &    $1.6957$ &       $d\indices*{*_{A_{5}}^{\tensor*[^{2}]{\hspace{-0.2em}A}{}_{1}^{}}_{\bar{A}_{4}}^{\tensor*[^{1}]{\hspace{-0.2em}B}{}_{2}^{}}_{\bar{A}_{1}}^{\tensor*[^{2}]{\hspace{-0.2em}B}{}_{1}^{}}}$ &    $0.7530$ \\[0.4em]
                                                                          $d\indices*{*_{A_{5}}^{\tensor*[^{2}]{\hspace{-0.2em}A}{}_{1}^{}}_{\bar{A}_{4}}^{B_{2}^{}}_{\bar{A}_{1}}^{B_{2}^{}}}$ &  $- 9.6680$ &                                        $d\indices*{*_{A_{5}}^{\tensor*[^{2}]{\hspace{-0.2em}A}{}_{1}^{}}_{\bar{A}_{4}}^{B_{2}^{}}_{\bar{A}_{1}}^{\tensor*[^{1}]{\hspace{-0.2em}B}{}_{2}^{}}}$ &    $8.0033$ &       $d\indices*{*_{A_{5}}^{\tensor*[^{2}]{\hspace{-0.2em}A}{}_{1}^{}}_{\bar{A}_{4}}^{\tensor*[^{1}]{\hspace{-0.2em}B}{}_{2}^{}}_{\bar{A}_{1}}^{\tensor*[^{1}]{\hspace{-0.2em}B}{}_{2}^{}}}$ &  $- 6.7499$ &                                                                                                          $d\indices*{*_{A_{5}}^{A_{2}^{}}_{\bar{A}_{4}}^{A_{2}^{}}_{\bar{A}_{1}}^{B_{1}^{}}}$ &   $11.1847$ &                                                                         $d\indices*{*_{A_{5}}^{A_{2}^{}}_{\bar{A}_{4}}^{A_{2}^{}}_{\bar{A}_{1}}^{\tensor*[^{1}]{\hspace{-0.2em}B}{}_{1}^{}}}$ &  $- 8.6838$ \\[0.4em]
                                                                          $d\indices*{*_{A_{5}}^{A_{2}^{}}_{\bar{A}_{4}}^{A_{2}^{}}_{\bar{A}_{1}}^{\tensor*[^{2}]{\hspace{-0.2em}B}{}_{1}^{}}}$ &  $- 3.7949$ &                                                                         $d\indices*{*_{A_{5}}^{A_{2}^{}}_{\bar{A}_{4}}^{B_{1}^{}}_{\bar{A}_{1}}^{\tensor*[^{1}]{\hspace{-0.2em}B}{}_{1}^{}}}$ &    $7.3284$ &                                                                         $d\indices*{*_{A_{5}}^{A_{2}^{}}_{\bar{A}_{4}}^{B_{1}^{}}_{\bar{A}_{1}}^{\tensor*[^{2}]{\hspace{-0.2em}B}{}_{1}^{}}}$ &    $3.6429$ &                                                                                                          $d\indices*{*_{A_{5}}^{A_{2}^{}}_{\bar{A}_{4}}^{B_{2}^{}}_{\bar{A}_{1}}^{B_{1}^{}}}$ &  $- 4.1433$ &                                                                         $d\indices*{*_{A_{5}}^{A_{2}^{}}_{\bar{A}_{4}}^{\tensor*[^{1}]{\hspace{-0.2em}B}{}_{2}^{}}_{\bar{A}_{1}}^{B_{1}^{}}}$ &    $3.7841$ \\[0.4em]
                                         $d\indices*{*_{A_{5}}^{A_{2}^{}}_{\bar{A}_{4}}^{\tensor*[^{1}]{\hspace{-0.2em}B}{}_{1}^{}}_{\bar{A}_{1}}^{\tensor*[^{2}]{\hspace{-0.2em}B}{}_{1}^{}}}$ &  $- 6.4073$ &                                                                         $d\indices*{*_{A_{5}}^{A_{2}^{}}_{\bar{A}_{4}}^{B_{2}^{}}_{\bar{A}_{1}}^{\tensor*[^{1}]{\hspace{-0.2em}B}{}_{1}^{}}}$ &    $6.5328$ &                                        $d\indices*{*_{A_{5}}^{A_{2}^{}}_{\bar{A}_{4}}^{\tensor*[^{1}]{\hspace{-0.2em}B}{}_{2}^{}}_{\bar{A}_{1}}^{\tensor*[^{1}]{\hspace{-0.2em}B}{}_{1}^{}}}$ &  $- 7.1608$ &                                                                         $d\indices*{*_{A_{5}}^{A_{2}^{}}_{\bar{A}_{4}}^{B_{2}^{}}_{\bar{A}_{1}}^{\tensor*[^{2}]{\hspace{-0.2em}B}{}_{1}^{}}}$ &   $10.9166$ &                                        $d\indices*{*_{A_{5}}^{A_{2}^{}}_{\bar{A}_{4}}^{\tensor*[^{1}]{\hspace{-0.2em}B}{}_{2}^{}}_{\bar{A}_{1}}^{\tensor*[^{2}]{\hspace{-0.2em}B}{}_{1}^{}}}$ &  $- 7.7727$ \\[0.4em]
                                                                          $d\indices*{*_{A_{5}}^{A_{2}^{}}_{\bar{A}_{4}}^{B_{2}^{}}_{\bar{A}_{1}}^{\tensor*[^{1}]{\hspace{-0.2em}B}{}_{2}^{}}}$ &  $- 0.9552$ &                                                                                                          $d\indices*{*_{A_{5}}^{B_{1}^{}}_{\bar{A}_{4}}^{B_{1}^{}}_{\bar{A}_{1}}^{B_{1}^{}}}$ &   $10.7756$ &                                                                         $d\indices*{*_{A_{5}}^{B_{1}^{}}_{\bar{A}_{4}}^{B_{1}^{}}_{\bar{A}_{1}}^{\tensor*[^{1}]{\hspace{-0.2em}B}{}_{1}^{}}}$ &  $- 2.5416$ &                                                                         $d\indices*{*_{A_{5}}^{B_{1}^{}}_{\bar{A}_{4}}^{B_{1}^{}}_{\bar{A}_{1}}^{\tensor*[^{2}]{\hspace{-0.2em}B}{}_{1}^{}}}$ &    $2.7674$ &                                        $d\indices*{*_{A_{5}}^{B_{1}^{}}_{\bar{A}_{4}}^{\tensor*[^{1}]{\hspace{-0.2em}B}{}_{1}^{}}_{\bar{A}_{1}}^{\tensor*[^{1}]{\hspace{-0.2em}B}{}_{1}^{}}}$ &  $- 0.0400$ \\[0.4em]
                                         $d\indices*{*_{A_{5}}^{B_{1}^{}}_{\bar{A}_{4}}^{\tensor*[^{1}]{\hspace{-0.2em}B}{}_{1}^{}}_{\bar{A}_{1}}^{\tensor*[^{2}]{\hspace{-0.2em}B}{}_{1}^{}}}$ &  $- 2.7555$ &                                                                         $d\indices*{*_{A_{5}}^{B_{2}^{}}_{\bar{A}_{4}}^{B_{1}^{}}_{\bar{A}_{1}}^{\tensor*[^{1}]{\hspace{-0.2em}B}{}_{1}^{}}}$ &  $- 6.0436$ &                                        $d\indices*{*_{A_{5}}^{\tensor*[^{1}]{\hspace{-0.2em}B}{}_{2}^{}}_{\bar{A}_{4}}^{B_{1}^{}}_{\bar{A}_{1}}^{\tensor*[^{1}]{\hspace{-0.2em}B}{}_{1}^{}}}$ &    $2.1282$ &                                        $d\indices*{*_{A_{5}}^{B_{1}^{}}_{\bar{A}_{4}}^{\tensor*[^{2}]{\hspace{-0.2em}B}{}_{1}^{}}_{\bar{A}_{1}}^{\tensor*[^{2}]{\hspace{-0.2em}B}{}_{1}^{}}}$ &  $- 6.4993$ &                                                                         $d\indices*{*_{A_{5}}^{B_{2}^{}}_{\bar{A}_{4}}^{B_{1}^{}}_{\bar{A}_{1}}^{\tensor*[^{2}]{\hspace{-0.2em}B}{}_{1}^{}}}$ &  $- 1.0368$ \\[0.4em]
                                         $d\indices*{*_{A_{5}}^{\tensor*[^{1}]{\hspace{-0.2em}B}{}_{2}^{}}_{\bar{A}_{4}}^{B_{1}^{}}_{\bar{A}_{1}}^{\tensor*[^{2}]{\hspace{-0.2em}B}{}_{1}^{}}}$ &  $- 5.3683$ &                                                                                                          $d\indices*{*_{A_{5}}^{B_{2}^{}}_{\bar{A}_{4}}^{B_{2}^{}}_{\bar{A}_{1}}^{B_{1}^{}}}$ &    $9.2645$ &                                                                         $d\indices*{*_{A_{5}}^{B_{2}^{}}_{\bar{A}_{4}}^{\tensor*[^{1}]{\hspace{-0.2em}B}{}_{2}^{}}_{\bar{A}_{1}}^{B_{1}^{}}}$ &  $- 5.9171$ &                                        $d\indices*{*_{A_{5}}^{\tensor*[^{1}]{\hspace{-0.2em}B}{}_{2}^{}}_{\bar{A}_{4}}^{\tensor*[^{1}]{\hspace{-0.2em}B}{}_{2}^{}}_{\bar{A}_{1}}^{B_{1}^{}}}$ &    $3.5999$ &       $d\indices*{*_{A_{5}}^{\tensor*[^{1}]{\hspace{-0.2em}B}{}_{1}^{}}_{\bar{A}_{4}}^{\tensor*[^{1}]{\hspace{-0.2em}B}{}_{1}^{}}_{\bar{A}_{1}}^{\tensor*[^{1}]{\hspace{-0.2em}B}{}_{1}^{}}}$ & $- 14.6626$ \\[0.4em]
        $d\indices*{*_{A_{5}}^{\tensor*[^{1}]{\hspace{-0.2em}B}{}_{1}^{}}_{\bar{A}_{4}}^{\tensor*[^{1}]{\hspace{-0.2em}B}{}_{1}^{}}_{\bar{A}_{1}}^{\tensor*[^{2}]{\hspace{-0.2em}B}{}_{1}^{}}}$ &    $1.8955$ &       $d\indices*{*_{A_{5}}^{\tensor*[^{1}]{\hspace{-0.2em}B}{}_{1}^{}}_{\bar{A}_{4}}^{\tensor*[^{2}]{\hspace{-0.2em}B}{}_{1}^{}}_{\bar{A}_{1}}^{\tensor*[^{2}]{\hspace{-0.2em}B}{}_{1}^{}}}$ &    $4.7158$ &                                        $d\indices*{*_{A_{5}}^{B_{2}^{}}_{\bar{A}_{4}}^{\tensor*[^{1}]{\hspace{-0.2em}B}{}_{1}^{}}_{\bar{A}_{1}}^{\tensor*[^{2}]{\hspace{-0.2em}B}{}_{1}^{}}}$ &    $1.1249$ &       $d\indices*{*_{A_{5}}^{\tensor*[^{1}]{\hspace{-0.2em}B}{}_{2}^{}}_{\bar{A}_{4}}^{\tensor*[^{1}]{\hspace{-0.2em}B}{}_{1}^{}}_{\bar{A}_{1}}^{\tensor*[^{2}]{\hspace{-0.2em}B}{}_{1}^{}}}$ &    $0.5575$ &                                                                         $d\indices*{*_{A_{5}}^{B_{2}^{}}_{\bar{A}_{4}}^{B_{2}^{}}_{\bar{A}_{1}}^{\tensor*[^{1}]{\hspace{-0.2em}B}{}_{1}^{}}}$ &    $0.0314$ \\[0.4em]
                                         $d\indices*{*_{A_{5}}^{B_{2}^{}}_{\bar{A}_{4}}^{\tensor*[^{1}]{\hspace{-0.2em}B}{}_{2}^{}}_{\bar{A}_{1}}^{\tensor*[^{1}]{\hspace{-0.2em}B}{}_{1}^{}}}$ &    $0.6866$ &       $d\indices*{*_{A_{5}}^{\tensor*[^{1}]{\hspace{-0.2em}B}{}_{2}^{}}_{\bar{A}_{4}}^{\tensor*[^{1}]{\hspace{-0.2em}B}{}_{2}^{}}_{\bar{A}_{1}}^{\tensor*[^{1}]{\hspace{-0.2em}B}{}_{1}^{}}}$ &  $- 0.2145$ &       $d\indices*{*_{A_{5}}^{\tensor*[^{2}]{\hspace{-0.2em}B}{}_{1}^{}}_{\bar{A}_{4}}^{\tensor*[^{2}]{\hspace{-0.2em}B}{}_{1}^{}}_{\bar{A}_{1}}^{\tensor*[^{2}]{\hspace{-0.2em}B}{}_{1}^{}}}$ &   $21.1081$ &                                                                         $d\indices*{*_{A_{5}}^{B_{2}^{}}_{\bar{A}_{4}}^{B_{2}^{}}_{\bar{A}_{1}}^{\tensor*[^{2}]{\hspace{-0.2em}B}{}_{1}^{}}}$ &    $2.2948$ &                                        $d\indices*{*_{A_{5}}^{B_{2}^{}}_{\bar{A}_{4}}^{\tensor*[^{1}]{\hspace{-0.2em}B}{}_{2}^{}}_{\bar{A}_{1}}^{\tensor*[^{2}]{\hspace{-0.2em}B}{}_{1}^{}}}$ &    $0.3261$ \\[0.4em]
        $d\indices*{*_{A_{5}}^{\tensor*[^{1}]{\hspace{-0.2em}B}{}_{2}^{}}_{\bar{A}_{4}}^{\tensor*[^{1}]{\hspace{-0.2em}B}{}_{2}^{}}_{\bar{A}_{1}}^{\tensor*[^{2}]{\hspace{-0.2em}B}{}_{1}^{}}}$ &  $- 5.7921$ &                                                                                                                       $d\indices*{*_{X_{1}}^{A_{1g}^{}}_{X_{2}}^{A_{1g}^{}}_{X}^{A_{1g}^{}}}$ & $- 10.7326$ &                                                                                                                         $d\indices*{*_{X_{1}}^{A_{1g}^{}}_{X_{2}}^{E_{g}^{}}_{X}^{E_{g}^{}}}$ &   $12.8641$ &                                                                                                                        $d\indices*{*_{X_{1}}^{A_{1g}^{}}_{X_{2}}^{A_{2u}^{}}_{X}^{E_{u}^{}}}$ &   $13.2602$ &                                                                                       $d\indices*{*_{X_{1}}^{A_{1g}^{}}_{X_{2}}^{A_{2u}^{}}_{X}^{\tensor*[^{1}]{\hspace{-0.2em}E}{}_{u}^{}}}$ & $- 15.6387$ \\[0.4em]
                                                                                                                        $d\indices*{*_{X_{1}}^{A_{1g}^{}}_{X_{2}}^{B_{1u}^{}}_{X}^{B_{1u}^{}}}$ & $- 11.7494$ &                                                                                                                         $d\indices*{*_{X_{1}}^{A_{1g}^{}}_{X_{2}}^{E_{u}^{}}_{X}^{E_{u}^{}}}$ &  $- 9.9416$ &                                                                                        $d\indices*{*_{X_{1}}^{A_{1g}^{}}_{X_{2}}^{E_{u}^{}}_{X}^{\tensor*[^{1}]{\hspace{-0.2em}E}{}_{u}^{}}}$ &   $11.6367$ &                                                       $d\indices*{*_{X_{1}}^{A_{1g}^{}}_{X_{2}}^{\tensor*[^{1}]{\hspace{-0.2em}E}{}_{u}^{}}_{X}^{\tensor*[^{1}]{\hspace{-0.2em}E}{}_{u}^{}}}$ & $- 14.3531$ &                                                                                                                          $d\indices*{*_{X_{1}}^{E_{g}^{}}_{X_{2}}^{E_{g}^{}}_{X}^{E_{g}^{}}}$ & $- 13.6350$ \\[0.4em]
                                                                                                                          $d\indices*{*_{X_{1}}^{A_{2u}^{}}_{X_{2}}^{E_{g}^{}}_{X}^{E_{u}^{}}}$ &   $13.0699$ &                                                                                        $d\indices*{*_{X_{1}}^{A_{2u}^{}}_{X_{2}}^{E_{g}^{}}_{X}^{\tensor*[^{1}]{\hspace{-0.2em}E}{}_{u}^{}}}$ & $- 18.9367$ &                                                                                                                        $d\indices*{*_{X_{1}}^{B_{1u}^{}}_{X_{2}}^{A_{2u}^{}}_{X}^{E_{g}^{}}}$ &   $12.9724$ &                                                                                                                         $d\indices*{*_{X_{1}}^{B_{1u}^{}}_{X_{2}}^{E_{g}^{}}_{X}^{E_{u}^{}}}$ &    $9.0770$ &                                                                                                                          $d\indices*{*_{X_{1}}^{E_{g}^{}}_{X_{2}}^{E_{u}^{}}_{X}^{E_{u}^{}}}$ &  $- 9.1131$ \\[0.4em]
                                                                        $\tensor*[^{1}]{d}{}\indices*{*_{X_{1}}^{E_{g}^{}}_{X_{2}}^{E_{u}^{}}_{X}^{\tensor*[^{1}]{\hspace{-0.2em}E}{}_{u}^{}}}$ &   $10.6899$ &                                                                                        $d\indices*{*_{X_{1}}^{B_{1u}^{}}_{X_{2}}^{E_{g}^{}}_{X}^{\tensor*[^{1}]{\hspace{-0.2em}E}{}_{u}^{}}}$ & $- 11.5334$ &                                                                       $\tensor*[^{0}]{d}{}\indices*{*_{X_{1}}^{E_{g}^{}}_{X_{2}}^{E_{u}^{}}_{X}^{\tensor*[^{1}]{\hspace{-0.2em}E}{}_{u}^{}}}$ &   $10.4270$ &                                                        $d\indices*{*_{X_{1}}^{E_{g}^{}}_{X_{2}}^{\tensor*[^{1}]{\hspace{-0.2em}E}{}_{u}^{}}_{X}^{\tensor*[^{1}]{\hspace{-0.2em}E}{}_{u}^{}}}$ & $- 16.0767$ &                                                                                                    $d\indices*{*_{\bar{A}_{1}}^{A_{1}^{}}_{\bar{A}_{2}}^{A_{1}^{}}_{\bar{A}_{5}}^{A_{1}^{}}}$ &   $10.1771$ \\[0.4em]
                                                                    $d\indices*{*_{\bar{A}_{1}}^{A_{1}^{}}_{\bar{A}_{2}}^{A_{1}^{}}_{\bar{A}_{5}}^{\tensor*[^{1}]{\hspace{-0.2em}A}{}_{1}^{}}}$ &   $12.5973$ &                                                                   $d\indices*{*_{\bar{A}_{1}}^{A_{1}^{}}_{\bar{A}_{2}}^{A_{1}^{}}_{\bar{A}_{5}}^{\tensor*[^{2}]{\hspace{-0.2em}A}{}_{1}^{}}}$ &    $2.4372$ &                                                                                                    $d\indices*{*_{\bar{A}_{1}}^{A_{1}^{}}_{\bar{A}_{2}}^{A_{1}^{}}_{\bar{A}_{5}}^{A_{2}^{}}}$ &   $14.6647$ &                                  $d\indices*{*_{\bar{A}_{1}}^{A_{1}^{}}_{\bar{A}_{2}}^{\tensor*[^{1}]{\hspace{-0.2em}A}{}_{1}^{}}_{\bar{A}_{5}}^{\tensor*[^{1}]{\hspace{-0.2em}A}{}_{1}^{}}}$ & $- 11.9249$ &                                  $d\indices*{*_{\bar{A}_{1}}^{A_{1}^{}}_{\bar{A}_{2}}^{\tensor*[^{1}]{\hspace{-0.2em}A}{}_{1}^{}}_{\bar{A}_{5}}^{\tensor*[^{2}]{\hspace{-0.2em}A}{}_{1}^{}}}$ &    $1.6419$ \\[0.4em]
                                                                    $d\indices*{*_{\bar{A}_{1}}^{A_{1}^{}}_{\bar{A}_{2}}^{\tensor*[^{1}]{\hspace{-0.2em}A}{}_{1}^{}}_{\bar{A}_{5}}^{A_{2}^{}}}$ &  $- 3.3018$ &                                                                   $d\indices*{*_{\bar{A}_{1}}^{A_{1}^{}}_{\bar{A}_{2}}^{\tensor*[^{1}]{\hspace{-0.2em}A}{}_{1}^{}}_{\bar{A}_{5}}^{B_{1}^{}}}$ &  $- 1.4939$ &                                  $d\indices*{*_{\bar{A}_{1}}^{A_{1}^{}}_{\bar{A}_{2}}^{\tensor*[^{1}]{\hspace{-0.2em}A}{}_{1}^{}}_{\bar{A}_{5}}^{\tensor*[^{1}]{\hspace{-0.2em}B}{}_{1}^{}}}$ &  $- 4.0709$ &                                  $d\indices*{*_{\bar{A}_{1}}^{A_{1}^{}}_{\bar{A}_{2}}^{\tensor*[^{1}]{\hspace{-0.2em}A}{}_{1}^{}}_{\bar{A}_{5}}^{\tensor*[^{2}]{\hspace{-0.2em}B}{}_{1}^{}}}$ &  $- 8.0771$ &                                                                   $d\indices*{*_{\bar{A}_{1}}^{A_{1}^{}}_{\bar{A}_{2}}^{\tensor*[^{1}]{\hspace{-0.2em}A}{}_{1}^{}}_{\bar{A}_{5}}^{B_{2}^{}}}$ &  $- 0.0034$ \\[0.4em]
                                   $d\indices*{*_{\bar{A}_{1}}^{A_{1}^{}}_{\bar{A}_{2}}^{\tensor*[^{1}]{\hspace{-0.2em}A}{}_{1}^{}}_{\bar{A}_{5}}^{\tensor*[^{1}]{\hspace{-0.2em}B}{}_{2}^{}}}$ &  $- 0.1387$ &                                  $d\indices*{*_{\bar{A}_{1}}^{A_{1}^{}}_{\bar{A}_{2}}^{\tensor*[^{2}]{\hspace{-0.2em}A}{}_{1}^{}}_{\bar{A}_{5}}^{\tensor*[^{2}]{\hspace{-0.2em}A}{}_{1}^{}}}$ &    $2.4595$ &                                                                   $d\indices*{*_{\bar{A}_{1}}^{A_{1}^{}}_{\bar{A}_{2}}^{\tensor*[^{2}]{\hspace{-0.2em}A}{}_{1}^{}}_{\bar{A}_{5}}^{A_{2}^{}}}$ &    $4.9239$ &                                                                   $d\indices*{*_{\bar{A}_{1}}^{A_{1}^{}}_{\bar{A}_{2}}^{\tensor*[^{2}]{\hspace{-0.2em}A}{}_{1}^{}}_{\bar{A}_{5}}^{B_{1}^{}}}$ &  $- 6.5173$ &                                  $d\indices*{*_{\bar{A}_{1}}^{A_{1}^{}}_{\bar{A}_{2}}^{\tensor*[^{2}]{\hspace{-0.2em}A}{}_{1}^{}}_{\bar{A}_{5}}^{\tensor*[^{1}]{\hspace{-0.2em}B}{}_{1}^{}}}$ &    $6.4743$ \\[0.4em]
                                   $d\indices*{*_{\bar{A}_{1}}^{A_{1}^{}}_{\bar{A}_{2}}^{\tensor*[^{2}]{\hspace{-0.2em}A}{}_{1}^{}}_{\bar{A}_{5}}^{\tensor*[^{2}]{\hspace{-0.2em}B}{}_{1}^{}}}$ &    $1.2736$ &                                                                   $d\indices*{*_{\bar{A}_{1}}^{A_{1}^{}}_{\bar{A}_{2}}^{\tensor*[^{2}]{\hspace{-0.2em}A}{}_{1}^{}}_{\bar{A}_{5}}^{B_{2}^{}}}$ &   $14.0383$ &                                  $d\indices*{*_{\bar{A}_{1}}^{A_{1}^{}}_{\bar{A}_{2}}^{\tensor*[^{2}]{\hspace{-0.2em}A}{}_{1}^{}}_{\bar{A}_{5}}^{\tensor*[^{1}]{\hspace{-0.2em}B}{}_{2}^{}}}$ &  $- 8.5987$ &                                                                                                    $d\indices*{*_{\bar{A}_{1}}^{A_{1}^{}}_{\bar{A}_{2}}^{A_{2}^{}}_{\bar{A}_{5}}^{A_{2}^{}}}$ &   $10.3692$ &                                                                                                    $d\indices*{*_{\bar{A}_{1}}^{A_{1}^{}}_{\bar{A}_{2}}^{A_{2}^{}}_{\bar{A}_{5}}^{B_{1}^{}}}$ &    $1.8392$ \\[0.4em]
                                                                    $d\indices*{*_{\bar{A}_{1}}^{A_{1}^{}}_{\bar{A}_{2}}^{A_{2}^{}}_{\bar{A}_{5}}^{\tensor*[^{1}]{\hspace{-0.2em}B}{}_{1}^{}}}$ &    $3.5624$ &                                                                   $d\indices*{*_{\bar{A}_{1}}^{A_{1}^{}}_{\bar{A}_{2}}^{A_{2}^{}}_{\bar{A}_{5}}^{\tensor*[^{2}]{\hspace{-0.2em}B}{}_{1}^{}}}$ &    $5.0108$ &                                                                                                    $d\indices*{*_{\bar{A}_{1}}^{A_{1}^{}}_{\bar{A}_{2}}^{A_{2}^{}}_{\bar{A}_{5}}^{B_{2}^{}}}$ &    $3.1395$ &                                                                   $d\indices*{*_{\bar{A}_{1}}^{A_{1}^{}}_{\bar{A}_{2}}^{A_{2}^{}}_{\bar{A}_{5}}^{\tensor*[^{1}]{\hspace{-0.2em}B}{}_{2}^{}}}$ &    $1.9313$ &                                                                                                    $d\indices*{*_{\bar{A}_{1}}^{A_{1}^{}}_{\bar{A}_{2}}^{B_{1}^{}}_{\bar{A}_{5}}^{B_{1}^{}}}$ & $- 13.5615$ \\[0.4em]
                                                                    $d\indices*{*_{\bar{A}_{1}}^{A_{1}^{}}_{\bar{A}_{2}}^{B_{1}^{}}_{\bar{A}_{5}}^{\tensor*[^{1}]{\hspace{-0.2em}B}{}_{1}^{}}}$ &    $7.4584$ &                                                                   $d\indices*{*_{\bar{A}_{1}}^{A_{1}^{}}_{\bar{A}_{2}}^{B_{1}^{}}_{\bar{A}_{5}}^{\tensor*[^{2}]{\hspace{-0.2em}B}{}_{1}^{}}}$ &    $0.2975$ &                                                                                                    $d\indices*{*_{\bar{A}_{1}}^{A_{1}^{}}_{\bar{A}_{2}}^{B_{2}^{}}_{\bar{A}_{5}}^{B_{1}^{}}}$ & $- 12.8487$ &                                                                   $d\indices*{*_{\bar{A}_{1}}^{A_{1}^{}}_{\bar{A}_{2}}^{\tensor*[^{1}]{\hspace{-0.2em}B}{}_{2}^{}}_{\bar{A}_{5}}^{B_{1}^{}}}$ &    $5.2543$ &                                  $d\indices*{*_{\bar{A}_{1}}^{A_{1}^{}}_{\bar{A}_{2}}^{\tensor*[^{1}]{\hspace{-0.2em}B}{}_{1}^{}}_{\bar{A}_{5}}^{\tensor*[^{1}]{\hspace{-0.2em}B}{}_{1}^{}}}$ &   $14.7333$ \\[0.4em]
                                   $d\indices*{*_{\bar{A}_{1}}^{A_{1}^{}}_{\bar{A}_{2}}^{\tensor*[^{1}]{\hspace{-0.2em}B}{}_{1}^{}}_{\bar{A}_{5}}^{\tensor*[^{2}]{\hspace{-0.2em}B}{}_{1}^{}}}$ &    $1.6088$ &                                                                   $d\indices*{*_{\bar{A}_{1}}^{A_{1}^{}}_{\bar{A}_{2}}^{B_{2}^{}}_{\bar{A}_{5}}^{\tensor*[^{1}]{\hspace{-0.2em}B}{}_{1}^{}}}$ &  $- 2.6820$ &                                  $d\indices*{*_{\bar{A}_{1}}^{A_{1}^{}}_{\bar{A}_{2}}^{\tensor*[^{1}]{\hspace{-0.2em}B}{}_{2}^{}}_{\bar{A}_{5}}^{\tensor*[^{1}]{\hspace{-0.2em}B}{}_{1}^{}}}$ &    $1.2421$ &                                  $d\indices*{*_{\bar{A}_{1}}^{A_{1}^{}}_{\bar{A}_{2}}^{\tensor*[^{2}]{\hspace{-0.2em}B}{}_{1}^{}}_{\bar{A}_{5}}^{\tensor*[^{2}]{\hspace{-0.2em}B}{}_{1}^{}}}$ &    $6.1106$ &                                                                   $d\indices*{*_{\bar{A}_{1}}^{A_{1}^{}}_{\bar{A}_{2}}^{B_{2}^{}}_{\bar{A}_{5}}^{\tensor*[^{2}]{\hspace{-0.2em}B}{}_{1}^{}}}$ &  $- 2.4576$ \\[0.4em]
                                   $d\indices*{*_{\bar{A}_{1}}^{A_{1}^{}}_{\bar{A}_{2}}^{\tensor*[^{1}]{\hspace{-0.2em}B}{}_{2}^{}}_{\bar{A}_{5}}^{\tensor*[^{2}]{\hspace{-0.2em}B}{}_{1}^{}}}$ &  $- 2.6292$ &                                                                                                    $d\indices*{*_{\bar{A}_{1}}^{A_{1}^{}}_{\bar{A}_{2}}^{B_{2}^{}}_{\bar{A}_{5}}^{B_{2}^{}}}$ &  $- 2.5822$ &                                                                   $d\indices*{*_{\bar{A}_{1}}^{A_{1}^{}}_{\bar{A}_{2}}^{B_{2}^{}}_{\bar{A}_{5}}^{\tensor*[^{1}]{\hspace{-0.2em}B}{}_{2}^{}}}$ &    $1.7682$ &                                  $d\indices*{*_{\bar{A}_{1}}^{A_{1}^{}}_{\bar{A}_{2}}^{\tensor*[^{1}]{\hspace{-0.2em}B}{}_{2}^{}}_{\bar{A}_{5}}^{\tensor*[^{1}]{\hspace{-0.2em}B}{}_{2}^{}}}$ &    $1.2008$ & $d\indices*{*_{\bar{A}_{1}}^{\tensor*[^{1}]{\hspace{-0.2em}A}{}_{1}^{}}_{\bar{A}_{2}}^{\tensor*[^{1}]{\hspace{-0.2em}A}{}_{1}^{}}_{\bar{A}_{5}}^{\tensor*[^{1}]{\hspace{-0.2em}A}{}_{1}^{}}}$ &  $- 0.4066$ \\[0.4em]
  $d\indices*{*_{\bar{A}_{1}}^{\tensor*[^{1}]{\hspace{-0.2em}A}{}_{1}^{}}_{\bar{A}_{2}}^{\tensor*[^{1}]{\hspace{-0.2em}A}{}_{1}^{}}_{\bar{A}_{5}}^{\tensor*[^{2}]{\hspace{-0.2em}A}{}_{1}^{}}}$ &  $- 7.3707$ &                                  $d\indices*{*_{\bar{A}_{1}}^{\tensor*[^{1}]{\hspace{-0.2em}A}{}_{1}^{}}_{\bar{A}_{2}}^{\tensor*[^{1}]{\hspace{-0.2em}A}{}_{1}^{}}_{\bar{A}_{5}}^{A_{2}^{}}}$ &    $3.1254$ & $d\indices*{*_{\bar{A}_{1}}^{\tensor*[^{1}]{\hspace{-0.2em}A}{}_{1}^{}}_{\bar{A}_{2}}^{\tensor*[^{2}]{\hspace{-0.2em}A}{}_{1}^{}}_{\bar{A}_{5}}^{\tensor*[^{2}]{\hspace{-0.2em}A}{}_{1}^{}}}$ &    $5.1465$ &                                  $d\indices*{*_{\bar{A}_{1}}^{\tensor*[^{1}]{\hspace{-0.2em}A}{}_{1}^{}}_{\bar{A}_{2}}^{\tensor*[^{2}]{\hspace{-0.2em}A}{}_{1}^{}}_{\bar{A}_{5}}^{A_{2}^{}}}$ &  $- 1.3917$ &                                  $d\indices*{*_{\bar{A}_{1}}^{\tensor*[^{1}]{\hspace{-0.2em}A}{}_{1}^{}}_{\bar{A}_{2}}^{\tensor*[^{2}]{\hspace{-0.2em}A}{}_{1}^{}}_{\bar{A}_{5}}^{B_{1}^{}}}$ &  $- 1.9174$ \\[0.4em]
  $d\indices*{*_{\bar{A}_{1}}^{\tensor*[^{1}]{\hspace{-0.2em}A}{}_{1}^{}}_{\bar{A}_{2}}^{\tensor*[^{2}]{\hspace{-0.2em}A}{}_{1}^{}}_{\bar{A}_{5}}^{\tensor*[^{1}]{\hspace{-0.2em}B}{}_{1}^{}}}$ &  $- 6.2116$ & $d\indices*{*_{\bar{A}_{1}}^{\tensor*[^{1}]{\hspace{-0.2em}A}{}_{1}^{}}_{\bar{A}_{2}}^{\tensor*[^{2}]{\hspace{-0.2em}A}{}_{1}^{}}_{\bar{A}_{5}}^{\tensor*[^{2}]{\hspace{-0.2em}B}{}_{1}^{}}}$ &  $- 2.1120$ &                                  $d\indices*{*_{\bar{A}_{1}}^{\tensor*[^{1}]{\hspace{-0.2em}A}{}_{1}^{}}_{\bar{A}_{2}}^{\tensor*[^{2}]{\hspace{-0.2em}A}{}_{1}^{}}_{\bar{A}_{5}}^{B_{2}^{}}}$ & $- 10.4293$ & $d\indices*{*_{\bar{A}_{1}}^{\tensor*[^{1}]{\hspace{-0.2em}A}{}_{1}^{}}_{\bar{A}_{2}}^{\tensor*[^{2}]{\hspace{-0.2em}A}{}_{1}^{}}_{\bar{A}_{5}}^{\tensor*[^{1}]{\hspace{-0.2em}B}{}_{2}^{}}}$ &    $3.4288$ &                                                                   $d\indices*{*_{\bar{A}_{1}}^{\tensor*[^{1}]{\hspace{-0.2em}A}{}_{1}^{}}_{\bar{A}_{2}}^{A_{2}^{}}_{\bar{A}_{5}}^{A_{2}^{}}}$ & $- 11.3964$ \\[0.4em]
                                                                    $d\indices*{*_{\bar{A}_{1}}^{\tensor*[^{1}]{\hspace{-0.2em}A}{}_{1}^{}}_{\bar{A}_{2}}^{A_{2}^{}}_{\bar{A}_{5}}^{B_{1}^{}}}$ &    $5.1774$ &                                  $d\indices*{*_{\bar{A}_{1}}^{\tensor*[^{1}]{\hspace{-0.2em}A}{}_{1}^{}}_{\bar{A}_{2}}^{A_{2}^{}}_{\bar{A}_{5}}^{\tensor*[^{1}]{\hspace{-0.2em}B}{}_{1}^{}}}$ &    $1.9193$ &                                  $d\indices*{*_{\bar{A}_{1}}^{\tensor*[^{1}]{\hspace{-0.2em}A}{}_{1}^{}}_{\bar{A}_{2}}^{A_{2}^{}}_{\bar{A}_{5}}^{\tensor*[^{2}]{\hspace{-0.2em}B}{}_{1}^{}}}$ &  $- 5.1911$ &                                                                   $d\indices*{*_{\bar{A}_{1}}^{\tensor*[^{1}]{\hspace{-0.2em}A}{}_{1}^{}}_{\bar{A}_{2}}^{A_{2}^{}}_{\bar{A}_{5}}^{B_{2}^{}}}$ &  $- 1.6231$ &                                  $d\indices*{*_{\bar{A}_{1}}^{\tensor*[^{1}]{\hspace{-0.2em}A}{}_{1}^{}}_{\bar{A}_{2}}^{A_{2}^{}}_{\bar{A}_{5}}^{\tensor*[^{1}]{\hspace{-0.2em}B}{}_{2}^{}}}$ &  $- 8.6795$ \\[0.4em]
                                                                    $d\indices*{*_{\bar{A}_{1}}^{\tensor*[^{1}]{\hspace{-0.2em}A}{}_{1}^{}}_{\bar{A}_{2}}^{B_{1}^{}}_{\bar{A}_{5}}^{B_{1}^{}}}$ &    $5.1046$ &                                  $d\indices*{*_{\bar{A}_{1}}^{\tensor*[^{1}]{\hspace{-0.2em}A}{}_{1}^{}}_{\bar{A}_{2}}^{B_{1}^{}}_{\bar{A}_{5}}^{\tensor*[^{1}]{\hspace{-0.2em}B}{}_{1}^{}}}$ &  $- 9.1824$ &                                  $d\indices*{*_{\bar{A}_{1}}^{\tensor*[^{1}]{\hspace{-0.2em}A}{}_{1}^{}}_{\bar{A}_{2}}^{B_{1}^{}}_{\bar{A}_{5}}^{\tensor*[^{2}]{\hspace{-0.2em}B}{}_{1}^{}}}$ &    $1.2593$ &                                                                   $d\indices*{*_{\bar{A}_{1}}^{\tensor*[^{1}]{\hspace{-0.2em}A}{}_{1}^{}}_{\bar{A}_{2}}^{B_{2}^{}}_{\bar{A}_{5}}^{B_{1}^{}}}$ &    $0.8558$ &                                  $d\indices*{*_{\bar{A}_{1}}^{\tensor*[^{1}]{\hspace{-0.2em}A}{}_{1}^{}}_{\bar{A}_{2}}^{\tensor*[^{1}]{\hspace{-0.2em}B}{}_{2}^{}}_{\bar{A}_{5}}^{B_{1}^{}}}$ &    $1.4312$ \\[0.4em]
  $d\indices*{*_{\bar{A}_{1}}^{\tensor*[^{1}]{\hspace{-0.2em}A}{}_{1}^{}}_{\bar{A}_{2}}^{\tensor*[^{1}]{\hspace{-0.2em}B}{}_{1}^{}}_{\bar{A}_{5}}^{\tensor*[^{1}]{\hspace{-0.2em}B}{}_{1}^{}}}$ &    $7.3288$ & $d\indices*{*_{\bar{A}_{1}}^{\tensor*[^{1}]{\hspace{-0.2em}A}{}_{1}^{}}_{\bar{A}_{2}}^{\tensor*[^{1}]{\hspace{-0.2em}B}{}_{1}^{}}_{\bar{A}_{5}}^{\tensor*[^{2}]{\hspace{-0.2em}B}{}_{1}^{}}}$ &    $7.5146$ &                                  $d\indices*{*_{\bar{A}_{1}}^{\tensor*[^{1}]{\hspace{-0.2em}A}{}_{1}^{}}_{\bar{A}_{2}}^{B_{2}^{}}_{\bar{A}_{5}}^{\tensor*[^{1}]{\hspace{-0.2em}B}{}_{1}^{}}}$ &    $3.1574$ & $d\indices*{*_{\bar{A}_{1}}^{\tensor*[^{1}]{\hspace{-0.2em}A}{}_{1}^{}}_{\bar{A}_{2}}^{\tensor*[^{1}]{\hspace{-0.2em}B}{}_{2}^{}}_{\bar{A}_{5}}^{\tensor*[^{1}]{\hspace{-0.2em}B}{}_{1}^{}}}$ &  $- 5.2838$ & $d\indices*{*_{\bar{A}_{1}}^{\tensor*[^{1}]{\hspace{-0.2em}A}{}_{1}^{}}_{\bar{A}_{2}}^{\tensor*[^{2}]{\hspace{-0.2em}B}{}_{1}^{}}_{\bar{A}_{5}}^{\tensor*[^{2}]{\hspace{-0.2em}B}{}_{1}^{}}}$ &    $3.4154$ \\[0.4em]
                                   $d\indices*{*_{\bar{A}_{1}}^{\tensor*[^{1}]{\hspace{-0.2em}A}{}_{1}^{}}_{\bar{A}_{2}}^{B_{2}^{}}_{\bar{A}_{5}}^{\tensor*[^{2}]{\hspace{-0.2em}B}{}_{1}^{}}}$ &    $3.0447$ & $d\indices*{*_{\bar{A}_{1}}^{\tensor*[^{1}]{\hspace{-0.2em}A}{}_{1}^{}}_{\bar{A}_{2}}^{\tensor*[^{1}]{\hspace{-0.2em}B}{}_{2}^{}}_{\bar{A}_{5}}^{\tensor*[^{2}]{\hspace{-0.2em}B}{}_{1}^{}}}$ &  $- 3.9657$ &                                                                   $d\indices*{*_{\bar{A}_{1}}^{\tensor*[^{1}]{\hspace{-0.2em}A}{}_{1}^{}}_{\bar{A}_{2}}^{B_{2}^{}}_{\bar{A}_{5}}^{B_{2}^{}}}$ &  $- 4.9251$ &                                  $d\indices*{*_{\bar{A}_{1}}^{\tensor*[^{1}]{\hspace{-0.2em}A}{}_{1}^{}}_{\bar{A}_{2}}^{B_{2}^{}}_{\bar{A}_{5}}^{\tensor*[^{1}]{\hspace{-0.2em}B}{}_{2}^{}}}$ &    $5.7850$ & $d\indices*{*_{\bar{A}_{1}}^{\tensor*[^{1}]{\hspace{-0.2em}A}{}_{1}^{}}_{\bar{A}_{2}}^{\tensor*[^{1}]{\hspace{-0.2em}B}{}_{2}^{}}_{\bar{A}_{5}}^{\tensor*[^{1}]{\hspace{-0.2em}B}{}_{2}^{}}}$ &  $- 3.3020$ \\[0.4em]
  $d\indices*{*_{\bar{A}_{1}}^{\tensor*[^{2}]{\hspace{-0.2em}A}{}_{1}^{}}_{\bar{A}_{2}}^{\tensor*[^{2}]{\hspace{-0.2em}A}{}_{1}^{}}_{\bar{A}_{5}}^{\tensor*[^{2}]{\hspace{-0.2em}A}{}_{1}^{}}}$ &  $- 3.7011$ &                                  $d\indices*{*_{\bar{A}_{1}}^{\tensor*[^{2}]{\hspace{-0.2em}A}{}_{1}^{}}_{\bar{A}_{2}}^{\tensor*[^{2}]{\hspace{-0.2em}A}{}_{1}^{}}_{\bar{A}_{5}}^{A_{2}^{}}}$ &    $3.3694$ &                                                                   $d\indices*{*_{\bar{A}_{1}}^{\tensor*[^{2}]{\hspace{-0.2em}A}{}_{1}^{}}_{\bar{A}_{2}}^{A_{2}^{}}_{\bar{A}_{5}}^{A_{2}^{}}}$ &  $- 0.2446$ &                                                                   $d\indices*{*_{\bar{A}_{1}}^{\tensor*[^{2}]{\hspace{-0.2em}A}{}_{1}^{}}_{\bar{A}_{2}}^{A_{2}^{}}_{\bar{A}_{5}}^{B_{1}^{}}}$ &  $- 5.7646$ &                                  $d\indices*{*_{\bar{A}_{1}}^{\tensor*[^{2}]{\hspace{-0.2em}A}{}_{1}^{}}_{\bar{A}_{2}}^{A_{2}^{}}_{\bar{A}_{5}}^{\tensor*[^{1}]{\hspace{-0.2em}B}{}_{1}^{}}}$ &    $4.3857$ \\[0.4em]
                                   $d\indices*{*_{\bar{A}_{1}}^{\tensor*[^{2}]{\hspace{-0.2em}A}{}_{1}^{}}_{\bar{A}_{2}}^{A_{2}^{}}_{\bar{A}_{5}}^{\tensor*[^{2}]{\hspace{-0.2em}B}{}_{1}^{}}}$ &    $5.4860$ &                                                                   $d\indices*{*_{\bar{A}_{1}}^{\tensor*[^{2}]{\hspace{-0.2em}A}{}_{1}^{}}_{\bar{A}_{2}}^{A_{2}^{}}_{\bar{A}_{5}}^{B_{2}^{}}}$ &    $4.7145$ &                                  $d\indices*{*_{\bar{A}_{1}}^{\tensor*[^{2}]{\hspace{-0.2em}A}{}_{1}^{}}_{\bar{A}_{2}}^{A_{2}^{}}_{\bar{A}_{5}}^{\tensor*[^{1}]{\hspace{-0.2em}B}{}_{2}^{}}}$ &  $- 0.8024$ &                                                                   $d\indices*{*_{\bar{A}_{1}}^{\tensor*[^{2}]{\hspace{-0.2em}A}{}_{1}^{}}_{\bar{A}_{2}}^{B_{1}^{}}_{\bar{A}_{5}}^{B_{1}^{}}}$ &  $- 0.0251$ &                                  $d\indices*{*_{\bar{A}_{1}}^{\tensor*[^{2}]{\hspace{-0.2em}A}{}_{1}^{}}_{\bar{A}_{2}}^{B_{1}^{}}_{\bar{A}_{5}}^{\tensor*[^{1}]{\hspace{-0.2em}B}{}_{1}^{}}}$ &  $- 5.0578$ \\[0.4em]
                                   $d\indices*{*_{\bar{A}_{1}}^{\tensor*[^{2}]{\hspace{-0.2em}A}{}_{1}^{}}_{\bar{A}_{2}}^{B_{1}^{}}_{\bar{A}_{5}}^{\tensor*[^{2}]{\hspace{-0.2em}B}{}_{1}^{}}}$ &  $- 5.7819$ &                                                                   $d\indices*{*_{\bar{A}_{1}}^{\tensor*[^{2}]{\hspace{-0.2em}A}{}_{1}^{}}_{\bar{A}_{2}}^{B_{2}^{}}_{\bar{A}_{5}}^{B_{1}^{}}}$ &    $2.1211$ &                                  $d\indices*{*_{\bar{A}_{1}}^{\tensor*[^{2}]{\hspace{-0.2em}A}{}_{1}^{}}_{\bar{A}_{2}}^{\tensor*[^{1}]{\hspace{-0.2em}B}{}_{2}^{}}_{\bar{A}_{5}}^{B_{1}^{}}}$ &    $0.1996$ & $d\indices*{*_{\bar{A}_{1}}^{\tensor*[^{2}]{\hspace{-0.2em}A}{}_{1}^{}}_{\bar{A}_{2}}^{\tensor*[^{1}]{\hspace{-0.2em}B}{}_{1}^{}}_{\bar{A}_{5}}^{\tensor*[^{1}]{\hspace{-0.2em}B}{}_{1}^{}}}$ &   $16.4719$ & $d\indices*{*_{\bar{A}_{1}}^{\tensor*[^{2}]{\hspace{-0.2em}A}{}_{1}^{}}_{\bar{A}_{2}}^{\tensor*[^{1}]{\hspace{-0.2em}B}{}_{1}^{}}_{\bar{A}_{5}}^{\tensor*[^{2}]{\hspace{-0.2em}B}{}_{1}^{}}}$ &    $5.1057$ \\[0.4em]
                                   $d\indices*{*_{\bar{A}_{1}}^{\tensor*[^{2}]{\hspace{-0.2em}A}{}_{1}^{}}_{\bar{A}_{2}}^{B_{2}^{}}_{\bar{A}_{5}}^{\tensor*[^{1}]{\hspace{-0.2em}B}{}_{1}^{}}}$ & $- 10.1465$ & $d\indices*{*_{\bar{A}_{1}}^{\tensor*[^{2}]{\hspace{-0.2em}A}{}_{1}^{}}_{\bar{A}_{2}}^{\tensor*[^{1}]{\hspace{-0.2em}B}{}_{2}^{}}_{\bar{A}_{5}}^{\tensor*[^{1}]{\hspace{-0.2em}B}{}_{1}^{}}}$ &    $6.9409$ & $d\indices*{*_{\bar{A}_{1}}^{\tensor*[^{2}]{\hspace{-0.2em}A}{}_{1}^{}}_{\bar{A}_{2}}^{\tensor*[^{2}]{\hspace{-0.2em}B}{}_{1}^{}}_{\bar{A}_{5}}^{\tensor*[^{2}]{\hspace{-0.2em}B}{}_{1}^{}}}$ &  $- 1.4810$ &                                  $d\indices*{*_{\bar{A}_{1}}^{\tensor*[^{2}]{\hspace{-0.2em}A}{}_{1}^{}}_{\bar{A}_{2}}^{B_{2}^{}}_{\bar{A}_{5}}^{\tensor*[^{2}]{\hspace{-0.2em}B}{}_{1}^{}}}$ & $- 13.6972$ & $d\indices*{*_{\bar{A}_{1}}^{\tensor*[^{2}]{\hspace{-0.2em}A}{}_{1}^{}}_{\bar{A}_{2}}^{\tensor*[^{1}]{\hspace{-0.2em}B}{}_{2}^{}}_{\bar{A}_{5}}^{\tensor*[^{2}]{\hspace{-0.2em}B}{}_{1}^{}}}$ &    $9.4068$ \\[0.4em]
                                                                    $d\indices*{*_{\bar{A}_{1}}^{\tensor*[^{2}]{\hspace{-0.2em}A}{}_{1}^{}}_{\bar{A}_{2}}^{B_{2}^{}}_{\bar{A}_{5}}^{B_{2}^{}}}$ &  $- 2.4830$ &                                  $d\indices*{*_{\bar{A}_{1}}^{\tensor*[^{2}]{\hspace{-0.2em}A}{}_{1}^{}}_{\bar{A}_{2}}^{B_{2}^{}}_{\bar{A}_{5}}^{\tensor*[^{1}]{\hspace{-0.2em}B}{}_{2}^{}}}$ &    $0.4903$ & $d\indices*{*_{\bar{A}_{1}}^{\tensor*[^{2}]{\hspace{-0.2em}A}{}_{1}^{}}_{\bar{A}_{2}}^{\tensor*[^{1}]{\hspace{-0.2em}B}{}_{2}^{}}_{\bar{A}_{5}}^{\tensor*[^{1}]{\hspace{-0.2em}B}{}_{2}^{}}}$ &    $1.7021$ &                                                                                                    $d\indices*{*_{\bar{A}_{1}}^{A_{2}^{}}_{\bar{A}_{2}}^{A_{2}^{}}_{\bar{A}_{5}}^{A_{2}^{}}}$ &   $11.6242$ &                                                                                                    $d\indices*{*_{\bar{A}_{1}}^{A_{2}^{}}_{\bar{A}_{2}}^{B_{1}^{}}_{\bar{A}_{5}}^{B_{1}^{}}}$ &    $5.0380$ \\[0.4em]
                                                                    $d\indices*{*_{\bar{A}_{1}}^{A_{2}^{}}_{\bar{A}_{2}}^{B_{1}^{}}_{\bar{A}_{5}}^{\tensor*[^{1}]{\hspace{-0.2em}B}{}_{1}^{}}}$ &    $6.4831$ &                                                                   $d\indices*{*_{\bar{A}_{1}}^{A_{2}^{}}_{\bar{A}_{2}}^{B_{1}^{}}_{\bar{A}_{5}}^{\tensor*[^{2}]{\hspace{-0.2em}B}{}_{1}^{}}}$ &  $- 2.8281$ &                                                                                                    $d\indices*{*_{\bar{A}_{1}}^{A_{2}^{}}_{\bar{A}_{2}}^{B_{2}^{}}_{\bar{A}_{5}}^{B_{1}^{}}}$ &  $- 6.0165$ &                                                                   $d\indices*{*_{\bar{A}_{1}}^{A_{2}^{}}_{\bar{A}_{2}}^{\tensor*[^{1}]{\hspace{-0.2em}B}{}_{2}^{}}_{\bar{A}_{5}}^{B_{1}^{}}}$ &    $4.5211$ &                                  $d\indices*{*_{\bar{A}_{1}}^{A_{2}^{}}_{\bar{A}_{2}}^{\tensor*[^{1}]{\hspace{-0.2em}B}{}_{1}^{}}_{\bar{A}_{5}}^{\tensor*[^{1}]{\hspace{-0.2em}B}{}_{1}^{}}}$ &    $3.0118$ \\[0.4em]
                                   $d\indices*{*_{\bar{A}_{1}}^{A_{2}^{}}_{\bar{A}_{2}}^{\tensor*[^{1}]{\hspace{-0.2em}B}{}_{1}^{}}_{\bar{A}_{5}}^{\tensor*[^{2}]{\hspace{-0.2em}B}{}_{1}^{}}}$ &    $9.1342$ &                                                                   $d\indices*{*_{\bar{A}_{1}}^{A_{2}^{}}_{\bar{A}_{2}}^{B_{2}^{}}_{\bar{A}_{5}}^{\tensor*[^{1}]{\hspace{-0.2em}B}{}_{1}^{}}}$ &    $6.3898$ &                                  $d\indices*{*_{\bar{A}_{1}}^{A_{2}^{}}_{\bar{A}_{2}}^{\tensor*[^{1}]{\hspace{-0.2em}B}{}_{2}^{}}_{\bar{A}_{5}}^{\tensor*[^{1}]{\hspace{-0.2em}B}{}_{1}^{}}}$ &  $- 4.4812$ &                                  $d\indices*{*_{\bar{A}_{1}}^{A_{2}^{}}_{\bar{A}_{2}}^{\tensor*[^{2}]{\hspace{-0.2em}B}{}_{1}^{}}_{\bar{A}_{5}}^{\tensor*[^{2}]{\hspace{-0.2em}B}{}_{1}^{}}}$ &    $9.9381$ &                                                                   $d\indices*{*_{\bar{A}_{1}}^{A_{2}^{}}_{\bar{A}_{2}}^{B_{2}^{}}_{\bar{A}_{5}}^{\tensor*[^{2}]{\hspace{-0.2em}B}{}_{1}^{}}}$ &    $0.6699$ \\[0.4em]
                                   $d\indices*{*_{\bar{A}_{1}}^{A_{2}^{}}_{\bar{A}_{2}}^{\tensor*[^{1}]{\hspace{-0.2em}B}{}_{2}^{}}_{\bar{A}_{5}}^{\tensor*[^{2}]{\hspace{-0.2em}B}{}_{1}^{}}}$ &    $0.7724$ &                                                                                                    $d\indices*{*_{\bar{A}_{1}}^{A_{2}^{}}_{\bar{A}_{2}}^{B_{2}^{}}_{\bar{A}_{5}}^{B_{2}^{}}}$ &    $9.7012$ &                                                                   $d\indices*{*_{\bar{A}_{1}}^{A_{2}^{}}_{\bar{A}_{2}}^{B_{2}^{}}_{\bar{A}_{5}}^{\tensor*[^{1}]{\hspace{-0.2em}B}{}_{2}^{}}}$ &  $- 6.0495$ &                                  $d\indices*{*_{\bar{A}_{1}}^{A_{2}^{}}_{\bar{A}_{2}}^{\tensor*[^{1}]{\hspace{-0.2em}B}{}_{2}^{}}_{\bar{A}_{5}}^{\tensor*[^{1}]{\hspace{-0.2em}B}{}_{2}^{}}}$ &    $9.1771$ &                                  $d\indices*{*_{\bar{A}_{1}}^{B_{1}^{}}_{\bar{A}_{2}}^{\tensor*[^{1}]{\hspace{-0.2em}B}{}_{1}^{}}_{\bar{A}_{5}}^{\tensor*[^{2}]{\hspace{-0.2em}B}{}_{1}^{}}}$ &  $- 2.8885$ \\[0.4em]
                                                                    $d\indices*{*_{\bar{A}_{1}}^{B_{2}^{}}_{\bar{A}_{2}}^{B_{1}^{}}_{\bar{A}_{5}}^{\tensor*[^{1}]{\hspace{-0.2em}B}{}_{1}^{}}}$ &    $6.4072$ &                                  $d\indices*{*_{\bar{A}_{1}}^{\tensor*[^{1}]{\hspace{-0.2em}B}{}_{2}^{}}_{\bar{A}_{2}}^{B_{1}^{}}_{\bar{A}_{5}}^{\tensor*[^{1}]{\hspace{-0.2em}B}{}_{1}^{}}}$ &  $- 3.9689$ &                                                                   $d\indices*{*_{\bar{A}_{1}}^{B_{2}^{}}_{\bar{A}_{2}}^{B_{1}^{}}_{\bar{A}_{5}}^{\tensor*[^{2}]{\hspace{-0.2em}B}{}_{1}^{}}}$ &    $5.0871$ &                                  $d\indices*{*_{\bar{A}_{1}}^{\tensor*[^{1}]{\hspace{-0.2em}B}{}_{2}^{}}_{\bar{A}_{2}}^{B_{1}^{}}_{\bar{A}_{5}}^{\tensor*[^{2}]{\hspace{-0.2em}B}{}_{1}^{}}}$ &  $- 4.8938$ &                                                                   $d\indices*{*_{\bar{A}_{1}}^{B_{2}^{}}_{\bar{A}_{2}}^{\tensor*[^{1}]{\hspace{-0.2em}B}{}_{2}^{}}_{\bar{A}_{5}}^{B_{1}^{}}}$ &    $0.8484$ \\[0.4em]
                                   $d\indices*{*_{\bar{A}_{1}}^{B_{2}^{}}_{\bar{A}_{2}}^{\tensor*[^{1}]{\hspace{-0.2em}B}{}_{1}^{}}_{\bar{A}_{5}}^{\tensor*[^{2}]{\hspace{-0.2em}B}{}_{1}^{}}}$ &    $1.3076$ & $d\indices*{*_{\bar{A}_{1}}^{\tensor*[^{1}]{\hspace{-0.2em}B}{}_{2}^{}}_{\bar{A}_{2}}^{\tensor*[^{1}]{\hspace{-0.2em}B}{}_{1}^{}}_{\bar{A}_{5}}^{\tensor*[^{2}]{\hspace{-0.2em}B}{}_{1}^{}}}$ &  $- 3.0630$ &                                  $d\indices*{*_{\bar{A}_{1}}^{B_{2}^{}}_{\bar{A}_{2}}^{\tensor*[^{1}]{\hspace{-0.2em}B}{}_{2}^{}}_{\bar{A}_{5}}^{\tensor*[^{1}]{\hspace{-0.2em}B}{}_{1}^{}}}$ &    $1.9333$ &                                  $d\indices*{*_{\bar{A}_{1}}^{B_{2}^{}}_{\bar{A}_{2}}^{\tensor*[^{1}]{\hspace{-0.2em}B}{}_{2}^{}}_{\bar{A}_{5}}^{\tensor*[^{2}]{\hspace{-0.2em}B}{}_{1}^{}}}$ &    $2.6136$ &                                                                                                                                                                                               &              \\
%
%
%

\end{longtable*}



\end{longtable*}

\clearpage

\bibliography{main}